\documentclass[12pt]{article}
\usepackage{cite,booktabs}
\usepackage{tikz}
\usepackage[english]{babel}
\usepackage{currfile}

\newlength{\abstractwidth}
\usepackage{indentfirst,hyperref,color,psfrag,verbatim}
\usepackage{amsmath,amssymb,amscd,mathrsfs,graphicx,dsfont,caption,
subcaption,mathtools,empheq,mathbbol}

\usepackage{listings,bm}
\usepackage{slashed}
\usepackage{ulem}
\usepackage{scalerel,stackengine}

\newcommand{\rmd}{\mathrm{d}}

\newcommand{\suminnt}{\sum \hspace{-3ex}\int}

\newcommand{\dfn}{\vcentcolon=}

\newcommand{\sk}{\smallskip\smallskip}

\newcommand{\zero}{\varnothing}

\DeclarePairedDelimiterX{\av}[1]{\langle}{\rangle}{#1}

\def\mathbi#1{\bm{ #1}}

\usepackage{lmodern}
\DeclareSymbolFont{euletters}{U}{zeur}{m}{n}
\DeclareMathSymbol\g{\mathalpha}{euletters}{`\g}
\def\enableeulerg
  {\edef\disableeulerg{\mathcode`g=\the\mathcode`\g}\mathcode`\g="8000 }
\def\disableeulerg{}
\begingroup\lccode`~=`g\lowercase{\endgroup\let~}\eulerg

\DeclareMathSymbol\n{\mathalpha}{euletters}{`\n}
\def\enableeulern
  {\edef\disableeulern{\mathcode`n=\the\mathcode`\n}\mathcode`\n="8000 }
\def\disableeulern{}
\begingroup\lccode`~=`n\lowercase{\endgroup\let~}\eulern

\DeclareMathSymbol\m{\mathalpha}{euletters}{`\m}
\def\enableeulerm
  {\edef\disableeulerm{\mathcode`m=\the\mathcode`\m}\mathcode`\m="8000 }
\def\disableeulerm{}
\begingroup\lccode`~=`m\lowercase{\endgroup\let~}\eulerm

\DeclareMathSymbol\B{\mathalpha}{euletters}{`\b}
\def\enableeulerb
  {\edef\disableeulerb{\mathcode`b=\the\mathcode`\B}\mathcode`\B="8000 }
\def\disableeulerb{}
\begingroup\lccode`~=`b\lowercase{\endgroup\let~}\eulerb

\DeclareFontFamily{OT1}{pzcbig}{}
\DeclareFontShape{OT1}{pzcbig}{e}{it}{<-> s* [1.1] pzcmi7t}{}
\DeclareMathAlphabet{\e}{OT1}{pzcbig}{e}{it}
\renewcommand{\thefootnote}{\fnsymbol{footnote}}
\renewcommand{\thanks}[1]{\footnote{#1}}
\newcommand{\starttext}{
\setcounter{footnote}{0}
\renewcommand{\thefootnote}{\arabic{footnote}}}

\begin{document}
\starttext
\setcounter{footnote}{0}

\begin{flushright}
{\tt MPP-2019-225}\\
{\tt LMU-ASC 37/19}
\end{flushright}

\baselineskip=18pt
\rightline{{\fontsize{0.40cm}{5.5cm}\selectfont{}}}
\vskip 0.3in

\begin{center}
{\LARGE {\bf Handle Operators in String Theory}}\\

\vskip 1cm

{\large \bf Dimitri Skliros$^{(a,b)}$ and Dieter L\"ust$^{(a,b)}$}
\vskip 0.2cm

{\sl (a) Max-Planck-Institut f\"ur Physik, Werner-Heisenberg-Institut, \\
F\"ohringer Ring 6, 80805 M\"unchen, Germany}

\vskip 0.08in

{\sl (b) Arnold Sommerfeld Center for Theoretical Physics,\\
Fakult\"at f\"ur Physik, Ludwig-Maximilians-Universit\"at M\"unchen,\\
 Theresienstr. 37, 80333 M\"unchen, Germany}
       
\vskip 0.18in
{\tt \small 
dp.skliros@gmail.com; dieter.luest@lmu.de}
\end{center}

\vskip .5cm
\begin{abstract}
We derive how to incorporate topological features of Riemann surfaces in string amplitudes by insertions of bi-local operators called `{\it handle operators}'. The resulting formalism is exact and globally well-defined in moduli space. After a detailed and pedagogical discussion of Riemann surfaces, complex structure deformations, global vs local aspects, boundary terms, an explicit choice of gluing-compatible and global (modulo U(1)) coordinates (termed `holomorphic normal coordinates'), finite changes in normal ordering, and factorisation of the path integral measure, we construct these handle operators explicitly. Adopting an offshell local coherent vertex operator basis for the latter, and gauge fixing invariance under Weyl transformations using holomorphic normal coordinates (developed by Polchinski), is particularly efficient. All physical loop amplitudes are gauge-invariant (BRST-exact terms decouple up to boundary terms in moduli space), and reparametrisation invariance is manifest, for arbitrary worldsheet curvature and topology (subject to the Euler number constraint). We provide a number of complementary viewpoints and consistency checks (including one-loop modular invariance, we compute all one- and two-point sphere amplitudes, glue two three-point sphere amplitudes to reproduce the exact four-point sphere amplitude, etc.). 
\end{abstract}

\setcounter{secnumdepth}{4}
\newpage
\tableofcontents

\newpage
\section{Introduction}\label{sec:intro}
A central question in high energy physics is to understand and get a working handle on the deep interplay between quantum gravity and corresponding emergence of spacetime geometry. String theory inevitably combines quantum mechanics and gravity, and so provides a context where, at least in principle, one has sharp tools to study this emergence. On the one hand we have the AdS/CFT correspondence \cite{Maldacena98} which relates quantum gravity in $d+1$ dimensions with AdS$_{d+1}$ boundary conditions to a well-defined CFT$_d$ path integral on a corresponding $d$-dimensional boundary spacetime. In principle, this correspondence can be used to define what we mean by a quantum theory of gravity, and, e.g., Ryu-Takayanagi-like formulas provide (to a certain extent) a means to reconstruct classical geometry in the bulk using CFT data \cite{RyuTakayanagi06}. Adding to this, there is the interesting proposal \cite{VanRaamsdonk10} that classically connected bulk spacetimes correspond to entangled states in the corresponding CFT, with various related ideas such as the ER $=$ EPR proposal \cite{MaldacenaSusskind13}. It is nevertheless difficult (if at all possible) to directly address certain questions within AdS/CFT, such as whether black holes have an interior. In particular, we only really have a concrete handle on boundary CFTs, with the corresponding bulk physics remaining largely mysterious (except in a supergravity or low energy limit). For example, one would like to generalise loop resummations of the type discussed in \cite{MaldacenaQi18,SaadShenkerStanford18,StanfordWitten19} to the full physical string theory.
\sk

More generally, it is desirable to get a more direct ``handle'' on bulk physics, which might more broadly be referred to as non-perturbative string theory. 
From a bulk perspective superstring perturbation theory seemingly provides the best starting point. The relevant framework is then associated to maps of Riemann surfaces into appropriate target spaces (determined by a choice of two-dimensional matter CFT's such that the total matter plus ghost central charge vanishes), and summing/integrating over all such (inequivalent) Riemann surfaces and target spaces subject to specified boundary conditions. 
Since the above program is particularly ambitious we will partition it into more manageable chunks, and focus here on developing an appropriate framework that might catalyse progress.
\sk

\begin{figure}
\begin{center}
\includegraphics[angle=0,origin=c,width=0.8\textwidth]{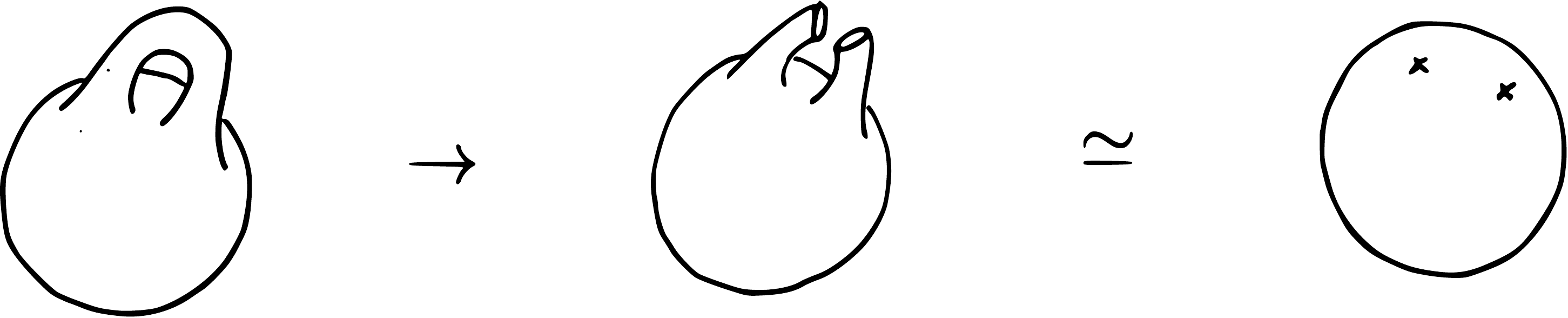}
\caption{An illustrative sketch of cutting open a non-trivial homology cycle of a Riemann surface, replacing it by a complete set of boundary states, and then mapping these states to local operators, $\mathscr{A}_a$, $\mathscr{A}^a$, using the operator-state correspondence. We can then dress every such handle operator, ${\Sigma\!\!\!\!\!\int}_a\mathscr{A}_a\mathscr{A}^a$, with three (complex) moduli, corresponding to two translation moduli for the resulting two local vertex operators and also a pinch/twist modulus.}\label{fig:pinch8}
\end{center}
\end{figure}

We will consider Riemann surfaces, $\Sigma$, from a viewpoint that will enable us to discuss cutting and gluing of worldsheet path integrals using appropriate handle operator insertions, see Fig.~\ref{fig:pinch8},
an idea that goes at least as far back as \cite{OoguriSakai87,Das88}; and in particular an early article by Polchinski \cite{Polchinski88}, work by Tseytlin \cite{Tseytlin90b,Tseytlin90} and recent work by Sen \cite{Sen15b} have been particularly influential for the developments in the current document. Some complementary work which includes some of the material needed to extend the current document to the superstring (which will be discussed elsewhere, beginning with \cite{Skliros20}) can also be found in the excellent papers by Atick, Moore and Sen \cite{AtickMooreSen88} and by Witten \cite{Witten12b,Witten12c} (as well as some important followups such as \cite{SenWitten15}). On the one hand, we would like to be able to cut open a worldsheet path integral across any given cycle, and possibly associate (pinch, twist and translation) moduli to this cycle. But we would like to be able to do so while keeping the moduli of the {\it remaining} Riemann surface fixed. This is a key point -- it corresponds to picking a slice in moduli space, $\mathcal{M}$, where the various moduli are (in a certain sense) ``as decoupled as possible'', and this places all string amplitudes of different genera on equal footing. The main hurdle to overcome  (which also underlies much of the formalism we develop) is to proceed in a manner that leads to a {\it globally} well-defined (on $\Sigma$ and $\mathcal{M}$) construction. So we need to study the relevant aspects of the deep interplay between local and global data, since most of the explicit formalism in string perturbation theory \cite{Polchinski_v1} is based on local considerations. 
\sk

On a parallel note, when cutting open (or gluing) a worldsheet path integral across a specified (trivial or non-trivial homology) cycle of $\Sigma$, it is convenient to insert a complete set of intermediate states in order to incorporate appropriate boundary conditions across the cut cycle. Using operator-state correspondence these states can be mapped to local vertex operators. Since states propagating through a given internal cycle of $\Sigma$ are generically offshell (and hence not BRST invariant), we need to proceed in a manner that keeps track of the associated frame dependence that arises \cite{Polchinski88,Nelson89}, while making sure that any such frame dependence cancels out and yields a well-defined S matrix. These intermediate states are also not conformal primaries, and so do not transform as tensors on patch overlaps.\footnote{Here we can imagine covering our Riemann surface with charts, with given holomorphic transition functions satisfying appropriate cocycle relations on triple chart overlaps.} How then (if we wish to associate their location to a modulus) can we ensure that the corresponding local operators can be integrated over the Riemann surface in a globally well-defined manner? A further difficulty is that these local operators are normal ordered, and so their transformation properties on patch overlaps are not the naive ones. Finally, although in CFT the energy-momentum tensor generates infinitesimal changes of coordinates (which includes a corresponding change in normal ordering), on chart overlaps we sometimes need {\it finite} changes of coordinates. So there are a number of subtle and deep obstacles to be overcome. The work of Polchinski \cite{Polchinski88} (also \cite{Polchinski87}) largely sets the necessary groundwork to overcome these obstacles, but since his analysis is in places perhaps exceedingly concise or opaque 
we will also review Polchinski's approach in detail.
\sk

Furthermore, in order to carry out this program explicitly we need to make a choice of basis for the aforementioned intermediate vertex operators. Alternatively, one can use a ``boundary state'' formalism, where the boundary state (which implements the correct boundary conditions across the cut cycle) is essentially the unit operator in the Hilbert space of interest. But this is not appropriate for our purposes because such boundary states do not {\it a priori} lead to a Hilbert space tensor product, which is in turn desirable if (in the case of separating degenerations) we are going to be able to compute the resulting amplitudes on either side of the cut cycle independently. Also, it is extremely convenient to work in terms of {\it local} operators, whereas boundary states are not local. Briefly, it is apparently efficient and possible to proceed in an exact manner if we use a local {\it coherent state} vertex operator basis, where, in addition to capturing the full tower of stringy states in one relatively simple and concise formula (which ensures the corresponding amplitudes will be non-perturbative in $\alpha'$), there are associated continuous quantum numbers that will presumably also make (some aspects of) emergence of classicality manifest. Of course, one would eventually like to sum over handle operators \cite{Tseytlin90b,Tseytlin90}.
\sk

Taking all of these considerations into account, we will (by a gentle and pedagogical sequence of steps) eventually end up with an exact expression for `handle operators' and a corresponding smooth gauge slice in moduli space.\footnote{The advantage of a {\it smooth} gauge slice in moduli space (as opposed to, say, a holomorphic slice) is (in addition to the fact that it is globally well-defined) clear also in the superstring generalisation \cite{Skliros20} of this work because there is no obstruction to a smooth splitting of supermoduli (as opposed to the obstruction to a holomorphic splitting discussed by Donagi and Witten \cite{DonagiWitten15}).} In the bosonic theory we can associate three complex moduli to every handle, but this is a choice. (There is also a slight distinction between cutting along trivial and non-trivial homology cycles, in that also non-level matched states propagate through the latter.) Regarding non-trivial homology cycles, we can increase the genus by increasing the number of handle operators (with an appropriate combinatorial factor to avoid overcounting). So all genera are treated on equal footing, and operator-product expansions (OPE) between handle operators and between handle operators and external vertex operators generate all boundaries of moduli space. The fact that this is possible is due to worldsheet duality (by which we mean associativity of the OPE and modular invariance) and has no obvious analogue in quantum field theory (and hence also no obvious analogue in string field theory\footnote{For recent developments in superstring field theory see \cite{deLacroixErbinKashyapSenVerma17,MoosavianZhou19}.} due to the ``field theory'' partitioning of moduli space). Or we can use these to glue amplitudes together, implement the Fischler-Susskind mechanism, etc. Related and particularly insightful developments, including the use of handle operators, have been discussed (but without the BRST machinery and including only leading contributions) in \cite{Tseytlin90b,Tseytlin90}. 
\sk

Incidentally, since we work in the BRST formalism and do not assume external vertex operators are primaries the handle operator construction we present certainly allows for insertions of offshell (as well as onshell) external vertex operators provided one adopts appropriate ``gluing-compatible'' coordinate charts \cite{PiusRudraSen14c,PiusRudraSen14b,Sen15b}. The point then is that although offshell amplitudes computed using different coordinate systems give different answers, they are all related by field redefinitions in the corresponding field theory while giving a unique answer for the onshell S matrix.
\sk

We present numerous consistency checks throughout. For example, we show that handle operator insertions lead to gauge-invariant amplitudes (at all loop orders). That is, BRST-exact contributions decouple up to boundary terms in moduli space as one expects. Another consistency check is to show that the formalism leads to modular-invariant amplitudes, a highly non-trivial statement since handle operators break manifest modular invariance. We demonstrate that modular invariance is present by an explicit one-loop example, whereby we glue a handle onto a sphere and extract the one-loop modular-invariant vacuum amplitude from it. This also checks consistency associated to cutting across non-trivial homology cycles. Regarding trivial homology cycles, a check we carry out is to reconstruct a four-point sphere amplitude by gluing together two three-point sphere amplitudes using an appropriate handle operator insertion. And indeed find that the expected four-point amplitude is precisely reproduced, to all orders in $\alpha'$. Regarding the path integral measure contributions, we derive these explicitly using both a metric viewpoint \cite{Polchinski88} and a holomorphic transition function viewpoint \cite{Nelson89,LaNelson90}. The latter is more efficient, but in both cases we pick a gauge slice in moduli space and in particular fix invariance under holomorphic reparametrisations by working with holomorphic normal coordinates. This choice fixes invariance under Weyl transformations but leaves reparametrisation invariance manifest and allows us to work with arbitrary (subject to global constraints) worldsheet curvature. Again, we find precise agreement between the metric and transition function approach.
\sk

The first six sections are almost entirely independent of the underlying 2D matter conformal field theory (CFT), so that (some or all of) the spacelike components in the matter sector could be associated to any unitary CFT (corresponding presumably to a general string background), the central charge always being such that the total matter plus ghost central charge vanishes. From Sec.~\ref{sec:GOCS} onwards we consider, for concreteness, `vanilla bosonic string theory', the best studied example of string theory. But perhaps the discussion will offer new perspectives, the point being that this simplest string theory allows one to go deeper than would be otherwise possible in a first attempt with a more elaborate string theory. In bosonic string theory the matter content is the target spacetime embedding, $x^{\nu}$, and the ghost fields, $b,c$. In the critical theory the superscript $\nu$ spans $D=26$ non-compact dimensions (or toroidal compactifications thereof, where $D=26-d_c$ when $d_c$ dimensions are compactified). In the vanilla theory $D$ will always denote the number of non-compact dimensions. 
\sk

Regarding underlying mathematics, there is perhaps a large gap between much of the string theory literature and corresponding Riemann surface theory. This is perhaps partly because conformal invariance and focus on lowest order amplitudes makes detailed understanding of Riemann surfaces somewhat unnecessary. But beyond tree-level and one-loop amplitudes, and if one needs to go offshell, conformal invariance does not play a particularly central role. We have tried to partly bridge this gap, in that we review the essential Riemann surface theory that we need in the main text in detail. We have a physicist audience in mind (but also attempt to be mathematically precise when there is possibility of confusion). We place emphasis on the transition from complex analysis on a plane to that on general Riemann surfaces, from local to global, but also endow surfaces with a metric at various instances allowing us to think of them as Riemannian manifolds. Our starting point is (in Sec.~\ref{sec:RS}) the very definition of a complex structure on a Riemann surface, and then we go on to discuss complex structure deformations from a variety of complementary viewpoints in detail. We also present an extensive discussion of the Euler characteristic, with and without boundaries, which serves as an excellent example to clarify the relation between local and global aspects at fixed complex structure. The techniques developed are applied extensively in the remaining article where we discuss how to cut open and glue path integrals, that in turn lead to the construction of explicit handle operators.
\sk

Regarding prerequisites, we have made an effort to make this work self-contained, in that it includes detailed reasoning and derivations, so as to be accessible to graduate students in string theory. This ambition has also contributed to its length, and there is a fine line between accessibility and conciseness. It is not clear we have succeeded in this respect. The organisation of material is such that readers with expertise in certain areas can skip corresponding sections after perhaps briefly skimming through them. The essential prerequisite is a working knowledge of Polchinski's textbook \cite{Polchinski_v1} (which in turn assumes working knowledge in differential geometry and quantum field theory), and a basic knowledge of complex analysis is certainly useful \cite{Ahlfors66}. We review most of what we need of Riemann surface theory, but excellent and complementary notes are \cite{BersRiemannSurfaces,Gunning66} and \cite{Bers81}. 
For complex manifold theory a similarly pedagogical and yet concise account is Chern's lectures \cite{Chern}, and also \cite{GunningRossi,MorrowKodaira}. We have made an effort (wherever possible) to adopt notation and conventions that are consistent with Polchinski's textbook \cite{Polchinski_v1}.

\section{Riemann Surfaces}\label{sec:RS}
In this section we set the scene by giving a detailed exposition of the Riemann surface theory we will be needing in order to discuss cutting and gluing of path integrals, that will in turn enable us to construct an exact and globally well-defined expression for generic handle operators. We will place particular emphasis on the link between local and global (on $\Sigma$ and $\mathcal{M}$) properties. The notion of a complex structure (and deformations thereof) plays a starring and fundamental role in perturbative string and Riemann surface theory, so we will explore and develop this concept in detail and from a variety of viewpoints.

\subsection{Fixed Complex Structure}\label{sec:CST}
Let us consider a genus-$\g$ closed Riemann surface, in particular a smooth, oriented and closed two-dimensional surface, $\Sigma$, that we endow with a corresponding open cover $\{U_1,U_2,\dots\}$ so that $\cup_m U_m=\Sigma$. We will eventually think of these surfaces as complex manifolds, but it will be convenient to initially regard $\Sigma$ as a real manifold. Every such manifold is locally diffeomorphic to an open subset of Euclidean space, $\mathbf{R}^2$. So we may primarily focus on a single coordinate chart, $(U,\sigma^a)$, with $\sigma^a$ (with $a=1,2$) a standard set of real coordinates mapping points in $U$ homeomorphically onto an open subset of $\mathbf{R}^{2}$ \cite{Tu11}. 

\subsubsection{Almost Complex Structure}\label{sec:NACS}
Let us build up to a useful and general definition of an almost complex structure on $\Sigma$. 
The surface $\Sigma$ has a `natural almost complex structure', $I$, which in turn has the following local realisation. 
To the chart $(U,\sigma^a)$ there corresponds a {\it holomorphic coordinate}, $z$, (and its complex conjugate, $\bar{z}\equiv z^*$) defined by:
\begin{equation}\label{eq:z=s1+is2}
\begin{aligned}
z&=\sigma^1+i\sigma^2\\
\bar{z}&=\sigma^1-i\sigma^2
\end{aligned}
\qquad{\rm and}\qquad
\begin{aligned}
\partial_{z}&=\tfrac{1}{2}(\partial_1-i\partial_2)\\
\partial_{\bar{z}}&=\tfrac{1}{2}(\partial_1+i\partial_2),
\end{aligned}
\end{equation}
satisfying, 
\begin{equation}
\partial_{\bar{z}}z=\partial_z\bar{z}=0.
\end{equation}
The quantity $\partial_a$ denotes differentiation with respect to the real coordinate $\sigma^a$. (An ``overbar'' throughout denotes complex conjugation.) These relations are useful and efficient if we are only interested in a single complex structure, since for any given complex structure there exist local coordinates as in (\ref{eq:z=s1+is2}). In string theory we are interested in a variety of distinct complex structures and their interrelations, and we will need to generalise the above. 
\sk

With this objective in mind, let us reformulate what we have just described. 
Construct a real matrix, $I$, with components,\footnote{Our convention for index placement in complex structure matrices is: $J_a^{\phantom{a}b}=\left(\begin{matrix} J_1^{\phantom{a}1} & J_1^{\phantom{a}2} \\ J_2^{\phantom{a}1} & J_2^{\phantom{a}2}\end{matrix}\right)$.}
\begin{equation}\label{eq:Imatrix}
I=
\Big(
   \begin{matrix} 
       0 & 1 \\
      -1 & 0 \\
   \end{matrix}
\Big),\qquad {\rm satisfying}\qquad I_a^{\phantom{a}b}I_b^{\phantom{a}c}=-\delta_a^{\phantom{a}c},
\end{equation}
in terms of which the holomorphicity condition, $\partial_{\bar{z}}z=0$, can equivalently be rewritten as:
\begin{equation}\label{eq:z=s1+is2-gen}
\big(\partial_a+iI_a^{\phantom{a}b}\partial_b\big)z(\sigma)=0.
\end{equation}
We refer to the components, $I_a^{\phantom{a}b}$, of the matrix in (\ref{eq:Imatrix}) as the components of the `natural almost complex structure' tensor with respect to the coordinates $\sigma^a$. The complex coordinate $z$ with the explicit realisation (\ref{eq:z=s1+is2}) is a holomorphic coordinate with respect to {\it this} complex structure $I$. 
\sk

Notice that given any solution $z(\sigma)$ to the differential equation (\ref{eq:z=s1+is2-gen}) we can construct another solution, $f(\sigma)$, as follows. We expand this complex function, $f(\sigma)=u(\sigma)+iv(\sigma)$, in terms of two real differentiable functions, $u(\sigma)$ and $v(\sigma)$, and demand that this be a solution to the differential equation (\ref{eq:z=s1+is2-gen}). Equating real and imaginary parts of the resulting relations yields the {\it Cauchy-Riemann equations},
\begin{equation}
\begin{aligned}
\partial_1u(\sigma)&=\partial_2v(\sigma)\\
\partial_2u(\sigma)&=-\partial_1v(\sigma),
\end{aligned}
\end{equation}
which (by Goursat's theorem) in turn guarantees that $f(\sigma)$ is holomorphic in $z(\sigma)$. So given a solution $z(\sigma)$ to (\ref{eq:z=s1+is2-gen}), any holomorphic function of $z(\sigma)$ will also be a solution, and in particular any holomorphic function of $z(\sigma)$ can serve as a `holomorphic coordinate with respect to the $I$ complex structure'. 
\sk

In equation (\ref{eq:z=s1+is2}) we have chosen to map the base point $\sigma=0$ to $z=0$. This is clearly not necessary, and we might more generally like to place the base point of the holomorphic coordinate system at a more generic point, perhaps at $\sigma=\sigma_1$. We could then define a more general holomorphic coordinate, $z_{\sigma_1}(\sigma)$, via,
\begin{equation}\label{eq:zs's=zs'-zs}
\begin{aligned}
z_{\sigma_1}(\sigma)&=z(\sigma)-z(\sigma_1)\\
&= ({\sigma}^{1}-\sigma_1^1)+i({\sigma}^{2}-\sigma_1^2),
\end{aligned}
\end{equation}
with $z(\cdot)$ as defined above. This base-point dependent coordinate in turn has the property $z_{\sigma_1}(\sigma_1)=0$ for every $\sigma^a_1$, so that the origin of the frame coordinate, $z_{\sigma_1}$, is carried along with the base point $\sigma_1$. In this case derivatives of the form $\partial_z$ or $\partial_{z_{\sigma_1}}$ will always implicitly mean differentiation with respect to $z_{\sigma_1}(\sigma)$. 
\sk

We emphasise from the outset that we will eventually need to abstract away from the simple expression for $z_{\sigma_1}(\sigma)$ given on the right-hand side of (\ref{eq:zs's=zs'-zs}). Because although such a simple expression exists (locally) for any given complex structure, such as the natural complex structure, $I$, represented by (\ref{eq:Imatrix}), we will need to decide how to incorporate the fact that general surfaces have Ricci curvature. The latter can in turn be distributed throughout the surface in a variety of ways subject to topological constraints (the Euler number): it can be encoded locally (e.g., uniformly or otherwise using a metric) or globally (using transition functions). Nevertheless, a property that does survive the abstraction is that there will {\it always} exist holomorphic coordinates, $z_{\sigma_1}(\sigma)$, with the property $z_{\sigma_1}(\sigma_1)=0$ (or $z_{\sigma}(\sigma)=0$ since there is nothing special about the chosen base point), but they need not take the simple form shown on the right-hand side in (\ref{eq:zs's=zs'-zs}). In particular, notice that in (\ref{eq:zs's=zs'-zs}) $z_{\sigma_1}(\sigma)$ depends holomorphically on both $\sigma$ and $\sigma_1$. More generally, as we will see, although $z_{\sigma_1}(\sigma)$ will always be holomorphic with respect to $\sigma$ it need not be holomorphic with respect to the base point $\sigma_1$. We will elaborate on this in Sec.~\ref{sec:SPUTF} where we discuss how to translate punctures across general Riemann surfaces with arbitrary local Ricci curvature. 
\sk

In the current section the base point will not play any role, so we will keep it implicit and drop it from the notation by writing $z_{\sigma_1}(\sigma)$ as $z(\sigma)$, with a similar remark applying to all other conformal coordinates we will introduce here.
\sk

Returning to the main theme of the current section, a general surface satisfying the properties set out in the opening paragraph of Sec.~\ref{sec:CST} (with the exception of $S^2$) has a variety of `inequivalent' complex structures. By `inequivalent' we mean complex structures not related by coordinate reparametrisations. The reason we reformulated the holomorphicity condition (\ref{eq:z=s1+is2}) as (\ref{eq:z=s1+is2-gen}) is that the latter immediately generalises the notion of a `holomorphic coordinate with respect to the natural complex structure $I$' to a `holomorphic coordinate with respect to a generic complex structure $J$'. Let us discuss how this comes about.
\sk

Notice that $I_a^{\phantom{a}b}$ maps tangent vector fields to tangent vector fields (on $\Sigma$) and satisfies $I_a^{\phantom{a}b}I_b^{\phantom{a}c}=-\delta_a^{\phantom{a}c}$. These properties (by definition) remain true for arbitrary complex structures, but we must in addition now allow for these complex structure matrix elements to become local functions, $I_a^{\phantom{a}b}(\sigma)$, on $\Sigma$. This can be deduced directly from (\ref{eq:z=s1+is2}) by performing a coordinate reparametrisation, $\sigma\mapsto \sigma'(\sigma)$, and noticing that the transformed components, $I_a^{\phantom{a}b}\rightarrow I_a^{'\phantom{i}b}(\sigma')$, become local functions even though $I_a^{\phantom{a}b}$ is constant. This suggests that a `generic' complex structure (i.e.~a complex structure that may or may not be related to $I_a^{\phantom{a}b}$ via a coordinate reparametrisation) will be a local function of $\sigma$.
\sk

Let us then denote such a generic complex structure by $J$. We then {\it define} a `holomorphic coordinate, $w$, with respect to complex structure $J$' by the following differential equation:
\begin{equation}\label{eq:w-gen}
\boxed{\big(\partial_a+iJ_a^{\phantom{a}b}\partial_b\big)w(\sigma)=0,\qquad {\rm with}\qquad J_a^{\phantom{a}b}J_b^{\phantom{a}c}=-\delta_a^{\phantom{a}c}}
\end{equation}
where now $J_a^{\phantom{a}b}(\sigma)$ is a local function on $U\subset\Sigma$. 
In order for this to be well-defined it must be that the sum of the two terms on the left-hand side of the first relation in (\ref{eq:w-gen}) transform covariantly under changes of coordinates, $\sigma\mapsto \sigma'(\sigma)$. This suggests we construct a {\it vector-valued} differential one form\footnote{Vector-valued forms were introduced by Nijenhuis, Eckmann and Fr\"olicher in the early 50s and studied in detail by Fr\"olicher and Nijenhuis in \cite{FrolicherNijenhuisI,FrolicherNijenhuis} (see also \cite{KodairaNirenbergSpencer} and \cite{BersRiemannSurfaces,Witten12c,D'HokerPhong15b} for complementary perspectives, and \cite{Tu17} for a modern discussion of vector-valued forms in general).}, 
\begin{equation}\label{eq:V-V J}
J=\rmd\sigma^aJ_a^{\phantom{a}b}(\sigma)\partial_b,
\end{equation}
and denoting an exterior derivative on $\Sigma$ by 
$
\rmd=\rmd\sigma^a\partial_a,
$ 
the differential equation in (\ref{eq:w-gen}) then takes on a manifestly coordinate-invariant form: $
(\rmd+iJ)w=0. 
$ 
This in turn implies an integrability condition, $\rmd J=0$, (recall that $\rmd^2=0$) whose component form reads,
$$
J_a^{\phantom{a}c}\partial_cJ_b^{\phantom{a}d}-J_b^{\phantom{a}c}\partial_cJ_a^{\phantom{a}d}-J_c^{\phantom{a}d}\partial_aJ_b^{\phantom{a}c}+J_c^{\phantom{a}d}\partial_bJ_a^{\phantom{a}c}=0.
$$
The quantity on the left-hand side, call it $J_{ab}^d$, is essentially (up to a convention-dependent overall factor) the component form of the Nijenhuis (or torsion) tensor \cite{KobayashiNomizu-II}. The statement $J_{ab}^d=0$ is an integrability condition for almost complex structure (by the Newlander-Nirenberg theorem), that is in turn identically satisfied in the case of interest of smooth closed two-dimensional surfaces. Since existence of an integrable almost complex structure is equivalent to existence of a complex structure, we will often omit the adjective `almost' and (somewhat imprecisely) refer to $J$ as a `complex structure' for brevity.  If furthermore $\Sigma$ admits a local holomorphic coordinate for $J$ around every point in $\Sigma$ then integrability ensures that these patch together to form a holomorphic atlas for $\Sigma$ which furthermore induces $J$. We will discuss this patching at fixed complex structure in detail in Sec.~\ref{sec:TFCR}, and a corresponding detailed example (the Euler characteristic) is discussed in Sec.~\ref{sec:EulerCh}.
\sk

For later reference, it is convenient to denote the set of all complex structures by:
$$
\mathscr{J}=\big\{\textrm{set of all complex structures $J$}\big\}.
$$
Moduli space will correspond to the quotient, 
$$
\mathcal{M}=\mathscr{J}/\mathscr{D}_{p_1,p_2,\dots},
$$
where $\mathscr{D}_{p_1,p_2,\dots}$ is the subset of diffeomorphisms, $\sigma\mapsto\sigma'(\sigma)$, that leave fixed a certain set of points, $\{p_1,p_2,\dots\}\in\Sigma$, at which vertex operators or (bi-local) handle operators are inserted.

\subsubsection*{Symmetries}
Notice that given any solution, $w(\sigma)$, to the defining differential equation in (\ref{eq:w-gen}) associated to complex structure $J(\sigma)$ in a given domain $U\subset \Sigma$, there is an invariance associated to:
\begin{equation}\label{eq:w->f(w)}
\boxed{w(\sigma)\mapsto w'(\sigma)=f(w(\sigma))}
\end{equation}
for any holomorphic function, $f(w(\sigma))$, since the holomorphic coordinate $w'(\sigma)$ will also be a solution to the same differential equation. This can be thought of as a residual gauge symmetry that leaves fixed the complex structure $J$. This residual symmetry can be fixed in a variety of ways (e.g., by identifying this holomorphic coordinate with a `holomorphic normal coordinate', that is in turn specified in the path integral by a special choice of Beltrami differential, but we elaborate on this in later sections). 
\sk

The invariance under holomorphic reparametrisations (\ref{eq:w->f(w)}) in turn provides the following alternative viewpoint: we can think of a {\it complex structure}, $J$, on $\Sigma$ as an {\it equivalence class}, $[w]$, of systems of local holomorphic coordinates. We can then reach a global construction by introducing holomorphic transition functions on patch overlaps and corresponding cocycle relations, as we discuss in detail in Sec.~\ref{sec:TFCR}. This provides an alternative definition of a complex structure on $\Sigma$, which in turn makes no explicit reference to an underlying real manifold. Much of the string theory literature emphasises this latter viewpoint since it is more efficient. But the latter viewpoint also obscures certain aspects, such as the fact that (as we discuss momentarily) holomorphic coordinates transform as {\it scalars} under reparametrisations, $\sigma\mapsto \sigma'(\sigma)$. This is a  useful result with far-reaching implications, since it is at the heart of making local composite (normal-ordered) operators (at least modulo U(1)) globally well-defined on $\Sigma$. (This is in particular at the heart of both conformal normal ordering \cite{Polchinski87} and Weyl normal ordering \cite{Polchinski88}, more about which later.)
\sk

So let us discuss the transformation property of a given holomorphic coordinate, $w_{\sigma_1}(\sigma)$, under coordinate reparametrisations, $\sigma\mapsto \sigma'(\sigma)$. We reintroduced an explicit base point, $\sigma_1$, at which $w_{\sigma_1}(\sigma_1)\equiv0$, since the holomorphic coordinate also depends on this. We will show that it transforms as a {\it scalar},
\begin{equation}\label{eq:w's'=ws}
\boxed{w_{\sigma_1}'(\sigma')=w_{\sigma_1}(\sigma)\quad \textrm{mod  U(1)}\qquad {\rm under}\qquad \sigma\mapsto\sigma'(\sigma),\qquad {\rm with}\qquad \sigma'(\sigma_1)\equiv \sigma_1}
\end{equation}
A few comments are in order:
\begin{itemize}
\item It is convenient to disentangle reparametrisations that leave the base point fixed from reparametrisations of the base point, $\sigma_1$. (Reparametrisations of the base point are considered in (\ref{eq:zsigma* scalar}) where we will see that the corresponding holomorphic coordinate also transforms as a scalar under reparametrisations of the base point.)
\item Following on from the above comment, when the base point $\sigma_1$ is associated to a modulus (which will be a case of interest) then the full path integral should be invariant under reparametrisations in moduli space \cite{Polchinski_v1} (which includes $\sigma_1\mapsto \sigma_1'(\sigma_1,\dots)$ where `\dots' denotes the remaining coordinates in moduli space), and in particular the full moduli space integrand should transform as a {\it density} (or top form). 
\item In \cite{Polchinski87} it was assumed that the conformal coordinate $w_{\sigma_1}(\sigma)$ transforms as a scalar under reparametrisations of both the base point, $\sigma_1$, and the point $\sigma$ at which the conformal coordinate is evaluated. This is guaranteed when the conformal coordinate, $w_{\sigma_1}(\sigma)$, is holomorphic with respect to {\it both} $\sigma$ and $\sigma_1$ (meaning that (\ref{eq:w-gen}) is satisfied even if we replace $\partial_a+iJ_a^{\phantom{a}b}\partial_b$ by the same quantity but with $\sigma_1$ replacing $\sigma$), which is in turn not satisfied generically. (A case where it is satisfied is (\ref{eq:zs's=zs'-zs}).) If one does assume holomorphicity in both $\sigma$ and $\sigma_1$ the price to pay is that charts are not guaranteed to glue together consistently and one must add Wu-Yang type terms on patch overlaps \cite{Nelson89}. This is because global {\it holomorphic} families of coordinates do not exist (even modulo U$(1)$) \cite{Nelson89}.\footnote{When we consider asymptotic state vertex operators as conformal primaries this distinction vanishes and one can assume the conformal coordinate transforms as a scalar with respect to both the base point and the point at which it is evaluated \cite{Polchinski87}, but for general handle operators we do not have the luxury of choosing a conformal primary basis.} Although $z_{\sigma_1}(\sigma)$ need not be holomorphic in $\sigma_1$, in Sec.~\ref{sec:SPUTF} we show that it also transforms as a scalar under reparametrisations of the form $\sigma_1\mapsto \sigma_1'(\sigma_1)$. A related discussion is given in Sec.~\ref{sec:comm dA-dA} and Sec.~\ref{sec:WuYang}.
\end{itemize}
For the time being we may now make the base point implicit again since it is to be considered inert under reparametrisations, and hence denote $w_{\sigma_1}$ by $w$ to lighten the notation (until the need arises to make it explicit again). 
\sk

The holomorphic coordinate, $w(\sigma)$, with respect to complex structure, $J$, satisfies the defining equation (\ref{eq:w-gen}),
\begin{equation}\label{eq:w-gen2}
\big(\partial_a+iJ_a^{\phantom{a}b}\partial_b\big)w(\sigma)=0.
\end{equation}
Under infinitesimal coordinate transformations,
\begin{equation}\label{eq:s->s'=s+e}
\sigma^a\mapsto \sigma^{'a}(\sigma)=\sigma^a+\epsilon^a(\sigma),
\end{equation}
requiring that the vector-valued one form, $J=\rmd\sigma^aJ_a^{\phantom{a}b}\partial_b$,  be globally-defined (i.e.~that it transforms as a tensor, $J'=J$, on patch overlaps) enforces the transformation, $J_a^{\phantom{a}b}(\sigma)\mapsto J_a^{'\phantom{i}b}(\sigma)=J_a^{\phantom{a}b}(\sigma)+\delta J_a^{\phantom{a}b}(\sigma)$, with,
\begin{equation}\label{eq:dJ-rep}
\boxed{\delta J_a^{\phantom{a}b}(\sigma)=\big(J_a^{\phantom{a}c}\partial_c\epsilon^b-J_c^{\phantom{a}b}\partial_a\epsilon^c-\epsilon^c\partial_cJ_a^{\phantom{a}b}\big)(\sigma)}
\end{equation}
A general variation of (\ref{eq:w-gen2}) in turn yields, 
\begin{equation}\label{eq:delta dw}
\delta\big(\partial_aw(\sigma)\big)+i\big(\delta J_a^{\phantom{a}b}(\sigma)\big)\partial_bw(\sigma)+iJ_a^{\phantom{a}b}(\sigma)\delta\big(\partial_bw(\sigma)\big)=0,
\end{equation}
so in the particular case that the variations are generated by reparametrisations (\ref{eq:s->s'=s+e}), and taking (\ref{eq:dJ-rep}) into account, it is not hard to see that (\ref{eq:delta dw}) is satisfied when $\delta w(\sigma)\equiv w'(\sigma)-w(\sigma)=-\epsilon^a\partial_aw(\sigma)$, which is the infinitesimal version of (\ref{eq:w's'=ws}). Therefore, successive compositions of infinitesimal reparametrisations will generate the finite transformation rule (\ref{eq:w's'=ws}), and so it is indeed consistent to assume $w(\sigma)$ transforms as a scalar under general coordinate reparametrisations, $\sigma\mapsto \sigma'(\sigma)$ (at least modulo U(1)). Regarding the remaining U(1), let us make the base point, $\sigma_1$, manifest again by writing $w(\sigma)$ as $w_{\sigma_1}(\sigma)$, and notice that if we replace $w_{\sigma_1}(\sigma)\rightarrow e^{i\theta(\sigma_1)}w_{\sigma_1}(\sigma)$, the defining differential equation (\ref{eq:w-gen2}) remains invariant and so the phase cannot be determined uniquely by this defining equation. In fact, by Hopf's index theorem the real phase $\theta(\sigma_1)$ cannot be gauged away globally, so $w_{\sigma_1}(\sigma)$ at best can be taken to transform as a scalar modulo U(1) phase rotations (we will elaborate on this in detail in Sec.~\ref{sec:WB}, see the discussion associated to (\ref{eq:Ipi=chi}) there).

\subsubsection{Metric Viewpoint and the Beltrami Equation}\label{sec:MVa}

Building on the formalism we have developed so far (taking into account the remarks in the opening paragraph of Sec.~\ref{sec:CST}), we now endow $\Sigma$ with a metric tensor associated to complex structure $J(\sigma)$,\footnote{We assume throughout that the associated line elements satisfy $\rmd s^2\geq0$, and also that $\det g_{ab}>0$.}
\begin{equation}\label{eq:ds2=gab}
\boxed{
g_J=g_{ab}(\sigma)\rmd\sigma^a \rmd\sigma^b}
\end{equation}
Introducing a metric enables us to think of $\Sigma$ as a Riemannian manifold. 
Although introducing a metric is not necessary it is sometimes convenient, since when we cut open and glue path integrals across various cycles of $\Sigma$ the intermediate steps in the calculation will depend on a choice of metric (or Weyl factor) even though this Weyl dependence always cancels out of physical observables. Introducing a metric also preserves reparametrisation invariance.
\sk

We have included a label $J$ on the left-hand side in (\ref{eq:ds2=gab}) to make explicit that the metric tensor depends on complex structure but have not included a corresponding label on the right-hand side. This is because there are two main complementary viewpoints here: we can either think of the components, $g_{ab}(\sigma)=g_{ab}^J(\sigma)$, as determining the complex structure with the coordinates, $\sigma^a$, fixed, or we can think of the coordinates as determining the complex structure, $\sigma^a=\sigma^a_J$, with the corresponding metric components, $g_{ab}(\sigma_J)$, considered fixed.\footnote{(A third possibility is that both coordinates and metric components depend on complex structure but we do not need to consider this case.)} We will explore both of these viewpoints.
\sk

Adopting the viewpoint that the metric components determine the complex structure with the corresponding coordinates considered fixed, we now claim that the complex structure components, $J_a^{\phantom{a}b}(\sigma)$, take the following explicit form in terms of metric components given in (\ref{eq:ds2=gab}), 
\begin{equation}\label{eq:J-metric}
\boxed{J_a^{\phantom{a}b}\dfn \frac{g_{ac}\varepsilon^{cb}}{\sqrt{g}},\qquad\textrm{satisfying}\qquad J_a^{\phantom{a}b}J_b^{\phantom{a}c}=-\delta_a^{\phantom{a}c},\qquad {\rm and}\qquad\left\{
\begin{aligned} 
\varepsilon^{12}&=-\varepsilon^{21}=1\\
\varepsilon^{11}&=-\varepsilon^{22}=0
\end{aligned}\right.}
\end{equation}
where $g=\det g_{ab}$. 
For easy reference, perhaps it is useful to also display $J$ in slightly more explicit terms,
\begin{equation}\label{eq:Jmatrix}
J= 
     \frac{1}{\sqrt{g}}\Big(\begin{matrix} 
      -g_{12} & g_{11} \\
      -g_{22} & g_{12} \\
   \end{matrix}\Big).
\end{equation}
Complex structure is of course invariant under Weyl rescalings, $g_{ab}(\sigma)\mapsto g_{ab}'(\sigma)=e^{\phi(\sigma)}g_{ab}(\sigma)$, in particular $
J|_{e^\phi g_{ab}}=J|_{g_{ab}}$. 
\sk

Let us understand why this complex structure, $J$, may be identified with the $J$ complex structure of the previous subsection. 
Recall that given a real set of standard local coordinates, $\sigma^a$, we defined a complex set by, 
\begin{equation}\label{eq:z=s1+is2,zbar=}
z=\sigma^1+i\sigma^2, \qquad {\rm and}\qquad \bar{z}=\sigma^1-i\sigma^2,
\end{equation}
and also that $z(\sigma)$ is a holomorphic coordinate with respect to a fixed natural complex structure $I$ (but not with respect to $J$ unless $J=I$). We wish to rewrite the $J$ complex structure metric (\ref{eq:ds2=gab}) in terms of $z,\bar{z}$ coordinates given in (\ref{eq:z=s1+is2,zbar=}) that are in turn holomorphic with respect to the $I$ complex structure,
\begin{equation}\label{eq:gab2}
\begin{aligned}
g_J
&=\frac{1}{4}\big(g_{11}-g_{22}-2ig_{12}\big)\rmd z^2+\frac{1}{4}\big(g_{11}-g_{22}+2ig_{12}\big)\rmd\bar{z}^2+\frac{1}{2}\big(g_{11}+g_{22}\big)\rmd z\rmd\bar{z}. 
\end{aligned}
\end{equation}
Since the corresponding line element $\rmd s^2\geq 0$ and the components in the first two terms in (\ref{eq:gab2}) are related by complex conjugation and the last term is real, we may define quantities $\rho,\mu_{\bar{z}}^{\phantom{z}z}$ (with $\rho(z,\bar{z})=|\rho|>0$ real and $\mu_{\bar{z}}^{\phantom{z}z}$ complex) by:
\begin{equation}\label{eq:gabmurho1}
\begin{aligned}
\boxed{g_J=\rho |\rmd z+\mu_{\bar{z}}^{\phantom{z}z} \rmd\bar{z}|^2}
\end{aligned}
\end{equation}

Following \cite{BersRiemannSurfaces}, setting (\ref{eq:gab2}) equal to (\ref{eq:gabmurho1}) determines the components $g_{ab}$ in terms of $\rho,\mu_{\bar{z}}^{\phantom{z}z}$,
\begin{subequations}\label{eq:rhomu-gab}
\begin{align}
&g_{11}=\rho(1+|\mu_{\bar{z}}^{\phantom{z}z}|^2+\mu_{\bar{z}}^{\phantom{z}z}+\mu_{z}^{\phantom{z}\bar{z}})\label{eq:g11rhomu}\\
&g_{22} =\rho(1+|\mu_{\bar{z}}^{\phantom{z}z}|^2-\mu_{\bar{z}}^{\phantom{z}z}-\mu_{z}^{\phantom{z}\bar{z}})\label{eq:g22rhomu}\\
&g_{12} =-i\rho(\mu_{\bar{z}}^{\phantom{z}z}-\mu_{z}^{\phantom{z}\bar{z}})\label{eq:g12rhomu}
\end{align}
\end{subequations}
which follows from elementary algebra, and $\mu_{z}^{\phantom{z}\bar{z}}\equiv (\mu_{\bar{z}}^{\phantom{z}z})^*$. 
Solving for $\rho$, there are two roots, $\rho_{\pm}=\frac{1}{4}\big(g_{11}+g_{22}\big)\pm\frac{1}{2}\sqrt{g_{11}g_{22}-g_{12}^2}$, 
leading to two expressions for the corresponding quantity, $\mu_{\pm}=(\mu_{\bar{z}}^{\phantom{z}z})_{\pm}$. 
It is not hard to show that only \cite{BersRiemannSurfaces} the positive root leads to an orientation-preserving change of coordinates (i.e.~with positive Jacobian), and we are only interested here in orientated surfaces. So writing $\rho_+$ as $\rho$ (and similarly $\mu_+$ as $\mu_{\bar{z}}^{\phantom{z}z}$), we can solve for $\mu_{\bar{z}}^{\phantom{z}z},\rho$ in terms of $g_{ab}$ and find,
\begin{equation}\label{eq:rho+}
\begin{aligned}
\rho &= \frac{1}{4}\Big({\rm Tr}\,g_{ab} +2\sqrt{\det \,g_{ab}}\Big),\qquad {\rm and}\qquad
\mu_{\bar{z}}^{\phantom{z}z} = \frac{\phantom{a}g_{11}-g_{22}+2ig_{12}\phantom{a}}{{\rm Tr}\,g_{ab} +2\sqrt{\det \,g_{ab}}}
\end{aligned}
\end{equation}
From the latter expression we also have that since $|\mu_{\bar{z}}^{\phantom{z}z}|^2\geq0$ and $\det \,g_{ab}>0$, the parametrisation (\ref{eq:gabmurho1}) requires, 
$
\frac{1}{2}{\rm Tr}\,g_{ab} \geq\sqrt{\det \,g_{ab}},
$ 
and since this is trivially true (given the components $g_{ab}$ are real) this in turn implies that only the following range is associated to orientation-preserving diffeomorphisms:
\begin{equation}\label{eq:|mu+|<1}
\boxed{0\leq|\mu_{\bar{z}}^{\phantom{z}z}|<1}
\end{equation}
This remains true for any real metric associated to an oriented surface. 
(Indeed, one can show that the complement, $|\mu_{\bar{z}}^{\phantom{z}z}|\geq1$, is associated to orientation-reversing diffeomorphisms, in which case the coefficient of $\sqrt{\det \,g_{ab}}$ in both expressions in (\ref{eq:rho+}) flips sign, the remaining terms being unaltered.) Saturating the first inequality in (\ref{eq:|mu+|<1}), namely setting $\mu_{\bar{z}}^{\phantom{z}z}=0$, automatically leads to a conformally-flat metric, where $g_{11}=g_{22}$ and $g_{12}=0$, and this in turn coincides with the $I$ complex structure.
\sk

Substituting (\ref{eq:rhomu-gab}) into (\ref{eq:Jmatrix}) leads to the following exact expression for the complex structure $J$ and determinant $\sqrt{g}=\sqrt{\det g_{ab}}$ in terms of the quantity $\mu_{\bar{z}}^{\phantom{a}z}$ (which we write as $\mu$ here in order to avoid unnecessary proliferation of indices) and $\rho$,
\begin{equation}\label{eq:Jmatrix-mu}
\boxed{
J= \frac{1}{1-|\mu|^2}
     \bigg(\begin{matrix} 
       i(\mu-\bar{\mu}) & 1+|\mu|^2+\mu+\bar{\mu}\\
      -1-|\mu|^2+\mu+\bar{\mu} & -i(\mu-\bar{\mu})\\
   \end{matrix}\bigg),\qquad \sqrt{g}=\rho\big(1-|\mu|^2\big)
}
\end{equation}

Let us substitute the exact explicit expression (\ref{eq:Jmatrix-mu}) for $J$ into the differential equation (\ref{eq:w-gen}) defining the $w(\sigma)$ holomorphic coordinate  with respect to the $J$ complex structure, taking also (\ref{eq:z=s1+is2,zbar=}) into account (according to which $\partial_1=\partial_z+\partial_{\bar{z}}$ and $\partial_2=i(\partial_z-\partial_{\bar{z}})$). We learn that the differential equation (\ref{eq:w-gen}) defining the $w$ holomorphic coordinate can equivalently be rewritten as follows,
\begin{equation}\label{eq:w-gen3}
\phantom{\qquad\textrm{(Betrami Equation)}\qquad}\boxed{\big(\partial_{\bar{z}}-\mu_{\bar{z}}^{\phantom{a}z}\partial_z\big)w(\sigma)=0}\qquad\textrm{(Betrami Equation)}\qquad
\end{equation}
We will assume throughout that $\partial_zw\neq0$. Remarkably, the expression (\ref{eq:w-gen3}) is exact despite the fact that $\mu_{\bar{z}}^{\phantom{a}z}$ and $J_{a}^{\phantom{b}b}$ are related non-linearly by (\ref{eq:Jmatrix-mu}). The differential equation (\ref{eq:w-gen3}) is called the {\it Beltrami equation} and will play a starring role in some of the following sections.
\sk

Geometrically speaking, and by analogy to (\ref{eq:V-V J}), the quantity $\mu_{\bar{z}}^{\phantom{z}z}$ might best be thought of as the component of a vector-valued $(0,1)$-form,
\begin{equation}\label{eq:beltramidifferential}
\mu=\rmd\bar{z}\mu_{\bar{z}}^{\phantom{z}z}\partial_z,
\end{equation}
so that $\mu_{\bar{z}}^{\phantom{z}z}\partial_z$ is a local section of the chiral half of the tangent bundle, $T\Sigma$, also denoted by $K^{-1}$, where $K$ is the chiral half of the cotangent or canonical bundle (also denoted by $T^*\Sigma$). The quantity $\mu$ is known as a {\it Beltrami differential} and $\mu_{\bar{z}}^{\phantom{z}z}$ the corresponding component in the $z$ coordinates. 
\sk

Let us make a few observations:
\begin{itemize}
\item  The coordinates $z,\bar{z}$ are holomorphic coordinates with respect to the natural complex structure, $I$. So (\ref{eq:w-gen3}) relates holomorphic coordinates $w$ and $z$ of (generically) distinct complex structures $J$ and $I$ respectively. We emphasise furthermore that $z$ is {\it not} a holomorphic coordinate with respect to the $J$ complex structure (unless $\mu=0$), since from (\ref{eq:w-gen3}) we see that $w$ is not a holomorphic function of $z$ when $\mu\neq0$. 
\item There are two points of view associated to (\ref{eq:w-gen3}), the first related to whether we wish to provide $w(\sigma)$ and read off $\mu_{\bar{z}}^{\phantom{z}z}$, and in the second case we can rather provide $\mu_{\bar{z}}^{\phantom{z}z}$ and then (at least locally) solve for $w(\sigma)$. Since $w(\sigma)$ is a solution to a first-order differential equation we must supply {\it boundary conditions} to obtain a unique solution.
\item The Beltrami equation is invariant under the holomorphic reparametrisation, $w\mapsto f(w)$, which is a residual symmetry of conformal gauge in the $J$ complex structure. Once we pick a gauge slide in moduli space, which may be phrased as picking an explicit expression for $\mu$, then we can also pick a {\it specific} corresponding solution, $w(z,\bar{z})$, to the Beltrami equation which fixes most of the residual symmetry of conformal gauge. This solution need not have the full symmetry of the original differential equation since the gauge slice generically fixes this symmetry. There is also a remaining ambiguous U(1) that causes some trouble -- we discuss this in various places throughout the document and primarily in Sec.~\ref{sec:WB} (where we relate it to Hopf's index theorem to exhibit its topological nature) and in Sec.~\ref{sec:HNC} (where we derive how it manifests itself when we fix the invariance, $w\mapsto f(w)$, of the Beltrami equation by choosing to work with holomorphic normal coordinates).
\item The differential equation (\ref{eq:w-gen3}) implies that we can if we wish forget entirely about the underlying real coordinate, $\sigma$, and instead replace $w(\sigma)$ by $w(z,\bar{z})$. This approach will indeed be useful at various points throughout, and is perhaps also the most common approach in the string theory literature \cite{DHokerPhong,Witten12c}.
\item That a solution to the Beltrami equation (\ref{eq:w-gen3}) always exists goes as far back as Gauss, at least in the case of real-analytic $\mu_{\bar{z}}^{\phantom{a}z}$, meaning that it has a power series expansion in $z,\bar{z}$. But real analyticity is sometimes too restrictive and in fact a generalised solution also exists provided $\mu_{\bar{z}}^{\phantom{a}z}$ satisfies a H\"older condition or even if it is simply measurable (see ch.~V in \cite{Alhfors}, Sec.~4 in \cite{Abikoff}, and also \cite{ChernIsothCoords55,BersRiemannSurfaces}). 
\end{itemize}

From the chain rule and the Beltrami equation (\ref{eq:w-gen3}) it follows immediately that 
$
\rmd w=\partial_zw\,(\rmd z+\mu_{\bar{z}}^{\phantom{a}z}\,\rmd\bar{z})
$, 
and so making use of this in (\ref{eq:gabmurho1}) and setting the latter equal to (\ref{eq:ds2=gab}) leads to the famous result,
\begin{equation}\label{eq:gab=rhomuz=w}
\begin{aligned}
g_{ab}(\sigma)\rmd\sigma^a \rmd\sigma^b&=\rho(z,\bar{z}) |\rmd z+\mu_{\bar{z}}^{\phantom{z}z} \rmd\bar{z}|^2\\
&= \rho(z,\bar{z}) |\partial_zw|^{-2} \rmd w\rmd\bar{w}\\
&\equiv \rho_0(w,\bar{w})\rmd w\rmd\bar{w},
\end{aligned}
\end{equation}
namely that, locally, all Riemann surfaces are conformally-flat. 
Since coordinate reparametrisations are by definition invertible in their domain of validity, given a solution $w(z,\bar{z})$ we can invert it (around an implicit base point, $\sigma=\sigma_1$) to construct $z(w,\bar{w})$ and so in the last equality we have defined:
\begin{equation}\label{eq:rho0rhozz}
\rho_0(w,\bar{w})\dfn \rho(z,\bar{z}) |\partial_zw(z,\bar{z})|^{-2},
\end{equation}
which may therefore be considered to be a function of $w,\bar{w}$ (and of the base point, $\sigma_1$). 
Notice that the various expressions (\ref{eq:gab=rhomuz=w}) for the metric represent the {\it same} complex structure, since they are related by a coordinate reparametrisations. Furthermore, for {\it every} complex structure there is a corresponding real coordinate expression for the metric (as on the left-hand side in (\ref{eq:gab=rhomuz=w})) and a corresponding conformally-flat metric (as on the right-hand side of the last equality in (\ref{eq:gab=rhomuz=w})). 

\subsubsection*{Symmetries}

Let us discuss the symmetries that leave fixed the line element (\ref{eq:gabmurho1}). Suppose there exist {\it smooth} vector fields, $v^z(z,\bar{z})\partial_z+v^{\bar{z}}(z,\bar{z})\partial_{\bar{z}}$, such that under, 
\begin{equation}\label{eq:z'=z+v(zzbar)}
z\mapsto z'=z+v^z(z,\bar{z}),
\end{equation}
the metric (\ref{eq:gabmurho1}) is invariant,
\begin{equation}\label{eq:gabmurho-rep}
\begin{aligned}
\rho(z,\bar{z}) |\rmd z+\mu(z,\bar{z}) \rmd\bar{z}|^2 = \rho'(z',\bar{z}') |\rmd z'+\mu'(z',\bar{z}') \rmd\bar{z}'|^2.
\end{aligned}
\end{equation}
We have written $\mu_{\bar{z}}^{\phantom{a}z}\equiv \mu(z,\bar{z})$ and similarly for the primed coordinates. Taylor expanding $\rho'(z',\bar{z}')$ and $\mu'(z',\bar{z}')$ around $z,\bar{z}$, writing $\rmd z'=\rmd z(1+\partial_zv)+\rmd\bar{z}\partial_{\bar{z}}v$ and similarly for the anti-chiral half, keeping terms up to linear order in $v$ and $\bar{v}$, and equating terms of equivalent tensor structures on left- and right-hand sides of (\ref{eq:gabmurho-rep}) yields:\footnote{Our conventions for covariant derivatives are spelt out in Appendix \ref{sec:CD}. Since these are defined to act on conformal tensors in $K^n$ (the $n$-fold tensor product of the canonical bundle) it might be helpful to recall that on account of metric compatibility, $\nabla_zg_{z\bar{z}}=0$, and $g^{z\bar{z}}g_{z\bar{z}}=1$ (note that $\rho\equiv 2g_{z\bar{z}}$), we have $\nabla_z\mu_{\bar{z}}^{\phantom{a}z}=\nabla_z(g^{z\bar{z}}g_{z\bar{z}}\mu_{\bar{z}}^{\phantom{a}z})=g_{z\bar{z}}\nabla_z(g^{z\bar{z}}\mu_{\bar{z}}^{\phantom{a}z})$, and then we can make use of the fact that $g^{z\bar{z}}\mu_{\bar{z}}^{\phantom{a}z}\partial_z\otimes\partial_z\in K^{-2}$ to evaluate the covariant derivative: $\nabla_z\mu_{\bar{z}}^{\phantom{a}z}=\partial_z\mu_{\bar{z}}^{\phantom{a}z}+\mu_{\bar{z}}^{\phantom{a}z}\partial_z\ln\rho$. Similar reasoning enables one to interpret all covariant derivatives appearing.}
\begin{equation}\label{eq:deltamumubarrho}
\begin{aligned}
\delta\mu_{\bar{z}}^{\phantom{a}z}&=-\big[(\nabla_z\mu_{\bar{z}}^{\phantom{a}z})+\nabla_{\bar{z}}-\mu_{\bar{z}}^{\phantom{a}z}\nabla_z\big]\big(v^z+\mu_{\bar{z}}^{\phantom{a}z}v^{\bar{z}}\big)\\
\delta\mu_{z}^{\phantom{a}\bar{z}}&=-\big[(\nabla_{\bar{z}}\mu_{z}^{\phantom{a}\bar{z}})+\nabla_{z}-\mu_{z}^{\phantom{a}\bar{z}}\nabla_{\bar{z}}\big]\big(v^{\bar{z}}+\mu_{z}^{\phantom{a}\bar{z}}v^{z}\big)\\
\delta \ln\rho&=-\big(\nabla_zv^z+\nabla_{\bar{z}}v^{\bar{z}}+\mu_{\bar{z}}^{\phantom{a}z}\nabla_zv^{\bar{z}}+\mu_{z}^{\phantom{a}\bar{z}}\nabla_{\bar{z}}v^{z}\big),\\
\end{aligned}
\end{equation}
where we have written $\delta\mu_{\bar{z}}^{\phantom{a}z}\equiv \mu'(z,\bar{z})-\mu(z,\bar{z})$ and $\delta\ln\rho\equiv \ln \rho'(z,\bar{z})- \ln \rho(z,\bar{z})$, so that both left- and right-hand sides are evaluated at $z,\bar{z}$. These variations are computed around a {\it finite} $\mu_{\bar{z}}^{\phantom{a}z},\mu_{z}^{\phantom{a}\bar{z}}$, and $\mu_{z}^{\phantom{a}\bar{z}}\equiv(\mu_{\bar{z}}^{\phantom{a}z})^*$. Notice that in the first two variations in (\ref{eq:deltamumubarrho}) we can replace all covariant derivatives by ordinary derivatives since the connection terms cancel out, e.g., $\delta\mu_{\bar{z}}^{\phantom{a}z}=-[(\partial_z\mu_{\bar{z}}^{\phantom{a}z})+\partial_{\bar{z}}-\mu_{\bar{z}}^{\phantom{a}z}\partial_z](v^z+\mu_{\bar{z}}^{\phantom{a}z}v^{\bar{z}})$, but they do not cancel in the first two terms of the last relation. For $v^z(z)$ holomorphic, $\delta\mu_{\bar{z}}^{\phantom{a}z}=\mu_{\bar{z}}^{\phantom{a}z}(\partial_zv^z-\partial_{\bar{z}}v^{\bar{z}})-(v^z\partial_z\mu_{\bar{z}}^{\phantom{a}z}+v^{\bar{z}}\partial_{\bar{z}}\mu_{\bar{z}}^{\phantom{a}z})$. An important special case of (\ref{eq:deltamumubarrho}) is when we look at reparametrisations, 
$$
z\mapsto z'=z+v^z(z,\bar{z}),
$$ 
around $\mu=0$ where it takes the form:
\begin{equation}\label{eq:deltamumubarrho2}
\begin{aligned}
\delta\mu_{\bar{z}}^{\phantom{a}z}\big|_{\mu=0}&=-\nabla_{\bar{z}}v^z\\
\delta\mu_{z}^{\phantom{a}\bar{z}}\big|_{\mu=0}&=-\nabla_{z}v^{\bar{z}}\\
\delta \ln\rho\big|_{\mu=0}&=-(\nabla_zv^z+\nabla_{\bar{z}}v^{\bar{z}}).\\
\end{aligned}
\end{equation}

Let us now return to the Beltrami equation (\ref{eq:w-gen3}). In terms of the quantity $\mu$ in (\ref{eq:beltramidifferential}), and writing $\bar{\partial}=\rmd\bar{z}\partial_{\bar{z}}$, $\partial=\rmd z\partial_z$, the Beltrami equation (\ref{eq:w-gen3}) takes on a manifestly conformally-invariant meaning:
$$
(\bar{\partial}-\mu)w=0.
$$
Since the quantity $\bar{\partial}-\mu$ is invariant under holomorphic reparametrisations, $z\mapsto z'(z)$ and $\bar{z}\mapsto \bar{z}'(\bar{z})$, the quantity $w(z,\bar{z})$ must transform as a {\it scalar} under such reparametrisations,
\begin{equation}\label{eq:w'=w hol z}
\boxed{w'(z',\bar{z}')=w(z,\bar{z}),\qquad {\rm with}\qquad z'=z'(z),\quad{\rm and}\quad \bar{z}'=\bar{z}'(\bar{z})}
\end{equation}
Clearly, the set of holomorphic reparametrisations, $z\mapsto z'(z)$, corresponds to only a small subset of the full set of reparametrisations, $\sigma\mapsto \sigma'(\sigma)$. In fact one can make a stronger statement. 
\sk

Under a general variation the Beltrami equation (\ref{eq:w-gen3}) reads,
$$
\delta(\partial_{\bar{z}}w)-(\delta\mu_{\bar{z}}^{\phantom{a}z})\partial_zw-\mu_{\bar{z}}^{\phantom{a}z}\delta(\partial_zw)=0,
$$
which is evaluated at a point $z,\bar{z}$. 
Under {\it arbitrary} reparametrisations (\ref{eq:z'=z+v(zzbar)}) in particular, making use of the explicit expression for $\delta\mu_{\bar{z}}^{\phantom{a}z}$ given in (\ref{eq:deltamumubarrho}), the analogue of the statement (\ref{eq:w's'=ws}) is that $w(z,\bar{z})$ transforms as a {\it scalar},
\begin{equation}\label{eq:w'=w  z,zbar}
w'(z',\bar{z}')=w(z,\bar{z}),\qquad {\rm with}\qquad z'=z'(z,\bar{z}),\quad{\rm and}\quad \bar{z}'=\bar{z}'(z,\bar{z})
\end{equation}
Clearly, holomorphic reparametrisations, $z'=z'(z)$, as given in (\ref{eq:w'=w hol z})  are a subset of general reparametrisations, $z'=z'(z,\bar{z})$. 

\subsubsection{Holomorphic Transition Functions and Cocycle Relations}\label{sec:TFCR}
We now generalise the local construction of the previous subsections to a global construction. 
\sk

\begin{figure}
\begin{center}
\includegraphics[angle=0,origin=c,width=0.8\textwidth]{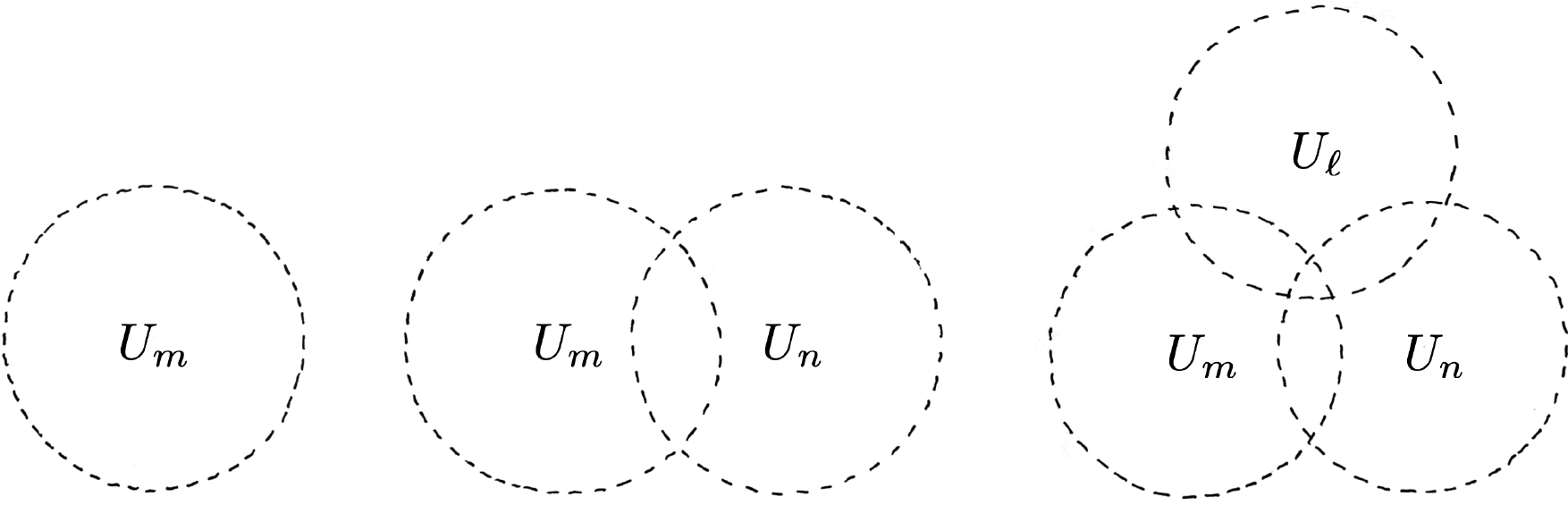}
\caption{Open sets, $U_m,U_n,U_\ell$, with corresponding double and triple patch intersections associated to charts $(U_m,z_m)$, $(U_n,z_n)$ and $(U_\ell,z_\ell)$. The transition functions (\ref{eq:transitionfuncs}) are only required to be satisfied on chart overlaps, such as $U_m\cap U_n$, and the cocycle relations (\ref{eq:cocycle}) are only required to be satisfied on triple overlaps, such as the central region of the third figure associated to $U_m\cap U_n\cap U_\ell$.}\label{fig:23overlaps}
\end{center}
\end{figure}

Let us consider an oriented genus-$\g$ closed Riemann surface,\footnote{We will sometimes write  $\Sigma_\g$ for a closed oriented genus-$\g$ Riemann surface (with unspecified number of punctures), $\Sigma_{\g,\n}$ for the corresponding genus-$\g$ surface with $\n$ punctures, and $\Sigma$ when we do not wish to specify either of the numbers $\g,\n$. In the presence of $\B$ (codimension-1) boundary components we might add a subscript `$\B$', e.g., $\Sigma_{\g,\n,\B}$.}  $\Sigma$, with fixed complex structure. Since any such surface is a complex manifold it may be fully specified by a collection of overlapping charts $\{(U_m,z_m)\}$ (such that $\mathscr{U}=\bigcup_m U_m$ is a cover for $\Sigma$) with (e.g., centred) coordinates that are one-to-one maps, $z_m:U_m\rightarrow \mathbf{C}$, with the special property that for every non-empty intersection $U_m\cap U_n$ the transition functions 
\begin{equation}\label{eq:transitionfuncs}
\boxed{f_{mn}\equiv z_m\circ z_n^{-1},\qquad {\rm or}\qquad z_m=f_{mn}(z_n)}
\end{equation}
are {\it biholomorphic} and therefore also invertible (on $U_m\cap U_n$). (Any one of these coordinates, $z_m$, can be identified with $w$ or $z$ of the previous section depending on whether we are in the $J$ or $I$ complex structure respectively, so this discussion is general.) The transition function $f_{mn}$ maps the chart associated to the open set $U_m\cap U_n\subset U_n$ to the chart associated to the open set $U_n\cap U_m\subset U_m$, the interpretation being that $z_m$ and $z_n$ represent the {\it same point} in $\Sigma$ if they are related as in (\ref{eq:transitionfuncs}) on $U_m\cap U_n$. See the first and second diagrams in Fig.~\ref{fig:23overlaps}. 
Note that $f_{mn}$ are holomorphic in $z_n$ but are not necessarily holomorphic in any remaining parameters that characterise the complex structure of $\Sigma$ (more about which later) \cite{Nelson89,LaNelson90}.\footnote{In the presence of boundaries, $\partial\Sigma$, \cite{Milnor} the essential modification is that in addition to the above there are also chart coordinates that map to the upper-half plane, $z_m:U_m\rightarrow \mathbf{H}$, and  boundaries are then mapped to the real line. The usual doubling trick \cite{Polchinski_v1} can then be adopted for local operators.} Having said that, it is possible to choose the $f_{mn}$ to be holomorphic in both $z_n$ and in all remaining parameters (or moduli) (on patch overlaps), and although this might seem desirable one must keep in mind that in this formulation one usually encounters Wu-Yang terms \cite{Polchinski87,Nelson89} when integrating by parts in moduli space (see Sec.~\ref{sec:WuYang} and the related discussion in Sec.~\ref{sec:comm dA-dA}), so that one cannot apply Stoke's theorem without first decomposing the surface into charts and then adding up the contributions from patch overlaps. If one does not adopt a slice where the  $f_{mn}$ are holomorphic in moduli one can find a parametrisation where the information contained in Wu-Yang terms is localised, so that integrals of total derivatives really do only get contributions from the boundary of moduli space \cite{Polchinski88}. (Again, this subtle point is discussed explicitly in Sec.~\ref{sec:WuYang}, see also Sec.~\ref{sec:comm dA-dA}.)
\sk

In general, the transition functions, $f_{mn}(z_n)$, being holomorphic in $z_n$ have a convergent and invertible power series expansion in $z_n$ on the relevant chart overlap, $U_m\cap U_n$.
\sk

\begin{figure}
\begin{center}
\includegraphics[angle=0,origin=c,width=1\textwidth]{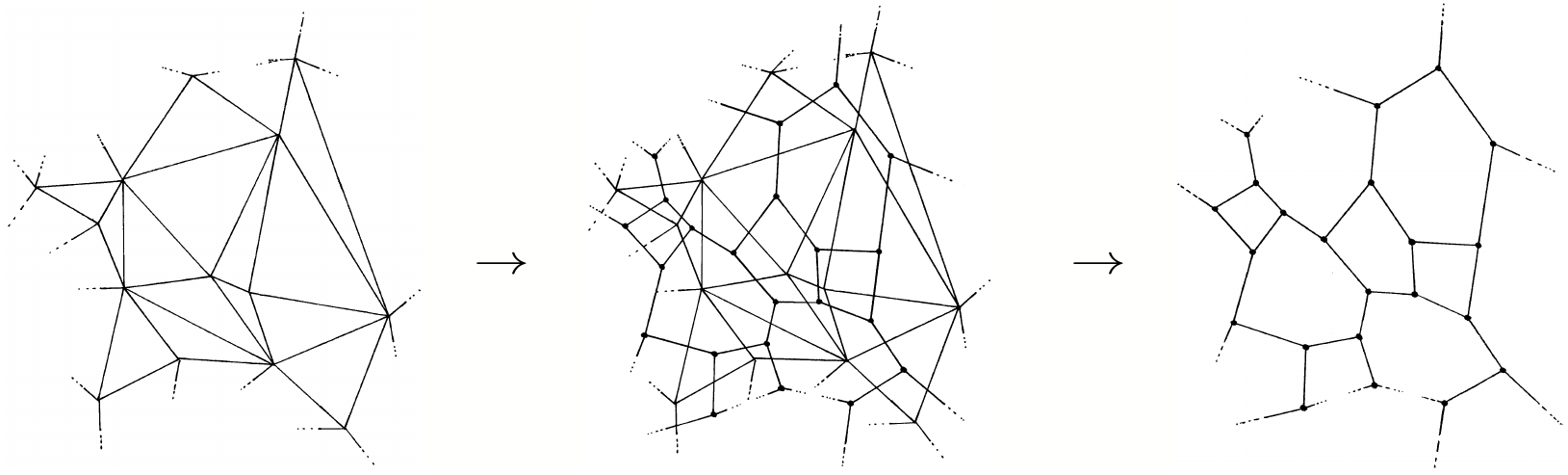}
\caption{
Given a generic triangulation (left figure) of a Riemann surface with generic $n$-point vertices ($n=3,4,\dots$) there is always a dual triangulation (right figure) where only 3-point vertices  appear. The dual triangulation is obtained (middle figure) from a generic triangulation by connecting nearest neighbour ``centre of masses'' (denoted by dots  `\,{\boldmath$\cdot$}') of every triangle (face) of the generic triangulation. One can then work with the dual triangulation and a corresponding cover $\mathscr{U}$, such that every double overlap, $U_m\cap U_n$, encloses an edge of the dual triangulation and every triple overlap, $U_m\cap U_n\cap U_\ell$, contains a 3-point vertex (`\,{\boldmath$\cdot$}') of the dual triangulation. Higher patch intersections do not appear in the dual triangulation (in 2 dimensions). The relation between the present figure and Fig.~\ref{fig:23overlaps} is made manifest in Fig.~\ref{fig:genericdualtriangles}.
}\label{fig:dualtriangulation}
\end{center}
\end{figure}

\begin{figure}
\begin{center}
\includegraphics[angle=0,origin=c,width=0.67\textwidth]{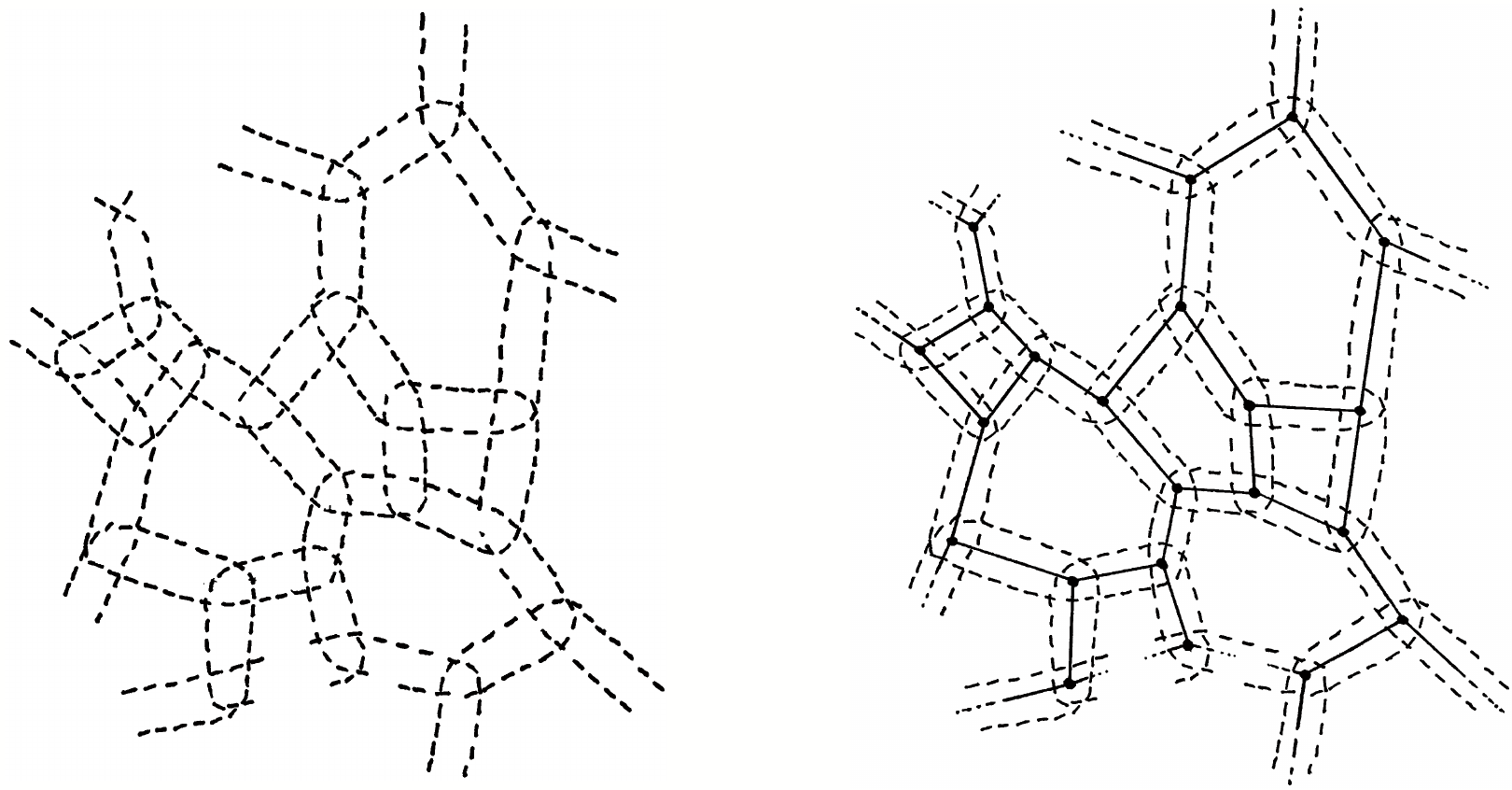}
\caption{Given a dual triangulation, such as that depicted in the third diagram of Fig.~\ref{fig:dualtriangulation}, one can construct an open cover, $\mathscr{U}=\{U_m\}$, of $\Sigma$ (left diagram), such that there is a one-to-one correspondence between every open set, $U_m$, (indicated by dashed lines) and a corresponding polygon, $V_m\subset U_m$, of the dual the triangulation (indicated by solid lines in the right diagram). The set of all polygons (which need not have ``straight'' edges) will throughout be denoted by $\mathscr{V}=\{V_m\}$. We also say that the set $\mathscr{V}=\{V_m\}$ provides a `cell decomposition' of $\Sigma$.}\label{fig:genericdualtriangles}
\end{center}
\end{figure}

To get a global description we must also satisfy {\it cocycle relations}. For example, on triple overlaps, $U_m\cap U_n\cap U_\ell\neq \zero$ (see the third diagram in Fig.~\ref{fig:23overlaps}), these take the form:
\begin{equation}\label{eq:cocycle}
\boxed{f_{mn}\circ f_{n\ell}=f_{m\ell},\qquad {\rm or}\qquad f_{mn}(f_{n\ell}(z_\ell))=f_{m\ell}(z_\ell)}
\end{equation}
satisfied for all labels $m,n,\ell$ of the cover. 
These cocycle relations provide (from a complex manifold viewpoint) the fundamental link between local and global data on a Riemann surface. We use the word `the' rather than `a' in the previous sentence because for a given complex structure the set of transition functions (\ref{eq:transitionfuncs}) supplemented by cocycle relations (\ref{eq:cocycle}) completely define the Riemann surface. 
In particular, depending on our choice of cover, there may also be cocycle relations associated to higher intersections, such as $f_{mn}\circ f_{n\ell}\circ f_{\ell k}=f_{mk}$ on $U_m\cap U_n\cap U_\ell \cap U_k\neq \zero$, etc., but actually one can always start from an arbitrary triangulation\footnote{That there always exists a triangulation on a general Riemann surface is known as {\it Rado's theorem} \cite{AhlforsSario}.}  of the surface (left diagram in Fig.~\ref{fig:dualtriangulation}), and then work on the dual cover (and corresponding dual triangulation, see the rightmost diagram in Fig.~\ref{fig:dualtriangulation}), where only triple patch intersections are encountered, so (\ref{eq:transitionfuncs}) and (\ref{eq:cocycle}) are not only necessary (if we want to think in terms of complex manifolds) but sufficient. 
This follows from the fact that Riemann surfaces are two-dimensional and does not hold (generically) for the corresponding moduli space of Riemann surfaces \cite{SenWitten15}.
Given such a dual triangulation, one can then associate open sets, $\{U_m\}$, (indicated by dashed lines in Fig.~\ref{fig:genericdualtriangles}) that enclose every simple polygon of the triangulation. 
\sk

From the cocycle condition (\ref{eq:cocycle}) one can deduce a number of properties. Suffice it for now to mention the following: taking $m=n=\ell$ in (\ref{eq:cocycle}) gives $f_{mm}\circ f_{mm}=f_{mm}$, and so $f_{mm}$ is the identity map,
\begin{equation}
f_{mm}=1,\qquad {\rm or}\qquad f_{mm}(z_m)=z_m.
\end{equation}
Secondly, given the map $f_{mn}$ is holomorphic (in $z_n$) it will also be invertible on $U_m\cap U_n$ provided $f_{mn}'(z_n)$ does not vanish on $U_m\cap U_n$. We will assume this condition is satisfied so that transition functions are invertible on the relevant chart overlaps. Defining the inverse, $f_{mn}^{-1}$, by $f_{mn}\circ f_{mn}^{-1}\equiv 1$ we can set $m=\ell$ in (\ref{eq:cocycle}) to find,
\begin{equation}\label{eq:fmnfnm=1}
f_{mn}\circ f_{nm}=1,\qquad{\rm or}\qquad f_{mn}(f_{nm}(z_m))=z_m,
\end{equation}
from which it follows that $f_{nm}^{-1}=f_{mn}$  (on $U_m\cap U_n\neq\zero$). 
\sk

Since we have singled out a particular set of charts (i.e.~an `atlas'), $\{(U_m,z_m)\}$, in order for the construction to be meaningful it is desirable to introduce an equivalence relation which is associated to a holomorphic change of charts, e.g.,
\begin{equation}\label{eq:wgz_mnot}
z_m\mapsto w_m=g_m(z_m),
\end{equation}
where the coordinates $z_m$ and $w_m$ map the same open set $U_m\subset \Sigma$ to different open subsets of $\mathbf{C}$ (or possibly $\mathbf{H}$, but this distinction will not be elaborated on here). In the $(U_m,w_m)$ chart suppose the transition functions are denoted by $h_{mn}\dfn w_m\circ w_n^{-1}$, so that the corresponding cocycle conditions on triple overlaps, $U_m\cap U_n\cap U_\ell\neq\zero$, take the form $h_{mn}\circ h_{n\ell}=h_{m\ell}$. So the new coordinate charts, $\{(U_m,w_m)\}$, and associated transition functions, $\{h_{mn}\}$, must satisfy:
\begin{equation}\label{eq:whmnnot}
\begin{aligned}
&w_m=h_{mn}(w_n)\\
&h_{m\ell}(w_\ell)=h_{mn}(h_{n\ell}(w_\ell)),
\end{aligned}
\end{equation}
for all $m,n,\ell$ labels of the cover, 
in precise analogy to (\ref{eq:transitionfuncs}) and (\ref{eq:cocycle}). In terms of transition functions, the required equivalence relation is furnished by:
\begin{equation}\label{eq:f'=g-1hg}
f'_{mn}(z_n) = g'_m(z_m)^{-1}h_{mn}'(w_n)g'_n(z_n),
\end{equation}
where primes denote derivatives with respect to the explicit arguments. The relation (\ref{eq:f'=g-1hg}) follows from substituting (\ref{eq:wgz_mnot}) into $w_m=h_{mn}(w_n)$ (for both $w_m,w_n$) and then substituting into the resulting relation on the left-hand side the transition function $z_m=f_{mn}(z_n)$ so that the same argument, $z_n$, is made manifest on both sides of the equation. Finally, one differentiates this with respect to $z_n$ and makes use of the above definitions to arrive at (\ref{eq:f'=g-1hg}). Notice that $f'_{mn}$ are the transition functions for the canonical bundle, $K\rightarrow \Sigma$, whose local sections are patched together via $\rmd z_m={f'}_{mn}\rmd z_n$ (with ${f'}_{mn}^{-1}=f'_{nm}$) subject to equivalence relations generated by (\ref{eq:f'=g-1hg}) and cocycle relations (\ref{eq:cocycle}). 
\sk

A choice of charts, the associated transition functions and cocycle relations (subject to the above equivalence relation) provide a {\it global} definition of all closed and oriented {\it Riemann surfaces}. Notice also that this definition does not require the introduction of a {\it Riemannian metric} (Sec.~\ref{sec:MVa}), even though (as we discuss in further detail below) it is convenient to introduce this concept in various contexts. Jumping ahead somewhat, we will see that it is convenient (but not necessary) to introduce a worldsheet metric to fix the residual invariance of conformal gauge (up to U(1)). This will prove convenient when we construct the map from fixed- to integrated-picture vertex operators\footnote{This includes offshell vertex operators arising from cutting open handles of the Riemann surface as well as external BRST-invariant or BRST-non-invariant vertex operators.}, where we will pick a gauge slice by a specific choice of Beltrami differential. But we will also phrase the resulting gauge slice in terms of holomorphic transition functions. This choice of gauge slice will also enable us to write integrated vertex operators in terms of covariant quantities that will automatically be globally-defined on $\Sigma$ or $\mathcal{M}$ (modulo U(1)). 

\subsection{Complex Structure Deformations: I}\label{sec:CSDI}
Having defined the essential notions associated to Riemann surfaces at fixed complex structure, we next focus on infinitesimal complex structure deformations or moduli deformations of $\Sigma$ (we use the two terms interchangeably), see e.g.~\cite{Kodaira,Polchinski_v1,VerlindeHphd,Witten12b}. The focus in particular will be on the interplay between local and global (on $\Sigma$ and on moduli space) data. In Sec.~\ref{sec:CSDa} we begin with a local discussion of complex structure deformations, in particular the defining equation for a holomorphic coordinate where the complex structure is explicit. In Sec.~\ref{sec:ITOCR} we connect to the global approach to complex structure deformations via \v{C}ech cohomology which follows directly from the defining properties of a Riemann surface (as laid out in the previous subsection). In Sec.~\ref{sec:CSDII} we present further complementary viewpoints, partly to exhibit the overall coherence of the subject and partly because complementary viewpoints will provide ``intellectual geodesics'' between technical tools we will be needing for explicit computation. In Sec.~\ref{sec:HNC} we make an explicit choice of coordinates (holomorphic normal coordinates) that are globally well-defined modulo U(1) and that will be used to translate points (and corresponding frames) across Riemann surfaces.  

\subsubsection{In Terms of the Beltrami Equation}\label{sec:CSDa}

We begin with a preliminary and {\it local} discussion of complex structure deformations based on the formalism we have developed in Sec.~\ref{sec:CST}, beginning from the defining equation of a holomorphic coordinate in a given fixed complex structure.
\sk

Let us in particular consider two holomorphic coordinates, $z(\sigma)$ and $w(\sigma)$, with respect to complex structures $I(\sigma)$ and $J(\sigma)$ respectively, with $\sigma^a$, $a=1,2$, an auxiliary real coordinate. From the defining equation (\ref{eq:w-gen}) these must satisfy:
\begin{equation}\label{eq:zw-IJ}
\begin{aligned}
\big(\partial_a+iI_a^{\phantom{a}b}(\sigma)\partial_b\big)z(\sigma)=0,\qquad {\rm and}\qquad
\big(\partial_a+iJ_a^{\phantom{a}b}(\sigma)\partial_b\big)w(\sigma)=0.
\end{aligned}
\end{equation}
Recall that these are essentially Beltrami equations (even though no reference is usually made to real coordinates in a Beltrami equation), and we shall refer to them as such.
\sk

Let us suppose now that the $J(\sigma)$ complex structure is infinitesimally close to the $I(\sigma)$ complex structure. That is, considering the underlying real coordinates, $\sigma^a$, {\it fixed} under small deformations, we can quite generally write:
\begin{equation}\label{eq:zw-IJinf}
\begin{aligned}
J_a^{\phantom{a}b}(\sigma)=I_a^{\phantom{a}b}(\sigma)+\delta I_a^{\phantom{a}b}(\sigma),\qquad {\rm and}\qquad
w(\sigma) = z(\sigma)+\delta z(\sigma),
\end{aligned}
\end{equation}
for the corresponding deformations. The latter relation relies only on a mild assumption of continuity, that small variations in complex structure induce small variations of the corresponding coordinate. Since both complex structure matrices satisfy the defining property, $J^2=I^2=-\mathds{1}$, we learn from the first relation in (\ref{eq:zw-IJinf}) that,
\begin{equation}\label{eq:IdI+dII=0}
I_a^{\phantom{a}b}\delta I_b^{\phantom{a}c}+\delta I_a^{\phantom{a}b}I_b^{\phantom{a}c}=0.
\end{equation}

Let us use these relations to compute the variation $\delta z(\sigma)$ induced by $\delta I_a^{\phantom{a}b}(\sigma)$. Substituting (\ref{eq:zw-IJinf}) into the second relation in (\ref{eq:zw-IJ}) and keeping terms up to linear order in the variations yields,
\begin{equation}\label{eq:csd-s}
\big(\partial_a+iI_a^{\phantom{a}b}(\sigma)\partial_b\big)\delta z(\sigma)+i\delta I_a^{\phantom{a}b}(\sigma)\partial_bz(\sigma)=0,
\end{equation}
where we also made use of the defining equation (\ref{eq:zw-IJ}) for the $z$ holomorphic coordinate. 
\sk

So far we have not made any assumption about which complex structure $I(\sigma)$ we are perturbing around in (\ref{eq:zw-IJinf}), other than the fact that $z$ is holomorphic with respect to the $I(\sigma)$ complex structure. Suppose now that the $I(\sigma)$ complex structure is `natural' with respect to the coordinates, $\sigma^a$, as defined in Sec.~\ref{sec:NACS} and in particular (\ref{eq:Imatrix}). With this choice, the consistency condition (\ref{eq:IdI+dII=0}) takes the form,
$$
\delta I_1^{\phantom{a}2}=\delta I_2^{\phantom{a}1},\qquad {\rm and}\qquad \delta I_1^{\phantom{a}1}=-\delta I_2^{\phantom{a}2},
$$
exposing that two of the four real components, $\delta I_a^{\phantom{a}b}$, are independent (which is true not only for the variations but also for the full complex structure tensor). 
Furthermore, making use of the fact that in the natural complex structure, $I$, we have the relations (\ref{eq:z=s1+is2}) and (\ref{eq:Imatrix}) we can rewrite the differential equation  (\ref{eq:csd-s}) in terms of $z,\bar{z}$ derivatives,
\begin{equation}\label{eq:dzbar dz = dI}
\boxed{\partial_{\bar{z}}\delta z(\sigma) = \frac{1}{2}\big(\delta I_1^{\phantom{a}2}+i\delta I_2^{\phantom{a}2}\big)(\sigma)}
\end{equation}
This equation may be viewed as a differential equation for the variation, $\delta z(\sigma)$, given a complex structure variation, $\delta I_a^{\phantom{a}b}(\sigma)$, which may or may not include variations induced by reparametrisations (\ref{eq:dJ-rep}). 
The relation (\ref{eq:dzbar dz = dI}) (and its various guises) will play a prominent role throughout the document.
\sk

Incidentally, for the reader that is wondering whether {\it infinitesimal} variations of complex structure (as in (\ref{eq:dzbar dz = dI})) are sufficient let us point out, if only briefly at this stage, that we are aiming to be able to compute the path integral measure associated to a given gauge slice of our choice in moduli space. Since this in turn amounts to computing a Jacobian for a change of coordinates, and since furthermore a Jacobian for the base space is the {\it same} as the Jacobian for the tangent space (see, e.g., \cite{DHokerPhong} for a string theory realisation of this) we can compute this Jacobian on the tangent space. The Jacobian on the tangent space is in turn entirely determined by an infinitesimal variation, and this is why infinitesimal variations of complex structure as in (\ref{eq:dzbar dz = dI}) are actually sufficient provided the moduli space integrand that we extract is globally well-defined. Furthermore, we are making use of the fact that {\it every} complex structure has a holomorphic coordinate with respect to which the complex structure is `natural', so that it is sufficient not only to consider infinitesimal variations but also to consider infinitesimal variations around the natural complex structure in particular.
\sk

We emphasise that $\partial_{\bar{z}}$ is a derivative with respect to the anti-holomorphic coordinate, $\bar{z}$, of the natural complex structure, $-I$, and that in this complex structure we have the relation, $\bar{z}=\sigma^1-i\sigma^2$. So (by a slight abuse of notation) we can therefore regard $\delta z(\sigma)$ and $\delta I_a^{\phantom{a}b}(\sigma)$ to be functions of $z,\bar{z}$, in particular,
\begin{equation}\label{eq:dzbar dz = dI zzbar}
\partial_{\bar{z}}\delta z(z,\bar{z}) = \frac{1}{2}\big(\delta I_1^{\phantom{a}2}+i\delta I_2^{\phantom{a}2}\big)(z,\bar{z})
\end{equation}

Clearly, when $\delta I_a^{\phantom{a}b}(\sigma)\neq0$ the deformation $\delta z(z,\bar{z})$ will not be holomorphic with respect to the $I$ complex structure holomorphic coordinate $z$, {\it even} if the variation was induced by a coordinate reparametrisation (\ref{eq:dJ-rep}). But the quantity $z(\sigma)+\delta z(\sigma)$ {\it is} a holomorphic coordinate with respect to the $J$ complex structure since it solves the second defining differential equation in (\ref{eq:zw-IJ}).
\sk

To also make further contact with much of the string theory literature \cite{Friedan82,DHokerPhong,Witten12c}, let us consider again two complex structures $I$ and $J$ that differ by a small variation as in (\ref{eq:zw-IJinf}), namely,
$$
J=I+\delta I,
$$ 
where the $I$ complex structure is the natural one with respect to the coordinates, $\sigma^a$, as defined in Sec.~\ref{sec:NACS} and in particular (\ref{eq:Imatrix}).  
Then, to {\it linear order} in $\mu$ we can extract an explicit expression for $\delta I$ in terms of $\mu,\bar{\mu}$ from (\ref{eq:Jmatrix-mu}),
\begin{equation}\label{eq:delta Jmatrix-mu}
   \delta I= \bigg(\begin{matrix}
      i&1\\
     1& -i\\
   \end{matrix}\bigg)\mu_{\bar{z}}^{\phantom{a}z}
   +
   \bigg(\begin{matrix} 
      -i&1\\
  1&i\\
   \end{matrix}\bigg)\mu_{z}^{\phantom{a}\bar{z}},
\end{equation}
and therefore in particular,
\begin{equation}\label{eq:mu=deltaI..}
\boxed{\mu_{\bar{z}}^{\phantom{a}z}=\frac{1}{2}\big(\delta I_1^{\phantom{a}2}+i\delta I_2^{\phantom{a}2}\big),\qquad {\rm and}\qquad \mu_{z}^{\phantom{a}\bar{z}}=\frac{1}{2}\big(\delta I_1^{\phantom{a}2}-i\delta I_2^{\phantom{a}2}\big)}
\end{equation}
So substituting this into (\ref{eq:dzbar dz = dI}) we learn that generic {\it small} deformations of complex structure $\delta I$ (and in particular $\mu_{\bar{z}}^{\phantom{a}z}$ via (\ref{eq:mu=deltaI..})) which preserve the complex structure $J$ can be associated to corresponding small deformations, $\delta z(\sigma)$, 
\begin{equation}\label{eq:dzbar dz = dI2}
\boxed{\partial_{\bar{z}}\delta z(\sigma) = \mu_{\bar{z}}^{\phantom{a}z}(\sigma)}
\end{equation}

\subsubsection{In Terms of Cocycle Relations}\label{sec:ITOCR}
We will in this section follow on from the discussion in Sec.~\ref{sec:NACS}, and continue to think of a {\it complex structure} on $\Sigma$ as an equivalence class of systems of local holomorphic coordinates on $\Sigma$ (recall in particular the last paragraph in Sec.~\ref{sec:NACS}). But we will take a slightly different viewpoint that will be much more closely based on the global construction of Sec.~\ref{sec:TFCR} whereby we will phrase things in terms of transition functions and cocycle relations. In particular, we will now regard a {\it complex structure deformation} of $\Sigma$ as the gluing of the {\it same} charts, $\{(U_m,z_m)\}$, via {\it different} identifications, $\{f_{mn}\}$. This is the approach pioneered (for general complex manifolds) by Kodaira and Spencer, see \cite{Kodaira} for a review, which is particularly efficient for certain applications. A major advantage of this approach is that it enables one to discuss deformations locally on $\Sigma$ keeping the remaining Riemann surface fixed, while also providing an explicit link to global data. By global data here we generically mean global on $\Sigma$. To be more precise, for a subset of complex structure deformations such as those associated to translating pinches across the surface (which might also arise from cutting open handles and using the operator-state correspondence to map the resulting states to local operators) the construction we develop will also be globally well-defined in {\it moduli space}.\footnote{In more generic cases the basic obstruction to integrating infinitesimal complex structure deformations to finite deformations is concisely discussed in \cite{Huybrechts}.} From this viewpoint the transition functions defined in (\ref{eq:transitionfuncs}) subject to (\ref{eq:cocycle}) (and perhaps higher cocycle relations associated to higher-point intersections, such as $f_{mn}\circ f_{n\ell}\circ f_{\ell k}\circ f_{km}=1$ when\footnote{We are using notation introduced in (\ref{eq:Umnellk-notation}) here.} $U_{mn\ell k}\neq\zero$, etc.) can be considered \cite{Kodaira,Polchinski_v1} to contain all moduli dependence. 
\sk

Pursuing this reasoning further, let us denote by $t=(t^1,t^2,\dots)$ a complete set of real or complex local coordinates on moduli space, $\mathcal{M}$, in which case the transition functions (\ref{eq:transitionfuncs}) may be written in more detail as follows,
\begin{equation}\label{eq:zmfmnzn}
z_m=f_{mn}(z_n,t).
\end{equation}
Note that $f_{mn}$ (for a fixed set of chart labels, $m,n$) may depend on only a subset of moduli or even no moduli at all depending on context. 
The consistency (cocycle) condition on triple overlaps (\ref{eq:cocycle}) demands that:
\begin{equation}\label{eq:coc}
f_{m\ell}(z_\ell,t)=f_{mn}(f_{n\ell}(z_\ell,t),t),
\end{equation}
and recall that $f_{\ell m}^{-1}=f_{m\ell}$. Following Kodaira and Spencer, see e.g.~\cite{Kodaira}, form the vector field:
\begin{equation}\label{eq:d/dt-dfn}
\frac{\partial}{\partial t}\equiv \sum_kc^k\frac{\partial}{\partial t^k},
\end{equation}
with unspecified components, $c^k$, 
and consider the derivation $\partial_tf_{m\ell}(z_\ell,t)$ (with $z_\ell$ held fixed) using (\ref{eq:coc}), multiply the resulting expression from the right by basis vectors, $\partial/\partial z_m$, and use the chain rule (to write $\frac{\partial z_m}{\partial z_n}|_{z_\ell}\frac{\partial }{\partial z_m}=\frac{\partial }{\partial z_n}$). The cocycle relation (\ref{eq:coc}) takes the form:
\begin{equation}\label{eq:coc1}
\begin{aligned}
\frac{\partial f_{m\ell}(z_\ell,t)}{\partial t}\Big|_{z_\ell}\frac{\partial }{\partial z_m}
&=\frac{\partial f_{n\ell}(z_\ell,t)}{\partial t}\Big\vert_{z_\ell}\frac{\partial }{\partial z_n}+\frac{\partial f_{mn}(z_n,t)}{\partial t}\Big\vert_{z_n}\frac{\partial }{\partial z_m},
\end{aligned}
\end{equation}
which leads us to define the following {\it local vector fields} on $\Sigma$, 
\begin{equation}\label{eq:phimn(t)}
\boxed{\phi_{mn}(t)\dfn \frac{\partial f_{mn}(z_n,t)}{\partial t}\Big\vert_{z_n}\frac{\partial }{\partial z_m}}
\end{equation}
in terms of which the cocycle condition reads, 
$
\phi_{n\ell}(t)-\phi_{m\ell}(t)+\phi_{mn}(t)=0.
$ 
Now since by definition $f_{mm}(z_m,t)=z_m$, which does not have any explicit dependence on $t$, differentiating with respect to $t$ keeping $z_m$ fixed we find that $\phi_{mm}(t)=0$. From the cocycle condition\footnote{Or by differentiating $f_{nm}(f_{mn}(z_n,t),t)=z_n$ with respect to $t$ at fixed $z_n$ and using the chain rule.} we then learn that (setting $\ell=m$), $\phi_{nm}(t)=-\phi_{mn}(t)$, and in particular taking also (\ref{eq:coc1}),  (\ref{eq:phimn(t)}) into account we arrive at \cite{Kodaira,Witten12b}:
\begin{equation}\label{eq:cocycles}
\boxed{
\begin{aligned}
&\phi_{n\ell}(t)+\phi_{\ell m}(t)+\phi_{mn}(t)=0\\
&\phi_{mn}(t)+\phi_{nm}(t)=0
\end{aligned},
\qquad {\rm on}\qquad U_m\cap U_n\cap U_\ell(\neq\zero)
}
\end{equation}
These are precisely \cite{Kodaira,Chern} the defining conditions for a 1-{\it cocycle}, and in particular $\{\phi_{mn}(t)\}$ is a 1-cocycle on $\Sigma$ with values in the sheaf $\mathcal{S}_t$ of holomorphic vector fields over $\Sigma$ associated to the deformation $t$. The set of 1-cocycles, $\{\phi_{mn}(t)\}$, subject to (\ref{eq:cocycles}) is conventionally denoted by $\mathcal{Z}^1(\Sigma,\mathcal{S}_t)$. 
\sk

If we were not working with a dual triangulation (see above) we would also have higher-point intersections, e.g.~4-point intersections, $U_{mn\ell k}$. The corresponding constraint arises from the cocycle condition, $f_{mn}\circ f_{n\ell}\circ f_{\ell k}\circ f_{km}=1$, and following precisely similar reasoning that led to (\ref{eq:cocycles}) yields:
\begin{equation}\label{eq:cocycles-q}
\begin{aligned}
&\phi_{n\ell}(t)+\phi_{\ell m}(t)+\phi_{mk}(t)+\phi_{kn}(t)=0
\end{aligned},
\qquad {\rm on}\qquad U_m\cap U_n\cap U_\ell\cap U_k(\neq\zero)
\end{equation}
with obvious generalisations for 5- or higher-point intersections. But notice that (\ref{eq:cocycles}) continues to hold on the triple intersection $U_{mn\ell}\subset U_{mn\ell k}$  regardless of the fact that there is an additional open set $U_k$, and so in particular $\phi_{n\ell}(t)+\phi_{\ell m}(t)=\phi_{nm}(t)$ and (\ref{eq:cocycles-q}) reduces to the original cocycle condition (\ref{eq:cocycles}). So higher-point intersections do not provide additional constraints that are not already contained in (\ref{eq:cocycles}). This observation is inline with the above discussion as to why working with dual triangulations is sufficient (recall Fig.~\ref{fig:dualtriangulation}), and will also be relevant below when we show that our expression for the path integral measure is independent of the cover $\mathscr{U}$. 

\subsubsection{Moduli Space and Equivalence Relations}\label{sec:MSER}
In order to associate such vector fields, $\phi_{mn}(t)$, with true moduli deformations we must also discuss how the $\{\phi_{mn}(t)\}$ change under holomorphic changes of coordinates, since any such change preserves complex structure. 
Let,
\begin{equation}\label{eq:wgz_m}
z_m\mapsto w_m=g_m(z_m,t),
\end{equation}
be such a holomorphic (in $z_m$) change of coordinates. The coordinates $z_m$ and $w_m$ map the {\it same} open set $U_m\subset \Sigma$ to different open subsets of $\mathbf{C}$. 
In terms of the collection $\{(U_m,w_m)\}$ of charts let the transition functions (analogous to (\ref{eq:transitionfuncs})) be denoted by $h_{mn}\dfn w_m\circ w_n^{-1}$, so that the corresponding cocycle conditions take the form $h_{mn}\circ h_{n\ell}=h_{m\ell}$ (analogous to (\ref{eq:cocycle})). That is, the new coordinate charts, $\{(U_m,w_m)\}$, and associated transition functions, $\{h_{mn}\}$, must satisfy:
\begin{equation}\label{eq:whmn}
\begin{aligned}
&w_m=h_{mn}(w_n,t)\\
&h_{m\ell}(w_\ell,t)=h_{mn}(h_{n\ell}(w_\ell,t),t).
\end{aligned}
\end{equation}
Given the relations (\ref{eq:whmn}), on chart overlaps, $U_m\cap U_n\neq\zero$, we can define vector fields $\varphi_{mn}(t)$ (analogous to $\phi_{mn}(t)$ above),
\begin{equation}\label{eq:varphimn(t)}
\varphi_{mn}(t)\dfn \frac{\partial h_{mn}(w_n,t)}{\partial t}\Big\vert_{w_n}\frac{\partial }{\partial w_m},
\end{equation}
and the same computation that led to (\ref{eq:cocycles}) yields:
\begin{equation}\label{eq:cocycles2}
\begin{aligned}
&\varphi_{n\ell}(t)+\varphi_{\ell m}(t)+\varphi_{mn}(t)=0\\
&\varphi_{mn}(t)+\varphi_{nm}(t)=0.
\end{aligned}
\end{equation}
That is, the set $\{\varphi_{mn}(t)\}$ also defines a 1-cocycle on $\Sigma$ with values in the sheaf $\mathcal{S}_t$ of holomorphic vector fields over $\Sigma$ associated to the deformation $t$. 
\sk

Now, from (\ref{eq:wgz_m}) and the first relation in (\ref{eq:whmn}), on chart overlaps $U_m\cap U_n\neq\zero$ we have $g_m(z_m,t)=h_{mn}(w_n,t)$, so that on account of (\ref{eq:zmfmnzn}) and (\ref{eq:wgz_m}):
\begin{equation}
\begin{aligned}
g_m(f_{mn}(z_n,t),t)&=h_{mn}(g_n(z_n,t),t).
\end{aligned}
\end{equation}
Differentiating this relation with respect to $t$ (keeping $z_n$ fixed), multiplying by the basis vector $\partial/\partial w_m$ from the right, using the chain rule (at fixed $t$), and making repeated use of the relations (\ref{eq:wgz_m}), (\ref{eq:whmn}) as necessary yields,
$$
\frac{\partial f_{mn}(z_n,t)}{\partial t}\Big|_{z_n}\frac{\partial}{\partial z_m}+\frac{\partial g_m(z_m,t)}{\partial t}\Big|_{z_m}\frac{\partial }{\partial w_m} = \frac{\partial g_n(z_n,t)}{\partial t}\Big|_{z_n}\frac{\partial }{\partial w_n}+\frac{\partial h_{mn}(w_n,t)}{\partial t}\Big|_{w_n}\frac{\partial }{\partial w_m}.
$$
If we define the vector:
\begin{equation}
\boxed{
\varphi_m(t)\dfn \frac{\partial g_m(z_m,t)}{\partial t}\Big|_{z_m}\frac{\partial }{\partial w_m}
}
\end{equation}
then from the preceding equation, and making use of the defining relations (\ref{eq:phimn(t)}) and (\ref{eq:varphimn(t)}),  we learn that under a holomorphic change of coordinates, $\{z_m,t\}\mapsto \{w_m,t\}$, on patch overlaps, $U_m\cap U_n$, the relation between the new holomorphic vector, $\varphi_{mn}(t)$, and the old, $\phi_{mn}(t)$, is:
\begin{equation}\label{eq:varphiphimn}
\boxed{
\phi_{mn}(t)=\varphi_{mn}(t)-\varphi_m(t)+\varphi_n(t)
}
\end{equation}
Since holomorphic coordinate transformations do not change the conformal class, the complex structure is also unchanged, and we should therefore consider moduli deformations generated by vector fields $\phi_{mn}(t)$ and $\varphi_{mn}(t)$ as being equivalent if they are related as in (\ref{eq:varphiphimn}). The relation (\ref{eq:varphiphimn}) is precisely analogous to (\ref{eq:f'=g-1hg}) which was for fixed complex structure moduli.  In the language of \v{C}ech cohomology the quantity $\varphi_m(t)-\varphi_n(t)$ is referred to as a 1-coboundary (with $\varphi_m,\varphi_n$ holomorphic in $U_m\cap U_n$). So there is an equivalence relation, $\varphi_{mn}(t)\sim\phi_{mn}(t)$, in particular:
\begin{equation}\label{eq:varphiphimn-eq}
\boxed{\phi_{mn}(t)\sim\phi_{mn}(t)-\phi_m(t)+\phi_n(t)}
\end{equation}
by which we should mod out if we wish to associate $\phi_{mn}(t)$ to a complex structure deformation. The collection of 1-cocycles, $\{\phi_{mn}(t)\}$, satisfying (\ref{eq:cocycles}) modulo the equivalence relation (\ref{eq:varphiphimn-eq}) (i.e.~modulo 1-coboundaries, $\phi_m(t)-\phi_n(t)$) is associated to elements of the sheaf cohomology group $H^1(\Sigma,\mathcal{S}_t)$ associated to the complex structure deformation $t$. Conversely, there is a `stability theorem of complex structures'  \cite{FrolicherNijenhuis57} whose essential content is that if $H^1(\Sigma,\mathcal{S}_t)=0$ the complex structures, $J_t$, and $J_{t+\delta t}$ (for sufficiently small $\delta t$) are equivalent. If we denote by $T\mathcal{M}$ the tangent bundle to moduli space, $\mathcal{M}$, and $T\mathcal{M}|_{\Sigma}$ its fibre at the point in $\mathcal{M}$ corresponding to $\Sigma$ then we have the identification \cite{Witten12c}: 
\begin{equation}\label{eq:TM=H1}
\boxed{T\mathcal{M}|_{\Sigma}=H^1(\Sigma,\mathcal{S}_t)}
\end{equation}

If we wish to be slightly pedantic, what we have derived is actually $H^1(\mathscr{U},\mathcal{S}_t)$ since we have not shown that the result is independent of the cover (or, more precisely, $H^1(\mathscr{N}(\mathscr{U}),\mathcal{S}_t)$, where $\mathscr{N}(\mathscr{U})$ is the {\it nerve} of the cover $\mathscr{U}$ \cite{Chern}, but modulo some comments in the following paragraph we will not need to elaborate on this distinction for what follows).
\sk

A technical requirement that is necessary to be satisfied is that in order to transition from $H^1(\mathscr{U},\mathcal{S}_t)$ to $H^1(\Sigma,\mathcal{S}_t)$ we should choose a  {\it good} cover, $\mathscr{U}$, which allows us to take a direct limit \cite{BottTu} to pass from the cohomology of the cover $\mathscr{U}$ to that of the manifold, $\Sigma$. Recall that a good cover, $\mathscr{U}$, is one for which \cite{BottTu} every open set $U_m$ in $\mathscr{U}$ and every (non-empty finite) intersection:
\begin{equation}\label{eq:Umnellk-notation}
\begin{aligned}
U_{mn}&=U_m\cap U_n\\
U_{mn\ell}&=U_m\cap U_n\cap U_\ell\\
U_{mn\ell k}&=U_m\cap U_n\cap U_\ell\cap U_k\\
&\,\,\,\vdots
\end{aligned}
\end{equation}
is contractible (diffeomorphic to an open disc of $\mathbf{C}$). E.g., if our choice of cover, $\mathscr{U}$, is such that there is an element $U_{mn}$ that is disconnected then this choice will not be associated to a good cover. For this reason we need at least three open sets to construct a good cover for the circle, $S^1$. A second example is to cover $S^2$ by two overlapping open sets, $\mathscr{U}=\{U_1,U_2\}$. In this case although $U_1$ and $U_2$ (which cover, say, the northern and southern hemispheres respectively) are contractible their intersection, $U_{12}$, is not since this forms a non-contractible equatorial band. In fact, the minimum number of sets we need to construct a good cover for $S^2$ is four, which can be achieved, e.g., by replacing one of the aforementioned two open sets, say $U_2$, by three overlapping open sets such as those in the third diagram in Fig.~\ref{fig:23overlaps} and then glue these to $U_1$ across an equatorial band using the transition functions induced from $z_1z_2=1$. This leads to a good cover for $S^2$ (with cocycle relations satisfied) as can be checked by explicit calculation.

\subsection{Complex Structure Deformations: II}\label{sec:CSDII}
To proceed further we will now consider deformations of complex structure from a metric viewpoint. The defining properties of a Riemann surface make no reference to the presence or existence of a metric, but it is sometimes nevertheless useful to endow a Riemann surface with a metric \cite{Polchinski88}. That the notion of a metric is not required also leads to considerable freedom in a choice of metric (subject to the constraint imposed by the Gauss-Bonnet theorem). E.g., it is even possible to take the metric to be singular along certain contours \cite{GiddingsMartinec86,D'HokerGiddings87,Polchinski88}, and this may be accomplished by a change in coordinates using a discontinuous local vector field to deform complex structure keeping the metric fixed\footnote{The essential ingredient here is not so much the `discontinuity' {\it per se} but rather that the change of coordinates is not invertible, otherwise it would generate a diffeomorphism.}. Another standard example is to endow $S^2$ with a metric and use a Weyl transformation to push all curvature to infinity, $\{\infty\}$, allowing one to use a flat metric for the whole of $S^2$ except at a point. 
\sk

The main purpose of introducing a metric in the current document is related to the fact that conformal gauge has a residual gauge invariance associated to holomorphic reparametrisations. We can fix this invariance (up to a residual U(1)) by using the metric to specify a gauge slice, and the specific choice of interest will in turn be equivalent to working with holomorphic normal coordinates. This choice leads to a gauge slice that fixes invariance under Weyl transformations, but reparametrisation invariance is actually preserved and becomes manifest if one chooses to bring to plain view an underlying fixed auxiliary coordinate system. It is particularly important to specify such a globally well-defined slice in moduli space, since handle operators represent {\it offshell} bi-local operators that are therefore not conformally-invariant. So if we are to insert operators into the path integral that are not conformally invariant we better make sure that the breaking is spontaneous and that the slice that implements it is globally well-defined. This is further developed in Sec.~\ref{sec:HNC}. In this section we proceed without specifying a gauge slice. 

\subsubsection{Deforming via a Change of Metric with Fixed Coordinates}
Suppose we consider an atlas $\mathscr{U}_I=\{(U^I_m,z_m)\}$ with associated complex structure that we shall denote by $I$. Now consider a second complex structure, denote it by $J$, with atlas $\mathscr{U}_J=\{(U^J_m,w_m)\}$. The two complex structures $I$ and $J$ will eventually be taken to be infinitesimally close, $J=I+\delta I$. In either of the two complex structures there exist \cite{ChernIsothCoords55,BersRiemannSurfaces} conformal coordinates, and therefore there are corresponding metric tensors, $g_I$, $g_J$, that when restricted to charts $U_m^I$, $U_m^J$ (for all $m$) are conformally flat:\footnote{We are being slightly more explicit here than we were in Sec.~\ref{sec:MVa}. And we use the shorthand notation $\rmd z\rmd\bar{z}=\frac{1}{2}(\rmd z\otimes \rmd\bar{z}+\rmd\bar{z}\otimes \rmd z)$.}
\begin{equation}\label{eq:dsJ2 dsI2}
g_I\big|_{U_m^I}=2g_{z_m\bar{z}_m}^{I}\rmd z_m\rmd\bar{z}_m,\qquad{\rm and}\qquad g_J\big|_{U_m^J}=2g_{w_m\bar{w}_m}^{J}\rmd w_m\rmd\bar{w}_m.
\end{equation}
The reader may wish to recall the discussion in Sec.~\ref{sec:MVa}.
\sk

Let us determine the metric in the $J$ complex structure but in the $z_m$ coordinates \cite{D'HokerPhong15b}. To achieve this, holomorphic reparametrisations will not be sufficient (since in that case $I=J$ and the two metrics in (\ref{eq:dsJ2 dsI2}) are in the same conformal class) so let us perform the more general (called `quasi-conformal', a term coined by L.V.~Alhfors, see e.g.~\cite{Alhfors}) coordinate transformation $w_m\mapsto z_m(w_m,\bar{w}_m)$, 
\begin{equation}\label{eq:ds2Jwwbar}
\begin{aligned}
g_J\big|_{U_m^J}&=2g_{w_m\bar{w}_m}^{J}\rmd w_m\rmd\bar{w}_m\\
&=2g_{w_m\bar{w}_m}^{J}|\partial_{z_m}w_m|^2\big(\rmd z_m+\mu_{\bar{z}_m}^{\phantom{a}z_m} \rmd\bar{z}_m\big)\big(\rmd\bar{z}_m+\mu_{z_m}^{\phantom{a}\bar{z}_m} \rmd z_m\big)\\
&=2g_{w_m\bar{w}_m}^{J}|\partial_{z_m}w_m|^2\Big((1+\mu_{\bar{z}_m}^{\phantom{a}z_m}\mu_{z_m}^{\phantom{a}\bar{z}_m})\rmd z_m\rmd\bar{z}_m+\mu_{z_m}^{\phantom{a}\bar{z}_m}\rmd z_m^2+\mu_{\bar{z}_m}^{\phantom{a}z_m} \rmd\bar{z}_m^2\Big)\\
\end{aligned}
\end{equation}
The quantity $\mu_{\bar{z}_m}^{\phantom{a}z_m}$ is defined by the {\it Beltrami equation} \cite{BersRiemannSurfaces},
\begin{equation}\label{eq:beltramieqnJI}
\boxed{
\big(\partial_{\bar{z}_m}-\mu_{\bar{z}_m}^{\phantom{a}z_m} \partial_{z_m}\big)w_m=0
}
\end{equation}
whose main interpretation was discussed in Sec.~\ref{sec:MVa}, and here we introduce yet another related viewpoint.
\sk

The coordinate transformations satisfying Beltrami's equation do not change the underlying complex structure of the surface, as hinted at by the fact that the right-hand sides of the first and third equalities in (\ref{eq:ds2Jwwbar}) are all equal to the metric tensor in the $J$ complex structure. So either of these expressions (\ref{eq:ds2Jwwbar}) for the metric all refer to a (subset of a) surface with complex structure $J$. Rather, one of the main purposes of such transformations is that they enable one to relate the conformal coordinates of different complex structures. So one can, e.g., work in the $J$ complex structure using the $I$ complex structure conformal coordinates, which is in fact what we shall do in places, and this is also what is achieved in the third equality in (\ref{eq:ds2Jwwbar}). Incidentally, by `$I$ complex structure conformal coordinates' we will always mean a specific set of coordinates in the equivalence class of conformally-flat metrics in a complex structure $I$; so in the first and second relations in (\ref{eq:dsJ2 dsI2}) we see the $I$ and $J$ complex structure conformal coordinates, $z_m$ and $w_m$, respectively. These are by no means unique, since we can perform arbitrary holomorphic reparametrisations while remaining within the class of conformally-flat metrics. 
\sk

As discussed in Sec.~\ref{sec:MVa}, the orientation-preserving diffeomorphisms $w_m\mapsto z_m(w_m,\bar{w}_m)$ are those for which $|\mu_{\bar{z}_m}^{\phantom{a}z_m}|<1$ (since from the Beltrami equation it follows that it is for this range that the Jacobian for this change of coordinates is positive), so this range will be the only case of interest since we only consider orientable (and in particular oriented) compact Riemann surfaces here. Note that since $z_m(w_m,\bar{w}_m)$ is an orientation-preserving diffeomorphism it has a unique inverse, $w_m(z_m,\bar{z}_m)$.\footnote{However, $\partial_{w_m}z_m$ and $\partial_{z_m}w_m$ are not inverses of each other but rather \cite{D'HokerPhong15b}:
\begin{equation}\label{eq:dzw dwz}
(\partial_{w_m}z_m)(\partial_{z_m}w_m)=\frac{1}{1-\mu_{\bar{z}_m}^{\phantom{a}z_m}\mu_{z_m}^{\phantom{a}\bar{z}_m}}
\end{equation}
which follows from the Beltrami equation and the chain rule.}
\sk

In the $z_m,\bar{z}_m$ coordinates, the $J$ complex structure metric tensor (\ref{eq:ds2Jwwbar}) is by definition of the general form:
\begin{equation}\label{eq:dsJ2-zm}
g_J\big|_{U_m^J}=2g^J_{z_m\bar{z}_m}\rmd z_m\rmd\bar{z}_m+g^J_{z_mz_m}\rmd z_m^2+g^J_{\bar{z}_m\bar{z}_m}\rmd\bar{z}_m^2,
\end{equation}
and so equating coefficients leads to the following identifications:
\begin{equation}\label{eq:gggJ}
\begin{aligned}
g^J_{z_m\bar{z}_m}&=g_{w_m\bar{w}_m}^{J}|\partial_{z_m}w_m|^2(1+\mu_{\bar{z}_m}^{\phantom{a}z_m}\mu_{z_m}^{\phantom{a}\bar{z}_m})\\
g^J_{z_mz_m}&=2g_{w_m\bar{w}_m}^{J}|\partial_{z_m}w_m|^2\mu_{z_m}^{\phantom{a}\bar{z}_m}\\
g^J_{\bar{z}_m\bar{z}_m}&=2g_{w_m\bar{w}_m}^{J}|\partial_{z_m}w_m|^2\mu_{\bar{z}_m}^{\phantom{a}z_m},
\end{aligned}
\end{equation}
and if we solve for $g_{w_m\bar{w}_m}^{J}|\partial_{z_m}w_m|^2$ in the first relation of (\ref{eq:gggJ}) (recalling that $|\mu_{z_m}^{\phantom{a}\bar{z}_m}|<1$) and substitute this into the second and third we find,
\begin{equation}
\begin{aligned}
g^J_{z_mz_m}&=\frac{2g^J_{z_m\bar{z}_m}}{1+\mu_{\bar{z}_m}^{\phantom{a}z_m}\mu_{z_m}^{\phantom{a}\bar{z}_m}}\mu_{z_m}^{\phantom{a}\bar{z}_m}\\
g^J_{\bar{z}_m\bar{z}_m}&=\frac{2g^J_{z_m\bar{z}_m}}{1+\mu_{\bar{z}_m}^{\phantom{a}z_m}\mu_{z_m}^{\phantom{a}\bar{z}_m}}\mu_{\bar{z}_m}^{\phantom{a}z_m}.
\end{aligned}
\end{equation}
Substituting these into (\ref{eq:dsJ2-zm}) leads to:
\begin{equation}\label{eq:dsJ2-zm2}
\begin{aligned}
g_J\big|_{U_m^J}&=2g^J_{z_m\bar{z}_m}\Big(\rmd z_m\rmd\bar{z}_m+\frac{\mu_{z_m}^{\phantom{a}\bar{z}_m}}{1+\mu_{\bar{z}_m}^{\phantom{a}z_m}\mu_{z_m}^{\phantom{a}\bar{z}_m}}\rmd z_m^2+\frac{\mu_{\bar{z}_m}^{\phantom{a}z_m}}{1+\mu_{\bar{z}_m}^{\phantom{a}z_m}\mu_{z_m}^{\phantom{a}\bar{z}_m}}\rmd\bar{z}_m^2\Big)\\
g_I\big|_{U_m^I}&=2g_{z_m\bar{z}_m}^{I}\rmd z_m\rmd\bar{z}_m,
\end{aligned}
\end{equation}
where in the second equality we also display the corresponding metric in the $I$ complex structure in the $z_m$ coordinates from (\ref{eq:dsJ2 dsI2}). Since $g^J_{z_m\bar{z}_m}$ and $g^I_{z_m\bar{z}_m}$ are local (and in fact real) non-vanishing functions of $z_m,\bar{z}_m$ we can define a quantity $e^{\nu}$ as the ratio:
\begin{equation}\label{eq:nuzmzbarmt}
\boxed{e^{\nu(z_m,\bar{z}_m,t)}\dfn (g^I_{z_m\bar{z}_m})^{-1}g^J_{z_m\bar{z}_m}}
\end{equation}
where $t=(t^1,t^2,\dots)$ is a local coordinate parametrising the complex structure deformation $I\rightarrow J$ such that $J_{t=0}=I$. 
Note that $\nu$ transforms as a scalar under diffeomorphisms or holomorphic reparametrisation, because each of the two ingredients, $g^J_{z_m\bar{z}_m}$, and $g^I_{z_m\bar{z}_m}$, having the same indices transform in the same manner. Therefore $\nu(z_m,\bar{z}_m,t)$ must be a globally-defined function on $\Sigma$ (at least for some finite range of $t$). In terms of $\nu$ (\ref{eq:dsJ2-zm2}) reads:
\begin{equation}\label{eq:dsJ2-zm3}
\begin{aligned}
g_J\big|_{U_m^J}&=2e^{\nu}g^I_{z_m\bar{z}_m}\Big(\rmd z_m\rmd\bar{z}_m+\frac{\mu_{z_m}^{\phantom{a}\bar{z}_m}}{1+\mu_{\bar{z}_m}^{\phantom{a}z_m}\mu_{z_m}^{\phantom{a}\bar{z}_m}}\rmd z_m^2+\frac{\mu_{\bar{z}_m}^{\phantom{a}z_m}}{1+\mu_{\bar{z}_m}^{\phantom{a}z_m}\mu_{z_m}^{\phantom{a}\bar{z}_m}}\rmd\bar{z}_m^2\Big)\\
g_I\big|_{U_m^I}&=2g_{z_m\bar{z}_m}^{I}\rmd z_m\rmd\bar{z}_m,
\end{aligned}
\end{equation}
so that these two relations allow us to read off finite deformations of complex structure of the metric in the $z_m$ coordinates, 
\begin{equation}\label{eq:Deltag-I}
\Delta g_I\dfn g_J-g_I,
\end{equation} 
where all terms are evaluated at the same point in the $z_m$ coordinates. The explicit components read:
\begin{equation}\label{eq:finitedeformations}
\begin{aligned}
(g^I_{z_m\bar{z}_m})^{-1}\Delta g_{z_m\bar{z}_m}^I &= e^{\nu}-1\\
\frac{1}{2}(g^I_{z_m\bar{z}_m})^{-1}\Delta g^I_{z_mz_m}&=\frac{e^{\nu}\mu_{z_m}^{\phantom{a}\bar{z}_m}}{1+\mu_{\bar{z}_m}^{\phantom{a}z_m}\mu_{z_m}^{\phantom{a}\bar{z}_m}}\\
\frac{1}{2}(g^I_{z_m\bar{z}_m})^{-1}\Delta g^I_{\bar{z}_m\bar{z}_m}&=\frac{e^{\nu}\mu_{\bar{z}_m}^{\phantom{a}z_m}}{1+\mu_{\bar{z}_m}^{\phantom{a}z_m}\mu_{z_m}^{\phantom{a}\bar{z}_m}}.
\end{aligned}
\end{equation}
We wish to emphasise that the first of these relations implies that a change of complex structure $I\rightarrow J$ also induces a Weyl transformation (that we are keeping track of) which is encoded in the choice of $\nu$. 
\sk

Now suppose the complex structure deformation is infinitesimal (and connected to the identity). Working with a set of (real or complex) coordinates in moduli space, $t=(t^1,t^2,\dots)$, and taking $\mu,\nu$ to be infinitesimal we may pick a gauge-slice in moduli space by specifying the components, $\mu_k,\nu_k$, defined by:
\begin{equation}\label{eq:munu}
\mu=\mu_k\rmd t^k,\qquad \nu=\nu_k\rmd t^k.
\end{equation}
Since $\mu$ is a vector-valued $(0,1)$-form on $\Sigma$ so is $\mu_k$ a vector-valued $(0,1)$-form on $\Sigma$. What (\ref{eq:munu}) also makes explicit is that both $\mu$ and $\nu$ are 1-forms in moduli space (we are not assuming the theory is Weyl-invariant \cite{Polchinski88}), and in particular we can regard them as (at least local) sections of the cotangent bundle $T^*\mathcal{M}\big|_{\Sigma}$. 
\sk

Summarising, given a reference conformal metric, $g_I$, according to (\ref{eq:finitedeformations}) we may specify an infinitesimal complex structure deformation, $I\rightarrow J=I+\delta I$, by the quantities:
\begin{equation}\label{eq:numumubar-metric2}
\boxed{
\begin{aligned}
\nu_k&=\frac{\partial }{\partial t^k}\ln g_{z_m\bar{z}_m}^I\big|_{z_m,\bar{z}_m}\\
\mu_{kz_m}^{\phantom{z_m}\bar{z}_m}&=\frac{1}{2}(g^I_{z_m\bar{z}_m})^{-1}\frac{\partial g_{z_mz_m}^I}{\partial t^k}\big|_{z_m,\bar{z}_m}\\
\mu_{k\bar{z}_m}^{\phantom{z_m}z_m}&=\frac{1}{2}(g^I_{z_m\bar{z}_m})^{-1}\frac{\partial g_{\bar{z}_m\bar{z}_m}^I}{\partial t^k}\big|_{z_m,\bar{z}_m}\\
\end{aligned}
}
\end{equation}
where we have indicated that the derivatives, $\partial/\partial t^k$, are evaluated at fixed chart coordinates, $z_m,\bar{z}_m$ as suggested by (\ref{eq:dsJ2-zm3}). 

\subsubsection{Deforming via a Change of Coordinates with Fixed Metric}\label{sec:DCCFM}
In (\ref{eq:numumubar-metric2}), as evident also from (\ref{eq:Deltag-I}) and (\ref{eq:finitedeformations}), we have parametrised a change in complex structure as a {\it change in metric} with {\it fixed coordinates}. It will be useful to also consider the alternative but equivalent viewpoint \cite{Polchinski_v1}, whereby we parametrise a change in complex structure as a {\it change of coordinates} with {\it fixed metric}. This statement is slightly ambiguous, but we will momentarily make it sharp. Let us first however explain {\it why} it is slightly ambiguous. The metric, $g_I$, as a tensor of course must change under a complex structure deformation, $g_I\mapsto g_J=g_I+\delta g_I$, since if it does not change, physical distances will remain unchanged and we will not be able to pinch cycles or translate punctures and pinches across Riemann surfaces.  To explain what is usually \cite{Polchinski88,Polchinski_v1} meant by the statement `change in coordinates keeping the metric fixed' let us consider a local patch, $U_m^I$, on $\Sigma$ with chart $(U_m^I,z_m)$. In the conformal class $[g_I]$ a conformal gauge metric takes the form: 
\begin{equation}\label{eq:gImetric}
g_I=\rho_I(z_m,\bar{z}_m)\rmd z_m\rmd\bar{z}_m.
\end{equation}
By `change of coordinates keeping the metric fixed' we will mean that it is often useful to regard a change of complex structure as a change of coordinates, $z_m\mapsto w_m=z_m+\delta z_m(z_m,\bar{z}_m)$, that induces a deformation $g_I\mapsto g_J=g_I+\delta g_I$, where \cite{Friedan82}:
\begin{equation}\label{eq:gJ=rhoI}
\begin{aligned}
g_J&=\rho_I(w_m,\bar{w}_m)\rmd w_m\rmd\bar{w}_m\\
&=\rho_I(z_m+\delta z_m,\bar{z}_m+\delta \bar{z}_m)d(z_m+\delta z_m)d(\bar{z}_m+\delta \bar{z}_m)\\
&=g_I+\rho_I(z_m,\bar{z}_m)\big[\nabla_{z_m}(\delta z_m)+\nabla_{\bar{z}_m}(\delta\bar{z}_m)\big]\rmd z_m\rmd\bar{z}_m\\
&\qquad+\rho_I(z_m,\bar{z}_m)\big[\nabla_{z_m}(\delta \bar{z}_m)\big]\rmd z_m^2+\rho_I(z_m,\bar{z}_m)\big[\nabla_{\bar{z}_m}(\delta z_m)\big]\rmd\bar{z}_m^2
\end{aligned}
\end{equation}
where $g_I$ is given in (\ref{eq:gImetric}). 
\sk

There are a few points worth emphasising before moving on:
\begin{itemize}
\item The first is that the change of coordinates, $z_m\mapsto w_m=z_m+\delta z_m(z_m,\bar{z}_m)$, need not be invertible, and since this is not meant to be a diffeomorphism this is permissible (see, e.g., \cite{D'HokerGiddings87} for a crystal clear demonstration and elaboration on some of the implications of this). For example, in addition to being a function of the moduli, $t$, the quantity $\delta z_m(z_m,\bar{z}_m)$ might be a {\it discontinuous} function of $z_m,\bar{z}_m$. In fact, the quantity $w_m(z_m,\bar{z}_m)$ need not even satisfy Beltrami's equation (\ref{eq:beltramieqnJI}).  For example, as already discussed above it is often convenient to allow the metric (and hence also the Beltrami differential) to become singular along certain contours or at isolated points \cite{D'HokerGiddings87,Polchinski88}. That such ``violent'' deformations are permissible might become even more plausible by recalling that the defining properties of a Riemann surface (reviewed in Sec.~\ref{sec:TFCR} and Sec.~\ref{sec:CSDI}) do not even require the introduction of a metric.  So one can instead phrase everything in terms of transition functions and cocycle relations that {\it are} required to be holomorphic (and hence well-behaved) on the relevant patch overlaps. 
\item The second point is that $\rho_I$ appears (and not $\rho_J$) in the first line in (\ref{eq:gJ=rhoI}), even though we are in the $J$ complex structure, and this is what we mean by `keeping the metric fixed' while deforming from $I$ to $J$: both metrics, $g_I,g_J$, are written in terms of the {\it same} conformal metric component, $\rho_I$, but with different conformal coordinates, $z_m,w_m$. From (\ref{eq:gJ=rhoI})  it is clear that $g_I\neq g_J$; the conformal classes, $[g_I]$ (with conformal coordinate $z_m$) and $[g_I+\delta g_I]$ (with conformal coordinate $w_m$) are different. (Upon interpreting the covariant derivatives appearing it is to be understood that $\delta z_m$ transforms {\it locally} as the component of a rank-$(-1)$ tensor (a vector) in the $[g_I]$ conformal class; the quantity $\delta z_m$ will not extend globally to a vector {\it unless} the complex structure remains unchanged under $z_m\mapsto w_m=z_m+\delta z_m$. Conversely, if the quantity $\delta z_m$ does transform as a vector globally then complex structure remains unchanged.) 
\end{itemize}

Summarising, an infinitesimal (possibly discontinuous and/or non-invertible) change of coordinates, 
\begin{equation}\label{eq:zm->wm}
z_m\mapsto w_m=z_m+\delta z_m(z_m,\bar{z}_m),
\end{equation}
induces the deformation:
\begin{equation}\label{eq:changecoordsmetricfixed}
\boxed{
\begin{aligned}
\delta \ln\rho_I\big|_{z_m} &= \nabla_{z_m}(\delta z_m)+\nabla_{\bar{z}_m}(\delta\bar{z}_m)\\
\rho_I^{-1}\delta g_{z_mz_m}^I\big|_{z_m}&=\nabla_{z_m}(\delta \bar{z}_m)\\
\rho_I^{-1}\delta g_{\bar{z}_m\bar{z}_m}^I\big|_{z_m}&=\nabla_{\bar{z}_m}(\delta z_m)
\end{aligned}
}
\end{equation}
where we noted that the metric in the initial and deformed complex structures are both evaluated in the $z_m$ coordinates, so that the variations are evaluated at fixed coordinates $z_m$. Note that, by definition, the covariant derivative $\nabla_{\bar{z}_m}(\delta z_m)$ does not require a connection (since $\delta z_m$ has an implicit ``upstairs'' index $z_m$ and hence is associated to an element of $K^{-1}$). So we can also replace it by $\partial_{\bar{z}_m}(\delta z_m)$, and similarly for the complex conjugate appearing in the second equality in (\ref{eq:changecoordsmetricfixed}).
\sk

Let us divert attention briefly to discuss how the procedure associated to the infinitesimal deformation of complex structures in (\ref{eq:gJ=rhoI}) generalises to the case of finite deformations, which  was discussed with great clarity in \cite{D'HokerPhong15b}. Using the notation developed in \cite{D'HokerPhong15b} and extracting only the relevant aspects for this subsection, the procedure involves a map $\imath_{z\leftarrow w}$ which in the current case of interest reads:
\begin{equation}
\imath_{z\leftarrow w}:\,\,\, \rho_I(w,\bar{w})\mapsto \rho_J(z,\bar{z})=\rho_I(w,\bar{w})|\partial_zw|^2.
\end{equation}
It is seen that taking $w=z+\delta z(z,\bar{z})$ results in:
\begin{equation}
\rho_J(z,\bar{z})=\rho_I(z,\bar{z})+\rho_I(z,\bar{z})\big[\nabla_{z}(\delta z)+\nabla_{\bar{z}}(\delta\bar{z})\big],
\end{equation}
in precise agreement with (\ref{eq:gJ=rhoI}) and (\ref{eq:changecoordsmetricfixed}) since $\delta \ln\rho_I\dfn (\rho_J-\rho_I)/\rho_I$. In terms of $\rho_J$ and when the Beltrami equation, $\partial_{\bar{z}}w=\mu_{\bar{z}}^{\phantom{a}z}\partial_zw$, is satisfied the full metric  in the $J$ complex structure for finite deformations reads,
$$
g_J = \rho_J(z,\bar{z})|\rmd z+\mu_{\bar{z}}^{\phantom{a}z}\rmd\bar{z}|^2,
$$
which reduces to the result (\ref{eq:gJ=rhoI}) of Friedan \cite{Friedan82}  in the case of infinitesimal deformations. We return to the case of infinitesimal deformations. 
\sk

The result (\ref{eq:changecoordsmetricfixed}) encodes how to deform complex structure by a change of coordinates (\ref{eq:zm->wm}) keeping the metric fixed. To connect this approach to the alternative viewpoint that resulted in (\ref{eq:numumubar-metric2}) (where we regarded a change in complex structure as a change of metric with fixed coordinates and transition functions) note that a change in coordinates  is determined by Beltrami's equation (\ref{eq:beltramieqnJI}). Recalling that (see Appendix~\ref{sec:DGlemma}) in a local simply connected patch there always exists a vector $v_m$ such that the (component of a) Beltrami differential reads:
\begin{equation}\label{eq:mu=dbarvm}
\mu_{\bar{z}_m}^{\phantom{a}z_m}=\partial_{\bar{z}_m}v_m^{z_m},\qquad \mu_{z_m}^{\phantom{a}\bar{z}_m}=\partial_{z_m}v_m^{\bar{z}_m},
\end{equation}
and concentrating on infinitesimal deformations  (\ref{eq:munu}), we learn that coordinate transformations of the form:
\begin{equation}\label{eq:w=z+v}
\begin{aligned}
w_m(z_m,\bar{z}_m)&=z_m+v_{km}^{z_m}(z_m,\bar{z}_m)\delta t^k+\mathcal{O}(v^2)\\
\bar{w}_m(z_m,\bar{z}_m)&=\bar{z}_m+v_{km}^{\bar{z}_m}(z_m,\bar{z}_m)\delta t^k+\mathcal{O}(v^2),
\end{aligned}
\end{equation}
indeed satisfy Beltrami's equation (\ref{eq:beltramieqnJI}) to the indicated order in $v_m$.   So we connect the two alternative viewpoints by identifying:
\begin{equation}\label{eq:dzm = vdt}
\delta z_m=v_{km}^{z_m}\delta t^k,\qquad \delta \bar{z}_m=v_{km}^{\bar{z}_m}\delta t^k,
\end{equation}
where $v_m^{z_m}=v_{km}^{z_m}\delta t^k$, 
from which it also follows that:
\begin{equation}\label{eq:v=dzdt}
v_{km}^{z_m}=\frac{\rmd z_m}{\rmd t^k},\qquad v_{km}^{\bar{z}_m}=\frac{\rmd\bar{z}_m}{\rmd t^k},
\end{equation}
which will be useful below. Notice these are {\it total} derivatives. 
\sk

Combining (\ref{eq:numumubar-metric2}) and (\ref{eq:changecoordsmetricfixed}) with the identification (\ref{eq:dzm = vdt}) we learn that consistency of the two viewpoints requires that in a given chart $(U_m^J,z_m)$:
\begin{equation}\label{eq:numumubar-dvs}
\boxed{
\begin{aligned}
\nu_k&=\nabla_{z_m}v_{km}^{z_m}+\nabla_{\bar{z}_m}v_{km}^{\bar{z}_m}\\
\mu_{kz_m}^{\phantom{z_m}\bar{z}_m}&=\nabla_{z_m} v_{km}^{\bar{z}_m}\\
\mu_{k\bar{z}_m}^{\phantom{z_m}z_m}&=\nabla_{\bar{z}_m} v_{km}^{z_m}\\
\end{aligned}
}
\end{equation}
These relations have the property that they do not make explicit reference to any metric, but they do nevertheless encode how complex structure deformations induce Weyl transformations. In the second and third relations in (\ref{eq:numumubar-dvs}) we can replace $\nabla_{\bar{z}_m} v_{km}^{z_m}$ by $\partial_{\bar{z}_m} v_{km}^{z_m}$ (and the corresponding complex conjugate relation) without loss of generality by the defining property of the $\partial_{\bar{z}}$ derivative that it does not require a connection when acting on tensors with solely holomorphic indices, $\phi_{zz\dots}\rmd z^n\in K^{\otimes n}$, with $K$ (as always) the canonical (or cotangent) bundle on $\Sigma$. So (\ref{eq:numumubar-dvs}) is consistent with (\ref{eq:mu=dbarvm}). Correspondingly, the covariant derivatives in the first relation in (\ref{eq:numumubar-dvs}) cannot be replaced by ordinary derivatives. In Sec.~\ref{sec:HNC} (and subsequent sections) we will make use of a deep and useful insight by Polchinski \cite{Polchinski88} that there exists a holomorphic reparametrisation (associated to a choice of metric) that enables one to write the relevant contributions associated to the path integral measure in terms of covariant derivatives of the Ricci scalar, 
$R_{(2)}$, (which encode any potential frame dependence) and holomorphic transition functions. 
\sk

It is instructive to compare (\ref{eq:numumubar-dvs}) to the corresponding relations (\ref{eq:deltamumubarrho}). The former deform complex structure away from $\mu=0$ whereas the latter correspond to reparametrisations around a generic finite value of $\mu$. Setting $\mu=0$ in (\ref{eq:deltamumubarrho}), see (\ref{eq:deltamumubarrho2}), one sees that they differ by overall minus signs.

\subsubsection{Relation to Transition Functions and Cocycle Relations}
In Sec.~\ref{sec:CSDI} we discussed how to deform complex structure and phrased everything in terms of the elementary data defining a Riemann surface, namely holomorphic transition functions and cocycle relations. That approach had the great advantage of being automatically globally well-defined on $\Sigma$. In this section we took an alternative (and more traditional in the string theory literature) approach where the discussion has been entirely local on $\Sigma$. To exhibit the underlying coherence of the subject let us end this subsection by discussing the precise relation between these two approaches. 
\sk

We primarily recall (\ref{eq:v=dzdt}), reproduced here for convenience:
\begin{equation}\label{eq:dzm/dt=vm}
\begin{aligned}
\frac{\rmd z_m}{\rmd t^k}=v_{km}^{z_m}\\
\end{aligned}
\end{equation}
which identifies the {\it total} derivative of the chart coordinate $z_m$ (of the open set $U_m^I$) with respect to a coordinate in moduli space, $t^k$, of our choice. Suppose there is a non-empty overlap $U_m^I\cap U_n^I$, where to $U_n^I$ there corresponds a chart coordinate $z_n$. From the discussion in Sec.~\ref{sec:RS}, see (\ref{eq:transitionfuncs}), we have that on the overlap these coordinates are identified using the holomorphic transition function, $z_m=f_{mn}(z_n)$. Furthermore, from the discussion in Sec.~\ref{sec:CSDI}, see (\ref{eq:zmfmnzn}), we know that we can consider the transition function to contain all moduli dependence \cite{Kodaira}, 
$
z_m=f_{mn}(z_n,t)
$. 
Therefore, substituting this into (\ref{eq:dzm/dt=vm}) yields:
\begin{equation}\label{eq:vm=dzm/dt}
\begin{aligned}
v_{km}^{z_m}(z_m,\bar{z}_m)&=\frac{\rmd f_{mn}(z_n,t)}{\rmd t^k}\\
&=\frac{\partial f_{mn}(z_n,t)}{\partial t^k}\big|_{z_n}+\frac{\partial f_{mn}(z_n,t)}{\partial z_n}\big|_{t}\frac{\rmd z_n}{\rmd t^k}
\end{aligned}
\end{equation}
However, in direct analogy to (\ref{eq:dzm/dt=vm}) we also know that in the $(U_n^I,z_n)$ chart,
\begin{equation}\label{eq:dzn/dt=vn}
\begin{aligned}
\frac{\rmd z_n}{\rmd t^k}=v_{kn}^{z_n}(z_n,\bar{z}_n),
\end{aligned}
\end{equation}
and so the last term in (\ref{eq:vm=dzm/dt}) is equivalent to,
$$
\frac{\partial f_{mn}(z_n,t)}{\partial z_n}\big|_{t}v_{kn}^{z_n}(z_n,\bar{z}_n) = \frac{\partial z_m}{\partial z_n}\big|_{t}v_{kn}^{z_n}(z_n,\bar{z}_n)=v_{kn}^{z_m}(z_m,\bar{z}_m),
$$
where we changed coordinates taking into account that (locally) $v_{kn}^{z_n}(z_n,\bar{z}_n)$ transforms as the component of a vector. Substituting the resulting relation into (\ref{eq:vm=dzm/dt}) therefore yields,
\begin{equation}\label{eq:vm-vn=dfmn/dt}
\boxed{
\begin{aligned}
v_{km}^{z_m}(z_m,\bar{z}_m)-v_{kn}^{z_m}(z_m,\bar{z}_m)=\frac{\partial f_{mn}(z_n,t)}{\partial t^k}\big|_{z_n}
\end{aligned}
}
\end{equation}
Recalling the notation (\ref{eq:phimn(t)}) we learn that what we have just computed is the component of the locally-defined holomorphic (in $z_n$) vector $\phi_{mn}(t^k)$ on $U_m^I\cap U_n^I\subset \Sigma$,
\begin{equation}\label{eq:phi_mn=vm-vn}
\begin{aligned}
\phi_{mn}(t^k)&=\frac{\partial f_{mn}(z_n,t)}{\partial t^k}\big|_{z_n}\frac{\partial}{\partial z_m}\\
&=\big(v_{km}^{z_m}-v_{kn}^{z_m}\big)\frac{\partial}{\partial z_m}\\
&\equiv v_{km}-v_{kn}.
\end{aligned}
\end{equation}
Since this is evaluated locally on $U_m\cap U_n$ notice that it is vital that the {\it locally}-defined vectors, $v_m,v_n$, do not agree on patch overlaps, for if they did agree on patch overlaps $\phi_{mn}$ would vanish and complex structure would remain unchanged. 
This identification (\ref{eq:phi_mn=vm-vn}) provides the link between {\it local and global data}, so that applying the analysis of Sec.~\ref{sec:CSDI} to the local discussion we have just presented ensures that we have arrived at a globally (on $\Sigma$) well-defined construction of infinitesimal complex structure deformations. Also, recall from the discussion in Sec.~\ref{sec:CSDI} that when $v_m,v_n$ are holomorphic in their arguments the quantity $v_m-v_n$ is (from the \v{C}ech cohomology viewpoint) a 1-coboundary by which we {\it mod out} to generate $H^1(\Sigma,\mathcal{S}_t)$, and the latter is in turn isomorphic to the tangent space of moduli space at $\Sigma$, $T\mathcal{M}|_{\Sigma}$. 

\subsection{Holomorphic Normal Coordinates}\label{sec:HNC}

\subsubsection{In Terms of a Metric}\label{sec:MV}
In \cite{Polchinski87} the notion of `{\it conformal normal ordering}' was introduced, which is to subtract self contractions from composite local operators using a holomorphic coordinate, which is in turn defined as a solution to the Beltrami equation for a given fixed complex structure. Given such a solution, $w(z,\bar{z})$, to the Beltrami equation (if only in an infinitesimal neighbourhood of a give point) there is an infinite set of solutions that can be generated from it, which are in turn generated by arbitrary holomorphic reparametrisations, $w\mapsto w'=f(w)$. So in order for the construction to be meaningful one must deal with the fact that the Beltrami equation does not completely determine the solution $w(z,\bar{z})$. In \cite{Polchinski87} this ambiguity was fixed by requiring that physical vertex operators be independent of the solution chosen in the normal ordering, and this in turn was shown to lead to the usual physical state conditions of primary vertex operators. In particular, the resulting vertex operators automatically satisfy the usual Virasoro constraints. So this procedure will not be appropriate for our purposes because the primary state conditions are clearly too strong for states propagating through handles (since these states are offshell). So we will need to fix this ambiguity in a different way.
\sk

In a followup paper \cite{Polchinski88} it was suggested to fix this ambiguity by singling out a specific solution (determined up to $U(1)$ frame rotations) of the Beltrami equation which has the property that it leads to a worldsheet that is `{\it as flat as possible}' at the location of a vertex operator, {\it independently} of where this vertex operator is inserted or translated to. This led to the notion of `{\it Weyl normal ordering}', which is to conformal normal order with this specific choice of conformal coordinate (where a metric is used to define a conformal frame). A virtue of this choice is that radially-normal ordered operators (in a free field realisation of the CFT of interest) are automatically Weyl normal ordered. We call these coordinates {\it holomorphic normal coordinates}. 
\sk

We adopt this terminology because these coordinates are obtained from a general conformal metric by a {\it holomorphic} change of coordinates, and they are furthermore analogous to Riemann {\it normal} coordinates (familiar from real differentiable manifolds \cite{Eisenhardt}, and used extensively in background-field expansions of non-linear sigma models \cite{Friedan80,Alvarez-GaumeFreedmanMukhi81,FradkinTseytlin85,Mukhi86,CallanGan86,Tseytlin87,HowePapadopoulosStelle88,Osborn90} for the target space metric). 
Furthermore, as in the case of Riemann normal coordinates for real manifolds, `holomorphic normal coordinates' are useful in making two-dimensional covariance manifest in string amplitudes, they serve to make amplitudes globally well-defined in  moduli space, and enable one to keep track of the frame dependence throughout a calculation. 
\sk

Going into further detail, the prescription is to make use of freedom in making holomorphic reparametrisations to assert that it is always possible to choose the metric $g_I|_{U_m^I}$ in (\ref{eq:dsJ2 dsI2}) such that it is ``as flat as possible'' at the point $p_m\in U_m^I$ (at which $z_m(p_m)=0$):
\begin{equation}\label{eq:Polchinskicoords}
\boxed{
\partial_{z_m}^ng^I_{z_m\bar{z}_m}(p_m)=\partial_{\bar{z}_m}^ng^I_{z_m\bar{z}_m}(p_m)=
\left\{
\begin{array}{l}
\begin{aligned}
&\tfrac{1}{2}\qquad \textrm{if $n=0$}\\
&0\qquad \textrm{if $n\geq1$}
\end{aligned}
\end{array}
\right.
}
\end{equation}
A proof that such a coordinate choice (\ref{eq:Polchinskicoords}) always exists can be found in Sec.~\ref{sec:PolCoords}, where it is also shown that the phase of the coordinate $z_m$ is not determined by the conditions (\ref{eq:Polchinskicoords}), and since this phase is not globally defined\footnote{This point is elaborated on in the following section in detail.} following \cite{Polchinski88} we will always take this phase to be integrated (or choose combinations of local operators that are independent of this phase). 
Notice from (\ref{eq:Polchinskicoords}) that all holomorphic and anti-holomorphic derivatives of $g^I_{z_m\bar{z}_m}$ at $p_m$ can be chosen to vanish, but mixed derivatives cannot be made to vanish by a coordinate choice since they involve the Ricci scalar which at any point in $U_m^I$ reads:
\begin{equation}\label{eq:R(2)}
\boxed{
R_{(2)}= -2(g^I_{z_m\bar{z}_m})^{-1}\partial_{z_m}\partial_{\bar{z}_m}\ln g^I_{z_m\bar{z}_m}
}
\end{equation}
From (\ref{eq:dsJ2 dsI2}) we can define the inverse metric component in the usual manner, i.e.~we denote it by raised indices: $g^{z_m\bar{z}_m}_Ig_{z_m\bar{z}_m}^I=1$. 
Since the only non-vanishing components of the metric-compatible and torsion-free (Christoffel) connection are $\Gamma_{z_mz_m}^{z_m}=\partial_{z_m}\ln g^I_{z_m\bar{z}_m}$ and $\Gamma_{\bar{z}_m\bar{z}_m}^{\bar{z}_m}=\partial_{\bar{z}_m}\ln g^I_{z_m\bar{z}_m}$, (\ref{eq:Polchinskicoords}) is equivalent to:
\begin{equation}\label{eq:Polchinskicoords-Gamma}
\partial_{z_m}^n\Gamma_{z_mz_m}^{z_m}\big|_{p_m}=\partial_{\bar{z}_m}^n\Gamma_{\bar{z}_m\bar{z}_m}^{\bar{z}_m}\big|_{p_m}=
\begin{aligned}
&0\quad \textrm{for}\quad n\geq0
\end{aligned}
\qquad {\rm and}\qquad g^I_{z_m\bar{z}_m}\big|_{p_m}=\tfrac{1}{2}
\end{equation}

It will be convenient to require that this change of coordinates does not move the point $p_m$, i.e.~$w_m(p_m)=z_m(p_m)=0$. This is ensured by demanding that:
\begin{equation}\label{eq:v(p)=0}
\boxed{
v_{km}^{z_m}(p_m)=v_{km}^{\bar{z}_m}(p_m)=0
}
\end{equation}
which may be enforced without loss of generality since we will absorb all translations of $p_m$ into corresponding transition functions with remaining neighbouring patches. These local vector components specify the quantities $\nu$ and $\mu$ appearing in (\ref{eq:numumubar-dvs}).
\sk

Taking also (\ref{eq:numumubar-metric2}) into account we have (in a general conformally-flat coordinate system):
\begin{equation}\label{eq:numumubar-metric4}
\begin{aligned}
\nu_k\big|_{U_m}&=\nabla_{z_m}v_{km}^{z_m}+\nabla_{\bar{z}_m}v_{km}^{\bar{z}_m}\\
\end{aligned}
\end{equation}
Let us then evaluate (\ref{eq:numumubar-metric4}) at $p_m$ in holomorphic normal coordinates where (\ref{eq:Polchinskicoords-Gamma}) holds. The connection terms vanish and we are therefore left with:
\begin{equation}
\nu_k(p_m) =\big(\partial_{z_m}v_{km}^{z_m}+\partial_{\bar{z}_m}v_{km}^{\bar{z}_m}\big)(p_m),
\end{equation}
and since the two terms on the right-hand side are related by complex conjugation,
\begin{equation}
\frac{1}{2}\nu_k(p_m) ={\rm Re}\big(\partial_{z_m}v_{km}^{z_m}\big)(p_m),
\end{equation}
or,
\begin{equation}\label{eq:dv=nu+gamma_k}
\boxed{\big(\partial_{z_m}v_{km}^{z_m}\big)(p_m)=\frac{1}{2}\big(\nu_k(p_m) +i\gamma_k(p_m)\big)}
\end{equation}
where we have defined:
\begin{equation}\label{eq:gammakdfn}
\gamma_k(p_m)\dfn 2{\rm Im}\big(\partial_{z_m}v_{km}^{z_m}\big)(p_m).
\end{equation}
The latter is associated to the phase ambiguity mentioned above and will drop out of string amplitude computations as we will see. 
We will also require expressions for higher derivatives, $\partial_{z_m}^{n+1}v_{km}^{z_m}(p_m)$, see (\ref{eq:Bintmub5}).\footnote{Our conventions for the Ricci scalar and covariant derivatives are specified in Appendix \ref{sec:CT}.} 
The first few read (note that $g_{z_m\bar{z}_m}(p_m)=1/2$):
\begin{equation}\label{eq:d2vd3vetc}
\begin{aligned}
\partial_{z_m}^2v_{km}^{z_m}(p_m)&=\big(\partial_{z_m}\nu_k-\partial_{\bar{z}_m}\mu_{kz_m}^{\phantom{aj}\bar{z}_m}\big)(p_m)\\
\partial_{z_m}^3v_{km}^{z_m}(p_m)&=\big(\partial_{z_m}^2\nu_k+\frac{1}{2}R_{(2)}\mu_{kz_m}^{\phantom{aa}\bar{z}_m}-\partial_{z_m}\partial_{\bar{z}_m}\mu_{kz_m}^{\phantom{aa}\bar{z}_m}\big)(p_m)\\
&\,\,\,\vdots
\end{aligned}
\end{equation}

The right-hand sides of the various terms in (\ref{eq:d2vd3vetc}) become somewhat unwieldy for a generic number of derivatives, but there is a vast simplification that occurs if we assume that the quantity $v_{km}^{z_m}$ is {\it real analytic} in the vicinity of the point $p_m$. In terms of real coordinates this means that it has a Taylor expansion. But it is more convenient to work in terms of complex coordinates, in which case we will interpret real analyticity\footnote{We would like to thank Simon Donaldson for a clarification on this point.} as meaning that it has a (convergent) power series expansion about $z_m,\bar{z}_m=0$ of the form:
\begin{equation}\label{eq:realanalyticvkm}
v_{km}^{z_m}(z_m,\bar{z}_m)=\sum_{a,b=0}^{\infty}\frac{z_m^a\bar{z}_m^b}{a!b!}\partial^a_{z_m}\partial_{\bar{z}_m}^bv_{km}^{z_m}(0,0).
\end{equation}
A simplifying implication of assuming such real analyticity is that {\it when} this expansion is substituted into the path integral measure we will see that only the $b=0$ terms in (\ref{eq:realanalyticvkm}) contribute, which in turn implies that only {\it holomorphic} terms contribute. In terms of the expansion (\ref{eq:d2vd3vetc}) this in turn implies that the only contributions that will actually contribute to the measure will be:
\begin{equation}\label{eq:dn+1v=vnnu}
\boxed{
\partial_{z_m}^{n+1}v_{km}^{z_m}(p_m)=\partial_{z_m}^n\nu_k(p_m)+\dots,\qquad n=1,2,\dots
}
\end{equation}
which in turn follows from the above statement and (\ref{eq:numumubar-metric4}). The dots `\dots' in (\ref{eq:dn+1v=vnnu}) indicate terms that are {\it not} holomorphic and that will consequently not contribute to the path integral measure as we will see in Sec.~\ref{sec:IToaM}. (Of course, the first term on the right-hand side is not holomorphic either but for real analytic $\nu_k$ it does, unlike the terms in `$\dots$', contain a term in a power series expansion that is holomorphic.)

\subsubsection{The Derivation}\label{sec:PolCoords}
In this section we derive explicitly the holomorphic change of variables starting from any given conformal metric that leads to a metric satisfying the properties (\ref{eq:Polchinskicoords}). 
We will also lighten the notation somewhat in this subsection, but at the cost of the notation here being somewhat inconsistent with the remaining document. 
\sk

Let us consider a local coordinate chart $(U,z)$ on a closed Riemann surface which is in turn endowed with a metric. As shown in Sec.~\ref{sec:MVa}, see in particular (\ref{eq:gab=rhomuz=w}), on this chart a general metric,
\begin{equation}\label{eq:gJ=gabsigma}
g_J=g_{ab}(\sigma)\rmd\sigma^a\rmd\sigma^b,
\end{equation}
can be written in the form:
\begin{equation}\label{eq:rho0dwdwbar}
g_J=\rho_0(w,\bar{w})\rmd w\rmd\bar{w}.
\end{equation}

We next make use of the freedom to perform arbitrary {\it holomorphic} reparametrisations to make a special choice of conformal coordinate, 
\begin{equation}\label{eq:w->zeta}
w,\bar{w}\mapsto \zeta(w),\bar{\zeta}(\bar{w}),
\end{equation}
in terms of which there will be a new local expression for the metric which is still equal to (\ref{eq:rho0dwdwbar}),\footnote{The quantities $\rho(z,\bar{z})$ and $\rho(\zeta,\bar{\zeta})$ in (\ref{eq:rhozzbar}) and (\ref{eq:rho0rhozz}) respectively are distinct. In the remaining subsection, by $\rho$ we will always mean the quantity $\rho(\zeta,\bar{\zeta})$ and not $\rho(z,\bar{z})$.}
\begin{equation}\label{eq:rhozzbar}
\boxed{g_J=\rho(\zeta,\bar{\zeta})\rmd\zeta \rmd\bar{\zeta}}
\end{equation}
In particular, setting (\ref{eq:rhozzbar}) equal to (\ref{eq:rho0dwdwbar}) requires that:
\begin{equation}\label{eq:lnrho}
\ln\rho(\zeta,\bar{\zeta})=\ln\rho_0(w,\bar{w})-\ln|\partial_w\zeta(w)|^2,
\end{equation}
in precise analogy to (\ref{eq:rho0rhozz}), but here $w(\zeta)$ is holomorphic in $\zeta$ whereas in (\ref{eq:rho0rhozz}) $w(z,\bar{z})$ was a solution to the general Beltrami equation. We emphasise that all metrics we have considered, namely (\ref{eq:rhozzbar}), (\ref{eq:rho0dwdwbar}), (\ref{eq:gabmurho1}) and (\ref{eq:gJ=gabsigma}) are equal in their common domain of validity since they are related by coordinate or conformal reparametrisations (and {\it not} Weyl rescalings).
\sk

Now comes a crucial point. Given any conformally-flat metric (\ref{eq:rho0dwdwbar}), $\rho_0(w,\bar{w})$, there exists a holomorphic change of coordinates (\ref{eq:w->zeta}) such that at a point $p\in U$ the new metric is ``as flat as possible'' \cite{Polchinski88}:
\begin{equation}\label{eq:PolchinskicoordsA}
\boxed{
\partial_\zeta^n\rho(\zeta,\bar{\zeta})|_p=\partial_{\bar{\zeta}}^n\rho(\zeta,\bar{\zeta})|_p=
\left\{
\begin{array}{l}
\begin{aligned}
&1\qquad \textrm{if $n=0$}\\
&0\qquad \textrm{if $n\geq1$}
\end{aligned}
\end{array}
\right.
}
\end{equation}
So all holomorphic and anti-holomorphic derivatives of $\rho(\zeta,\bar{\zeta})$ at $p$ can be chosen to vanish, but mixed derivatives cannot be made to vanish by a coordinate choice since they involve the Ricci scalar,
$$
R_{(2)}= -4\rho^{-1}\partial_\zeta\partial_{\bar{\zeta}}\ln \rho(\zeta,\bar{\zeta}).
$$
Since the only non-vanishing components of the Christofel connection in the metric (\ref{eq:rhozzbar}) are $\Gamma_{\zeta\zeta}^\zeta=\partial_\zeta\ln \rho(\zeta,\bar{\zeta})$ and $\Gamma_{\bar{\zeta}\bar{\zeta}}^{\bar{\zeta}}=\partial_{\bar{\zeta}}\ln \rho(\zeta,\bar{\zeta})$, (\ref{eq:PolchinskicoordsA}) is equivalent to:
\begin{equation}\label{eq:Polchinskicoords-GammaA}
\boxed{
\partial_\zeta^n\Gamma_{\zeta\zeta}^\zeta\big|_p=\partial_{\bar{\zeta}}^n\Gamma_{\bar{\zeta}\bar{\zeta}}^{\bar{\zeta}}\big|_p=
\begin{aligned}
&0\quad \textrm{for}\quad n\geq0
\end{aligned}
\qquad {\rm and}\qquad \rho(\zeta,\bar{\zeta})|_p=1
}
\end{equation}

Before we set out to derive the explicit holomorphic coordinate transformation $w\mapsto \zeta$ that makes (\ref{eq:Polchinskicoords-GammaA}) possible given an arbitrary conformally-flat metric $ds^2=\rho_0(w,\bar{w})\rmd w\rmd\bar{w}$, let us discuss (if only briefly) why it is useful. Recall that the covariant derivative of a scalar (say the Ricci scalar for concreteness) in complex coordinates satisfies,
\begin{equation}
\nabla_\zeta R_{(2)} = \partial_\zeta R_{(2)},
\end{equation}
where (as is common in the physics literature we work in terms of components). Since the covariant derivative of a scalar transforms as a conformal tensor of rank $1$ taking a second covariant derivative of the above yields,
\begin{equation}\label{eq:Nabla2R}
\nabla_\zeta^2 R_{(2)} = \partial_\zeta^2 R_{(2)}-\Gamma_{\zeta\zeta}^\zeta \partial_\zeta R_{(2)},
\end{equation}
whereas a third covariant derivative yields,
\begin{equation}\label{eq:Nabla3R}
\nabla_\zeta^3 R_{(2)} = \partial_\zeta\big(\partial_\zeta^2 R_{(2)}-\Gamma_{\zeta\zeta}^\zeta \partial_\zeta R_{(2)}\big)-2\Gamma_{\zeta\zeta}^\zeta \big(\partial_\zeta^2 R_{(2)}-\Gamma_{\zeta\zeta}^\zeta \partial_\zeta R_{(2)}\big),
\end{equation}
where we have taken into account that every additional covariant derivative increases the rank of the tensor by 1. In particular, recall that the covariant derivative of a rank $n$ tensor, $\varphi=\varphi_{\zeta\zeta\dots}\rmd\zeta^n$, reads: 
$$
\nabla^{(n)}_\zeta\varphi_{\zeta\zeta\dots} 
= \partial_\zeta\varphi_{\zeta\zeta\dots}-n\Gamma_{zz}^z\varphi_{\zeta\zeta\dots}.
$$ 
So it is clear that if we work in holomorphic normal coordinates and are interested in evaluating the resulting expressions such as (\ref{eq:Nabla2R}) and (\ref{eq:Nabla3R}) at $p$ where (\ref{eq:Polchinskicoords-GammaA}) is valid, the $n^{\rm th}$ covariant derivative of the Ricci scalar evaluated at $p$ takes the simple form:
\begin{equation}\label{eq:NablanR-Polch}
\boxed{\nabla_\zeta^n R_{(2)}(p) = \partial_\zeta^n R_{(2)}(p)}
\end{equation}
So this choice of coordinates enables one to replace ordinary derivatives by covariant derivatives, and since an expression written in terms of covariant quantities is independent of coordinate system we see that the resulting expressions are valid in any coordinate system. So holomorphic normal coordinates are analogous to Riemann normal coordinates for real manifolds, and the latter have in turn played a major role in the study of non-linear sigma models where such coordinates are systematically used in the target space (a short list of early references regarding the use of Riemann normal coordinates is \cite{Friedan80,Alvarez-GaumeFreedmanMukhi81,FradkinTseytlin85,Mukhi86,CallanGan86,Tseytlin87,HowePapadopoulosStelle88,Osborn90}). 
\sk

We next prove existence of a holomorphic coordinate chart $(U,\zeta)$ satisfying (\ref{eq:PolchinskicoordsA}) given any initial holomorphic chart $(U,w)$ with metric $\rho_0(w,\bar{w})$ (with $\zeta(w)$ holomorphic in $w$). First note that $\zeta(w)$ is invertible around a given (implicit) base point since it is analytic, so $w(\zeta)$ always exists locally. For $n\geq1$ the defining properties (\ref{eq:PolchinskicoordsA}) are equivalent to $\partial_w^n\rho(\zeta,\bar{\zeta})|_p=\partial_{\bar{w}}^n\rho(\zeta,\bar{\zeta})|_p=0$ as one can show by differentiating (\ref{eq:lnrho}) and making use of (\ref{eq:PolchinskicoordsA}). Therefore, from (\ref{eq:lnrho}), taking into account (\ref{eq:PolchinskicoordsA}), we have that at $p$ and for $n\geq1$ (using Faa di Bruno's formula):
\begin{equation}\label{eq:dwnlnrho0}
\begin{aligned}
\partial_w^n\ln\rho_0(w,\bar{w})&=\sum_{k=1}^n (-)^{k+1}(k-1)!B_{n,k}(\partial_w^2\zeta/\partial_w\zeta,\partial_w^3\zeta/\partial_w\zeta,\dots,\partial_w^{n-k+2}\zeta/\partial_w\zeta)\\
\end{aligned}
\end{equation}
The quantities $B_{n,k}(a_1,\dots,a_{n-k+1})$ are {\it Bell polynomials} \cite{Riordan58,Andrews}. 
Since the quantity,
$$
x_n\dfn \sum_{k=1}^n (-)^{k+1}(k-1)!B_{n,k}(y_1,y_2,\dots,y_{n-k+1}),
$$
can be inverted using a standard Bell polynomial identity,
\begin{equation}
\begin{aligned}
y_n&=\sum_{k=1}^nB_{n,k}(x_1,x_2,\dots,x_{n-k+1})\\
&=B_{n}(x_1,x_2,\dots,x_n)
\end{aligned}
\end{equation}
it follows that we can invert (\ref{eq:dwnlnrho0}),
\begin{equation}
\begin{aligned}
\partial_w^{n+1}\zeta(p)&=B_{n}\big(\partial_w\ln\rho_0,\dots,\partial_w^{n}\ln\rho_0\big)\partial_w\zeta(p),\qquad n\geq1
\end{aligned}
\end{equation}
where $B_n(a_1,\dots,a_n)$ are {\it complete Bell polynomials} \cite{Riordan58,Andrews}. 
Evidently, although this relation was derived for $n\geq1$ it also holds for $n=0$ since $B_0=1$. This relation then determines $\zeta(w)$ by Taylor expansion around $p$.
\sk

So we have shown that given any metric, $\rho_0(w,\bar{w})$, in a local chart associated to a local patch $U\in \Sigma$ we can always find a frame $(U,\zeta)$, obtained from the conformally-flat frame $(U,w)$ by a holomorphic reparametrisation, $w\mapsto w(\zeta)$, such that at a point $p\in U$ the metric is `as flat as possible', i.e.~satisfies (\ref{eq:PolchinskicoordsA}), and when expanded around $p\in U$, at a point $p'\in U$ it takes the form:
\begin{equation}\label{eq:zetap'p}
\begin{aligned}
&\zeta_p(p')=\sum_{n=0}^{\infty}\frac{1}{n!}(w_p(p')-w_p(p))^n\partial_w^n\zeta_p(p)\\
&\quad=\zeta_p(p)+\sum_{n=1}^{\infty}\frac{1}{n!}(w_p(p')-w_p(p))^nB_{n-1}\big(\partial_w\ln\rho_0(p),\dots,\partial_w^{n-1}\ln\rho_0(p)\big)\partial_w\zeta_p(p)
\end{aligned}
\end{equation}
where by $\partial_w^n\zeta_p(p)$ we really mean $(\partial^n\zeta_p(p')/\partial w_p(p')^n)|_{p'=p}$.

We have not yet enforced the $n=0$ condition in the defining relations (\ref{eq:PolchinskicoordsA}) of the holomorphic coordinate transformation $w,\bar{w}\mapsto \zeta(w),\bar{\zeta}(\bar{w})$. From (\ref{eq:lnrho}) we see that this determines the magnitude $|\partial_w\zeta|$ but not the phase, $\alpha$, where:
$$
\partial_w\zeta=e^{i\alpha}|\partial_w\zeta|,
$$
and so according to (\ref{eq:lnrho}) we have:
$$
\partial_w\zeta_p(p)=e^{i\alpha(p)}\sqrt{\rho_0}(p),
$$
and note furthermore that $\rho_0$ is real and positive definite. As shown in Sec.~\ref{sec:WB}, the phase $\alpha$ is in general not globally-defined, the obstruction being the Euler number, and so to circumvent this one can \cite{Polchinski88} always take this phase to be integrated in string amplitude computations or consider combinations of operators that are independent of this phase. We have:
\begin{equation}\label{eq:zetap'pbx}
\begin{aligned}
\zeta_p(&p')=\sum_{n=0}^{\infty}\frac{1}{n!}(w_p(p')-w_p(p))^n\partial_w^n\zeta_p(p)\\
\end{aligned}
\end{equation}
If we wish to take the coordinate $\zeta$ to be centred at $p$ then we can enforce $\zeta_p(p)=0$. We are also free to choose $w_p(p)$. 
Taking the above considerations into account the expansion (\ref{eq:zetap'pbx}) for the holomorphic (in $p'\in U$, it is {\it not} holomorphic in $p\in U$ since $\rho_0(p)$ is not holomorphic) coordinate $\zeta_p(p')$ takes the form:
\begin{equation}\label{eq:zetap'p2}
\boxed{
\begin{aligned}
\,\,\zeta_p(p')&=\zeta_p(p)+e^{i\alpha(p)}\sqrt{\rho_0(p)}\sum_{n=0}^{\infty}\frac{1}{(n+1)!}B_{n}\big(\partial_w^{s}\ln\rho_0(p)\big)(w_p(p')-w_p(p))^{n+1}\!\!
\end{aligned}
}
\end{equation}
where we also shifted $n\rightarrow n+1$. Note that $\zeta_p(p')$ is a different function for each $p$. Unless otherwise stated we adopt the convention to centre the frame such that $\zeta_p(p)=0$. 
\sk

To summarise, given any conformally-flat metric, $g_J=\rho_0(w,\bar{w})\rmd w\rmd\bar{w}$, with centred conformal coordinate, $w$, the holomorphic coordinate transformation $w\mapsto \zeta(w)$ given by (\ref{eq:zetap'p2}) gives rise to a new conformal metric, $g_J=\rho(\zeta,\bar{\zeta})\rmd\zeta \rmd\bar{\zeta}$, which (at a distinguished point $p\in\Sigma$) satisfies the properties (\ref{eq:PolchinskicoordsA}). This holomorphic coordinate transformation, $w\mapsto \zeta(w)$, leaves the phase, $\alpha(p)$, of $\zeta$ undetermined. 
We refer to the centred system of conformal coordinates, $(\zeta,\bar{\zeta})$, as `{\it holomorphic normal coordinates}'. (This is not standard terminology but it seems appropriate given the close analogy with Riemann normal coordinates.) {\it Unless} specified otherwise, throughout the entire document these coordinates will be denoted by $(z_m,\bar{z}_m)$ (equivalently $(z_{\sigma_m}(\sigma),\bar{z}_{\sigma_m}(\sigma))$ when we want to specify the base point, $\sigma_m$, and frame in terms of an underlying auxiliary coordinate, $\sigma$, and such that $z_{\sigma_m}(\sigma_m)\equiv \bar{z}_{\sigma_m}(\sigma_m)\equiv 0$), but we will try to always clarify this within the various subsections to avoid confusion.

\subsubsection{Shifting Punctures with Transition Functions}\label{sec:SPUTF}
Recall from Sec.~\ref{sec:TFCR} and Sec.~\ref{sec:CSDI} that we may think of a Riemann surface as a complex manifold, which is in turn fully specified by an {\it atlas}, i.e.~a collection of holomorphic charts, 
\begin{equation}\label{eq:Umzm}
\{(U_m,z_m),(U_n,z_n),\dots\},
\end{equation}
such that $\mathscr{U}=\bigcup_\ell U_\ell$ is a cover for $\Sigma$, with holomorphic chart coordinates corresponding to one-to-one maps, $z_m:U_m\rightarrow \mathbf{C}$, such that on every non-empty intersection $U_m\cap U_n$ we identify $z_n$ with $z_m$ provided they are related by biholomorphic (and hence invertible) transition functions,
\begin{equation}\label{eq:transitionfuncsX}
z_m=f_{mn}(z_n;t).
\end{equation}
In this formulation (recall Sec.~\ref{sec:CSDI}) although the transition functions (\ref{eq:transitionfuncsX}) are holomorphic in $z_n$ (for every $m,n$) they are not necessarily holomorphic in the moduli, $t=(t^1,t^2,\dots)$. Recalling the discussion in Sec.~\ref{sec:ITOCR}, in particular (\ref{eq:phimn(t)}), in this section we will derive an explicit expression for the vector,\footnote{Since only the first derivative of $f_{mn}(z_n;t)$ with respect to $t$ appears in string amplitudes we need only consider first order variations thereof.)}
$$
\phi_{mn}(t)=\frac{\partial f_{mn}(z_n,t)}{\partial t}\Big\vert_{z_n}\frac{\partial }{\partial z_m},
$$
in the specific case that the zeros or base points, $z_m=0$ and $z_n=0$, of the two holomorphic charts $(U_m,z_m)$ and $(U_n,z_n)$ are related by a {\it translation}. Roughly speaking, we wish to associate a modulus $t^k$ to this translation. A slightly more precise statement is that we wish to associate a modulus to the {\it base point} with respect to which the relevant holomorphic coordinate, say $z_n$, has been defined. This will then allow us (in later sections) to define a local operator at, say, $z_n=0$ (and normal-ordered with respect to the $(U_n,z_n)$ chart), and the transition function under consideration will then be used to construct the corresponding path integral measure that is associated to a gauge slice in moduli space that generates translations of this operator across a Riemann surface. 
\sk

One way of making the geometrical aspects of these introductory statements precise is to introduce an auxiliary {\it fixed} coordinate system, $\sigma^a$, with respect to which the aforementioned specified charts can be defined, more about which momentarily.\footnote{The introduction of an auxiliary coordinate system is not necessary, but introducing it here is useful since it does offer geometrical insight while tying together various results from preceding sections. In Sec.~\ref{sec:TP} we translate punctures (and by extension `handle operators') without introducing an auxiliary coordinate. The resulting amplitudes are of course the same as those derived using an auxiliary coordinate system.} We do not want to restrict the scalar curvature, $R_{(2)}$, of the underlying surface in any way, since this will then lead to a construction that is automatically globally well-defined in moduli space (while avoiding the need \cite{Polchinski87,Nelson89} to introduce Wu-Yang-type contributions across boundaries of the open sets associated to the cover $\mathscr{U}$ of interest \cite{Polchinski88}). A simple way to proceed will be to identify $z_m,z_n$ with holomorphic normal coordinates (see Sec.~\ref{sec:PolCoords}).
\sk

\begin{figure}
\begin{center}
\includegraphics[angle=0,origin=c,width=0.9\textwidth]{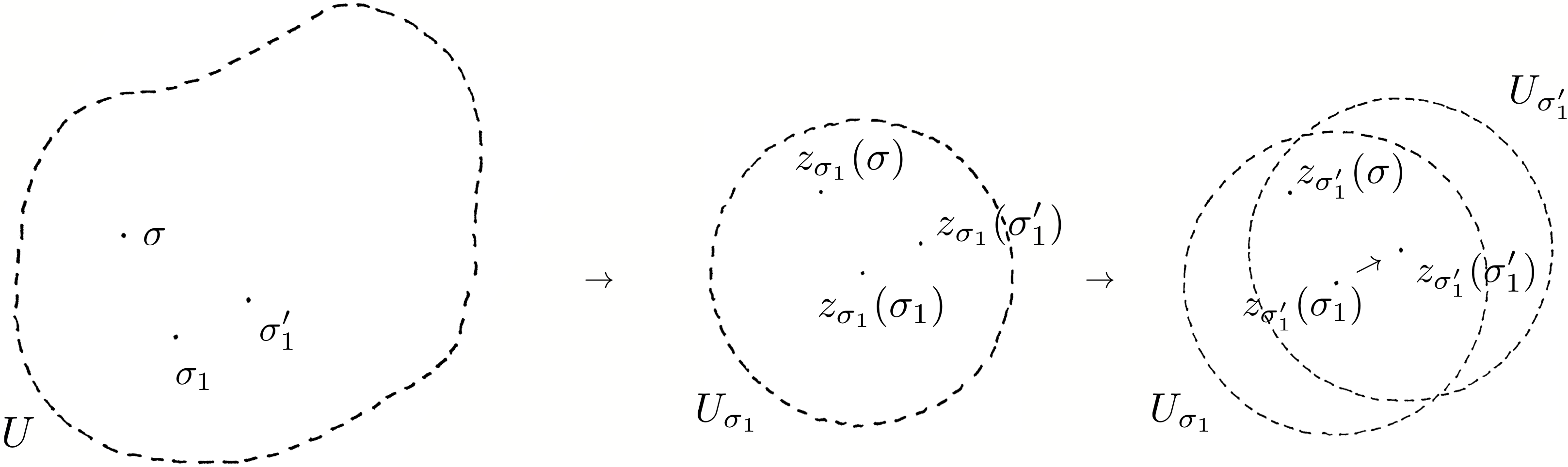}
\caption{\underline{\it Left sketch}: Three (real) coordinate points, $\sigma,\sigma_1,\sigma_1'$, specified by a chart $(U,\sigma)$, associated to an open set $U\subset\Sigma$ and chosen as follows. We pick two arbitrary coordinate points $\sigma,\sigma_1$ within the chart and construct a third, $\sigma_1'$, such that: $\sigma_1'\equiv \sigma_1+\delta\sigma_1$, with $\delta \sigma_1$ infinitesimal. \underline{\it Middle sketch}: The same three points specified in terms of a holomorphic chart, $(U_{\sigma_1},z_{\sigma_1})$, that is centred at $\sigma_1$, i.e.~$z_{\sigma_1}(\sigma_1)\dfn0$. \underline{\it Right sketch}:  The same three points specified in terms of a shifted holomorphic chart, $(U_{\sigma_1'},z_{\sigma_1'})$, that is centred at $\sigma_1'$, i.e.~$z_{\sigma_1'}(\sigma_1')\dfn0$. All three points are within the overlap $U_{\sigma_1'}\cap U_{\sigma_1}\neq\zero$.}\label{fig:trans}
\end{center}
\end{figure}

So let us consider a real {\sl auxiliary coordinate chart} (associated to an atlas with smooth and real transition functions), 
$$
(U,\sigma^a),
$$
(with $a=1,2$) of $\Sigma$, so that $\sigma$ maps arbitrary points of an open set $U\subset\Sigma$ to $\mathbf{R}^2$. Being a real coordinate chart, $(U,\sigma)$ is not an element of the holomorphic atlas (\ref{eq:Umzm}), but we will make contact with the latter momentarily. 
We can use this chart $(U,\sigma)$ to single out two arbitrary coordinate points, $\sigma^a,\sigma_1^a$, and then choose a third $\sigma_1^{'a}$ such that:
$$
\sigma_1'=\sigma_1+\delta \sigma_1,
$$ 
with $\delta \sigma_1$ ``small''. (A corresponding diagram is shown in the left sketch in Fig.~\ref{fig:trans}.) Let us then construct two overlapping open sets, $U_{\sigma_1}$ and $U_{\sigma_1'}$, (with a convenient labelling that is explained in the next paragraph) such that $U_{\sigma_1}\cup U_{\sigma_1'}\subset U$ while the three aforementioned points, $\sigma^a,\sigma_1^a$ and $\sigma_1^{'a}$, are coordinate points associated to the (non-empty) overlap, $U_{\sigma_1}\cap U_{\sigma_1'}\subset U$. Generalising, we  similarly construct an atlas of such auxiliary coordinate systems, and require that the corresponding transition functions on patch overlaps be {\sl smooth} (but for notational simplicity we can keep this implicit for now).
\sk

We now want to make the link between the real smooth atlas of the previous paragraph to the holomorphic charts in (\ref{eq:Umzm}) (and the corresponding notation adopted in Sec.~\ref{sec:TFCR}, \ref{sec:ITOCR}, \ref{sec:MSER} and Sec.~\ref{sec:CSDII}). 
To do so, we want to use the aforementioned two open sets $U_{\sigma_1}$ and $U_{\sigma_1'}$ to define two holomorphic charts, 
$$
(U_{\sigma_1},z_{\sigma_1}),\qquad{\rm and}\qquad (U_{\sigma_1'},z_{\sigma_1'}),
$$ 
with holomorphic coordinates $z_{\sigma_1}$ and $z_{\sigma_1'}$ respectively. Loosely speaking, we want to identify {\it these} with elements of the holomorphic atlas (\ref{eq:Umzm}), but a more precise statement is the following. Given an auxiliary chart $(U,\sigma)$ with $\sigma:U\rightarrow \mathbf{R}^2$ (associated to a real atlas with smooth transition functions), a chart $(U_1,z_1)$ with $z_1:U_1\rightarrow \mathbf{C}$ (associated to a complex atlas with holomorphic transition functions), and assuming that $U\cap U_1\subset\Sigma$ is non-empty, we define the chart $(U_{\sigma_1},z_{\sigma_1})$ via the composition, 
$$
z_{\sigma_1}\dfn z_1\circ \sigma^{-1},\qquad \textrm{such that}\qquad z_{\sigma_1}(\sigma_1)=0.
$$
Consistency with our convention that $z_1(p_1)=0$ determines $\sigma_1=\sigma(p_1)$. Notice that $z_{\sigma_1}$ is well-defined on $U\cap U_1$, so we might as well define, $U_{\sigma_1}\dfn U\cap U_1$. From this viewpoint it is clear that $z_1(p)=z_{\sigma_1}(\sigma)$, it being understood that $\sigma\equiv\sigma(p)$, 
so for all intents and purposes we can use these two viewpoints interchangeably:
\begin{equation}\label{eq:Us1zs1=U1z1}
\boxed{(U_{\sigma_1},z_{\sigma_1}(\sigma))\leftrightarrow (U_{1},z_{1}(p)),\qquad {\rm and}\qquad p_1\leftrightarrow\sigma_1,\quad p\leftrightarrow \sigma}
\end{equation}
with $p,p_1\in U\subset \Sigma$ abstractly denoting points on the Riemann surface and $\sigma^a,\sigma_1^a$ (respectively) denoting auxiliary coordinate representations of these same points. 
\sk

We can proceed in an entirely analogous manner to define the neighbouring shifted chart, $(U_{\sigma_1'},z_{\sigma_1'})$. This too can be defined using the chart $(U,\sigma)$;  in terms of a ``sufficiently close'' shifted holomorphic chart $(U_{1'},z_{1'})$ in the holomorphic atlas (\ref{eq:Umzm}), we define $z_{\sigma_1'}\dfn z_{1'}\circ \sigma^{-1}$. Again, we might as well take $U_{\sigma_1'}=U\cap U_{1'}$, and also $U_{\sigma_1}\cup U_{\sigma_1'}\subset U$ and $U_{\sigma_1}\cap U_{\sigma_1'}$ non-empty  as shown in Fig.~\ref{fig:trans}. 
A point to emphasise is that the subscripts in $U_{\sigma_1}$ and $U_{\sigma_1'}$ denote the auxiliary coordinate representation of the {\it base points} at which the holomorphic coordinates vanish: 
$$
z_{\sigma_1}(\sigma)\big|_{\sigma=\sigma_1}\equiv z_{\sigma_1}(\sigma_1)\dfn 0,\qquad {\rm and}\qquad z_{\sigma_1'}(\sigma)\big|_{\sigma=\sigma_1'}\equiv z_{\sigma_1'}(\sigma_1')\dfn 0.
$$

Suppose then that a generic point $\sigma$ in the two holomorphic charts, $(U_{\sigma_1},z_{\sigma_1})$ and $(U_{\sigma_1'},z_{\sigma_1'})$, is represented by $z_{\sigma_1}(\sigma)$ and $z_{\sigma_1'}(\sigma)$ respectively. In the language of Sec.~\ref{sec:NACS}, and to make the link to corresponding complex structures, $J$ and $J'$, see (\ref{eq:w-gen}), that these coordinates are holomorphic means precisely that they satisfy the differential equations:
\begin{equation}\label{eq:z-genX}
\big(\partial_a+iJ_a^{\phantom{a}b}\partial_b\big)z_{\sigma_1}(\sigma)=0,\qquad{\rm and}\qquad \big(\partial_a+i{J'}_{\!\!a}^{\!\phantom{a}b}\partial_b\big)z_{\sigma_1'}(\sigma)=0,
\end{equation} 
which follows from the defining property of an almost complex structure as being a tangent space endomorphism that squares to minus one. 
So to different points within (say) chart $(U_{\sigma_1},z_{\sigma_1})$ and for {\it fixed} base point, $\sigma_1$, there corresponds a holomorphic coordinate, $z_{\sigma_1}(\sigma)$, for every point $\sigma$ in the relevant domain. 
\sk

We can now make one of the key statements in the opening paragraph of this section precise: we wish to identify the {\it base point} $\sigma_1$ with one of the {\it moduli}, $t=(\sigma_1^a,\dots)$, whose variations, $\delta\sigma_1=\sigma_1'-\sigma_1$,  generate complex structure deformations. This variation will correspondingly be identified with a variation, $\delta t$. To proceed, notice that since there is non-empty overlap $U_{\sigma_1}\cap U_{\sigma_1'}$ we may by definition construct a holomorphic transition function, $f_{\sigma_1'\sigma_1}$,
\begin{equation}\label{eq:zsigmafss}
\boxed{
\begin{aligned}
z_{\sigma_1'}(\sigma)&=f_{\sigma_1'\sigma_1}(z_{\sigma_1}(\sigma);\sigma_1)\\
&\equiv z_{\sigma_1}(\sigma)+\delta z_{\sigma_1}(\sigma),
\end{aligned}
}
\end{equation}
where $\sigma :U_{\sigma_1}\cap U_{\sigma_1'}\rightarrow \mathbf{R}^2$. Since $z_{\sigma_1'}(\sigma)$ and $z_{\sigma_1}(\sigma)$ are both holomorphic coordinate representations of the same point $\sigma$ but centred or based at different points $\sigma_1'$ and $\sigma_1$ respectively, by one of the defining properties of transition functions for complex manifolds (see Sec.~\ref{sec:TFCR}), the quantity,
\begin{equation}\label{eq:dzs*}
\phantom{\quad{\rm (passive)}}\boxed{
\delta z_{\sigma_1}(\sigma)\equiv z_{\sigma_1'}(\sigma)-z_{\sigma_1}(\sigma),\qquad {\rm with}\qquad \sigma_1'\dfn \sigma_1+\delta\sigma_1
}
\end{equation}
will be a {\it holomorphic} function of $z_{\sigma_1}(\sigma)$. Notice that this variation (\ref{eq:dzs*}) is ``passive'', in that it corresponds to shifting the location of the {\it frame} (parametrised by $\sigma_1\mapsto \sigma_1'$) keeping the coordinate, $\sigma$, {\it fixed}. (It will occasionally be convenient to switch to an ``active'' variation, whereby we parametrise the variation as a shift in $\sigma$ keeping the location, $\sigma_1$, of the frame fixed, more about which below.) Furthermore, since this variation (\ref{eq:dzs*}) is defined at a fixed and generic point $\sigma$, we can also set $\sigma=\sigma_1$ if we so please and (\ref{eq:dzs*}) then provides meaning to the variation $\delta z_{\sigma_1}(\sigma_1)$ despite the fact that $z_{\sigma_1'}(\sigma_1')\equiv0\equiv z_{\sigma_1}(\sigma_1)$. 
\sk

Elaborating on this last statement, a quantity that will play a particularly important role will be the variation, $\delta z_{\sigma_1}(\sigma_1)$, generated by the base-point  shift $\sigma_1\mapsto \sigma_1'=\sigma_1+\delta\sigma_1$. We can use (\ref{eq:dzs*}) to write $\delta z_{\sigma_1}(\sigma_1)$ in terms of the variation, $\delta\sigma_1$, as follows. Since we have two holomorphic charts, $(U_{\sigma_1},z_{\sigma_1})$ and $(U_{\sigma_1'},z_{\sigma_1'})$, we can read off the ``coordinate distance in moduli space'', $\sigma_1'-\sigma_1$, with respect to each of these directly from Fig.~\ref{fig:trans}. In the $(U_{\sigma_1},z_{\sigma_1})$ chart the infinitesimal quantity $\sigma_1'-\sigma_1$ is mapped to:
\begin{equation}\label{eq:active}
\begin{aligned}
z_{\sigma_1}(\sigma_1')-z_{\sigma_1}(\sigma_1)&=z_{\sigma_1}(\sigma_1+\delta \sigma_1)-z_{\sigma_1}(\sigma_1)\\
&\simeq z_{\sigma_1}(\sigma_1)+\delta \sigma_1^a\frac{\partial z_{\sigma_1}(\sigma)}{\partial \sigma^a}\Big|_{\sigma=\sigma_1}-z_{\sigma_1}(\sigma_1)\\
&=\delta \sigma_1^a\frac{\partial z_{\sigma_1}(\sigma)}{\partial \sigma^a}\Big|_{\sigma=\sigma_1}
\end{aligned}
\end{equation}
Similarly, in the $(U_{\sigma_1'},z_{\sigma_1'})$ chart the quantity $\sigma_1'-\sigma_1$ is mapped to:
\begin{equation}\label{eq:zs'*s'*-zs'*s*}
\begin{aligned}
z_{\sigma_1'}(\sigma_1')-z_{\sigma_1'}(\sigma_1)
&=\delta \sigma_1^a\frac{\partial z_{\sigma_1'}(\sigma)}{\partial \sigma^a}\Big|_{\sigma=\sigma_1}
\end{aligned}
\end{equation}
Recalling that $z_{\sigma_1'}(\sigma_1')=z_{\sigma_1}(\sigma_1)=0$, we then evaluate (\ref{eq:dzs*}) at $\sigma=\sigma_1$ by making use of (\ref{eq:zs'*s'*-zs'*s*}):
\begin{equation}\label{eq:active}
\begin{aligned}
\delta z_{\sigma_1}(\sigma_1)&= z_{\sigma_1'}(\sigma_1)-z_{\sigma_1}(\sigma_1)\\
&= \Big(z_{\sigma_1'}(\sigma_1')-\delta \sigma_1^a\frac{\partial z_{\sigma_1'}(\sigma)}{\partial \sigma^a}\Big|_{\sigma=\sigma_1}\Big)-z_{\sigma_1}(\sigma_1)\\
&= -\delta \sigma_1^a\frac{\partial z_{\sigma_1'}(\sigma)}{\partial \sigma^a}\Big|_{\sigma=\sigma_1}\\
\end{aligned}
\end{equation}
with an important minus sign. 
Taking into account (\ref{eq:dzs*}) once more we learn that:
\begin{equation}\label{eq:dz*=-ds*e}
\boxed{
\delta z_{\sigma_1}(\sigma_1)=-\delta \sigma_1^a\frac{\partial z_{\sigma_1}(\sigma)}{\partial \sigma^a}\Big|_{\sigma=\sigma_1}
}
\end{equation}
where note that the variation on the left-hand side is to be interpreted as above, namely we first compute the variation at {\it fixed} $\sigma$ as in (\ref{eq:dzs*}) and then evaluate it at $\sigma=\sigma_1$. 
\sk

Regarding the derivative on the right-hand side of (\ref{eq:dz*=-ds*e}) we primarily emphasise that by construction $\sigma$ and $\sigma_1$ are independent. In arriving at (\ref{eq:dz*=-ds*e}) we neglected terms of higher order in the variation since we {\it only} need the infinitesimal case\footnote{The reason we only need infinitesimal variations boils down to a standard result in differential geometry that we can compute the Jacobian associated to a change of coordinates (that in our case is associated to choosing a specific gauge slice in the path integral measure encoded in the $\hat{B}_k$ insertions, so it is essentially the Fadeev-Popov determinant) in either the tangent space, $T\mathcal{M}$, or base space, $\mathcal{M}$, the result being independent of this choice. By considering only first order variations we implicitly have in mind that we will be computing the contribution to the Fadeev-Popov determinant associated to translating punctures or handle operators in the {\it tangent space}, $T\mathcal{M}$. We then use this same expression for the $\hat{B}_k$ insertions to integrate over $\mathcal{M}$.} where $\delta\sigma_1\rightarrow0$. The quantity:
\begin{equation}\label{eq:framefield2}
\boxed{
e_a^{z_{\sigma_1}}(\sigma_1)\dfn \frac{\partial z_{\sigma_1}(\sigma)}{\partial \sigma^a}\Big|_{\sigma=\sigma_1}=-\frac{\partial z_{\sigma_1}(\sigma_1)}{\partial \sigma_1^a}
}
\end{equation}
will play the role of a {\it frame field} \cite{Polchinski88}. The first equality is a definition whereas the second equality follows from the chain rule applied to the left-hand side in (\ref{eq:dz*=-ds*e}). Taking into account that this variation, $\delta z_{\sigma_1}(\sigma_1)$, is evaluated at {\it fixed} $\sigma$ as indicated in (\ref{eq:dzs*}), we could more precisely write the quantity appearing on the right-hand side of the second equality in (\ref{eq:framefield2}) as follows,
$$
\frac{\partial z_{\sigma_1}(\sigma_1)}{\partial \sigma_1^a}\equiv \frac{\partial z_{\sigma_1}(\sigma)}{\partial \sigma_1^a}\Big|_{\sigma=\sigma_1}.
$$

Regarding the frame field (\ref{eq:framefield2}), when we evaluate (\ref{eq:z-genX}) at $\sigma=\sigma_1$ we see that it satisfies,
$$
e_a^{z_{\sigma_1}}(\sigma_1)+iJ_a^{\phantom{b}b}(\sigma_1)e_b^{z_{\sigma_1}}(\sigma_1)=0,
$$
and since $J_a^{\phantom{b}b}(\sigma)$ transforms as the components of a $(1,1)$ tensor at any point $\sigma$ under $\sigma\mapsto \hat{\sigma}(\sigma)$, so will $J_a^{\phantom{b}b}(\sigma_1)$ transform in the same manner under reparametrisations of the {\it base point}, $\sigma_1\mapsto \hat{\sigma}_1(\sigma_1)$, in particular,
$$
J_a^{\phantom{b}b}(\sigma_1)\mapsto \hat{J}_a^{\phantom{b}b}(\hat{\sigma}_1)=\frac{\partial \sigma_1^c}{\partial \hat{\sigma}_1^a}\frac{\partial \hat{\sigma}_1^b}{\partial \sigma_1^d}J_c^{\phantom{b}d}(\sigma_1),\qquad{\rm under}\qquad \sigma_1\mapsto \hat{\sigma}_1(\sigma_1).
$$
From the above two displayed equations we therefore learn that the quantities $e_a^{z_{\sigma_1}}(\sigma_1)$ transform as the components of a one form,
\begin{equation}\label{eq:eaz-reparam}
e_a^{z_{\sigma_1}}(\sigma_1)\mapsto e_a^{z_{\hat{\sigma}_1}}(\hat{\sigma}_1)=\frac{\partial \sigma_1^b}{\partial \hat{\sigma}_1^a}e_b^{z_{\sigma_1}}(\sigma_1),\qquad{\rm under}\qquad \sigma_1\mapsto \hat{\sigma}_1(\sigma_1).
\end{equation}
But in order for this to be true, and according to the {\it second} equality in (\ref{eq:framefield2}), it must be that $z_{\sigma_1}(\sigma_1)$ transforms as a {\it scalar} under such reparametrisations,
\begin{equation}\label{eq:zsigma* scalar}
\boxed{
z_{\sigma_1}(\sigma_1)\mapsto z_{\hat{\sigma}_1}(\hat{\sigma}_1)=z_{\sigma_1}(\sigma_1),\qquad{\rm under}\qquad \sigma_1\mapsto \hat{\sigma}_1(\sigma_1)
}
\end{equation}
So the statement that $z_{\sigma_1}(\sigma)$ is a centred coordinate satisfying $z_{\sigma_1}(\sigma_1)=0$ has coordinate invariant meaning.  The reparametrisations $\sigma_1\mapsto \hat{\sigma}(\sigma_1)$ inherited from $\sigma\mapsto \hat{\sigma}(\sigma)$ will be associated to reparametrisations in moduli space when we associate the base point $\sigma_1$ (at which a local operator is inserted) to a modulus. See also the discussion below, the paragraph containing (\ref{eq:dzv=ds1s1}). These discussions are at the heart of the reparametrisation invariance of the gauge slice in moduli space that we are constructing.
\sk

It will also be useful to consider the measure generated by variations (\ref{eq:dz*=-ds*e}) of the base point. Taking also the definition in (\ref{eq:framefield2}) into account,
\begin{equation}\label{eq:dzsigma*-measure}
\begin{aligned}
\rmd^2z_{\sigma_1}(\sigma_1)&\equiv i\rmd z_{\sigma_1}(\sigma_1)\wedge \rmd\bar{z}_{\sigma_1}(\sigma_1) \\
&=i\rmd\sigma_1^1\wedge \rmd\sigma_1^2\big(e_1^{z_{\sigma_1}}(\sigma_1)e_2^{\bar{z}_{\sigma_1}}(\sigma_1)-e_2^{z_{\sigma_1}}(\sigma_1)e_1^{\bar{z}_{\sigma_1}}(\sigma_1)\big)\\
&=\rmd\sigma_1^1\wedge \rmd\sigma_1^2\,2\sqrt{\det g_{ab}(\sigma_1)}\\
&\equiv \rmd^2\sigma_1 2\sqrt{g(\sigma_1)},
\end{aligned}
\end{equation}
where the first equality is a standard convention \cite{Polchinski_v1} that we use throughout for complex coordinates.\footnote{Notice that reparametrisation invariance of the measure in (\ref{eq:dzsigma*-measure}) is manifest on the right-hand side of the last equality and indicates that the left-hand side, which by (\ref{eq:dzv=ds1s1}) we write as $\rmd^2z_{v_1}$, is also to be considered a reparametrisation-invariant measure. This is consistent with the above comments because $z_{\sigma_1}(\sigma_1)$ or $z_v$ transforms as a scalar as indicated in (\ref{eq:zsigma* scalar}).} The overall sign in the last two equalities is chosen by hand such that area elements are positive, and we defined the quantity:
\begin{equation}\label{eq:gabzs*}
\boxed{
g_{ab}(\sigma_1)\dfn \frac{1}{2}\big(e_a^{z_{\sigma_1}}(\sigma_1)e_b^{\bar{z}_{\sigma_1}}(\sigma_1)+e_a^{\bar{z}_{\sigma_1}}(\sigma_1)e_b^{z_{\sigma_1}}
(\sigma_1)\big)
}
\end{equation}
which can be identified with the components of a {\it metric tensor} in an auxiliary coordinate system associated to the frame field $e_a^{z_{\sigma_1}}(\sigma_1)$.\footnote{The factor of $1/2$ in (\ref{eq:gabzs*}) is associated to taking the ``tangent space metric'' in complex structure, $I$, to be $g_{z_{\sigma_1}\bar{z}_{\sigma_1}}^I(\sigma_1)=1/2=\rho_I(\sigma_1)/2$, and as we discuss below we will require that this remains true under shifts or reparametrisations of the base point, $\sigma_1$.} It also transforms as such under general reparametrisations (which follows from (\ref{eq:eaz-reparam})), and in particular the right-hand side of the last equality in (\ref{eq:dzsigma*-measure}) is clearly invariant under general reparametrisations $\sigma_1\mapsto \hat{\sigma}_1(\sigma_1)$, and therefore so must the left-hand side be. The latter is precisely the statement that $z_{\sigma_1}(\sigma_1)$ transforms as a scalar as in (\ref{eq:zsigma* scalar}). 
Notice that the frame field or metric (\ref{eq:gabzs*}) contains only a small amount of information about the frame, $z_{\sigma_1}(\sigma)$, namely its first derivative \cite{Polchinski88}.
\sk

Let us as usual define the inverse metric by $g^{ab}g_{bc}\dfn\delta^a_{\phantom{a}c}$, and since $g_{ab}$ and hence also the inverse, $g^{ab}$, are symmetric under $a\leftrightarrow b$, it follows from (\ref{eq:gabzs*}) that,\footnote{These two relations are equivalent (respectively) to: $g^{z_{\sigma_1}\bar{z}_{\sigma_1}}_I(\sigma_1)=2$ and $g^{z_{\sigma_1}z_{\sigma_1}}_I(\sigma_1)=0=g^{\bar{z}_{\sigma_1}\bar{z}_{\sigma_1}}_I(\sigma_1)$.}
\begin{equation}\label{eq:gab relations}
\begin{aligned}
e_a^{z_{\sigma_1}}(\sigma_1)g^{ab}(\sigma_1&)e_b^{\bar{z}_{\sigma_1}}(\sigma_1)=2,\\
e_a^{z_{\sigma_1}}(\sigma_1)g^{ac}(\sigma_1)e_c^{z_{\sigma_1}}(\sigma_1)=&\,0=e_a^{\bar{z}_{\sigma_1}}(\sigma_1)g^{ac}(\sigma_1)e_c^{\bar{z}_{\sigma_1}}(\sigma_1).
\end{aligned}
\end{equation}
The last two relations in (\ref{eq:gab relations}) follow from (\ref{eq:gabzs*}) after multiplying left- and right-hand sides of the latter by $e_c^{z_{\sigma_1}}$, contracting both sides by $g^{ac}$ and making use of the first relation in (\ref{eq:gab relations}). All quantities in (\ref{eq:gab relations}) are evaluated at $\sigma_1$.
\sk

Another aspect that is now completely transparent in (\ref{eq:dzsigma*-measure}) is that Weyl rescalings of the metric, $g_{ab}(\sigma_1)\mapsto e^{\delta\phi(\sigma_1)}g_{ab}(\sigma_1)$, can be absorbed into {\it holomorphic} reparametrisations, $z_{\sigma_1}\mapsto w_{\sigma_1}(z_{\sigma_1})$ that leave fixed the base point $\sigma=\sigma_1$. We will discuss Weyl rescalings in Sec.~\ref{sec:HTFWR} in detail.
\sk

After this brief interlude on frame fields, we now wish to compute the variation $\delta z_{\sigma_1}(\sigma)$ in (\ref{eq:dzs*}) (under $\sigma_1\mapsto \sigma_1+\delta\sigma_1$ at fixed $\sigma$) of the full transition function (\ref{eq:zsigmafss}). But we do not yet have enough information to determine it uniquely: it remains to specify how the gauge slice is to vary as we shift the base point $\sigma_1$. We specify this by primarily identifying $z_{\sigma_1}(\sigma)$ with a {\it holomorphic normal coordinate} (corresponding to a metric that is `as flat as possible' at the base point $\sigma_1$), and then the dependence on the gauge slice as we shift $\sigma_1$ will be completely determined (up to an immaterial phase) if we require that translations of the base point, $\sigma_1\mapsto \sigma_1'=\sigma_1+\delta\sigma_1$, {\it preserve this gauge slice}. 
\sk

To implement the statements in the preceding paragraph, let us recall the defining relations of holomorphic normal coordinates (\ref{eq:PolchinskicoordsA}), which in the notation of the current section read,
\begin{equation}\label{eq:PolchinskicoordsAX}
\partial_{z_{\sigma_1}}^n\rho_I(z_{\sigma_1},\bar{z}_{\sigma_1})\big|_{\sigma=\sigma_1}=\partial_{\bar{z}_{\sigma_1}}^n\rho_I(z_{\sigma_1},\bar{z}_{\sigma_1})\big|_{\sigma=\sigma_1}=
\left\{
\begin{array}{l}
\begin{aligned}
&1\qquad \textrm{if $n=0$}\\
&0\qquad \textrm{if $n\geq1$}
\end{aligned}
\end{array}
\right.
\end{equation}
where by $z_{\sigma_1}$ we actually mean $z_{\sigma_1}(\sigma)$ (and similarly $\bar{z}_{\sigma_1}=\bar{z}_{\sigma_1}(\sigma)$). So the derivative $\partial_{z_{\sigma_1}}$ appearing in (\ref{eq:PolchinskicoordsAX}) is with respect to $z_{\sigma_1}(\sigma)$ and subsequently evaluated at $\sigma=\sigma_1$, and similarly for the anti-holomorphic counterpart. 
We next draw from the formalism developed in Sec.~\ref{sec:DCCFM}, in particular (\ref{eq:changecoordsmetricfixed}) which gives the change in metric under arbitrary coordinate variations, repeated here with the notation appropriately adapted to the current context,
\begin{equation}\label{eq:changecoordsmetricfixedX}
\begin{aligned}
\delta \ln\rho_I\big|_{z_{\sigma_1}} &= \nabla_{z_{\sigma_1}}\delta z_{\sigma_1}+\nabla_{\bar{z}_{\sigma_1}}\delta\bar{z}_{\sigma_1}\\
\rho_I^{-1}\delta g_{z_{\sigma_1} z_{\sigma_1}}^I\big|_{z_{\sigma_1}}&=\nabla_{z_{\sigma_1}}\delta \bar{z}_{\sigma_1}\\
\rho_I^{-1}\delta g_{\bar{z}_{\sigma_1}\bar{z}_{\sigma_1}}^I\big|_{z_{\sigma_1}}&=\nabla_{\bar{z}_{\sigma_1}}\delta z_{\sigma_1},
\end{aligned}
\end{equation}
where all relations are evaluated at a generic point $\sigma$, so that as above $z_{\sigma_1}$ stands for $z_{\sigma_1}(\sigma)$. On the left-hand sides in (\ref{eq:changecoordsmetricfixedX}) we make explicit that the variations are evaluated at fixed $z_{\sigma_1}(\sigma)$. We will associate the depicted variations with variations in the base point $\sigma_1$ (which we think of as a modulus). 
Since (according to (\ref{eq:zsigmafss})) $\delta z_{\sigma_1}$ is holomorphic in $z_{\sigma_1}$ it follows that $\nabla_{z_{\sigma_1}}\delta \bar{z}_{\sigma_1}=\nabla_{\bar{z}_{\sigma_1}}\delta z_{\sigma_1}=0$ for all $\sigma$. Secondly, if the gauge slice associated to holomorphic normal coordinates is to be preserved {\it at} the location of the puncture it must be that also the variation $\delta \ln\rho_I\big|_{z_{\sigma_1}}$ vanishes {\it at} $\sigma=\sigma_1$ (but it need not vanish at a more generic point $\sigma\neq \sigma_1$). 
\sk

Taking these observations into account let us then consider the first relation in (\ref{eq:changecoordsmetricfixedX}), 
$
\delta \ln\rho_I|_{z_{\sigma_1}}= \nabla_{z_{\sigma_1}}\delta z_{\sigma_1}+\nabla_{\bar{z}_{\sigma_1}}\delta\bar{z}_{\sigma_1},
$ 
differentiate it with respect to $z_{\sigma_1}(\sigma)$ $n-1$ times (for $n\geq2$) taking into account (\ref{eq:zsigmafss}) (according to which $\partial_{z_{\sigma_1}}\delta\bar{z}_{\sigma_1}=0$ at all points $\sigma$), and evaluate the resulting quantity at $\sigma=\sigma_1$ taking into account (\ref{eq:PolchinskicoordsAX}),
\begin{equation}\label{eq:0=dn-1dzs}
\begin{aligned}
0&=\partial_{z_{\sigma_1}}^{n-1}\big(\nabla_{z_{\sigma_1}}\delta z_{\sigma_1}+\nabla_{\bar{z}_{\sigma_1}}\delta\bar{z}_{\sigma_1}\big)\big|_{\sigma=\sigma_1}\\
&=\partial_{z_{\sigma_1}}^n\delta z_{\sigma_1}(\sigma_1)-\frac{1}{4}\delta \bar{z}_{\sigma_1}(\sigma_1)\partial_{z_{\sigma_1}}^{n-2}R_{(2)}(\sigma_1),\\
\end{aligned}
\end{equation}
where we expanded out the covariant derivatives (see Appendix \ref{sec:CD}) taking into account that $\delta z_{\sigma_1}$ transforms as a vector, and made use of the following expression for the Ricci scalar, $R_{(2)}(\sigma)=-4\rho_I^{-1}\partial_{z_{\sigma_1}}\partial_{\bar{z}_{\sigma_1}}\ln\rho_I(\sigma)$. For clarity, note that the quantity $\partial_{z_{\sigma_1}}^n\delta z_{\sigma_1}(\sigma_1)$ is to be interpreted as $\partial_{z_{\sigma_1}(\sigma)}^n\delta z_{\sigma_1}(\sigma)|_{\sigma=\sigma_1}$ and similarly $\partial_{z_{\sigma_1}}^{n-2}R_{(2)}(\sigma_1)\equiv \partial_{z_{\sigma_1}(\sigma)}^{n-2}R_{(2)}(\sigma)|_{\sigma=\sigma_1}$. 
\sk

Next multiply the resulting expression in (\ref{eq:0=dn-1dzs}) by $\frac{1}{n!}z_{\sigma_1}(\sigma)^{n}$, sum over $n=2,3,\dots$, and rearrange, 
\begin{equation}\label{eq:n=2deltazss}
\begin{aligned}
\sum_{n=2}^\infty\frac{1}{n!}z_{\sigma_1}(\sigma)^{n}\partial_{z_{\sigma_1}}^n\delta z_{\sigma_1}(\sigma_1)=\delta \bar{z}_{\sigma_1}(\sigma_1)\frac{1}{4}\sum_{n=2}^\infty\frac{1}{n!}z_{\sigma_1}(\sigma)^{n}\nabla_{z_{\sigma_1}}^{n-2}R_{(2)}(\sigma_1),
\end{aligned}
\end{equation}
where we took into account (\ref{eq:PolchinskicoordsAX}) to replace the partial derivatives of the Ricci scalar by covariant derivatives. 
We can now include the $n=0,1$ terms in the sum on the left-hand side provided we subtract these terms out again, since then the resulting sum over $n=0,1,\dots$ is the {\it Taylor expansion} of $\delta z_{\sigma_1}(\sigma)$. Shifting also the summation index on the right-hand side in (\ref{eq:n=2deltazss}) we learn that:\begin{equation}\label{eq:deltazss'}
\begin{aligned}
\delta z_{\sigma_1}(\sigma)=\delta z_{\sigma_1}(\sigma_1)+z_{\sigma_1}(\sigma)\partial_{z_{\sigma_1}}\delta z_{\sigma_1}(\sigma_1)+\delta \bar{z}_{\sigma_1}(\sigma_1)\frac{1}{4}\sum_{n=1}^\infty\frac{1}{(n+1)!}z_{\sigma_1}(\sigma)^{n+1}\nabla_{z_{\sigma_1}}^{n-1}R_{(2)}(\sigma_1).
\end{aligned}
\end{equation}
Notice that the right-hand side of (\ref{eq:deltazss'}) is {\it holomorphic} in $z_{\sigma_1}(\sigma)$ but is generically {\it not} holomorphic in $\sigma_1$. The relation (\ref{eq:deltazss'}) describes how the holomorphic coordinate, $z_{\sigma_1}(\sigma)$, changes under variations of the base point, $\delta z_{\sigma_1}(\sigma_1),\delta\bar{z}_{\sigma_1}(\sigma_1)$,  and therefore determines how the holomorphic chart coordinate must vary in order to preserve the gauge slice and remain a holomorphic normal coordinate. So this ensures the metric remains `as flat as possible' at $\sigma_1$ while the puncture is translated across the surface. 
\sk

The first equality in (\ref{eq:0=dn-1dzs}) is also valid when $n=1$. So we see that $\partial_{z_{\sigma_1}}\delta z_{\sigma_1}(\sigma_1)+\partial_{\bar{z}_{\sigma_1}}\delta \bar{z}_{\sigma_1}(\sigma_1)=0$, implying that the real part of $\partial_{z_{\sigma_1}}\delta z_{\sigma_1}(\sigma_1)$ is zero. That is, the variation $\delta z_{\sigma_1}(\sigma_1)$ also generates an overall phase (which appears in the second term on the right-hand side of (\ref{eq:deltazss'})), and in particular (to leading order in the variation),
\begin{equation}\label{eq:zsds(s')}
\boxed{
\begin{aligned}
z_{\sigma_1+\delta\sigma_1}&(\sigma)=e^{i{\rm Im}\,\partial_{z_{\sigma_1}}\delta z_{\sigma_1}(\sigma_1)}\\
&\times\Big(z_{\sigma_1}(\sigma)+\delta z_{\sigma_1}(\sigma_1)+\delta \bar{z}_{\sigma_1}(\sigma_1)\frac{1}{4}\sum_{n=1}^\infty\frac{1}{(n+1)!}\big(\nabla_{z_{\sigma_1}}^{n-1}R_{(2)}(\sigma)\big)\Big|_{\sigma=\sigma_1}\!\!\!z_{\sigma_1}(\sigma)^{n+1}\Big)
\end{aligned}
}
\end{equation}
This phase, $e^{i{\rm Im}\,\partial_{z_{\sigma_1}}\delta z_{\sigma_1}(\sigma_1)}$, is base-point dependent, as was the analogous phase $e^{i\alpha(p)}$  in (\ref{eq:zetap'p2}) (or $e^{i\alpha(\sigma_1)}$ in the notation of the current section). Recall that the latter phase is arbitrary, it is undetermined by our gauge choice and furthermore (as discussed in Sec.~\ref{sec:WB}) it is generically not globally well-defined, the obstruction being the Euler number. So by adjusting $\alpha(\sigma_1)$ we can change ${\rm Im}\,\partial_{z_{\sigma_1}}\delta z_{\sigma_1}(\sigma_1)$, and since the former is arbitrary also the latter will not be meaningful. We circumvent this by always taking the phase of the frame coordinate to be integrated or by considering combinations that are independent of this phase as suggested in \cite{Polchinski88}.  Dropping this phase, (\ref{eq:deltazss'}) and its complex conjugate take the form:
\begin{equation}\label{eq:deltazss'2}
\boxed{
\begin{aligned}
\delta z_{\sigma_1}(\sigma)\big|_{\sigma}&=\delta z_{\sigma_1}(\sigma_1)+\delta \bar{z}_{\sigma_1}(\sigma_1)\frac{1}{4}\sum_{n=1}^\infty\frac{1}{(n+1)!}\big(\nabla_{z_{\sigma_1}}^{n-1}R_{(2)}(\sigma)\big)\Big|_{\sigma=\sigma_1}\!\!\!z_{\sigma_1}(\sigma)^{n+1}\\
\delta \bar{z}_{\sigma_1}(\sigma)\big|_{\sigma}&=\delta \bar{z}_{\sigma_1}(\sigma_1)+\delta z_{\sigma_1}(\sigma_1)\frac{1}{4}\sum_{n=1}^\infty\frac{1}{(n+1)!}\big(\nabla_{\bar{z}_{\sigma_1}}^{n-1}R_{(2)}(\sigma)\big)\Big|_{\sigma=\sigma_1}\!\!\!\bar{z}_{\sigma_1}(\sigma)^{n+1}\\
\end{aligned}
}
\end{equation}
We emphasise these variations are evaluated at a generic and fixed coordinate point $\sigma$. Also, as emphasised above the right-hand sides of the first and second relations are holomorphic in $z_{\sigma_1}(\sigma)$ and anti-holomorphic in $\bar{z}_{\sigma_1}(\sigma)$ respectively.
\sk

According to (\ref{eq:zsigmafss}), taking into account that by definition $f_{\sigma_1\sigma_1}$ is the identity map, $f_{\sigma_1\sigma_1}(z_{\sigma_1}(\sigma))\equiv z_{\sigma_1}(\sigma)$, we learn that, $\delta z_{\sigma_1}(\sigma)=\delta f_{\sigma_1'\sigma_1}(z_{\sigma_1}(\sigma))$, and therefore we can read off from (\ref{eq:deltazss'2}) that,
\begin{equation}\label{eq:dfss'/ds}
\boxed{
\begin{aligned}
\frac{\partial f_{\sigma_1'\sigma_1}(z_{\sigma_1}(\sigma))}{\partial z_{\sigma_1}(\sigma_1)}\Big|_{z_{\sigma_1}(\sigma)}&=1\\
\frac{\partial f_{\sigma_1'\sigma_1}(z_{\sigma_1}(\sigma))}{\partial \bar{z}_{\sigma_1}(\sigma_1)}\Big|_{z_{\sigma_1}(\sigma)}&=\frac{1}{4}\sum_{n=1}^\infty\frac{1}{(n+1)!}\big(\nabla_{z_{\sigma_1}}^{n-1}R_{(2)}(\sigma)\big)\Big|_{\sigma=\sigma_1}\!\!\!z_{\sigma_1}(\sigma)^{n+1}
\end{aligned}
}
\end{equation}
where we indicate that these derivatives are evaluated at fixed holomorphic chart coordinate, $z_{\sigma_1}(\sigma)$. These relations give the variation of the transition function with respect to the moduli $z_{\sigma_1}(\sigma_1),\bar{z}_{\sigma_1}(\sigma_1)$ at fixed $z_{\sigma_1}(\sigma)$. 
\sk

Tracing back the definitions one notices that the derivatives on the left-hand side in (\ref{eq:dfss'/ds}) are associated to the variation $\delta z_{\sigma_1}(\sigma_1)=(z_{\sigma_1'}(\sigma)-z_{\sigma_1}(\sigma))|_{\sigma=\sigma_1}$ (and the complex conjugate). So this is a ``passive variation'', i.e. we are shifting the base point of the frame, $\sigma_1\mapsto \sigma_1'$, keeping the coordinate, $\sigma$, where the frame is evaluated fixed. We can instead shift the coordinate keeping the frame fixed. This just introduces a minus sign because as one verifies using the above (in particular, (\ref{eq:active}), (\ref{eq:active}) and (\ref{eq:dz*=-ds*e})):
\begin{equation}\label{eq:activepassive}
\delta z_{\sigma_1}(\sigma_1)=(z_{\sigma_1'}(\sigma)-z_{\sigma_1}(\sigma))|_{\sigma=\sigma_1}=-(z_{\sigma_1}(\sigma_1')-z_{\sigma_1}(\sigma_1))\equiv -\delta z_{v_1}.
\end{equation}
In later sections it will occasionally be convenient to work in terms of the `{\it active variation}', $\delta z_{v_1}$, and we will then use the following shorthand notation:
\begin{equation}\label{eq:dzv=ds1s1}
\boxed{
\delta z_{v_1}\equiv -\delta z_{\sigma_1}(\sigma_1),\qquad \delta \bar{z}_{v_1}\equiv-\delta \bar{z}_{\sigma_1}(\sigma_1),\qquad {\rm and}\qquad 
\frac{\partial}{\partial z_{v_1}}\equiv -\frac{\partial}{\partial z_{\sigma_1}(\sigma_1)},
}
\end{equation}
where the first two correspond to translation moduli, and the notation in the third relation  appears, e.g, in (\ref{eq:d/dzv1=-d/dzs1s1}). Notice that, since $z_{\sigma_1}(\sigma)$ transforms as a scalar under reparametrisations, $\sigma\mapsto \hat{\sigma}(\sigma)$, from (\ref{eq:activepassive}) is follows immediately that $\delta z_{\sigma_1}(\sigma_1)$ will also transform as a scalar under such reparametrisations, and therefore so will $\delta z_{v_1}(\sigma_1)$ and $z_{v_1}(\sigma_1)$ transform as scalars under reparametrisations of the underlying auxiliary coordinate (this complements the above discussion associated to (\ref{eq:zsigma* scalar})). (Notice that since $z_{v_1},\bar{z}_{v_1}$ is an `active' coordinate it is not centred at the puncture -- the latter is identified with the value of the coordinate point; so locally we can regard, e.g., the Ricci scalar or a vertex operator as being inserted at $z_{v_1},\bar{z}_{v_1}$.)
\sk

We will need these relations (\ref{eq:dfss'/ds}) to implement this global (modulo U(1)) gauge slice in the path integral measure, in particular when we construct the map from fixed- to integrated-picture handle operators and vertex operators.  A guiding principle underlying this formalism has been to fix as much symmetry as possible without encountering any topological obstructions while keeping reparametrisation invariance manifest. The price to pay has been to obscure holomorphicity in moduli space, but this is necessary if we want to have a manifestly global construction that avoids Wu-Yang type boundary terms on patch overlaps \cite{Polchinski87,Polchinski88,Nelson89}. More precisely, this gauge slice associated to adopting holomorphic normal coordinates gauge fixes invariance under Weyl transformations, and we will discuss this further in Sec.~\ref{sec:HTFWR}, but reparametrisation invariance remains intact according to the discussion in the preceding paragraph. 
\sk

\subsubsection{Holomorphic Transformations from Weyl Rescalings}\label{sec:HTFWR}
We would like to understand how Weyl rescalings affect the gauge slice associated to holomorphic normal coordinates, $z_{\sigma_1}(\sigma)$, and recall that by definition, $z_{\sigma_1}(\sigma_1)\equiv 0$. Suppose we have two metrics, related by Weyl rescaling,
\begin{equation}\label{eq:edphig,g ab}
e^{\delta\phi(\sigma)}g_{ab}(\sigma)\rmd\sigma^a \rmd\sigma^b,\qquad{\rm and}\qquad g_{ab}(\sigma)\rmd\sigma^a \rmd\sigma^b,
\end{equation}
with $\delta\phi(\sigma)$ small. From the equality of (\ref{eq:gJ=gabsigma}) and (\ref{eq:rhozzbar}) we have,
\begin{equation}\label{eq:gab=rhozeta}
g_{ab}(\sigma)\rmd\sigma^a \rmd\sigma^b=\rho(z_{\sigma_1},\bar{z}_{\sigma_1})\rmd z_{\sigma_1}\rmd\bar{z}_{\sigma_1},
\end{equation}
where all quantities on the right-hand side are evaluated at $\sigma$, in particular $z_{\sigma_1}=z_{\sigma_1}(\sigma)$. Furthermore, since $z_{\sigma_1}(\sigma)$ are  holomorphic normal coordinates $\rho(z_{\sigma_1},\bar{z}_{\sigma_1})$ satisfies the defining relation (\ref{eq:PolchinskicoordsA}) at $\sigma=\sigma_1$, namely:
\begin{equation}\label{eq:hol norm coords dfn}
\partial_{z_{\sigma_1}}^n\rho(z_{\sigma_1},\bar{z}_{\sigma_1})\big|_{\sigma=\sigma_1}=\partial_{\bar{z}_{\sigma_1}}^n\rho(z_{\sigma_1},\bar{z}_{\sigma_1})\big|_{\sigma=\sigma_1}=
\left\{
\begin{array}{l}
\begin{aligned}
&1\qquad \textrm{if $n=0$}\\
&0\qquad \textrm{if $n\geq1$}
\end{aligned}
\end{array}
\right.
\end{equation}
Notice that {\it at} $\sigma=\sigma_1$ we can identify the metric components, $g_{ab}(\sigma_1)$, in (\ref{eq:edphig,g ab}) and the corresponding components in (\ref{eq:gabzs*}), see also (\ref{eq:dzsigma*-measure}). 
\sk

We now wish to associate Weyl rescalings, $\delta\phi(\sigma)$, in (\ref{eq:edphig,g ab}) to a {\it holomorphic} change of frame,
\begin{equation}\label{eq:zs*zs*'Weyl}
z_{\sigma_1}(\sigma)\mapsto w_{\sigma_1}(\sigma)=z_{\sigma_1}(\sigma)+\delta z_{\sigma_1}(z_{\sigma_1}(\sigma)),
\end{equation}
keeping the metric components, $\rho(z_{\sigma_1}(\sigma),\bar{z}_{\sigma_1}(\sigma))$, and coordinate $\sigma$ {\it fixed}. In particular, we want to demand that:
\begin{equation}\label{eq:edphiwz}
\begin{aligned}
e^{\delta\phi(\sigma)}g_{ab}(\sigma)\rmd\sigma^a \rmd\sigma^b&=\rho(w_{\sigma_1},\bar{w}_{\sigma_1})\rmd w_{\sigma_1}\rmd\bar{w}_{\sigma_1}\\
&=\rho(z_{\sigma_1},\bar{z}_{\sigma_1})\rmd z_{\sigma_1}\rmd\bar{z}_{\sigma_1}+\big(\nabla_{z_{\sigma_1}} \delta z_{\sigma_1}+\nabla_{\bar{z}_{\sigma_1}}\delta \bar{z}_{\sigma_1}\big)\rho(z_{\sigma_1},\bar{z}_{\sigma_1})\rmd z_{\sigma_1}\rmd\bar{z}_{\sigma_1},
\end{aligned}
\end{equation}
where in going from the first to the second equality we took into account that at a generic point $\sigma$, $\nabla_{\bar{z}_{\sigma_1}} \delta z_{\sigma_1}=\nabla_{z_{\sigma_1}}\delta \bar{z}_{\sigma_1}=0$. So taking $\delta\phi(\sigma)$ to be small and on account of (\ref{eq:gab=rhozeta}) we learn that (\ref{eq:edphiwz}) implies the relation,
\begin{equation}\label{eq:dphi=Ddz+Dbardbarz}
\phantom{\qquad\textrm{(Weyl Rescalings)}}\boxed{\delta\phi(\sigma)=\big(\nabla_{z_{\sigma_1}} \delta z_{\sigma_1}+\nabla_{\bar{z}_{\sigma_1}}\delta \bar{z}_{\sigma_1}\big)(\sigma)}\qquad\textrm{(Weyl Rescalings)}
\end{equation}
We think of this as a first-order differential equation determining (the real part of) $\delta z_{\sigma_1}(\sigma)$ in terms of $\delta\phi(\sigma)$. We will solve this differential equation subject to the following boundary condition: we require that the variation, $\delta z_{\sigma_1}(\sigma)$, generated by the Weyl transformation vanishes at the base point, $\sigma=\sigma_1$:
\begin{equation}\label{eq:dzs*=0}
\phantom{\qquad\textrm{(Weyl Rescalings)}}
\boxed{\delta z_{\sigma_1}(\sigma_1)=\delta\bar{z}_{\sigma_1}(\sigma_1)=0}
\qquad\textrm{(Weyl Rescalings)}
\end{equation}
but of course the derivatives of these variations need not vanish at the base point. According to (\ref{eq:zs*zs*'Weyl}) this ensures that Weyl transformations do not shift punctures at $\sigma_1$ -- the Weyl-transformed frame, $w_{\sigma_1}$, remains centred at $\sigma_1$. As discussed in Sec.~\ref{sec:SPUTF} (and in later sections) we wish to associate such variations, $\delta z_{\sigma_1}(\sigma_1)$ and $\delta\bar{z}_{\sigma_1}(\sigma_1)$, with {\it complex structure deformations}, so that (\ref{eq:dzs*=0}) should be interpreted as specifying that (for the gauge slice of interest) complex structure deformations are transverse to Weyl transformations.
\sk

We will solve the above differential equation by constructing a Taylor series solution. 
Let us expand the covariant derivatives in (\ref{eq:dphi=Ddz+Dbardbarz}), and hit the left- and right-hand sides with $\partial^{n-1}_{z_{\sigma_1}}$ with $n\geq2$ (where differentiation is with respect to $z_{\sigma_1}(\sigma)$). Evaluating the resulting relation at the base point, $\sigma=\sigma_1$, we learn that,
\begin{equation}\label{eq:dn-1dphi}
\begin{aligned}
\big(\partial^{n-1}_{z_{\sigma_1}}\delta\phi(\sigma)\big)\big|_{\sigma=\sigma_1}&=\big(\partial^{n-1}_{z_{\sigma_1}}\nabla_{z_{\sigma_1}} \delta z_{\sigma_1}(\sigma)+\partial^{n-1}_{z_{\sigma_1}}\nabla_{\bar{z}_{\sigma_1}}\delta \bar{z}_{\sigma_1}(\sigma)\big)\big|_{\sigma=\sigma_1}\\
&=\big(\partial^{n}_{z_{\sigma_1}}\delta z_{\sigma_1}(\sigma)
\big)\big|_{\sigma=\sigma_1}\\
\end{aligned}
\end{equation}
where we made use of the defining relations of holomorphic normal coordinates (\ref{eq:hol norm coords dfn}), that Weyl transformations leave the frame coordinate associated to the base point unchanged (\ref{eq:dzs*=0}), and that $\delta \bar{z}_{\sigma_1}(\bar{z}_{\sigma_1}(\sigma))$ is anti-holomorphic. Let us now multiply left- and right-hand sides in (\ref{eq:dn-1dphi}) by $\frac{1}{n!}z_{\sigma_1}(\sigma)^{n}$ and sum over $n=2,3,\dots$ We then identify the right-hand side of the resulting relation with the Taylor expansion of $\delta z_{\sigma_1}(\sigma)$ with the $n=0,1$ terms subtracted out. The $n=0$ subtraction term vanishes by (\ref{eq:dzs*=0}), so rearranging the resulting equation we hence learn that:
\begin{equation}\label{eq:dz from Weyl}
\delta z_{\sigma_1}(\sigma)=\big(\partial_{z_{\sigma_1}}\delta z_{\sigma_1}(\sigma)
\big)\big|_{\sigma=\sigma_1}z_{\sigma_1}(\sigma)+
\sum_{n=1}^\infty \frac{1}{(n+1)!}\big(\partial^{n}_{z_{\sigma_1}}\delta\phi(\sigma)\big)\big|_{\sigma=\sigma_1}z_{\sigma_1}(\sigma)^{n+1}
\end{equation}

According to (\ref{eq:dphi=Ddz+Dbardbarz}) and (\ref{eq:hol norm coords dfn}) we can also rewrite the first term on the right-hand side in terms of $\delta \phi(\sigma_1)$, 
\begin{equation}\label{eq:dz from Weyl2}
\begin{aligned}
\delta z_{\sigma_1}(\sigma)&=\frac{1}{2}\big(\delta \phi(\sigma_1)+i\delta\beta(\sigma_1)
\big)z_{\sigma_1}(\sigma)\\
&\qquad+
\sum_{n=1}^\infty \frac{1}{(n+1)!}\big(\partial^{n}_{z_{\sigma_1}}\delta\phi(\sigma)\big)\big|_{\sigma=\sigma_1}z_{\sigma_1}(\sigma)^{n+1},
\end{aligned}
\end{equation}
where (taking into account that covariant and partial derivatives are interchangeable at $\sigma=\sigma_1$) we have defined the phase,
$$
\delta\beta(\sigma_1)\dfn {\rm Im}\,\nabla_{z_{\sigma_1}}\delta z_{\sigma_1}(\sigma)\big|_{\sigma=\sigma_1}.
$$
Once again, since the phase of the holomorphic normal coordinate is arbitrary, recall the discussion associated to the base-point dependent phase $\alpha(p)$ in (\ref{eq:zetap'p2}), the quantity $\delta\beta(\sigma_1)$ in (\ref{eq:dz from Weyl2}) can be neglected provided we always take the overall phase to be integrated, recall the related discussion in (\ref{eq:zsds(s')}). 
\sk

Given a set of {\it holomorphic normal coordinates}, $z_{\sigma_1}(\sigma)$, we can now display the full result for the holomorphic transformation, $z_{\sigma_1}(\sigma)\mapsto w_{\sigma_1}(\sigma)$, generated by a Weyl rescaling, $\delta\phi(\sigma)$, of the auxiliary metric,
$$
\boxed{
g_{ab}(\sigma)\mapsto e^{\delta\phi(\sigma)}g_{ab}(\sigma)
}
$$
On account of (\ref{eq:zs*zs*'Weyl}) and (\ref{eq:dz from Weyl2}), to leading order in the variation,
\begin{equation}\label{eq:dz from Weyl3}
\boxed{
\begin{aligned}
z_{\sigma_1}(\sigma)\mapsto w_{\sigma_1}(\sigma)&= e^{\frac{1}{2}(\delta \phi(\sigma_1)+i\delta\beta(\sigma_1))}\Big[z_{\sigma_1}(\sigma)+
\sum_{n=1}^\infty \frac{1}{(n+1)!}\big(\nabla^{n}_{z_{\sigma_1}}\delta\phi(\sigma_1)\big)z_{\sigma_1}(\sigma)^{n+1}\Big]
\end{aligned}
}
\end{equation}
where $\nabla^{n}_{z_{\sigma_1}}\delta\phi(\sigma_1)\equiv (\nabla^{n}_{z_{\sigma_1}(\sigma)}\delta\phi(\sigma))|_{\sigma=\sigma_1}$, and we took into account that covariant and partial derivatives evaluated at $\sigma=\sigma_1$ are equivalent in holomorphic normal coordinates. 
Notice that the right-hand side in (\ref{eq:dz from Weyl3}) is {\it holomorphic} in $z_{\sigma_1}(\sigma)$, and the variation is such that the defining equation (\ref{eq:hol norm coords dfn}) for the gauge slice associated to holomorphic normal coordinates is preserved up to a rescaling of the metric. Furthermore, it is manifest that $w_{\sigma_1}(\sigma_1)=z_{\sigma_1}(\sigma_1)=0$ by construction, so that these Weyl transformations leave fixed the base point $\sigma=\sigma_1$. So we can use these relations to unravel how general local operators transform under Weyl transformations. This is carried out in Sec.~\ref{sec:LOUWR}.

\subsection{Warmup: The Euler Characteristic}\label{sec:EulerCh}
In this section we begin with a warmup calculation (associated to the Euler characteristic) that will make certain aspects of the interplay between local and global data transparent. Recall in particular that the Euler characteristic of a Riemann surface is purely topological (and hence sensitive to global data).
\sk

More precisely, we will recast the differential-geometric representation of the Euler characteristic, namely (\ref{eq:GaussBonnet1}), in terms of the fundamental data defining a Riemann surface, namely holomorphic transition functions subject to cocycle relations (as defined in Sec.~\ref{sec:TFCR}). The approach we adopt will have a close analogue in our study of the worldsheet path integral measure. To add further context, note in particular that when we cut open a compact Riemann surface across a separating (or non-separating) cycle we end up with two (or one) Riemann surfaces {\it with boundaries}. This emergence of a boundary when cutting open path integrals propagates through to the path integral measure (via appearance of boundary terms in the latter in the gauge slice of interest) that we will develop fully. This section is therefore a warmup for what is to come. 
\sk

Consider a genus-$\g$ oriented Riemann surface, $\Sigma$, possibly with boundary, $\partial\Sigma$, that is in turn comprised of $\B$ disconnected components, such that $\Sigma$ is endowed with metric tensor $g$. The {\it Euler characteristic}, $\chi(\Sigma)$, is the basic topological invariant of this manifold, see e.g.~\cite{Chern67}, defined by:\footnote{A more general definition is to write the Euler characteristic as an alternating sum of {\it Betti numbers}, $b_i$, as $\chi(M)=\sum_{i=1}^{\rm dim M}(-)^ib_i$. In particular, the numbers, $\mathbf{v},\mathbf{e}$ and $\mathbf{f}$ individually depend on the triangulation, whereas every individual number $b_i$ is itself a topological invariant.}
\begin{equation}\label{eq:chi=vef}
\boxed{\chi(\Sigma)\dfn\mathbf{v}-\mathbf{e}+\mathbf{f}}
\end{equation}
Here we imagine subdividing $\Sigma$ into a union of polygons, $V_m$, $m=1,\dots,\mathbf{f}$, such that any two $V_m$ either have no point, or one vertex, or a whole side in common, and we denote the number of {\it interior vertices} by $\mathbf{v}$ and the number of {\it interior edges} by $\mathbf{e}$. Whether we also include boundary {\it boundary} edges and vertices in the counting or not does not affect the Euler characteristic because the two resulting surfaces are of the same homotopy type (e.g., the Euler characteristic of a point is equal to that of a line). So the Euler characteristic is independent of whether boundaries are considered part of the surface or not.
\sk

For example, a disc, ${\rm D}$, is topologically equivalent to a square, cell-decomposed into four regions such that there are two diametrically opposite lines connecting the four corners of the square which in turn meet at a vertex at the centre of the square. Then the counting would be such that $(\mathbf{v},\mathbf{e},\mathbf{f})=(1,4,4)$, because the edges and vertices on the square boundary are not {\it interior}, leading to the standard result $\chi({\rm D})=1$. But as follows from the comment in the last sentence of the preceding paragraph, the result for the disc would have been the same had we rather included {\it all} vertices, edges and faces in the counting in which case $(\mathbf{v},\mathbf{e},\mathbf{f})=(5,8,4)$. 
\sk

Such vertices, edges and faces defined in the paragraph containing (\ref{eq:chi=vef}) are depicted in the second diagram in Fig.~\ref{fig:tripleoverlapsUV}, but they need not be as regular as depicted there: more generally, we may identify the vertices, edges and faces with those associated to a dual triangulation (briefly discussed in Sec.~\ref{sec:TFCR}), see the third diagram in Fig.~\ref{fig:dualtriangulation} and Fig.~\ref{fig:genericdualtriangles} (on p.~\pageref{fig:genericdualtriangles}). We also do not require the edges be ``straight'' in any sense, despite the fact that they are depicted as being straight in the various figures (such as Fig.~\ref{fig:dualtriangulation},\ref{fig:genericdualtriangles},\ref{fig:tripleoverlapsUV}); the underlying Riemann surface is also generically curved of course. The Euler characteristic (\ref{eq:chi=vef}) of $\Sigma$ is well-known to take the form: 
\begin{equation}\label{eq:chi=2-2g-b}
\chi(\Sigma)=2-2\g-\B.
\end{equation}
The primary focus of this section is on the differential-geometric definition of $\chi(\Sigma)$ given by the {\it Gauss-Bonnet theorem}:
\begin{equation}\label{eq:GaussBonnet1}
\boxed{\chi(\Sigma)=\frac{i}{2\pi}\int_\Sigma \mathcal{R}+\frac{1}{2\pi}\int_{\partial \Sigma}\rmd s\, k_{\rm g}(s)+\frac{1}{2\pi}\sum_i(\pi-\alpha_i)}
\end{equation}
but we will take a slightly different approach from the usual one \cite{Chern67,Tu17}, and in particular we take an approach that will be directly applicable to (and that will provide some additional insight in) the path integral measure in later sections, in particular Sec.~\ref{sec:TPIM}. 
\sk

The quantity $\mathcal{R}$ in (\ref{eq:GaussBonnet1}) is the curvature tensor of $\Sigma$, $k_{\rm g}(s)$ is the geodesic curvature of $\partial\Sigma$, and $\pi-\alpha_i$ are the exterior angles at the vertices of $\partial\Sigma$ (if the latter is only {\it sectionally} smooth rather than smooth; if it is smooth the third term on the right-hand side of (\ref{eq:GaussBonnet1}) is zero). All of these quantities are defined below. The three terms on the right-hand side of (\ref{eq:GaussBonnet1}) correspond respectively to the surface curvature, line curvature and point curvature of $\Sigma$, so that $\chi(\Sigma)$ is interpreted as the total curvature of $\Sigma$ \cite{Chern}. The remarkable property of (\ref{eq:GaussBonnet1}) is that it provides a link between global and local data.

\subsubsection{No Boundaries}\label{sec:NB}
Let us begin with the case when $\partial \Sigma$ is empty so that $\Sigma$ is a {\it closed} oriented Riemann surface. According to (\ref{eq:GaussBonnet1}), the Euler characteristic is given in terms of the integrated Ricci curvature tensor, $\mathcal{R}$, \cite{Chern67}:
\begin{equation}\label{eq:chiA}
\boxed{\chi(\Sigma) =\frac{i}{2\pi}\int_{\Sigma} \mathcal{R}}
\end{equation}
which famously takes the form $\chi(\Sigma)=2-2\g$, 
for all $\g=0,1,\dots$, and this result is independent of the local geometry. 
Since it is locally always possible to work in a conformally-flat coordinate chart \cite{BersRiemannSurfaces}, we may endow $\Sigma$ with a metric, $g$, which in a local chart $(U,z)$ with conformal coordinate $z$ takes the general form,
$$
g=\rho(z,\bar{z})\rmd z\rmd\bar{z},
$$ 
with $\rho(z,\bar{z})$ a real and positive definite function of the complex coordinates, $z,\bar{z}$. In this chart the curvature tensor then reads explicitly \cite{Chern},
\begin{equation}\label{eq:Rtensor}
\mathcal{R}=R_{z\bar{z}}\rmd z\wedge \rmd\bar{z}=\big[\!-\partial_z\partial_{\bar{z}}\ln \rho(z,\bar{z})\big]\rmd z\wedge \rmd\bar{z}
\end{equation}
For many purposes it is convenient to work with a metric, but as we have already noted (and as we also elaborate on next) this is (at least far from the boundary of moduli space) not necessary. Since $\chi(\Sigma)$ is topological the moduli space of Riemann surfaces will play no role in this section; it is to be understood that we are sufficiently far from any boundary of moduli space.
\sk

\begin{figure}
\begin{center}
\includegraphics[angle=0,origin=c,width=0.75\textwidth]{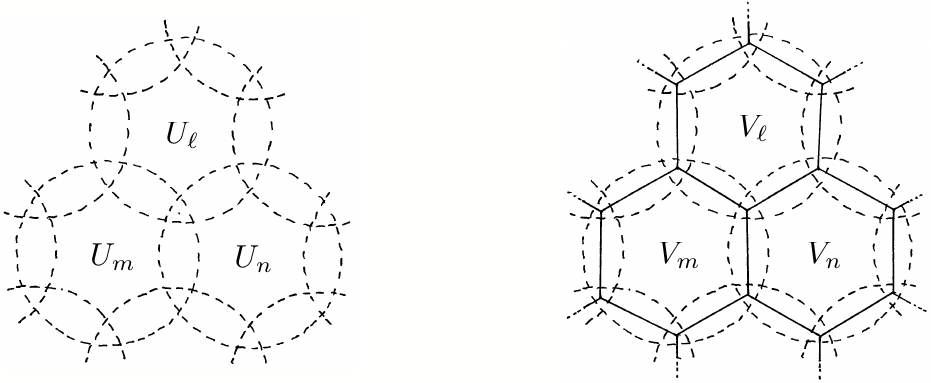}
\caption{The first diagram shows part of an open cover, $\mathscr{U}=\{U_m,U_n,\dots\}$, of a Riemann surface with a finite number of open sets (three of which are displayed, $U_m,U_n,U_\ell$) such that they overlap on double, $\{U_m\cap U_n,\dots\}$, triple, $\{U_m\cap U_n\cap U_\ell,\dots\}$, (or, not explicitly shown higher) patch intersections. In the second diagram we introduce a cell decomposition $\mathscr{V}=\{V_m\}$ with $V_m\subset U_m$ for every $m$ with the solid lines denoting the boundaries of the cells, $V_m$. Using a partition of unity, $\sum_m\lambda_m=1$, subordinate to the open cover $\mathscr{U}$, and the corresponding decomposition $\mathscr{V}$, allows one to replace integrals over $\Sigma$ by a sum of integrals over $\{V_m\}$: $\int_\Sigma = \int_\Sigma \sum_m\lambda_m=\sum_m\int_{V_m}$ (independently of the fact that $\int_\Sigma\lambda_m$ has support on $U_m$ which is a larger domain than $V_m$). 
}\label{fig:tripleoverlapsUV}
\end{center}
\end{figure}

\begin{figure}
\begin{center}
\includegraphics[angle=0,origin=c,width=0.87\textwidth]{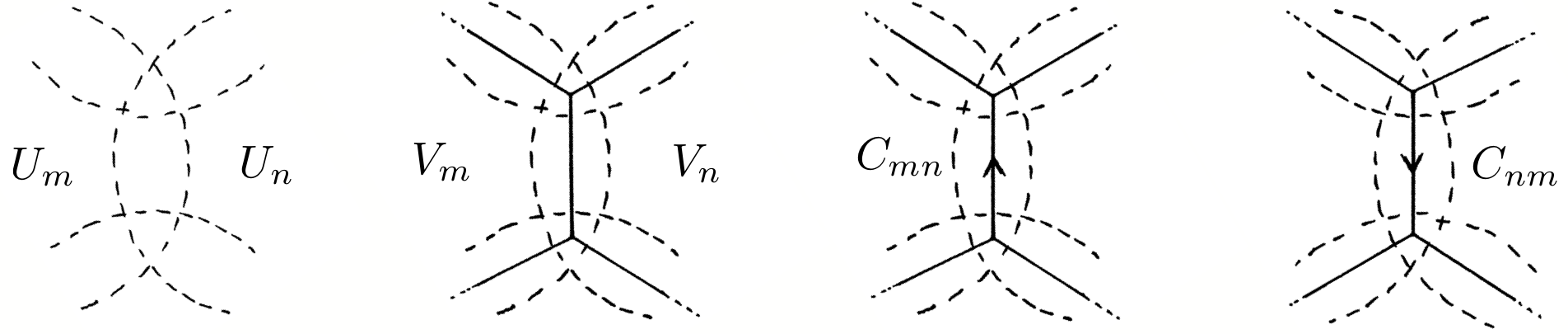}
\caption{A zoomed in version of Fig.~\ref{fig:tripleoverlapsUV}, displaying not only the double intersection $U_m\cap U_n$ and the corresponding partitioning into non-overlapping cells, $V_m$, $V_n$, but also the oriented segment of the common boundary, $C_{mn}$, of $V_m$ and $V_n$ which is oriented in an anti-clockwise sense with respect to $V_m$.}\label{fig:segment}
\end{center}
\end{figure}

Recall in particular that in Sec.~\ref{sec:TFCR} and Sec.~\ref{sec:CSDI} we defined Riemann surfaces and their complex structure deformations using only charts and (holomorphic) transition functions on chart overlaps, in particular without introducing the notion of a metric. So this implies we should be able to express $\chi(\Sigma)$ in terms of charts and holomorphic transition functions alone, namely in terms of the fundamental data that defines a Riemann surface. Indeed, independently of whether we have a good cover or not, we will begin by showing that the integral (\ref{eq:chiA}) (in the absence of boundaries) can be recast into the form \cite{Chern,BottTu}:
\begin{equation}\label{eq:Euler}
\boxed{
\begin{aligned}
\chi(\Sigma)
&=\frac{i}{2\pi}\sum_{(mn)}\int_{C_{mn}}\!\!\rmd z_m\partial_{z_m}\ln f'_{nm}(z_m),
\end{aligned}
}
\end{equation}
where the sum is over all {\it pairs} of overlapping patches, $U_{m}\cap U_n\neq \varnothing$, and the oriented contour $C_{mn}$ traverses the overlap $U_m\cap U_n$ counterclockwise with respect to $U_m$. Furthermore, $C_{mn}$ is either a closed contour (if $U_{m}\cap U_n\neq \zero$ is homeomorphic to an annulus), or it begins and ends on higher (i.e.~triple or higher) patch overlaps where other contours from the complete set $\{C_{mn}\}$ begin or end. E.g., the contour $C_{mn}$ may be identified with the common boundary of the sets $V_m$ and $V_n$ displayed (as a solid line) in Fig.~\ref{fig:tripleoverlapsUV} (with the orientation depicted in Fig.~\ref{fig:segment}). 
A prime denotes a partial derivative with respect to the (explicit) argument, e.g.~$f_{nm}'(z_m)\equiv \partial_{z_m}f_{nm}(z_m)$, keeping any implicit arguments (such as moduli) fixed. Taking into account (\ref{eq:transitionfuncs}), the relation $f_{nm}'(z_m)=f_{mn}'(z_n)^{-1}$ which follows from differentiating (\ref{eq:fmnfnm=1}), and $C_{mn}=-C_{nm}$, it is seen that the summand in (\ref{eq:Euler}) is symmetric in $m$ and $n$. 
\sk

To go from (\ref{eq:chiA}) to (\ref{eq:Euler}) is an instructive exercise, and we will go through it in detail since we apply similar reasoning to the path integral measure in later sections. Schematically, we introduce a partition of unity\footnote{Recall that \cite{GunningRossi} a {\it partition of unity} subordinate to a locally finite open covering $\{U_m\}$ of $\Sigma$ is a collection of non-negative $C^{\infty}$ functions $\{\lambda_m:\Sigma\rightarrow \mathbf{R}\}$, such that $\lambda_m(p)=0$ in an open neighbourhood of $\Sigma\setminus U_m$ and $\sum_m\lambda_m(p)=1$ at every $p\in\Sigma$. Here `locally finite' ensures that the sum over $m$ at any point $p\in\Sigma$ is finite.} \cite{GunningRossi}, $\sum_m\lambda_m=1$, subordinate to the open cover $\mathscr{U}=\{U_m,U_n,\dots\}$, and introduce a cell decomposition $\mathscr{V}=\{V_m,V_n,\dots\}$ with $V_m\subset U_m$ (for all $m$) such that every $V_m$ is bounded by a union of curves $C_m=C_{mn_1}\cup C_{mn_2}\cup\dots$ (that are either disjoint or meet at triple or higher intersections) as shown in Fig.~\ref{fig:tripleoverlapsUV} and Fig.~\ref{fig:segment}. More precisely, we subdivide $\Sigma$ into a union of polygons, $\mathscr{V}=\{V_m\}$, such that: 
\begin{itemize}\label{itemize-Vms}
\item[(1)] each $V_m$ lies in one coordinate chart (in particular, $V_m\subset U_m$, with chart $(U_m,z_m)$); 
\item[(2)] any two polygons, $V_m,V_n$, have either no point, or one vertex, or a whole edge, in common;
\item[(3)] the $V_m$ are coherently oriented with $\Sigma$, so that every {\it interior edge} has different senses induced by the two polygons of which it is a side (see the third and fourth diagram in Fig.~\ref{fig:segment}). 
\item[(4)] denote the cardinality of $\mathscr{V}$ (equivalently, the number of {\it faces} or polygons in $\{V_m\}$) by $\mathbf{f}$, the number of {\it interior edges} (i.e.~the number of double overlaps, $U_m\cap U_n$) by $\mathbf{e}$, and the number of {\it interior vertices} (i.e.~the number of triple or higher overlaps, $U_m\cap U_n\cap U_\ell$) by $\mathbf{v}$. (So $\mathbf{v}$ and $\mathbf{e}$ count vertices and edges that are {\it not} on a boundary, $\partial\Sigma$.)
\end{itemize}
We have phrased these properties of $\mathscr{V}$ in such a manner that they remain valid independently of whether $\Sigma$ has a boundary, $\partial \Sigma$, or not. Notice that the number of edges, $\mathbf{f}$, is the same as the number of elements in the set $\{C_{mn}\}$ (which is partially why we adopted this method of counting in the definition (\ref{eq:chi=vef}) of the Euler characteristic).
\sk

We may then make a special choice for the parameters defining the $\{\lambda_m\}$ by demanding that $\lambda_m=1$ in $V_m$ and $\lambda_m=0$ in $\Sigma\setminus V_m$ (so that every $\lambda_m$ is a characteristic function on $V_m$)\footnote{This is not quite what we do (in particular, we do not need to make a specific choice for the partition of unity, see below) but the result is the same. Note also that although the characteristic function, $\lambda_m$, of $V_m$ is not strictly speaking $C^\infty$ (one of the defining properties of a partition of unity), one can imagine constructing it using an appropriate $C^\infty$ function by a limiting procedure \cite{GunningRossi}.}. Under the mild assumption that a limit exists such that the boundaries $\{\partial V_m\}$ contribute measure zero to the integral over $\Sigma$, we can then use Stoke's theorem to arrive at (\ref{eq:Euler}). 
\begin{figure}
\begin{center}
\includegraphics[angle=0,origin=c,width=0.8\textwidth]{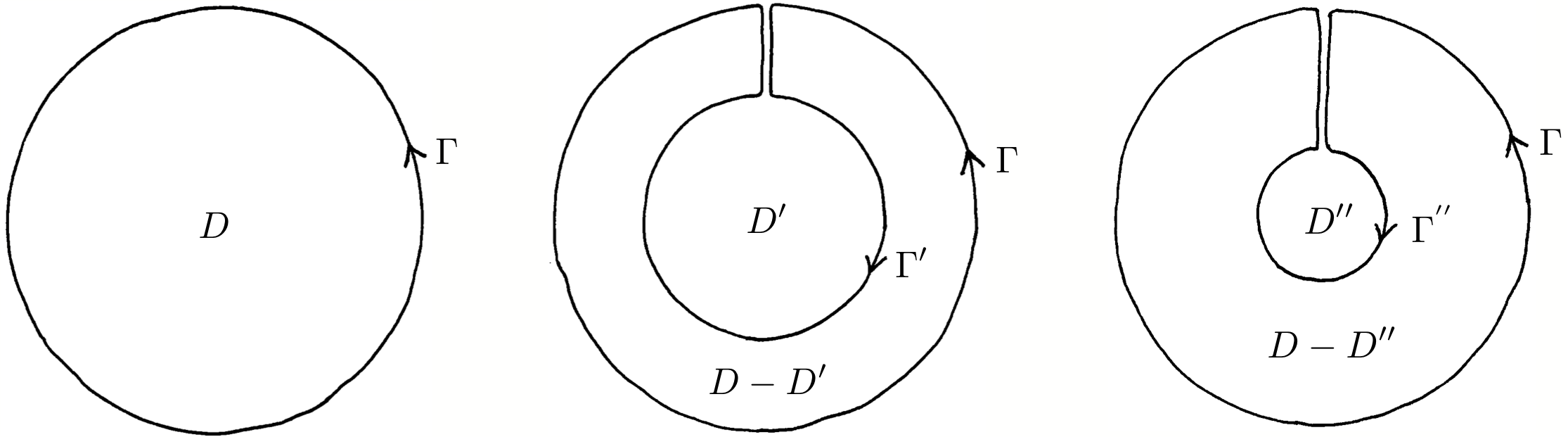}
\caption{Integration contours associated to various simply-connected domains.}\label{fig:Greens}
\end{center}
\end{figure}
\sk

Let us now exhibit these steps in detail. We consider the integral representation (\ref{eq:chiA}) and insert a partition of unity subordinate to $\mathscr{U}$,
\begin{equation}\label{eq:chi}
\begin{aligned}
\chi(\Sigma) &=\frac{i}{2\pi}\int_{\Sigma} \mathcal{R}\\
&=\frac{i}{2\pi}\int_{\Sigma} \big(\sum_m\lambda_m\big)\mathcal{R}\\
&=\frac{i}{2\pi}\sum_m\int_{U_m} \mathcal{R}\lambda_m.
\end{aligned}
\end{equation} 
It is sometimes convenient to work with an explicit partition of unity, $\{\lambda_m\}$, such as the choice mentioned above (which can in principle be constructed along the lines discussed in \cite{GunningRossi}), but in fact we will not need to do so. In fact, let us primarily show how to rearrange the sum over $m$ in (\ref{eq:chi}) such that for any globally-defined top form $\omega$ (such as $\frac{i}{2\pi}\mathcal{R}$ in the above example):
\begin{equation}\label{eq:nopartitionofunity}
\boxed{\sum_m\int_{U_m}\omega\lambda_m=\sum_m\int_{V_m} \omega}
\end{equation}
We will derive (\ref{eq:nopartitionofunity}) next and show that it holds independently of a choice of partition of unity. Readers willing to accept the validity of (\ref{eq:nopartitionofunity}) without a detailed reasoning may wish to skip to the beginning of the paragraph containing (\ref{eq:chix}). 
\sk

\begin{figure}
\begin{center}
\includegraphics[angle=0,origin=c,width=0.25\textwidth]{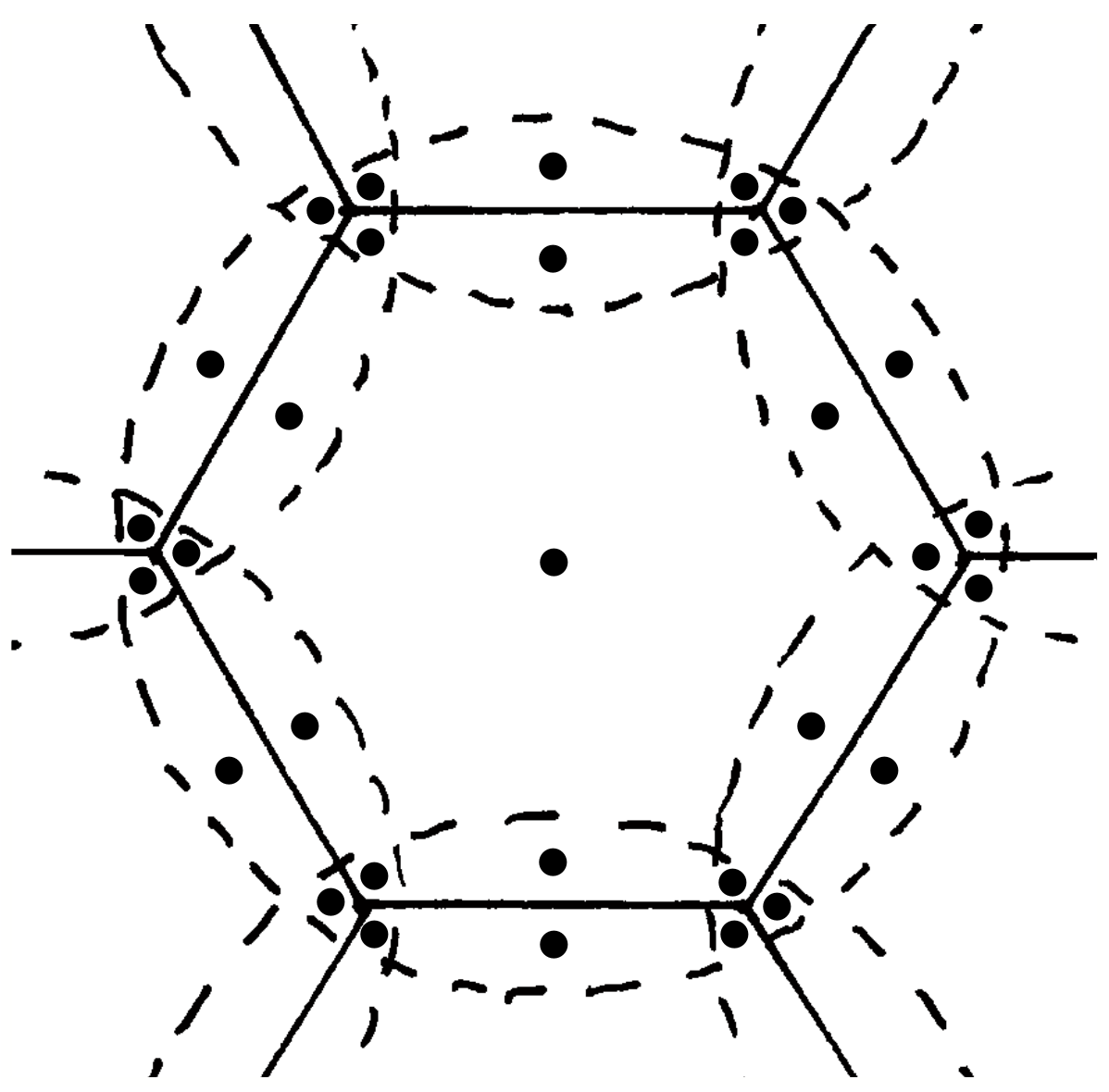}
\caption{This figure is used in the derivation of (\ref{eq:nopartitionofunity}). It  displays the decomposition of any one of the open sets of the open cover $\mathscr{U}$ (say the specific set $U_m$) into (in this example) 31 distinct cells, either one of which (for ease of visual identification) contains a single dot `\,{\boldmath$\cdot$}' while being bounded by lines (irrespective of whether they are dashed or solid).}\label{fig:UVdots}
\end{center}
\end{figure}

Consider\footnote{A sketch of a proof along these lines was explained to DS by Lee Mosher.} an element, $U_m$, of the cover $\mathscr{U}$ (see Fig.~\ref{fig:tripleoverlapsUV}). We then partition $U_m$ into cells such as those depicted in Fig.~\ref{fig:UVdots} where the solid lines indicate the decomposition of interest $\mathscr{V}$ that we are aiming for to obtain (\ref{eq:nopartitionofunity}). The choice of $\mathscr{V}$ is such that the four properties listed above are satisfied. 
In the example displayed in Fig.~\ref{fig:UVdots} there are $31$ (and more generally, say, $\kappa_m$) distinct cells (within the set $U_m$), let us denote them by $U_m^{(1)},\dots,U_m^{(\kappa_m)}$, either one of which contains a single dot `\,{\boldmath$\cdot$}\,' (for visual identification purposes) while being bounded by solid and/or dashed lines. Since every $U_m$ is diffeomorphic to a disc (or perhaps an annulus if we do not consider a good cover) there exists a single coordinate system on it and we may use ordinary integration theory to decompose the integral over $U_m$ into a sum of integrals over the specified cells:
$$
\int_{U_m}\omega\lambda_m=\sum_{a=1}^{\kappa_m}\int_{U_m^{(a)}}\omega\lambda_m.
$$
The fact that this is always possible follows from the defining property of an ordinary integral, namely that it can be extracted from a Riemann sum using an appropriate limiting procedure. We then repeat this procedure for every $U_m$ of the cover $\mathscr{U}$ and sum over $m$,
\begin{equation}\label{eq:summol}
\sum_m\int_{U_m}\omega\lambda_m=\sum_m\sum_{a=1}^{\kappa_m}\int_{U_m^{(a)}}\omega\lambda_m.
\end{equation}
Let us denote the full collection of such cells associated to the entire manifold $\Sigma$ by $\mathscr{W}=\{U_m^{(a)},U_n^{(b)},\dots\}$. 
Since there are non-empty patch overlaps, such as $U_m\cap U_n$ or $U_m\cap U_n\cap U_\ell$, etc., there exist elements in $\mathscr{W}$ that have different labels but represent nevertheless {\it identical} cells in $\Sigma$. For example, on the triple overlap $U_m\cap U_n\cap U_\ell$ (see Fig.\ref{fig:tripleoverlapsUV}) all three sets $U_m^{(1)}$, $U_n^{(1)}$ and $U_\ell^{(1)}$ in $\mathscr{W}$ might coincide with $U_n\cap U_\ell\cap V_m$. It is therefore useful to construct a set, $\mathscr{D}=\{D_k\}$, (not necessarily ordered) whose elements are the cells in $\mathscr{W}$ with the additional rule that whenever a cell is repeated in $\mathscr{W}$ (in the above sense) only a single copy of it is retained in $\mathscr{D}$. In particular, there exists a projection (a surjective map) $\pi$ that takes every element of $\mathscr{W}$ to an element in $\mathscr{D}$, and, conversely, every element in $\mathscr{D}$ has at least one identical element in $\mathscr{W}$. In the above example all three cells $U_m^{(1)}$, $U_n^{(1)}$ and $U_\ell^{(1)}$ are mapped to a single cell, $U_n\cap U_\ell\cap V_m$ (which we could call $D_1$), under $\pi$. By construction therefore we can use $\pi$ to give an equivalent expression to (\ref{eq:summol}),
\begin{equation}\label{eq:summol2}
\sum_m\int_{U_m}\omega\lambda_m=\sum_m\sum_{a=1}^{\kappa_m}\int_{\pi(U_m^{(a)})}\omega\lambda_m.
\end{equation}
(Note that the number of terms appearing in the double sums on the right-hand sides of both (\ref{eq:summol}) and (\ref{eq:summol2}) is the same.) 
To see how this works explicitly let us consider the terms in the sums on the right-hand side of (\ref{eq:summol2}) associated to the specific cells in the above example, namely the cells $U_m^{(1)}$, $U_n^{(1)}$ and $U_\ell^{(1)}$ associated to $D_1=U_n\cap U_\ell\cap V_m$. The corresponding terms in the double sum in (\ref{eq:summol2}) take the form:
\begin{equation}\label{eq:intpiUintD1}
\begin{aligned}
&\int_{\pi(U_m^{(1)})}\omega\lambda_m+\int_{\pi(U_n^{(1)})}\omega\lambda_n+\int_{\pi(U_\ell^{(1)})}\omega\lambda_\ell\\
&\qquad=\int_{D_1}\omega\big(\lambda_m+\lambda_n+\lambda_\ell\big)\\
&\qquad=\int_{D_1}\omega.
\end{aligned}
\end{equation}
The remaining terms in the sums in (\ref{eq:summol2}) are associated to the complement of $U_n\cap U_\ell\cap V_m$ in $\Sigma$. In the last equality in (\ref{eq:intpiUintD1}) we took into account that $\sum_m\lambda_m(p)=1$ for every point $p\in\Sigma$ and in particular therefore for every $p\in D_1$ since $D_1\subset \Sigma$. Proceeding in the same manner for all remaining terms in the double sum in (\ref{eq:summol2}) we learn that,
\begin{equation}\label{eq:summol3}
\sum_m\int_{U_m}\omega\lambda_m=\sum_k\int_{D_k}\omega.
\end{equation}
Finally, every element $V_m\subset \mathscr{V}$ is by construction a union of a subset of distinct elements in $\mathscr{D}$, so we may rearrange the sum over $k$ in (\ref{eq:summol3}) to finally obtain precisely (\ref{eq:nopartitionofunity}) which is what we set out to show.
\sk

Applying the identity (\ref{eq:nopartitionofunity}) to the case of interest (\ref{eq:chi}) we learn that the Euler characteristic takes the form:
\begin{equation}\label{eq:chix}
\begin{aligned}
\chi(\Sigma) &=\frac{i}{2\pi}\sum_m\int_{V_m} \mathcal{R}.
\end{aligned}
\end{equation} 
In the local chart associated to $V_m\subset U_m$ we have the explicit local coordinate representation for the curvature that is read off from (\ref{eq:Rtensor}), in particular,
\begin{equation}\label{eq:chi(S)temp}
\begin{aligned}
\chi(\Sigma)
&=\frac{1}{2\pi}\sum_m\int_{V_m} \rmd^2z_m(-\partial_{z_m}\partial_{\bar{z}_m}\ln \rho_{m})\\
&=\frac{1}{2\pi i}\sum_m\oint_{\partial V_m} \!\!\!\rmd z_m(-\partial_{z_m}\ln \rho_{m}).
\end{aligned}
\end{equation}
where in going from the first to the second equality we used Green's theorem; we display the latter for concreteness since we will be making use of it throughout \cite{BersRiemannSurfaces}:
\begin{equation}\label{eq: Greens theorem2}
\frac{1}{2\pi}\int_D\rmd^2z\,\partial_{\bar{z}}w(z,\bar{z})=\frac{1}{2\pi i}\oint_{\Gamma}\rmd z\, w(z,\bar{z}).
\end{equation} 
We have included a short discussion of this in Appendix \ref{sec:GT}.
\sk

Green's theorem (\ref{eq: Greens theorem2}) is valid for complex $C^1$ functions $w(z,\bar{z})$, $D$ is a simply connected region in $\Sigma$ bounded by a sufficiently smooth curve $\Gamma=\partial D$ with the standard orientation depicted in the first diagram in Fig.~\ref{fig:Greens} (or the second or third diagram when $D$ is replaced by $D-D'$ or $D-D''$ respectively). Incidentally, note that (crucially) (\ref{eq: Greens theorem2}) is also valid for a more general surface (such as an oriented Riemann surface, $\Sigma$, with boundary and arbitrary topology) {\it provided} the integrands are {\it globally well-defined} on $\Sigma$. In many of the integrals of interest this is {\it not} the case, and we will see in detail how to deal with this. 
Taking the complex conjugate of (\ref{eq: Greens theorem2}) (with $w^*\equiv\bar{w}$), replacing $\bar{w}$ with a new $C^1$ complex function $\tilde{w}(z,\bar{z})$ (that is {\it not} necessarily the complex conjugate of $w(z,\bar{z})$) and taking a linear combination with (\ref{eq: Greens theorem2}) yields:
\begin{equation}\label{eq: Greens theorem full}
\boxed{\frac{1}{2\pi}\int_D\rmd^2z\,\big{(}\partial_{\bar{z}}w-\partial_z\tilde{w}\big{)}=\frac{1}{2\pi i}\oint_{\Gamma}\big(\rmd z\,w+\rmd\bar{z}\tilde{w}\big)}
\end{equation}
Of course, this is in agreement with the more conventional expression, $\int_{D}\rmd \varphi=\int_{\partial D}\varphi$, if we take $\varphi=w\rmd z+\tilde{w}\rmd\bar{z}$ (note that $\rmd = \rmd z\partial_z+\rmd\bar{z}\partial_{\bar{z}}$). Here and throughout we use the conventions: $\oint \rmd\bar{z}/\bar{z}=-2\pi i$, $\oint \rmd z/z=2\pi i$ (so that under complex conjugation we have the effective rule ``$(\oint)^*=\oint$'') and $\rmd^2z\equiv i\rmd z\wedge \rmd\bar{z}$. In passing, note that (since $\rho_m$ is real) we could use these integral identities to write (\ref{eq:chi(S)temp}) more democratically between chiral and anti-chiral halves,
\begin{equation}\label{eq:chi(S)tempdemo}
\begin{aligned}
\frac{i}{2\pi}\int_{\Sigma} \mathcal{R}&=\frac{1}{4\pi i}\sum_m\oint_{\partial V_m} \!\!\!(-\partial \ln \rho_{m}+\bar{\partial} \ln \rho_{m}).
\end{aligned}
\end{equation}
This will become relevant when we consider Riemann surfaces with boundaries in the following section. In particular, note that (\ref{eq:chi(S)tempdemo}) remains valid in the presence of true boundaries, $\partial\Sigma$, (but is then not a topological invariant unless boundary terms are added, more about which later). For now we focus on the equivalent chiral representation (\ref{eq:chi(S)temp}) with $\partial\Sigma=\zero$. 
\sk

The resulting contour integrals in (\ref{eq:chi(S)temp}) are along the boundaries of the $V_m$, see the second diagram in Fig.~\ref{fig:tripleoverlapsUV} and the four zoomed-in diagrams in Fig.~\ref{fig:segment} (all of which emphasise different aspects of the local setup), with the standard counterclockwise orientation with respect to $V_m$ inherited by the orientation of $\Sigma$. Every such boundary, $\partial V_m$, consists of one or more segments $\{C_{mn},C_{m\ell},\dots\}$ that either close or meet at triple or higher intersections\footnote{As already mentioned, we will usually work directly on the dual cover where there are only triple intersections (but also allow for annular overlaps where a given element $C_{mn}$ might be closed). We will also consider higher-point intersections when we consider the cover independence of the path integral measure below.}. So the segments $\{C_{mn},C_{m\ell},\dots\}$ are curves that traverse patch intersections $\{U_{mn},U_{m\ell},\dots\}$ as seen in the figure that begin and end at the vertices contained in $\{U_{mn\ell},\dots\}$. 
To every contribution along $C_{mn}\subset\partial V_m$ there will clearly be a corresponding contribution from the curve $C_{nm}\subset \partial V_n$ (notice the interchange of indices) that arises from the boundary integral along $\partial V_n$. (This is {\it not} true in the presence of boundaries; when $\partial\Sigma\neq\zero$ there remain {\it unpaired} boundary terms, as we elaborate on in the following subsection.) 
This in turn implies that the sum over $m$ in (\ref{eq:chi(S)temp}) can therefore be rewritten as a sum over {\it pairs}, $(mn)$, of curves as follows,\footnote{Although the contour integral along every element $\partial V_m$ in (\ref{eq:chi(S)tempdemo}) was compact (homeomorphic to a circle), the contour integral appearing in (\ref{eq:chi(S)temp2}) for every pair $(mn)$ may be either a circle or a segment, and in the latter case a union of segments corresponds to the original compact contour integrals appearing in (\ref{eq:chi(S)tempdemo}).}
\begin{equation}\label{eq:chi(S)temp2}
\begin{aligned}
\chi(\Sigma)
&=\frac{1}{2\pi i}\sum_{(mn)}\Big(\int_{C_{mn}} \!\!\!\rmd z_m(-\partial_{z_m}\ln \rho_{m})+\int_{C_{nm}} \!\!\!\rmd z_n(-\partial_{z_n}\ln \rho_{n})\Big).
\end{aligned}
\end{equation}
But both of these integrals are defined within the double intersection $U_{mn}$, and therefore the two relevant coordinate charts $(U_m,z_m)$ and $(U_n,z_n)$ will be related on $U_{mn}$ by a holomorphic transition function (\ref{eq:transitionfuncs}), namely 
$
z_n=f_{nm}(z_m),
$ 
as discussed in detail in Sec.~\ref{sec:TFCR}. The quantity $\partial =\rmd z_n\partial_{z_n}=\rmd z_m\partial_{z_m}$ since this is invariant under holomorphic changes of coordinate, and since the $C_{nm}$ contour is precisely equivalent to the $C_{mn}$ contour with opposite orientation (\ref{eq:chi(S)temp2}) also equals,
\begin{equation}\label{eq:chi(S)temp3}
\begin{aligned}
\chi(\Sigma)&=\frac{1}{2\pi i}\sum_{(mn)}\int_{C_{mn}} \!\!\!\rmd z_m\partial_{z_m}\big(\!-\ln \rho_{m}+\ln \rho_{n}\big).
\end{aligned}
\end{equation}
Finally, since the metric tensor is invariant under holomorphic reparametrisations, on $U_{mn}$:
$$
\rho_{m}(z_m,\bar{z}_m)\rmd z_m\rmd\bar{z}_m=\rho_{n}(z_n,\bar{z}_n)\rmd z_n\rmd\bar{z}_n\quad\Rightarrow \quad \rho_{n}=\rho_{m}|f_{nm}'(z_m)|^{-2}.
$$
Given that $f_{nm}(z_m)$ is holomorphic in $U_{mn}$, when we substitute this expression for the metric into (\ref{eq:chi(S)temp3}) the metric and anti-holomorphic dependences cancel out and we are left with:
\begin{equation}\label{eq:chi(S)}
\begin{aligned}
\chi(\Sigma)
&=\frac{i}{2\pi}\sum_{(mn)}\int_{C_{mn}} \!\!\!\rmd z_m\partial_{z_m}\!\ln f_{nm}'(z_m),
\end{aligned}
\end{equation}
which is precisely the desired result (\ref{eq:Euler}) that we set out to prove. 
\sk

Let us also check that the result (\ref{eq:chi(S)}) is invariant under holomorphic changes of coordinates (\ref{eq:wgz_mnot}), namely\footnote{We have not fixed the invariance under Weyl transformations here, which is why invariance under such holomorphic reparametrisations is still present. Recall the discussion associated to holomorphic normal coordinates in Sec.~\ref{sec:HNC} and in particular Sec.~\ref{sec:HTFWR}.} $z_m\mapsto w_m=g_m(z_m)$. Taking into account the transformation law for derivatives of transition functions (\ref{eq:f'=g-1hg}) (exchanging $m\leftrightarrow n$) we learn that (\ref{eq:chi(S)}) takes the form,
\begin{equation}\label{eq:chi(S)w}
\begin{aligned}
\chi(\Sigma)&=\frac{i}{2\pi}\sum_{(mn)}\int_{C_{mn}} \!\!\!\rmd z_m\partial_{z_m}\!\ln \big[g'_n(z_n)^{-1}h_{nm}'(w_m)g'_m(z_m)\big]\\
&=\frac{i}{2\pi}\sum_{(mn)}\int_{C_{mn}} \!\!\!\rmd w_m\partial_{w_m}\!\ln h_{nm}'(w_m)-\frac{i}{2\pi}\sum_{(mn)}\int_{C_{mn}} \!\!\!\rmd z_m\partial_{z_m}\!\ln \big[g'_n(z_n)g'_m(z_m)^{-1}\big]\\
\end{aligned}
\end{equation}
where we took into account that $\partial\equiv \rmd z_m\partial_{z_m}=\rmd w_m\partial_{w_m}$. 
That the last term on the right-hand side in (\ref{eq:chi(S)w}) indeed vanishes is shown in (\ref{eq:chi(S)Xw2}) (where in the absence of true boundaries the quantity $\{{\rm b.v.}\}=0$), and so (\ref{eq:chi(S)}) is indeed well-defined and independent of the choice of coordinates; in particular, we can also write:
\begin{equation}\label{eq:chi(S)w2}
\begin{aligned}
\chi(\Sigma)&=\frac{i}{2\pi}\sum_{(mn)}\int_{C_{mn}} \!\!\!\rmd w_m\partial_{w_m}\!\ln h_{nm}'(w_m).
\end{aligned}
\end{equation}
Notice also that the last term in (\ref{eq:chi(S)w}) does {\it not} vanish trivially, but rather that each integral over $C_{mn}$ contributes two boundary terms and it is the {\it sum} of all boundary terms that vanish. Therefore, that (\ref{eq:chi(S)}) is invariant under holomorphic changes of coordinates requires global information. This is a common feature of using such a ``holomorphic formalism'' and remains true for the string path integral (an example can be found in the Appendix in \cite{Polchinski87} and also in \cite{Nelson89}).\footnote{Incidentally, below we discuss how to make such boundary contributions effectively {\it local} in $\Sigma$ in the case of interest that is the ghost contributions to the path integral measure. The trick is to rewrite expressions in the holomorphic formalism in terms of covariant quantities that transform as tensors on patch overlaps, in which case any boundary contributions vanish trivially by virtue of the fact that tensors, $T$, (rather than tensor components) restricted to open sets, $T_m=T|_{U_m}$, are equal, $T_m(p)=T_n(p)$, on chart overlaps, $p\in U_m\cap U_n$.}
\sk

The result (\ref{eq:chi(S)}) is a special case of a much more general result \cite{BottTu}. 
To highlight the geometrical interpretation, let us also mention that the set of all quantities $\mathrm{h}_{nm}(z_m)\dfn f_{nm}'(z_m)^{-1}$ (satisfying $\mathrm{h}_{mn}(z_n)=\mathrm{h}_{nm}(z_m)^{-1}$) for all $m,n$ is precisely \cite{Chern} the set of holomorphic transition functions for the tangent bundle, $T\Sigma\rightarrow\Sigma$; on patch overlaps, $U_m\cap U_n\neq\varnothing$, the bases transform as:
$$
\partial_{z_n}=\mathrm{h}_{nm}(z_m)\partial_{z_m},\qquad {\rm with}\qquad \mathrm{h}_{nm}(z_m)\dfn f_{nm}'(z_m)^{-1},
$$
in terms of which, writing $\rmd=\rmd z_m\partial_{z_m}+\rmd\bar{z}_m\partial_{\bar{z}_m}$ for the total differential on $\Sigma$ (at fixed complex structure) expanded in terms of the $(U_m,z_m)$ chart coordinates,
$$
\chi(\Sigma)=\frac{i}{2\pi}\sum_{(mn)}\int_{C_{mn}}\!\!\rmd\ln \mathrm{h}_{nm}^{-1}.
$$

An elementary consistency check of (\ref{eq:chi(S)}) is to consider the case $\Sigma=S^2$, cover it with two charts, $\{(U_{\rm S},z_{\rm S}), (U_{\rm N},z_{\rm N})\}$, with $U_{\rm S}\cap U_{\rm N}\neq\zero$ spanning an equatorial band of finite width and with transition function $z_{\rm N}=f_{\rm NS}(z_{\rm S})=1/z_{\rm S}$ on $U_{\rm S}\cap U_{\rm N}$. (Here `${\rm N}$' and `${\rm S}$' stand for `North' and `South' respectively.) Applying (\ref{eq:chi(S)}) to this context indeed yields,
\begin{equation}\label{eq:chi(S2)}
\begin{aligned}
\chi(S^2)
&=\frac{i}{2\pi}\oint_{C_{\rm SN}} \!\!\!\rmd z_{\rm S}\partial_{z_{\rm S}}\!\ln f_{\rm NS}'(z_1)\\
&=\frac{i}{2\pi}\oint_{C_{\rm SN}} \!\!\!\rmd z_{\rm S}\partial_{z_{\rm S}}\!\ln \frac{-1}{z_{\rm S}^2}\\
&=\frac{i}{2\pi}\oint_{C_{\rm SN}} \!\!\!\rmd z_{\rm S}\frac{1}{z_{\rm S}}(-2)\\
&=2,
\end{aligned}
\end{equation}
as expected from the general result at arbitrary genus (\ref{eq:chi=2-2g-b}). In the last equality we took into account that the contour, $C_{\rm SN}$, is (as discussed) always to be interpreted as being counterclockwise from the viewpoint of the $(U_{\rm S},z_{\rm S})$ chart. At higher genus and focusing on a single handle, we may use plumbing fixture to introduce a handle by singling out two cells, $V_1,V_2$, and promoting them to annuli by removing circles of radius $|z_1|=|q^{1/2}|=|z_2|$ and then declare that $z_1$ is to be identified with $z_2$ if they are related by the transition function $z_1z_2=q$ (with $q$ a complex number such that $|q|<1$). In particular, we may declare that if $|z_1|<|q^{1/2}|$ we are to use the $z_2$ coordinate induced by the transition function, and vice versa if $|z_2|<|q^{1/2}|$ we are to use the $z_1$ coordinate. This creates a handle. By an explicit calculation that is entirely analogous to (\ref{eq:chi(S2)}) one finds that this handle contribution to $\chi(\Sigma)$ is precisely $-2$ because the orientation of the contour $C_{12}$ in this case is {\it opposite} to that in (\ref{eq:chi(S2)}), namely to $C_{\rm SN}$. Introducing additional handles is equivalent to carrying out the same procedure in other charts (assuming a sufficient number of charts has been introduced and that we are at a generic point in moduli space), and so one finds that every handle contributes an integer $-2$. Combining this with the result (\ref{eq:chi(S2)}) for a sphere leads to a genus-$\g$ compact Riemann surface with Euler characteristic $\chi(\Sigma)=2-2\g$, as one would expect. Notice that (in the absence of physical boundaries) there is no topological obstruction to taking the {\it remaining} transition functions (i.e.~those not associated to handles or the inversion (\ref{eq:chi(S2)})) to be linear, $f_{mn}(z_n)=a_{mn}z_n+b_{mn}$ with $a_{mn},b_{mn}\in \mathbf{C}$. 
This computation is simple because $\chi$ is topological, but in more general contexts complex structure deformations introduce additional structure.
\sk

Incidentally, the procedure outlined above can be used \cite{Polchinski87,Nelson89} to compute integrals of total derivatives of quantities that are not globally-defined on $\Sigma$, and in the physics literature this goes at least as far back as the work of Wu and Yang \cite{WuYang75}. An underlying motivation for going through the above in detail is that we apply a procedure analogous to the derivation of (\ref{eq:chi(S)})  from (\ref{eq:chiA}) to the path integral measure in Sec.~\ref{sec:TPIM}. But first we will need to consider the corresponding generalisation associated to Riemann surfaces with boundaries.

\subsubsection{With Boundaries}\label{sec:WB}
\begin{figure}
\begin{center}
\includegraphics[angle=0,origin=c,width=0.46\textwidth]{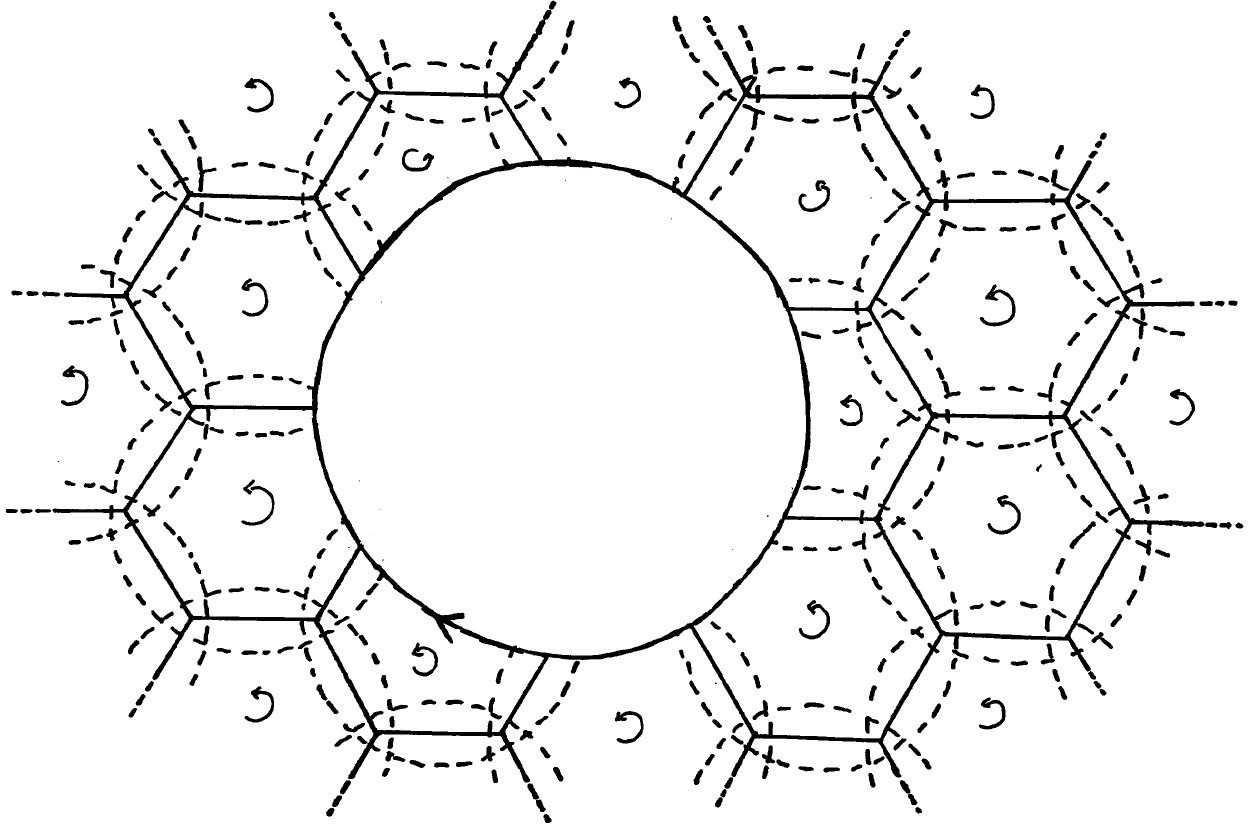}\qquad
\includegraphics[angle=0,origin=c,width=0.46\textwidth]{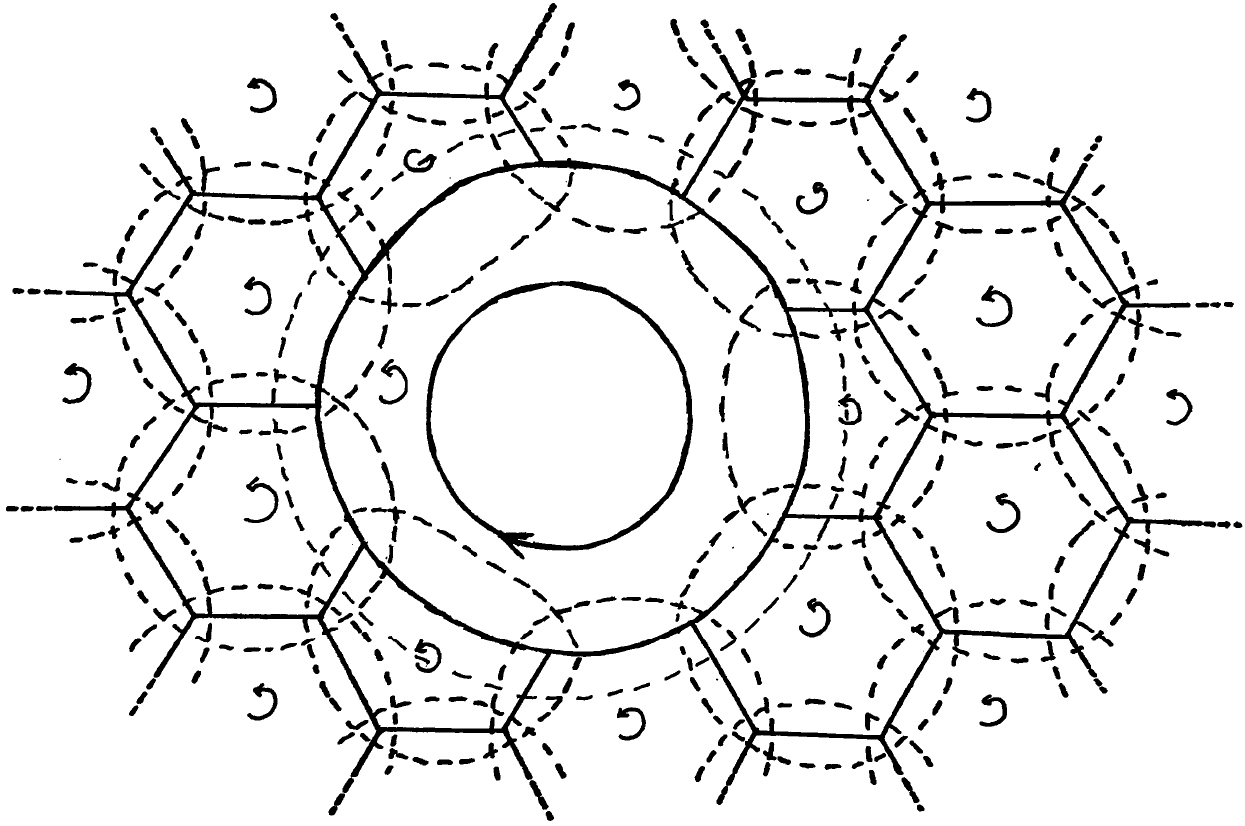}
\caption{Two smooth physical boundaries of a surface, $\Sigma$, their corresponding covers, chart overlaps, cell decompositions (interpreted as in Fig.~\ref{fig:23overlaps} and Fig.~\ref{fig:genericdualtriangles}), and canonical orientations induced by the orientation of $\Sigma$. The first diagram is associated to a {\sl good cover}, since it only contains contractible charts (diffeomorphic to discs), whereas the second is {\sl not} associated to a good cover since the boundary chart is topologically an annulus. Complementary diagrams corresponding to the first and second diagrams here are shown in the first and second diagrams respectively in Fig.~\ref{fig:goodboundaryshrink}.}\label{fig:holeorient}
\end{center}
\end{figure}

We now generalise the above calculation of the differential-geometric derivation of the Euler characteristic, $\chi(\Sigma)$, of an oriented genus-$\g$ Riemann surface, $\Sigma$, to that in the presence of a boundary, $\partial \Sigma$, consisting of $\B$ disconnected components (either of which is topologically $S^1$). The Euler characteristic is now given by Gauss-Bonnet theorem (\ref{eq:GaussBonnet1}), repeated here for convenience\footnote{Of numerous available expositions of the Gauss-Bonnet theorem, Chern's exposition \cite{Chern67} (in conjunction with \cite{ChernChenLam}) is in particular highly recommended, as is Tu's complementary and recent exposition \cite{Tu17}. Our approach will differ from these in several respects, the emphasis here being on a procedure that will also work best for the path integral measure later on.},
\begin{equation}\label{eq:GaussBonnet}
\boxed{\chi(\Sigma)=\frac{i}{2\pi}\int_\Sigma \mathcal{R}+\frac{1}{2\pi}\int_{\partial \Sigma}\rmd s\, k_{\rm g}(s)+\frac{1}{2\pi}\sum_i(\pi-\alpha_i)}
\end{equation}
where $k_{\rm g}(s)$ is the geodesic curvature of $\partial\Sigma$ (see below), $\pi-\alpha_i$ are the exterior angles at the vertices of $\partial\Sigma$ (if the latter is only {\it sectionally} smooth rather than smooth). The three terms on the right-hand side of (\ref{eq:GaussBonnet}) correspond respectively to the surface curvature, line curvature and point curvature of $\Sigma$, so that $\chi(\Sigma)$ is interpreted as the total curvature of $\Sigma$ \cite{Chern}. If the boundary $\partial \Sigma$ contains no vertices (i.e.~if $\partial\Sigma$ is smooth, such as that depicted in Fig.~\ref{fig:holeorient}) then the last term in (\ref{eq:GaussBonnet}) vanishes since every $\alpha_i=\pi$. The quantities $k_{\rm g}(s)$ and $\alpha_i$ will be defined much more explicitly below.
\sk

The expression for the Euler characteristic (\ref{eq:GaussBonnet}) for a surface with boundaries differs from the corresponding expression (\ref{eq:chiA}) of the previous section by the addition of the second and third terms in the former. But let us understand in detail {\it why} the new terms are required in order to reproduce a well-defined expression for $\chi(\Sigma)$. This will in turn guide us as we search for the correct boundary contributions associated to cutting open the worldsheet path integral measure across various cycles of $\Sigma$. 
\sk

\begin{figure}
\begin{center}
\includegraphics[angle=0,origin=c,width=0.95\textwidth]{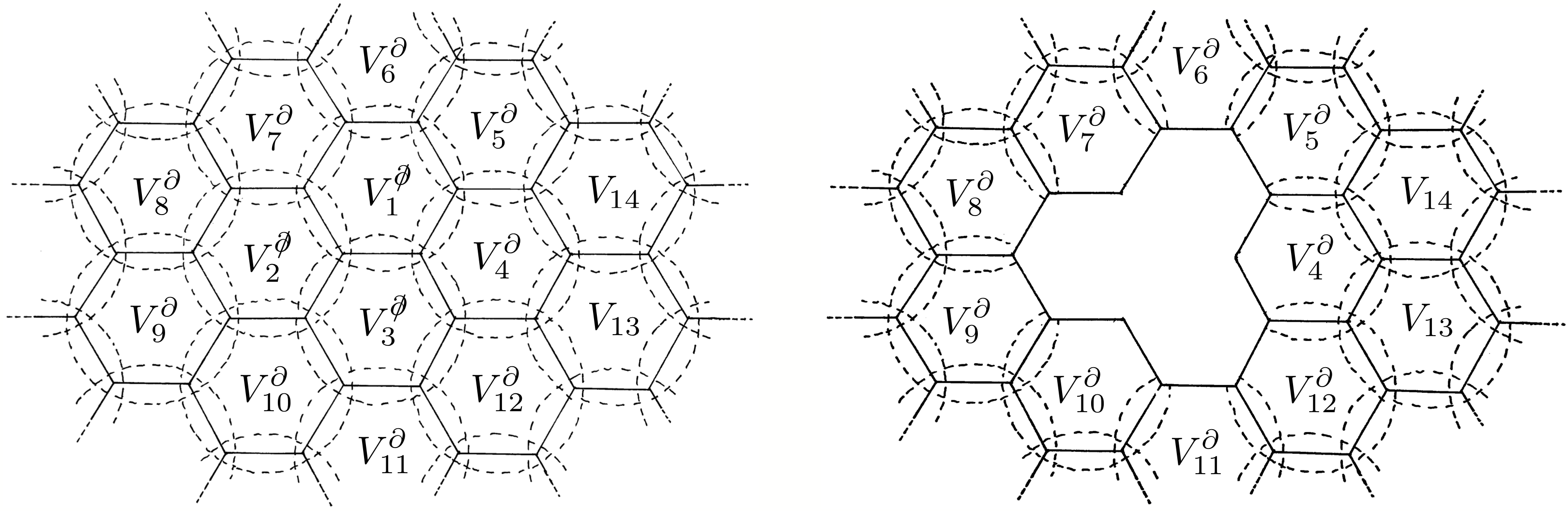}
\caption{Creating a boundary on $\Sigma$. In the first diagram part of a compact oriented Riemann surface, $\Sigma$, is depicted with a choice of cell decomposition, $\mathscr{V}$ (as defined in Sec.~\ref{sec:NB}). We then label which cells we wish to remove, call them $\{V_m^{\partial\!\!\!\slash}\}$, and which cells will acquire a physical boundary when the depicted cells in the set $\{V_m^{\partial\!\!\!\slash}\}$ are removed. We label the cells that will acquire a physical boundary by $\{V_n^\partial\}$. Only a subset of the resulting edges of these latter cells will acquire a physical boundary. Finally, we leave the labelling of the cells not adjacent to a boundary unchanged. In the second figure we have removed the sets $\{V_m^{\partial\!\!\!\slash}\}$ resulting in a physical boundary, $\partial\Sigma$. The orientation of $\partial\Sigma$ is that inherited by the orientation of $\Sigma$, see the first diagram in Fig.~\ref{fig:holeorient}.}\label{sec:holes2}
\end{center}
\end{figure}

Let us begin by considering the integral over the curvature tensor in the {\it presence} of boundaries,
$$
\frac{i}{2\pi}\int_\Sigma \mathcal{R}.
$$
The discussion leading to (\ref{eq:chi(S)tempdemo}) in the previous section may be carried over to this context without modification, so we may immediately write down:
\begin{equation}\label{eq:chi(S)tempdemo2}
\begin{aligned}
\frac{i}{2\pi}\int_{\Sigma} \mathcal{R}&=\frac{1}{4\pi i}\sum_m\oint_{\partial V_m} \!\!\!(-\partial \ln \rho_{m}+\bar{\partial} \ln \rho_{m}).
\end{aligned}
\end{equation}
The fact that there are boundaries, $\partial\Sigma\neq\zero$, can implemented by simply omitting terms in the sum over $m$ (the sum over cells), which would correspond to the fact that certain polygons in the cell decomposition are simply absent when $\partial\Sigma\neq\zero$. The relevant procedure is outlined in Fig.~\ref{sec:holes2}. 
In particular, making this explicit we primarily {\it relabel} the elements of the set $\mathscr{V}$ as defined in the previous subsection:
$$
\mathscr{V}=\{V_1,V_2,\dots\}\equiv \big\{V_m,V_n^{\partial},V_\ell^{\partial\!\!\!\slash}\big\},
$$
and introduce boundaries by simply deleting the subset $\{V_\ell^{\partial\!\!\!\slash}\}$. 
Then, the resulting reduced set corresponds to a cell decomposition of a Riemann surface with boundaries,
\begin{equation}\label{eq:Vpartial}
\mathscr{V}_{\partial}=\big\{V_m,V_n^{\partial}\big\},
\end{equation}
the notation being such that {\it every} edge of any one element in $\{V_m\}$ coincides with an edge of another element in $\mathscr{V}_{\partial}$, whereas any one element in $\{V_n^{\partial}\}$ has at least one edge  that is not in common with another element in $\mathscr{V}_{\partial}$ and at least one edge that {\it is} in common with that of another element in $\mathscr{V}_{\partial}$. This is illustrated in Fig~\ref{sec:holes2}. The union of the cells in $\{V_n^{\partial}\}$ correspond topologically to an annulus (or a set of annuli when there are multiple disconnected boundary components in $\partial\Sigma$, only one of which is shown in Fig~\ref{sec:holes2}). 
\sk

Restricting to the atlas associated to (\ref{eq:Vpartial}), we make the physical boundary contributions explicit in (\ref{eq:chi(S)tempdemo2}) using the above definitions,
\begin{equation}\label{eq:chi(S)tempdemo3}
\begin{aligned}
\frac{i}{2\pi}\int_{\Sigma} \mathcal{R}&=\frac{1}{4\pi i}\sum_m\oint_{\partial V_m} \!\!\!(-\partial \ln \rho_{m}+\bar{\partial} \ln \rho_{m})+\\
&\qquad+\frac{1}{4\pi i}\sum_m\oint_{\partial V_m^{\partial}}(-\partial \ln \rho_{m}+\bar{\partial} \ln \rho_{m}).
\end{aligned}
\end{equation}

Next consider the set $\{\partial V_m^{\partial}\}$. Every element, $\partial V_m^{\partial}$, is topologically a circle. We decompose the collection of all such elements into {\it physical} boundary segments or edges, $\mathscr{C}=\{C_\ell\}$, and then the remaining segments are all interior edges that in turn coincide with corresponding edges of neighbouring cells. In the notation of the previous subsection, namely the notation explained in the paragraphs leading to (\ref{eq:chi(S)temp3}), the entire set of distinct edges (with the standard orientation explained there) correspond to those associated to {\it pairs}, $(mn)$, labelled by $\{C_{mn}\}$, and the physical edges that constitute the physical boundary labelled by $\{C_\ell\}$. That is, the complete set of distinct edges associated to a cell decomposition of a Riemann surface with boundary is:
$$
\big\{C_{mn},C_\ell\big\},
$$
which in turn enables us to rewrite (\ref{eq:chi(S)tempdemo3}) (using the same reasoning that led to (\ref{eq:chi(S)temp3})) as follows,
\begin{equation}\label{eq:chi(S)temp4}
\begin{aligned}
\frac{i}{2\pi}\int_{\Sigma} \mathcal{R}&=\frac{1}{4\pi i}\sum_{(mn)}\int_{C_{mn}} \!\!\!\big(\!-\partial\ln \rho_{m}+\partial\ln \rho_{n}\big)-\frac{1}{4\pi i}\sum_{(mn)}\int_{C_{mn}} \!\!\!\big(-\bar{\partial} \ln \rho_{m}+\bar{\partial} \ln \rho_{n}\big)\\
&\qquad +\frac{1}{4\pi i}\sum_{\ell}\int_{C_\ell}(-\partial \ln \rho_{\ell}+\bar{\partial} \ln \rho_{\ell}).
\end{aligned}
\end{equation}
Since the metric components, $\rho(z,\bar{z})$, are real we can make a change of variables to rewrite the result in terms of a chiral half only,
\begin{equation}\label{eq:chi(S)temp5}
\begin{aligned}
\frac{i}{2\pi}\int_{\Sigma} \mathcal{R}&=\frac{1}{2\pi i}\sum_{(mn)}\int_{C_{mn}} \!\!\!\big(\!-\partial\ln \rho_{m}+\partial\ln \rho_{n}\big)-\frac{1}{2\pi i}\sum_{\ell}\int_{C_\ell}\partial \ln \rho_{\ell}.
\end{aligned}
\end{equation}
So the sum over $\ell$ is over all segments, $C_\ell$, (i.e.~edges) that make up the physical boundary components of $\Sigma$,
$$
\partial\Sigma=\bigcup_\ell C_\ell.
$$ 
For example, if there was only a single boundary, and that was identified with the boundary shown in Fig.~\ref{sec:holes2}, the total number of elements in the set $\{C_\ell\}$ would be 12, and therefore the sum over $\ell$ in (\ref{eq:chi(S)temp5}) would be over these 12 terms. 
\sk

Finally, since the integral associated to the sum over pairs $(mn)$ is along {\it interior} edges, identical reasoning that led to (\ref{eq:chi(S)}) applies here too for these terms, so that (\ref{eq:chi(S)temp5}) may be written directly in terms of transition functions on patch overlaps,
\begin{equation}\label{eq:chi(S)X}
\boxed{
\begin{aligned}
\frac{i}{2\pi}\int_{\Sigma} \mathcal{R}
&=\frac{i}{2\pi}\sum_{(mn)}\int_{C_{mn}}\!\!\! \partial\ln f_{nm}'(z_m)-\frac{1}{2\pi i}\sum_{\ell}\int_{C_\ell}\partial \ln \rho_{\ell}(z_\ell,\bar{z}_\ell)
\end{aligned}
}
\end{equation}
This is the result of including boundaries on $\Sigma$ simply by {\it omitting} cells, as depicted in Fig.~\ref{sec:holes2}. 
But we must be very careful, because whether the right-hand side in (\ref{eq:chi(S)X}) is well-defined or not depends crucially on whether we are considering a {\it good cover} or {\it not}. In particular, although the first term on the right-hand side seemingly depends entirely on the data that defines a Riemann surface (namely holomorphic transition functions on patch overlaps), the last term on the right-hand side is an integral of a {\it connection} (see (\ref{eq:Gammazzz})) so it is not obvious whether it is well-defined under holomorphic reparametrisations. We will show that the right-hand side depends on the choice of cover.
\sk

Let us understand this last point in detail, and primarily examine how the right-hand side in (\ref{eq:chi(S)X}) transforms under a holomorphic reparametrisation,
$$
z_m\mapsto w_m(z_m),
$$
in every chart, $(U_m,z_m)$, of the cover. We will show that it is invariant {\it when} we choose a good cover, but if we allow for a cover that contains sets, $U_m$, that are diffeomorphic to annuli rather than discs then the right-hand side in (\ref{eq:chi(S)X}) depends on the choice of coordinates.
\sk

Recall that in the original chart coordinates the transition functions were given by $z_n=f_{nm}(z_m)$ on patch overlaps $U_m\cap U_n$ (on which the contour $C_{mn}$ in (\ref{eq:chi(S)X}) is defined, for every pair $(mn)$). Let us then suppose that in the new holomorphic coordinates, $w_n$, we have new transition functions, $w_n=h_{nm}(w_m)$, so that if we perform a holomorphic reparametrisation, $z_n\mapsto w_n(z_n)$, in every chart, taking into account that $z_n=f_{nm}(z_m)$, we learn that,
$$
w_n\big(f_{nm}(z_m)\big) = h_{nm}\big(w_m(z_m)\big).
$$
Let us then differentiate this relation with respect to $z_m$ and use the chain rule to arrive at,
$$
h_{nm}'(w_m)=(\partial_{z_n}w_n)f_{nm}'(z_m)(\partial_{z_m}w_m)^{-1}.
$$
This is simply the result we arrived at in (\ref{eq:f'=g-1hg}) using slightly differently phrasing. 
Likewise, the metric components in the new coordinates in terms of the old read, for every $\ell$,
$$
\ln\tilde{\rho}_{\ell}(w_\ell,\bar{w}_\ell) = \ln\rho_\ell(z_\ell,\bar{z}_\ell)-\ln|\partial_{z_\ell}w_\ell|^{2}.
$$

Next, we make use of these relations in (\ref{eq:chi(S)X}) to find that the right-hand side of (\ref{eq:chi(S)X}) in the new coordinates is given by (omitting the overall factor, $\frac{i}{2\pi}$),
\begin{equation}\label{eq:chi(S)Xw}
\begin{aligned}
&\sum_{(mn)}\int_{C_{mn}}\!\!\! \partial\ln h_{nm}'(w_m)+\sum_{\ell}\int_{C_\ell}\partial \ln \tilde{\rho}_{\ell}(w_\ell,\bar{w}_\ell)\\
&\qquad=\sum_{(mn)}\int_{C_{mn}}\!\!\! \partial\ln f_{nm}'(z_m)+\sum_{\ell}\int_{C_\ell}\partial \ln \rho_{\ell}(z_\ell,\bar{z}_\ell)+\\
&\quad\qquad +\sum_{(mn)}\int_{C_{mn}}\!\!\! \partial\ln \big[(\partial_{z_n}w_n)(\partial_{z_m}w_m)^{-1}\big]-\sum_{\ell}\int_{C_\ell}\partial \ln (\partial_{z_\ell}w_\ell).
\end{aligned}
\end{equation}

Now comes the distinction between a good cover and a more general cover. Let us suppose initially that we {\it do} have a good cover. Then, the contours $C_{mn}$ are along interior edges with either both endpoints at triple interior vertices, or with one endpoint at an interior triple vertex and one endpoint at a triple boundary vertex. Correspondingly, the contours $C_\ell$ are along boundary edges with endpoints on triple  boundary vertices, as illustrated in the second diagram in Fig.~\ref{sec:holes2} or the {\it first} diagram in Fig.~\ref{fig:holeorient}. If (\ref{eq:chi(S)X}) was well-defined the two terms in the last line on the right-hand side in (\ref{eq:chi(S)Xw}) would be zero. To study these two terms let us define the function:
$$
f_m(z_m)\dfn \ln\partial_{z_m}w_m.
$$ 
Given that by assumption we are initially considering a good cover (so that all patches and patch intersections are simply connected) then at each endpoint of the contour, $C_{mn}$, there is a point $z_{mn\ell}:p\mapsto \mathbf{C}$ (with $p\subset U_m\cap U_n\cap U_\ell$) at which three interior contours meet, and {\it also} points where interior contours meet boundary vertices, and in particular the first integral in the last line in (\ref{eq:chi(S)Xw}) therefore takes the form:
\begin{equation}\label{eq:chi(S)Xw2}
\begin{aligned}
&\sum_{(mn)}\int_{C_{mn}}\!\!\! \partial\ln \big[(\partial_{z_n}w_n)(\partial_{z_m}w_m)^{-1}\big]\\
&=\sum_{(mn)}
\int_{C_{mn}}\!\!\! \partial\big(f_n(z_n) -f_m(z_m)\big)\\
&=\sum_{(mn\ell)}\Big\{
\big[f_n(z_{\hat{n}m\ell})-f_m(z_{n\hat{m}\ell})\big]+\big[f_m(z_{n\hat{m}\ell})-f_\ell(z_{nm\hat{\ell}})\big]+\big[f_\ell(z_{nm\hat{\ell}})-f_n(z_{\hat{n}m\ell})\big]\Big\}+\{{\rm b.v.}\}\\
&=0+\{{\rm b.v.}\}.
\end{aligned}
\end{equation}
A ``hat'', `$\hat{\phantom{n}}$', denotes the coordinate system in which the point $z_{mn\ell}$ is defined. The remaining quantity, $\{{\rm b.v.}\}$, on the right-hand side encodes the total contribution associated to {\it boundary vertices} from this term, and this precisely cancels the last term in (\ref{eq:chi(S)Xw}), since:
$$
\{{\rm b.v.}\}=\sum_{\ell}\int_{C_\ell}\partial f_\ell(z_\ell)=\sum_{\ell}\int_{C_\ell}\partial \ln (\partial_{z_\ell}w_\ell).
$$
Regarding the interior vertices' contribution that cancel out in (\ref{eq:chi(S)Xw2}), the statement (\ref{eq:chi(S)Xw2}) is precisely the statement that given a set of charts with cocycle relations, $f_{mn}\circ f_{n\ell}\circ f_{\ell m}=1$, under a holomorphic change of coordinates the new cocycle relations, $g_{mn}\circ g_{n\ell}\circ g_{\ell m}=1$, are trivially satisfied when they are satisfied in the initial system of coordinates. 
So we have therefore shown that for a {\it good cover}:
\begin{equation}\label{eq:chi(S)Xwzz}
\begin{aligned}
&\sum_{(mn)}\int_{C_{mn}}\!\!\! \partial\ln h_{nm}'(w_m)+\sum_{\ell}\int_{C_\ell}\partial \ln \tilde{\rho}_{\ell}(w_\ell,\bar{w}_\ell)\\
&\qquad=\sum_{(mn)}\int_{C_{mn}}\!\!\! \partial\ln f_{nm}'(z_m)+\sum_{\ell}\int_{C_\ell}\partial \ln \rho_{\ell}(z_\ell,\bar{z}_\ell),
\end{aligned}
\end{equation}
and therefore that the right-hand side in (\ref{eq:chi(S)X}) is invariant under holomorphic changes of coordinates.
\sk

If we did {\it not} choose a good cover however, and in particular considered a cover such  that boundary components are as in the {\it second} diagram in Fig.~\ref{fig:holeorient}, the quantity $\{{\rm b.v.}\}$ on the right-hand side of (\ref{eq:chi(S)Xw2}) would be {\it zero}.  (More generally it would not precisely cancel the last term on the right-hand side in (\ref{eq:chi(S)Xw})). This is because if we allowed a subset of $\{U_m\}$ to correspond to annuli (such that the ``interior'' boundaries of these annuli are identified with the boundaries of $\Sigma$ and the ``exterior'' boundaries are connected to the remaining bulk of $\Sigma$), there would be no contour $C_{mn}$ that connects an interior vertex to a boundary vertex. Therefore, there would be no way to cancel all contributions coming from the term $\sum_{\ell}\int_{C_\ell}\partial \ln (\partial_{z_\ell}w_\ell)$ on the right-hand side of (\ref{eq:chi(S)Xw}), and so the right-hand side in (\ref{eq:chi(S)X}) would be ill-defined if we required invariance under arbitrary holomorphic reparametrisations. We could evade this by specifying boundary conditions on the metric, and then demanding invariance under only a subset of holomorphic reparametrisations (which would perhaps be mostly trivial at a boundary component). But then the answer for the integral $\frac{i}{2\pi}\int_{\Sigma} \mathcal{R}$ would depend  on detailed information about the surface and it would be sensitive to data that is not purely topological.) 
\sk

In fact, we {\it will} be working with such a cover (as in the {\it second} diagram in Fig.~\ref{fig:holeorient}) in string perturbation theory. In the latter case we will be interested in creating ``fictitious'' boundaries by cutting open internal handles of Riemann surfaces. And we will there be interested in slices in moduli space that decouple complex structure moduli ``as much as possible'' (we elaborate on this in detail in following sections). As we will see, a good way to proceed is to include appropriate boundary terms, the analogue of which in the present example of the Euler characteristic is the following. Namely, if we allow for charts that are topologically annuli with the inner circles identified with boundaries of $\Sigma$, and demand invariance with respect to the full set of holomorphic reparametrisations, we should add boundary terms to (\ref{eq:chi(S)X}),
\begin{equation}\label{eq:chi(S)Xquifckfix}
\begin{aligned}
\frac{i}{2\pi}\int_{\Sigma} \mathcal{R}+\frac{1}{2\pi i}\sum_{\ell}\oint_{C_\ell}\partial \ln \rho_{\ell}(z_\ell,\bar{z}_\ell)
&=\frac{i}{2\pi}\sum_{(mn)}\int_{C_{mn}}\!\!\! \partial\ln f_{nm}'(z_m),
\end{aligned}
\end{equation}
where (as indicated in the notation $\oint_{C_\ell}$ as opposed to $\int_{C_\ell}$) for this choice of cover the contours $C_{\ell}$ are now {\it closed}, because by assumption there is a single coordinate system (which covers an annulus) for every boundary. So therefore the number of elements in the sum over $\ell$ in (\ref{eq:chi(S)Xquifckfix}) now equals the number of boundaries, $\B$. 
According to the above, both the left- and right-hand sides in (\ref{eq:chi(S)Xquifckfix}) are then invariant under holomorphic reparametrisations (when we consider a cover such that boundary components are as in the {\it second} diagram in Fig.~\ref{fig:holeorient}).  Although this is not the whole story in the case of the Euler characteristic (we elaborate on this below). Another way of phrasing this is that the left- and right-hand sides of this  precise expression (\ref{eq:chi(S)Xquifckfix}) are well-defined when all boundaries that we create on $\Sigma$ are associated to interior boundaries of open sets $\{U_m\}$ that are in turn diffeomorphic to annuli; this case is also depicted in the middle and right-most diagrams in Fig.~\ref{fig:goodboundaryshrink} (as opposed to the first diagram which might be part of a good cover). 
\sk

The necessary boundary terms in the current example are, more generally, taken care of by the inclusion of geodesic curvature (and a point curvature if the boundary is only sectionally smooth rather than smooth). This latter viewpoint has the advantage that it is independent of the choice of cover, so that the full Euler characteristic will be explicitly topological for all covers. Let us discuss this in detail next.
\sk

Consider an oriented Riemann surface, $\Sigma$, with or without boundary. To specify a coordinate representation of a given point $p\in \Sigma$ we (as usual) primarily construct an atlas, $\mathscr{U}=\{(U_m,z_m)\}$, and then pick a chart, call it $(U_m,z_m)$, such that $p\in U_m$. This then provides a coordinate representation for $p$ defined by $p\mapsto z_m(p)$, which we simply denote by $z_m\in \mathbf{C}$ (with appropriate restrictions for boundary charts). Since the surface is 2-dimensional we need two numbers to specify the coordinate of $p$, which we can take to be $(z_m,\bar{z}_m)$ in the specified chart. Suppose now that at every point $p\in \Sigma$ we construct a unit tangent space, $\mathrm{B}_p$ (we denote the collection of all such unit tangent spaces by $\mathrm{B}$), comprised at any one point $p$ of unit tangent vectors, 
$$
\xi_m=\xi^{z_m}_m\partial_{z_m}+\xi^{\bar{z}_m}_m\partial_{\bar{z}_m}.
$$
These are unit tangent vectors with respect to the inner product induced by the metric, $g$,  e.g.~in the chart $(U_m,z_m)$ we may go to conformal gauge where $g=\rho_m(z_m,\bar{z}_m)\rmd z_m\rmd\bar{z}_m$, in which case, 
\begin{equation}\label{eq:rhoxixi=1}
\rho_m(z_m,\bar{z}_m)\xi_m^{z_m}\xi_m^{\bar{z}_m}=1.
\end{equation}
Since the tangent space at $p$ is flat this assignment is well-defined ({\it locally} in $\Sigma$ there is no obstruction to requiring a tangent vector have unit length, more about which below). Taking into account also that $\xi_m^{\bar{z}_m}=(\xi_m^{z_m})^*$ we find the following general solution to (\ref{eq:rhoxixi=1}):
\begin{equation}\label{eq:xi_m}
\xi_m^{z_m}(z_m,\bar{z}_m,\theta_m)=\rho_m(z_m,\bar{z}_m)^{-\frac{1}{2}}e^{i\theta_m},\qquad \xi_m^{\bar{z}_m}(z_m,\bar{z}_m,\theta_m)=\rho_m(z_m,\bar{z}_m)^{-\frac{1}{2}}e^{-i\theta_m},
\end{equation}
and because $\rho_m$ is real and positive definite the square root is well-defined. 
So to fully specify an element in $\mathrm{B}$ we actually need three coordinates, $(z_m,\bar{z}_m,\theta_m)$, and in particular therefore $\mathrm{B}$ is a {\it three}-dimensional space. What we have just described is a local trivialisation, $\mathbf{C}\times U(1)$, of the unit tangent or frame bundle $\mathrm{B}\rightarrow \Sigma$ with $\theta_m$ playing the role of a fibre coordinate over $p\in U_m\subset\Sigma$. Given the unit tangent vector $\xi_m$, let the quantity
$$
\eta_m=\eta_m^{z_m}\partial_{z_m}+\eta_m^{\bar{z}_m}\partial_{\bar{z}_m},
$$
be the unique unit tangent vector that is orthogonal to $\xi_m$ such that $\eta_m$ is obtained from $\xi_m$ by rotating the latter in a counterclockwise sense in the $(U_m,z_m)$ chart by an angle $\pi/2$, i.e.~$
\eta_m=i\xi_m.
$ 
An explicit expression for the corresponding components follows immediately:
$$
\eta_m^{z_m}(z_m,\bar{z}_m,\theta_m)=i\rho_m(z_m,\bar{z}_m)^{-\frac{1}{2}}e^{i\theta_m},\qquad \eta_m^{\bar{z}_m}(z_m,\bar{z}_m,\theta_m)=-i\rho_m(z_m,\bar{z}_m)^{-\frac{1}{2}}e^{-i\theta_m}.
$$
Note that $\xi_m\cdot \xi_m=\eta_m\cdot \eta_m=1$ and $\eta_m\cdot\xi_m=0$ with respect to the inner product induced by $g$ as in (\ref{eq:rhoxixi=1}). If we assign tangent vectors, $\xi_m,\eta_m$, to every point $p\in U_m$ we obtain local unit vector fields, and since furthermore the pair, $(\xi_m,\eta_m)_p$, at any point $p$ forms a basis for the tangent space $T_pU_m$ we have in fact defined a {\it local} frame field. We emphasise that these unit tangent vectors are only defined locally and generically {\it do not} extend to the entire Riemann surface, the obstruction being the Euler number. But they do extend to the entire Riemann surface modulo U(1). Since this is a central point that will play a role below let us elaborate further and explain this in detail. 
\sk

\begin{figure}
\begin{center}
\includegraphics[angle=0,origin=c,width=0.35\textwidth]{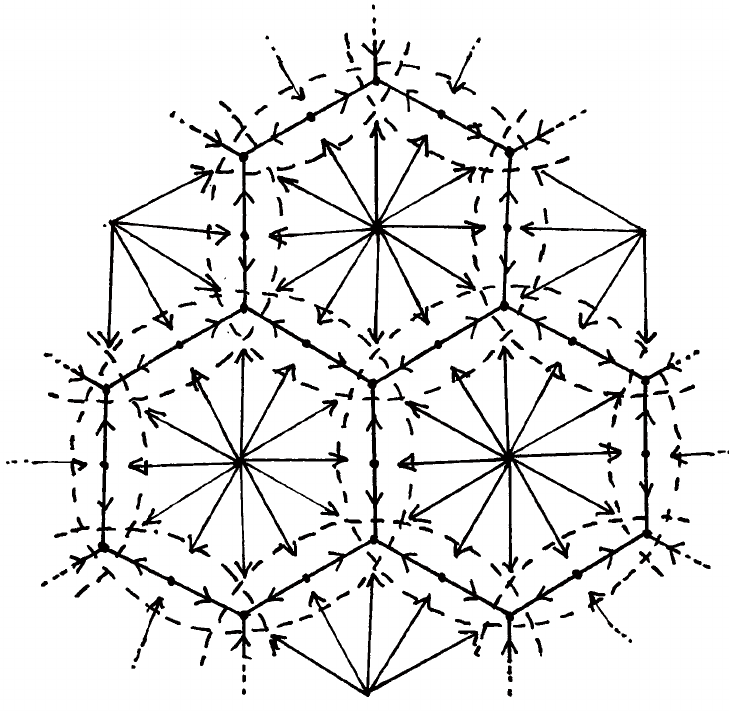}
\caption{Part of the cell decomposition of a Riemann surface, $\Sigma$, also depicted in Fig.~\ref{fig:tripleoverlapsUV}, but now endowed with a smooth {\it globally}-defined vector field, $v$, (consistent with Hopf's index theorem). The arrows indicate flow lines of the vector field, such that every vertex, edge and face contain a single zero (termed  `singularity' by mathematicians) of $v$, with indices $+1$, $-1$ and $+1$ respectively. Adding all indices yields $\sum_iI_{p_i}=\mathbf{v}-\mathbf{e}+\mathbf{f}$ which by definition (\ref{eq:chi=vef}) equals the Euler characteristic, $\chi(\Sigma)$. }\label{sec:flows}
\end{center}
\end{figure}

A {\it zero} (usually referred to as a {\it singular point} \cite{ChernChenLam} in the math literature) of a smooth vector field, $v$, on $\Sigma$ is a point $p_i\in\Sigma$ such that $v(p_i)=0$. The properties of a vector field, $v$, near a singular point, $p_i$, can be quite rich \cite{Chern67}, but the property of interest here is summarised by {\it Hopf's index theorem} \cite{BottTu} which states that the sum of the indices of a vector field on a compact oriented manifold is the Euler characteristic of the manifold \cite{Milnor,Chern67,BottTu},
\begin{equation}\label{eq:Ipi=chi}
\boxed{\sum_iI_{p_i}=\chi(\Sigma)}
\end{equation}
This is simply the higher-genus generalisation of the {\it hairy ball theorem}. 
So the sum is over all $i$ corresponding to isolated zeros of the given vector field, $v(p_i)=0$. Informally speaking, the {\it index}, $I_{p_i}$, of a vector field around a singular point, $p_i\in\Sigma$, is in turn an integer equal to the number of times the ``arrow'' of the vector field twists around itself as we traverse (in an anti-clockwise sense) a simple closed curve which contains that particular singular point (and no other singular point). A more precise statement is given below, see the reasoning leading up to (\ref{eq:Ipi}), but for illustration see also Fig.~\ref{sec:flows} for an example of how this theorem can be used to draw the flow diagram of a {\it globally}-defined smooth vector field on a general Riemann surface. 
Anti-clockwise twisting (i.e.~positive with respect to the orientation of $\Sigma$) contributes positive index and clockwise twisting contributes negative index \cite{Chern67}. Clearly, since the Euler characteristic is a topological invariant, globally-defined {\it unit} tangent vectors do not exist unless the Euler characteristic vanishes. If however rather than requiring unit tangent vectors to be globally defined we define them locally and allow {\it phase} discontinuities across patch overlaps, then the sum of these discontinuities can `mimic' an index such that the sum over such indices adds up to give the Euler characteristic required for a globally consistent setup. In the string theory literature, explicit implications of this were perhaps first been pointed out by Polchinski \cite{Polchinski87,Polchinski88} (with related discussion found in \cite{FriedanShenker87} and later elaborated on in \cite{Nelson89,LaNelson90}), who phrased this rather cryptically as `the phase of a local coordinate is not globally defined, the obstruction being the Euler number'. The above explanation is, presumably, the origin of the requirement \cite{Polchinski88} that the corresponding ill-defined phase always be taken to be integrated (or that only combinations of operators are considered that are insensitive to this phase). This is, e.g., directly related to the $b_0-\tilde{b}_0=0$ constraint (as an operator equation), more about which below. We refer to this inability to define the phase of a local coordinate globally on $\Sigma$ as the `U(1) {\it ambiguity}'. This U(1) ambiguity will manifest itself in numerous places throughout the document.
\sk

These frame fields, $\xi,\eta$, lead us to define a {\it connection form}, $\varphi$, on $\mathrm{B}$, in particular a real and linear differential 1-form given by,\footnote{When the tangent vector $\xi$ is identified with a tangent vector to the boundary $\partial\Sigma$ (and is also coherently oriented with the standard orientation inherited by that of $\Sigma$, see e.g., Fig.~\ref{fig:Greens}) then the normal vector $\eta$ is {\it inwards-pointing}; in the notation of Polchinski \cite{Polchinski_v1}, $\xi^a=t^a$ and $\eta^a=-n^a$, see Ex.~1.3 in \cite{Polchinski_v1} and related comments on p.~56,57,90 there.} 
\begin{equation}\label{eq:varphiDxieta}
\varphi \dfn g_{ab}(D\xi^a)\eta^b,
\end{equation}
with $D$ the exterior covariant derivative (or covariant differential) \cite{Chern67,Tu17}. Written out in components in the $(U_m,z_m)$ chart (using that $g_{z_m\bar{z}_m}=\frac{1}{2}\rho_m$ and $g_{z_mz_m}=0$), we have\footnote{The quantity $D\xi_m^{z_m}$ is not to be confused with the covariant derivative, $\nabla \xi_m^{z_m}$, which is turn reads: $\nabla\xi_m^{z_m}=\partial\xi_m^{z_m}+(\partial\ln\rho_m)\xi_m^{z_m}$ (with $\nabla=\rmd z_m\nabla_{z_m}$ and $\overline{\nabla}=\rmd\bar{z}_m\nabla_{\bar{z}_m}$). (See Appendix \ref{sec:CD}.)} $D\xi_m^{z_m}=\rmd\xi_m^{z_m}+(\partial\ln\rho_m)\xi_m^{z_m}$ and\footnote{To derive $D\xi_m^{\bar{z}_m}=(D\xi_m^{z_m})^*$ one makes use of metric compatibility, $Dg_{z_m\bar{z}_m}=0$, that we raise and lower indices with the metric, $\xi_m^{\bar{z}_m}=g^{z_m\bar{z}_m}\xi_{mz_m}$, and also $D\xi_{mz_m}=\rmd\xi_{mz_m}-(\partial\ln\rho_m)\xi_{mz_m}$. Note that $\rho_m$ is independent of $\theta_m$.} $D\xi_m^{\bar{z}_m}=(D\xi_m^{z_m})^*$ and in particular therefore,
\begin{equation}
\begin{aligned}
\varphi_m &= \frac{1}{2}\rho_m\big[\rmd\xi_m^{z_m}+(\partial\ln\rho_m)\xi_m^{z_m}\big]\eta_m^{\bar{z}_m}+{\rm c.c.},
\end{aligned}
\end{equation}
where in the $(U_m,z_m)$ chart $\partial=\rmd z_m\partial_{z_m}$ (with $\bar{\partial}$ the complex conjugate). Making use of the explicit expressions for $\xi_m^{z_m},\eta_m^{\bar{z}_m}$, in terms of $\rho_m$ and $\theta_m$ provided above we can recast the connection form, $\varphi_m$, as:
\begin{equation}\label{eq:varphi-covdiff}
\boxed{\varphi_m=\rmd\theta_m-i(\partial-\bar{\partial})\ln\rho_m^{1/2}}
\end{equation}
As noted above this is a differential form on $B$, the $(z_m,\bar{z}_m,\theta_m)$ being local trivialising coordinates. Taking the exterior derivative of $\varphi_m$ leads immediately to the fundamental relation:\footnote{We are being somewhat naive in this discussion; the sharp way to phrase these developments is in terms of a {\it frame bundle}, but placing the current discussion in the latter context would require introducing additional machinery that we will not be needing below. For a crystal clear discussion along those lines see \cite{Tu17}.}
\begin{equation}\label{eq:varphi=-iR}
\boxed{\rmd \varphi_m=-i\mathcal{R}}
\end{equation}
where we made use of the explicit expression for the curvature tensor, $\mathcal{R}$,  given in terms of $\rho_m$ in (\ref{eq:Rtensor}). $\mathcal{R}$ is of course globally defined, and therefore so is $\rmd \varphi_m$ globally defined, despite the fact that $\varphi_m$ is only locally defined.
\sk

Let us elaborate on this last point. Suppose that we rotate the frame by an angle $\alpha_m$ keeping $p$ fixed. We then obtain  rotated (but still mutually orthogonal and unit-normalised) frame fields, $\xi_m^{'z_m}=e^{i\alpha_m}\xi_m^{z_m}$, $\eta_m^{'z_m}=e^{i\alpha_m}\eta_m^{z_m}$, and by (\ref{eq:xi_m}) and (\ref{eq:varphi-covdiff}) a new connection form, $\varphi_m'$,
\begin{equation}\label{eq:varphi'=da+varphi}
\varphi_m'=\rmd\alpha_m+\varphi_m.
\end{equation}
Since $p$ is fixed this rotation induces motion along a fibre by action of the group, $\mathrm{G}={\rm U}(1)$. In particular, $\varphi'_m=\varphi_m\circ G$ is also a differential form on $\mathrm{B}$. From (\ref{eq:varphi=-iR}) and (\ref{eq:varphi'=da+varphi}) we obtain $\rmd\varphi_m'=-i\mathcal{R}$, so that the curvature tensor is invariant under rotations of frame and since furthermore on patch overlaps, $U_m\cap U_n$, we have $\mathcal{R}=\mathcal{R}_m=\mathcal{R}_n$, it should therefore indeed be regarded as a globally-defined 2-form on $\Sigma$. 
\sk

Above we explained that unit tangent vectors that extend to the entire manifold $\Sigma$ do not exist unless $\chi=0$, in that there is a topological obstruction that enforces a given vector field to vanish at a discrete set of points. Let us suppose that $p_i\in\Sigma$ is such a singular point of $\xi,\xi'$. In the {\it vicinity} of $p_i$ (or more generally on $\Sigma\setminus \{p_1,p_2,\dots\}$ if $\xi,\xi'$ has a multitude of singular points in the manifold) and at fixed complex structure the tangent vector $\xi$ is well-defined, so we can integrate (\ref{eq:varphi'=da+varphi}) along the contour $\Gamma_i+\Gamma_i'$ (or $\Gamma+\Gamma_i''$) indicated in Fig.~\ref{fig:Greens}, such that $D$ is centred at this singular point, $p_i$, and such that there are no other singular points within $D$,
$$
\oint_{\Gamma_i+\Gamma_i'}\varphi_m'=\oint_{\Gamma_i+\Gamma_i'}\rmd\alpha_m+\oint_{\Gamma_i+\Gamma_i'}\varphi_m
$$
Making use of Green's theorem (\ref{eq: Greens theorem full}), we write the contour integrals of $\varphi_m,\varphi_m'$ as integrals over $D$, so that taking into account that $\rmd\varphi_m=\rmd\varphi_m'=-i\mathcal{R}$ yields,
\begin{equation}
\oint_{\Gamma_i} \rmd\alpha_m=\oint_{-\Gamma_i'}\rmd\alpha_m=\oint_{-\Gamma_i''}\rmd\alpha_m,
\end{equation}
from which it follows that this contour integral is invariant under continuous deformations of the simple closed contour containing $p_i$ and changes sign when we reverse its orientation. Since $\Gamma_i$ (or $\Gamma_i',\Gamma_i''$) are simple closed contours (i.e.~Jordan curves) the angle $\alpha_m$ must return to itself as we traverse such a contour {\it up to} an integer multiple of $2\pi$ (since the tangent vector is single-valued). This integer multiple is precisely the {\it index} of $\xi$ around the singular point, $p_i$,
\begin{equation}\label{eq:Ipi}
I_{p_i}\equiv \frac{1}{2\pi}\oint_{\Gamma_i} \rmd\alpha_m.
\end{equation}

Consider now a closed {\it curve}, $\gamma:[0,1]\rightarrow \Sigma$, which we denote by $C$, and parametrise it by proper time (or arc length \cite{Tu17}), so that $\gamma(0)=\gamma(1)$ represent the same point in $\Sigma$. In order to avoid cumbersome notation however we might also write somewhat imprecisely $\gamma(s)=s$, with $s$ a parameter taking values in the unit interval. Let us focus on a local patch, $U_m$, where the line element, $\rmd s^2=\rho_m \rmd z_m\rmd\bar{z}_m$. Then, a coordinate parametrisation of the curve, $C$, in the subset $U_m$ would be $(z_m(s),\bar{z}_m(s))$, and we may construct a local and continuous (but not necessarily smooth) section, $\sigma_m$, over $U_m$ by also assigning an angle $\theta_m$ to every point $(z_m,\bar{z}_m)$, i.e.,
$$
\sigma_m:(z_m,\bar{z}_m)\mapsto (z_m,\bar{z}_m,\theta_m).
$$ 
So if we suppose that the domain of $\sigma_m$ is that determined by the curve of interest, $C$, (i.e.~if we consider the composition $\sigma_m\circ \gamma$) then we obtain a map from the curve arclength to a point in the fibre over $\gamma(s)$, namely $s\mapsto(z_m(s),\bar{z}_m(s),\theta_m(s))$, and this enables us to assign a smooth unit vector field $\xi_m(s)$ along $C$ with components,
$$
\frac{\rmd z_m}{\rmd s}=\xi_m^{z_m}(s),\qquad \frac{\rmd\bar{z}_m}{\rmd s}=\xi_m^{\bar{z}_m}(s),
$$ 
where the $s$-dependence of the angle, $\theta_m(s)$, is such that the resulting vector field, $\xi_m(s)$, is tangent to $C$ for some appropriate parameter range for $s$. 
Accordingly, since we want to integrate $\varphi_m$ in  (\ref{eq:varphi-covdiff}) (which lives on $B_m$) along the curve, $C$, (which lives in $U_m$) we can pull it back to $U_m$ using the aforementioned section,
\begin{equation}\label{eq:varphi=()ds}
(\sigma_m\circ\gamma)^*\varphi_m=\Big(\frac{\rmd\theta_m}{\rmd s}-i\frac{\rmd z_m}{\rmd s}\partial_{z_m}\ln\rho_m^{\frac{1}{2}}+i\frac{\rmd\bar{z}_m}{\rmd s}\partial_{\bar{z}_m}\ln\rho_m^{\frac{1}{2}}\Big)\rmd s.
\end{equation}
The quantity in the parenthesis on the right-hand side in (\ref{eq:varphi=()ds}) is called the {\it geodesic curvature}, $k_{\rm g}(s)$,
\begin{equation}\label{eq:kg(s)}
\begin{aligned}
k_{\rm g}(s) 
&= \frac{\rmd\theta_m}{\rmd s}-i\xi_m^{z_m}(s)\partial_{z_m}\ln\rho_m^{\frac{1}{2}}+i\xi_m^{\bar{z}_m}(s)\partial_{\bar{z}_m}\ln\rho_m^{\frac{1}{2}}
\end{aligned}
\end{equation}
According to the Gauss-Bonnet theorem (\ref{eq:GaussBonnet}) we should integrate $k_{\rm g}(s)$ over the curve, $C$, parametrised by $s$ and choose the curve such that the integral is along the boundary, $C=\partial\Sigma$, and the vector $\xi$ is tangent to $C$ with an orientation (see Fig.~\ref{fig:holeorient}) induced by that of $\Sigma$. If it so happens that the curve spans various charts we should sum over the contribution from each chart (while ensuring that we do not integrate over any one subset of $C$ more than once). 
\sk

Having gone through the relevant reasoning in detail it is now convenient to keep some of these details (in particular the pullback by $\sigma_m\circ\gamma$) implicit and instead write the result of integrating (\ref{eq:kg(s)}) over $\partial \Sigma$ as a sum of integrals as follows,
\begin{equation}
\begin{aligned}
\frac{1}{2\pi}\int_{\partial \Sigma}\rmd s\,k_{\rm g}(s)
&=\sum_{\ell}\frac{1}{2\pi}\int_{C_\ell}\rmd\theta_\ell-\frac{1}{4\pi i}\sum_{\ell}\int_{C_\ell}\big(-\partial\ln\rho_\ell+\bar{\partial}\ln\rho_\ell\big)
\end{aligned}
\end{equation}
Clearly, since $\rho_\ell$ is real the last two integrals are equal and we can equivalently write the result as an integral over the chiral half,
\begin{equation}\label{eq:geodesiccurvatureint}
\boxed{
\begin{aligned}
\frac{1}{2\pi}\int_{\partial \Sigma}\rmd s\,k_{\rm g}(s)
&=\sum_{\ell}\frac{1}{2\pi}\int_{C_\ell}\rmd\theta_\ell+\frac{1}{2\pi i}\sum_{\ell}\int_{C_\ell}\partial\ln\rho_\ell(z_\ell,\bar{z}_\ell)
\end{aligned}
}
\end{equation}

Referring back to the Gauss-Bonnet theorem (\ref{eq:GaussBonnet}), we see that when we add the surface and line curvatures given by (\ref{eq:chi(S)X}) and (\ref{eq:geodesiccurvatureint}) respectively the term, $-\frac{1}{2\pi i}\sum_{\ell}\int_{C_\ell}\partial\ln\rho_\ell(z_\ell,\bar{z}_\ell)$, precisely cancels out and we are left with,
\begin{equation}\label{eq:R+kg}
\begin{aligned}
\frac{i}{2\pi}\int_{\Sigma} \mathcal{R}+\frac{1}{2\pi}\int_{\partial \Sigma}\rmd s\,k_{\rm g}(s)
&=\frac{i}{2\pi}\sum_{(mn)}\int_{C_{mn}}\!\!\! \partial\ln f_{nm}'(z_m)+\sum_{\ell}\frac{1}{2\pi}\int_{C_\ell}\rmd\theta_\ell
\end{aligned}
\end{equation}
Notice that the geodesic curvature term provides precisely the necessary boundary term that appeared already in the second term on the left-hand side in (\ref{eq:chi(S)Xquifckfix}). In addition, the Gaussian curvature provides a contribution associated to `point curvature' that we discuss momentarily. Recall that the first term on the right-hand side of (\ref{eq:R+kg}) will be invariant under holomorphic reparametrisations only if we pick a cover such that the contours, $C_{mn}$, do not meet a boundary vertex, e.g., if we pick a cover for which boundaries are created by deleting discs contained within a given subset of charts, $\{U_m\}$, so that the boundary components are connected to the remaining surface only via annuli as in the middle and right-most diagrams in Fig.~\ref{fig:goodboundaryshrink} (in which case there is no patch overlap $U_m\cap U_n$ in the ``near vicinity'' of any one boundary edge).
\sk

\begin{figure}
\begin{center}
\includegraphics[angle=0,origin=c,width=0.4\textwidth]{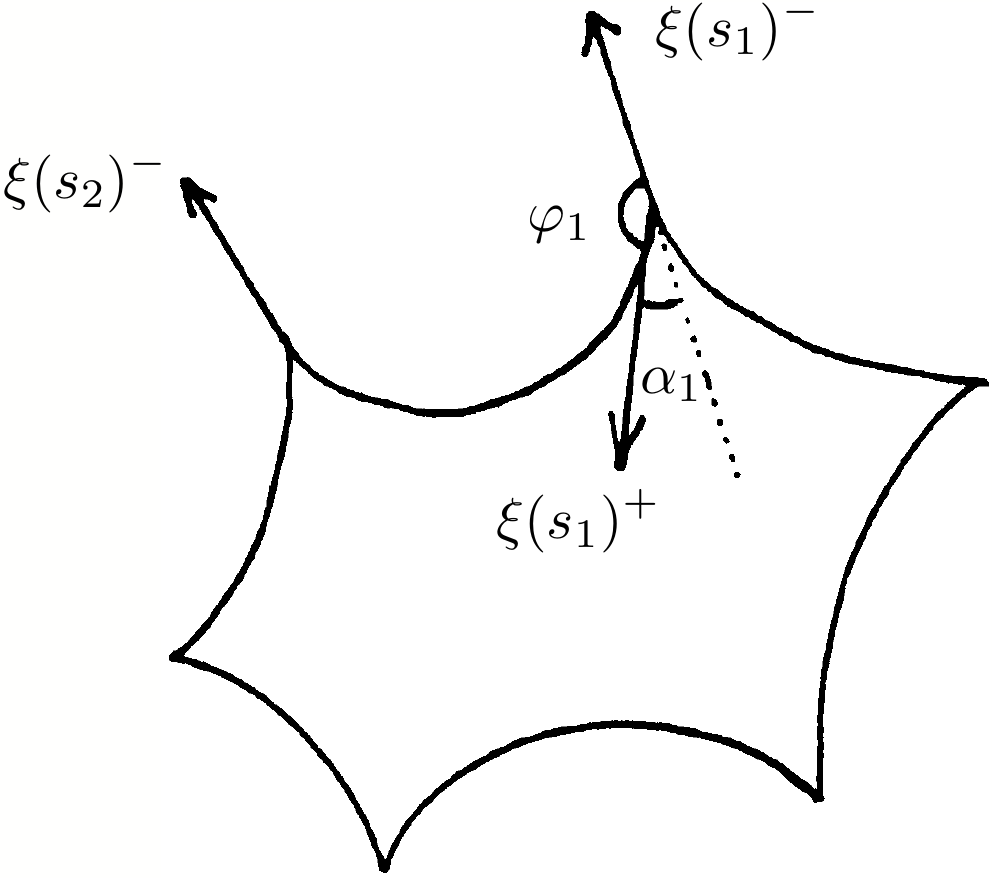}\qquad\quad
\includegraphics[angle=0,origin=c,width=0.36\textwidth]{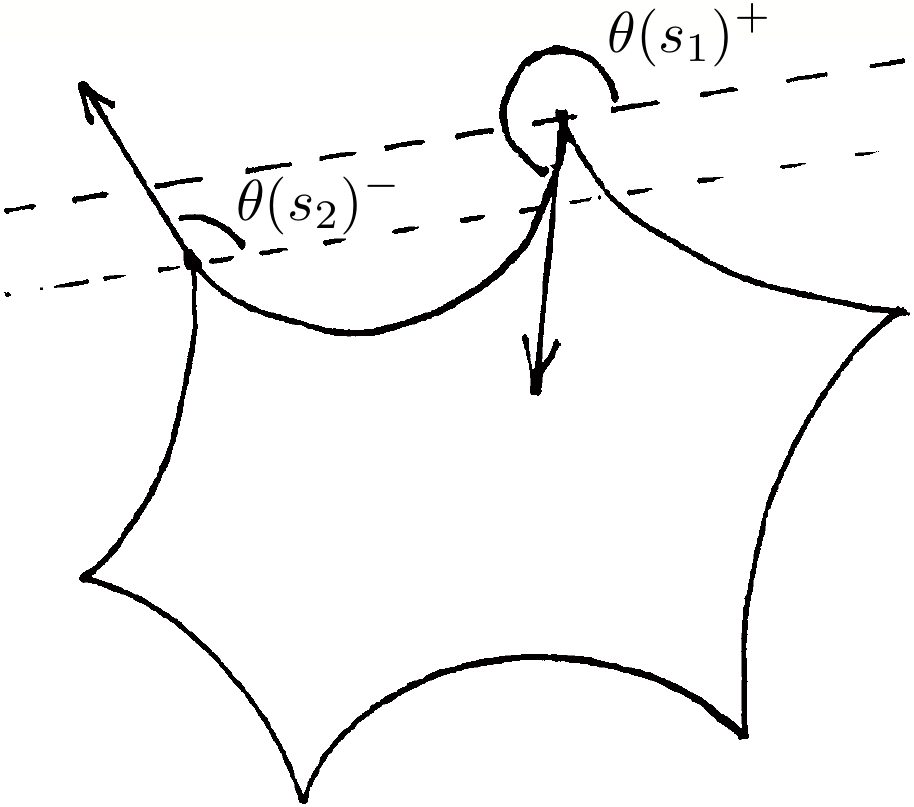}\,\,
\caption{Illustration of the various tangent vectors and angles that define the rotation index (\ref{eq:2pigamma}). It is to be understood that the bulk of the surface $\Sigma$ (of which the spiky closed curve is a boundary) corresponds to the ``outside'' of the bounded region. The dashed lines in the diagram on the right denote an arbitrary fixed frame with respect to which the angle $\theta(s_i)^{\pm}$ is measured. (In this example $m=6$.)}\label{fig:spikes}
\end{center}
\end{figure}

The {\it second} term on the right-hand side in (\ref{eq:R+kg}) gives rise to a sum over angles over the sectionally-smooth physical boundaries, where the angle, $\theta_\ell(s)$, parametrises the direction of the tangent vector with respect to a fixed frame (only differences in angles enter so further information about this fixed frame cancels out) for every open set $C_\ell$. This angle, $\theta_\ell(s)$, is furthermore only well-defined at points where the tangent vector along the boundary is single-valued. To incorporate the case where the tangent vector along the boundary is not necessarily single-valued suppose rather that $\Sigma$ has $\B$ {\it sectionally-smooth} boundaries, so that $\partial\Sigma$ is comprised of $\B$ disconnected components.\footnote{(In this case the angle $\theta_\ell(s)$ is ill-defined, in particular double-valued, at a discrete set of points, so we adopt the notation $\theta_\ell(s_i)^{\pm}$ at these points to distinguish the two, corresponding to the angle between the fixed frame and the vectors $\xi(s_i)^{\pm}$, see below.)} 
Every such component is a sectionally-smooth simple closed curve. Consider one such simple closed curve and let $s_i$, $i=1,\dots,m$, be the arc length measured from a vertex $A_0$ to vertex $A_i$, such that $s_m$ is the arc length of the curve (with a condition $A_{m+1}\equiv A_1$ that ensures the curve is closed). At the $i^{\rm th}$ vertex there are two unit tangent vectors, $\xi(s_i)^-$ and $\xi(s_i)^+$, tangent respectively to the smooth arcs $A_{i-1}A_i$ and $A_{i}A_{i+1}$ at $s_i$, and we let the angle from the tangent $\xi(s_i)^-$ to $\xi(s_i)^+$ be\footnote{The angle $\varphi_i$ is positive when the vector $\xi(s_i)^-$ is rotated in a sense, $\circlearrowleft$, that is coherently induced by the orientation, $\circlearrowleft$, of $\Sigma$ to reach $\xi(s_i)^+$ as indicated in Fig.~\ref{fig:spikes}.} $\varphi_i$, $-\pi<\varphi_i<\pi$, referred to as the {\it exterior angle} \cite{Chern67} or {\it jump angle} \cite{Tu17} at the $i^{\rm th}$ vertex. See Fig.~\ref{fig:spikes}. 
In addition to the set of jump angles, $\{\varphi_i\}$, there is also the set of angles, $\theta(s_i)^--\theta(s_{i-1})^+$, corresponding to the angle from the tangent vector $\xi(s_{i-1})^+$ to the vector $\xi(s_i)^-$  (with respect to a fixed frame) after parallel transporting the former to the latter. It is a theorem \cite{Chern67,Tu17} that the sum of all these angles equals $2\pi\gamma$, with $\gamma$ the {\it rotation index},
\begin{equation}\label{eq:2pigamma}
2\pi\gamma=\sum_i^m\big[\theta(s_i)^--\theta(s_{i-1})^+\big]+\sum_{i=1}^m\varphi_i,\qquad \gamma=\pm1.
\end{equation}
We will not prove this here, but it is not hard to convince oneself of the result (\ref{eq:2pigamma}) since the jump angles simply ``fill the gaps'' in angle differences not counted in the first term on the right-hand side (a detailed proof can be found in \cite{Chern67,Tu17}). In our context of interest every closed contour appearing in the last term in (\ref{eq:R+kg}) has an orientation induced by that of $\Sigma$, and since we have introduced boundaries by introducing ``holes'' in $\Sigma$ it follows that $\gamma=-1$ for every boundary component. The above was for a single connected boundary component, but now we sum over {\it all} boundary components of $\Sigma$. The left hand side of (\ref{eq:2pigamma}) then becomes $-2\pi\B$, the first term on the right-hand side takes the form $\sum_{\ell}\int_{C_\ell}\rmd\theta_\ell$ (which coincides with the corresponding term in (\ref{eq:R+kg})), and the second term on the right-hand side of (\ref{eq:2pigamma}) also results in a simple sum over all jump angles, denoted briefly by $\sum_i\varphi_i$ where the range of the sum is now over {\it all} vertices in every component of every boundary. (In Fig.~\ref{fig:spikes}, e.g., the number of vertices is $m=6$.) In particular therefore,
\begin{equation}\label{eq:2pigamma2}
\sum_{\ell}\frac{1}{2\pi}\int_{C_\ell}\rmd\theta_\ell=-\B-\frac{1}{2\pi}\sum_i\varphi_i.
\end{equation}
It is traditional to write the exterior angles, $\varphi_i$, in terms of the complementary {\it interior angles}, $\alpha_i$, defined by $\alpha_i\dfn \pi-\varphi_i$. Summarising, (\ref{eq:R+kg}) takes the form:
\begin{equation}\label{eq:R+kgX}
\boxed{
\begin{aligned}
\frac{i}{2\pi}\int_{\Sigma} \mathcal{R}+\frac{1}{2\pi}\int_{\partial \Sigma}\rmd s\,k_{\rm g}(s)+\frac{1}{2\pi}\sum_i(\pi-\alpha_i)&=\frac{i}{2\pi}\sum_{(mn)}\int_{C_{mn}}\!\!\! \partial\ln f_{nm}'(z_m)-\B.
\end{aligned}
}
\end{equation}

Let us consider now surfaces with only {\it smooth} boundaries, and in particular with boundaries such as that shown in the second diagram in Fig.~\ref{fig:holeorient}, where boundary charts are topologically equivalent to annuli, so that no interior edge reaches any given true boundary of $\Sigma$. 
For such a smooth boundary the sum $\sum_i(\pi-\alpha_i)=0$ since then the interior angles, $\{\alpha_i\}$, at the vertices are then all equal to $\pi$: locally, a smooth curve is by definition ``straight''. 
\sk

The left-hand side of (\ref{eq:R+kgX}) is precisely the Euler characteristic, quoted in (\ref{eq:GaussBonnet1}), which in turn famously takes the form (\ref{eq:chi=2-2g-b}), namely $\chi(\Sigma)=2-2\g-\B$. Setting the latter relation equal to the right-hand side of (\ref{eq:R+kgX}), we recover the relation that we obtained in the previous subsection associated to Riemann surfaces without boundary:
\begin{equation}\label{eq:2-2g=fmn}
\boxed{\frac{i}{2\pi}\sum_{(mn)}\int_{C_{mn}}\!\!\! \partial\ln f_{nm}'(z_m)=2-2\g}
\end{equation}
This is an important consistency check, and it must be the case since as mentioned above the contours, $C_{mn}$, do not meet the boundary (we are not considering a good cover here), so that boundaries are effectively invisible to (\ref{eq:2-2g=fmn}). 
The contours, $C_{mn}$, are still over {\it all} patch overlaps but there are no patch overlaps that coincide with boundaries. 
\sk

Secondly, we have shown that in the presence of boundary components the integral $\frac{i}{2\pi}\int_{\Sigma} \mathcal{R}$ (or more precisely the right-hand side in (\ref{eq:chi(S)X})) is not conformally-invariant unless we consider a `good cover' and one needs to add boundary terms to make it conformally-invariant. The left-hand side in (\ref{eq:R+kgX}) is well-defined independently of the choice of cover. From (\ref{eq:chi(S)X}), making use of the explicit expression for the curvature tensor given in (\ref{eq:Rtensor}), and taking into account (\ref{eq:2-2g=fmn}), we learn that:
\begin{equation}\label{eq:R+Boundary=fmn}
\boxed{
\begin{aligned}
\frac{1}{2\pi}\int_{\Sigma}\rmd^2z\,\partial_{\bar{z}}\partial_z\ln \rho(z,\bar{z})-\frac{1}{2\pi i}\int_{\partial\Sigma}\rmd z\partial_z \ln \rho(z,\bar{z})
&=\frac{1}{2\pi i}\sum_{(mn)}\int_{C_{mn}}\!\!\! \partial\ln f_{nm}'(z_m)
\end{aligned}
}
\end{equation}
is also well-defined when $\rho(z,\bar{z})$ is evaluated in the relevant chart inherited by the atlas associated to our chosen cover (which is such that the contours, $C_{mn}$, never reach boundary vertices). Any ill-defined terms on the left-hand side in (\ref{eq:R+Boundary=fmn}) cancel between the first and second integrals, so that the result (on the right-hand side) is given entirely in terms of the fundamental data defining the Riemann surface, namely holomorphic transition functions satisfying appropriate cocycle relations (that were in turn defined in Sec.~\ref{sec:TFCR}). 
\sk

On a related note, and just as importantly, notice that if we naively applied Green's theorem (\ref{eq: Greens theorem2}) to (\ref{eq:R+Boundary=fmn}) we would conclude that the left-hand side in the latter is zero -- this is clearly not the case, the obstruction being precisely the Euler characteristic. Of course, Green's theorem does not apply directly to (\ref{eq:R+Boundary=fmn}) because the integrand in the second term on the left-hand side is {\it not} globally defined (it does not transform as a tensor on patch overlaps), which is to be expected since the integrand is essentially a connection.
\sk

In string perturbation theory we encounter such integrands (i.e.~local sections that do not extend to global sections so that that the underlying bundles are non-trivial), and in this section we have derived and shown very explicitly one way to deal with this subtle issue, even if we do not adopt a good cover. Namely we decompose the manifold into charts, and we can also allow for annular charts in the vicinity of boundaries. On any one such chart the corresponding integral is well-defined and we can use Green's or Stoke's theorem. We then glue these together paying careful attention to any boundary terms that arise. We will of course go through the relevant reasoning for the path integral measure in detail below.
\sk

Before closing this section it will be useful (for what follows) to also display an equivalent expression to (\ref{eq:R+Boundary=fmn}). Let us suppose that $\Sigma$ has only a single boundary component that is (in the spirit of the above comments) in turn created by {\it removing} a disc $D'$ of radius $|z_1|=r$ from the cell $D\equiv V_1\subset U_1$ (in the chart $(U_1,z_1)$), and that the resulting range of $z_1$ is now $r<|z_1|<1$. So $z_1$ is an annulus coordinate on, $D-D'$, see Fig.~\ref{fig:Greens} on p.~\pageref{fig:Greens}. The boundary of the annulus associated to $|z_1|=r$ is therefore identified with $\partial\Sigma$ (but with opposite orientation with respect to the $z_1$ coordinate) and (\ref{eq:R+Boundary=fmn}) may (somewhat imprecisely) then be written as:
\begin{equation}\label{eq:R+Boundary=fmn2}
\begin{aligned}
\frac{1}{2\pi}\int_{|z_1|>r}\rmd^2z_1\,\partial_{\bar{z}_1}\partial_{z_1}\ln \rho_1(z_1,\bar{z}_1)+\frac{1}{2\pi i}\int_{|z_1|=r}\rmd z_1\partial_{z_1} \ln \rho_1(z_1,\bar{z}_1)
&=\frac{1}{2\pi i}\sum_{(mn)}\int_{C_{mn}}\!\!\! \partial\ln f_{nm}'(z_m)
\end{aligned}
\end{equation}
Using Green's theorem (\ref{eq: Greens theorem2}) (see Fig.~\ref{fig:Greens}) it is an elementary exercise to show that:
\begin{equation}\label{eq:R+Boundary=fmn3}
\begin{aligned}
\frac{1}{2\pi i}\int_{|z_1|=1}\rmd z_1\partial_{z_1} \ln \rho_1&=\frac{1}{2\pi}\int_{1>|z_1|>r}\rmd^2z_1\,\partial_{\bar{z}_1}\partial_{z_1}\ln \rho_1+\frac{1}{2\pi i}\int_{|z_1|=r}\rmd z_1\partial_{z_1} \ln \rho_1.
\end{aligned}
\end{equation}
Now since the left-hand side of the latter relation is independent of $r$ so is the right-hand side. Adding terms associated to the remaining Riemann surface ($|z_1|>1$) to the left- and right-hand sides in (\ref{eq:R+Boundary=fmn3}) implies immediately that (\ref{eq:R+Boundary=fmn2}) is independent of the radius, $r$. An analogous observation in the context of the path integral measure below will allow us to transition from a path integral with a boundary of radius $r$ to a path integral with a puncture, precisely by considering such a cover and demonstrating the same $r$-independence which allows us to take the limit $r\rightarrow 0$. This is the essence of why path integrals factorise in string theory, and more precisely `why' we can represent intermediate states by a pair of {\it local} offshell vertex operators (whose quantum numbers are summed/integrated over).\footnote{It may be of interest to notice the parallel reasoning presented in \cite{Polchinski88} associated to eqn.~(23) there and the reasoning explained below eqn.~(25) there, but we will come to this below.} 
\sk

This pair of local vertex operators will correspond to the {\it handle operators} we are aiming to construct. Since we need to integrate over complex structures, we need to primarily specify a gauge slice in moduli space, and then unravel how the corresponding handle operators change as we move along this gauge slice. Since different gauge slices in moduli space give rise to different path integral measures, we will begin (in the following section) by studying the path integral measure (first in general and then for a specific globally well-defined mod U(1) slice) in detail. 

\section{The Path Integral Measure}\label{sec:TPIM}
In this section we consider the path integral measure in detail, paying careful attention to boundary contributions arising from cutting open the path integral across various (trivial or non-trivial) homology cycles. To be precise, the quantity of interest is the path integral measure associated to integrating over worldsheet metrics (in particular complex structures modulo diffeomorphisms), but we take the corresponding Fadeev-Popov ghost or BRST outlook throughout \cite{Polchinski_v1,Witten12c}. In this formulation the measure associated to integrating over metrics becomes (at a fixed loop order) a finite dimensional (moduli space) integral with an appropriate Jacobian factor that determines a gauge slice in moduli space. This Jacobian factor is in turn written in terms of ghosts and transition functions of the underlying surface, and takes different forms depending on which cycles we wish to cut open or pinch and how we wish to distribute or partition the various moduli of the underlying surface. To orient the discussion, we begin with a subsection on a very brief discussion of the full path integral before focusing on the measure.

\subsection{Full Path Integral I: `Fixed Picture'}\label{sec:FPIFP}
The general starting point is the following expression for (the connected part of) generic closed string S-matrix elements \cite{Polchinski_v1}, $S_{\g,\n}\equiv S_{\g}(1;\dots;\n)$, at genus $\g$ and with $\n$ fixed-picture vertex operator insertions,\footnote{The title of this subsection is meant to reflect the fact that all external vertex operators are in `fixed picture'. Although this is the natural picture for external vertex operators \cite{Witten12c}, there are cases where the integrated-picture S matrix is more useful (e.g., when the number of vertex operators is smaller than the number of conformal Killing vectors admitted by the surface); this is discussed in Sec.~\ref{sec:TEC}.}  $\hat{\mathscr{V}}_j$, 
\begin{equation}\label{eq:fullpathintegral}
\boxed{
\begin{aligned}
S_{\g,\n}&=e^{-\chi(\Sigma_\g)\Phi}\int_{\mathcal{M}_{\g,\n}}\frac{\rmd^{2\m}\tau}{n_R}\Big\langle\prod_{k=1}^\m\hat{B}_{\tau^k}\hat{B}_{\bar{\tau}^k}\prod_{j=1}^\n\hat{\mathscr{V}}_j\Big\rangle_{\Sigma_\g}
\end{aligned}
}
\end{equation}
The measure contributions, $\hat{B}_{\tau^k}\hat{B}_{\bar{\tau}^k}$, are discussed and defined in Sec.~\ref{sec:IToTF:I}-\ref{sec:SHUTF}. We are working in the BRST formalism, a handful of early references being \cite{GiddingsMartinec86,Martinec86b}, see also \cite{Mansfield87}, and especially Witten's recent analysis \cite{Witten12c} and Polchinski's textbook \cite{Polchinski_v1}. (Corresponding early references for the path integral measure in terms of determinants are \cite{MooreNelson86,DHokerPhong86}, see also \cite{Alvarez83} and \cite{Knizhnik86,Knizhnik89}).  The correlation function, $\langle \dots\rangle_{\Sigma_\g}$, denotes a path integral over (chiral and anti-chiral) ghost and matter fields,
$$
\boxed{\langle \dots\rangle_{\Sigma_\g}\dfn \int \mathcal{D}(b,c,\dots)\,e^{-I(b,c,\dots)}(\dots)}
$$
where we suppress the matter contributions (so that the discussion in this section remains true for arbitrary string backgrounds provided the total central charge vanishes), and the relevant (Euclidean worldsheet and spacetime signature) action \cite{Polchinski_v1} is denoted by $I(b,c,\dots)$. 
The quantity $\chi(\Sigma_\g) = 2-2\g$ is the standard Euler characteristic, see Sec.~\ref{sec:EulerCh}, and $\m$ is the total number of  (complex) moduli,
\begin{equation}
\m=
\left\{
\begin{aligned}
&0\\
&\n-3\\
&1\\
&\n\\
&3\g-3+\n
\end{aligned}
\right.
\qquad 
\begin{aligned}
&(\textrm{for $\g=0$ and $\n=0,1,2$})\\
&(\textrm{for $\g=0$ and $\n\geq3$})\\
&(\textrm{for $\g=1$ and $\n=0$})\\
&(\textrm{for $\g=1$ and $\n\geq1$})\\
&(\textrm{for $\g\geq2$ and $\n\geq0$})
\end{aligned}
\end{equation}
Notice that if $\n\geq3$ the number of moduli is always $3\g-3+\n$, for all $\g=0,1,2,\dots$, which one can think of as being associated to the $\n-3$ moduli of a sphere ($\g=0$) with $\n\geq3$ vertex operators plus 3 moduli for every handle. In fact, if we decide to represent all $\g$ handles by insertions of handle operators on a sphere this counting can be given precise meaning (more about which later).
\sk

In writing down (\ref{eq:fullpathintegral}) we have defined:
\begin{equation}\label{eq:d2mtau}
\begin{aligned}
\rmd^{2\m}\tau &\dfn \rmd^2\tau^1 \dots \rmd^2\tau^\m\\
&=i\rmd\tau^1\wedge \rmd\bar{\tau}^1\wedge \dots \wedge i\rmd\tau^m\wedge \rmd\bar{\tau}^\m,
\end{aligned}
\end{equation}
with $i^2=-1$, 
whereas the positive integer $n_R$ in (\ref{eq:fullpathintegral}) corresponds to the (finite) order of any residual discrete group of symmetries not fixed by the conformal gauge choice \cite{Polchinski_v1}. The full expression for S-matrix elements associated to $\n$ asymptotic vertex operator insertions is obtained from (\ref{eq:fullpathintegral}) by summing over topologies,\footnote{We might sometimes denote the genus $\g=0$ (sphere) contribution, $S_{0,\n}$, by $S_{S^2}(1;\dots,\n)$, and the genus $\g=1$ (torus) contribution, $S_{1,\n}$, by $S_{T^2}(1;\dots,\n)$. In the absence of vertex operators (where $\n=0$) the latter will be denoted by $Z_{T^2}$ (or $Z_{\g}$ at arbitrary genus, $\g$).}
$$
S_\n=\sum_{\g=0}^\infty S_{\g,\n}
$$
We adopt covariant normalisation for external onshell vertex operators in a mass eigenstate basis, whereby the tree-level ($\g=0$) contribution to two-point amplitudes in the special case of $D$ flat non-compact dimensions reads \cite{ErbinMaldacenaSkliros19}:
\begin{equation}\label{eq:2ptonshell}
S_{S^2}(1;2)=2k_1^0(2\pi)^{D-1}\delta^{D-1}(\mathbf{k}_1-\mathbf{k}_2)\delta_{j_1,j_2},
\end{equation}
and $j_1,j_2$ denote any remaining quantum numbers other than the momenta, $k_1,k_2$. (See also \cite{SekiTakahashi19} for an interesting suggestion about how to arrive at (\ref{eq:2ptonshell}) in the BRST formulation, at least for open strings\footnote{The closed string calculation in \cite{SekiTakahashi19} is perhaps not entirely satisfactory because external states are not annihilated by $b_0-\tilde{b}_0$; see the associated comment regarding the necessity of this condition in Sec.~\ref{sec:EVO} (combined with the fact that $S^2$ has positive curvature).  The derivation outlined in Sec.~\ref{sec:TEC}, which is essentially a hybrid of the Polyakov and BRST approach, works in all cases for sufficiently generic elements in the BRST cohomology.}) The defining properties of external fixed-picture physical vertex operators, $\hat{\mathscr{V}}_j$, and some comments on their offshell extension are discussed in the following subsection.
\sk\sk

\underline{\it A note on notation}: it is often useful to think in terms of real moduli rather than complex moduli, and/or often rather than $\hat{B}_{\tau^k}$ or $\hat{B}_{\bar{\tau}^k}$ we will write $\hat{B}_{k}$, or $\hat{B}_{t^k}$, or $\hat{B}_q$, or $\hat{B}_{z_v}$, etc., depending on context.  It is always to be understood that the normalisation of the full path integral is chosen to be consistent with the expressions in terms of complex moduli (\ref{eq:fullpathintegral}) and (\ref{eq:d2mtau}), and the normalisation of the full amplitude is in turn always fixed by unitarity and/or factorisation \cite{Weinberg85,Polchinski_v1}. In fact, the normalisation of the full S matrix is already fixed by (\ref{eq:2ptonshell}), but it took a while for this to be understood (a related discussion is in \cite{ErbinMaldacenaSkliros19}).
\sk

\subsection{External Vertex Operators}\label{sec:EVO}
Let us briefly discuss the conditions that must be satisfied by vertex operators used in the path integral (\ref{eq:fullpathintegral}). We begin with the onshell case and then discuss the offshell extension.

\subsubsection*{BRST-Invariant Vertex Operators}
The simplest context is when external vertex operators do not receive quantum corrections. Quantum corrections can be induced by mass renormalisation \cite{Seiberg87,Sen88,PiusRudraSen14c,PiusRudraSen14b} when we consider vertex operators whose mass is not protected by a symmetry.  Quantum corrections can also be induced by background shifts via the Fischler-Susskind mechanism \cite{FischlerSusskind86a,FischlerSusskind86b}, see in particular \cite{Polchinski88,LaNelson90,Tseytlin90b}, Sec.7 in \cite{Witten12c}, and the more recent dedicated study \cite{PiusRudraSen14}, whereby in the presence of non-vanishing massless tadpoles one needs to destroy tree-level conformal invariance (including the tree-level Virasoro physical state conditions below) in order to restore it at loop-level. This is because \cite{CallanGan86} although tadpole cancellation deforms the background equations of motion, the first variation of the deformed background equations of motion determine the corrected physical state conditions for external vertex operators. Finally, there is typically also wavefunction renormalisation \cite{Weinberg85b}. So when all such corrections are absent, one can \cite{Polchinski_v1,Witten12c} identify asymptotic (fixed-picture) vertex operators, $\hat{\mathscr{V}}_j$, with conformal primaries of ghost number two and conformal weight $(h,\tilde{h})=(0,0)$. More precisely, in the absence of such quantum corrections one can always choose:
\begin{equation}\label{eq:Vhat=ccV}
\hat{\mathscr{V}}_j=\tilde{c}c\mathscr{V}_j,
\end{equation}
where $\mathscr{V}_j$ is a {\it matter}\footnote{More generally, and also below, vertex operators without a ``hat'', $\hat{\phantom{a}}$, are in integrated picture (but are not necessarily constructed solely out of matter contributions).} primary of weight $(1,1)$. This condition is equivalent to the Virasoro physical state conditions, which is to choose $\mathscr{V}_j$ in (\ref{eq:Vhat=ccV}) to satisfy:
\begin{equation}\label{eq:L0LnV=0}
L_0\mathscr{V}_j=\mathscr{V}_j,\qquad {\rm and}\qquad L_n\mathscr{V}_j=0,\qquad {(n\geq1)},
\end{equation}
with similar relations for the anti-chiral halves. 
\sk

When it is possible to use such primaries as asymptotic states things become extremely simple, because these vertex operators, $\hat{\mathscr{V}}_j$, transform as {\it scalars} on patch overlaps and are hence manifestly globally-defined on any Riemann surface. That is, given two coordinate charts, $(U_m,z_m)$ and $(U_n,z_n)$, then {\it if} $p\in U_m\cap U_n\neq\zero$ there will be a corresponding holomorphic transition function $z_m(p)=f_{mn}(z_n(p))$, and the corresponding normal-ordered vertex operators at $p$ in the $z_m$ and $z_n$ charts respectively will be equal:
$$
\hat{\mathscr{V}}^{(z_m)}(p)=\hat{\mathscr{V}}^{(z_n)}(p),\qquad {\rm with}\qquad z_m=f_{mn}(z_n).
$$

It will be useful for the generalisation that follows to note also that total (i.e.~matter plus ghost) conformal weight $(h,\tilde{h})$ vertex operators of the specific form (\ref{eq:Vhat=ccV}) also satisfy:
\begin{equation}\label{eq:BRSTVhatV=0,dW}
\begin{aligned}
Q_B\cdot \hat{\mathscr{V}}_j &= \big(h\partial c -\tilde{h}\bar{\partial} \tilde{c} \big)\hat{\mathscr{V}}_j\\
Q_B\cdot \mathscr{V}_j& = \partial (c\mathscr{V}_j)+\bar{\partial} (\tilde{c}\mathscr{V}_j)+h\partial c\,\mathscr{V}_j+\tilde{h}\bar{\partial} \tilde{c}\,\mathscr{V}_j,
\end{aligned}
\end{equation}
as can be checked explicitly using the explicit expressions (\ref{eq:Vhat=ccV}) and (\ref{eq:L0LnV=0}), as well as the expression for the BRST charge (\ref{eq:QB}). 
So fixed picture vertex operators of the form (\ref{eq:Vhat=ccV}) are BRST-invariant when $(h,\tilde{h})=(0,0)$, whereas the matter (or more generally {\it integrated-picture}) vertex operators, $\mathscr{V}_j$, are BRST-invariant when $(h,\tilde{h})=(0,0)$ {\it up to a total derivative}. 
\sk

For the reasons given above we need to be able to go offshell \cite{Sen15b}. 
With this objective in mind it is necessary \cite{Sen15b} to abandon the identification of external vertex operators, $\hat{\mathscr{V}}_j$, with conformal primaries (\ref{eq:Vhat=ccV}) and (\ref{eq:L0LnV=0}), and sufficient to rather elevate the Virasoro physical state conditions to the more general notion of {\it BRST invariance} \cite{Polchinski88}. More precisely, an appropriate generalisation is the following. 
\sk

We primarily elevate the above requirement of conformal invariance to BRST invariance. In particular, 
the fixed-picture asymptotic state vertex operators, $\hat{\mathscr{V}}_j$, used in the path integral (\ref{eq:fullpathintegral}), will be required (before quantum corrections) to be annihilated by the BRST charge, $Q_B$, (modulo BRST-exact terms), and it is also necessary that they be annihilated by $b_0-\tilde{b}_0$:
\begin{equation}\label{eq:QBVhat=0=b0b0Vhat}
\boxed{
\hat{\mathscr{V}}_j\in \ker Q_B/{\rm Image} \,Q_B,\qquad{\rm and}\qquad (b_0-\tilde{b}_0)\hat{\mathscr{V}}_j =0
}
\end{equation}
But they need not satisfy the Virasoro conditions above. We also choose {\it not} to restrict to definite ghost number (the usual choice \cite{Sen15b} being to restrict to ghost-number two for closed strings), since as we will see having indefinite ghost number is not only natural but is also efficient if one wishes to capture arbitrarily excited string states with one fairly simple local operator. Finally, in addition to (\ref{eq:QBVhat=0=b0b0Vhat}) we can also require that external vertex operators satisfy the {\it Siegel gauge} condition:
\begin{equation}\label{eq:b0+b0tilde=0}
\boxed{
(b_0+\tilde{b}_0)\hat{\mathscr{V}}_j =0
}
\end{equation}

That such a projection (\ref{eq:b0+b0tilde=0}) exists\footnote{It has recently been demonstrated \cite{Sen20} that open string zero modes on a D-instanton (that are not associated with the collective coordinates of the D-instanton) can lead to a breakdown of  Siegel gauge. We are focusing here on the closed string, and we are not aware of such a breakdown of Siegel gauge in the closed string.} follows immediately from the explicit expression for handle operators that we derive, e.g., in Sec.~\ref{sec:GWDCSII}, but the underlying reasoning is actually independent of the matter CFT. In particular, and jumping ahead slightly, the precise reasoning leading to the additional condition (\ref{eq:b0+b0tilde=0}) amounts to the fact that it is possible to pick a slice in moduli space whereby we associate {\it both} pinch and twist moduli to every internal handle (whether it is homologically trivial or not) in amplitudes of arbitrary genus. In that case, the corresponding handle operators are effectively annihilated by both $b_0$ and $\tilde{b}_0$, see (\ref{eq:eq:Ahatprime}), due to the contribution, $\hat{B}_q\hat{B}_{\bar{q}}\propto b_0\tilde{b}_0$, from the path integral measure associated to this gauge slice (see also Sec.~\ref{sec:PH}). Therefore, since external physical states (due to unitarity) correspond to only a {\it subset} of the full set of offshell states associated to handle operators (which in addition contain unphysical and BRST-exact contributions when taken onshell) it is immediate that we can always enforce (\ref{eq:b0+b0tilde=0}) without loss of generality. 
\sk

The statements in the previous paragraph are for onshell physical vertex operators that can be used as asymptotic states in string path integrals (\ref{eq:fullpathintegral}). Before discussing the offshell generalisation there are two crucial points we should elaborate on:
\begin{itemize}
\item $\boxed{L_0-\tilde{L}_0=0}$: The physical state conditions (\ref{eq:QBVhat=0=b0b0Vhat}), due to the identity $\{Q_B,b_0-\tilde{b}_0\}=L_0-\tilde{L}_0$, automatically imply fixed-picture vertex operators, $\hat{\mathscr{V}}_j$, are annihilated by $L_0-\tilde{L}_0$. This ensures that they are insensitive to phase rotations of the local frame used to define them: $\hat{\mathscr{V}}_j^{(z_j)}(p_j)=\hat{\mathscr{V}}_j^{(z_j/e^{i\theta})}(p_j)$. The reason as to why this is a fundamental requirement is because this phase, $e^{i\theta(p_j)}$, is not globally well-defined as we discussed in Sec.~\ref{sec:WB}; the corresponding manifestation in terms of holomorphic normal coordinates was derived in Sec.~\ref{sec:PolCoords}.\footnote{Perhaps we should emphasise that this phase depends only on the base point, $p_j$ (which in an auxiliary coordinate system is denoted by $\sigma_j$), with respect to which the frame coordinate, $z_{\sigma_j}(\sigma)$, is defined and it does not depend on $\sigma$. So for a given base point (i.e.~a given frame) it is a global U(1) but becomes a local U(1) under base-point shifts. The conclusion of the above discussion is that it is not possible to define $\theta(\sigma_j)$ at every point $\sigma_j\in\Sigma$, and in particular that it must be ill-defined at a discrete set of points in $\Sigma$ when the Euler characteristic, $\chi(\Sigma)\neq0$.} Since this is an important point let us briefly summarise the relevant conclusion of the discussion in Sec.~\ref{sec:WB}. In order to be able to define such a phase one should be able to determine relative angles of tangent vectors on a general Riemann surface. If a tangent vector vanishes at certain points on $\Sigma$ angles cannot be defined at these points using this tangent vector. So in particular one needs to be able to define {\it non-vanishing}  tangent vectors {\it globally} on $\Sigma$ in order to define such a phase globally. But this is impossible by Hopf's index theorem (\ref{eq:Ipi=chi}) unless the Euler characteristic vanishes, $\chi(\Sigma)=0$. So such a phase is not globally well-defined, and the obstruction is topological, namely the Euler number. So requiring external vertex operators be insensitive to this phase, $\hat{\mathscr{V}}_j^{(z_j)}=\hat{\mathscr{V}}_j^{(z_j/e^{i\theta})}$, i.e., demanding they be annihilated by $L_0-\tilde{L}_0$, removes this U(1) ambiguity. 
\item $\boxed{b_0-\tilde{b}_0=0}$: That the $b_0-\tilde{b}_0$ condition in (\ref{eq:QBVhat=0=b0b0Vhat}) is necessary was elegantly (but formally) explained on general grounds by Witten in Sec.~2.4.4 in \cite{Witten12c}, which essentially summarises a result of Nelson's short paper \cite{Nelson89}. The $b_0-\tilde{b}_0$ condition is also adopted in string field theory, but the motivation there is that \cite{Zwiebach93} it is not known how to write down a kinetic term and interaction terms for string fields unless they are annihilated by $b_0-\tilde{b}_0$ and $L_0-\tilde{L}_0$. In fact, one can also derive it by explicit calculation and independently of string field theory. We do so in Sec.~\ref{sec:IPVO}, where the essential conclusion will be that if a fixed-picture vertex operator,\footnote{The quantities $\hat{B}_{z_v}\hat{B}_{\bar{z}_v}$ are the path integral measure contributions that translate fixed- to integrated-picture vertex operators \cite{Polchinski88}, and a careful and explicit derivation using holomorphic normal coordinates is one of the main objectives of this document, see Sec.~\ref{sec:TP} and Sec.~\ref{sec:SHUTF} for a derivation using a metric and transition functions respectively.} $\hat{\mathscr{V}}_j$, satisfies $Q_B\hat{\mathscr{V}}_j=0$, then the corresponding {\it integrated-picture} vertex operator, $\mathbb{V}_j=\rmd^2z_v\hat{B}_{z_v}\hat{B}_{\bar{z}_v}\hat{\mathscr{V}}_j$, satisfies,
\begin{equation}\label{eq:QBVint=0}
\boxed{
Q_B\hat{\mathscr{V}}_j=0\qquad\Rightarrow\qquad Q_B\mathbb{V}_j =i\mathcal{R}(b_0-\tilde{b}_0)\hat{\mathscr{V}}_j+\rmd \hat{\mathbb{W}}_j
}
\end{equation}
where $\mathcal{R}$ is the curvature tensor of $\Sigma$ (see Appendix \ref{sec:C}) and $\rmd \hat{\mathbb{W}}_j$ is a total derivative of a local normal-ordered operator.\footnote{In Sec.~\ref{sec:IPVO} the derivation of (\ref{eq:QBVint=0}) was carried out for arbitrary local operators $\hat{\mathscr{A}}_a^{(z)}$ (such as those associated to handle operators) but in the current section we are only interested in the subset $\hat{\mathscr{V}}_j\subset \{\hat{\mathscr{A}}_a^{(z)}\}$ which can be identified with external vertex operators. Also, the derivative is {\it outside} the normal ordering, so we could have more precisely written $\rmd :\!\hat{\mathbb{W}}_j\!:$ on the right-hand side in (\ref{eq:QBVint=0}). Derivatives do not commute with conformal normal ordering in general, and it is because the derivative is outside the normal ordering that we can freely integrate by parts using Stoke's theorem.} This relation (\ref{eq:QBVint=0}) is derived using holomorphic normal coordinates which corresponds to choosing a (global, modulo the immaterial U(1) phase $e^{i\theta(p_j)}$ commented on above) non-holomorphic (with respect to shifts of the base point, $\sigma_j$, of the frame) section of the bundle \cite{Nelson89,Witten12c} $\hat{\mathscr{P}}_{\g,\n}\rightarrow \mathcal{M}_{\g,\n}$ and hence leads to a globally well-defined construction. Indeed, as we show by explicit computation in Sec.~\ref{sec:WuYang}, there are no `Wu-Yang' boundary terms when using holomorphic normal coordinates to choose a section of $\hat{\mathscr{P}}_{\g,\n}\rightarrow \mathcal{M}_{\g,\n}$, (as briefly commented on already in \cite{Polchinski88,Nelson89}) and hence one can integrate by parts the quantity $\rmd \hat{\mathbb{W}}_j$ and pick up contributions only from the boundary of moduli space (as opposed to also picking up ``fictitious'' boundary contributions from patch overlaps). Therefore, from (\ref{eq:QBVint=0}) we see that $Q_B\mathbb{V}_j$ will decouple up to a total derivative {\it provided} $(b_0-\tilde{b}_0)\hat{\mathscr{V}}_j=0$. The fact that the curvature, $\mathcal{R}$, appears is reminiscent of the fact that the obstruction is {\it global} (and this is also consistent with Hopf's index theorem, see Sec.~\ref{sec:WB} and (\ref{eq:Ipi=chi}) there). To conclude, the $b_0-\tilde{b}_0=0$ constraint in (\ref{eq:QBVhat=0=b0b0Vhat}) is derived from the requirement that when fixed-picture vertex operators are BRST-invariant then {\it integrated vertex operators should be BRST-closed up to total derivatives}:
\begin{equation}\label{eq:QVhat=0,QV=dW}
\boxed{
\left.
\begin{aligned}
Q_B\hat{\mathscr{V}}_j&=0\\
Q_B\mathbb{V}_j&=\rmd \hat{\mathbb{W}}_j
\end{aligned}\right\}
\qquad\Rightarrow \qquad(b_0-\tilde{b}_0)\hat{\mathscr{V}}_j=0
}
\end{equation}
The analogy between (\ref{eq:QVhat=0,QV=dW}) and (\ref{eq:BRSTVhatV=0,dW}) is immediate when $h=\tilde{h}=0$ is enforced in the latter. 
\end{itemize}

Let us also point out that the $b_0-\tilde{b}_0=0$ and $L_0-\tilde{L}_0=0$ constraints on fixed-picture vertex operators are in general  independent {\it unless} these vertex operators are BRST-invariant. This is due to the relation,
$$
L_0-\tilde{L}_0=\{Q_B,b_0-\tilde{b}_0\},
$$
which holds as an operator statement inside the path integral. Notice in particular that $(b_0-\tilde{b}_0)\hat{\mathscr{V}}_j=0$ implies that also $(L_0-\tilde{L}_0)\hat{\mathscr{V}}_j=0$ if and only if $Q_B\hat{\mathscr{V}}_j=0$. 

\subsubsection*{Offshell Extension}
Having recast the physical state conditions as above the extension to an {\it offshell description} is essentially immediate. Namely, we are to drop the BRST-invariance condition from (\ref{eq:QBVhat=0=b0b0Vhat}) but retain the following conditions \cite{PiusRudraSen14c,PiusRudraSen14b,PiusRudraSen14,Sen15b}:
\begin{equation}\label{eq:L0L0tildeV=0=b0b0Vhat}
\boxed{
(L_0-\tilde{L}_0)\hat{\mathscr{V}}_j =0,\qquad{\rm and}\qquad (b_0\pm\tilde{b}_0)\hat{\mathscr{V}}_j =0
}
\end{equation}
and then one writes down a complete set of states satisfying these constraints. These states can be used to compute offshell amplitudes. 
For the reasons explained above, we can still require the Siegel gauge conditions (\ref{eq:b0+b0tilde=0}). As discussed in \cite{PiusRudraSen14c,PiusRudraSen14b,PiusRudraSen14,Sen15b} it is essential that a {\it gluing compatibility} condition is satisfied. This requirement demands that close to a boundary of moduli space where a punctured Riemann surface, $\Sigma$, used for computing an amplitude can be represented by two separate punctured Riemann surfaces, $\Sigma_1$ and $\Sigma_2$, glued together using plumbing fixture, the choice of local coordinates at the external punctures of $\Sigma$ must agree with those induced from the choice of local coordinates at the punctures of $\Sigma_1$ and $\Sigma_2$. Although the resulting offshell amplitudes then depend on the choice of local coordinates, the corresponding renormalised masses and S-matrix elements are independent of this choice \cite{PiusRudraSen14c,PiusRudraSen14b}. Also, the different offshell amplitudes that can be obtained by using different sets of coordinates are all related by field redefinitions in the corresponding string field theory.\footnote{DS thanks Ashoke Sen for explaining this to him.}
\sk

In the following subsections we discuss the measure contribution, $\prod_{k=1}^\m\hat{B}_{\tau^k}\hat{B}_{\bar{\tau}^k}$, in detail and from various viewpoints, focusing in particular on local and global aspects and various gauge slices in moduli space (sections of $\hat{\mathscr{P}}_{\g,\n}\rightarrow \mathcal{M}_{\g,\n}$), namely explicit realisations with explicit choices of coordinates (associated to choosing holomorphic normal coordinates and generic worldsheet curvature, $\mathcal{R}$) induced by cutting open the path integral across different cycles. 

\subsection{In Terms of Transition Functions I: `No Boundaries'}\label{sec:IToTF:I}
We will now focus on the path integral measure contribution associated to integrating over worldsheet metrics (in the BRST formulation) in detail. Namely, it remains to specify  the insertions:
\begin{equation}\label{eq:prodBBk}
\prod_{k=1}^\m\hat{B}_{\tau^k}\hat{B}_{\bar{\tau}^k},
\end{equation}
in (\ref{eq:fullpathintegral}), in relation to cutting and gluing path integrals across various (trivial or non-trivial) homology cycles, and in particular to choose a convenient gauge slice. 
We will not define these insertions quite yet. Suffice it to say for now that the quantities (\ref{eq:prodBBk}) are closely related to the natural pairing \cite{GiddingsMartinec86,Martinec86b} between a Beltrami differential, $\mu=\rmd\bar{z}\mu_{\bar{z}}^{\phantom{a}z}\partial_z$, (and its complex conjugate $\bar{\mu}=\rmd z\mu_{z}^{\phantom{a}\bar{z}}\partial_{\bar{z}}$), and the Grassmann-odd anti-ghost, $b=b_{zz}\rmd z^2$, (and its anti-chiral half, $\tilde{b}=\tilde{b}_{\bar{z}\bar{z}}\rmd\bar{z}^2$),
\begin{equation}\label{eq:Bintmub0}
B=\frac{1}{2\pi}\int_{\Sigma}\rmd^2z\,\big(\mu_{\bar{z}}^{\phantom{a}z}b_{zz}+\mu_z^{\phantom{a}\bar{z}}\tilde{b}_{\bar{z}\bar{z}}\big).
\end{equation}
The quantities $\mu,b$ are tensors and so this integral is globally-defined. Notice that the quantities in (\ref{eq:prodBBk}) (in addition to a subscript) have a `$\hat{\phantom{B}}$' to distinguish them from (\ref{eq:Bintmub0}). Since we wish to cut open the path integral across various cycles we also need to consider (\ref{eq:Bintmub0}) in the presence of boundaries.
\sk

In particular, it will be useful to expose certain {\it boundary terms} in (\ref{eq:Bintmub0}). The role of these boundary terms is twofold:
\begin{itemize}
\item[(1)] They will enable us to take into account that the ``shape of a puncture'' where a local operator is inserted (associated perhaps to a handle that is cut open via operator-state correspondence) {\it changes} as it is translated across the relevant Riemann surface due to Ricci curvature. Furthermore, we will carry out the computation for general Ricci curvature, so that the resulting insertions are covariant \cite{CallanGan86,Polchinski88}.
\item[(2)] They will enable us to pick a globally well-defined (modulo U(1)) gauge slice in moduli space where the various moduli are ``{\it as decoupled as possible}''. E.g., measure contributions associated to {\it translations of handles} (or local operators after having cut open these handles) and those associated to {\it pinching and twisting} will be independent. 
\end{itemize}


We will rely heavily on the formalism developed in Sec.~\ref{sec:RS}.  As discussed there the chiral half, $\mu_{\bar{z}}^{\phantom{i}z}$, is best thought of as the component of a vector-valued $(0,1)$-form on $\Sigma$, and\footnote{As also discussed below in detail, Beltrami differentials deform complex structure only modulo shifts \cite{Witten12c},  $\mu_{\bar{z}}^{\phantom{i}z}\mapsto \mu_{\bar{z}}^{\phantom{i}z}+\partial_{\bar{z}}u^z$, for $u=u^z\partial_z$ a globally-defined vector field on $\Sigma$ or a locally-defined holomorphic vector. When $u^z\partial_z$ is locally or globally defined but holomorphic in $z$ the assertion is obvious. When $u^z\partial_z$ is not holomorphic (e.g.~it may instead be smooth) but globally-defined the assertion requires global information and will be discussed below. (Perhaps we should mention also that we are not assuming the equation of motion, $\partial_{\bar{z}}b_{zz}=0$, is satisfied in particular the left-hand side will be non-vanishing at a discrete set of points and whether it contributes will then depend on what operator is inserted at these points.)} is associated to complex structure deformations of $\Sigma$. Conventions regarding the chiral component, $b_{zz}$, are spelt out in Appendix \ref{sec:CFTC}. This operator, $b_{zz}$, has a Laurent expansion in any one chart. Whether the poles associated to this Laurent expansion contribute or not depends on whether there are other operator insertions on the chart where it is inserted (we will make this entirely explicit momentarily). 
\sk

We will see that the aforementioned boundary terms (which are somewhat to those required to make the Euler characteristic a topological invariant in Sec.~\ref{sec:EulerCh}) are reflected here (from one viewpoint) in the poles of the Laurent expansion just mentioned. This will make sharp the statement that given a Riemann surface with boundaries (and corresponding states on these boundaries) we can pick a gauge slice that enables us to shrink these boundaries to points, and recast the same surface as a Riemann surface with punctures (and corresponding local vertex operators inserted at these). This is usually referred to as the {\it operator-state correspondence} \cite{Polchinski_v1}, which will in turn play a starring role when we cut open path integrals.\footnote{To make further preliminary contact between the current section and the calculation of Sec.~\ref{sec:EulerCh} let us further note from the outset that the operator-state correspondence is intimately related to the fact that the result (\ref{eq:R+Boundary=fmn}) is independent of the radii of the various boundary components, so we can shrink them to points resulting in a conformally-equivalent punctured surface. There will be an analogue of this for the path integral measure.}
\sk

Let us consider a closed oriented Riemann surface, $\Sigma$, with no ``extended'' boundary component (by which we mean that $\Sigma$ may have punctures, so, technically, $\Sigma$ will be allowed to have co-dimension-2 boundaries but no co-dimension-1 boundaries yet). 
As just alluded to, locally $b_{zz}, \tilde{b}_{\bar{z}\bar{z}}$ have Laurent expansions around any given base point as in (\ref{eq:phi_n modes Laurent}) with corresponding modes as in (\ref{eq:phi_n modes}). In particular, given a cover $\mathscr{U}=\bigcup_m U_m$ of $\Sigma$, see Sec.~\ref{sec:TFCR}, with charts $\{(U_m,z_m)\}$ and centred coordinates in each chart, $z_m(p_m)=0$ (we may somewhat imprecisely think of this as defining the point $p_m\in U_m\subset\Sigma$), in any one such chart we have:
\begin{equation}\label{eq:bzmexps}
\begin{aligned}
b_{z_mz_m}(z)&=\sum_{n\in\mathbf{Z}}\frac{b_n^{(z_m)}}{z^{n+2}}\\
b_n^{(z_m)}&=\frac{1}{2\pi i}\oint \frac{\rmd z}{z}z^{n+2}b_{z_mz_m}(z).
\end{aligned}
\end{equation}
As discussed in Appendix \ref{sec:LO} and \ref{sec:PMO}, $b_n^{(z_m)}\equiv b_n^{(z_m)}(p_m)$ which indicates that the mode is based at the same point $p_m\in\Sigma$ at which the frame is based. The contour integral is around a small circle containing the image of  $p_m$ in the $z_m$ chart coordinates with the standard orientation. There are similar relations for the anti-chiral halves,
\begin{equation}\label{eq:bzmexpstilde}
\begin{aligned}
\tilde{b}_{\bar{z}_m\bar{z}_m}(\bar{z})&=\sum_{n\in\mathbf{Z}}\frac{\tilde{b}_n^{(\bar{z}_m)}}{\bar{z}^{n+2}}\\
\tilde{b}_n^{(\bar{z}_m)}&=-\frac{1}{2\pi i}\oint \frac{\rmd\bar{z}}{\bar{z}}\bar{z}^{n+2}\tilde{b}_{\bar{z}_m\bar{z}_m}(\bar{z}).
\end{aligned}
\end{equation}
Note also that $z$ is the coordinate of a given point $p\in U_m\subset \Sigma$ (in the $z_m$ frame coordinates), in particular $z=z_m(p)$, and it is to be understood that, e.g., $b_{z_mz_m}(z)$ is the field evaluated at $z$ in the $z_m$ chart coordinates.\footnote{A clarification is already in order. The location of the poles in these Laurent expansions act as ``placeholders'' for any potential punctures (where external vertex operators or even local operators associated to string loops might inserted) that {\it might} be present (in the radial quantisation sense). This will become entirely transparent in what follows.}
\sk

Furthermore, by the Dolbeault-Grothendieck lemma \cite{Chern}, see  Appendix~\ref{sec:DGlemma}, given a {\it good} cover $\mathscr{U}$ of $\Sigma$, in any one chart $(U_m,z_m)$ there exists a locally-defined vector $v_m=v_m^{z_m}\partial_{z_m}$, $\bar{v}_m=\bar{v}_m^{\bar{z}_m}\partial_{\bar{z}_m}$ (with $\bar{v}_m\equiv v_m^*$ the complex conjugate) such that when $\mu$ is restricted to a simply-connected open set $U_m$:
\begin{equation}\label{eq:mu=dvzm}
\boxed{
\mu_{\bar{z}_m}^{\phantom{a}z_m}\big|_{U_m}=\partial_{\bar{z}_m}v_m^{z_m},\qquad {\rm and}\qquad \mu_{z_m}^{\phantom{a}\bar{z}_m}\big|_{U_m}=\partial_{z_m}\bar{v}_m^{\bar{z}_m}
}
\end{equation}
We have added a subscript $m$ to $v^{z_m}_m,\bar{v}_m^{\bar{z}_m}$ to indicate that (\ref{eq:mu=dvzm}) only holds locally, in particular within $U_m$. It is crucial that (\ref{eq:mu=dvzm}) only holds locally, since otherwise $B$  in (\ref{eq:Bintmub0}) would be mostly (but not necessarily entirely) trivial; we make this statement sharp below.
\sk

Taking into account the information in the preceding two paragraphs we wish to evaluate (\ref{eq:Bintmub0}).  We will proceed as in our warmup calculation in  Sec.~\ref{sec:EulerCh}, starting from the chiral half. ``Boundary'' contributions will take on a different guise here, in terms of punctures. Again, we insert a partition of unity, $\sum_m\lambda_m=1$, subordinate to the open cover $\mathscr{U}$ and again perform a cell decomposition, $\mathscr{V}=\{V_m\}$, allowing us to replace integrals over $\Sigma$ by a sum of integrals over (non-overlapping) cells $\{V_m\}$, recall Fig.~\ref{fig:tripleoverlapsUV} (on p.~\pageref{fig:tripleoverlapsUV}) and the discussion thereabouts. Namely, the steps analogous to (\ref{eq:chi}) and (\ref{eq:chix}) read:\footnote{We could have equivalently written $b_{z_mz_m}$ as $b^{(z_m)}(p)$, where $z_m(p)$ corresponds to the argument of integration in the $z_m$ frame.}
\begin{equation}\label{eq:Bintmub1}
\begin{aligned}
\frac{1}{2\pi}\int_{\Sigma}\rmd^2z\,\mu_{\bar{z}}^{\phantom{a}z}b_{zz} 
&=\frac{1}{2\pi}\int_{\Sigma}\rmd^2z\,\mu_{\bar{z}}^{\phantom{a}z}b_{zz}\big(\sum_m\lambda_m\big) \\
&=\frac{1}{2\pi}\sum_m\int_{V_m} \rmd^2z_m\,\mu_{\bar{z}_m}^{\phantom{a}z_m}b_{z_mz_m},
\end{aligned}
\end{equation}
where we made use of the general result (\ref{eq:nopartitionofunity}) derived there, which in turn implies that (\ref{eq:Bintmub1}) is {\it independent} of a specific choice of partition of unity, though this is not manifest. The sum over $m$ is over all cells $V_m\subset U_m$ of the cover $\mathscr{U}$.
It is useful to have the diagrams of Fig.~\ref{fig:tripleoverlapsUV}  in mind. 
\sk

Next, taking into account that $V_m\subset U_m$ is simply-connected (diffeomorphic to a disc), we make use of Dolbeault-Grothendieck lemma which led to (\ref{eq:mu=dvzm}) to rewrite (\ref{eq:Bintmub1}) as follows,
\begin{equation}\label{eq:Bintmub2}
\begin{aligned}
\frac{1}{2\pi}\int_{\Sigma}\rmd^2z\,\mu_{\bar{z}}^{\phantom{a}z}b_{zz} 
&=\frac{1}{2\pi}\sum_m\int_{V_m} \rmd^2z_m\,\big(\partial_{\bar{z}_m}v_m^{z_m}\big)b_{z_mz_m}\\
&=\frac{1}{2\pi}\sum_m\int_{V_m} \rmd^2z_m\,\partial_{\bar{z}_m}\big(v_m^{z_m}b_{z_mz_m}\big)-\frac{1}{2\pi}\sum_m\int_{V_m} \rmd^2z_m\,v_m^{z_m}\partial_{\bar{z}_m}b_{z_mz_m}\\
&=\frac{1}{2\pi i}\sum_m\oint_{\partial V_m} \rmd z_m\,v_m^{z_m}b_{z_mz_m}-\frac{1}{2\pi}\sum_m\int_{V_m} \rmd^2z_m\,v_m^{z_m}\partial_{\bar{z}_m}b_{z_mz_m},
\end{aligned}
\end{equation}
where in the second equality we integrated by parts, and in the last equality we used Green's theorem (\ref{eq: Greens theorem2a}) for the first term (the orientation of $\partial V_m$ being, as usual, counterclockwise with respect to $V_m$, which is in turn the standard orientation inherited from that of $\Sigma$). 
\sk

Let us consider now the second term on the right-hand side of the last equality in (\ref{eq:Bintmub2}). According to the Laurent expansion in (\ref{eq:bzmexps}), the components $b_{z_mz_m}$ are holomorphic in $U_m$ except at $z_m=0$, and in particular we pick up delta function contributions:\footnote{The notation $\partial_{\bar{z}_m}b_{z_mz_m}(z_m)$ is really shorthand for $\partial_{\bar{z}}b_{z_mz_m}(z)|_{z=z_m}$, where in this context $(\dots)|_{z=z_m}$ means `evaluate $(\dots)$ at $z=z_m$'.}
\begin{equation}\label{eq:dbarbzmzm}
\partial_{\bar{z}_m}b_{z_mz_m}(z_m)=\sum_{n\geq-1}\frac{(-)^{n+1}}{(n+1)!}b_n^{(z_m)}2\pi\,\partial_{z_m}^{n+1}\delta^2(z_m),
\end{equation}
where we took into account that derivatives commute and also $\partial_{\bar{z}_m}z_m^{-1}=2\pi\delta^2(z_m)$, which in turn is a direct consequence of the generalised Cauchy integral formula and Green's theorem.\footnote{Given a $C^1$ complex function $w(z,\bar{z})$ and the notation of Fig.~\ref{fig:Greens}, the `{\it generalised Cauchy integral formula}' at a point $\zeta,\bar{\zeta}$ in $D$ reads \cite{BersRiemannSurfaces}:
\begin{equation}\label{eq:genCauchIntForm}
w(\zeta,\bar{\zeta})=\frac{1}{2\pi i}\oint_\Gamma \,\rmd z\frac{w(z,\bar{z})}{z-\zeta}-\frac{1}{2\pi}\int_D\rmd^2z\frac{\partial_{\bar{z}}w(z,\bar{z})}{z-\zeta}.
\end{equation}
A derivation is included in Appendix \ref{sec:GCIF}. 
Integrating by parts in the second integral of the right-hand side in (\ref{eq:genCauchIntForm}) and using Green's theorem (\ref{eq: Greens theorem2a}), the resulting boundary term cancels the first integral on the right-hand side in (\ref{eq:genCauchIntForm}). Then, on account of the defining property of the Dirac delta function, $w(\zeta,\bar{\zeta})=\int \rmd^2z\,w(z,\bar{z})\delta^2(z-\zeta)$, the above yields precisely $\partial_{\bar{z}}z^{-1}=2\pi\delta^2(z)$.
} 
Only the range $n\geq-1$ survives in the corresponding Laurent expansion, because the terms with $n<-1$ are holomorphic and are therefore annihilated by the anti-chiral derivative, and this determines the corresponding range in the sum over $n$ in (\ref{eq:dbarbzmzm}). 
\sk

We substitute (\ref{eq:dbarbzmzm}) into the last term on the right-hand side in (\ref{eq:Bintmub2}) and integrate by parts to remove the derivatives from the delta function. This integration by parts does not produce  additional boundary terms because the delta function only has support at $z_m=0$ (which is within $V_m$). All in all, we learn that (\ref{eq:Bintmub2}) is equal to:
\begin{equation}\label{eq:Bintmub3}
\frac{1}{2\pi i}\sum_m\oint_{\partial V_m} \rmd z_m\,v_m^{z_m}b_{z_mz_m}-\sum_m\sum_{n\geq-1}\frac{1}{(n+1)!}\partial_{z_m}^{n+1}v_m^{z_m}(p_m)b_n^{(z_m)}(p_m),
\end{equation}
and we could have equivalently written $v_m^{z_m}(0)$ or $v_m^{z_m}(z,\bar{z})|_{z=\bar{z}=0}$ for the same quantity $v_m^{z_m}(p_m)$ since $z_m(p_m)=0$. 
Including the contribution from the anti-chiral half, which is entirely analogous, we learn that the quantity $B$ defined in (\ref{eq:Bintmub0}) reads:
\begin{equation}\label{eq:Bintmub4}
\begin{aligned}
\frac{1}{2\pi}\int_{\Sigma}\rmd^2z\,\big(&\mu_{\bar{z}}^{\phantom{a}z}b_{zz}+\mu_z^{\phantom{a}\bar{z}}\tilde{b}_{\bar{z}\bar{z}}\big)=\sum_m\frac{1}{2\pi i}\oint_{\partial V_m} \big(\rmd z_m\,v_m^{z_m}b_{z_mz_m}-\rmd\bar{z}_m\,\bar{v}_m^{\bar{z}_m}\tilde{b}_{\bar{z}_m\bar{z}_m}\big)\\
&\qquad-\sum_m\sum_{n\geq-1}\frac{1}{(n+1)!}\big(\partial_{z_m}^{n+1}v_m^{z_m}(p_m)b_n^{(z_m)}(p_m)+\partial_{\bar{z}_m}^{n+1}\bar{v}_m^{\bar{z}_m}(p_m)\tilde{b}_n^{(\bar{z}_m)}(p_m)\big).
\end{aligned}
\end{equation}
Upon interpreting this result it is to be understood that (\ref{eq:mu=dvzm}) provides the bridge between the globally-defined Beltrami differentials, $\mu$, and the locally defined vectors $v_m$ for every chart labelled by $m$. 
\sk

It is important to emphasise that the derivation leading to (\ref{eq:Bintmub4}) assumed we are considering a {\it good cover}, because this is where the Dolbeault-Grothendieck lemma leading to (\ref{eq:mu=dvzm}) applies.
\sk

It will be useful to now rewrite (\ref{eq:Bintmub4}) to make manifest the fundamental data (namely holomorphic transition functions satisfying appropriate cocycle relations) that defines our Riemann surface of interest. 
\sk

Proceeding by direct analogy to the derivation leading from (\ref{eq:chi(S)temp}) to (\ref{eq:chi(S)temp3}) in Sec.~\ref{sec:NB}, let us consider the contour integral terms in (\ref{eq:Bintmub4}). For every $m$, the contours $\partial V_m$ are along the boundaries of the $V_m$ and are coherently oriented with the orientation of $\Sigma$ (i.e.~in a counterclockwise sense with respect to $V_m$), see Fig.~\ref{fig:tripleoverlapsUV} and Fig.~\ref{fig:segment}. Every such boundary $\partial V_m$ consists of one or more segments $\{C_{mn},C_{m\ell},\dots\}$ that meet at triple or higher intersections.\footnote{Eventually it will be useful to depart from the requirement of a `good cover', so that we allow for annular chart overlaps, where the $C_{mn}$ will also be allowed to close.} The segments $\{C_{mn},C_{m\ell},\dots\}$ are then curves that traverse patch intersections $\{U_{mn},U_{m\ell},\dots\}$ in the figure that begin and end at the vertices contained in $\{U_{mn\ell},\dots\}$. To every contribution along $C_{mn}$ arising from $\partial V_m$ there is a contribution with opposite orientation from the curve $C_{nm}(=-C_{mn})$ that arises from the boundary integral along $\partial V_n$, so the sum over $m$ in the contour integral terms in (\ref{eq:Bintmub4}) can therefore be rewritten as a sum over {\it pairs}, $(mn)$, of curves as follows (see Fig.~\ref{fig:segment}),
\begin{equation}\label{eq:Bintmub5}
\begin{aligned}
\frac{1}{2\pi}\int_{\Sigma}&\rmd^2z\,\big(\mu_{\bar{z}}^{\phantom{a}z}b_{zz}+\mu_z^{\phantom{a}\bar{z}}\tilde{b}_{\bar{z}\bar{z}}\big)=\\
&=\sum_{(mn)}\frac{1}{2\pi i}\int_{C_{mn}} \Big[\rmd z_m\,\big(v_m^{z_m}-v_n^{z_m}\big)b_{z_mz_m}-\rmd\bar{z}_m\,\big(\bar{v}_m^{\bar{z}_m}-\bar{v}_n^{\bar{z}_m}\big)\tilde{b}_{\bar{z}_m\bar{z}_m}\Big]\\
&\quad-\sum_m\sum_{n\geq-1}\frac{1}{(n+1)!}\Big[\partial_{z_m}^{n+1}v_m^{z_m}(p_m)b_n^{(z_m)}(p_m)+\partial_{\bar{z}_m}^{n+1}\bar{v}_m^{\bar{z}_m}(p_m)\tilde{b}_n^{(\bar{z}_m)}(p_m)\Big].
\end{aligned}
\end{equation}

Let us pick a gauge slice in moduli space such that the Beltrami differential can be expanded as in (\ref{eq:munu}), where $\{t^1,t^2,\dots\}$ are an appropriate set of coordinates in moduli space, $\mu=\mu_k\rmd t^k$, (although we will not yet specify the precise coordinate choices). Taking also (\ref{eq:mu=dvzm}) into account we can write the components, $\mu_k$, in the chart $(U_m,z_m)$ more precisely as $\mu_{k\bar{z}_m}^{\phantom{aa}z_m}|_{U_m}=\partial_{\bar{z}_m}v_{km}^{z_m}$. This enables us to make use of (\ref{eq:vm-vn=dfmn/dt}), according to which the terms associated to sums over pairs, $(mn)$, on the right-hand side of (\ref{eq:Bintmub5}) can be written entirely in terms of the fundamental data defining the Riemann surface, namely the holomorphic transition functions, $f_{mn}(z_n,t)$,
\begin{equation}\label{eq:Bintmub5bx}
\begingroup\makeatletter\def\f@size{11}\check@mathfonts
\def\maketag@@@#1{\hbox{\m@th\large\normalfont#1}}%
\begin{aligned}
&\frac{1}{2\pi}\int_{\Sigma}\rmd^2z\,\big(\mu_{k\bar{z}}^{\phantom{aa}z}b_{zz}+\mu_{kz}^{\phantom{aa}\bar{z}}\tilde{b}_{\bar{z}\bar{z}}\big)=\\
&=\sum_{(mn)}\frac{1}{2\pi i}\int_{C_{mn}} \bigg(\rmd z_m\,\frac{\partial f_{mn}(z_n,t)}{\partial t^k}\big|_{z_n}b_{z_mz_m}-\rmd\bar{z}_m\,\frac{\partial \bar{f}_{mn}(\bar{z}_n,\bar{t})}{\partial t^k}\big|_{\bar{z}_n}\tilde{b}_{\bar{z}_m\bar{z}_m}\bigg)\\
&\quad-\sum_m\sum_{n\geq-1}\frac{1}{(n+1)!}\Big[\partial_{z_m}^{n+1}v_{km}^{z_m}(p_m)b_n^{(z_m)}(p_m)+\partial_{\bar{z}_m}^{n+1}\bar{v}_{km}^{\bar{z}_m}(p_m)\tilde{b}_n^{(\bar{z}_m)}(p_m)\Big]\\
\end{aligned}
\endgroup
\end{equation}

The sum over $m$ on the right-hand side of (\ref{eq:Bintmub5bx}) is such that the $m^{\rm th}$ term in the summand is evaluated at the point $p_m\in U_m$ where the coordinate system of the $(U_m,z_m)$ chart is centred (recall that $z_m(p_m)=0$). {\it If} there is no other operator insertion at $p_m$ then these terms vanish, since by (\ref{eq:bn-annihilation-creation}) the $b_{-1},b_0$ and $b_{n}$ for $n\geq1$ all annihilate the ${\rm SL}(2,\mathbf{C})$ vacuum, and more precisely the contours that define these modes in the contour integral representation in (\ref{eq:bzmexps}) can then all be shrunk to zero without obstruction and the corresponding contour integrals vanish. 
\sk

The terms in the sum over $m$ on the right-hand side of (\ref{eq:Bintmub5bx}) cannot be written in terms of the fundamental data defining the Riemann surface, and furthermore these terms are not invariant under coordinate transformations. Since we wish to obtain a covariant expression for the measure which depends entirely on the defining properties of Riemann surfaces, one way to proceed is to specify a gauge slice (that should implicitly be attributed to a choice of Beltrami differential, $\mu$) which explicitly cancels these coordinate-dependent terms. In particular, this choice can be implemented by adding {\it boundary terms} such that the gauge-dependent terms are precisely cancelled, $B_k\rightarrow \hat{B}_k=B_k+B_{k}^{\rm (b.t.)}$, 
\begin{equation}\label{eq:Bintmub5bxx}
\begingroup\makeatletter\def\f@size{11}\check@mathfonts
\def\maketag@@@#1{\hbox{\m@th\large\normalfont#1}}%
\boxed{
\begin{aligned}
\hat{B}_k&=\frac{1}{2\pi}\int_{\Sigma}\rmd^2z\,\big(\mu_{k\bar{z}}^{\phantom{aa}z}b_{zz}+\mu_{kz}^{\phantom{aa}\bar{z}}\tilde{b}_{\bar{z}\bar{z}}\big)+\sum_m\sum_{n\geq-1}\frac{1}{(n+1)!}\bigg[\partial_{z_m}^{n+1}v_{km}^{z_m}b_n^{(z_m)}(p_m)+\partial_{\bar{z}_m}^{n+1}\bar{v}_{km}^{\bar{z}_m}\tilde{b}_n^{(\bar{z}_m)}(p_m)\bigg]\\
&=\sum_{(mn)}\frac{1}{2\pi i}\int_{C_{mn}} \bigg(\rmd z_m\,\frac{\partial f_{mn}(z_n,t)}{\partial t^k}\Big|_{z_n}b_{z_mz_m}-\rmd\bar{z}_m\,\frac{\partial \bar{f}_{mn}(\bar{z}_n,\bar{t})}{\partial t^k}\Big|_{\bar{z}_n}\tilde{b}_{\bar{z}_m\bar{z}_m}\bigg)\\
\end{aligned}
}
\endgroup
\end{equation}
This is identified with that appearing in the path integral (\ref{eq:fullpathintegral}) {\it when} the Riemann surfaces under consideration are closed and oriented. Either of the two expressions on the right-hand side in (\ref{eq:Bintmub5bxx}) depend solely on the fundamental data defining the Riemann surface, namely holomorphic transition functions, $f_{mn}(z_n,t)$, on patch overlaps $U_m\cap U_n$. 
\sk

But we are not done yet since we have not yet completely specified the gauge slice of interest in moduli space. In particular, we still need to specify how to cut open handles and replace them with handle operators, and we need to ensure that the resulting slice is globally well-defined in moduli space.
\sk

The significance of the boundary terms on the right-hand side of the first equality in (\ref{eq:Bintmub5bxx}) will emerge when this operator, $\hat{B}_k$, acts on (either offshell or onshell) vertex operators associated to asymptotic states or corresponding operators that arise within a handle/cycle of a Riemann surface. 
The result (\ref{eq:Bintmub5bxx}) is of central importance in string perturbation theory. 
\sk

The first equality in (\ref{eq:Bintmub5bxx}) is a slight generalisation of Polchinski's result, namely eqn.~(26) in \cite{Polchinski88}, associated to his (so-called \cite{Nelson89}) $\hat{b}$ prescription. To be precise, (\ref{eq:Bintmub5bxx}) does {\it not} rely on a specific choice of conformal gauge metric (contrary to eqn.~(26)  in \cite{Polchinski88} which is valid for the choice of metric given in eqn.~(16) there).\footnote{In fact, there is a relative sign difference in this expression compared to that in \cite{Polchinski88}. Nevertheless, we will see that there is an additional sign difference below that compensates this so that there is in fact perfect agreement with equations (28) and (30) in \cite{Polchinski88}. (The difference can be attributed to the distinction between active and passive coordinate frames and so has no physical significance.)} Secondly, we have not assumed that $v_m$ is real analytic, but only that it is complex-valued and $C^\infty$ at $p_m$. Nevertheless, the approach of Polchinski, using a metric on $\Sigma$ compatible with its complex structure to single out a conformal frame at a point $p_m\in\Sigma$, a procedure referred to \cite{Polchinski88} as {\it Weyl normal ordering}, is extremely useful. And furthermore, his {\it specific} choice \cite{Polchinski88} of choosing a metric to define a conformal coordinate system that is ``as flat as possible'' at a point $p_m$ will be indispensable since it provides an explicit globally well-defined (modulo an immaterial U(1) phase) and covariant slice in moduli space, so we will discuss it in further detail below. This is to be contrasted with Nelson's approach \cite{Nelson89} based on holomorphicity which requires Wu-Yang type boundary terms on patch overlaps that must in turn be constructed on a case-by-case basis -- Polchinski's approach effectively determines all these Wu-Yang contributions automatically and localises them.\footnote{To elaborate a bit further, Polchinski's specific choice of metric plays a role analogous to using Riemann normal coordinates for real manifolds, which has in turn proved extremely valuable in non-linear $\sigma$-model studies in non-trivial string backgrounds (a very short list of early references being \cite{Friedan80,Alvarez-GaumeFreedmanMukhi81,FradkinTseytlin85,Mukhi86,CallanGan86,Tseytlin87,HowePapadopoulosStelle88,Osborn90}). } 
This procedure will be equivalent to adopting holomorphic normal coordinates which fixes invariance under holomorphic reparametrisations (and therefore also Weyl transformations), but reparametrisation invariance will be manifest.
\sk

Although the second equality in (\ref{eq:Bintmub5bxx}) makes it manifest that the left-hand side is well-defined, it does nevertheless naively seem to rely on the explicit choice of cover, $\mathscr{U}$, so we will need to show that the result is actually independent of this choice.  This will be done in Sec.~\ref{sec:CIoM}. In Sec.~\ref{sec:TP} we use the first equality to translate pinches, handles and vertex operators across a Riemann surface; in Sec.~\ref{sec:SHUTF} we use the second equality for the same purpose as an independent consistency check of the result and also to provide a complementary viewpoint. We are developing both approaches. The transition function approach will be more efficient.

\subsection{Cover-Independence of Measure}\label{sec:CIoM}
We next show that the expression for the path integral measure  (\ref{eq:Bintmub5bxx}) is independent of the cover, $\mathscr{U}$, of the Riemann surface $\Sigma$, since this is certainly not explicit. To do so we will need to appeal to global data. For this computation it will be simplest to work with the {\it second} equality in (\ref{eq:Bintmub5bxx}), namely:
\begin{equation}\label{eq:Bintmub5bxxfmn}
\begin{aligned}
\hat{B}_k
&=\sum_{(mn)}\frac{1}{2\pi i}\int_{C_{mn}} \bigg(\rmd z_m\,\frac{\partial f_{mn}(z_n,t)}{\partial t^k}\Big|_{z_n}b_{z_mz_m}-\rmd\bar{z}_m\,\frac{\partial \bar{f}_{mn}(\bar{z}_n,\bar{t})}{\partial t^k}\Big|_{\bar{z}_n}\tilde{b}_{\bar{z}_m\bar{z}_m}\bigg)\\
\end{aligned}
\end{equation}

Let us then consider a specific non-empty triple patch intersection, $U_m\cap U_n\cap U_\ell$, and in particular a point we shall (in this subsection) call $p$, 
$$
p\in U_m\cap U_n\cap U_\ell\neq\zero,
$$
where three boundary components in $\{\partial V_m,\partial V_n,\partial V_\ell\}$ meet, see Fig.~\ref{fig:tripleoverlapsUV}. The three boundary segments (with orientation such that, e.g., the three contours are incoming towards $p$ although the result does not depend on this) are denoted by: $C_{mn},C_{\ell m},C_{n\ell}$, where we continue to use the notation developed above. Let us label the corresponding coordinate of $p$ in each of the three charts by:
$$
z_{\hat{m}n\ell}\dfn z_m(p),\qquad z_{m\hat{n}\ell}\dfn z_n(p),\qquad {\rm and}\qquad z_{mn\hat{\ell}}\dfn z_\ell(p).
$$

We must primarily verify that the measure (\ref{eq:Bintmub5bxxfmn}) does not depend on the point $p$ since (after applying the same reasoning to {\it every} triple intersection) this will primarily ensure that the path integral measure is independent of continuous deformations of the cover. It will be sufficient to focus on the chiral half since identical reasoning applies also to the anti-chiral half. Let us then make explicit the (chiral half of the) term in (\ref{eq:Bintmub5bxxfmn}) that potentially depends on $p$:
\begin{equation}\label{eq:Bintmub7}
\begin{aligned}
\hat{B}_k&=\frac{1}{2\pi i}\int_{C_{mn}} \rmd z_m\,\frac{\partial f_{mn}(z_n,t)}{\partial t^k}\big|_{z_n}b_{z_mz_m}+\frac{1}{2\pi i}\int_{C_{\ell m}} \rmd z_\ell\,\frac{\partial f_{\ell m}(z_m,t)}{\partial t^k}\big|_{z_m}b_{z_\ell z_\ell}\\
&\qquad +\frac{1}{2\pi i}\int_{C_{n\ell}} \rmd z_n\,\frac{\partial f_{n\ell}(z_\ell,t)}{\partial t^k}\big|_{z_\ell}b_{z_nz_n} +\textrm{(anti-chiral half)}+\textrm{(terms independent of $p$)},
\end{aligned}
\end{equation}
equivalently,
\begin{equation}\label{eq:Bintmub7b}
\begin{aligned}
\hat{B}_k&=\frac{1}{2\pi i}\int^{z_{\hat{m}n\ell}} \rmd z_m\,\frac{\partial f_{mn}(z_n,t)}{\partial t^k}\big|_{z_n}b_{z_mz_m}+\frac{1}{2\pi i}\int^{z_{mn\hat{\ell}}} \rmd z_\ell\,\frac{\partial f_{\ell m}(z_m,t)}{\partial t^k}\big|_{z_m}b_{z_\ell z_\ell}\\
&\qquad +\frac{1}{2\pi i}\int^{z_{m\hat{n}\ell}} \rmd z_n\,\frac{\partial f_{n\ell}(z_\ell,t)}{\partial t^k}\big|_{z_\ell}b_{z_nz_n} +\textrm{(anti-chiral half)}+\textrm{(terms independent of $p$)},
\end{aligned}
\end{equation}
where we need not specify the lower limits on the line integrals since these are $p$-independent. 
Furthermore, since we are only interested on the possible $p$ dependence we can partition the three line integrals into contributions associated to segments {\it inside} the triple overlap, $U_{mn\ell}$, plus contributions coming from {\it outside} the triple overlap. Terms associated to the outside of the triple overlap can (and will) be absorbed into the `(terms independent of $p$)' in (\ref{eq:Bintmub7b}), since this conveniently allows us to rewrite the $p$-dependent terms in (\ref{eq:Bintmub7b}) in terms of a single chart coordinate, say $z_m$. We then use the fact that $\frac{\partial f_{\ell m}(z_m,t)}{\partial t^k}\big|_{z_m}$ transforms as the component of the locally-defined vector, $\phi_{\ell m}(t^k)$ recall (\ref{eq:phimn(t)}), in particular,
\begin{equation*}
\begin{aligned}
\frac{\partial f_{\ell m}(z_m,t)}{\partial t^k}\big|_{z_m}&=v^{z_\ell}_{k\ell}-v^{z_\ell}_{km}\\
&=\frac{\partial z_\ell}{\partial z_m}\Big|_{t^k}\big(v^{z_m}_{k\ell}-v^{z_m}_{km}\big)\\
&=-\frac{\partial z_\ell}{\partial z_m}\Big|_{t^k}\frac{\partial f_{m\ell}(z_\ell,t)}{\partial t^k}\Big|_{z_\ell},\\
\end{aligned}
\end{equation*}
with a similar manipulation for the integrand associated to the $z_n$ integral in (\ref{eq:Bintmub7b}). Then, taking also into account that $b_{(m)}\dfn b_{z_mz_m}\rmd z_m\otimes \rmd z_m$ is invariant under holomorphic reparametrisations\footnote{Recall from Sec.~\ref{sec:RS} that transition functions on chart overlaps are by definition (since Riemann surfaces are complex manifolds) {\it holomorphic} in the chart coordinates.}, so that $b_{(m)}=b_{(n)}=b_{(\ell)}$ on the relevant triple overlap, we arrive at the following equivalent expression to (\ref{eq:Bintmub7b}):
\begin{equation}\label{eq:Bintmub7c}
\begin{aligned}
\hat{B}_k&=\frac{1}{2\pi i}\int^{z_{m}(p)} \Big(\phi_{mn}(t^k)-\phi_{m\ell}(t^k)+\phi_{n\ell}(t^k)\Big)b_{(m)} \\
&\qquad+\textrm{(anti-chiral half)}+\textrm{(terms independent of $p$)}.
\end{aligned}
\end{equation}
We see immediately, on account of the cocycle relations (\ref{eq:cocycles}), namely $\phi_{n\ell}(t)+\phi_{\ell m}(t)+\phi_{mn}(t)=0$ and $\phi_{m\ell}(t)=-\phi_{\ell m}(t)$, that the $p$ dependence cancels out entirely since the explicit integrand in (\ref{eq:Bintmub7c}) vanishes identically. Conversely, requiring that the $p$ dependence cancels out implies the cocycle relations (\ref{eq:cocycles}), thus providing an alternative derivation of the latter.
\sk

In the above derivation we singled out a specific triple intersection point $p\in U_{mn\ell}$ but clearly the above reasoning is independent of this choice, and therefore the right-hand side of the equality in (\ref{eq:Bintmub5bxxfmn}) is independent of (at least) {\it continuous} deformations of the (good) cover $\mathscr{U}$ of $\Sigma$ that we started from.
\sk

Modulo a caveat mentioned below, it remains to check the independence of the path integral measure on {\it discontinuous} changes of cover. In particular, the triple intersection containing $p$ might now be associated to a 4-point intersection. The corresponding derivation that led to (\ref{eq:Bintmub7c}) now yields:
\begin{equation}\label{eq:Bintmub7d}
\begin{aligned}
\hat{B}_k=\frac{1}{2\pi i}\int^{z_{m}(p)} \Big(\phi_{mn}(t^k)+\phi_{n\ell}(t^k)+\phi_{\ell k}(t^k)+\phi_{km}(t^k)\Big)b_{(m)} +\textrm{(terms independent of $p$)}.
\end{aligned}
\end{equation}
But this also vanishes identically due to the 4-point cocycle relation (\ref{eq:cocycles-q}) (which is in turn guaranteed to be satisfied when the basic 3-point cocycle relation (\ref{eq:cocycles}) is satisfied), and therefore the expression for the path integral measure (\ref{eq:Bintmub7d}) is still independent of $p$. Proceeding similarly for higher-point intersections and the remaining patch overlaps of the cover will clearly yield the same outcome. In particular, since any two covers $\mathscr{U}$, $\mathscr{U}'$ of a Riemann surface, $\Sigma$, can only differ in the number and locations of the open sets that comprise them we have shown that the expression (\ref{eq:Bintmub5bxxfmn}) for the path integral measure is independent of the cover. 
\sk

There is an important caveat however. The derivation that led to (\ref{eq:Bintmub7}) assumed that $\mathscr{U}$ is a {\it good cover} \cite{BottTu}, and the above procedure of checking cover independence did not deviate from this assumption. For example, we did not allow for annular chart overlaps, and so we have not demonstrated that the expression (\ref{eq:Bintmub7}) for the path integral measure is correct if we do not use a good cover. The reason we required a good cover in the first place is twofold:
\begin{itemize}
\item[{\bf (1)}] We made use of the Dolbeault-Grothendieck lemma (which led to the Beltrami differential relation (\ref{eq:mu=dvzm})). The proof of this lemma (Appendix \ref{sec:DGlemma}) requires that the explicit open set, say $U_m$, on which the lemma is used is diffeomorphic to a disc. 
\item[{\bf (2)}] As mentioned in Sec.~\ref{sec:MSER}, in order to ensure that we are constructing a measure in moduli space we need to \cite{GunningRossi,Chern} take a direct limit to transition from the cohomology group, $H^1(\mathscr{N}(\mathscr{U}),\mathcal{S}_t)$, of the nerve of the cover, $\mathscr{N}(\mathscr{U})$, to the cohomology group, $H^1(\Sigma,\mathcal{S}_t)$, of the manifold, $\Sigma$. Because it is the {\it latter} \cite{Witten12c} that corresponds to the fibre of the tangent bundle,\footnote{Note that we can construct the path integral measure on either the tangent space, $T\mathcal{M}$, or base space, $
\mathcal{M}$, and since the Jacobian for a change of variables (which in our case is the Fadeev-Popov determinant in terms of ghosts) is independent of this choice it suffices to construct it on $T\mathcal{M}$.} $T\mathcal{M}|_{\Sigma}$, corresponding to $\Sigma$, in particular $T\mathcal{M}|_{\Sigma}=H^1(\Sigma,\mathcal{S}_t)$. This direct limit which takes us from the cohomology of the cover to that of the manifold requires that the cover, $\mathscr{U}$, be `good'.
\end{itemize}
The above reasoning allows us to conclude that (\ref{eq:Bintmub7}) is invariant if we replace the good cover $\mathscr{U}$ by another {\it good cover} $\mathscr{U}'$. But we have not shown that we can relax the condition that the cover be good. 
\sk

Looking ahead, if we are going to be able to use plumbing fixture \cite{Sen15b} to pinch and twist trivial or non-trivial homology cycles it is easiest to proceed if we also allow for a subset of annular overlaps and charts and in particular open sets, $U_m'=U_m\setminus {\rm D}$ (where ${\rm D}$ is a disc that is removed from the open set $U_m$ to obtain $U_m'$), that are diffeomorphic to annuli in addition to allowing for open sets diffeomorphic to discs. One might try to justify using such annuli for gluing by demanding that the corresponding U(1) rotation symmetry of the annulus always be taken to be integrated. So one could imagine beginning with a good cover, and then replacing a subset of charts with annuli might be thought of as being justified if one integrates or averages over the location of the (at least three) charts of the underlying good cover which every annulus replaces (see Fig.~\ref{fig:sep-nonsep}--\ref{fig:goodannuliX}, and also Fig.~\ref{fig:pinchtranslate}). This is presumably sensible because of the variety of gauge slices in moduli space that we have at our disposal. Also, the resulting amplitudes will depend solely on the complex structure and not on details about the specific coordinate parametrisation or frame associated to these annuli. 
\sk

Incidentally, when we cut open a trivial or non-trivial homology cycle we will see that we can associate translation moduli to the resulting two states that are produced by the cutting procedure. Furthermore, our chosen slice in moduli space will be such that the measure contributions associated to these translation moduli are insensitive to the ``size'' of the cut circle on which these states are defined, so we can take a zero-radius limit whereby these circles are reduced to punctures without affecting resulting amplitudes. This makes the operator-state correspondence (whereby we map arbitrary offshell states produced in loops to local operators) possible. This is also closely associated to the comment that we are interested in a gauge slice that ``decouples moduli as much as possible''. And when this is the case, as we have already seen in the previous subsection, we {\it can} apply the Dolbeault-Grothendieck lemma, because the resulting boundary terms that we have added essentially subtract out the contribution to the measure from the puncture. We will see this in detail below. So the above comment on annular charts only really plays a role for pinch/twist moduli, and in particular therefore we should retain the complex structure information if we replace the three or more overlapping charts with annuli, because the information we need to encode in every annulus transition function is a single plumbing-fixture modulus \cite{Fay}. 
\sk

In fact, this procedure will pass all of the consistency checks that we carry out (most notably one-loop modular invariance and worldsheet duality, which probes the consistency of this procedure in the cases of cutting across non-trivial and trivial homology cycles respectively).
\sk

This U(1) ambiguity (see also the discussion in Sec.~\ref{sec:WB}) takes on different guises throughout the document. We will follow Polchinski's suggestion \cite{Polchinski88} which is to either take this phase to be integrated or to consider combinations of operators that are independent of this phase. (Incidentally, taking this phase to be integrated does {\it not} mean that non-level-matched states are projected out, except in the case of trivial homology cycles. For non-trivial homology cycles non-level-matched states contribute to amplitudes.)

\subsection{In Terms of Transition Functions II: `With Boundaries'}\label{sec:IToTF:II}
\begin{figure}
\begin{center}
\includegraphics[angle=0,origin=c,width=0.7\textwidth]{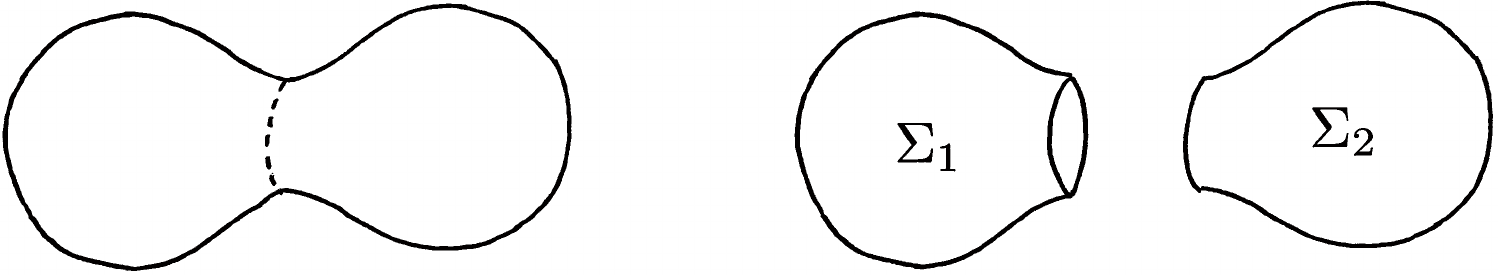}
\caption{When we cut open a compact oriented Riemann surface (first diagram) across a given cycle (indicated by a dashed line) we end up with two oriented Riemann surfaces with boundaries (second diagram). The boundaries can in turn be shrunk to points leading to punctures. In the corresponding path integral one inserts a resolution of unity to connect the boundaries and enforce appropriate boundary conditions (ensuring that contours associated to conserved charges such as $Q_B,L_n,\mathscr{O}_n,\dots$ can be deformed from one Riemann surface to the other without obstruction). Shrinking the boundaries to points corresponds to replacing the states in the resolution of unity by local (offshell) vertex operators (with a corresponding sum over their continuous/discrete quantum numbers).}\label{fig:SigmaCut2}
\end{center}
\end{figure}

We now generalise the above calculation of Sec.~\ref{sec:IToTF:I} which led to an explicit globally-defined expression for $\hat{B}_k$ (associated to an oriented compact genus-$\g$ Riemann surface, $\Sigma$) to that in the presence of a boundary, $\partial \Sigma$, consisting of $\B$ disconnected components. In particular, since our objective is to develop a prescription for cutting open path integrals (using a coherent state basis) we must address the elementary fact that cutting along a cycle introduces {\it boundary components}, see Fig.~\ref{fig:SigmaCut2}, and this will be reflected in the path integral measure.
\sk

We will proceed by direct analogy to the analysis of Sec.~\ref{sec:WB}. Let us again begin by considering the natural pairing between a Beltrami differential, $\mu$, and the $b$-ghost (\ref{eq:Bintmub0}), but in the {\it presence} of boundaries,
\begin{equation}\label{eq:Bintmub0z}
B=\frac{1}{2\pi}\int_{\Sigma}\rmd^2z\,\big(\mu_{\bar{z}}^{\phantom{a}z}b_{zz}+\mu_z^{\phantom{a}\bar{z}}\tilde{b}_{\bar{z}\bar{z}}\big).
\end{equation}
The discussion leading to (\ref{eq:Bintmub4}) in Sec.~\ref{sec:IToTF:I} may be carried over to this context essentially without modification, so we may immediately write down:
\begin{equation}\label{eq:Bintmub4z}
\begin{aligned}
\frac{1}{2\pi}\int_{\Sigma}\rmd^2z\,\big(&\mu_{\bar{z}}^{\phantom{a}z}b_{zz}+\mu_z^{\phantom{a}\bar{z}}\tilde{b}_{\bar{z}\bar{z}}\big)=\sum_m\frac{1}{2\pi i}\oint_{\partial V_m} \big(\rmd z_m\,v_m^{z_m}b_{z_mz_m}-\rmd\bar{z}_m\,\bar{v}_m^{\bar{z}_m}\tilde{b}_{\bar{z}_m\bar{z}_m}\big)\\
&\qquad-\sum_m\sum_{n\geq-1}\frac{1}{(n+1)!}\big(\partial_{z_m}^{n+1}v_m^{z_m}(p_m)b_n^{(z_m)}(p_m)+\partial_{\bar{z}_m}^{n+1}\bar{v}_m^{\bar{z}_m}(p_m)\tilde{b}_n^{(\bar{z}_m)}(p_m)\big).
\end{aligned}
\end{equation}
That there are boundaries, $\partial\Sigma\neq\zero$, will (as in Sec.~\ref{sec:WB}) primarily be implemented by simply omitting terms in the sum over $m$ (the sum over cells), which would correspond to the fact that certain polygons in the cell decomposition are simply absent when $\partial\Sigma\neq\zero$, as outlined in Fig.~\ref{sec:holes2}. We will then discover that simply omitting cells is not well-defined, and we will also discover the boundary terms that make it well-defined.
\sk

Making this explicit, we again {\it relabel} the elements of the set,  
$
\mathscr{V}=\{V_1,V_2,\dots\}\equiv \big\{V_m,V_n^{\partial},V_\ell^{\partial\!\!\!\slash}\big\},
$ 
and introduce boundaries by simply deleting the subset $\{V_\ell^{\partial\!\!\!\slash}\}$. The resulting reduced set will be identical to that in (\ref{eq:Vpartial}), namely:
\begin{equation}\label{eq:Vpartialz}
\mathscr{V}_{\partial}=\big\{V_m,V_n^{\partial}\big\},
\end{equation}
where the relevant notation is explained below (\ref{eq:Vpartial}), with a corresponding illustration of the relevant setup in Fig~\ref{sec:holes2}. 
\sk

Restricting to the atlas associated to (\ref{eq:Vpartialz}), we make the physical boundary contributions explicit in (\ref{eq:Bintmub4z}) using the above definitions,
\begin{equation}\label{eq:Bintmub5z}
\begin{aligned}
\frac{1}{2\pi}\int_{\Sigma}\rmd^2z&\,\big(\mu_{\bar{z}}^{\phantom{a}z}b_{zz}+\mu_z^{\phantom{a}\bar{z}}\tilde{b}_{\bar{z}\bar{z}}\big)=\sum_m\frac{1}{2\pi i}\oint_{\partial V_m} \big(\rmd z_m\,v_m^{z_m}b_{z_mz_m}-\rmd\bar{z}_m\,\bar{v}_m^{\bar{z}_m}\tilde{b}_{\bar{z}_m\bar{z}_m}\big)\\
&\qquad-\sum_m\sum_{n\geq-1}\frac{1}{(n+1)!}\big(\partial_{z_m}^{n+1}v_m^{z_m}(p_m)b_n^{(z_m)}(p_m)+\partial_{\bar{z}_m}^{n+1}\bar{v}_m^{\bar{z}_m}(p_m)\tilde{b}_n^{(\bar{z}_m)}(p_m)\big)\\
&\qquad+\sum_m\frac{1}{2\pi i}\oint_{\partial V_m^{\partial}} \big(\rmd z_m\,v_m^{z_m}b_{z_mz_m}-\rmd\bar{z}_m\,\bar{v}_m^{\bar{z}_m}\tilde{b}_{\bar{z}_m\bar{z}_m}\big).
\end{aligned}
\end{equation}
The sum in the second line on the right-hand side is over all elements of the set (\ref{eq:Vpartialz}) and the summand is evaluated at the origins of the corresponding charts. 
Considering the subset $\{\partial V_m^{\partial}\}$, every element, $\partial V_m^{\partial}$, is topologically a circle. We then decompose the collection of all such elements into the set of {\it physical} boundary segments or edges, $\mathscr{C}=\{C_\ell\}$, and then the remaining segments are all interior edges that in turn coincide with corresponding edges of neighbouring cells. Using the notation explained in the paragraphs leading up to (\ref{eq:chi(S)temp3}), the entire set of distinct edges (with the standard orientation explained there) correspond to those associated to {\it pairs}, $(mn)$, labelled by $\{C_{mn}\}$ (with positive orientation with respect to the first entry in the subscript), and the set of physical edges, $\{C_\ell\}$, that constitute the physical boundary, $\partial\Sigma$. That is, the complete set of distinct edges associated to a cell decomposition of a Riemann surface with boundary is:
$$
\big\{C_{mn},C_\ell\big\},
$$
which in turn enables us to rewrite (\ref{eq:Bintmub5z}) (using the same reasoning that led to (\ref{eq:chi(S)temp4})) as follows,
\begin{equation}\label{eq:Bintmub6z}
\begin{aligned}
\frac{1}{2\pi}&\int_{\Sigma}\rmd^2z\,\big(\mu_{\bar{z}}^{\phantom{a}z}b_{zz}+\mu_z^{\phantom{a}\bar{z}}\tilde{b}_{\bar{z}\bar{z}}\big)=\\
&\qquad=\sum_{(mn)}\frac{1}{2\pi i}\int_{C_{mn}} \Big[\rmd z_m\,\big(v_m^{z_m}-v_n^{z_m}\big)b_{z_mz_m}-\rmd\bar{z}_m\,\big(\bar{v}_m^{\bar{z}_m}-\bar{v}_n^{\bar{z}_m}\big)\tilde{b}_{\bar{z}_m\bar{z}_m}\Big]\\
&\qquad\quad-\sum_m\sum_{n\geq-1}\frac{1}{(n+1)!}\big(\partial_{z_m}^{n+1}v_m^{z_m}(p_m)b_n^{(z_m)}(p_m)+\partial_{\bar{z}_m}^{n+1}\bar{v}_m^{\bar{z}_m}(p_m)\tilde{b}_n^{(\bar{z}_m)}(p_m)\big)\\
&\quad\qquad+\sum_\ell\frac{1}{2\pi i}\int_{C_\ell} \big(\rmd z_\ell\,v_\ell^{z_\ell}b_{z_\ell z_\ell}-\rmd\bar{z}_\ell\,\bar{v}_\ell^{\bar{z}_\ell}\tilde{b}_{\bar{z}_\ell\bar{z}_\ell}\big).
\end{aligned}
\end{equation}
The sum over pairs $(mn)$ is over interior edges, whereas the sum over $\ell$ is over all segments, $C_\ell$, (i.e.~edges) that make up the physical boundary of $\Sigma$,
$$
\partial\Sigma=\bigcup_\ell C_\ell.
$$ 
For example, if there was only a single boundary, and that was identified with the boundary shown in Fig.~\ref{sec:holes2}, the total number of elements in the set $\{C_\ell\}$ (i.e.~the number of terms in the sum over $\ell$) in (\ref{eq:Bintmub6z}) would be 12. If instead we considered the setup shown in the first diagram of Fig.~\ref{fig:goodboundaryshrink} then the sum over $\ell$ would be over the 3 depicted boundary segments. 
\sk

Finally, since the contours $\{C_{mn}\}$ are all along {\it interior} edges the relation (\ref{eq:vm-vn=dfmn/dt}) (linking the locally-defined vectors to transition functions) remains valid so that (\ref{eq:Bintmub6z}) may be written directly in terms of transition functions on patch overlaps,
\begin{equation}\label{eq:Bintmub7z}
\begingroup\makeatletter\def\f@size{11}\check@mathfonts
\def\maketag@@@#1{\hbox{\m@th\large\normalfont#1}}%
\begin{aligned}
B_k&=\frac{1}{2\pi}\int_{\Sigma}\rmd^2z\,\big(\mu_{k\bar{z}}^{\phantom{aa}z}b_{zz}+\mu_{kz}^{\phantom{aa}\bar{z}}\tilde{b}_{\bar{z}\bar{z}}\big)\\
&=\sum_{(mn)}\frac{1}{2\pi i}\int_{C_{mn}} \bigg(\rmd z_m\,\frac{\partial f_{mn}(z_n,t)}{\partial t^k}\big|_{z_n}b_{z_mz_m}-\rmd\bar{z}_m\,\frac{\partial \bar{f}_{mn}(\bar{z}_n,\bar{t})}{\partial t^k}\big|_{\bar{z}_n}\tilde{b}_{\bar{z}_m\bar{z}_m}\bigg)\\
&\quad-\sum_m\sum_{n\geq-1}\frac{1}{(n+1)!}\Big[\partial_{z_m}^{n+1}v_{km}^{z_m}(p_m)b_n^{(z_m)}(p_m)+\partial_{\bar{z}_m}^{n+1}\bar{v}_{km}^{\bar{z}_m}(p_m)\tilde{b}_n^{(\bar{z}_m)}(p_m)\Big]\\
&\quad\qquad+\sum_\ell\frac{1}{2\pi i}\int_{C_\ell} \big(\rmd z_\ell\,v_{k\ell}^{z_\ell}b_{z_\ell z_\ell}-\rmd\bar{z}_\ell\,\bar{v}_{k\ell}^{\bar{z}_\ell}\tilde{b}_{\bar{z}_\ell\bar{z}_\ell}\big)\\
\end{aligned}
\endgroup
\end{equation}
where we have found it convenient to work in terms of the components, $\mu_k$ and $\{v_{km}^{z_m}\}$, recall (\ref{eq:munu}) and the infinitesimal version (which is all one needs) of (\ref{eq:dzm = vdt}). The expression (\ref{eq:Bintmub7z}) for the ghost insertions in the path integral measure is the result of trying to include boundaries on $\Sigma$ simply by {\it omitting} cells, as depicted in Fig.~\ref{sec:holes2}. Note also that (\ref{eq:Bintmub7z}) is precisely the analogue of (\ref{eq:chi(S)X}), an important difference being that the boundary terms appearing in the sum over $\ell$ in the latter transform as connections whereas the boundary terms in the sum over $\ell$ in (\ref{eq:Bintmub7z}) transform as one-forms in the vicinity of the various contours, $C_\ell$.
\sk

The comments below (\ref{eq:Bintmub5bx}) apply directly also to (\ref{eq:Bintmub7z}), the only difference here being the additional boundary terms in the sum over $\ell$ that are present in (\ref{eq:Bintmub7z}) but absent in (\ref{eq:Bintmub5bx}). As in the `no codimension-1 boundary' case (\ref{eq:Bintmub5bxx}) we will add boundary terms to (\ref{eq:Bintmub7z}) to subtract out the frame-dependent contributions in the sum over $m$. The terms in the sum over $\ell$ are perfectly well-defined (as mentioned the integrand transforms as a one form\footnote{Incidentally, one can also take the chiral half of the terms in the sum over $\ell$ to be {\it holomorphic} in the vicinity of the $C_\ell$ contours, and although this is used, e.g., in string field theory, it will not be desirable here because we are searching for a slice that will be globally well-defined in moduli space.} in the vicinity of the contours $C_\ell$), but it will be convenient to add boundary terms to both sides of (\ref{eq:Bintmub7z}) to cancel these also, since (as we will explain momentarily) this will enable us to ``decouple moduli as much as possible''. These boundary terms can be attributed to a partial choice of gauge slice which make these measure contributions insensitive to the Beltrami differential along the contours $C_\ell$. This will in turn enable the pinch/twist moduli and corresponding translation moduli of handle operators to ``decouple''. In particular, we will replace $B_k\rightarrow \hat{B}_k=B_k+B_k^{\rm (b.t.)}$, where:
\begin{equation}\label{eq:Bintmub8z}
\begingroup\makeatletter\def\f@size{11}\check@mathfonts
\def\maketag@@@#1{\hbox{\m@th\large\normalfont#1}}%
\boxed{
\begin{aligned}
\hat{B}_k&=\frac{1}{2\pi}\int_{\Sigma}\rmd^2z\,\big(\mu_{k\bar{z}}^{\phantom{aa}z}b_{zz}+\mu_{kz}^{\phantom{aa}\bar{z}}\tilde{b}_{\bar{z}\bar{z}}\big)-\sum_{\ell\in\partial \Sigma}\frac{1}{2\pi i}\int_{C_\ell} \big(\rmd z_\ell\,v_{k\ell}^{z_\ell}b_{z_\ell z_\ell}-\rmd\bar{z}_\ell\,\bar{v}_{k\ell}^{\bar{z}_\ell}\tilde{b}_{\bar{z}_\ell\bar{z}_\ell}\big)\\
&\qquad +\sum_m\sum_{n\geq-1}\frac{1}{(n+1)!}\Big[\partial_{z_m}^{n+1}v_{km}^{z_m}(p_m)b_n^{(z_m)}(p_m)+\partial_{\bar{z}_m}^{n+1}\bar{v}_{km}^{\bar{z}_m}(p_m)\tilde{b}_n^{(\bar{z}_m)}(p_m)\Big]\\
&=\sum_{(mn)}\frac{1}{2\pi i}\int_{C_{mn}} \bigg(\rmd z_m\,\frac{\partial f_{mn}(z_n,t)}{\partial t^k}\big|_{z_n}b_{z_mz_m}-\rmd\bar{z}_m\,\frac{\partial \bar{f}_{mn}(\bar{z}_n,\bar{t})}{\partial t^k}\big|_{\bar{z}_n}\tilde{b}_{\bar{z}_m\bar{z}_m}\bigg)
\end{aligned}
}
\endgroup
\end{equation}
and we also made it explicit that the sum over $\ell$ is over all contours that comprise the physical boundary of $\Sigma$. 
This expression generalises (\ref{eq:Bintmub5bxx}) and can be used to define the ghost contributions to the path integral measure in the presence of boundaries {\it and} punctures. Crucially, the second equality in (\ref{eq:Bintmub8z}) shows that $\hat{B}_k$ depends entirely on the fundamental data defining the Riemann surface. We emphasise that the contours $\{C_{mn}\}$ only traverse {\it internal} edges of any given cell decomposition, since this is where there are patch overlaps.
\sk

That the result on the right-hand side of the second equality in (\ref{eq:Bintmub8z}) is identical in form to that on the right-hand side of the second equality in (\ref{eq:Bintmub5bxx}), {\it even though} the former was derived in the presence of physical boundaries whereas the latter was derived in the absence of physical boundaries is precisely what we expect. This also happens in the simple example of the Euler characteristic, recall in particular the discussion associated to (\ref{eq:2-2g=fmn}) and the remaining paragraphs in that section. 
\sk

The second equality in (\ref{eq:Bintmub8z}) in conjunction with the full path integral expression (\ref{eq:fullpathintegral}) shows that if the path integral is to be invariant under reparametrisations of the moduli, $t\mapsto \hat{t}(t)$ (or $\tau\mapsto \hat{\tau}(\tau)$ in a holomorphic construction) then the transition functions must (at least on patch overlaps) transform as {\it scalars} under such reparametrisations. But since $z_m=f_{mn}(z_n,t)$ we see that $z_m$ must also transform as a scalar, and hence so must $z_n$, since this must be true for all charts of the cover. That is, on {\it patch overlaps},\footnote{XXX}
\begin{equation}\label{eq:f=fhat}
\boxed{
f_{mn}(z_n,t) = \hat{f}_{mn}(\hat{z}_n(\hat{t}),\hat{t}(t)),\qquad{\rm and}\qquad z_m(t)=\hat{z}_m(\hat{t}(t))
}\end{equation}
This is a key equation which is closely related to making amplitudes globally  well-defined.
\sk

Incidentally, the derivation leading to (\ref{eq:Bintmub8z}) holds for {\it any} conformal tensor whose chiral half has weight $(2,0)$ and anti-chiral half has weight $(0,2)$ which has a Laurent expansion at $p_m$, not just $b_{zz}\rmd z^2$ and $\tilde{b}_{\bar{z}\bar{z}}\rmd\bar{z}^2$. Therefore, in particular, it also holds for the total energy-momentum tensor, $T=T_{zz}\rmd z^2$ and $\tilde{T}=\tilde{T}_{\bar{z}\bar{z}}\rmd\bar{z}^2$ (see Appendix \ref{sec:CFTC} and in particular (\ref{eq:phi_n modes Laurent}) and Table \ref{tab:weightsmodes} on p.~\pageref{tab:weightsmodes}) with vanishing central charge, $c=0$. That is, from (\ref{eq:Bintmub8z}) we can construct a globally well-defined operator, call it $\hat{D}_k$:
\begin{equation}\label{eq:Tintmub8z}
\begingroup\makeatletter\def\f@size{11}\check@mathfonts
\def\maketag@@@#1{\hbox{\m@th\large\normalfont#1}}%
\boxed{
\begin{aligned}
\hat{D}_k&=\frac{1}{2\pi}\int_{\Sigma}\rmd^2z\,\big(\mu_{k\bar{z}}^{\phantom{aa}z}T_{zz}+\mu_{kz}^{\phantom{aa}\bar{z}}\tilde{T}_{\bar{z}\bar{z}}\big)-\sum_\ell\frac{1}{2\pi i}\int_{C_\ell} \big(\rmd z_\ell\,v_{k\ell}^{z_\ell}T_{z_\ell z_\ell}-\rmd\bar{z}_\ell\,\bar{v}_{k\ell}^{\bar{z}_\ell}\tilde{T}_{\bar{z}_\ell\bar{z}_\ell}\big)\\
&\qquad +\sum_m\sum_{n\geq-1}\frac{1}{(n+1)!}\Big[\partial_{z_m}^{n+1}v_{km}^{z_m}(p_m)L_n^{(z_m)}(p_m)+\partial_{\bar{z}_m}^{n+1}\bar{v}_{km}^{\bar{z}_m}(p_m)\tilde{L}_n^{(\bar{z}_m)}(p_m)\Big]\\
&=\sum_{(mn)}\frac{1}{2\pi i}\int_{C_{mn}} \bigg(\rmd z_m\,\frac{\partial f_{mn}(z_n,t)}{\partial t^k}\big|_{z_n}T_{z_mz_m}-\rmd\bar{z}_m\,\frac{\partial \bar{f}_{mn}(\bar{z}_n,\bar{t})}{\partial t^k}\big|_{\bar{z}_n}\tilde{T}_{\bar{z}_m\bar{z}_m}\bigg)
\end{aligned}
}
\endgroup
\end{equation}

This expression (\ref{eq:Tintmub8z}) will play a vital role when we show that BRST-exact terms decouple from the path integral up to boundary terms in moduli space. This is due to the relation, $\{Q_B,\hat{B}_k\}=\hat{D}_k$, and as we will see this in turn gives rise to total derivatives of normal-ordered operators. A related comment is that unlike in dimensional regularisation \cite{Polchinski88}, conformal (or more precisely Weyl) normal ordering does {\it not} generically commute with derivatives \cite{Polchinski87}, $\partial_k\!:\!(\dots)\!:_z\,\,\neq\,\, :\!\partial_k(\dots)\!:_z$ on curved surfaces. When acting on normal-ordered operators, the quantity $\hat{D}_k$ will generate a derivative that is {\it outside} the normal ordering. The boundary terms on the right-hand side of the first equality in (\ref{eq:Tintmub8z}) ensure that one indeed ends up with total derivatives in moduli space. We will discuss this in detail in Sec.~\ref{sec:comm dA-dA}, see also the related discussion in Sec.~\ref{sec:WuYang}, and also the dedicated section on gauge invariance, Sec.~\ref{sec:BRST-AC}.

\subsection{In Terms of Transition Functions III: `Emergent Boundaries'}\label{sec:MWEB}
We will in this section consider the measure contribution to the path integral (associated to a Riemann surface without physical boundaries), and in particular we would like to understand how the boundary terms in (\ref{eq:Bintmub8z}) (that were ``conjectured'' to be present in the previous section) emerge via the cutting procedure. We will also make it manifest that this choice of boundary terms ``decouples moduli as much as possible''.
\sk

To proceed we will primarily consider cutting the path integral associated to a Riemann surface without boundary across a trivial homology cycle, as in Fig.~\ref{fig:SigmaCut2}. The left-hand side of the figure is a closed Riemann surface (with at most punctures but no codimension-1 boundaries), and so the path integral measure is a product of insertions of the form (\ref{eq:Bintmub5bxx}). Let us then choose a good cover, $\mathscr{U}$, for the uncut surface $\Sigma$ such that,
$$
\mathscr{U}=\mathscr{U}_1\cup \mathscr{U}_2,
$$
where $\mathscr{U}_1$ and $\mathscr{U}_2$ are corresponding covers for the surfaces $\Sigma_1$ and $\Sigma_2$ respectively, as shown on the right-hand side of Fig.~\ref{fig:SigmaCut2}. But there is a still an overlap region, $\mathscr{U}_1\cap \mathscr{U}_2$. In order to decouple the two sides of the cut as much as possible we will again construct a cell decomposition of $\Sigma$, but such that each of the two the resulting surfaces, $\Sigma_1$ and $\Sigma_2$, correspond to surfaces with holes as shown in the {\it second} diagram of Fig.~\ref{sec:holes2} (but they can also be smooth as shown in the first diagram in Fig.~\ref{fig:holeorient}). We then still denote the boundary segments of $\Sigma_1$ and $\Sigma_2$ by $C_\ell$ as in the previous section, the only difference being that now whether $C_\ell$ is part of a boundary curve of $\Sigma_1$ or $\Sigma_2$ depends on whether $\ell\in \partial\Sigma_1$  or $\ell\in\partial\Sigma_2$ respectively. So proceeding precisely as in the previous section the quantity $B$ defined in (\ref{eq:Bintmub0}) corresponding to cutting open the path integral shown in Fig.~\ref{fig:SigmaCut2} takes the form,
\begin{equation}\label{eq:Bintmub8}
\begin{aligned}
\frac{1}{2\pi}&\int_{\Sigma}\rmd^2z\,\big(\mu_{\bar{z}}^{\phantom{a}z}b_{zz}+\mu_{z}^{\phantom{a}\bar{z}}\tilde{b}_{\bar{z}\bar{z}}\big)=\\
&=\bigg[\frac{1}{2\pi}\int_{\Sigma_1}\rmd^2z\,\big(\mu_{\bar{z}}^{\phantom{a}z}b_{zz}+\mu_z^{\phantom{a}\bar{z}}\tilde{b}_{\bar{z}\bar{z}}\big)-\sum_{\ell\in\partial \Sigma_1}\frac{1}{2\pi i}\int_{C_\ell} \big(\rmd z_\ell\,v_\ell^{z_\ell}b_{z_\ell z_\ell}-\rmd\bar{z}_\ell\,\bar{v}_\ell^{\bar{z}_\ell}\tilde{b}_{\bar{z}_\ell\bar{z}_\ell}\big)\bigg]\\
&\qquad+\sum_{\ell,\ell'\in \mathscr{U}_1\cap\mathscr{U}_2}\frac{1}{2\pi i}\int_{C_{\ell\ell'}} \Big[\rmd z_\ell\,\big(v_\ell^{z_\ell}-v_{\ell'}^{z_\ell}\big)b_{z_\ell z_\ell}-\rmd\bar{z}_\ell\,\big(\bar{v}_\ell^{\bar{z}_\ell}-\bar{v}_{\ell'}^{\bar{z}_\ell}\big)\tilde{b}_{\bar{z}_\ell\bar{z}_\ell}\Big]+\\
&\quad+\bigg[\frac{1}{2\pi}\int_{\Sigma_2}\rmd^2z\,\big(\mu_{\bar{z}}^{\phantom{a}z}b_{zz}+\mu_z^{\phantom{a}\bar{z}}\tilde{b}_{\bar{z}\bar{z}}\big)-\sum_{\ell\in\partial \Sigma_2}\frac{1}{2\pi i}\int_{C_\ell} \big(\rmd z_\ell\,v_\ell^{z_\ell}b_{z_\ell z_\ell}-\rmd\bar{z}_\ell\,\bar{v}_\ell^{\bar{z}_\ell}\tilde{b}_{\bar{z}_\ell\bar{z}_\ell}\big)\bigg]\\
\end{aligned}
\end{equation}
The left-hand side is the integral over the uncut surface, $\Sigma$, whereas the right-hand side is comprised of three terms, corresponding to the three brackets\footnote{We use the mostly standard nomenclature: `braces'$=\{\dots\}$, `bracket'$=[\dots]$, `parenthesis'$=(\dots)$.}, the first of which depends {\it entirely} on local data in $\Sigma_1$ while being {\it independent} of boundary conditions along $\partial \Sigma_1$ (because these boundary contributions are subtracted out as shown and therefore cancel out), the second bracket depends entirely on data across the overlap, $\mathscr{U}_1\cap\mathscr{U}_2$, (to which we always associate twist moduli but also pinch moduli if we wish), and the third bracket depends entirely on local data in $\Sigma_2$ and which is also independent of boundary conditions on  $\partial \Sigma_2$.
\sk

We have already seen in (\ref{eq:Bintmub5bxx}) that the left-hand side in (\ref{eq:Bintmub8}) associated to the full uncut surface, $\Sigma$, requires additional boundary terms to cancel the coordinate dependence at punctures\footnote{Incidentally, it is {\it these} subtractions or boundary terms that enable one to insert arbitrary BRST-invariant (or even BRST non-invariant) external vertex operators into the string path integral rather than only Virasoro primaries.}, and (since we are considering a good cover) this will also be true for the corresponding cells on the right-hand side: the interior points, $p_m$, of every cell at which local coordinates, $z_m$, are centred, $z_m(p_m)=0$, is unaffected by the fact that the surface is cut open to produce the two surfaces, $\Sigma_1,\Sigma_2$. So adding the appropriate boundary terms that remove this coordinate dependence at punctures on the left- and right-hand sides in (\ref{eq:Bintmub8}) results precisely in:
\begin{equation}\label{eq:Bintmub8f}
\begingroup\makeatletter\def\f@size{11}\check@mathfonts
\def\maketag@@@#1{\hbox{\m@th\large\normalfont#1}}%
\begin{aligned}
\frac{1}{2\pi}&\int_{\Sigma}\rmd^2z\,\big(\mu_{\bar{z}}^{\phantom{a}z}b_{zz}+\mu_{z}^{\phantom{a}\bar{z}}\tilde{b}_{\bar{z}\bar{z}}\big)-\sum_{m\in\Sigma}\sum_{n\geq-1}\frac{1}{(n+1)!}\big(\partial_{z_m}^{n+1}v_m^{z_m}(p_m)b_n^{(z_m)}(p_m)+\partial_{\bar{z}_m}^{n+1}\bar{v}_m^{\bar{z}_m}(p_m)\tilde{b}_n^{(\bar{z}_m)}(p_m)\big)\\
&=\bigg[\frac{1}{2\pi}\int_{\Sigma_1}\rmd^2z\,\big(\mu_{\bar{z}}^{\phantom{a}z}b_{zz}+\mu_z^{\phantom{a}\bar{z}}\tilde{b}_{\bar{z}\bar{z}}\big)-\sum_{\ell\in\partial \Sigma_1}\frac{1}{2\pi i}\int_{C_\ell} \big(\rmd z_\ell\,v_\ell^{z_\ell}b_{z_\ell z_\ell}-\rmd\bar{z}_\ell\,\bar{v}_\ell^{\bar{z}_\ell}\tilde{b}_{\bar{z}_\ell\bar{z}_\ell}\big)\\
&\qquad-\sum_{m\in\Sigma_1}\sum_{n\geq-1}\frac{1}{(n+1)!}\big(\partial_{z_m}^{n+1}v_m^{z_m}(p_m)b_n^{(z_m)}(p_m)+\partial_{\bar{z}_m}^{n+1}\bar{v}_m^{\bar{z}_m}(p_m)\tilde{b}_n^{(\bar{z}_m)}(p_m)\big)\bigg]\\
&\quad\qquad+\sum_{\ell,\ell'\in \mathscr{U}_1\cap\mathscr{U}_2}\frac{1}{2\pi i}\int_{C_{\ell\ell'}} \Big[\rmd z_\ell\,\big(v_\ell^{z_\ell}-v_{\ell'}^{z_\ell}\big)b_{z_\ell z_\ell}-\rmd\bar{z}_\ell\,\big(\bar{v}_\ell^{\bar{z}_\ell}-\bar{v}_{\ell'}^{\bar{z}_\ell}\big)\tilde{b}_{\bar{z}_\ell\bar{z}_\ell}\Big]+\\
&\quad+\bigg[\frac{1}{2\pi}\int_{\Sigma_2}\rmd^2z\,\big(\mu_{\bar{z}}^{\phantom{a}z}b_{zz}+\mu_z^{\phantom{a}\bar{z}}\tilde{b}_{\bar{z}\bar{z}}\big)-\sum_{\ell\in\partial \Sigma_2}\frac{1}{2\pi i}\int_{C_\ell} \big(\rmd z_\ell\,v_\ell^{z_\ell}b_{z_\ell z_\ell}-\rmd\bar{z}_\ell\,\bar{v}_\ell^{\bar{z}_\ell}\tilde{b}_{\bar{z}_\ell\bar{z}_\ell}\big)\\
&\qquad-\sum_{m\in\Sigma_2}\sum_{n\geq-1}\frac{1}{(n+1)!}\big(\partial_{z_m}^{n+1}v_m^{z_m}(p_m)b_n^{(z_m)}(p_m)+\partial_{\bar{z}_m}^{n+1}\bar{v}_m^{\bar{z}_m}(p_m)\tilde{b}_n^{(\bar{z}_m)}(p_m)\big)\bigg].
\end{aligned}
\endgroup
\end{equation}

Recall that the Beltrami differential $\mu=\mu_{z}^{\phantom{a}\bar{z}}\rmd z\otimes \partial_{\bar{z}}$ can be thought of as a one form in moduli space. So in particular for, say, real coordinates, $\{t^k\}=(t^1,\dots,t^{2\m})$:
$$
\mu_{z}^{\phantom{a}\bar{z}} = \mu_{kz}^{\phantom{aa}\bar{z}}\rmd t^k,\qquad {\rm and}\qquad \mu_{\bar{z}}^{\phantom{a}z} = \mu_{k\bar{z}}^{\phantom{aa}z}\rmd t^k,\qquad k=1,\dots,2\m,
$$
with an implicit sum over $k$, 
and for complex coordinates, $\{t^k\}=(\tau^1,\bar{\tau}^1,\dots,\tau^\m,\bar{\tau}^\m)$,
$$
\mu_{z}^{\phantom{a}\bar{z}} = \mu_{\tau^kz}^{\phantom{aaa}\bar{z}}\rmd\tau^k+\mu_{\bar{\tau}^kz}^{\phantom{aaa}\bar{z}}\rmd\bar{\tau}^k,\qquad {\rm and}\qquad \mu_{\bar{z}}^{\phantom{a}z} = \mu_{\tau^k\bar{z}}^{\phantom{aaa}z}\rmd\tau^k+\mu_{\bar{\tau}^k\bar{z}}^{\phantom{aaa}z}\rmd\bar{\tau}^k,\qquad k=1,\dots,\m,
$$
subject to the conditions\footnote{Although it is \cite{BelavinKnizhnik86} possible to choose a local holomorphic slice such that $\mu_{\bar{\tau}^kz}^{\phantom{aaa}\bar{z}}= \mu_{\tau^k\bar{z}}^{\phantom{aaa}z}=0$ it will be easier to construct a globally well-defined measure in moduli space if we rather {\it do not} require that the Beltrami differential, $\mu$, be meromorphic in $\tau^k$.} $(\mu_{\tau^k\bar{z}}^{\phantom{aaa}z})^*=\mu_{\bar{\tau}^kz}^{\phantom{aaa}\bar{z}}$ and $(\mu_{\bar{\tau}^k\bar{z}}^{\phantom{aaa}z})^*=\mu_{\tau^kz}^{\phantom{aaa}\bar{z}}$. Similar reasoning applies to the local vector components, $\mu_{z}^{\phantom{a}\bar{z}}|_{U_m}=\partial_{z_m}v_m^{\bar{z}_m}$, etc. So we may then decompose the measure contributions as follows,
\begin{equation}\label{eq:B=Bkdtk}
\hat{B}=\hat{B}_{t^k}\rmd t^k,\qquad {\rm or}\qquad \hat{B}=\hat{B}_{\tau^k}\rmd\tau^k+\hat{B}_{\bar{\tau}^k}\rmd\bar{\tau}^k,
\end{equation}
depending on whether we are interested in real or complex moduli space coordinates respectively (and with appropriate implicit sums over corresponding ranges of $k$). So we see that (\ref{eq:Bintmub8f}) is precisely the statement that:
\begin{equation}\label{eq:Bkfactor}
\boxed{\hat{B}_k^{\Sigma} =\hat{B}_k^{\Sigma_1}+\hat{B}_k^{\mathscr{U}_1\cap\mathscr{U}_2}+\hat{B}_k^{\Sigma_2}}
\end{equation}
where in the latter $k$ can denote any real or complex modulus of our choice, 
each term of which is precisely given by the term we derived above, see (\ref{eq:Bintmub8z}), but we also included appropriate superscripts here to indicate the underlying Riemann surfaces on which the various quantities are evaluated. 
\sk

A crucial point now is that (for a {\it given} modulus labelled by $k$) the quantities $\hat{B}_k^{\Sigma_1}$, $\hat{B}_k^{\mathscr{U}_1\cap\mathscr{U}_2}$ and $\hat{B}_k^{\Sigma_2}$, can be taken to be {\it independent}, since they are each defined on distinct regions of the corresponding Riemann surfaces with no overlapping region of validity. And in particular the derivation leading to (\ref{eq:Bkfactor}) shows that we can pick a slice in moduli space such that this decomposition, for a given $k$, is ``orthogonal'', so that the various moduli will therefore be ``as decoupled as possible'', which is what we set out to show. In particular, for a {\it given} $k$ only {\it one} of the three explicit terms in (\ref{eq:Bkfactor}) will be non-vanishing. 
\sk

Taking also (\ref{eq:vm-vn=dfmn/dt}) (linking the locally-defined vectors to transition functions) into account, what we have shown is that the quantities $\hat{B}_k$ in (\ref{eq:Bintmub8z}) are precisely the correct elementary building blocks to be used to construct the full measure. We can, in particular, now choose gauge slices such that if, say, we consider pinch/twist moduli, $(q,\bar{q})$, of a handle only the quantities $\hat{B}_q^{\mathscr{U}_1\cap\mathscr{U}_2}$ and $\hat{B}_{\bar{q}}^{\mathscr{U}_1\cap\mathscr{U}_2}$ in (\ref{eq:Bkfactor}) are non-vanishing. Or if we wish to associate translations (within $\Sigma_1$) of, say, the boundary $\partial \Sigma_1$, (to which we associate moduli $(z_{v_1},\bar{z}_{v_1})$) then only the terms $\hat{B}_{z_{v_1}}^{\Sigma_1}$ and $\hat{B}_{\bar{z}_{v_1}}^{\Sigma_1}$ in (\ref{eq:Bkfactor}) are non-vanishing, etc. Notice also that the  compensating boundary terms (such as those in (\ref{eq:Bintmub8f}) associated to $\ell\in\partial \Sigma_1$, or $\ell\in \partial \Sigma_2$, etc.) are such that the terms $\hat{B}_k^{\Sigma_1}$ and $\hat{B}_k^{\Sigma_2}$ in (\ref{eq:Bkfactor}) are {\it independent} of the boundary conditions across the cut cycles, so that we can freely shrink these cycles to {\it points}, reducing them to punctures. It therefore makes sense to associate moduli $z_{v_j},\bar{z}_{v_j}$ to the location of these punctures (in precisely the same way as we associate moduli to the location of external fixed-picture vertex operators to translate them to integrated picture), and we do so explicitly from two independent viewpoints in Sec.~\ref{sec:TP} and Sec.~\ref{sec:SHUTF}.
\sk

The derivation leading to (\ref{eq:Bkfactor}) assumed we cut the path integral across  a single separating (or trivial) homology cycle, but the derivation is entirely general. To every cycle that we cut open we are to include compensating boundary contributions (such as the explicit sums\footnote{If the boundaries are not smooth there will generically be required additional compensating boundary terms of higher codimension (as we already saw in the famous Euler characteristic example); a general and related discussion associated to boundary terms can be found in \cite{Sen19}. All these cut cycles that we are interested in here are to be considered smooth. For the specific integrals under consideration here boundary contributions (additional to the ones we have included already) are generically not required, but one must proceed carefully.} over $\ell\in\partial \Sigma_1$, or $\ell\in \partial \Sigma_2$, etc.), as well as a term of the form $\hat{B}_k^{\mathscr{U}_1\cap\mathscr{U}_2}$ for every handle. This remains true for non-trivial homology cycles also, and the precise expressions derived above remain valid. 
\sk

\begin{figure}
\begin{center}
\includegraphics[width=0.6\textwidth]{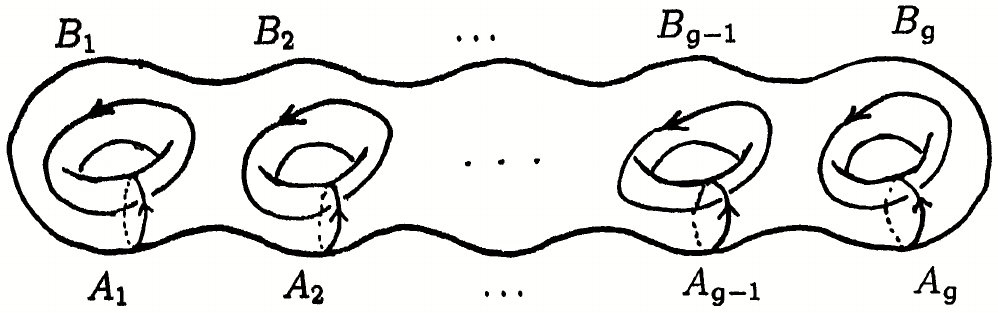}
\caption{A sketch of a canonical intersection homology basis, $A_I,B_I$, with $I=1,\dots,\g$, for  $2\g$ (non-trivial) homology cycles of a genus-$\g$ closed and oriented Riemann surface associated to the first homology group $H_1(\Sigma,\mathbf{Z})$. This corresponds to an abelian subgroup of the mapping class group, $\mathrm{Mod}_{\g,\n}$, which (when $\n=0$) one can think of as being generated by $2\g+1$ Dehn twists 
around non-separating simple closed curves on $\Sigma_\g$. }
\label{fig:homology}
\end{center}
\end{figure}

To elaborate further on this last point, suppose that we now consider a genus-$\g$ Riemann surface, $\Sigma_\g$, and a corresponding canonical intersection basis, $A_I,B_I$ with $I=1,\dots,\g$, of homology cycles, as shown in Fig.~\ref{fig:homology}. Let us cut across all $A_I$ cycles. 
The above observations then allow us, e.g., to consider such a genus-$\g$ closed Riemann surface, $\Sigma_\g$, with or without external punctures, and represent the corresponding path integral as a path integral over a {\it sphere}, $S^2$, with $2\g$ insertions of the form $\hat{B}_{z_{v_j}}^{S^2}\hat{B}_{\bar{z}_{v_j}}^{S^2}$, $j=1,\dots, 2\g$, (which will translate the two states on the two sides of the cut to integrated picture) and $\g$ insertions of the form $\hat{B}_{q^I}^{\rm handle}\hat{B}_{\bar{q}^I}^{\rm handle}$, $I=1,\dots,\g$ (each one of which takes the form $\hat{B}_k^{\mathscr{U}_1\cap\mathscr{U}_2}$ as above and can be associated to pinch/twist moduli of the corresponding $A_I$-cycle handles corresponding the the magnitude/phase of the corresponding coordinates, $q^I$, respectively). This full set of coordinates (if there are external vertex operators we include their position moduli also) could be thought of as parametrising Teichm\"uller space, $\mathrm{T}_{\g,\n}$, or more precisely as being associated to a global section of a principal bundle over Teichm\"uller space (where moving along a given fibre is generated by reparametrisations). Incidentally, since Teichm\"uller space is the {\it universal} cover of moduli space, $\mathcal{M}_{\g,\n}$, of genus-$\g$ closed Riemann surfaces with $\n$ marked points, the fundamental group of $\mathrm{T}_{\g,\n}$ is trivial and hence global sections exist. Since furthermore different choices of sections are therefore homotopic (at least up to boundary terms), making a particular choice leads to a globally well defined construction.\footnote{Recall also that phases of local coordinates are taken to be integrated. One must eventually restrict to the quotient, $\mathcal{M}_{\g,\n}=\mathrm{T}_{\g,\n}/\mathrm{Mod}_{\g,\n}$, where $\mathrm{Mod}_{\g,\n}$ is the mapping class group associated to a closed oriented genus-$\g$ Riemann surface with $\n$ punctures. Discussing an explicit parametrisation of the space $\mathcal{M}_{\g,\n}$ in sufficient detail (in relation to the handle operator approach) is clearly important but we cannot do it justice here. Some relevant foundational material with references to classical results on the mapping class group (e.g.,~by Lickorish, Mumford, Humphries and Dehn, etc.) can be found in \cite{FarbMargalit12}, see also \cite{Bers81}.} Perhaps we should also emphasise that by a theorem of Earle (see Appendix A in \cite{AtickMooreSen88} for a review) there is no global {\it holomorphic} section of the space of conformal structures over Teichmuller space, but (mod U(1)) there is no such obstruction for a {\it smooth} global section (which is the case of interest in the current document). So for every one of these aforementioned Teichm\"uller or moduli coordinates,
$$
t^k=\{z_{v_1},\bar{z}_{v_1},\dots, z_{v_{2\g}},\bar{z}_{v_{2\g}},q^1,\bar{q}^1,\dots,q^\g,\bar{q}^\g\},
$$
we have an insertion of the form:
\begin{equation}\label{eq:BkfactorS2g}
\boxed{\hat{B}_k^{\Sigma_\g} =\hat{B}_k^{S^2}+\sum_{\rm handles}\hat{B}_k^{\rm handles}}
\end{equation}
and, more precisely, for every one of these aforementioned moduli there will only be a {\it single} term in (\ref{eq:BkfactorS2g}) that is non-vanishing. Precisely because we have decoupled the various moduli ``as much as possible''. So in complex moduli space coordinates, on account of (\ref{eq:B=Bkdtk}) we can represent the entire path integral measure contribution as follows,
$$
\frac{1}{n_R}\frac{i^\m}{(2\m)!}\underbrace{\hat{B}\wedge \dots\wedge \hat{B}}_{2\m} = \frac{1}{n_R} \prod_{k=1}^\m \rmd^{2}\tau^k\hat{B}_{\tau^k}\hat{B}_{\bar{\tau}^k},\qquad \rmd^2\tau^k\equiv i\rmd\tau^k\wedge \rmd\bar{\tau}^k,
$$
where the integer $n_R$ accounts for any residual discrete unfixed symmetries \cite{Polchinski_v1}, 
and in the particular slice associated to cutting open all $A_I$ cycles this reduces to:
$$
\frac{1}{n_R}\Big(\prod_{j=1}^{2\g}\rmd^2z_{v_j}\hat{B}_{z_{v_j}}^{S^2}\hat{B}_{\bar{z}_{v_j}}^{S^2}\Big)\Big(\prod_{I=1}^{\g}\rmd^2q^I\hat{B}_{q^I}^{\rm handle}\hat{B}_{\bar{q}^I}^{\rm handle}\Big),
$$
which accounts for $3\g$ complex moduli. Of course one must be careful to account for the number of conformal Killing vectors, and the total number of moduli must be consistent with the Riemann-Roch index theorem \cite{Polchinski_v1}. 
\sk

If, e.g., we wish to also insert $\n$ fixed-picture external vertex operators (and if it so happens that $\n\geq3$) then we can translate $\n-3$ of these external vertex operators to integrated picture and the full measure contribution would be:
$$
\frac{1}{n_R}\Big(\prod_{j=1}^{2\g+\n-3}\rmd^2z_{v_j} \hat{B}_{z_{v_j}}^{S^2}\hat{B}_{\bar{z}_{v_j}}^{S^2}\Big)\Big(\prod_{I=1}^{\g}\rmd^2q^I\hat{B}_{q^I}^{\rm handle}\hat{B}_{\bar{q}^I}^{\rm handle}\Big),
$$
which again holds for arbitrary genus $\g=0,1,\dots$. 
If the case of interest was instead $0\leq\n<3$ (again for any genus $\g$) then we can  unfix the conformal Killing group symmetries and use the integrated picture S matrix derived in Sec.~\ref{sec:TEC} which remains true for arbitrary $\n,\g=0,1,2,\dots$. The total measure contribution would then be:
$$
\Big(\prod_{j=1}^{2\g+\n}\rmd^2z_{v_j}\hat{B}_{z_{v_j}}^{S^2}\hat{B}_{\bar{z}_{v_j}}^{S^2}\Big)\Big(\prod_{I=1}^{\g}\rmd^2q^I\hat{B}_{q^I}^{\rm handle}\hat{B}_{\bar{q}^I}^{\rm handle}\Big)\frac{1}{V_R}\frac{\prod_{a=1}^\kappa\,\,\tilde{c}^{(w_{\sigma_a})} c^{(w_{\sigma_a})}(\sigma_a)}{|\det \psi_a^{(w_{\sigma_b})}(\sigma_b)|^2},
$$
where as we will discuss in Sec.~\ref{sec:TEC} the extra factor compensates for the unfixed conformal Killing group symmetries (with $V_R$ the volume of the conformal Killing group). Notice that the (complex) number of conformal Killing vectors is $\kappa=3$ for {\it all} genera,
$$
\kappa \equiv \dim_{\mathbf{C}}\textrm{PSL(2,$\mathbf{C}$)}= 3,
$$
since we have reduced the genus-$\g$ path integral to a path integral on a Riemann sphere, $S^2$, with handle operator insertions. 
\sk

In the following sections we w ill specify in much greater detail how to single out specific and convenient globally well-defined gauge slices in moduli space (modulo a remaining U(1)).

\subsection{Measure in Terms of a Metric}\label{sec:IToaM}
Returning to the path integral measure contribution (\ref{eq:Bintmub8z}) in the presence of boundaries, there is an important special case that we need to develop (and that will be used to convert fixed-picture offshell vertex operators to integrated picture). This is to adopt the `Weyl normal ordering' prescription of Polchinski \cite{Polchinski88} whereby a special choice of metric is made that is ``as flat as possible'' at the location of a puncture, as described in Sec.~\ref{sec:MV}. The virtue of this choice is that the resulting measure will be globally well-defined in moduli space. In this section we work in terms of a metric.
\sk

\begin{figure}
\begin{center}
\includegraphics[angle=0,origin=c,width=0.85\textwidth]{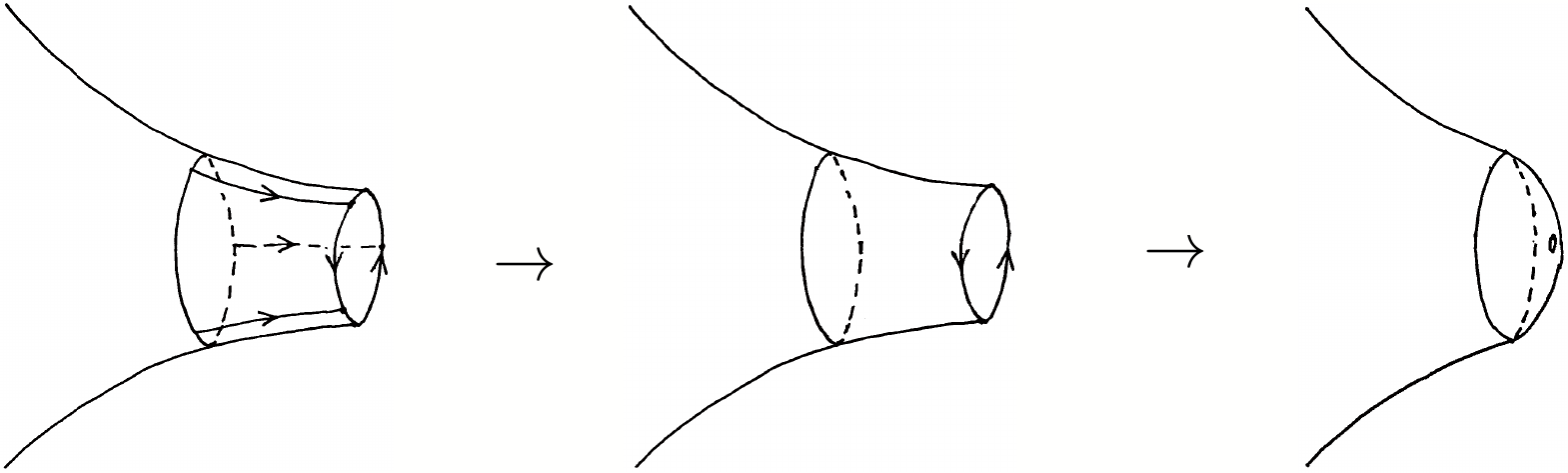}
\caption{Part of a Riemann surface, $\Sigma$, with a boundary. In the first diagram we show the part of a cell decomposition of $\Sigma$ that is adjacent to the boundary. Three open sets are drawn explicitly, each one of which (as well as their intersections) are diffeomorphic to open discs (corresponding to a `good cover'). In the second diagram we have replaced the three overlapping open sets with an annulus (with a single coordinate), denoted by $V_\ell'$ in the main text, and in the third diagram we take the inner radius, $r$, of the annulus to zero, $r\rightarrow 0$, whereby the corresponding chart becomes diffeomorphic to a disc with a puncture.}\label{fig:goodboundaryshrink}
\end{center}
\end{figure}

Let us suppose that $\Sigma$ has a single boundary, $\B=1$. (This is for notational convenience, in that any additional boundaries play no role in what follows; we generalise again to multiple boundaries below.) Then, {\it if} we can arrange for the set of locally-defined vectors, $v_{k\ell}$, associated to that specific boundary of $\Sigma$ to be equal, $v_{k\ell}=v_{k\ell'}$, on patch overlaps, $U_\ell\cap U_{\ell'}$, where $C_\ell$ and $C_{\ell'}$ denote different segments of the single boundary, then we can replace the sum over contour integrals in the first equality in (\ref{eq:Bintmub8z}) associated to that specific boundary by a single contour integral,\footnote{It has been argued that this is nevertheless effectively correct if we average or take the corresponding phase of the annulus coordinate to be integrated \cite{Polchinski88}.  We would like to thank Ashoke Sen for a discussion along these lines. See also the comments in the remaining paragraphs after (\ref{eq:Bintmub7d}) of Sec.~\ref{sec:CIoM}.\label{foot:annulusphase}}
\begin{equation}\label{eq:Cellcontourint}
\begin{aligned}
\hat{B}_k&=\frac{1}{2\pi}\int_{\Sigma}\rmd^2z\,\big(\mu_{k\bar{z}}^{\phantom{aa}z}b_{zz}+\mu_{kz}^{\phantom{aa}\bar{z}}\tilde{b}_{\bar{z}\bar{z}}\big)-\frac{1}{2\pi i}\oint_{C_\ell} \big(\rmd z_\ell\,v_{k\ell}^{z_\ell}b_{z_\ell z_\ell}-\rmd\bar{z}_\ell\,\bar{v}_{k\ell}^{\bar{z}_\ell}\tilde{b}_{\bar{z}_\ell\bar{z}_\ell}\big)\\
&\qquad +\sum_m\sum_{n\geq-1}\frac{1}{(n+1)!}\Big[\partial_{z_m}^{n+1}v_{km}^{z_m}(p_m)b_n^{(z_m)}(p_m)+\partial_{\bar{z}_m}^{n+1}\bar{v}_{km}^{\bar{z}_m}(p_m)\tilde{b}_n^{(\bar{z}_m)}(p_m)\Big]\\
\end{aligned}
\end{equation}
where (unlike as in (\ref{eq:Bintmub8z}) where $C_\ell$ was a {\it segment} of a boundary) now $C_\ell$ is a {\it closed} contour. This is precisely the kind of modification that was required in order to ensure that the Euler characteristic does not depend on details about shape or size of the various boundaries (see Sec.~\ref{sec:WB}) when the cover we considered was not a `good cover', and in particular when we allowed for charts diffeomorphic to annuli to construct boundaries as we are doing here. So this corresponds to transitioning from the first to the second diagram in Fig.~\ref{fig:goodboundaryshrink}. 
When this is the case we can take the contour integral (\ref{eq:Cellcontourint}) to be along the inner boundary of an annulus in a chart $(U_\ell',z_\ell)$, where $U_\ell'\equiv U_\ell \setminus {\rm D}_\ell$ is the annulus, i.e.~the set associated to removing a disc ${\rm D}_\ell$ from the chart open set $U_\ell$. We can take the centre of ${\rm D}_\ell$ to be the point $p_\ell$ at which $z_\ell(p_\ell)=0$. Let us suppose furthermore that in the chart $(U_\ell',z_\ell)$ there is a corresponding cell, $V_\ell'$, such that restricting the range of $z_\ell$ to $V_\ell'$  is equivalent to taking $r<|z_\ell|<1$. The ``outer'' boundary of the annulus (identified with the outer boundary of $V_\ell'$) is then identified with $|z_\ell|=1$, and it is this boundary that is connected to the remaining Riemann surface (which is in turn associated to $|z_\ell|>1$). Incidentally, the sum over $m$ in the last line in (\ref{eq:Cellcontourint}) is over all cells that are ``outside'' the annulus $V_\ell'$ (corresponding to the region $|z_1|>1$). The inner boundary of $V_\ell'$ (or, equivalently, of $U_\ell'$) is identified with the range of $z_\ell$ subject to $|z_\ell|=r$, so that $r$ is the corresponding radius of the boundary in this coordinate chart, see the second diagram in Fig.~\ref{fig:goodboundaryshrink}. In terms of this data the quantity $\hat{B}_k$ in (\ref{eq:Cellcontourint}) takes the form:
\begin{equation}\label{eq:Cellcontourint2}
\begin{aligned}
\hat{B}_k&=\frac{1}{2\pi}\int_{|z_\ell|>r}\rmd^2z_\ell\,\big(\mu_{k\bar{z}_\ell}^{\phantom{aa}z_\ell}b_{z_\ell z_\ell}+\mu_{kz_\ell}^{\phantom{aa}\bar{z}_\ell}\tilde{b}_{\bar{z}_\ell\bar{z}_\ell}\big)+\frac{1}{2\pi i}\oint_{|z_\ell|=r} \big(\rmd z_\ell\,v_{k\ell}^{z_\ell}b_{z_\ell z_\ell}-\rmd\bar{z}_\ell\,\bar{v}_{k\ell}^{\bar{z}_\ell}\tilde{b}_{\bar{z}_\ell\bar{z}_\ell}\big)\\
&\qquad +\sum_m\sum_{n\geq-1}\frac{1}{(n+1)!}\Big[\partial_{z_m}^{n+1}v_{km}^{z_m}(p_m)b_n^{(z_m)}(p_m)+\partial_{\bar{z}_m}^{n+1}\bar{v}_{km}^{\bar{z}_m}(p_m)\tilde{b}_n^{(\bar{z}_m)}(p_m)\Big]\\
\end{aligned}
\end{equation}
where taking into account that the orientation of the contour, $C_\ell$, is opposite to that induced by the $z_\ell$ chart coordinate the contour integral has opposite sign to that in (\ref{eq:Cellcontourint}). Applying identical reasoning here to that associated to the relations (\ref{eq:R+Boundary=fmn2}) and (\ref{eq:R+Boundary=fmn3}) we learn immediately that the quantity $\hat{B}_k$ is, crucially, {\it independent of $r$} (unless the integrand depends explicitly on $r$). This will allow us to take a limit $r\rightarrow 0$ in (\ref{eq:Cellcontourint2}) without affecting the left-hand side, $\hat{B}_k$. The second and third diagrams in Fig.~\ref{fig:goodboundaryshrink} denote respectively the surface before and after the limit $r\rightarrow 0$ has been taken.
\sk

To proceed further it will prove convenient to make one more assumption: that $v_{k\ell}^{z_\ell}$ is real analytic\footnote{Also here, DS would like to thank Simon Donaldson for a related comment.} within $|z_\ell|\leq r+\epsilon$, (with $|\epsilon|\ll1$) which in particular we will take to mean that it has a power series expansion of the form (\ref{sec:MV}), repeated here for convenience:
\begin{equation}\label{eq:realanalyticvkm2}
v_{k\ell}^{z_\ell}(z_\ell,\bar{z}_\ell)=\sum_{a,b=0}^{\infty}\frac{z_\ell^a\bar{z}_\ell^b}{a!b!}\partial^a_{z_\ell}\partial_{\bar{z}_\ell}^bv_{k\ell}^{z_\ell}(0,0).
\end{equation}
Writing $v_{k\ell}^{z_\ell}(0,0)\equiv v_{k\ell}^{z_\ell}(p_\ell)$, we then substitute this expansion and the Laurent expansions for the $b$-ghosts (\ref{eq:bzmexps}) and (\ref{eq:bzmexpstilde}) into (\ref{eq:Cellcontourint2}) and evaluate the contour integrals associated to $|z_\ell|=r$ explicitly, leading to:
\begin{equation}\label{eq:Cellcontourint3}
\begin{aligned}
\hat{B}_k&=\frac{1}{2\pi}\int_{|z_\ell|>r}\rmd^2z_\ell\,\big(\mu_{k\bar{z}_\ell}^{\phantom{aa}z_\ell}b_{z_\ell z_\ell}+\mu_{kz_\ell}^{\phantom{aa}\bar{z}_\ell}\tilde{b}_{\bar{z}_\ell\bar{z}_\ell}\big)+\\
&\qquad+\sum_{a,b=0}^\infty \frac{r^{2b}}{a!b!}\big(\partial^a_{z_\ell}\partial_{\bar{z}_\ell}^bv_{k\ell}^{z_\ell}(p_\ell)\big)b_{a-b-1}^{(z_\ell)}(p_\ell)+\sum_{a,b=0}^\infty \frac{r^{2b}}{a!b!}\big(\partial^b_{z_\ell}\partial_{\bar{z}_\ell}^a\bar{v}_{k\ell}^{\bar{z}_\ell}(p_\ell)\big)\tilde{b}_{a-b-1}^{(\bar{z}_\ell)}(p_\ell)\\
&\qquad +\sum_m\sum_{n\geq-1}\frac{1}{(n+1)!}\Big[\partial_{z_m}^{n+1}v_{km}^{z_m}(p_m)b_n^{(z_m)}(p_m)+\partial_{\bar{z}_m}^{n+1}\bar{v}_{km}^{\bar{z}_m}(p_m)\tilde{b}_n^{(\bar{z}_m)}(p_m)\Big]\\
\end{aligned}
\end{equation}

Since (\ref{eq:Cellcontourint2}) and hence also (\ref{eq:Cellcontourint3}) are independent of $r$ we will take the limit $r\rightarrow 0$ in the latter. The first integral goes over to the ordinary integral over the entire Riemann surface, $\Sigma$, whereas the terms in the sums over $a,b$ localise on $b=0$, with the remarkable implication that only holomorphic (and anti-holomorphic) terms in the power series expansion (\ref{eq:realanalyticvkm2}) (and it's anti-holomorphic counterpart) contribute to $\hat{B}_k$,
\begin{equation}\label{eq:Cellcontourint4a}
\begin{aligned}
\hat{B}_k&=\frac{1}{2\pi}\int_{\Sigma}\rmd^2z\,\big(\mu_{k\bar{z}}^{\phantom{aa}z}b_{z z}+\mu_{kz}^{\phantom{aa}\bar{z}}\tilde{b}_{\bar{z}\bar{z}}\big)+\\
&\qquad+\sum_{a=0}^\infty \frac{1}{a!}\big(\partial^a_{z_\ell}v_{k\ell}^{z_\ell}(p_\ell)\big)b_{a-1}^{(z_\ell)}(p_\ell)+\sum_{a=0}^\infty \frac{1}{a!}\big(\partial_{\bar{z}_\ell}^a\bar{v}_{k\ell}^{\bar{z}_\ell}(p_\ell)\big)\tilde{b}_{a-1}^{(\bar{z}_\ell)}(p_\ell)\\
&\qquad +\sum_m\sum_{n\geq-1}\frac{1}{(n+1)!}\Big[\partial_{z_m}^{n+1}v_{km}^{z_m}(p_m)b_n^{(z_m)}(p_m)+\partial_{\bar{z}_m}^{n+1}\bar{v}_{km}^{\bar{z}_m}(p_m)\tilde{b}_n^{(\bar{z}_m)}(p_m)\Big]\\
\end{aligned}
\end{equation}
But then the sums over $a$ are precisely of the same form as the contribution of a puncture (the last line in (\ref{eq:Cellcontourint4a})), so that the result of summing over $a$ can be absorbed into the latter and we obtain the result:
\begin{equation}\label{eq:Cellcontourint4}
\begin{aligned}
\hat{B}_k&=\frac{1}{2\pi}\int_{\Sigma}\rmd^2z\,\big(\mu_{k\bar{z}}^{\phantom{aa}z}b_{z z}+\mu_{kz}^{\phantom{aa}\bar{z}}\tilde{b}_{\bar{z}\bar{z}}\big)+\\
&\qquad +\sum_m\sum_{n\geq-1}\frac{1}{(n+1)!}\Big[\partial_{z_m}^{n+1}v_{km}^{z_m}(p_m)b_n^{(z_m)}(p_m)+\partial_{\bar{z}_m}^{n+1}\bar{v}_{km}^{\bar{z}_m}(p_m)\tilde{b}_n^{(\bar{z}_m)}(p_m)\Big]\\
\end{aligned}
\end{equation}

There is however one difference between the contribution of a puncture to $\hat{B}_k$ and that of a boundary whose radius is shrunk to zero, namely the derivation of the latter makes manifest that only holomorphic (or anti-holomorphic) contributions in $v_{km}^{z_m}$ (or $\bar{v}_{km}^{\bar{z}_m}$) contribute to $\hat{B}_k$ {\it when}: 
\begin{itemize}
\item[(a)] these local vector components, $v_{km}^{z_m}$ (or $\bar{v}_{km}^{\bar{z}_m}$), are taken to be real analytic as in (\ref{eq:realanalyticvkm2}) 
\item[(b)] we can cover the annulus with a {\it single} coordinate chart, $(U_\ell',z_\ell)$
\end{itemize}
Under these assumptions, the boundary contributions in (\ref{eq:Cellcontourint4}) clearly preserve complex structure (since holomorphic vectors preserve complex structure) and are therefore solely associated to conformal changes of frame as briefly mentioned in \cite{Polchinski88}. So making an explicit choice of conformal frame will also lead to a specific choice for these vectors, more about which momentarily.
\sk

It is hopefully clear that under assumptions (a) and (b) the derivation leading to (\ref{eq:Cellcontourint4}) holds for an arbitrary number of boundary components provided they (and any other features or insertions) are ``sufficiently far'' on the Riemann surface. Assumption (b) is actually quite subtle and it is certainly not obvious whether it is a good assumption: the Dolbeault-Grothendieck lemma (Appendix~\ref{sec:DGlemma}) only guarantees that Beltrami differentials take the form $\mu_{\bar{z}}^{\phantom{aa}z}=\partial_{\bar{z}}v^z$ within a domain that is diffeomorphic to a {\it disc}, whereas the above rather requires that it take this form in an {\it annular} region. Furthermore, covering an annulus with a single coordinate chart does {\it not} lead to a good cover, so in particular it is not obvious that the correct moduli space is captured -- recall the discussion below (\ref{eq:TM=H1}). This is nevertheless the standard approach \cite{Polchinski88,LaNelson90,Witten12c,Polchinski_v1}, and we will adopt it also, but it is a point that certainly deserves further study (since we are not aware of this having been elaborated on in any sufficient detail in the string theory literature\footnote{Recall the footnote on p.~\ref{foot:annulusphase}.}). The justification we adopt for proceeding in this manner is that the essential distinction between `an annulus with a single coordinate chart' and `an annulus covered by three charts' is effectively removed if we take the phase of the annulus chart coordinate to be integrated or averaged, or if we consider combinations of operators that are insensitive to this phase. We will always take this to be the case (for both trivial and non-trivial homology cycles), and this is related to the ambiguous U(1) that we discuss in numerous places throughout the document, see, e.g., the discussion following (\ref{eq:Ipi=chi}).
\sk

We will now finally make a special choice of frame, namely that introduced by Polchinski \cite{Polchinski88}, whereby the corresponding metric is taken to be ``as flat as possible'' at the location of a puncture. This is equivalent to adopting holomorphic normal coordinates, see Sec.~\ref{sec:HNC}. To implement this, all that is needed (since the notation has been kept consistent) is to substitute the results of Sec.~\ref{sec:MV} into (\ref{eq:Cellcontourint4}). So let us gather these results (\ref{eq:v(p)=0}), (\ref{eq:dv=nu+gamma_k}) and (\ref{eq:dn+1v=vnnu}), repeated here for convenience,
\begin{equation}\label{eq:Polchinskivs}
\begin{aligned}
v_{km}^{z_m}(p_m)&=v_{km}^{\bar{z}_m}(p_m)=0\\
\partial_{z_m}v_{km}^{z_m}(p_m)&=\frac{1}{2}\big(\nu_k(p_m) +i\gamma_k(p_m)\big)\\
\partial_{z_m}^{n+1}v_{km}^{z_m}(p_m)&=\partial_{z_m}^n\nu_k(p_m)+\dots,\qquad n=1,2,\dots
\end{aligned}
\end{equation}
and since the dots `\dots' in the last relation denote contributions that do not contain purely holomorphic terms they drop out (according to the discussion below (\ref{eq:realanalyticvkm2})). So $\hat{B}_k$ takes the form:
\begin{equation}\label{eq:Cellcontourint5nu}
\boxed{
\begin{aligned}
\hat{B}_k
&=\frac{1}{2\pi}\int_{\Sigma}\rmd^2z\,\big(\mu_{k\bar{z}}^{\phantom{aa}z}b_{z z}+\mu_{kz}^{\phantom{aa}\bar{z}}\tilde{b}_{\bar{z}\bar{z}}\big)+\\
&\qquad +\sum_m\frac{1}{2}\Big[\nu_k(p_m) \big(b_0^{(z_m)}+\tilde{b}_0^{(\bar{z}_m)}\big)(p_m)+i\gamma_k(p_m)\big(b_0^{(z_m)}-\tilde{b}_0^{(\bar{z}_m)}\big)(p_m)\Big]\\
&\qquad +\sum_m\sum_{n\geq1}\frac{1}{(n+1)!}\Big[\partial_{z_m}^n\nu_k(p_m)b_n^{(z_m)}(p_m)+\partial_{\bar{z}_m}^n\nu_k(p_m)\tilde{b}_n^{(\bar{z}_m)}(p_m)\Big]
\end{aligned}
}
\end{equation}

This completes the derivation of eqn (26) in \cite{Polchinski88}, thus furnishing the link between Polchinski's approach of using a metric to define a conformal frame and the corresponding expression in terms of transition functions in the second equality in (\ref{eq:Bintmub8z}). Note that the $\gamma_k$ term in (\ref{eq:Cellcontourint5nu}) is absent in \cite{Polchinski88} since eqn (26) there is applied to cutting internal cycles (separating or non-separating degenerations) whereby the phase of the coordinate of the annulus is either identified with a modulus (and is hence integrated over) or combinations of operators are considered that are independent of this phase -- this is related to comment (b) above. (Note that offshell and onshell external states are required to be annihilated by $L_0-\tilde{L}_0$ and $b_0-\tilde{b}_0$, but this need not be the case for offshell states associated to factorisation.\footnote{Incidentally, although as discussed in Sec.~\ref{sec:EVO} offshell {\it external} states may always be taken to be annihilated by $L_0-\tilde{L}_0$ and $b_0-\tilde{b}_0$, for internal states the first quantised approach is actually different from string field theory \cite{deLacroixErbinKashyapSenVerma17} in this respect: non-level matched internal states (which are required to be present by modular invariance) are actually absorbed into SFT vertices, whereas in the first quantised formalism they are part of the offshell Hilbert space. We thank Edward Witten and Ashoke Sen for related remarks.}). 
\sk

The virtue of phrasing $\hat{B}_k$ in terms of $\mu_k$ and $\nu_k$ as in (\ref{eq:Cellcontourint5nu}) will become manifest in Sec.~\ref{sec:TP} where we use it to provide the map between fixed- and integrated-picture vertex operators. Briefly, we will see that this then leads to integrated-picture vertex operators that are written directly in terms of covariant quantities (e.g., covariant derivatives of the curvature tensor), so that although a special choice of metric was made here the result is independent of this choice.  (This is analogous to the use of Riemann normal coordinates in the case of real manifolds, where a special coordinate system is chosen at an intermediate step which allows one to express various quantities in terms of fully covariant objects.) 
\sk

There is an entirely analogous expression to (\ref{eq:Cellcontourint5nu}) for the total energy-momentum tensor. Comparing (\ref{eq:Tintmub8z}) to the expression for $\hat{B}_k$ given in (\ref{eq:Bintmub8z}), and then making use of (\ref{eq:Cellcontourint5nu}), we learn that:
\begin{equation}\label{eq:TCellcontourint5nu}
\boxed{
\begin{aligned}
\hat{D}_k
&=\frac{1}{2\pi}\int_{\Sigma}\rmd^2z\,\big(\mu_{k\bar{z}}^{\phantom{aa}z}T_{z z}+\mu_{kz}^{\phantom{aa}\bar{z}}\tilde{T}_{\bar{z}\bar{z}}\big)+\\
&\qquad +\sum_m\frac{1}{2}\Big[\nu_k(p_m) \big(L_0^{(z_m)}+\tilde{L}_0^{(\bar{z}_m)}\big)(p_m)+i\gamma_k(p_m)\big(L_0^{(z_m)}-\tilde{L}_0^{(\bar{z}_m)}\big)(p_m)\Big]\\
&\qquad +\sum_m\sum_{n\geq1}\frac{1}{(n+1)!}\Big[\partial_{z_m}^n\nu_k(p_m)L_n^{(z_m)}(p_m)+\partial_{\bar{z}_m}^n\nu_k(p_m)\tilde{L}_n^{(\bar{z}_m)}(p_m)\Big]
\end{aligned}
}
\end{equation}
(Note that since we are using holomorphic normal coordinates we can equivalently replace $\partial_{z_m}^n$ by covariant derivatives, $\nabla_{z_m}^n$ since the result is evaluated at $p_m$.)
As mentioned above, this quantity (\ref{eq:TCellcontourint5nu}) will appear when establishing that BRST-exact states decouple from string amplitudes. In particular, the quantities (\ref{eq:TCellcontourint5nu}) and (\ref{eq:Cellcontourint5nu}) are related by:
\begin{equation}\label{eq:D=QB}
\boxed{
\hat{D}_k
=\big\{Q_B,\hat{B}_k\big\}
}
\end{equation}
where $Q_B$ is the BRST charge (note that the BRST current (\ref{eq:jB}) transforms as a {\it tensor} on patch overlaps). The relation (\ref{eq:D=QB}) may be established using (\ref{eq:QbLn commutators}) after writing the integrals over $\Sigma$ as a sum of integrals over cells, then using the Dolbeault-Grothendieck lemma to rewrite the Beltrami differential in terms of locally-defined vectors, and finally writing the resulting integrals as contour integrals using Greens theorem.
\sk

As already mentioned, when applied to translations (\ref{eq:TCellcontourint5nu}) provides the relation between the derivative of a normal-ordered quantity and the normal-ordered derivative of that quantity (since, unlike in dimensional regularisation, conformal normal ordering does not commute with derivatives when curvature is encoded in local information). We explore this in detail in Sec.~\ref{sec:comm dA-dA}. 

\subsection{Pinching \& Twisting Handles with Transition Functions}\label{sec:PH}
We will use plumbing fixture \cite{Vafa87,Polchinski_v1} to study a degeneration of the form (\ref{eq:completeness}). Let us select a cycle in $\Sigma$ that is associated to a separating degeneration, and (keeping moduli fixed) cut across this cycle so as to produce two Riemann surfaces, $\Sigma_1$ and $\Sigma_2$, each with a single boundary, see Fig.~\ref{fig:SigmaCut2}. 
\begin{figure}
\begin{center}
\includegraphics[angle=0,origin=c,width=0.85\textwidth]{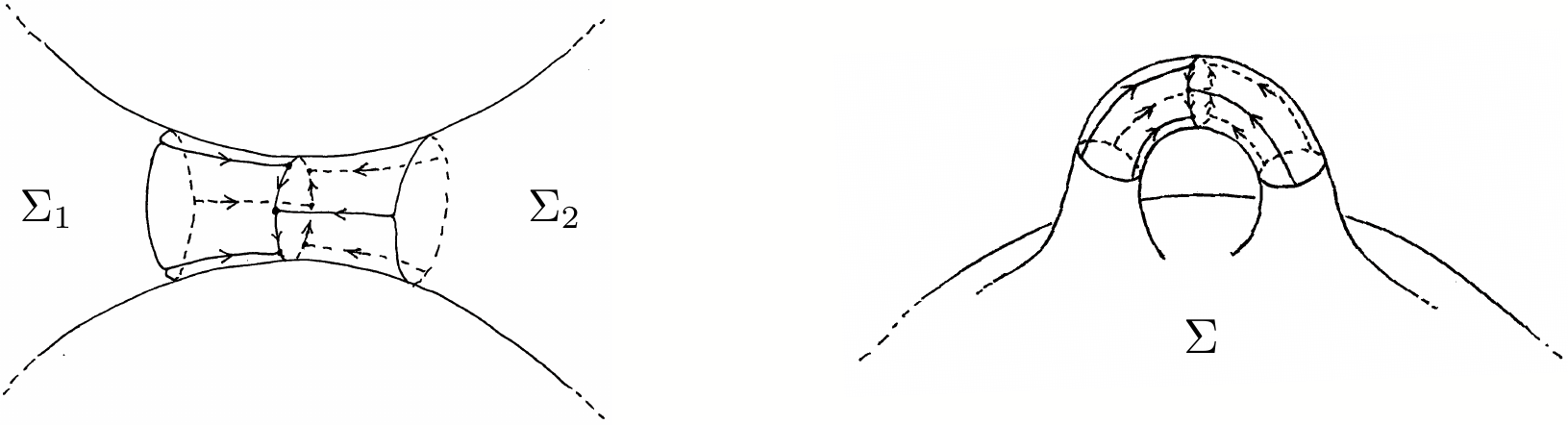}
\caption{Separating and non-separating degenerations of Riemann surfaces. The vicinity of the pinch regions are depicted as well as the corresponding cell decompositions (recall Fig.~\ref{fig:tripleoverlapsUV}) associated to a `good cover'. The Riemann surface is cut along the closed cycle, which is the cycle bounded by two annuli on either of the two sides of the pinch. Each of the two annuli, also depicted in Fig.~\ref{fig:goodannuliV}, are in turn covered by three overlapping open sets whose cell decomposition (rather than the corresponding open cover) is shown. Notice that only 3-point vertices appear since we are working with a `dual triangulation' (recall Fig.~\ref{fig:dualtriangulation},\ref{fig:genericdualtriangles}).}\label{fig:sep-nonsep}
\end{center}
\end{figure}
\begin{figure}
\begin{center}
\includegraphics[angle=0,origin=c,width=0.65\textwidth]{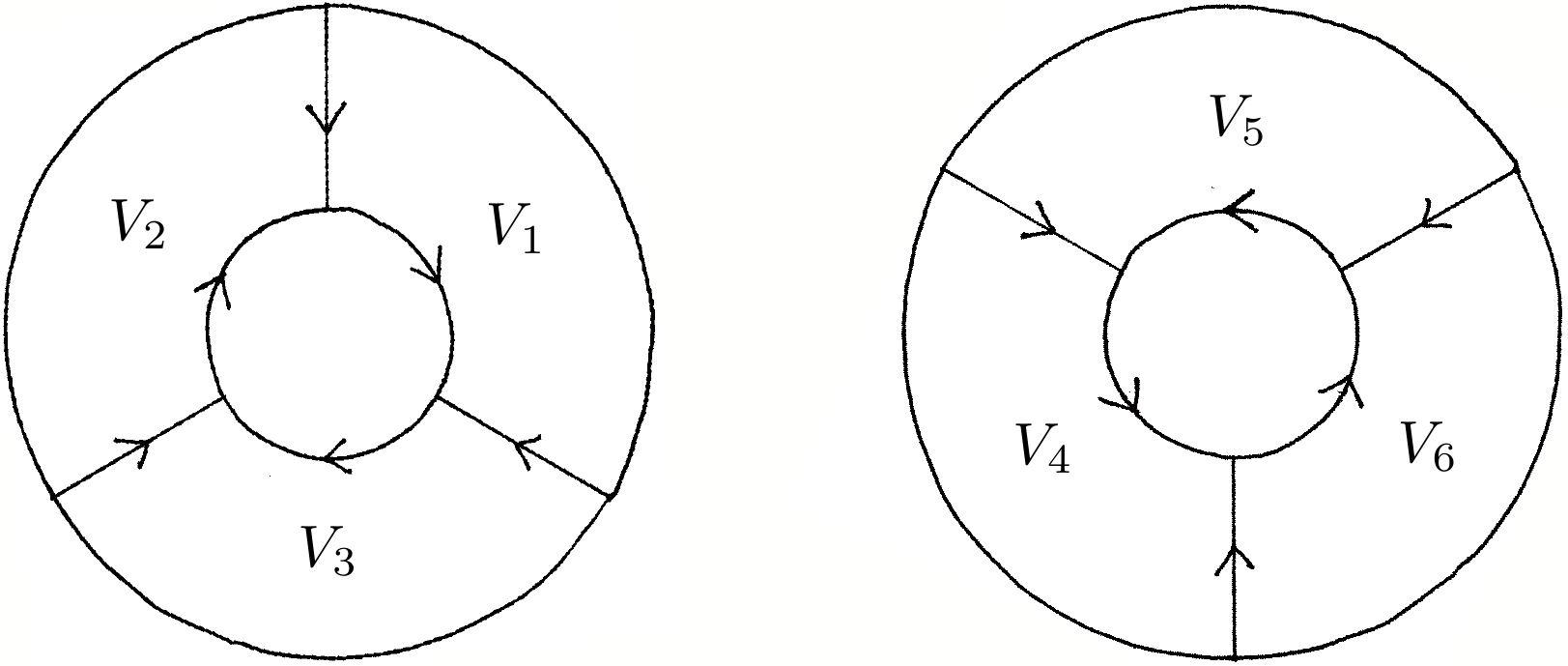}
\caption{Two annuli associated to either a separating or a non-separating degeneration of a Riemann surface, see Fig.~\ref{fig:sep-nonsep}. The inner boundaries of the two annuli are identified with the central cycle shown in either the first or second diagrams in Fig.~\ref{fig:sep-nonsep}. The three straight segments in the first diagram (starting from the top one and continuing counterclockwise) are identified with the contours $C_{12}$, $C_{23}$ and $C_{31}$, whereas those in the second diagram (starting from the top left one and continuing counterclockwise) are identified with $C_{56}$, $C_{46}$ and $C_{65}$. The {\rm six} arc segments along the inner boundary in the first diagram starting from the top left one and continuing counterclockwise are $C_{25}$, $C_{26}$, $C_{36}$, $C_{34}$, $C_{14}$ and $C_{15}$. That there are six may be understood by either inspecting Fig.~\ref{fig:sep-nonsep}, or by inverting the second diagram here (via the corresponding transition functions) and gluing it to the first along the inner boundary of the latter (as shown in Fig.~\ref{fig:goodannuliX}).}\label{fig:goodannuliV}
\end{center}
\end{figure}
\begin{figure}
\begin{center}
\includegraphics[angle=0,origin=c,width=0.85\textwidth]{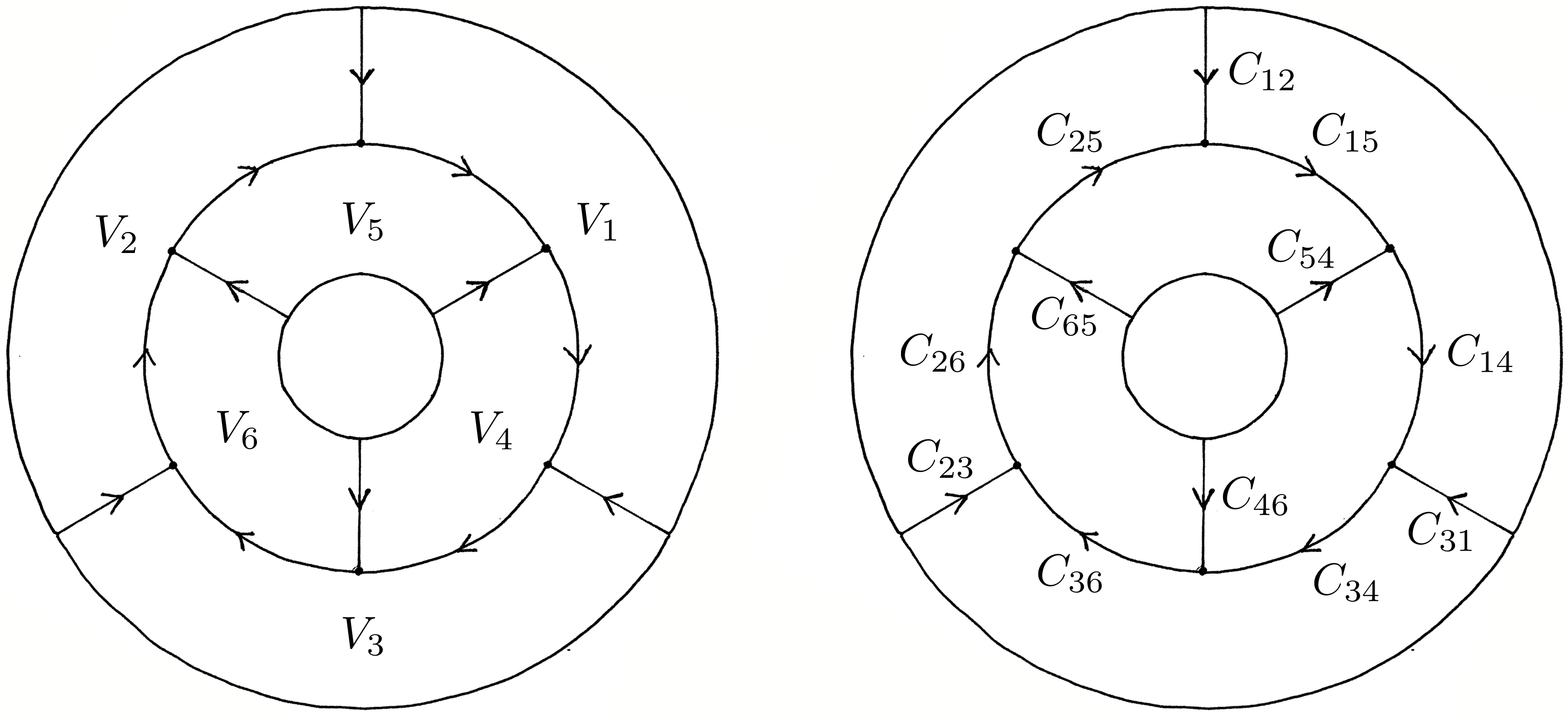}
\caption{Part of a cell decomposition of a good cover (of a dual triangulation) of an annulus (left) and the corresponding contour notation for moduli integrals (right). These six cells correspond to those in Fig.~\ref{fig:sep-nonsep} and Fig.~\ref{fig:goodannuliV}. In the main text, and modulo the caveats spelt out in Sec.\ref{sec:CIoM}, we use the simpler setup of retaining a single chart for every annulus.}\label{fig:goodannuliX}
\end{center}
\end{figure}
In order to reconstruct $\Sigma$ given $\Sigma_1$ and $\Sigma_2$ (keeping moduli fixed) we may endow $\Sigma_1$ with a local chart $(U_1,z_1)$ such that the region $|z_1|<(1-\epsilon)|q|^{1/2}$ is absent due to the aforementioned cutting procedure, where $\epsilon$ is a small parameter and $q$ a complex number such that $|q|^{1/2}$ roughly corresponds to the `narrowness of the neck' at the cut. Similarly, we endow $\Sigma_2$ with local chart $(U_2,z_2)$ such that the region $|z_2|<(1-\epsilon)|q|^{1/2}$ is again absent due to the aforementioned cut. We can now reconstruct $\Sigma$ by identifying pairs of points $z_1,z_2$ (on $\Sigma_1,\Sigma_2$ respectively) if on $U_1\cap U_2$ they are related by the following transition function:
\begin{equation}\label{eq:z1z2=q}
\boxed{z_1z_2=q,\qquad {\rm with}\qquad z_1=f_{12}(z_2,q)}
\end{equation}
So $q,\bar{q}$ correspond to coordinates in moduli space associated to this specific degeneration. 
The remaining transition functions associated to all other patch overlaps will be taken to be independent of $q,\bar{q}$. This corresponds to freedom in picking a slice in moduli space. However, this construction does depend on the choice of coordinates; e.g., from (\ref{eq:z1z2=q}) it follows that we can absorb $q$ into a redefinition of $z_1,z_2$, by defining $z_1'=z_1q^{-1/2}$ and $z_2'=z_2q^{-1/2}$, so that $q=1$ gives an equivalent surface. (In (\ref{eq:completeness}) we made the choice $q=1$ which corresponds to the clearest starting point, but taking $q<1$ will eventually be convenient to discuss degeneration with fixed coordinate patches). So one must check that the corresponding amplitudes do not depend on such a choice of coordinates. 
\sk

The annular region (to order $\mathcal{O}(\epsilon)$):
\begin{equation}\label{eq:sewingregion}
U_1\cap U_2=\left\{z_1,z_2\,\bigg|\,
\begin{aligned}
&(1-\epsilon)|q|^{1/2}<|z_1|<(1+\epsilon)|q|^{1/2}\\
&(1-\epsilon)|q|^{1/2}<|z_2|<(1+\epsilon)|q|^{1/2}
\end{aligned}
\right\}
\end{equation}
corresponds to the sewn region. In the $q\rightarrow 0$ limit the handle pinches, and for non-zero $q$ we have smooth interpolation between metrics $\rho_1(z_1,\bar{z}_1)\rmd z_1\rmd\bar{z}_1$ and $\rho_2(z_2,\bar{z}_2)\rmd z_2\rmd\bar{z}_2$ in the overlap region, in particular on $U_1\cap U_2$ where: $\rho_1(z_1,\bar{z}_1)=\rho_2(z_2,\bar{z}_2)|z_2|^4/|q|^2$. 
\sk

We wish to write down the corresponding path integral measure contributions for this degeneration. The general formula for every modulus $t^k$ (which here we identify with $q,\bar{q}$) was given in (\ref{eq:Bintmub8z}), and given the transition function (\ref{eq:z1z2=q}) it is clearly convenient to use the explicit form for $\hat{B}_k$ given in the {\it second} equality in (\ref{eq:Bintmub8z}). We may in particular associate the sewing region (\ref{eq:sewingregion}) to the region traversed by a contour $C_{12}$ in (\ref{eq:Bintmub8z}) which we may in turn identify with the dotted line in Fig.~\ref{fig:SigmaCut2} (with the orientation induced by that of $\Sigma_1$).  So for this degeneration we learn that, taking into account the transition function (\ref{eq:z1z2=q}), and that the remaining transition functions are independent of $q$ (so that only the pair $(mn)=(1,2)$ survives in the sum over pairs),
\begin{equation}\label{eq:Bintpinchq}
\begin{aligned}
\hat{B}_q
&=\frac{1}{2\pi i}\int_{C_{12}} \bigg(\rmd z_1\,\frac{\partial f_{12}(z_2,q)}{\partial q}\big|_{z_2}b_{z_1z_1}-\rmd\bar{z}_1\,\frac{\partial \overline{f_{12}(z_2,q)}}{\partial q}\big|_{\bar{z}_2}\tilde{b}_{\bar{z}_1\bar{z}_1}\bigg)\\
&=-\frac{b_0^{(z_1)}(p_1)}{q}=-\frac{b_0^{(z_2)}(p_2)}{q}
\end{aligned}
\end{equation}
Similarly, for the anti-chiral half we have, $\hat{B}_{\bar{q}}=-\tilde{b}_0^{(z_1)}(p_1)/\bar{q}=-\tilde{b}_0^{(z_2)}(p_2)/\bar{q}$. 
\sk

So, including the integration measure, the total ghost insertion associated to pinching and twisting a separating or non-separating degeneration is given by:
\begin{equation}\label{eq:BqBqbar}
\boxed{
\int \rmd^2q\hat{B}_q\hat{B}_{\bar{q}}=\int \rmd^2q\frac{b_0^{(z_1)}\tilde{b}_0^{(z_1)}(p_1)}{q\bar{q}}=\int \rmd^2q\frac{b_0^{(z_2)}\tilde{b}_0^{(z_2)}(p_2)}{q\bar{q}},\qquad \rmd^2q\equiv idq\wedge \rmd\bar{q}
}
\end{equation}
This is an operator equation, and so holds inside the path integral. Also, the operand will generically be $q,\bar{q}$-dependent as follows from the gluing relation (\ref{eq:z1z2=q}).  An important point is that the $b_0,\tilde{b}_0$ operators appearing (\ref{eq:BqBqbar}) are based either in the $z_1$ or $z_2$ frame (we are free to choose which one since in fact $b_0^{(z_1)}=b_0^{(z_2)}\equiv b_0$, as follows from (\ref{eq:gluingmodes=0bosonic})). These modes encircle the operator that is inserted at the origin of either of these two frames (with any remaining operators outside of the contour). 
\sk

Since we do not always need to pick a gauge slice in moduli space where twisting and pinching a handle are both regarded as moduli deformations, let us also display the above ghost insertions in terms of real coordinates, $r,\theta$. We write $q=re^{i\theta}$ and $\bar{q}=re^{-i\theta}$, and the transition function is still given by (\ref{eq:z1z2=q}) with this identification, so the corresponding insertions take the form:
\begin{equation}\label{eq:Bintpinchr}
\begin{aligned}
\hat{B}_r
&=\frac{1}{2\pi i}\int_{C_{12}} \bigg(\rmd z_1\,\frac{\partial f_{12}(z_2,q)}{\partial r}\big|_{z_2}b_{z_1z_1}-\rmd\bar{z}_1\,\frac{\partial \overline{f_{12}(z_2,q)}}{\partial r}\big|_{\bar{z}_2}\tilde{b}_{\bar{z}_1\bar{z}_1}\bigg)\\
&=-\frac{1}{r}\big(b_0^{(z_1)}+\tilde{b}_0^{(z_1)}\big)(p_1)=-\frac{1}{r}\big(b_0^{(z_2)}+\tilde{b}_0^{(z_2)}\big)(p_2)\\
\end{aligned}
\end{equation}
and,
\begin{equation}\label{eq:Bintpinchtheta}
\begin{aligned}
\hat{B}_\theta
&=\frac{1}{2\pi i}\int_{C_{12}} \bigg(\rmd z_1\,\frac{\partial f_{12}(z_2,q)}{\partial \theta}\big|_{z_2}b_{z_1z_1}-\rmd\bar{z}_1\,\frac{\partial \overline{f_{12}(z_2,q)}}{\partial \theta}\big|_{\bar{z}_2}\tilde{b}_{\bar{z}_1\bar{z}_1}\bigg)\\
&=-i\big(b_0^{(z_1)}-\tilde{b}_0^{(z_1)}\big)(p_1)=-i\big(b_0^{(z_2)}-\tilde{b}_0^{(z_2)}\big)(p_2),
\end{aligned}
\end{equation}
in terms of which the full insertion analogous to (\ref{eq:Bintpinchq}) would be:
$$
\int i\rmd r\wedge \rmd\theta\,\hat{B}_r\hat{B}_\theta = \int (2r\rmd r\wedge \rmd\theta)\frac{b_0\tilde{b}_0}{r^2},
$$
which is precisely equal to (\ref{eq:BqBqbar}). (The curious factor of $i$ on the left-hand side appears in the change of variables because the $\hat{B}_r,\hat{B}_\theta$ anticommute.) So we can also consider gauge slices where the insertion: 
$$
i\int \rmd\theta\hat{B}_\theta=\int \rmd\theta\,\big(b_0^{(z_1)}-\tilde{b}_0^{(z_1)}\big)(p_1)=\int \rmd\theta\,\big(b_0^{(z_2)}-\tilde{b}_0^{(z_2)}\big)(p_2),
$$ 
appears when we cut open a given handle and not $\int \rmd r\hat{B}_r$. Also, rotating $z_1$ (or $z_2$) by a phase, $z_2\rightarrow z_2e^{-i\theta}$, or rescaling, $z_2\rightarrow z_2/r$, leaves $b_0,\tilde{b}_0$ invariant, 
$$
b_0^{(z_1/q)} =b_0^{(z_1)}=b_0^{(z_2/q)} =b_0^{(z_2)},\qquad {\rm for}\qquad q=re^{i\theta}.
$$

Let us add some comments on the range of integration over $q,\bar{q}$ or $r,\theta$. If the handle we cut open is associated to a {\it trivial-homology cycle} we can take\footnote{DPS thanks Ashoke Sen for a related comment here.},
\begin{equation}\label{eq:qqbar-rangeZ}
\theta\in [0,2\pi),\qquad {\rm and}\qquad r\in(0,\Lambda],
\end{equation}
where $\Lambda=\max(|z_1z_2|) $ is chosen such that the size of the pinch increases up to the closest other local (or smeared out) operator insertion (which might also be associated to a modulus). This corresponds also to the radius of convergence of the OPE. Using furthermore the fact that the local operators of the given handle operator will have well-defined scaling dimension (as we will show explicitly in Sec.~\ref{sec:GOCS}) it follows from the gluing relation, $z_1z_2=q$, that we can always absorb $\Lambda$ into a rescaling of the frame coordinates, $z_1$ and/or $z_2$, so that the range (\ref{eq:qqbar-rangeZ}) can always be replaced by:
\begin{equation}\label{eq:qqbar-range}
\theta\in [0,2\pi),\qquad {\rm and}\qquad r\in(0,1],\qquad\textrm{(separating degeneration)}
\end{equation}
We will consider an explicit example in Sec.~\ref{sec:VSG}, the relevant equations there being (\ref{eq:zu=q}), (\ref{eq:|q| range VS}), (\ref{eq:zz'uu'qq'}) and also (\ref{eq:FeynmannProp}). This simple result is possible because modular transformations leave trivial homology cycles invariant.
\sk

Cutting along a {\it non-trivial homology cycle} the range of integration of the pinch and twist moduli requires more work. Modular invariance is mostly spontaneously broken if we choose to insert $\g$ $A_I$-cycle (integrated-picture) handle operators (with $I=1,\dots,\g$) on a genus-$\g$ Riemann surface. This is analogous to the D'Hoker and Phong approach\footnote{A review and extension to arbitrary external states and constant fixed toroidally-compactified backgrounds is contained in \cite{SklirosCopelandSaffin17}.} \cite{D'HokerPhong89} (and Dijkgraaf, E.~Verlinde and H.~Verlinde \cite{DijkgraafVerlindeVerlinde88} and also \cite{VerlindeVerlinde87}) of fixing the loop momenta in amplitudes of arbitrary genus (supplemented by the Belavin and Knizhnik  theorem \cite{BelavinKnizhnik86} for the chiral splitting of the ghost determinants). This certainly leaves the subgroup generated by $\theta\rightarrow \theta+2\pi$ manifest (so that we should integrate over $\theta\in [0,2\pi)$ to avoid overcounting), and clearly there is a remaining discrete symmetry associated to interchange of any two handle operators, leading to an overall factor of $1/\g!$. Since the local operators out of which the handle operator is constructed will not be Hermitian (after rotating to Lorentzian signature) there should be no multiplicative factor of $1/2^\g$. 
\sk

It is important to unravel the precise moduli space over which we should integrate. In general this is subtle, but we will show by explicit computation in Sec.\ref{sec:MI} that at one loop the resulting integrand with one handle operator insertion is indeed modular invariant, so that to avoid overcounting one should restrict to a fundamental domain of SL(2,$\mathbf{Z})/\mathbf{Z}_2$. The precise domain of integration at higher loops is also essentially determined by modular invariance, but the result is more complicated because the domain of integration of any one handle operator moduli will depend on where in moduli space remaining handle operators (and vertex operators) are. It will presumably be convenient to decouple these domains of integration by an appropriate unfolding of moduli space,\footnote{DS thanks Juan Maldacena for a suggestion along these lines.} but to even begin to unravel this one needs to primarily derive how the mapping class group \cite{Bers81,FarbMargalit12} acts on amplitudes with handle operator insertions.

\subsection{Shifting Handles using a Metric}\label{sec:TP}
Recall that we derived two equivalent expressions for the ghost contributions to the path integral measure, see the two equalities in (\ref{eq:Bintmub8z}) on p.~\pageref{eq:Bintmub8z}. The first of these takes a metric viewpoint whereas the second relies solely on holomorphic transition functions and cocycle relations, namely on the defining properties of a Riemann surface. We here consider the first of these approaches and derive the explicit expression for the particular case of moduli associated to translating vertex operators across a Riemann surface.  That is, we will work out the map from fixed-picture vertex operators to integrated picture with emphasis on obtaining a globally (on $\Sigma$ and $\mathcal{M}$) well-defined expression. In Sec.~\ref{sec:SHUTF} we derive the same result, but rather than using a metric we will work directly in terms of transition functions. (The latter approach is much more efficient, and the reader can also skip directly to Sec.~\ref{sec:SHUTF}.)
\sk

In particular, in this section we will use holomorphic normal coordinates to fix invariance under holomorphic reparametrisations (which include Weyl transformations), and  evaluate the path integral measure contribution associated  to picking a slice in moduli space that associates `translation moduli' to handle operators. An advantage (compared to the choice of Nelson \cite{Nelson89}) is that the resulting expression will be {\it explicitly} covariant and globally well-defined in moduli space (modulo U(1)), while also preserving explicit reparametrisation invariance\footnote{To be precise, reparametrisation invariance will be more explicit in Sec.~\ref{sec:SHUTF}, but the results of the current section and Sec.~\ref{sec:SHUTF} are entirely equivalent.} (so that it also remains valid in any coordinate system). In contrast, the holomorphic slice chosen in \cite{Nelson89} is not globally defined, even modulo U(1), so that one needs in the latter case to add Wu-Yang boundary contributions from patch overlaps to get a global description. 
\sk

Our starting point here will be the expression for $\hat{B}_k$ given in (\ref{eq:Cellcontourint5nu}) (which, recall, was derived from (\ref{eq:Bintmub8z})).  Detailed discussion along similar lines can be found in \cite{Witten12c}, but here we take into account some additional boundary terms that will enable us in addition to discuss: (1) the case where the vertex operators are general BRST-invariant operators (rather than only primaries as in \cite{Witten12c}); (2) general offshell vertex operators arising from cutting open the path integral across specific cycles of $\Sigma$ and using the operator-state correspondence (in which case the analysis of \cite{Witten12c} only applies to cycles for which only the imaginary part is considered, in the Cutkosky sense\footnote{This is because it is only physical states that contribute to the imaginary part of a cut handle, whereas the real part also propagates offshell and hence unphysical degrees of freedom.}). So our analysis here is much closer to that of Polchinski \cite{Polchinski88}. 
\sk

In Sec.~\ref{sec:CSDII} we considered general infinitesimal deformations of complex structure from a metric viewpoint. Following on from that discussion, let us suppose we consider a compact Riemann surface, $\Sigma$, and a corresponding cover, $\{U_m^I\}$, (associated to some complex structure, $I$) such that $\cup_m U_m^I=\Sigma$, with non-trivial double and triple overlaps as shown in Fig.~\ref{fig:genericdualtriangles} (see also Fig.~\ref{fig:23overlaps}). Every open set $U_m^I$ is homeomorphic to a disc with at most a puncture inserted. Now to every such cover we can associate an atlas, $\mathscr{U}_I=\{(U_m^I,z_m)\}$, such that $z_m:U_m^I\rightarrow \mathbf{C}$ provides a holomorphic coordinate for the region $U_m^I$, with corresponding holomorphic transition functions on patch overlaps and cocycle relations as discussed in Sec.~\ref{sec:RS}. Let us be a bit more explicit,
\begin{equation}\label{eq:U_I-trans}
\mathscr{U}_I=\{(U_m^I,z_m)\} = \{(U_1^I,z_1),(U_2^I,z_2),(U_3^I,z_3),\dots\}.
\end{equation}
We then single out a specific chart, $(U_1^I,z_1)$, and arbitrarily assign to a point, $p_1\in U_1^I$, the origin of the coordinate system, $z_1(p_1)=0$. 
\sk

\begin{figure}
\begin{center}
\includegraphics[angle=0,origin=c,width=0.75\textwidth]{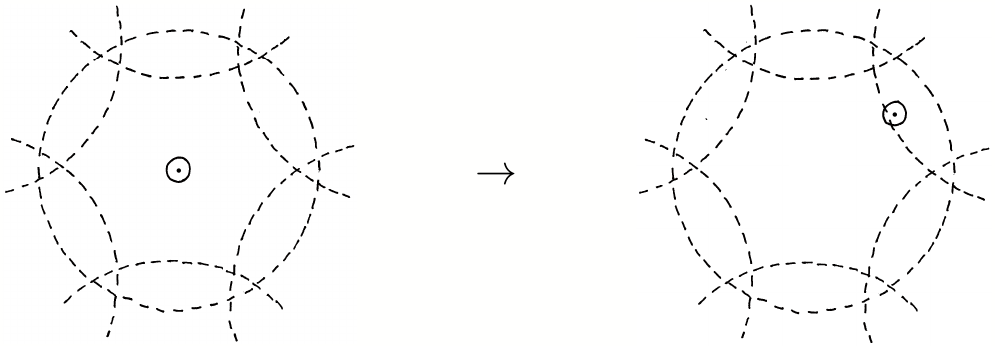}
\caption{A diagram of a complex structure deformation, $I\rightarrow J=I+\delta I$, that leaves the entire Riemann surface fixed except that a point $p_1\in U_1^I$ (with $z_1(p_1)=0$) (left diagram) is translated to a point $p_1\in U_1^J$ (with $w_1(p_1)=z_v$) (right diagram). The point $p_1\in U_1^J$ (as viewed from the initial complex structure, $I$) corresponds to a point $p_1'\in U_1^I$, such that $z_1(p_1')=z_v$. We emphasise that we translate the physical point $p_1$, so that it makes sense to say that after translation the point $p_1$ is in the patch of the new complex structure, $U_1^J$.}\label{fig:opensetstranslate}
\end{center}
\end{figure}
Let us suppose now that there is a puncture at $p_1$, i.e.~a hole in $\Sigma$ centred at $p_1$ whose radius as measured in the $z_1$ coordinate, $|z_1|=r$, is sent to zero.  To be precise,
\begin{quote}
\textit{We are free to (and will) send $r\rightarrow 0$ in terms that are independent of $r$ in the path integral measure. This allows us to treat external punctures and ``punctures'' arising from cutting open the path integral along an internal cycle of the Riemann surface on an equal footing.\footnote{This is so even though properties of external vertex operators and those arising from cutting handles will be different in general.} So in fact $r$ might even correspond to a {\it modulus} of our Riemann surface associated to pinching a cycle.}
\end{quote}
Clearly, it is perfectly consistent to send $r\rightarrow0$ in terms independent of $r$ and this is what we shall do. The relevant point is that we pick a gauge slice in moduli space such that the measure contributions, $\hat{B}_k$, associated to {\it translating} a pinched handle across the surface (as opposed to measure contributions associated to {\it pinching} or {\it twisting} a handle) are independent of $r$ -- recall the discussion in Sec.~\ref{sec:IToaM} -- so we may (and will) in fact take the limit $r\rightarrow 0$ in {\it these} particular terms that are associated to translation moduli (but not for pinch or twist moduli). In fact, these particular terms are precisely the relevant terms needed to implement the operator-state correspondence for vertex operators in integrated picture. This allows us to speak about ``local (or bi-local) operator insertions'', even though these insertions might arise from state(s) associated to cutting open a cycle of a Riemann surface whose radius corresponds to a modulus.\footnote{In case the reader has a hard time digesting the implications of this observation (or the underlying reason as to why it is true) they are advised to read on since we will try to clarify it in great detail. The relevant clue was provided in Sec.~\ref{sec:IToaM} where the $r$-independence was discussed. Note that such $r$-independence is only present when the boundary terms mentioned there are included. It is because these boundary terms were omitted in Witten's analysis \cite{Witten12c} (see equation (2.58) there) that the analysis there only applies to asymptotic primary vertex operators, or to vertex operators associated to pinches {\it when} only the imaginary part (in the Cutkosky sense \cite{Cutkosky60,PiusSen16}) of the cut cycle is considered.\label{foot:WittenNoBoundaryTerms}}
\sk

Furthermore, as far as the ghost contributions, $\hat{B}_k$, associated to translation moduli in the path integral measure are concerned, whether there is a puncture at $p_1$ or not is specified by whether there is (using the operator-state correspondence) a local vertex operator or the vacuum (i.e.~the unit operator) at $p_1$. In this section we are interested in the case where there is a local vertex operator inserted at $p_1\in U_1^I$ and when there is no other operator insertion within $U_1^I$ -- we want to understand how to translate this point $p_1$ across the surface. In particular, we want to associate to the motion of the puncture, $p_1$, a complex structure deformation, $I\rightarrow J=I+\delta I$. On general grounds then, in the $J$ complex structure there will by definition be a corresponding holomorphic atlas,
\begin{equation}\label{eq:U_J-trans}
\mathscr{U}_J=\{(U_m^J,w_m)\} = \{(U_1^J,w_1),(U_2^J,w_2),(U_3^J,w_3),\dots\}.
\end{equation}
where every $w_m$ is a holomorphic coordinate with respect to the $J$ complex structure, as was every $z_m$ a holomorphic coordinate with respect to the $I$ complex structure in (\ref{eq:U_I-trans}). 
\sk

This is of course entirely general, in that it holds for {\it any} small deformation, $I\rightarrow I+\delta I$. Now we wish to pick a gauge slice in moduli space where $I\rightarrow I+\delta I$ leaves the {\it entire} Riemann surface invariant {\it except} that the point $p_1$ is translated to a new point $p_1'\in U_1^I$. We can accomplish this by choosing the atlas, $\mathscr{U}_J$, in the $J$ complex structure as follows,
\begin{figure}
\begin{center}
\includegraphics[angle=0,origin=c,width=0.45\textwidth]{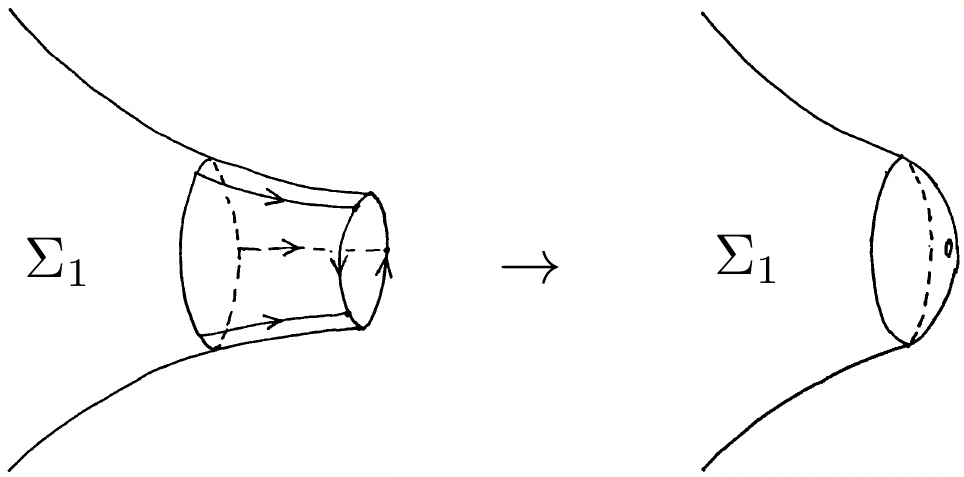}
\caption{Cutting open a Riemann surface (and a corresponding worldsheet path integral) across a cycle introduces two boundaries, one of which is shown in the first diagram (see also Fig.~\ref{fig:sep-nonsep}). One can pick a gauge slice in moduli space whereby the path integral measure factorises into a term that associates the ``size'' and ``twist'' of the cut cycle to a complex modulus, such that the remaining terms are independent of the size (or radius), $r$, of the cut cycle. In the latter terms we can freely take $r\rightarrow 0$ (second diagram), allowing one to replace the (three, in the depicted example) partially overlapping charts by a single coordinate chart. The resulting puncture may then be translated across the Riemann surface in this fixed chart as depicted in Fig.~\ref{fig:opensetstranslate}.}\label{fig:pinchtranslate}
\end{center}
\end{figure}
\begin{equation}\label{eq:U_J-trans2}
\begin{aligned}
\mathscr{U}_J&=\{(U_1^J,w_1),(U_2^J,w_2),(U_3^J,w_3),\dots\}\\
&=\big\{\big(U_1^J,z_1+\theta(r-|z_1|)\delta z_v\big),(U_2^I,z_2),(U_3^I,z_3),\dots\big\}\big|_{r\rightarrow0}\\
\end{aligned}
\end{equation}
so that we demand that $(U_m^J,w_m)=(U_m^I,z_m)$ for all $m\neq1$, whereas for $m=1$ (including also the complex conjugate), 
\begin{equation}\label{eq:w1zwzv}
\boxed{
\begin{aligned}
w_1&=z_1+\theta(r-|z_1|)\delta z_v\\
\bar{w}_1&=\bar{z}_1+\theta(r-|z_1|)\delta \bar{z}_v
\end{aligned}
\qquad \textrm{(with $r\ll1$)}
}
\end{equation}
and $r$ is the radius that is to be taken to zero in order to ensure that only the point $p_1$ is translated leaving the remaining Riemann surface invariant. The relation (\ref{eq:w1zwzv}) should {\it not} be interpreted as a transition function since: (1) transition functions are holomorphic (for fixed moduli parameters) whereas (\ref{eq:w1zwzv}) is not; (2) $U_J$ and $U_I$ represent different complex structures.
\sk

Rather, (\ref{eq:w1zwzv}) represents the relation between holomorphic coordinates of two different complex structures. Since $z_1$ and $w_1$ are holomorphic parameters in their respective charts associated to complex structures, $I$ and $J$, it must be that $z_v$ is {\it not} a holomorphic parameter\footnote{Already here (and also below) we diverge slightly from Witten's derivation \cite{Witten12c}, see the comment below (2.52) in \cite{Witten12c}. The correspondence with the notation in \cite{Witten12c} is: $(w_1,z_1,z_v)_{\rm here}=(z,w,\mathbf{z})_{\rm there}$, and rather than taking $f(|w|)$ (defined there) to be a general monotonically decreasing function we choose to use the Heaviside function since we {\it only} wish to translate a single point, $p_1$, while keeping the remaining surface fixed. (Recall also footnote \ref{foot:WittenNoBoundaryTerms} on p.~\pageref{foot:WittenNoBoundaryTerms}.)} since the Heaviside function with argument $|z_1|$ appears in (\ref{eq:w1zwzv}). The discontinuity in (\ref{eq:w1zwzv}) ensures that $z_1\mapsto w_1$ is also not a diffeomorphism, since diffeomorphisms are by definition injective and surjective smooth maps with a smooth inverse. We could have smoothed this discontinuity out if we so wished (as done in \cite{Witten12c}) but the map would still not be a diffeomorphism: the essential point is this change of coordinates is {\it not invertible}. Viewing the complement of diffeomorphisms as complex structure deformations we are led to interpret this map as a complex structure deformation. To be precise, the map $I\rightarrow J=I+\delta I$ is {\it non-injective} and {\it non-surjective}: since (after taking the limit, $r\rightarrow0$) there are {\it two} points in $p_1,p_1'\in U_1^I$ that map to the {\it same} point $p_1\in U_1^J$ the map is non-injective (in particular both $z_1(p_1)=0$ and $z_1(p_1')$ are mapped to $w(p_1)=z_v$), and since there is also the point $p_0$ (defined by $w(p_0)=0$) that is not in the codomain of the map it is also non-surjective. 
\sk

Dissecting the notion of a diffeomorphism and complex structure deformations in this manner is useful since it highlights a multitude of ways available to us of deforming complex structures. We may in particular construct a map and simply drop one or more of the defining properties of a diffeomorphism -- with some care, the resulting map can be identified with a complex structure deformation. Choosing to parametrise the specific deformation (\ref{eq:w1zwzv}) by a discontinuous change of variables is evidently not necessary but it is certainly convenient and as general as it needs to be.
\sk

The discontinuity induced by the Heaviside function in (\ref{eq:w1zwzv}) corresponds to taking the metric to be singular along the contour $|z_1|=r$ (as we discuss below). Since the metric is not even part of the defining properties of a Riemann surface there is no obstruction to letting it become singular provided this does not interfere with global constraints (namely the cocycle relations (\ref{eq:cocycle}) or (\ref{eq:cocycles}) and hence also the Euler characteristic (\ref{eq:GaussBonnet1})). Indeed, the transition functions defining the Riemann surface remain holomorphic on patch overlaps for any one fixed complex structure so these are unaffected by the discontinuity. Furthermore, it is clear that the Euler characteristic is also not affected by the discontinuity since it is evaluated in a {\it fixed} complex structure -- the discontinuity does not appear in either complex structure (since we can work with the $z_m$ or $w_m$ coordinates in the $I$ or $J$ complex structures respectively). So the Heaviside function in (\ref{eq:w1zwzv}) for the entire domain in, $z_1:U_1^I\rightarrow \mathbf{C}$, is not only well-defined (with some care), but it will also streamline what follows.
\sk

We wish to identify the complex parameter $z_v$ with the coordinate in moduli space that parametrises the location of the point $p_1$, and we take it {\it initially} to be such that the translated point stays within the patch $U_1^J$, see Fig.~\ref{fig:opensetstranslate} and Fig.~\ref{fig:pinchtranslate}. The contribution to the path integral that we wish to evaluate will then be $\hat{B}_{z_v}\hat{B}_{\bar{z}_v}$, where according to (\ref{eq:Cellcontourint5nu}):
\begin{equation}\label{eq:Cellcontourint5nutrans}
\begin{aligned}
\hat{B}_{z_v}
&=\frac{1}{2\pi}\int_{\Sigma}\rmd^2z\,\big(\mu_{{z_v}\bar{z}}^{\phantom{aaa}z}b_{z z}+\mu_{{z_v}z}^{\phantom{aaa}\bar{z}}\tilde{b}_{\bar{z}\bar{z}}\big)+\\
&\qquad +\sum_m\frac{1}{2}\Big[\nu_{z_v}(p_m) \big(b_0^{(z_m)}+\tilde{b}_0^{(\bar{z}_m)}\big)(p_m)+i\gamma_{z_v}(p_m)\big(b_0^{(z_m)}-\tilde{b}_0^{(\bar{z}_m)}\big)(p_m)\Big]\\
&\qquad +\sum_m\sum_{n\geq1}\frac{1}{(n+1)!}\Big[\partial_{z_m}^n\nu_{z_v}(p_m)b_n^{(z_m)}(p_m)+\partial_{\bar{z}_m}^n\nu_{z_v}(p_m)\tilde{b}_n^{(\bar{z}_m)}(p_m)\Big]
\end{aligned}
\end{equation}
with a similar relation for the anti-chiral half, $\hat{B}_{\bar{z}_v}$, obtained from (\ref{eq:Cellcontourint5nutrans}) by the replacement $z_v\rightarrow \bar{z}_v$ keeping everything else fixed. Recalling the general relation between the holomorphic coordinates, $z_m$ and $w_m$, associated to complex structures $I$ and $J=I+\delta I$ respectively, given in (\ref{eq:w=z+v}),
\begin{equation}\label{eq:w=z+vtrans}
\begin{aligned}
w_m(z_m,\bar{z}_m)&=z_m+v_{km}^{z_m}(z_m,\bar{z}_m)\delta t^k+\mathcal{O}(v^2)\\
\bar{w}_m(z_m,\bar{z}_m)&=\bar{z}_m+v_{km}^{\bar{z}_m}(z_m,\bar{z}_m)\delta t^k+\mathcal{O}(v^2),
\end{aligned}
\qquad \textrm{(for all $m$)}
\end{equation}
and comparing (\ref{eq:w=z+vtrans}) with (\ref{eq:U_J-trans2}) we see that for the deformation of interest (i.e.~that associated to translating the point $p_1$ keeping the remaining Riemann surface fixed by identifying $z_v,\bar{z}_v$ with the corresponding modulus, namely $(t^{z_v},t^{\bar{z}_v})=(z_v,\bar{z}_v)$) all of the $v_{z_vm}^{z_m}$, $v_{z_vm}^{\bar{z}_m}$, (and their derivatives) are identically zero except for a subset of the $m=1$ terms:
\begin{equation}\label{eq:vvbarvanish}
\begin{aligned}
\big(v_{z_v1}^{z_1},v_{z_v2}^{z_2},\dots\big)&=\big(v_{z_v1}^{z_1},0,\dots\big)\\
\big(v_{\bar{z}_v1}^{z_1},v_{\bar{z}_v2}^{z_2},\dots\big)&=\big(0,0,\dots\big)\\
\end{aligned}\quad,
\quad 
\begin{aligned}
\big(v_{z_v1}^{\bar{z}_1},v_{z_v2}^{\bar{z}_2},\dots\big)&=\big(0,0,\dots\big)\\
\big(v_{\bar{z}_v1}^{\bar{z}_1},v_{\bar{z}_v2}^{\bar{z}_2},\dots\big)&=\big(v_{\bar{z}_v1}^{\bar{z}_1},0,\dots\big),
\end{aligned}
\end{equation}
and the $m=1$ terms read in particular,
\begin{equation}\label{eq:vzv1=theta}
\boxed{
\begin{aligned}
v_{z_v1}^{z_1}(z_1,\bar{z}_1)&=\theta(r-|z_1|)\\
v_{\bar{z}_v1}^{z_1}(z_1,\bar{z}_1)&=0
\end{aligned}\quad,\quad
\begin{aligned}
v_{z_v1}^{\bar{z}_1}(z_1,\bar{z}_1)&=0\\
v_{\bar{z}_v1}^{\bar{z}_1}(z_1,\bar{z}_1)&=\theta(r-|z_1|)
\end{aligned}
}
\end{equation}
But then from the first relation in (\ref{eq:numumubar-dvs}) it follows that all terms in the sums over $m$ in (\ref{eq:Cellcontourint5nutrans}) vanish except for the $m=1$ term, so the latter simplifies to:
\begin{equation}\label{eq:Cellcontourint5nutrans2}
\begin{aligned}
\hat{B}_{z_v}
&=\frac{1}{2\pi}\int_{\Sigma}\rmd^2z\,\big(\mu_{{z_v}\bar{z}}^{\phantom{aaa}z}b_{z z}+\mu_{{z_v}z}^{\phantom{aaa}\bar{z}}\tilde{b}_{\bar{z}\bar{z}}\big)+\\
&\qquad +\frac{1}{2}\Big[\nu_{z_v}(p_1) \big(b_0^{(z_1)}+\tilde{b}_0^{(z_1)}\big)(p_1)+i\gamma_{z_v}(p_1)\big(b_0^{(z_1)}-\tilde{b}_0^{(z_1)}\big)(p_1)\Big]\\
&\qquad +\sum_{n\geq1}\frac{1}{(n+1)!}\Big[\partial_{z_1}^n\nu_{z_v}(p_1)b_n^{(z_1)}(p_1)+\partial_{\bar{z}_1}^n\nu_{z_v}(p_1)\tilde{b}_n^{(z_1)}(p_1)\Big].
\end{aligned}
\end{equation}
We will evaluate the operator (\ref{eq:Cellcontourint5nutrans2}) next, beginning from the $\nu_{z_v}$-dependent terms. 
\sk

Let us appeal to the formalism developed in Sec.~\ref{sec:CSDII}. (We have tried to use a consistent notation throughout the entire document.) In that section we discussed how to parametrise deformation of complex structures from a metric viewpoint. Let us then consider the metric in the $J$ complex structure in the $U_1^J$ patch given in (the first equality in) (\ref{eq:gJ=rhoI}), repeated here for convenience,
\begin{equation}\label{eq:gJtransl}
g_J=\rho_I(w_1,\bar{w}_1)\rmd w_1\rmd\bar{w}_1
\end{equation}
where we have noted that in $U_1^J$ the holomorphic coordinate is $w_1$. Now since $\rho_I(w_1,\bar{w}_1)$ is the {\it same} function as that appearing in the $I$ complex structure, namely $\rho_I(z_1,\bar{z}_1)$, the only difference being that we have replaced $z_1$ by $w_1$ in the latter (recall the extensive discussion above and below (\ref{eq:gJ=rhoI}) on this point),  it follows that $\rho_I(w_1,\bar{w}_1)$ depends on the moduli $z_v,\bar{z}_v$ only via its dependence on $w_1,\bar{w}_1$. This is essentially why there is the subscript $I$ in $\rho_I(w_1,\bar{w}_1)$ rather than $J=I+\delta I$, {\it even though} (confusingly) the full metric tensor $g_J$ is in the $J$ complex structure as emphasised on the left-hand side in (\ref{eq:gJtransl}). Recall that this is what we meant by `parametrising a complex structure deformation by a change of coordinates keeping the metric fixed'. 
\sk

It will also be convenient to allow for a small (so that a single coordinate patch will be sufficient) but {\it finite} deformation. Rather than introduce new notation we will then infer from (\ref{eq:w1zwzv}) that for finite deformations:
\begin{equation}\label{eq:w1zwzvfinite}
\boxed{
\begin{aligned}
w_1&=z_1+\theta(r-|z_1|)z_v\\
\bar{w}_1&=\bar{z}_1+\theta(r-|z_1|)\bar{z}_v
\end{aligned}\quad,\quad
\textrm{(with $r\ll1$)}
}
\end{equation}
where we have taken into account that $\theta(r-|z_1|)$ and $z_1$ are independent of the modulus $z_v,\bar{z}_v$, and we have chosen the integration constant such that $z_v=0$ (when $z_1=0$) is identified with $w_1=0$. 
\sk

Given (\ref{eq:w1zwzvfinite}), it then should clearly be the case that if we differentiate $\ln\rho_I(w_1,\bar{w}_1)$ with respect to $z_v$ keeping both $w_1,\bar{w}_1$ fixed then the resulting quantity should vanish:
\begin{equation}\label{eq:ddzvrhoI}
\frac{\rmd }{\rmd z_v}\ln\rho_I(w_1,\bar{w}_1)\big|_{w_1,\bar{w}_1}=0,
\end{equation}
because the entire $z_v$ dependence is in $w_1$ via (\ref{eq:w1zwzvfinite}). As far as {\it this} derivative is concerned we need only consider the case $|z_1|<r$ (because in order to evaluate (\ref{eq:Cellcontourint5nutrans2}), jumping ahead slightly, we will need to evaluate it at $p_1$, i.e.~at $z_1=0$, {\it before} taking $r\rightarrow0$), so that the Heaviside function in (\ref{eq:w1zwzvfinite}) is equal to unity here. Therefore, on account of (\ref{eq:w1zwzvfinite}) we can regard $z_1+z_v,\bar{z}_1+\bar{z}_v$ as being the arguments of $\rho_I(w_1,\bar{w}_1)$ (which we denote by $\rho_I(z_1+z_v)\equiv \rho_I(z_1+z_v,\bar{z}_1+\bar{z}_v)$ for brevity). Using the chain rule we can then rewrite (\ref{eq:ddzvrhoI}) as follows,
\begin{equation}\label{eq:ddzvrhoI2}
\begin{aligned}
\frac{\partial}{\partial z_v}\ln\rho_I(z_1&+z_v)\big|_{z_1,\bar{z}_1}+\frac{\partial}{\partial z_1}\ln\rho_I(z_1+z_v)\big|_{\bar{z}_1,z_v}\frac{\partial z_1}{\partial z_v}\big|_{w_1,\bar{w}_1}\\
&+\frac{\partial}{\partial \bar{z}_1}\ln\rho_I(z_1+z_v)\big|_{z_1,z_v}\frac{\partial \bar{z}_1}{\partial z_v}\big|_{w_1,\bar{w}_1}=0,
\end{aligned}
\end{equation}
where $z_v$ and $\bar{z}_v$ (for $|z_1|<r$) are interpreted as independent variables, which is in turn why a term proportional to $\partial \bar{z}_v/\partial z_v$ is absent in (\ref{eq:ddzvrhoI2}). Since on account of (\ref{eq:w1zwzvfinite}) for $|z_1|<r$ the last term on the left-hand side in (\ref{eq:ddzvrhoI2}) vanishes, and in fact:
$$
\frac{\partial z_1}{\partial z_v}\big|_{w_1,\bar{w}_1}=-1,\qquad \frac{\partial \bar{z}_1}{\partial z_v}\big|_{w_1,\bar{w}_1}=0,\qquad {\rm when}\qquad |z_1|<r,
$$
we learn that the requirement (\ref{eq:ddzvrhoI}) or (\ref{eq:ddzvrhoI2}) is equivalent to:
\begin{equation}\label{eq:ddzvrhoI3}
\frac{\partial}{\partial z_v}\ln\rho_I(z_1+z_v)\big|_{z_1,\bar{z}_1}=\frac{\partial}{\partial z_1}\ln\rho_I(z_1+z_v)\big|_{z_v,\bar{z}_v},\qquad {\rm when}\qquad |z_1|<r.
\end{equation}
We emphasise that this is valid in a finite region (within the disc, $|z_1|<r$, in particular), which includes $z_1=0$. We have also taken this into account when specifying what is held fixed when the various partial derivatives are taken (to be precise, the quantities $z_1$ and $\bar{z}_1$ are independent, as are $z_v$ and $\bar{z}_v$, when $|z_1|<r$), and we will also make use of this below since this allows us to take multiple derivatives of (\ref{eq:ddzvrhoI3}) with respect to $z_1$ before evaluating it at $z_1(p_1)=0$. 
\sk

In Sec.~\ref{sec:CSDI} and Sec.~\ref{sec:CSDII} we denoted by $t=(t^1,t^2,\dots)$ a generic local coordinate system in moduli space. As mentioned above, we here identify a pair of these coordinates, $t^{z_v},t^{\bar{z}_v}$, with the location of the puncture in the $J$ complex structure, that is: 
$$
(t^{z_v},t^{\bar{z}_v})=(z_v,\bar{z}_v). 
$$
Taking this into account, let us then substitute (\ref{eq:ddzvrhoI3}) back into the first relation in (\ref{eq:numumubar-metric2}) (displayed on p.~\pageref{eq:numumubar-metric2}), recalling that $\rho_I=2g^I_{z_1\bar{z}_1}$. That is, in the $J$ complex structure (written in terms of the $I$ complex structure coordinates, $z_1$) we have the relations:
\begin{equation}\label{eq:numumubar-metric3}
\begin{aligned}
\nu_{z_v}(z_1+z_v)&=\frac{\partial }{\partial z_v}\ln\rho_I(z_1+z_v)\big|_{z_1,\bar{z}_1}\\
&=\frac{\partial }{\partial z_1}\ln\rho_I(z_1+z_v)\big|_{z_v,\bar{z}_v}\\
\end{aligned}
,\qquad
\begin{aligned}
\nu_{\bar{z}_v}(z_1+z_v)&=\frac{\partial }{\partial \bar{z}_v}\ln\rho_I(z_1+z_v)\big|_{z_1,\bar{z}_1}\\
&=\frac{\partial }{\partial \bar{z}_1}\ln\rho_I(z_1+z_v)\big|_{z_v,\bar{z}_v}\\
\end{aligned}
\end{equation}
provided $|z_1|<r$. Notice in particular that we can freely interchange $\partial_{z_1}$ with $\partial_{z_v}$ derivatives of $\ln\rho_I(z_1+z_v)$ provided $|z_1|<r$. 
\sk

We may then use the relation (\ref{eq:numumubar-metric3}) to rewrite the derivatives appearing in the last line in (\ref{eq:Cellcontourint5nutrans2}) in terms of derivatives with respect to $z_v,\bar{z}_v$ at fixed frame coordinate, $z_1,\bar{z}_1$,
\begin{equation}\label{eq:dz1lnrhoI}
\begin{aligned}
\partial_{z_1}^n\nu_{z_v}(z_1+z_v)&=\partial_{z_v}^{n+1}\ln\rho_I(z_1+z_v)\big|_{z_1,\bar{z}_1}\\
\partial_{\bar{z}_1}^n\nu_{z_v}(z_1+z_v)&=\partial_{\bar{z}_v}^n\partial_{z_v}\ln\rho_I(z_1+z_v)\big|_{z_1,\bar{z}_1},
\end{aligned}
\end{equation}
Evaluating (\ref{eq:dz1lnrhoI}) at $p_1$ taking into account that $z_1(p_1)=0$ yields,
\begin{equation}\label{eq:dz1lnrhoI2}
\begin{aligned}
\partial_{z_1}^n\nu_{z_v}(p_1)
&=\partial_{z_v}^{n+1}\ln\rho_I(z_v)\big|_{z_1,\bar{z}_1}\\
\partial_{\bar{z}_1}^n\nu_{z_v}(p_1)
&=\partial_{\bar{z}_v}^n\partial_{z_v}\ln\rho_I(z_v)\big|_{z_1,\bar{z}_1}
\end{aligned}
,\qquad ( n\geq0)
\end{equation}

We next make use of the choice of metric (\ref{eq:Polchinskicoords}) of Polchinski that was introduced in Sec.~\ref{sec:MV}. But before we carry on let us divert momentarily before implementing this choice into (\ref{eq:dz1lnrhoI2}) in order to briefly elaborate on a subtle point that can lead to confusion. In the analysis of Sec.~\ref{sec:MV} it would seem that in order to make the choice of metric (\ref{eq:Polchinskicoords}) we referred to a specific complex structure, $I$, with corresponding chart $(U_1^I,z_1)$ (we take the patch associated to $m=1$ here which is the focus of the current section). And that consequently the choice of metric (\ref{eq:Polchinskicoords}) only remains valid at $z_1=0$. Although it is true that the choice of metric holds at $z_1=0$, the relation (\ref{eq:Polchinskicoords}) and all the subsequent relations that follow from it (such as the boxed relations in that subsection) also remain valid in a deformed complex structure $J=I+\delta I$ (with a corresponding open set $U_1^J$) when the latter differs from $I$ only in that (when $r\rightarrow0$) the point $p_1$ has been translated across $U_1^I$ as we have discussed in the current subsection leaving the entire {\it remaining} Riemann surface unchanged. Making the explicit chart coordinate, $z_1$, manifest, the relation (\ref{eq:Polchinskicoords}) in the $I$ complex structure in the chart $(U_1^I,z_1)$ reads $\partial_{z_1}^ng^I_{z_1\bar{z}_1}(0)=0$, whereas in the $J$ complex structure in the chart $(U_1^J,w_1)$ (but written in terms of the $z_1$ coordinate) it reads:
\begin{equation}\label{eq:prelude g}
\begin{aligned}
&\partial_{z_1}^n\rho_I(z_1+z_v)\big|_{z_1=0}=\partial_{z_v}^n\rho_I(z_1+z_v)\big|_{z_1=0}=
\left\{
\begin{array}{l}
\begin{aligned}
&1\qquad \textrm{if $n=0$}\\
&0\qquad \textrm{if $n\geq1$}
\end{aligned}
\end{array}
\right.\\
&\partial_{\bar{z}_1}^n\rho_I(z_1+z_v)\big|_{z_1=0}=\partial_{\bar{z}_v}^n\rho_I(z_1+z_v)\big|_{z_1=0}=
\left\{
\begin{array}{l}
\begin{aligned}
&1\qquad \textrm{if $n=0$}\\
&0\qquad \textrm{if $n\geq1$}
\end{aligned}\end{array}
\right.\\
\end{aligned}
\end{equation}
where it might be helpful to recall that $w_1=z_1+z_v$ when $|z_1|<r$ (for some sufficiently small radius $r$). Similarly, the relation for the Ricci scalar corresponding to (\ref{eq:R(2)}) 
in the $J$ complex structure reads:
\begin{equation}\label{eq:prelude}
\begin{aligned}
R_{(2)}(z_1+z_v)&= -4\rho_I(z_1+z_v)^{-1}\partial_{z_v}\partial_{\bar{z}_v}\ln \rho_I(z_1+z_v).
\end{aligned}
\end{equation}
Given this holds in a finite region $|z_1|<r$ and not only at $p_1$ we can also consider derivatives of (\ref{eq:prelude}) and then evaluate it at $p_1$. 
Since it is the physical point $p_1$ that has been actively translated, these relations are still eventually evaluated at $p_1$ where $z_1=0$, but now the explicit arguments will correspond to the modulus $z_v$. 
Again, it may come across as slightly confusing that the $I$ complex structure label and coordinates appear in the relations (\ref{eq:prelude g}) and (\ref{eq:prelude}) even though we are in the $J$ complex structure, but as mentioned the reason that this must be the case can be traced back to (\ref{eq:gJ=rhoI}) (where this was discussed in detail). 
\sk

Taking these comments into account, we next evaluate the two expressions in (\ref{eq:dz1lnrhoI2}) using the choice of metric (\ref{eq:prelude g}) (associated to holomorphic normal coordinates) and find that the first expression vanishes,
\begin{equation}\label{eq:dz1lnrhoI3a}
\begin{aligned}
\partial_{z_1}^n\nu_{z_v}(p_1)
&=\partial_{z_v}^{n+1}\ln\rho_I(z_v)\big|_{z_1,\bar{z}_1}=0\\
\end{aligned}
,\qquad ( n\geq0)
\end{equation}
whereas the second expression in (\ref{eq:dz1lnrhoI2}) vanishes only for $n=0$. For $n\geq1$ the latter can primarily be massaged into the form:
\begin{equation}\label{eq:dbarvup1}
\begin{aligned}
\partial_{\bar{z}_1}^n\nu_{z_v}(p_1)
&=\partial_{\bar{z}_v}^n\partial_{z_v}\ln\rho_I(z_v)\big|_{z_1,\bar{z}_1}\\
&=-\frac{1}{4}\partial_{\bar{z}_v}^{n-1}\Big[\rho_I(z_v)\big(-4\rho_I(z_v)^{-1}\partial_{\bar{z}_v}\partial_{z_v}\ln\rho_I(z_v)\big)\Big]\big|_{z_1,\bar{z}_1}\\
\end{aligned}\qquad(n\geq1)
\end{equation}
Now since in holomorphic normal coordinates purely anti-holomorphic (or purely holomorphic) derivatives of $\rho_I(z_v)$ vanish at $p_1$, and since $\rho_I(z_v)=1$, on account of the above expression for the Ricci scalar (\ref{eq:prelude}) we learn that in holomorphic normal coordinates (\ref{eq:dbarvup1}) is equivalent to,
\begin{equation}\label{eq:dbarvup2}
\begin{aligned}
\partial_{\bar{z}_1}^n\nu_{z_v}(p_1)
&=-\frac{1}{4}\partial_{\bar{z}_v}^{n-1}R_{(2)}(z_v)\big|_{z_1,\bar{z}_1}\qquad(n\geq1)
\end{aligned}
\end{equation}
Given that in holomorphic normal coordinates purely (anti-)holomorphic derivatives of the connection, $\Gamma_{z_1z_1}^{z_1}(z_v)=\partial_{z_v}\ln\rho_I(z_v)$, always vanish when evaluated at $p_1$ (see the reasoning leading up to (\ref{eq:NablanR-Polch}) in Sec.~\ref{sec:PolCoords}) we learn that (\ref{eq:dbarvup2}) can be written entirely in terms of {\it covariant} quantities,
\begin{equation}\label{eq:dbarvup3}
\boxed{
\begin{aligned}
\nu_{z_v}(p_1)&=0\\
\partial_{z_1}^n\nu_{z_v}(p_1)&=0\\
\,\,\partial_{\bar{z}_1}^n\nu_{z_v}(p_1)
&=-\frac{1}{4}\nabla_{\bar{z}_v}^{n-1}R_{(2)}(z_v)\big|_{z_1,\bar{z}_1}\end{aligned}\quad,\quad (n\geq1)
}
\end{equation}
where in the first two relations we reproduce (\ref{eq:dz1lnrhoI3a}) for later reference. Crucially, the first two relations in (\ref{eq:dbarvup3}) imply that variations in $z_v$ {\it preserve} the gauge fixing associate to the choice of holomorphic normal coordinates. So as we shift a puncture across a general Riemann surface the metric will {\it remain} `as flat as possible' at the puncture.
\sk

Since the right-hand side of the third relation in (\ref{eq:dbarvup3}) is explicitly independent of $z_1,\bar{z}_1$ we will drop the fact that it is evaluated at fixed $z_1,\bar{z}_1$ from the notation in what follows.  
Written in this manner it is clear that the right-hand side of the third relation in (\ref{eq:dbarvup3}) is valid in {\it any} coordinate system, even though we chose to work in the specific coordinate system introduced in Sec.~\ref{sec:MV} and Sec.~\ref{sec:PolCoords}. This is analogous to the use of Riemann normal coordinates for real manifolds and is, consequently, particularly powerful. 
For example, local operators written entirely in terms of covariant quantities such as (\ref{eq:dbarvup3}) will transform as (components of) conformal tensors (modulo U(1)) on patch overlaps.\footnote{A complementary viewpoint is that the frame coordinates transform as scalars under reparametrisations of an underlying auxiliary coordinate, $\sigma$, or reparametrisations of the base point, $\sigma=\sigma_1$, in terms of which what we have been calling $z_1$ is identified with $z_{\sigma_1}(\sigma)$ and $z_1=0$ corresponds to $z_{\sigma_1}(\sigma_1)=0$. Moduli variations, $\delta z_v$, are identical to $-\delta z_{\sigma_1}(\sigma_1)$. The relations (\ref{eq:numumubar-metric3}) can also be used to replace $\nabla_{\bar{z}_v}^{n-1}R_{(2)}(z_v)\big|_{z_1,\bar{z}_1}$ in (\ref{eq:dbarvup3}) by $\nabla_{\bar{z}_1}^{n-1}R_{(2)}(z_1+z_v)\big|_{z_v,\bar{z}_v}$ with the latter evaluated at $z_1=0$. In terms of the auxiliary coordinate the latter is equivalent to $\nabla_{\bar{z}_{\sigma_1}}^{n-1}R_{(2)}(\sigma)\big|_{\sigma=\sigma_1}$, where the covariant derivatives are with respect to $\bar{z}_{\sigma_1}(\sigma)$ and then evaluated at $\sigma=\sigma_1$. So on account of (\ref{eq:w's'=ws}) and (\ref{eq:zsigma* scalar}) reparametrisation invariance is perhaps clearer in terms of an auxiliary coordinate system, see also the discussion in the following section. But the results are equivalent.} 
\sk

We must also verify that the left-hand sides in the {\it first two} relations in (\ref{eq:dbarvup3}) also vanish in every coordinate system, since this is not immediate (the second relation is not a tensor equation). The first relation in (\ref{eq:dbarvup3}) indeed holds in every coordinate system since $\nu_{z_v}$ is a globally-defined function on $\Sigma$ (i.e.~a scalar with respect to diffeomorphisms) as discussed below (\ref{eq:nuzmzbarmt}), see also (\ref{eq:munu}). The second relation also holds in every coordinate system since in holomorphic normal coordinates it is true that $\partial_{z_1}^n\nu_{z_v}(p_1)=\nabla_{z_1}^n\nu_{z_v}(p_1)$, and since this is the $n^{\rm th}$ covariant derivative of a scalar and covariant derivatives of scalars are tensors, it follows that the second relation in (\ref{eq:dbarvup3}) indeed holds in every conformal coordinate system. Summarising, all relations in (\ref{eq:dbarvup3}) hold in a general coordinate system and lead to a globally-defined construction. 
\sk

Let us now consider the $\gamma_{z_v}(p_1)$ factor in (\ref{eq:Cellcontourint5nutrans2}). Recall primarily that this was defined in (\ref{eq:gammakdfn}). This quantity will never contribute to the path integral provided: (1) we always identify the phase of the frame coordinate, $z_1$, with a modulus; or (2) when the vertex operator on which  $\hat{B}_{z_v}$ acts is annihilated by $(b_0^{(z_1)}-\tilde{b}_0^{(z_1)})(p_1)$ (when $\hat{B}_{z_v}$ is associated to translating an external state across the Riemann surface). In the first case, that the $\gamma_{z_v}(p_1)$ factor in (\ref{eq:Cellcontourint5nutrans2}) does not contribute is because in this case there is always an additional factor of $\hat{B}_{\theta}$ in the path integral measure, see (\ref{eq:Bintpinchtheta}), and $(b_0^{(z_1)}-\tilde{b}_0^{(z_1)})^2=0$ by the Grassmann-odd nature of these modes. In the latter case it is trivially true that the $\gamma_{z_v}(p_1)$ drops out. We may consequently drop this term from the following formulas. 
\sk

Summarising, let us substitute (\ref{eq:dbarvup3}) into (\ref{eq:Cellcontourint5nutrans2}) to find the following expression for the $\hat{B}_{z_v}$ operator, namely the measure contribution associated to translating pinches or local operators across a Riemann surface with corresponding moduli $z_v,\bar{z}_v$,
\begin{equation}\label{eq:Cellcontourint5nutrans3}
\begin{aligned}
\hat{B}_{z_v}
&=\frac{1}{2\pi}\int_{\Sigma}\rmd^2z\,\big(\mu_{{z_v}\bar{z}}^{\phantom{aaa}z}b_{z z}+\mu_{{z_v}z}^{\phantom{aaa}\bar{z}}\tilde{b}_{\bar{z}\bar{z}}\big)-\frac{1}{4}\sum_{n\geq1}\frac{1}{(n+1)!}\nabla_{\bar{z}_v}^{n-1}R_{(2)}(z_v)\tilde{b}_n^{(z_1)}(p_1).
\end{aligned}
\end{equation}

It remains to evaluate the first term in this equation associated to the integral over $\Sigma$. We evaluate this by introducing a partition of unity satisfying $\sum_m\lambda_m=1$, as discussed extensively in Sec.~\ref{sec:EulerCh} and then make use of (\ref{eq:nopartitionofunity}) in order to write the integral over $\Sigma$ as a sum of integrals over cells, $\mathscr{V}=\{V_m\}$. That is,
\begin{equation}\label{eq:mumubarintegrala}
\begin{aligned}
\frac{1}{2\pi}\int_{\Sigma}\rmd^2z\,\big(\mu_{{z_v}\bar{z}}^{\phantom{aaa}z}b_{z z}+\mu_{{z_v}z}^{\phantom{aaa}\bar{z}}\tilde{b}_{\bar{z}\bar{z}}\big)&=\sum_m\frac{1}{2\pi}\int_{U_m}\rmd^2z_m\,\big(\mu_{{z_v}\bar{z}_m}^{\phantom{aaa}z_m}b_{z_m z_m}\lambda_m+\mu_{{z_v}z_m}^{\phantom{aaa}\bar{z}_m}\tilde{b}_{\bar{z}_m\bar{z}_m}\lambda_m\big)\\
&=\sum_m\frac{1}{2\pi}\int_{V_m}\rmd^2z_m\,\big(\mu_{{z_v}\bar{z}_m}^{\phantom{aaa}z_m}b_{z_m z_m}+\mu_{{z_v}z_m}^{\phantom{aaa}\bar{z}_m}\tilde{b}_{\bar{z}_m\bar{z}_m}\big).
\end{aligned}
\end{equation}
We then consider every integral in the sum over cells, labelled by $m$, separately. In any one such cell we will assume the Dolbeault-Grothendieck lemma holds so that the (components of the) Beltrami differentials take the form:
$$
\mu_{{z_v}\bar{z}_m}^{\phantom{aaa}z_m}=\partial_{\bar{z}_m}v_{z_vm}^{z_m},\qquad \mu_{{z_v}z_m}^{\phantom{aaa}\bar{z}_m}=\partial_{z_m}v_{z_vm}^{\bar{z}_m}.
$$
For the deformation of interest the local vectors $v_{z_vm}^{z_m}, v_{z_vm}^{\bar{z}_m}$ satisfy (\ref{eq:vvbarvanish}) and (\ref{eq:vzv1=theta}), so that all terms in the sum over $m$ in (\ref{eq:mumubarintegrala}) vanish except for the chiral half of the $m=1$ term,
\begin{equation}\label{eq:mumubarintegral}
\begin{aligned}
\frac{1}{2\pi}\int_{\Sigma}\rmd^2z\,\big(\mu_{{z_v}\bar{z}}^{\phantom{aaa}z}b_{z z}+\mu_{{z_v}z}^{\phantom{aaa}\bar{z}}\tilde{b}_{\bar{z}\bar{z}}\big)
&=\frac{1}{2\pi}\int_{V_1}\rmd^2z_1\,\partial_{\bar{z}_1}v_{z_v1}^{z_1}b_{z_1 z_1}\\
&=\lim_{r\rightarrow 0}\frac{1}{2\pi}\int_{V_1}\rmd^2z_1\,\partial_{\bar{z}_1}\theta(r-|z_1|)b_{z_1 z_1}\\
\end{aligned}
\end{equation}
We have also noted that we may take $r\rightarrow0$ {\it after} having evaluated the integral since $\hat{B}_{z_v}$ is independent of $r$ as discussed extensively above. 
\sk

A derivative of the Heaviside function appearing in (\ref{eq:mumubarintegral}) is by definition a delta function,
\begin{equation}\label{eq:partialzbartheta}
\begin{aligned}
\partial_{\bar{z}_1}\theta(r-|z_1|) 
&= -\delta(r-|z_1|)\partial_{\bar{z}_1}|z_1|\\
\end{aligned}
\end{equation}
To proceed further, when it is not immediate what the derivative of a complex-valued function, $f(z,\bar{z})$, (in the sense of real analysis) of one complex variable is we always interpret it in the sense of Pompeiu \cite{Pompeiu}:
\begin{equation}\label{eq:Pompeiu}
(\partial_{\bar{z}}f)(z_0,\bar{z}_0)\dfn \lim_{\epsilon\rightarrow 0}\frac{1}{2\pi i\epsilon^2}\oint_{\partial\Delta(z_0,\epsilon)}\rmd z\,f(z,\bar{z}),
\end{equation}
where $\Delta(z_0,\epsilon)$ is a disc in the $z,\bar{z}$ coordinate system centred at $z=z_0$ of radius $\epsilon$ and $\partial \Delta(z_0,\epsilon)$ its boundary with the standard orientation so that $\oint_{\partial\Delta(z_0,\epsilon)}\rmd z/(z-z_0)=2\pi i$. Applying this to the case of interest, we take $f(z,\bar{z})=|z|$ and find,
\begin{equation}\label{eq:partial|z|}
\partial_{\bar{z}_1}|z_1|=\left\{
\begin{aligned}
&0\qquad \textrm{if $z_1=0$}\\
&\frac{1}{2}\frac{\bar{z}_1}{|z_1|}\qquad\textrm{if $z_1\neq0$}
\end{aligned}\right.
\end{equation}
Notice that (\ref{eq:Pompeiu}) makes sense even when $f(z,\bar{z})$ is not differentiable, but this plays no role here since by (\ref{eq:partialzbartheta}) (and given that $z_1$ in (\ref{eq:mumubarintegral}) is integrated) only the $|z_1|=r$ case is of interest. In particular, (\ref{eq:partialzbartheta}) reads,
\begin{equation}\label{eq:partialzbartheta2}
\begin{aligned}
\partial_{\bar{z}_1}\theta(r-|z_1|) 
&= -\delta(r-|z_1|)\frac{1}{2}\frac{\bar{z}_1}{|z_1|},
\end{aligned}
\end{equation}
and substituting this into (\ref{eq:mumubarintegral}) yields,
\begin{equation}\label{eq:mumubarintegral2}
\begin{aligned}
\frac{1}{2\pi}\int_{\Sigma}\rmd^2z\,\big(\mu_{{z_v}\bar{z}}^{\phantom{aaa}z}b_{z z}+\mu_{{z_v}z}^{\phantom{aaa}\bar{z}}\tilde{b}_{\bar{z}\bar{z}}\big)
&=-\lim_{r\rightarrow 0}\frac{1}{2\pi}\int_{V_1}\rmd^2z_1\,\delta(r-|z_1|)\frac{1}{2}\frac{\bar{z}_1}{|z_1|}b_{z_1 z_1}.
\end{aligned}
\end{equation}
We next make use of the fact that $b_{z_1 z_1}$ has a Laurent expansion (within $V_1$)  centred at $z_1(p_1)\equiv0$,
\begin{equation}\label{eq:bz1z1z}
b_{z_1z_1}(z)=\sum_{n\in\mathbf{Z}}\frac{b_n^{(z_1)}(p_1)}{z^{n+2}},\qquad{\rm with}\qquad z\dfn z_1(p_1'),
\end{equation}
where $p_1'$ is a point on the manifold at which the local operator, $b_{z_1z_1}$, is evaluated, 
and change the integration variables, $(z_1,\bar{z}_1)\mapsto (\rho,\theta)$, by  $z_1=\rho e^{i\theta}$ and $\bar{z}_1=\rho e^{-i\theta}$ (with range $0\leq\rho\leq r+\Lambda$ and $0\leq \theta<2\pi$, and $\Lambda$ any positive constant). The measure takes the form $\rmd^2z_1\equiv i\rmd z_1\wedge \rmd\bar{z}_1=2\rho \rmd\rho \wedge \rmd\theta$, so that the integral (\ref{eq:mumubarintegral2}) is then trivially evaluated and yields:
\begin{equation}\label{eq:mumubarintegral3}
\begin{aligned}
\frac{1}{2\pi}\int_{\Sigma}\rmd^2z\,\big(\mu_{{z_v}\bar{z}}^{\phantom{aaa}z}b_{z z}+\mu_{{z_v}z}^{\phantom{aaa}\bar{z}}\tilde{b}_{\bar{z}\bar{z}}\big)&=-\lim_{r\rightarrow0}\sum_{n\in\mathbf{Z}}r^{-n-1}\delta_{n+1,0}b_n^{(z_1)}(p_1)\\
&=-b_{-1}^{(z_1)}(p_1).
\end{aligned}
\end{equation}

We next substitute (\ref{eq:mumubarintegral3}) into the expression for the $\hat{B}_{z_v}$ operator given in (\ref{eq:Cellcontourint5nutrans3}) to find that the final expression for the path integral measure contribution associated to the complex structure deformation $\delta z_v,\delta \bar{z}_v$ that translates fixed picture vertex operators to integrated picture reads:
\begin{equation}\label{eq:hatBzvR}
\begin{aligned}
\hat{B}_{z_v}
&=-b_{-1}^{(z_1)}(p_1)-\frac{1}{4}\sum_{n\geq1}\frac{1}{(n+1)!}\nabla_{\bar{z}_v}^{n-1}R_{(2)}(z_v)\tilde{b}_n^{(z_1)}(p_1)\\
\hat{B}_{\bar{z}_v}
&=-\tilde{b}_{-1}^{(z_1)}(p_1)-\frac{1}{4}\sum_{n\geq1}\frac{1}{(n+1)!}\nabla_{z_v}^{n-1}R_{(2)}(z_v)b_n^{(z_1)}(p_1)
\end{aligned}
\end{equation}
where in the second equality we also display the corresponding relation for the $\bar{z}_v$ modulus which is completely analogous. 
\sk

Since $b_{z_1z_1}(\rmd z_1)^2$ transforms as a rank-2 conformal tensor under conformal transformations, {\it when} $|z_1|<r$ the relation to the $J$ complex structure coordinates reads $w_1=z_1+z_v$, and is such that $b_{z_1z_1}(\rmd z_1)^2=b_{w_1w_1}(\rmd w_1)^2$. Therefore, since $\rmd z_1=\rmd w_1$ the corresponding modes satisfy,
\begin{equation}\label{eq:bnz1=bnw1}
b_n^{(z_1)}(p_1) = b_n^{(w_1)}(p_1)\qquad \Rightarrow \qquad b_n^{(z_1)}(0) = b_n^{(w_1)}(z_v),
\end{equation}
where,
$$
b_n^{(w_1)}(z_v) = \oint \frac{\rmd w}{2\pi i(w-z_v)}(w-z_v)^{n+2}b_{w_1w_1}(w),\qquad {\rm with}\qquad w\dfn w_1(p_1'),
$$
and (according to (\ref{eq:bz1z1z})) $p_1'$ is now interpreted as a point on the manifold along the contour at which the integrand is evaluated, and the contour integral integrates over all such points. (Note that we imagine taking the contour such that $|z_1|<r$ and {\it then} take $r\rightarrow 0$.) 
In terms of the $J$ complex structure coordinates, taking into account (\ref{eq:bnz1=bnw1}), we could equivalently have written (\ref{eq:hatBzvR}) as,
\begin{equation}\label{eq:hatBzvw1}
\boxed{
\begin{aligned}
\hat{B}_{z_v}
&=-b_{-1}^{(w_1)}(p_1)-\frac{1}{4}\sum_{n\geq1}\frac{1}{(n+1)!}\nabla_{\bar{z}_v}^{n-1}R_{(2)}(z_v)\tilde{b}_n^{(w_1)}(p_1)\\
\hat{B}_{\bar{z}_v}
&=-\tilde{b}_{-1}^{(w_1)}(p_1)-\frac{1}{4}\sum_{n\geq1}\frac{1}{(n+1)!}\nabla_{z_v}^{n-1}R_{(2)}(z_v)b_n^{(w_1)}(p_1)
\end{aligned}
}
\end{equation}
The relations (\ref{eq:hatBzvR}) are consistent with \cite{Polchinski88}. Here we have also discussed the detailed and very subtle reasoning that leads to this result.
\sk

Let us also discuss how to apply this result (\ref{eq:hatBzvw1}) in explicit computations. When there is a fixed-picture vertex operator, $\hat{\mathscr{A}}_a^{(z_1)}(p_1)$, inserted at $p_1$ in the $z_1$ frame coordinate, and when we wish to associate its location, $p_1$,  to a modulus, the measure contribution of the path integral provides the integrated-picture vertex operator,
\begin{equation}\label{eq:intvertop1}
\mathscr{A}_a^{(z_1)}\dfn \int \rmd^2z_v\hat{B}_{z_v}\hat{B}_{\bar{z}_v}\hat{\mathscr{A}}_a^{(z_1)}(p_1)
\end{equation}
where the domain of integration depends on whether this vertex operator arises from cutting open a handle of $\Sigma$ or whether $\hat{\mathscr{A}}_a^{(z_1)}(p_1)$ is an asymptotic state vertex operator, but in general it also depends on the remaining features on the surface. Adhering to standard convention \cite{Polchinski_v1}, for integrated vertex operators, $\mathscr{A}_a^{(z_1)}$, we omit the hat `$\hat{\phantom{A}}$' from $\hat{\mathscr{A}}_a^{(z_1)}$.
\sk

The quantity $\hat{B}_{z_v}\hat{B}_{\bar{z}_v}\hat{\mathscr{A}}_a^{(z_1)}(p_1)$ is evaluated in the $J$ complex structure where locally the chart coordinates, $(U_1^J,w_1)$, are used and recall that:
$$
z_1(p_1)=0,\qquad {\rm and}\qquad w_1(p_1)=z_v.
$$
So to make $z_v$ more explicit we could transform the $z_1$ frame vertex operators, $\hat{\mathscr{A}}_a^{(z_1)}(p_1)$, into the corresponding $w_1$ frame vertex operators, $\hat{\mathscr{A}}_a^{(w_1)}(p_1)$. {\it If} we can show that these vertex operators transform as {\it conformal tensor components under rigid translations}, $z_1\mapsto w_1=z_1+z_v$ (for $|z_1|\ll r$) then by the standard transformation law of a tensor\footnote{Recall that the defining property of a conformal tensor component, $\mathscr{O}^{(z)}(p)$, is that under (anti-)holomorphic transformations $z\mapsto w(z)$ (and $\bar{z}\mapsto \bar{w}(\bar{z})$), 
$$
\mathscr{O}^{(z)}(p)\rmd z^h\rmd\bar{z}^{\tilde{h}} =\mathscr{O}^{(w)}(p)\rmd w^h\rmd\bar{w}^{\tilde{h}} 
$$} under rigid translations (since $\rmd w_1=\rmd z_1$) (and making the coordinates, $z_1(p_1)$, of the location, $p_1$, at which the vertex operator is evaluated explicit),
$$
\hat{\mathscr{A}}_a^{(z_1)}(p_1)\equiv \hat{\mathscr{A}}_a^{(z_1)}(z_1(p_1))=\hat{\mathscr{A}}_a^{(z_1)}(0) = \hat{\mathscr{A}}_a^{(w_1)}(w_1(p_1)) =\hat{\mathscr{A}}_a^{(w_1)}(z_v)\equiv \hat{\mathscr{A}}_a^{(w_1)}(p_1).
$$
It is indeed true that the handle operators we will consider transform as tensors under rigid shifts of frame, because of the following two reasons: (1) they transform naively as such when we neglect normal ordering; (2) the change induced by a change in normal ordering is zero for rigid shifts -- the latter follows from the general result for changes in normal ordering (\ref{eq:CNOxbc-w<->z}) with (\ref{eq:DDDw<->z}) evaluated at $w_1(z)=z+z_v$ (and similarly $w_1(z')=z'+z_v$), see (\ref{eq:CNOxbc-w<->z-zero}). 
So applying this to (\ref{eq:intvertop1}) we learn that given a set of fixed picture vertex operators, $\hat{\mathscr{A}}_a^{(z_1)}(p_1)$, the corresponding {\it integrated-picture vertex operators} are also given by:
\begin{equation}\label{eq:intvertop}
\boxed{\mathscr{A}_a^{(w_1)}\dfn\int \rmd^2z_{v_1}\hat{B}_{z_{v_1}}\hat{B}_{\bar{z}_{v_1}}\hat{\mathscr{A}}_a^{(w_1)}(z_{v_1})}
\end{equation}
where the ghost insertions are given by (\ref{eq:hatBzvw1}). (For later convenience we replaced $z_v$ by $z_{v_1}$.) 
Notice that the modes in $\hat{B}_{z_{v_1}},\hat{B}_{\bar{z}_{v_1}}$ are centred at the vertex operator insertions, for every value of the modulus, $z_{v_1},\bar{z}_{v_1}$. 
\sk

We have set things up such that the modes hitting $\hat{\mathscr{A}}_a^{(w_1)}(z_{v_1})$ in (\ref{eq:intvertop}) remain centred at the local vertex operator, and this remains true for all $z_{v_1},\bar{z}_{v_1}$. So this also remains true while $\hat{\mathscr{A}}_a^{(w_1)}(z_{v_1})$ is integrated over the entire Riemann surface of interest. 
\sk

The result (\ref{eq:intvertop}) is a general and important result (and the normalisation is precisely that required by unitarity in our conventions, although we have not yet specified the normalisation of $\hat{\mathscr{A}}_a^{(w_1)}(z_{v_1})$). Integrating over $z_v,\bar{z}_v$ corresponds to integrating over complex structures $J$ (with respect to the fixed $I$ complex structure) keeping all {\it remaining Riemann surface moduli fixed}. The subtlety of integrating infinitesimal complex structure deformations to corresponding finite deformations, from this viewpoint, corresponds to unravelling the {\it range} of the integral over $z_v,\bar{z}_v$ in (\ref{eq:intvertop}). Clearly, this will depend on the various neighbouring features on the surface. 
\sk

When the insertion (\ref{eq:intvertop1}) or (\ref{eq:intvertop})  arises from cutting open a handle one must distinguish between two cases depending on whether the degeneration results in a separating or a non-separating degeneration -- the expressions (\ref{eq:intvertop1}) and (\ref{eq:intvertop}) remain true in either case since it is entirely local (it is the {\it range} of the $z_v$ integral and the corresponding transition functions and cocycle relations that distinguish the two cases). When there are also pinch and twist moduli for the cut handle there are additional differences, in that only level-matched states propagate through separating degenerations, whereas for non-separating degenerations also non-level-matched states propagate -- we discuss this in detail in following sections.
\sk

To the computation of $\hat{B}_{z_v}$ (associated to choosing a slice in moduli space that associates translations of points and handle operators to moduli variations) there corresponds an entirely analogous expression for the energy-momentum tensor contribution (\ref{eq:TCellcontourint5nu}), the only difference being that the $\gamma_{z_v}(p_m)$ term does not necessarily decouple unless what it acts on is annihilated by $L_0^{(z_1)}-\tilde{L}_0^{(\bar{z}_1)}$:
\begin{equation}\label{eq:hatDzvR}
\begin{aligned}
\hat{D}_{z_{v}}
&=-L_{-1}^{(z_1)}(p_1)+i\gamma_{z_v}(p_1)\big(L_0^{(z_1)}-\tilde{L}_0^{(\bar{z}_1)}\big)(p_1)-\frac{1}{4}\sum_{n\geq1}\frac{1}{(n+1)!}\nabla_{\bar{z}_v}^{n-1}R_{(2)}(z_v)\tilde{L}_n^{(z_1)}(p_1)\\
\hat{D}_{\bar{z}_{v}}
&=-\tilde{L}_{-1}^{(z_1)}(p_1)-i\gamma_{z_v}(p_1)\big(L_0^{(z_1)}-\tilde{L}_0^{(\bar{z}_1)}\big)(p_1)-\frac{1}{4}\sum_{n\geq1}\frac{1}{(n+1)!}\nabla_{z_v}^{n-1}R_{(2)}(z_v)L_n^{(z_1)}(p_1),
\end{aligned}
\end{equation}
with similar relations for the $J$ complex structure coordinates, $w_1$.
\sk

Let us study the $\gamma_{z_v}(p_m)$ term in further detail. We will show that it vanishes in a given chart in holomorphic normal coordinates, but also that it is ambiguous since it is not guaranteed to vanish in all charts. 
From the defining relation (\ref{eq:gammakdfn}), identifying the modulus $t^{z_v}$ with $z_v$ and setting $m=1$, yields,
$$
\gamma_{z_v}(p_1)= 2{\rm Im}\big(\partial_{z_1}v_{z_v1}^{z_1}\big)(p_1).
$$
The quantity $\partial_{z_1}v_{z_v1}^{z_1}$ in holomorphic normal coordinates is also equal to $\nabla_{z_1}v_{z_v1}^{z_1}$, but notice that this is not globally-defined since the quantity $v_{z_v1}^{z_1}$ is only defined in the chart $(U_1^I,z_1)$ (or in $(U_1^J,w_1)$). Nevertheless, we can evaluate it locally in holomorphic normal coordinates and we do so next. 
We take into account that $v_{z_v1}^{z_1}$ is a Heaviside function, see (\ref{eq:vzv1=theta}), and learn that:
\begin{equation}\label{eq:gammazv}
\begin{aligned}
\gamma_{z_v}(p_1)
&= 2{\rm Im}\big(\partial_{z_1}\theta(r-|z_1|)\big)(p_1)\\
&=-2{\rm Im}\big(\delta(r-|z_1|)\partial_{z_1}|z_1|\big)(p_1).
\end{aligned}
\end{equation}
The derivative is in turn given by,
\begin{equation}
\partial_{z_1}|z_1|=\left\{
\begin{aligned}
&0\qquad \textrm{if $z_1=0$}\\
&\frac{1}{2}\frac{z_1}{|z_1|}\qquad\textrm{if $z_1\neq0$}.
\end{aligned}\right.
\end{equation}
which is derived using (\ref{eq:Pompeiu}), see (\ref{eq:partial|z|}). Since we wish to evaluate (\ref{eq:gammazv}) at $p_1$ where $z_1(p_1)=0$, but we also need to take the limit $r\rightarrow0$, it seems there is a potential ambiguity depending on whether we first take $r\rightarrow 0$ and then set $z_1=0$ or whether we first set $z_1=0$ and then take $r\rightarrow0$. To resolve this we can imagine smoothing the delta function (since the Heaviside function was used for convenience rather than necessity, we can imagine replacing it by a smooth monotonic function). Then it is clear what the correct limit is for (\ref{eq:gammazv}), namely we should smooth out the delta function and set $z_1=0$ since the delta function singularity is removable. This then implies that in holomorphic normal coordinates:
\begin{equation}\label{eq:gamma=0}
\gamma_{z_v}(p_1)=0,
\end{equation}
and we deduce that (\ref{eq:hatDzvR}) on $(U_1^I,z_1)$ reduces to,
\begin{equation}\label{eq:hatDzv2R}
\boxed{
\begin{aligned}
\hat{D}_{z_{v}}
&=-L_{-1}^{(z_1)}(p_1)-\frac{1}{4}\sum_{n\geq1}\frac{1}{(n+1)!}\nabla_{\bar{z}_v}^{n-1}R_{(2)}(z_v)\tilde{L}_n^{(z_1)}(p_1)\\
\hat{D}_{\bar{z}_{v}}
&=-\tilde{L}_{-1}^{(z_1)}(p_1)-\frac{1}{4}\sum_{n\geq1}\frac{1}{(n+1)!}\nabla_{z_v}^{n-1}R_{(2)}(z_v)L_n^{(z_1)}(p_1).
\end{aligned}
}
\end{equation}
Clearly though, since $v_{z_v1}^{z_1}$ is only defined in the chart $(U_1^I,z_1)$, the statement (\ref{eq:gamma=0}) not guaranteed to be satisfied throughout the entire manifold. This is related to the U(1) ambiguity discussed, e.g., in Sec.~\ref{sec:WB}.
\sk

It is also useful to have at hand the (anti-)commutator of $\hat{B}_{z_v}$ with the BRST charge, $Q_B^{(z_1)}$. Using (\ref{eq:QbLn commutators}), namely,
\begin{equation}\label{eq:QbLn commutators2}
\big\{Q_{\rm B}^{(z_1)},b_n^{(z_1)}\big\} = L_n^{(z_1)},\qquad \big\{Q_{\rm B}^{(z_1)},\tilde{b}_n^{(z_1)}\big\} = \tilde{L}_n^{(z_1)},
\end{equation}
it is immediate to show that:
\begin{equation}\label{eq:QBBDzv}
\boxed{
\big\{Q_B^{(z_1)},\hat{B}_{z_v}\big\}=\hat{D}_{z_v}
}
\end{equation}
where we made use of (\ref{eq:hatDzv2R}) and (\ref{eq:hatBzvR}). 
The fact that this relation follows immediately should be clear since both $\hat{B}_{z_v}$ and $\hat{D}_{z_v}$ are linear in their respective generators. Note that this operator acts on normal-ordered operators. In particular, the operator appearing on the right-hand side in (\ref{eq:QBBDzv}) will generate a derivative with respect to the base point of the holomorphic normal coordinate that is {\it outside} the normal ordering, whereas $L_{-1}$ generates a normal-ordered derivative of the operator on which it acts. We show this in Sec.~\ref{sec:comm dA-dA}, see also Sec.~\ref{sec:WuYang} and Sec.~\ref{sec:BRST-AC} for related comments. 

\subsection{Shifting Handles using Transition Functions}\label{sec:SHUTF}
In the previous section we derived an explicit expression for the path integral measure that determines (up to U(1)) a gauge slice in moduli space that translates a fixed-picture offshell vertex operator to integrated-picture. We found it convenient to adopt holomorphic normal coordinates, since this leads automatically to a globally well-defined measure (modulo U(1)). This fixes Weyl invariance (since holomorphic normal coordinates do) but reparametrisation invariance (generated by reparametrisations of the auxiliary real coordinates, $\sigma_1^a$, specifying the location of the vertex operator) remains unbroken\footnote{This was not manifest in the previous section but will become manifest in the current section.}. Furthermore, in the previous section we worked exclusively in terms of a metric on $\Sigma$, leading to the explicit representation (\ref{eq:hatBzvw1}). In particular, we derived this by making use of the {\it first} equality in (\ref{eq:Bintmub8z}). In this section we discuss how to proceed when we work entirely in terms of holomorphic transition functions on patch overlaps, namely the defining data of a Riemann surface when we think of the latter as a complex manifold. That is, the starting point of interest is now the {\it second} equality in (\ref{eq:Bintmub8z}). This latter approach is much more efficient as we will see.
\sk

So let us begin from the second equality in (\ref{eq:Bintmub8z}), and evaluate the right-hand side of the latter in the case where the moduli, $t^k$, of interest are identified with two worldsheet translation moduli, $z_v,\bar{z}_v$, namely:
\begin{equation}\label{eq:Bintmub8z2}
\begin{aligned}
\hat{B}_{z_v}
&=\sum_{(mn)}\frac{1}{2\pi i}\int_{C_{mn}} \!\bigg(\rmd z_m\,\frac{\partial f_{mn}(z_n;t)}{\partial z_v}\big|_{z_n}b_{z_mz_m}-\rmd\bar{z}_m\,\frac{\partial \overline{f_{mn}(z_n;t)}}{\partial z_v}\big|_{\bar{z}_n}\tilde{b}_{\bar{z}_m\bar{z}_m}\bigg),
\end{aligned}
\end{equation}
with an analogous expression for $\hat{B}_{\bar{z}_v}$, obtained from $\hat{B}_{z_v}$ by replacing $\partial_{z_v}$ by $\partial_{\bar{z}_v}$. 
By definition (recall Sec.~\ref{sec:RS} and in particular Sec.~\ref{sec:CST} and Sec.~\ref{sec:CSDI}), the transition functions $z_m=f_{mn}(z_n;t)$ are holomorphic in $z_n$ (for every $m,n$), but are not necessarily holomorphic in the moduli, $t=(z_v,\bar{z}_v,\dots)$. Following on from the discussion of Sec.~\ref{sec:SPUTF}, generically, every such transition function can depend on various moduli, but we will pick a slice in moduli space whereby the $(m,n)=(\sigma_1',\sigma_1)$ term with corresponding holomorphic transition function, 
$$
f_{mn}(z_n;t)=f_{\sigma_1'\sigma_1}(z_{\sigma_1}(\sigma),\sigma_1),
$$ 
in the sum over $m,n$ in (\ref{eq:Bintmub8z2}) depends solely (in addition to $z_{\sigma_1}(\sigma)$) on the specific moduli associated to the location of the base point, $\sigma_1$, of the frame, $z_{\sigma_1}(\sigma)$, which in terms of complex coordinates is parametrised by $z_{\sigma_1}(\sigma_1)$ (or $z_v$ with $\delta z_v\equiv -\delta z_{\sigma_1}(\sigma_1)$) and their complex conjugates and does not depend on any other moduli. Recall that $z_{\sigma_1}(\sigma)$ are in particular {\it holomorphic normal coordinates}. Furthermore, the gauge slice will be such that all remaining transition functions with $(m,n)\neq(\sigma_1',\sigma_1)$ are independent of the translation moduli $z_v,\bar{z}_v$, at least for some finite range of the latter.
\sk

Up to an irrelevant overall phase (associated to the U(1) ambiguity discussed at various points, e.g., Sec.~\ref{sec:WB}), the transition function of interest takes the form (\ref{eq:zsds(s')}), namely,
\begin{equation}\label{eq:zsds(s')x}
\begin{aligned}
z_{\sigma_1'}(\sigma)&=f_{\sigma_1'\sigma_1}(z_{\sigma_1}(\sigma),\sigma_1)\\
&=z_{\sigma_1}(\sigma)-\delta z_{v_1}-\delta \bar{z}_{v_1}\frac{1}{4}\sum_{n=1}^\infty\frac{1}{(n+1)!}\big(\nabla_{z_{\sigma_1}}^{n-1}R_{(2)}(\sigma)\big)\Big|_{\sigma=\sigma_1}\!\!\!z_{\sigma_1}(\sigma)^{n+1},
\end{aligned}
\end{equation}
so that in particular the ghost insertion (\ref{eq:Bintmub8z2}) we need to evaluate takes the form,
\begin{equation}\label{eq:Bintmub8z2x}
\begin{aligned}
\hat{B}_{z_v}
&=\frac{1}{2\pi i}\int_{C_{\sigma_1'\sigma_1}} \bigg(\rmd z_{\sigma_1'}\,\frac{\partial f_{\sigma_1'\sigma_1}(z_{\sigma_1}(\sigma),\sigma_1)}{\partial z_v}\big|_{z_{\sigma_1}}b_{z_{\sigma_1'}z_{\sigma_1'}}-\rmd\bar{z}_{\sigma_1'}\,\frac{\partial \overline{f_{\sigma_1'\sigma_1}(z_{\sigma_1}(\sigma),\sigma_1)}}{\partial z_v}\big|_{\bar{z}_{\sigma_1}}\tilde{b}_{\bar{z}_{\sigma_1'}\bar{z}_{\sigma_1'}}\bigg).\\
\end{aligned}
\end{equation}
To evaluate this we make a holomorphic change of variables, $z_{\sigma_1'}(\sigma)\mapsto z_{\sigma_1}(\sigma)$, using the transition function (\ref{eq:zsds(s')x}), taking into account that at {\it fixed} $z_{v_1},\bar{z}_{v_1}$, 
$$
\Big(\frac{\partial z_{\sigma_1'}}{\partial z_{\sigma_1}}\Big|_{z_{v_1},\bar{z}_{v_1}}\Big)^{-1}=\frac{\partial z_{\sigma_1}}{\partial z_{\sigma_1'}}\Big|_{z_{v_1},\bar{z}_{v_1}}=1.
$$
This result relies on the explicit expression (\ref{eq:zsds(s')x}) (and does not hold for generic transition functions). 
Furthermore, the contours before and after the change of variables are homotopically (or isotopically in $\Sigma$) equivalent, so that $C_{\sigma_1'\sigma_1}=C_{\sigma_1\sigma_1'}$, both of which encircle the images (under the maps $z_{\sigma_1'}$ and $z_{\sigma_1}$) of both $\sigma_1'$ and $\sigma_1$. That is, the contour orientations are the same before and after the change of variables, $z_{\sigma_1'}(\sigma)\mapsto z_{\sigma_1}(\sigma)$. So in particular (\ref{eq:Bintmub8z2x}) can be rewritten as,
\begin{equation}\label{eq:Bintmub8z2xx}
\begin{aligned}
\hat{B}_{z_v}
&=\frac{1}{2\pi i}\int_{C_{\sigma_1\sigma_1'}} \bigg(\rmd z_{\sigma_1}\,\frac{\partial f_{\sigma_1'\sigma_1}(z_{\sigma_1}(\sigma),\sigma_1)}{\partial z_v}\big|_{z_{\sigma_1}}b_{z_{\sigma_1}z_{\sigma_1}}-\rmd\bar{z}_{\sigma_1}\,\frac{\partial \overline{f_{\sigma_1'\sigma_1}(z_{\sigma_1}(\sigma),\sigma_1)}}{\partial z_v}\big|_{\bar{z}_{\sigma_1}}\tilde{b}_{\bar{z}_{\sigma_1}\bar{z}_{\sigma_1}}\bigg).
\end{aligned}
\end{equation}
The reason for making this change of variables is that now the integrand is explicitly a (anti-)holomorphic function of $z_{\sigma_1}(\sigma)$ (or $\bar{z}_{\sigma_1}(\sigma)$). Evaluating the two contour integrals making use of (\ref{eq:zsds(s')x}) and that the $z_{\sigma_1}(\sigma)$ chart is centred at $\sigma=\sigma_1$ the result is immediate,
\begin{equation}\label{eq:Bintmub8z2xxx}
\boxed{
\begin{aligned}
\hat{B}_{z_v}&=-b_{-1}^{(z_{\sigma_1})}(\sigma_1)-\frac{1}{4}\sum_{n=1}^\infty\frac{1}{(n+1)!}\big(\nabla_{\bar{z}_{\sigma_1}}^{n-1}R_{(2)}(\sigma)\big)\Big|_{\sigma=\sigma_1}\!\!\!\tilde{b}_n^{(z_{\sigma_1})}(\sigma_1)\\
\hat{B}_{\bar{z}_v}&=-\tilde{b}_{-1}^{(z_{\sigma_1})}(\sigma_1)-\frac{1}{4}\sum_{n=1}^\infty\frac{1}{(n+1)!}\big(\nabla_{z_{\sigma_1}}^{n-1}R_{(2)}(\sigma)\big)\Big|_{\sigma=\sigma_1}\!\!\!b_n^{(z_{\sigma_1})}(\sigma_1)
\end{aligned}
}
\end{equation}
where in the second line we also displayed the anti-chiral contribution associated to moduli deformations $\delta\bar{z}_{v_1}$. 
This operator acts on local operators inserted at $\sigma_1$ and constructed using the frame $z_{\sigma_1}(\sigma)$. 
\sk

As discussed in the previous section, the product, $\hat{B}_{z_v}\hat{B}_{\bar{z}_v}$,  translates fixed-picture vertex operators to integrated picture in the gauge slice associated to picking holomorphic normal coordinates that leads to a slice that is ``as flat as possible'' where the local operator is inserted. In particular, (\ref{eq:intvertop1}) continues to hold but perhaps it is useful to make explicit that (taking (\ref{eq:dzsigma*-measure}) into account and that $\delta z_v=-\delta z_{\sigma_1}(\sigma_1)$) we can also write it as follows,
\begin{equation}\label{eq:intvertop1sigma_1}
\boxed{
\mathscr{A}_a^{(z_{\sigma_1})}\dfn \int \rmd^2\sigma_12\sqrt{\det g_{ab}(\sigma_1)}\hat{B}_{z_v}\hat{B}_{\bar{z}_v}\hat{\mathscr{A}}_a^{(z_{\sigma_1})}(\sigma_1)
}
\end{equation}
Since $z_{\sigma_1}(\sigma_1)$ transforms as a scalar under reparametrisations, $\sigma_1\mapsto \hat{\sigma}_1(\sigma_1)$, recall (\ref{eq:zsigma* scalar}), and since the quantity $\hat{B}_{z_v}\hat{B}_{\bar{z}_v}\hat{\mathscr{A}}_a^{(z_{\sigma_1})}(\sigma_1)$ is constructed entirely out of $z_{\sigma_1}(\sigma_1)$ so does the quantity $\hat{B}_{z_v}\hat{B}_{\bar{z}_v}\hat{\mathscr{A}}_a^{(z_{\sigma_1})}(\sigma_1)$ transform as a scalar under such reparametrisations. So that this operator is globally well-defined is from this viewpoint immediate.
\sk

Regarding the corresponding transformation rule under local Weyl rescalings,
$$
g_{ab}(\sigma)\mapsto e^{\delta\phi(\sigma)}g_{ab}(\sigma),
$$
first note from Sec.~\ref{sec:HTFWR} and in particular (\ref{eq:dz from Weyl3}) that Weyl rescalings of the metric induce a holomorphic change of coordinates, $z_{\sigma_1}\mapsto w_{\sigma_1}(z_{\sigma_1})$. Up to an overall phase and for operators inserted at $\sigma=\sigma_1$ this transformation reads,
$$
z_{\sigma_1}(\sigma)\mapsto w_{\sigma_1}(\sigma)=z_{\sigma_1}(\sigma)+
\sum_{n=1}^\infty \frac{1}{(n+1)!}\big(\nabla^{n}_{z_{\sigma_1}}\delta\phi(\sigma)\big)\big|_{\sigma=\sigma_1}z_{\sigma_1}(\sigma)^{n+1}.
$$
Recall also that $w_{\sigma_1}(\sigma_1)=z_{\sigma_1}(\sigma_1)=0$ by construction, so that these Weyl transformations leave fixed the base point $\sigma=\sigma_1$. 
But since the operators (\ref{eq:intvertop1sigma_1}) are normal ordered such a holomorphic change of coordinates does not correspond to the naive one, essentially because of the change in normal ordering in addition to the naive change of coordinates. Choosing a slice in moduli space by picking holomorphic normal coordinates centred at the local vertex operator insertions spontaneously breaks the invariance under holomorphic changes of coordinates. So in fact since we have chosen a slice in moduli space we do not need to have invariance under holomorphic changes of coordinates. But it is useful nevertheless to think carefully about how the operators we  construct (and in particular the handle operators) transform under Weyl transformations since this allows us to keep track of this Weyl dependence of amplitudes (which must of course explicitly cancel out of full amplitudes when asymptotic vertex operators are taken onshell). So from the above comments it is useful to primarily understand how to change normal ordering by a holomorphic change of coordinates since this is what Weyl transformations induce. We study normal ordering and changes in normal ordering in path integral language in Sec.~\ref{sec:NOII}. For more on Weyl transformations of local operators defined using holomorphic normal coordinates see Sec.~\ref{sec:LOUWR}.
\sk

The result (\ref{eq:Bintmub8z2xxx}) is equivalent to that derived in the previous section, see (\ref{eq:hatBzvR}), which is a good consistency check, but the derivation here used only the defining properties of a Riemann surface and a choice of holomorphic chart coordinates (which in turn fixes Weyl invariance and leaves reparametrisation invariance intact). In the previous section we rather took a metric viewpoint throughout the calculation. It is clearly more efficient (and perhaps more transparent) to work in terms of holomorphic transition functions as done in this section.
\sk

In Sec.~\ref{sec:mainSCS} we apply the results of Sec.~\ref{sec:RS} and Sec.~\ref{sec:TPIM} on Riemann surfaces and the path integral measure to cutting and gluing string amplitudes using an offshell coherent state basis. The formalism we have developed enables us to proceed in a globally well-defined manner, so that the corresponding analysis is exact. 

\subsection{Full Path Integral II: `Integrated Picture'}\label{sec:TEC}
In this section we tie up a loose end which is associated to some exceptional cases related to string amplitudes with ``too few'' external vertex operators here. This is not necessary for most of the remaining document (except in Sec.~\ref{sec:EA}) and unless specifically of interest can be omitted on a first reading.
\sk

When the number, $\n$, of external vertex operators is not sufficient to saturate the number of conformal Killing vectors (in particular the cases $\n=0,1,2$ for $\g=0$ and $\n=0$ for $\g=1$) care is required, and it is safest to translate the external vertex operators in the path integral into integrated-picture. In particular, the derivation leading to the path integral $S_{\g,\n}$ in (\ref{eq:fullpathintegral}) generically (but not always\footnote{Examples where the standard gauge fixing does not break down so that one can directly apply (\ref{eq:fullpathintegral}) is when one uses external offshell vertex operators, $\mathscr{A}_a^{(z_1)}$ and their duals $\mathscr{A}^a_{(z_2)}$, of indefinite ghost number, such as those constructed in the current document. We compute $S_{0,1},S_{0,2}$ and $S_{1,0}$ explicitly using these offshell vertex operators in Sec.~\ref{sec:EA}.}) can break down in the cases $S_{0,0},S_{0,1},S_{0,2}$ and $S_{1,0}$. A related observation (including a corresponding cure) for the tree-level two-point amplitude, $S_{0,2}$, was discussed in \cite{ErbinMaldacenaSkliros19} in the Polyakov formalism. There is an easy trick (see Sec.~7.3 in \cite{Polchinski_v1}) to deal with the case $(\g,\n)=(1,0)$, and we will here generalise that derivation (if only briefly) and present a procedure that allows one to compute all these amplitudes, $S_{0,0},S_{0,1},S_{0,2}$ and $S_{1,0}$, in the BRST approach, {\it provided} the following two mild assumptions are satisfied:
\begin{itemize}
\item[{\bf(A)}] The OPE's of $c,\tilde{c}$ with the fixed-picture external vertex operators, $\hat{\mathscr{V}}_j$, vanish at coincident points;
\item[{\bf(B)}] The OPE's of $c,\tilde{c}$ with the integrated-picture external vertex operators, $\mathscr{V}_j$, are  non-singular. 
\end{itemize}
If these OPE's are singular the following procedure may work but this has not been shown. Under these conditions, ${\bf(A)}$ and ${\bf(B)}$, we start from the full path integral (\ref{eq:fullpathintegral}), taking the number of external vertex operators, $\n$, to be initially {\it greater or equal} to the (complex) number, $\kappa$, of {\it conformal Killing vectors} (CKV) on $\Sigma$,
\begin{equation}
\kappa\dfn( \#_{\mathbf{C}}\textrm{ CKV})=
\left\{
\begin{aligned}
&3\qquad \textrm{if $\Sigma=S^2$}\\
&1\qquad \textrm{if $\Sigma=T^2$}\\
&0\qquad \textrm{if $\Sigma=\Sigma_{\g>1}$}
\end{aligned}
\right.
\end{equation}
The procedure then consists of the following steps: 
\begin{itemize}
\item[{\bf 1.}] Undo the conformal Killing group gauge fixing by translating all external vertex operators to integrated picture. 
\item[{\bf2.}] {\it Then} explicitly set $\n=0,1,2,\dots$, to the value of interest in the resulting expression. 
\item[{\bf3a.}] In some cases (such as for\footnote{This was discussed in Sec.~7.3 in \cite{Polchinski_v1}, and we will generalise that argument here.} $S_{1,0}$) it is possible to leave some of the residual symmetry unfixed and rather divide explicitly by the corresponding volume. This works well when the residual volume is finite. If this condition is not satisfied, proceed to step 3b.
\item[{\bf3b.}] Fadeev-Popov gauge fix using gauge fixing functions that are ``appropriate'' for the specific amplitude of interest.
\end{itemize}
This procedure can be carried out for all amplitudes, $S_{\g,\n}$, (i.e.~for any positive integers, $\g,\n=0,1,\dots$), and (unlike in \cite{Witten12c}) also applies to cases where the kinematics is not necessarily generic. To elaborate briefly on this last point, the obstruction discussed in \cite{Witten12c} in the case of exceptional kinematics is evaded by working with more general representatives in the BRST cohomology class of interest that are not conformal primaries. 
The first step, ${\bf 1.}$, replaces the factor $1/n_R$ in (\ref{eq:fullpathintegral}) by $1/V_R$, where $V_R$ is the volume of the resulting residual unfixed symmetry. 
Regarding the use of the adjective `appropriate' in  {\bf 3b.}, note for example that it does not make sense to pick gauge-fixing functions that fix the position of three vertex operators when there are only two present \cite{ErbinMaldacenaSkliros19}. 
\sk

The starting point is the full gauge-fixed (up to a residual discrete symmetry accounted for by the factor $1/n_R$) path integral (\ref{eq:fullpathintegral}), repeated here for convenience:
\begin{equation}\label{eq:fullpathintegralX}
\begin{aligned}
S_{\g,\n}&=e^{-\chi(\Sigma_\g)\Phi}\int_{\mathcal{M}_{\g,\n}}\frac{\rmd^{2\m}\tau}{n_R}\Big\langle\prod_{k=1}^\m\hat{B}_{\tau^k}\hat{B}_{\bar{\tau}^k}\prod_{j=1}^\n\hat{\mathscr{V}}_j^{(z_{\sigma_j})}\Big\rangle_{\Sigma_\g},
\end{aligned}
\end{equation}
where we here made explicit that the fixed-picture vertex operators, $\hat{\mathscr{V}}_j^{(z_{\sigma_j})}$, are defined using frame coordinates, $z_{\sigma_j}$, and inserted at $\sigma_j$ at which $z_{\sigma_j}(\sigma_j)=0$. 
To implement the first step, ${\bf 1.}$, we pick a subset of vertex operators, $\{\hat{\mathscr{V}}_a^{(z_{\sigma_a})}\}\subset \{\hat{\mathscr{V}}_j^{(z_{\sigma_j})}\}$, such that $a=1,\dots,\kappa$ (taking into account that we initially consider the case $\n\geq\kappa$), and observe that:
\begin{equation}\label{eq:ccBB=BBcc+1+..}
\tilde{c}_1^{(z_{\sigma_a})} c_1^{(z_{\sigma_a})}\hat{B}_{z_{v_a}}\hat{B}_{\bar{z}_{v_a}} =1+\hat{B}_{z_{v_a}}\hat{B}_{\bar{z}_{v_a}}  \tilde{c}_1^{(z_{\sigma_a})} c_1^{(z_{\sigma_a})}+\hat{B}_{\bar{z}_{v_a}} \tilde{c}_1^{(z_{\sigma_a})}+\hat{B}_{z_{v_a}}c_1^{(z_{\sigma_a})},
\end{equation}
where we used the explicit expressions for the measure contributions (\ref{eq:Bintmub8z2xxx}) that translate fixed-picture to integrated picture. The relation (\ref{eq:ccBB=BBcc+1+..}) holds as an operator statement inside the path integral. Assumption ${\bf (A)}$ can then be used to infer that,
\begin{equation}\label{eq:Vhat=ccBBVhat}
:\!\hat{\mathscr{V}}_a^{(z_{\sigma_a})}\!\!:_{z_{\sigma_a}}\,\, = \,\,:\!\tilde{c}_1^{(z_{\sigma_a})} c_1^{(z_{\sigma_a})}\hat{B}_{z_{v_a}}\hat{B}_{\bar{z}_{v_a}} \hat{\mathscr{V}}_a^{(z_{\sigma_a})}\!\!:_{z_{\sigma_a}}\,\,,
\end{equation}
since the three remaining terms on the right-hand side in (\ref{eq:ccBB=BBcc+1+..}) then vanish when acting on $\hat{\mathscr{V}}_a^{(z_{\sigma_a})}$. 
We made explicit the normal ordering, and this relation holds 
for all external vertex operators (and in particular the $\kappa$ vertex operators labelled by `$a$' out of the full set $\{\hat{\mathscr{V}}_j^{(z_{\sigma_j})}\}$ appearing in (\ref{eq:fullpathintegralX})). Assumption ${\bf (B)}$ then allows us to factorise the right-hand side in (\ref{eq:Vhat=ccBBVhat}),
\begin{equation}
:\!\hat{\mathscr{V}}_a^{(z_{\sigma_a})}\!\!:_{z_{\sigma_a}}\,\, = \,\,:\!\tilde{c}_1^{(z_{\sigma_a})} c_1^{(z_{\sigma_a})}\!:_{z_{\sigma_a}}\,\,:\!\hat{B}_{z_{v_a}}\hat{B}_{\bar{z}_{v_a}} \hat{\mathscr{V}}_a^{(z_{\sigma_a})}\!\!:_{z_{\sigma_a}}\,\,,
\end{equation}
since the OPE's are by assumption non-singular. In fact, given the OPE's are non-singular we can also use the operator-state correspondence to rewrite the quantity $:\!\tilde{c}_1^{(z_{\sigma_a})} c_1^{(z_{\sigma_a})}\!:_{z_{\sigma_a}}$ as a local operator. 
\begin{equation}
:\!\hat{\mathscr{V}}_a^{(z_{\sigma_a})}\!\!:_{z_{\sigma_a}}\,\, = \,\,:\!\tilde{c}^{(z_{\sigma_a})} c^{(z_{\sigma_a})}\!:_{z_{\sigma_a}}\,\,:\!\hat{B}_{z_{v_a}}\hat{B}_{\bar{z}_{v_a}} \hat{\mathscr{V}}_a^{(z_{\sigma_a})}\!\!:_{z_{\sigma_a}}\,\,.
\end{equation}

The factor $:\!\tilde{c}^{(z_{\sigma_a})} c^{(z_{\sigma_a})}\!:_{z_{\sigma_a}}$ transforms as a $(-1,-1)$ conformal primary under holomorphic reparametrisations, $z_{\sigma_a}\mapsto w_{\sigma_a}(z_{\sigma_a})$, in particular,
$$
:\!\tilde{c}^{(z_{\sigma_a})} c^{(z_{\sigma_a})}\!:_{z_{\sigma_a}}\,\, = \Big|\frac{\partial z_{\sigma_a}}{\partial w_{\sigma_a}}\Big|^2 :\!\tilde{c}^{(w_{\sigma_a})} c^{(w_{\sigma_a})}\!:_{w_{\sigma_a}}.
$$
This reparametrisation might simply correspond to a shift, and we do {\it not} require that the $w_{z_{\sigma_a}}$ frame is centred at $\sigma_a$. Making this change of variables in a product over $a=1,\dots,\kappa$ in the above relation we learn that,
\begin{equation}\label{eq:prodaVdzdw}
\prod_{a=1}^\kappa:\!\hat{\mathscr{V}}_a^{(z_{\sigma_a})}\!\!:_{z_{\sigma_a}}\,\, = \Big(\prod_{a=1}^\kappa \big|\frac{\partial z_{\sigma_a}}{\partial w_{\sigma_a}}\big|^2\Big)\,\Big(\prod_{a=1}^\kappa\,\,:\!\tilde{c}^{(w_{\sigma_a})} c^{(w_{\sigma_a})}\!:_{w_{\sigma_a}}\Big)\,\,\Big(\prod_{a=1}^\kappa:\!\hat{B}_{z_{v_a}}\hat{B}_{\bar{z}_{v_a}} \hat{\mathscr{V}}_a^{(z_{\sigma_a})}\!\!:_{z_{\sigma_a}}\,\,\Big).
\end{equation}
Let us denote conformal Killing vector (components) in the $z_{\sigma_a}$ frame by: $\psi_i^{(z_{\sigma_a})}(\sigma_j)$, with a corresponding expression for the $w_{\sigma_a}$-frame expression, and $i,j=1,\dots,\kappa$. These transform in the same way as do the $c$-ghost insertions under holomorphic reparametrisations, and in particular therefore:
\begin{equation}\label{eq:dzdw=dets}
\prod_{a=1}^\kappa \big|\frac{\partial z_{\sigma_a}}{\partial w_{\sigma_a}}\big|^2=\left|\frac{\det \psi_a^{(z_{\sigma_b})}(\sigma_b)}{\det \psi_a^{(w_{\sigma_b})}(\sigma_b)}\right|^2
\end{equation}
where the determinants are over the indices $a,b$. 
\sk

Choosing a gauge slice for the integral over moduli in (\ref{eq:fullpathintegralX}) such that we associate $\n-\kappa$ of the measure contributions to promoting $\n-\kappa$ fixed-picture vertex operators to integrated picture, we can rearrange the path integrand in (\ref{eq:fullpathintegralX}) as follows,
\begin{equation}\label{eq:prodrearrange}
\begin{aligned}
\prod_{k=1}^\m\hat{B}_{\tau^k}\hat{B}_{\bar{\tau}^k}\prod_{j=1}^\n\hat{\mathscr{V}}_j^{(z_{\sigma_j})} = \prod_{k=1}^{\m_\g}\hat{B}_{\tau^k}\hat{B}_{\bar{\tau}^k}\prod_{j=1}^{\n-\kappa}\hat{B}_{z_{v_j}}\hat{B}_{\bar{z}_{v_j}}\hat{\mathscr{V}}_j^{(z_{\sigma_j})}\prod_{a=1}^\kappa\hat{\mathscr{V}}_a^{(z_{\sigma_a})},
\end{aligned}
\end{equation}
where we defined:
\begin{equation}
\m_\g =
\left\{
\begin{aligned}
&0\qquad\textrm{when $\g=0$}\\
&1\qquad\textrm{when $\g=1$}\\
&3\g-3\qquad \textrm{when $\g>1$}.
\end{aligned}
\right.
\end{equation}

We now substitute (\ref{eq:dzdw=dets}) into (\ref{eq:prodaVdzdw}), which is in turn substituted into (\ref{eq:prodrearrange}), leading to the following representation for the path integrand,
\begin{equation}\label{eq:prodrearrange2}
\begin{aligned}
&\prod_{k=1}^\m\hat{B}_{\tau^k}\hat{B}_{\bar{\tau}^k}\prod_{j=1}^\n\hat{\mathscr{V}}_j^{(z_{\sigma_j})} = \prod_{k=1}^{\m_\g}\hat{B}_{\tau^k}\hat{B}_{\bar{\tau}^k}\prod_{j=1}^{\n-\kappa}\hat{B}_{z_{v_j}}\hat{B}_{\bar{z}_{v_j}}\hat{\mathscr{V}}_j^{(z_{\sigma_j})}\\
&\qquad\qquad\qquad\times
\Big|\frac{\det \psi_a^{(z_{\sigma_b})}(\sigma_b)}{\det \psi_a^{(w_{\sigma_b})}(\sigma_b)}\Big|^2\,\Big(\prod_{a=1}^\kappa\,\,\tilde{c}^{(w_{\sigma_a})} c^{(w_{\sigma_a})}\Big)\,\,\Big(\prod_{a=1}^\kappa \hat{B}_{z_{v_a}}\hat{B}_{\bar{z}_{v_a}} \hat{\mathscr{V}}_a^{(z_{\sigma_a})}\Big)\\
&\qquad= \Big(\prod_{k=1}^{\m_\g}\hat{B}_{\tau^k}\hat{B}_{\bar{\tau}^k}\Big)\Big|\frac{\det \psi_a^{(z_{\sigma_b})}(\sigma_b)}{\det \psi_a^{(w_{\sigma_b})}(\sigma_b)}\Big|^2\,\Big(\prod_{a=1}^\kappa\,\,\tilde{c}^{(w_{\sigma_a})} c^{(w_{\sigma_a})}\Big)\,\,\Big(\prod_{j=1}^\n \hat{B}_{z_{v_j}}\hat{B}_{\bar{z}_{v_j}} \hat{\mathscr{V}}_j^{(z_{\sigma_j})}\Big)\\
\end{aligned}
\end{equation}
where in the second equality we gathered the $\n$ integrated-picture vertex operators into a single product. We now substitute (\ref{eq:prodrearrange2}) into the path integral expression (\ref{eq:fullpathintegralX}), multiply left- and right-hand sides in the resulting expression by $n_R|\psi_a^{(z_{\sigma_b})}(\sigma_b)|^{-2}$,
\begin{equation}\label{eq:fullpathintegralX2}
\begin{aligned}
&n_R\big|\det\psi_a^{(z_{\sigma_b})}(\sigma_b)\big|^{-2}S_{\g,\n}=\\
&\qquad=e^{-\chi(\Sigma_\g)\Phi}\int_{\mathcal{M}_{\g,\n}}\rmd^{2\m_\g}\tau \int \rmd^{2(\n-\kappa)}z_{v}\Big\langle\prod_{k=1}^{\m_\g}\hat{B}_{\tau^k}\hat{B}_{\bar{\tau}^k}\frac{\prod_{a=1}^\kappa\,\,\tilde{c}^{(w_{\sigma_a})} c^{(w_{\sigma_a})}}{|\det \psi_a^{(w_{\sigma_b})}(\sigma_b)|^2}\prod_{j=1}^\n \hat{B}_{z_{v_j}}\hat{B}_{\bar{z}_{v_j}} \hat{\mathscr{V}}_j^{(z_{\sigma_j})}\Big\rangle_{\Sigma_\g},
\end{aligned}
\end{equation}
and integrate left- and right-hand sides in (\ref{eq:fullpathintegralX2}) over the remaining $\kappa$ vertex operators with measure $\rmd^{2\kappa}z_v$. Rearranging the resulting expression we are led to the following result for the {\it full path integral in integrated picture}:
\begin{equation}\label{eq:fullpathintegralX3}
\begingroup\makeatletter\def\f@size{11}\check@mathfonts
\def\maketag@@@#1{\hbox{\m@th\large\normalfont#1}}%
\boxed{
\begin{aligned}
&S_{\g,\n}=e^{-\chi(\Sigma_\g)\Phi}\int_{\mathcal{M}_{\g,\n}}\frac{\rmd^{2\m_\g}\tau}{V_R} \Big\langle\prod_{k=1}^{\m_\g}\hat{B}_{\tau^k}\hat{B}_{\bar{\tau}^k}\frac{\prod_{a=1}^\kappa\,\,\tilde{c}^{(w_{\sigma_a})} c^{(w_{\sigma_a})}(\sigma_a)}{|\det \psi_a^{(w_{\sigma_b})}(\sigma_b)|^2}\prod_{j=1}^\n \int_{\Sigma} \rmd^2z_{v_j}\hat{B}_{z_{v_j}}\hat{B}_{\bar{z}_{v_j}} \hat{\mathscr{V}}_j^{(z_{\sigma_j})}\Big\rangle_{\Sigma_\g}
\end{aligned}
}
\endgroup
\end{equation}
where we defined the quantity $V_R$ which is precisely the residual {\it volume of the conformal Killing group} (including the residual discrete symmetry factor, $n_R$):
\begin{equation}\label{eq:VR}
\boxed{V_R\dfn n_R\int_{\Sigma^{\otimes \kappa}} \rmd^{2\kappa}z_{v}\big|\det\psi_a^{(z_{\sigma_b})}(\sigma_b)\big|^{-2}}
\end{equation}
where $\rmd^{2\kappa}z_v\equiv \rmd^2z_{v_1}\dots \int \rmd^2z_{v_\kappa}$. Taking (\ref{eq:dzsigma*-measure}) into account, and that $\delta z_{v_j}=-\delta z_{\sigma_j}(\sigma_j)$, we could have also written: 
$
\rmd^{2}z_{v_j}=\rmd^2\sigma_j2\sqrt{\det g(\sigma_j)}.
$ 
\sk

Notice that in order for this procedure to make sense the holomorphic reparametrisation, $w_{\sigma_a}(z_{\sigma_a})$, for all $a=1,\dots,\kappa$, must be such that it remains {\it fixed} under variations, $\delta z_{v_a}\equiv -\delta z_{\sigma_a}(\sigma_a)$. In particular, a {\it general} holomorphic reparametrisation that does {\it not} leave fixed the base point at a generic point, $\sigma$ in the auxiliary system of coordinates can be taken to have the general form:
\begin{equation}\label{eq:CKGwz}
w_{\sigma_a}(\sigma) = z_{\sigma_a}(\sigma)+\sum_{n=1}^\infty \epsilon_n(\sigma_a)z_{\sigma_a}(\sigma)^{n+1}+v_a(\sigma_a),
\end{equation}
for $\epsilon_n(\sigma_a)$ (and for every $n=1,2,\dots$) an unspecified function of $\sigma_a$. Then the necessary requirement is $\delta w_{\sigma_a}(\sigma_a) = 0$, i.e., 
$
\delta z_{\sigma_a}(\sigma_a)=-\delta v_a(\sigma_a)
$ since $z_{\sigma_a}(\sigma)|_{\sigma=\sigma_a}\equiv0$. 
In particular, this then ensures that under variations $\delta z_{v_a}\equiv -\delta z_{\sigma_a}(\sigma_a)$,
$$
\delta \Big(\tfrac{\prod_{a=1}^\kappa\,\,\tilde{c}^{(w_{\sigma_a})} c^{(w_{\sigma_a})}(\sigma_a)}{|\det \psi_a^{(w_{\sigma_b})}(\sigma_b)|^2} \Big)= 0,
$$
because the quantity in the parenthesis is constructed entirely out of the frame coordinates $w_{\sigma_a},\bar{w}_{\sigma_a}$ evaluated at $\sigma=\sigma_a$.
\sk

The main point now is that (\ref{eq:fullpathintegralX3}) makes sense for {\it any} number of external vertex operator insertions, $\n=0,1,2,\dots$. This is to be contrasted with the path integral (\ref{eq:fullpathintegralX}) with vertex operators in fixed picture where the conformal Killing group symmetries were already gauge-fixed. 
\sk

If the residual volume, $V_R$, is finite (such as in the case $\Sigma=T^2$) then it is convenient to allow for the overcounting and compute $V_R$ explicitly. When $V_R$ is infinite one can again take (\ref{eq:fullpathintegralX3}) as the starting point and evaluate it using the Fadeev-Popov method, which is to choose a convenient set of $\kappa$ (or 2$\kappa$ for the real case) gauge-fixing functions, $F_a$, (subject to the condition that these are {\it not} invariant under the group of the residual symmetries) and insert the quantity:
\begin{equation}\label{eq:FP}
1=\Delta\int \mathcal{D}g\,\prod_{a=1}^{2\kappa} \delta(F_a^g)
\end{equation}
into the path integral (\ref{eq:fullpathintegralX3}), where one formally identifies $V_R$ (as given in (\ref{eq:VR})) with $\int \mathcal{D}g$, corresponding to the integral over the volume of the residual gauge group. The quantity $\Delta$ is then the Fadeev-Popov determinant (which {\it is} gauge-invariant) that can be calculated by this defining equation (\ref{eq:FP}) after specifying the gauge-fixing functions. This procedure was discussed recently in \cite{ErbinMaldacenaSkliros19} (in the Polyakov formalism) where it was shown that the $S_{0,2}$ path integral for onshell external states is equal to the covariant free particle result (\ref{eq:2ptonshell}). Here we have generalised this procedure to the more general BRST formalism and to arbitrary amplitudes.
\sk

Let us consider the torus example in further detail, since we will be needing the result in Sec.~\ref{sec:MI}. So we wish to compute $V_R$ in the case $\Sigma=T^2$. The torus has one conformal Killing vector, corresponding to rigid translations across the $A$ and $B$ cycles, so the residual group is $G=$U(1)$\times$U(1), and using the above notation $\kappa=1$. Since the torus is flat we can choose a basis for the corresponding conformal Killing vectors such that $\psi_1^{(z_{\sigma_1})}=\alpha$, with $\alpha$ an arbitrary complex number. It is convenient (and permissible globally since the torus is flat) to choose the holomorphic reparametrisation (\ref{eq:CKGwz}) such that all the $\epsilon_n=0$, in which case the change of variables implies that $\psi_1^{(z_{\sigma_1})}=\psi_1^{(w_{\sigma_1})}$, since these transform as vector components (and are hence invariant under rigid shifts). Adopting a convenient set of coordinates with the usual identifications, $z_v\sim z_v+1$ and $z_v\sim z_v+\tau$, where $\tau=\tau_1+i\tau_2$ is the single modulus of $T^2$, we learn that:
$$
V_R = n_R|\alpha|^{-2}\int \rmd^2z_v = n_R|\alpha|^{-2}2\tau_2.
$$
Substituting this into (\ref{eq:fullpathintegralX3}), noting that there is a residual discrete $\mathbf{Z}_2$ symmetry associated to invariance under $z_v\rightarrow -z_v$, which sets $n_R=2$ \cite{Polchinski86}, that the Euler characteristic of the torus, $\chi(T^2)=0$, and $\m_\g|_{\g=1}=1$, we see that,
\begin{equation}\label{eq:fullpathintegralX4a}
\begingroup\makeatletter\def\f@size{11}\check@mathfonts
\def\maketag@@@#1{\hbox{\m@th\large\normalfont#1}}%
\boxed{
\begin{aligned}
&S_{1,\n}=\int_{\mathcal{M}_{1,\n}}\frac{\rmd^{2}\tau}{4\tau_2} \Big\langle\hat{B}_{\tau}\hat{B}_{\bar{\tau}}\tilde{c}^{(w_{\sigma_1})} c^{(w_{\sigma_1})}\prod_{j=1}^\n \int_{\Sigma} \rmd^2z_{v_j}B_{z_{v_j}}B_{\bar{z}_{v_j}} \hat{\mathscr{V}}_j^{(z_{\sigma_j})}\Big\rangle_{T^2}
\end{aligned}
}
\endgroup
\end{equation}
where we also took into account (from the above discussion) that the explicit determinant in (\ref{eq:fullpathintegralX3}) also equals $|\det \psi_1^{(w_{\sigma_1})}(\sigma_1)|^2=|\alpha|^2$. The path integral (\ref{eq:fullpathintegralX4a}) is also valid when $\n=0$ where it is customary to denote it by $Z_{T^2}$,
\begin{equation}\label{eq:fullpathintegralX4}
\begin{aligned}
Z_{T^2}=\int_{\mathcal{M}_{1,0}}\frac{\rmd^{2}\tau}{4\tau_2} \Big\langle\hat{B}_{\tau}\hat{B}_{\bar{\tau}}\tilde{c}^{(w_{\sigma_1})} c^{(w_{\sigma_1})}\Big\rangle_{T^2},
\end{aligned}
\end{equation}
and $\mathcal{M}_{1,0}$ is a fundamental domain of SL(2,$\mathbf{Z})/\mathbf{Z}_2$, e.g.,
$$
\mathcal{M}_{1,0} = \big\{\tau,\bar{\tau}\,\,\big|-\tfrac{1}{2}\leq \tau_2\leq\tfrac{1}{2},|\tau|\geq1\big\},
$$
which corresponds to integrating over all tori with distinct complex structures. 
In Sec.~\ref{sec:MI} we evaluate (\ref{eq:fullpathintegralX4}) explicitly using an $A$-cycle handle operator insertion to rewrite it as a correlator on $S^2$, and demonstrate that the result is indeed modular invariant (which corresponds to invariance of the integrand of (\ref{eq:fullpathintegralX4}) under $\tau\mapsto -1/\tau$ and $\tau\mapsto \tau+1$).

\section{Normal Ordering of Composite Operators}\label{sec:NOII}

In this subsection we will study how generic normal-ordered composite operators in a free CFT change under {\it finite} changes of the normal-ordering prescription. To place the discussion into context, if one wishes to construct a Riemann surface by gluing together charts using holomorphic transition functions as discussed in Sec.~\ref{sec:TFCR}, then one must confront the fact that as local operators are translated across the surface one must transform these local composite operators across patch overlaps. Since these composite operators are normal-ordered there is also a (typically {\it finite}) change in the normal ordering {\it prescription} that is induced by this holomorphic transition function. As a result, the transformation properties of composite operators are not the naive ones and it is non-trivial to ensure that one ends up with a globally well-defined construction. In particular, these composite operators (e.g., handle operators) do not in general transform as tensors on patch overlaps, so how can one ensure that one ends up with a globally well-defined construction? 
\sk

One viewpoint that we adopt in this article is to adopt Polchinski's resolution \cite{Polchinski88}, which is to consider an auxiliary (e.g.~real) fixed coordinate system, $\sigma^a$, and glue together charts using reparametrisations, $\sigma\mapsto \hat{\sigma}(\sigma)$. One then adopts holomorphic normal coordinates, $z_{\sigma_1}(\sigma)$, which have the property that they transform as scalars (modulo U(1)) on patch overlaps, recall (\ref{eq:w's'=ws}) and (\ref{eq:zsigma* scalar}). One can then normal order using these coordinates and then the corresponding composite local operators (which are constructed then entirely out of $z_{\sigma_1}(\sigma)$ and evaluated at $\sigma=\sigma_1$) will also transform as scalars on patch overlaps (modulo U(1)), thus enabling one to extend their definition to the entire Riemann surface in a globally well-defined manner. The point of this is then that one does not need to change the normal ordering prescription across patch overlaps, and so is a significant simplification. Using holomorphic normal coordinates to normal order operators also has the effect of gauge fixing invariance under Weyl transformations. And (much like Riemann normal coordinates on real manifolds) it also preserves covariance, in that one can (and we will) proceed without making a choice for the underlying 2D Riemannian or Ricci curvature.
\sk

Having said that, if one wishes (for whatever reason)  to change the normal ordering prescription of a given composite operator one would like to understand how to carry this out, even if the change in normal ordering is {\it finite}. E.g., it may sometimes be desirable to pullback the various handle operators onto a sphere using, e.g., SL(2,$\mathbf{C}$), or it may be desirable to study the transformation properties of the path integral measure and handle operators, and/or of external offshell vertex operators, under Weyl transformations. In this section we discuss how to change the normal ordering prescription (after defining such a prescription) when it is desirable to do so; nevertheless, for the reasons mentioned in the opening two paragraphs of this section the reader can perhaps skip this section upon a first reading and revisit it as necessary.
\sk

We will adopt an approach that is perhaps somewhat different from the usual CFT approach. In particular, emphasis will be placed on disentangling the following notions that are usually combined into a single notion in standard CFT literature,
\begin{itemize}
\item[{\bf (a)}]  {\it Change of normal ordering keeping coordinates fixed}
\item[{\bf (b)}]  {\it Change of coordinates keeping normal ordering fixed}
\item[{\bf (c)}] {\it Change of coordinates}
\end{itemize}
the first of which we will explore in detail in Sec.~\ref{sec:HCNO}. This distinction between {\bf (a)} and {\bf (b)} is closely related to the distinction between primary and non-primary operators. But in order to set the groundwork let us briefly recall some standard material regarding normal ordering. 

\subsection{Preliminary Remarks and CFT Approach}

Let us begin the discussion with a very simple example, that will in turn highlight some relevant features as well as the relation to the usual CFT way of thinking. The general prescription for normal ordering composite operators and finite changes in normal ordering will be discussed in the following two subsections.
\sk

Suppose that we consider a local operator of the form:
\begin{equation}\label{eq:d2xeipx-ex}
\mathscr{A}^{(z_1)}=\,:\!\partial^2x\,e^{ip\cdot x(0)}\!:_{z_1}
\end{equation}
This operator is certainly not primary and so one expects it will transform non-trivially under a change of frame. We will deconstruct the notion of normal ordering here, before making the transition to the CFT approach (which naturally generates infinitesimal changes of normal ordering). 
\sk

An elementary viewpoint is to primarily think of the depicted normal ordering in (\ref{eq:d2xeipx-ex}) as the operator $\partial^2x\,e^{ip\cdot x(0)}$ placed at $z_1=0$ in the $z_1$ frame coordinate minus all possible self contractions using point-splitting and (\ref{eq:opes}) for the subtractions \cite{Polchinski87}:
\begin{equation}\label{eq:d2xepx}
:\!\partial^2x\,e^{ip\cdot x(0)}\!:_{z_1}\equiv \lim_{z_1\rightarrow 0}\Big(\partial^{2}x\,e^{ip\cdot x(0)}-\frac{i\alpha'p}{2}\frac{1}{z_1^2}e^{ip\cdot x(0)}\Big)e^{-\frac{\alpha'p^2}{4}\ln z_1}
\end{equation}
where the overall exponential factor subtracts the self contractions in the exponential, $e^{ip\cdot x(0)}$, whereas the $1/z_1^2$ term subtracts the contractions of $\partial^2x$ with the exponential. 
\sk

Let us then consider the infinitesimal holomorphic change of frame, 
\begin{equation}\label{eq:infholz1}
z_1\mapsto z_1'=z_1+\sum_{n=0}^{\infty}\varepsilon_nz_1^{n+1},
\end{equation}
in terms of which we have similarly,
\begin{equation}\label{eq:d2xepx'}
:\!\big(\partial^2x\,e^{ip\cdot x(0)}\big)'\!:_{z_1'}\equiv \lim_{z_1'\rightarrow 0}\Big(\big(\partial^{2}x\,e^{ip\cdot x(0)}\big)'-\frac{i\alpha'p}{2}\frac{1}{z_1^{'2}}\big(e^{ip\cdot x(0)}\big)'\Big)e^{-\frac{\alpha'p^2}{4}\ln z_1'}.
\end{equation}
The primes denote that the corresponding quantities are evaluated using $z_1'$ frame coordinates. 
Using (\ref{eq:infholz1}) that defines $z_1'$ in terms of $z_1$, we are to interpret the quantities appearing explicitly on the right-hand side of (\ref{eq:d2xepx'}) as follows,
\begin{equation}\label{eq:eipxd2x}
\begin{aligned}
&\big(e^{ip\cdot x(0)}\big)'=e^{ip\cdot x(0)}\\
&\big(\partial^{2}x(0)\big)'=\partial^2x(0)-2\varepsilon_0\partial^2x(0)-2\varepsilon_1\partial x(0)\\
\end{aligned}
\end{equation}
with corresponding relations for $1/z_1'^2$ and $e^{-\frac{\alpha'p^2}{4}\ln z_1'}$ that are read off from (\ref{eq:infholz1}) (and we keep terms up to linear order in the $\varepsilon_n$). Notice that the quantity $e^{ip\cdot x(0)}$ in (\ref{eq:eipxd2x}) transforms as a {\it scalar} (which reflects the general principle that operators that are not normal ordered transform naively, i.e.~as their classical counterparts, under changes of coordinates). So substituting these four relations into (\ref{eq:d2xepx'}), making use of (\ref{eq:d2xepx}) and the related expressions, 
\begin{equation}
\begin{aligned}
:\!\partial x\,e^{ip\cdot x(0)}\!:_{z_1}&\equiv \lim_{z_1\rightarrow 0}(\partial x\,e^{ip\cdot x(0)}-\frac{i\alpha'p}{2}\frac{1}{z_1}e^{ip\cdot x(0)})e^{-\frac{\alpha'p^2}{4}\ln z_1}\\
:\!e^{ip\cdot x(0)}\!:_{z_1}&=\lim_{z_1\rightarrow 0}e^{ip\cdot x(0)}e^{-\frac{\alpha'p^2}{4}\ln z_1},
\end{aligned}
\end{equation}
yields the following result for the $z_1'$-frame operator $\mathscr{A}^{(z_1')}$ associated to the $z_1$-frame operator $\mathscr{A}^{(z_1)}$ in (\ref{eq:d2xeipx-ex}):
\begin{equation}\label{eq:d2xepx'2}
\begin{aligned}
:\!\big(\partial^2x\,e^{ip\cdot x(0)}\big)'\!:_{z_1'}&=\,:\!\partial^2x\,e^{ip\cdot x(0)}\!:_{z_1}-\varepsilon_0\Big(\frac{\alpha'p^2}{4}+2\Big)\,:\!\partial^2x\,e^{ip\cdot x(0)}\!:_{z_1}\\
&\qquad-\varepsilon_12:\!\partial x\,e^{ip\cdot x(0)}\!:_{z_1}-\varepsilon_2(-i\alpha'p)\!:\!e^{ip\cdot x(0)}\!:_{z_1}.
\end{aligned}
\end{equation}

Making use of the explicit expressions for the Virasoro generators in terms of contour integrals involving the energy-momentum tensor, it is then a simple calculation to show that the right-hand side in (\ref{eq:d2xepx'2}) is precisely equal to,
$$
:\!\partial^2x\,e^{ip\cdot x(0)}\!:_{z_1}-\sum_{n=0}^{\infty}\big(\varepsilon_nL_{n}^{(z_1)}+\bar{\varepsilon}_n\tilde{L}_{n}^{(z_1)}\big):\!\partial^2x\,e^{ip\cdot x(0)}\!:_{z_1},
$$
Notice for example that the $L_0^{(z_1)}$ eigenvalue of $:\!\!\partial^2x\,e^{ip\cdot x(0)}\!\!:_{z_1}$ is the conformal weight $h=\frac{\alpha'p^2}{4}+2$ as one also reads off from (\ref{eq:d2xepx'2}). 
\sk

The above short calculation motivates the following general prescription \cite{Polchinski_v1} to calculating how a local composite operator changes under an {\it infinitesimal} change of frame. Given a chart, $(U_1,z_1)$, with a local operator, $\hat{\mathscr{A}}_a^{(z_1)}$, inserted at the origin $z_1=0$, under a holomorphic change of frame that leaves the origin invariant we have the standard CFT result \cite{Polchinski_v1},
\begin{equation}\label{eq:z'z}
\boxed{
\begin{aligned}
z_1&\mapsto w_1=z_1+\sum_{n=0}^{\infty}\varepsilon_nz_1^{n+1},\\
:\!\hat{\mathscr{A}}_a^{(z_1)}(p_1)\!:_{z_1}\,\,\mapsto \,\,:\!\hat{\mathscr{A}}_a^{(w_1)}(p_1)\!:_{w_1}&=\,\,:\!\hat{\mathscr{A}}_a^{(z_1)}(p_1)\!:_{z_1}-\sum_{n=0}^{\infty}\big(\varepsilon_nL_{n}^{(z_1)}+\bar{\varepsilon}_n\tilde{L}_{n}^{(z_1)}\big):\!\hat{\mathscr{A}}_a^{(z_1)}(p_1)\!:_{z_1}
\end{aligned}
}
\end{equation}
with $\varepsilon_n$ small. Notice that this transformation changes {\it both} the coordinates and the normal ordering, and as mentioned above in the following subsections we will find it useful to disentangle these notions.  
\sk

Also, let us point out that (\ref{eq:z'z}) implies (since all quantities are evaluated at the origin of the frame coordinates) the standard result that only primary fields of vanishing weights are independent of frame. Another comment is that (\ref{eq:z'z}) remains true if we also extend the sums over $n$ appearing to all integers, and it also remains true if we replace $p_1$ by a more generic point $p$ provided it is understood that the contours appearing in the mode operators enclose $z_1(p)$ (or $\bar{z}_1(p))$.
\sk

What is also immediately evident is that the prescription (\ref{eq:z'z}) for a change of frame is directly applicable only for infinitesimal changes; for finite changes one must either compose such infinitesimal transformations and exponentiate or proceed differently. Since finite changes of frame do play an important role (e.g., we can define Riemann surfaces using holomorphic transition functions, in which case changes of frame are generically not small on patch overlaps) we will present a more general procedure below, in particular (\ref{eq:CNOxbc-w<->z}). 
\sk

Indeed, looking ahead, the approach we present in Sec.~\ref{sec:HCNO} and in particular (\ref{eq:CNOxbc-w<->z}) yields under $z_1\mapsto w_1(z_1)$ almost immediately the much more general result,
\begin{equation}\label{eq:d2xepx'2b}
\begin{aligned}
:\!\partial_{z_1}^2&x_{(z_1)}e^{ik\cdot x_{(z_1)}(z)}\!:_{w_1}=\\
&=\Big[:\!\partial_{z_1}^2x_{(z_1)}e^{ik\cdot x_{(z_1)}(z)}\!:_{z_1}+\frac{i\alpha'k}{6}\Big(\frac{\partial_{z_1}^3w_1}{\partial_{z_1}w_1}-\frac{1}{4}\Big(\frac{\partial_{z_1}^2w_1}{\partial_{z_1}w_1}\Big)^2\Big):\!e^{ik\cdot x_{(z_1)}(z)}\!:_{z_1}\Big](\partial_{z_1}w_1)^{-\alpha'k^2/4},
\end{aligned}
\end{equation}
which reduces to (\ref{eq:d2xepx'2}) when $w_1(z_1)=z_1+\varepsilon_0z_1+\varepsilon_1z_1^2+\varepsilon_2z_1^3+\dots$ (after setting $z=0$ and changing coordinates naively on the left-hand side using that when acting on scalars, $\partial_{z_1}^2 = (\partial_{z_1}^2w_1)\partial_{w_1}+(\partial_{z_1}w_1)^2\partial_{w_1}^2$). So this motivates the developments in the following subsections where we generalise the usual CFT prescription (\ref{eq:z'z}) (which is best suited for infinitesimal changes) to finite holomorphic changes of normal ordering.

\subsection{Conformal Normal Ordering}\label{sec:CNO}
So as discussed with a simple example in the previous subsection, a {\it normal ordering prescription} is a prescription for subtracting infinities arising from self contractions within (usually one or more composite) operators. Of the various approaches (such as dimensional regularisation, zeta function regularisation, reparametrisation-invariant short-distance cutoff, etc.) there is a particularly convenient choice that works for arbitrary vertex operators while preserving most of the naive reparametrisation-invariance of a ``naively'' constructed vertex operator. This is the `conformal normal ordering' (CNO) scheme \cite{Polchinski87} (and the related notion \cite{Polchinski88} of `Weyl normal ordering' where a ``preferred'' frame or holomorphic coordinate is used, in particular a {\it holomorphic normal coordinate}, see Sec.~\ref{sec:HNC}). So Weyl normal ordering is a special case of CNO. A slight variant of CNO that will be appropriate for our purposes (and which extends the analysis of \cite{Polchinski87} to the BRST context) is the following. 
\sk

Let us recall from Sec.~\ref{sec:MVa} the Beltrami equation (\ref{eq:w-gen3}),
\begin{equation}\label{eq:beltramieqnbx}
\big(\partial_{\bar{z}}-\mu_{\bar{z}}^{\phantom{a}z}\partial_z\big)w(z,\bar{z})=0,
\end{equation}
which defines what we mean by a conformal coordinate, $w(z,\bar{z})$, associated to a complex structure $J$, which is in turn given in terms of a coordinate, $z$. Notice that the coordinate $z$ is also a conformal (or holomorphic) coordinate but in a different (when $\partial_{\bar{z}}w\neq0$) complex structure that we have called $I$ in Sec.~\ref{sec:MVa}. So a solution to the Beltrami equation allows us to express the conformal coordinate of one complex structure in terms of the conformal coordinate of a different complex structure. 
\sk

As discussed in Sec.~\ref{sec:MVa} the Beltrami equation (\ref{eq:beltramieqnbx}) is invariant under the following two independent sets of (anti-)holomorphic reparametrisations,
\begin{subequations}\label{eq:Beltr-invariances}
\begin{align}
w\mapsto w'(w)\qquad &\Rightarrow \qquad w'(z,\bar{z})=w'(w(z,\bar{z})),\label{eq:w->f}\\
z\mapsto z'(z)\qquad&\Rightarrow\qquad w'(z'(z),\bar{z}'(\bar{z}))=w(z,\bar{z}).\label{eq:z->z'}
\end{align}
\end{subequations}
Recall furthermore that invariance of the Beltrami equation under (\ref{eq:z->z'}) is the residual gauge symmetry associated to going to conformal gauge, which is in turn a subset of the full set of real reparametrisations, $\sigma\mapsto \sigma'(\sigma)$. But in conformal gauge there is still a large local symmetry, namely holomorphic reparametrisations (\ref{eq:w->f}). Using holomorphic normal coordinates fixes this latter invariance and the resulting normal ordering prescription was termed Weyl normal ordering in \cite{Polchinski88}. We will develop both of these types of normal ordering in parallel, since the defining equations are essentially of identical form.
\sk

The idea underlying conformal normal ordering is to work with a Riemann surface in, say, complex structure $J$, to normal order by subtracting self contractions using the conformal coordinate $w(z,\bar{z})$, and to associate the naive transformation property of the corresponding vertex operator to holomorphic reparametrisations $z\mapsto z'(z)$. Because (\ref{eq:z->z'}) then guarantees that the transformation properties of the normal ordered vertex operator will be the same (possibly up to a Weyl rescaling) as the naive transformation property of the vertex operator. So if a vertex operator is constructed to be naively invariant under holomorphic reparametrisations $z\mapsto z'(z)$ then so will the conformal normal-ordered vertex operator be invariant (up to a Weyl rescaling). This in turn guarantees that if we use holomorphic transition functions to glue charts together, i.e.~to go from a local description to a global description, then we will end up with a globally well-defined local operator (up to a Weyl rescaling that in turn enables us to keep precise track of the Weyl dependence in intermediate steps of the computation).
\sk

We take the relations (\ref{eq:opes}) as our starting point, which we will call `$z_1$ normal ordering in $z_1$ frame coordinates', 
\begin{equation}\label{eq:opes2}
\begingroup\makeatletter\def\f@size{11}\check@mathfonts
\def\maketag@@@#1{\hbox{\m@th\large\normalfont#1}}%
\boxed{
\begin{aligned}
&x^{\mu}_{(z_1)}(p')x^{\nu}_{(z_1)}(p) = -\frac{\alpha'}{2}\eta^{\mu\nu}\ln|z_1(p')-z_1(p)|^2+:\!x^{\mu}_{(z_1)}(p')x^{\nu}_{(z_1)}(p)\!:_{z_1}\\
&b^{(z_1)}(p')c^{(z_1)}(p) = \frac{1}{z_1(p')-z_1(p)}+:\!b^{(z_1)}(p')c^{(z_1)}(p)\!:_{z_1}\\
\end{aligned}
}
\endgroup
\end{equation}
It is also possible to use the auxiliary coordinate system, $\sigma^a$, to specify the abstract notation for the points $p',p$, since then it becomes manifest (taking into account the comments in Sec.~\ref{sec:MVa}) that since (perhaps up to an immaterial phase) $z_1(\sigma)$ transforms as a {\it scalar} under general reparametrisations, $\sigma\mapsto \sigma'(\sigma)$, normal ordered operators can be extended to globally well-defined operators.
\sk
 
These relations (\ref{eq:opes2}) {\it define} what we mean by `$z_1$ normal ordering in $z_1$ frame coordinates'. In a free theory, such as the case of interest here\footnote{For the corresponding case for composite operators in interacting theories see  \cite{Zimmermann73a,Zimmermann73b} and \cite{EllisMavromatosSkliros15}.}, Wick's theorem gives all the self contractions and we therefore have, succinctly, that for any functional, $\hat{\mathscr{A}}^{(z_1)}(x,b,c)$, of $x_{(z_1)}^\mu(p)$, $b^{(z_1)}(p)$ and $c^{(z_1)}(p)$,\footnote{Note that this is equivalent to Polchinski's {\it conformal normal ordering}  \cite{Polchinski87}, generalised here to include ghost contributions; see also equation (2.2.7) in \cite{Polchinski_v1}. The relation between the two approaches stems from a comment on p.~152 in Coleman's book `Aspects of Symmetry' and is elaborated on in \cite{EllisMavromatosSkliros15}.} 
\begin{equation}\label{eq:CNOxbc-z1}
\begingroup\makeatletter\def\f@size{11}\check@mathfonts
\def\maketag@@@#1{\hbox{\m@th\large\normalfont#1}}%
\boxed{
\begin{aligned}
&:\hat{\mathscr{A}}^{(z_1)}(x,b,c,\tilde{b},\tilde{c})\!:_{z_1} \,\,= \hat{\mathscr{A}}^{(z_1)}(\delta_K,\delta_I,\delta_J,\delta_{\tilde{I}},\delta_{\tilde{J}})\\
&\quad\times\exp\Bigg(-\frac{1}{2}\int_{z'}\int_z\,K^{(z_1)}(z')\cdot K^{(z_1)}(z)\,G_{\rm m}^{(z_1)}(z',z)+\int_zK^{(z_1)}(z)\!\cdot x_{(z_1)}(z)\Bigg)\\
&\quad\times \exp\Bigg(\int_{z'}\int_z\,I^{(z_1)}(z')\cdot J^{(z_1)}(z)\,G_{\rm g}^{(z_1)}(z',z)+\int_zI^{(z_1)}(z)b^{(z_1)}(z)+\int_zJ^{(z_1)}(z)c^{(z_1)}(z)\Bigg)\\
&\quad\times \exp\Bigg(\int_{z'}\int_z\,\tilde{I}^{(z_1)}(z')\cdot \tilde{J}^{(z_1)}(z)\,\bar{G}_{\rm g}^{(z_1)}(z',z)+\int_z\tilde{I}^{(z_1)}(z)\tilde{b}^{(z_1)}(z)+\int_z\tilde{J}^{(z_1)}(z)\tilde{c}^{(z_1)}(z)\Bigg)\Bigg|_{K=I=J=0}\\
\end{aligned}
}
\endgroup
\end{equation}
where we write $\int_z\dfn \int \rmd^2z\sqrt{g}$, etc., and $\delta_K$, $\delta_I$ and $\delta_J$ then denote covariant functional derivatives with respect to the corresponding sources with normalisation, $\int_z\delta_{I^{(z_1)}(z')}I^{(z_1)}(z)=1$. The source $K$ is commuting and transforms as a scalar under conformal transformations, whereas $I$ and $J$ are anticommuting and transform as tensor components of weights $(h,\tilde{h})=(-2,0)$ and $(1,0)$ respectively. We have also defined the quantities:
\begin{equation}\label{eq:GGGz1}
\begin{aligned}
G_{\rm m}^{(z_1)}(z',z)&\dfn -\frac{\alpha'}{2}\ln|z_1(p')-z_1(p)|^2\\
G_{\rm g}^{(z_1)}(z',z)&\dfn \frac{1}{z_1(p')-z_1(p)}\\
\bar{G}_{\rm g}^{(z_1)}(z',z)&\dfn \frac{1}{\bar{z}_1(p')-\bar{z}_1(p)},\qquad {\rm with}\qquad z'=z_1(p'),\quad z=z_1(p).
\end{aligned}
\end{equation}

A comment on notation. Throughout this article we denote fixed-picture coherent state vertex operators in the $z_1$ frame as, $\hat{\mathscr{A}}_a^{(z_1)}(p_1)$, (and similarly for their duals, $\hat{\mathscr{A}}^a_{(z_2)}(p_2)$), more about which later. We could have also written the same quantity as $:\!\hat{\mathscr{A}}_a^{(z_1)}(p_1)\!:_{z_1}$ to emphasise that normal ordering is also carried out in the $z_1$ frame coordinates. In particular,
\begin{equation}\label{eq:Az1normalorderingnotation}
\boxed{\hat{\mathscr{A}}_a^{(z_1)}(p_1)\equiv\,\, :\!\hat{\mathscr{A}}_a^{(z_1)}(p_1)\!:_{z_1},\qquad \hat{\mathscr{A}}^a_{(z_2)}(p_2)\equiv\,\, :\!\hat{\mathscr{A}}^a_{(z_2)}(p_2)\!:_{z_2}}
\end{equation}
So the notation is such that when the normal ordering is not explicitly exhibited (as in the left-hand sides of (\ref{eq:Az1normalorderingnotation})) then it is always carried out in the same frame as the frame of the coordinates, namely $z_1$ (or $z_2$) in this example (as exhibited on the right-hand sides in (\ref{eq:Az1normalorderingnotation})). Since in this section we also discuss changes in the normal ordering prescription with coordinates held fixed the more explicit notation on the right-hand sides in (\ref{eq:Az1normalorderingnotation}) will be more convenient.
\sk

Incidentally, (although we are jumping ahead slightly) it is perhaps useful to mention in passing the relation between the normal ordering prescription (\ref{eq:CNOxbc-z1}) used to defined local operators and the corresponding local operators obtained in the Polyakov treatment of defining local operators, see p.~102-107 in Polchinski's textbook \cite{Polchinski_v1}. The case of interest is when the frame coordinate in (\ref{eq:GGGz1}) is identified with a holomorphic normal coordinate. The two approaches differ in renormalisation scheme, and in particular using Polchinski's classification characterised by a parameter $\gamma$ (see equation (3.6.15) in \cite{Polchinski_v1}) the conformal normal ordering defined above using holomorphic normal coordinates corresponds to $\gamma=-1$ (rather than $\gamma=-2/3$ which instead corresponds to subtracting self contractions using geodesic distance, see equation (3.6.6) there). One can extract all the relevant formulas such as the analogues of those on p.~105 in  \cite{Polchinski_v1} by replacing equations (3.6.5) and (3.6.6) by (\ref{eq:CNOxbc-z1}) and (\ref{eq:GGGz1}) respectively. One must be a little careful because although the equation of motion does not hold in geodesic normal ordering (see equation (3.6.18) in \cite{Polchinski_v1}), it does hold in Weyl normal ordering (modulo contact terms \cite{Polchinski88}). E.g., in Weyl normal ordering (and when $z$ is a holomorphic normal coordinate) we have relations such as:
$$
\,:\!4\big(\partial_{z}\partial_{\bar{z}}X^\mu\big) e^{ik\cdot X}(\sigma_1)\!:_z\,=\,:\!\big(\nabla^2X^\mu\big) e^{ik\cdot X}(\sigma_1)\!:_z\,=0,
$$
with $\nabla_a$ a covariant derivative with respect to $\sigma^a$ with $a=1,2$ (which can be related to the $z,\bar{z}$ covariant derivative by means of the zweibein or frame field), and,
\begin{equation}\label{eq:repl3.6.18Pol}
\nabla_a:\!\nabla^aX^\mu e^{ik\cdot X}(\sigma_1)\!:_z\,=\,:\!\nabla_aX^\mu\,\nabla^a e^{ik\cdot X}(\sigma_1)\!:_z+\frac{i\alpha'\gamma}{4}k^\mu R_{(2)}(\sigma_1):\!e^{ik\cdot X}(\sigma_1)\!:_z\,
\end{equation}
where $\gamma=-1$. 
The latter exhibits the sense in which equation (3.6.18) in \cite{Polchinski_v1} can be understood, despite the fact that in Weyl normal ordering the equation of motion holds. Namely, it is ``as if'' we have used the product rule for differentiation, but rather the above relation follows from the fact that differentiation and Weyl normal ordering do not commute. The general commutator of differentiation with Weyl normal ordering is derived in Sec.~\ref{sec:comm dA-dA} and (\ref{eq:repl3.6.18Pol}) follows immediately from the result of that section. Further details along these lines can be found in \cite{SklirosPhysStackEx}. 
\sk

In the next section we discuss how to (and the effect of) changing the above normal ordering prescription by a holomorphic reparametrisation.

\subsection{Finite Change in Normal Ordering}\label{sec:HCNO}
The $z_1$ coordinate used in the previous subsection (and throughout) is clearly not special, and when there are multiple overlapping charts, $\{(U_m,z_m)\}$, this is particularly important. So (as we have discussed at various points) in order to obtain a meaningful construction we should provide an appropriate equivalence relation associated to holomorphic reparametrisations, $z_1\mapsto w_1(z_1)$, which in turn preserve complex structure. In particular, we could have used any holomorphic coordinate, $w_1$, to define normal ordering. That is, in direct analogy to (\ref{eq:opes2}),
\begin{equation}\label{eq:opes_w1}
\begin{aligned}
&x^{\mu}_{(w_1)}(p')x^{\nu}_{(w_1)}(p) = -\frac{\alpha'}{2}\eta^{\mu\nu}\ln|w_1(p')-w_1(p)|^2+:\!x^{\mu}_{(w_1)}(p')x^{\nu}_{(w_1)}(p)\!:_{w_1}\\
&b^{(w_1)}(p')c^{(w_1)}(p) = \frac{1}{w_1(p')-w_1(p)}+:\!b^{(w_1)}(p')c^{(w_1)}(p)\!:_{w_1}\\
\end{aligned}
\end{equation}
Taking into account the transformation properties (\ref{eq:chiralprimary2}) of the various local tensor operators, namely,\footnote{The generalisation of what we discuss here to general $b,c$ systems with weights $(\lambda,1-\lambda)$ is immediate (with an appropriate minus sign for commuting fields \cite{Friedan86,FriedanShenkerMartinec85}). The case of interest here is bosonic string theory where we have taken $\lambda=2$ with $b,c$ anticommuting.} $b^{(w_1)}(p)(\partial_{z_1}w_1)^2=b^{(z_1)}(p)$, $c^{(w_1)}(p)(\partial_{z_1}w_1)^{-1}=c^{(z_1)}(p)$, and that $x_{(w_1)}(p)$ (and functionals thereof) transforms as a {\it scalar}, $x_{(w_1)}(p)=x_{(z_1)}(p)$, {\it when} the normal ordering is held fixed, we learn that (\ref{eq:opes_w1}) further implies:
\begin{equation}\label{eq:opes_w1fixedz1}
\boxed{
\begin{aligned}
&x^{\mu}_{(z_1)}(p')x^{\nu}_{(z_1)}(p) = -\frac{\alpha'}{2}\eta^{\mu\nu}\ln|w_1(z_1(p'))-w_1(z_1(p))|^2+:\!x^{\mu}_{(z_1)}(p')x^{\nu}_{(z_1)}(p)\!:_{w_1}\\
&b^{(z_1)}(p')c^{(z_1)}(p) = \frac{(\partial_{z_1}w_1(z_1(p')))^2(\partial_{z_1}w_1(z_1(p)))^{-1}}{w_1(z_1(p'))-w_1(z_1(p))}+:\!b^{(z_1)}(p')c^{(z_1)}(p)\!:_{w_1}\\
\end{aligned}
}
\end{equation}
The left-hand sides are  independent of $w_1(z_1)$, and we will take the transition from the relations (\ref{eq:opes2}) to (\ref{eq:opes_w1fixedz1}) to {\it define} what we mean by `{\it change in normal ordering keeping coordinates fixed}'. This is a central notion when transitioning from a local to a global construction, because the various charts required to describe the full Riemann surface are related by holomorphic transition functions. 
\sk

The analogue of (\ref{eq:CNOxbc-z1}) is immediate, namely we will define `{\it $w_1$ normal ordering in $z_1$ chart coordinates}' for any operator, $\hat{\mathscr{A}}^{(z_1)}(x,b,c,\tilde{b},\tilde{c})$, by:
\begin{equation}\label{eq:CNOxbc-w1}
\begingroup\makeatletter\def\f@size{11}\check@mathfonts
\def\maketag@@@#1{\hbox{\m@th\large\normalfont#1}}%
\begin{aligned}
&:\hat{\mathscr{A}}^{(z_1)}(x,b,c,\tilde{b},\tilde{c})\!:_{w_1} \,\,= \hat{\mathscr{A}}^{(z_1)}(\delta_K,\delta_I,\delta_J,\delta_{\tilde{I}},\delta_{\tilde{J}})\\
&\quad\times\exp\Bigg(-\frac{1}{2}\int_{z'}\int_z\,K^{(z_1)}(z')\cdot K^{(z_1)}(z)\,G_{\rm m}^{(w_1;z_1)}(z',z)+\int_zK^{(z_1)}(z)\!\cdot x_{(z_1)}(z)\Bigg)\\
&\quad\times \exp\Bigg(\int_{z'}\int_z\,I^{(z_1)}(z')\cdot J^{(z_1)}(z)\,G_{\rm g}^{(w_1;z_1)}(z',z)+\int_zI^{(z_1)}(z)b^{(z_1)}(z)+\int_zJ^{(z_1)}(z)c^{(z_1)}(z)\Bigg)\\
&\quad\times \exp\Bigg(\int_{z'}\int_z\,\tilde{I}^{(z_1)}(z')\cdot \tilde{J}^{(z_1)}(z)\,\bar{G}_{\rm g}^{(w_1;z_1)}(z',z)+\int_z\tilde{I}^{(z_1)}(z)\tilde{b}^{(z_1)}(z)+\int_z\tilde{J}^{(z_1)}(z)\tilde{c}^{(z_1)}(z)\Bigg)\Bigg|_{K=I=J=0}\\
\end{aligned}
\endgroup
\end{equation}
where we read off the following quantities from (\ref{eq:opes_w1fixedz1}):
\begin{equation}\label{eq:GGGw1}
\begin{aligned}
G_{\rm m}^{(w_1;z_1)}(z',z)&\dfn -\frac{\alpha'}{2}\ln|w_1(z')-w_1(z)|^2\\
G_{\rm g}^{(w_1;z_1)}(z',z)&\dfn \frac{(\partial_{z'}w_1(z'))^2(\partial_zw_1(z))^{-1}}{w_1(z')-w_1(z)}\\
\bar{G}_{\rm g}^{(w_1;z_1)}(z',z)&\dfn \frac{(\partial_{\bar{z}'}\bar{w}_1(\bar{z}'))^2(\partial_{\bar{z}}\bar{w}_1(\bar{z}))^{-1}}{\bar{w}_1(\bar{z}')-\bar{w}_1(\bar{z})},\qquad {\rm with}\qquad z'=z_1(p'),\quad z=z_1(p).
\end{aligned}
\end{equation}
We have derived (\ref{eq:CNOxbc-w1}) by simply {\it replacing} all $z_1$ frame coordinates in the exponential in (\ref{eq:CNOxbc-z1}) by $w_1$ frame coordinates, and then {\it transforming} coordinates back to the $z_1$ frame taking into account the transformation properties of the measures and sources. This is what we mean by `change in normal ordering keeping coordinates fixed', and notice that the operator $\hat{\mathscr{A}}^{(z_1)}(\delta_K,\delta_I,\delta_J,\delta_{\tilde{I}},\delta_{\tilde{J}})$ is unaffected by this sequence of steps. 
Also, setting $w_1(z_1)=z_1$ yields the $z_1$ normal ordering relations (\ref{eq:CNOxbc-z1}) since:
$$
G_{\rm m}^{(z_1;z_1)}(z',z)=G_{\rm m}^{(z_1)}(z',z),\qquad {\rm and}\qquad G_{\rm g}^{(z_1;z_1)}(z',z)=G_{\rm g}^{(z_1)}(z',z).
$$

When explicitly applying the prescription (\ref{eq:CNOxbc-w1}) to general operators (e.g., to unravel how offshell general coherent state vertex operators transform under {\it finite} changes of normal ordering) one encounters expressions involving some number of derivatives of (\ref{eq:GGGw1}). The relevant expressions for the matter sector (and the corresponding limits, $z'\rightarrow z$, if one uses point splitting to define the local operators) may be extracted from:
\begin{equation}\label{eq:dndmGwz}
\begin{aligned}
\partial_{z'}^n\partial_{z}^mG_{\rm m}^{(w_1;z_1)}(z',z)&=\sum_{k=1}^n\sum_{\ell=1}^m(-)^k(k+\ell-1)!B_{n,k}\big(\tfrac{\partial_{z'}^sw_1(z')}{w_1(z')-w_1(z)}\big)B_{m,\ell}\big(\tfrac{\partial_{z}^sw_1(z)}{w_1(z')-w_1(z)}\big)\\
\partial_{z'}^nG_{\rm m}^{(w_1;z_1)}(z',z)&=\sum_{k=1}^n(-)^k(k-1)!B_{n,k}\big(\tfrac{\partial_{z'}^sw_1(z')}{w_1(z')-w_1(z)}\big)\\
\partial_{z}^mG_{\rm m}^{(w_1;z_1)}(z',z)&=\sum_{\ell=1}^m(\ell-1)!B_{m,\ell}\big(\tfrac{\partial_{z}^sw_1(z)}{w_1(z')-w_1(z)}\big),
\end{aligned}
\end{equation}
where $n,m\geq1$, and the quantities, $B_{n,k}(a_s)\dfn B_{n,k}(a_1,a_2,\dots,a_{n-k+1})$, etc., are Bell polynomials (with the useful properties $B_{n,k}(a_sb^sc)=b^nc^kB_{n,k}(a_s)$ and $B_{n,k}(s!)=\binom{n-1}{k-1}\frac{n!}{k!}$). There are corresponding expressions for the ghost sector (which can be calculated using Fa\'a di Bruno's formula \cite{Riordan58}). The relevant Taylor expansions around $z'=z$ are (for every $s=1,2,\dots$), 
\begin{equation}
\begin{aligned}
\frac{\partial_{z'}^sw_1(z')}{w_1(z')-w_1(z)}&=\frac{1}{z'-z}\Bigg(\frac{1+\sum_{a=1}^\infty\frac{1}{a!}(z'-z)^a\frac{\partial_z^{a+s}w_1(z)}{\partial_zw_1(z)}}{1+\sum_{b=1}^\infty\frac{1}{(b+1)!}(z'-z)^b\frac{\partial_z^{b+1}w_1(z)}{\partial_zw_1(z)}}\Bigg)\\
\frac{\partial_{z}^sw_1(z)}{w_1(z')-w_1(z)}&=\frac{1}{z'-z}\Bigg(\frac{\frac{\partial_z^{s}w_1(z)}{\partial_zw_1(z)}}{1+\sum_{b=1}^\infty\frac{1}{(b+1)!}(z'-z)^b\frac{\partial_z^{b+1}w_1(z)}{\partial_zw_1(z)}}\Bigg)
\end{aligned}
\end{equation}

Finally, since the left-hand sides of (\ref{eq:opes2}) and (\ref{eq:opes_w1fixedz1}) are actually equal so must the right-hand sides be, in particular,
\begin{equation}\label{eq:opes_w1fixedz1xx}
\begingroup\makeatletter\def\f@size{11}\check@mathfonts
\def\maketag@@@#1{\hbox{\m@th\large\normalfont#1}}%
\begin{aligned}
:\!x^{\mu}_{(z_1)}(p')x^{\nu}_{(z_1)}(p)\!:_{w_1}&=\,\,
:\!x^{\mu}_{(z_1)}(p')x^{\nu}_{(z_1)}(p)\!:_{z_1}+\frac{\alpha'}{2}\eta^{\mu\nu}\ln\Big|\frac{w_1(z_1(p'))-w_1(z_1(p))}{z_1(p')-z_1(p)}\Big|^2\\
:\!b^{(z_1)}(p')c^{(z_1)}(p)\!:_{w_1}&=\,\,:\!b^{(z_1)}(p')c^{(z_1)}(p)\!:_{z_1}-\Big(\frac{(\partial_{z_1}w_1(z_1(p')))^2(\partial_{z_1}w_1(z_1(p)))^{-1}}{w_1(z_1(p'))-w_1(z_1(p))}-\frac{1}{z_1(p')-z_1(p)}\Big)\\
\end{aligned}
\endgroup
\end{equation}
So by analogy to the above, e.g.~(\ref{eq:CNOxbc-w1}), we can directly write down how the same operators are related in different normal ordering prescriptions (again keeping coordinates fixed),
\begin{equation}\label{eq:CNOxbc-w<->z}
\begingroup\makeatletter\def\f@size{11}\check@mathfonts
\def\maketag@@@#1{\hbox{\m@th\large\normalfont#1}}%
\boxed{
\begin{aligned}
&:\hat{\mathscr{A}}^{(z_1)}(x,b,c,\tilde{b},\tilde{c})\!:_{w_1} \,\,= \,\,:\!\hat{\mathscr{A}}^{(z_1)}(\delta_K,\delta_I,\delta_J,\delta_{\tilde{I}},\delta_{\tilde{J}})\\
&\quad\times\exp\Bigg(-\frac{1}{2}\int_{z'}\int_z\,K^{(z_1)}(z')\cdot K^{(z_1)}(z)\,\Delta_{\rm m}^{(w_1;z_1)}(z',z)+\int_zK^{(z_1)}(z)\!\cdot x_{(z_1)}(z)\Bigg)\\
&\quad\times \exp\Bigg(\int_{z'}\int_z\,I^{(z_1)}(z')\cdot J^{(z_1)}(z)\,\Delta_{\rm g}^{(w_1;z_1)}(z',z)+\int_zI^{(z_1)}(z)b^{(z_1)}(z)+\int_zJ^{(z_1)}(z)c^{(z_1)}(z)\Bigg)\\
&\quad\times \exp\Bigg(\int_{z'}\int_z\,\tilde{I}^{(z_1)}(z')\cdot \tilde{J}^{(z_1)}(z)\,\bar{\Delta}_{\rm g}^{(w_1;z_1)}(z',z)+\int_z\tilde{I}^{(z_1)}(z)\tilde{b}^{(z_1)}(z)+\int_z\tilde{J}^{(z_1)}(z)\tilde{c}^{(z_1)}(z)\Bigg)\Bigg|_{K=I=J=0}\!\!\!\!\!\!\!\!\!\!\!\!\!\!\!\!\!\!\!\!\!\!:_{z_1}\,\,\,\,\,\,\,\,\,\,\,\,\,\,\,\,\,\,\\
\end{aligned}
}
\endgroup
\end{equation}
This generalises equation (2.7.14) on p.~60 in \cite{Polchinski_v1}. 
The quantities $\Delta_{\rm m},\Delta_{\rm g}$ and $\bar{\Delta}_{\rm g}$ are read off from (\ref{eq:opes_w1fixedz1xx}) and explicitly read:
\begin{equation}\label{eq:DDDw<->z}
\boxed{
\begin{aligned}
\Delta_{\rm m}^{(w_1;z_1)}(z',z)&\dfn -\frac{\alpha'}{2}\ln\Big|\frac{w_1(z')-w_1(z)}{z'-z}\Big|^2\\
\Delta_{\rm g}^{(w_1;z_1)}(z',z)&\dfn \frac{(\partial_{z'}w_1(z'))^2(\partial_{z}w_1(z))^{-1}}{w_1(z')-w_1(z)}-\frac{1}{z'-z}\\
\bar{\Delta}_{\rm g}^{(w_1;z_1)}(z',z)&\dfn \frac{(\partial_{\bar{z}'}\bar{w}_1(\bar{z}'))^2(\partial_{\bar{z}}\bar{w}_1(\bar{z}))^{-1}}{\bar{w}_1(\bar{z}')-\bar{w}_1(\bar{z})}-\frac{1}{\bar{z}'-\bar{z}}
\end{aligned}
}
\end{equation}
where as above, $z'=z_1(p')$, and $z=z_1(p)$. Notice that these three quantities are non-singular at $z'=z$.
\sk

In practice, when the local operators, $\hat{\mathscr{A}}^{(z_1)}$, of interest are those associated to handle operators then they involve exponentials. In applying the change in normal ordering identity (\ref{eq:CNOxbc-w<->z}), the following relation is therefore useful:
\begin{equation}\label{eq:exp YdK}
e^{\int_z Y(z)\delta_K}e^{H(K)} = e^{\sum_{N=0}^\infty\frac{1}{N!}\int_{z_1,\dots,z_N}Y(z_1)\dots Y(z_N)\delta_{K_1}\dots \delta_{K_N}H(K)}.
\end{equation}
The quantity $Y(z)$ could in general be identified with any Grassmann-even local operator of interest, e.g., $Y_\mu(z)=i\e{p}_\mu\delta^2(z-z_1(p))$, or $Y_\mu(z) =\delta^2(z-z_1(p))\sum_nA_{n\mu}\partial_z^n $, etc. For $Y(z)$ and $\delta_K$ Grassmann-odd the same identity applies but we are to include an overall coefficient, $(-)^{\frac{N(N-1)}{2}}$ in the summand of the right-hand side in (\ref{eq:exp YdK}),
\begin{equation}\label{eq:exp YdI}
e^{\int_z Y(z)\delta_I}e^{H(I)} = e^{\sum_{N=0}^\infty\frac{1}{N!}(-)^{\frac{N(N-1)}{2}}\int_{z_1,\dots,z_N}Y(z_1)\dots Y(z_N)\delta_{I_1}\dots \delta_{I_N}H(I)}.
\end{equation}
This is derived from (\ref{eq:exp YdK}) by first grouping together the Grassmann-even factors, $Y\delta_K$, on the right-hand side and {\it then} promoting $Y$ and $\delta_K$ to Grassmann-odd quantities and rearranging (keeping track of the resulting minus signs). 
The relations (\ref{eq:exp YdK}) and (\ref{eq:exp YdI}) (and combinations thereof) can be used, e.g., to change normal ordering prescription in any coherent state.
\sk

{\it If} it so happens that the change in normal ordering, $z_1\mapsto w_1(z_1)$, is such that the quantities (\ref{eq:DDDw<->z}) vanish 
then (\ref{eq:CNOxbc-w<->z}) reduces to,
\begin{equation}\label{eq:CNOxbc-w<->z-zero}
\begin{aligned}
&\Delta_{\rm m}^{(w_1;z_1)}=\Delta_{\rm g}^{(w_1;z_1)}=0\qquad\Rightarrow\qquad:\hat{\mathscr{A}}^{(z_1)}(x,b,c,\tilde{b},\tilde{c})\!:_{w_1} \,\,= \,\,:\hat{\mathscr{A}}^{(z_1)}(x,b,c,\tilde{b},\tilde{c})\!:_{z_1}
\end{aligned}
\end{equation}
This is true, e.g., in the case of rigid shifts, $w_1(z_1)=z_1+z_v$, so for rigid shifts arbitrary normal-ordered local operators transform as do the corresponding local operators when one neglects the normal ordering.
\sk

Another special case of (\ref{eq:CNOxbc-w<->z}) is when we wish to pullback handle operators to a stereographic projection of the sphere onto a plane, which may be accomplished by using SL(2,$\mathbf{C}$). Since a generic SL(2,$\mathbf{C}$) transformation, $z\mapsto w(z)$, is of the form,
$$
w(z)=\frac{az+b}{cz+d},\qquad {\rm with}\qquad a,b,c,d\in\mathbf{C},\quad ad-bc=1,
$$
it follows that the quantities (\ref{eq:DDDw<->z}) that are to be used in (\ref{eq:CNOxbc-w<->z}) to change normal ordering keeping coordinates fixed simplify to,
\begin{equation}\label{eq:DDDw<->z SL2C}
\begin{aligned}
\Delta_{\rm m}^{(w_1;z_1)}(z',z)&= \frac{\alpha'}{2}\ln\big|(cz'+d)(cz+d)\big|^2\\
\Delta_{\rm g}^{(w_1;z_1)}(z',z)&= \frac{1}{z'-z}\Big[\Big(\frac{cz+d}{cz'+d}\Big)^3-1\Big]\\
\bar{\Delta}_{\rm g}^{(w_1;z_1)}(z',z)&=  \frac{1}{\bar{z}'-\bar{z}}\Big[\Big(\frac{\bar{c}\bar{z}+\bar{d}}{\bar{c}\bar{z}'+\bar{d}}\Big)^3-1\Big]
\end{aligned}
\end{equation}
Notice that $a$ and $b$ (subject to $ad-bc=1$) cancel out.
\sk

Disentangling holomorphic changes in normal ordering from holomorphic changes in coordinates, namely the definition (\ref{eq:CNOxbc-w1}) or (\ref{eq:CNOxbc-w<->z}), is extremely useful in practice. 
Let us discuss a few simple but illustrative examples next.

\subsection{Simple Examples}
In Sec.~\ref{sec:HCNO} we discussed how to change normal ordering keeping coordinates fixed. Let us discuss some simple but illustrative examples.

\subsubsection*{\underline{Example 1:} Energy-Momentum Operator}

This is the most important case we will consider, since it is one of the simplest examples of a local operator that is not a conformal tensor (unless the corresponding central charge vanishes). And because coherent state vertex operators are not conformal tensors either, by studying the energy-momentum operator under {\it finite} changes of frame we can learn about the corresponding finite changes of frame for general local operators. This in turn is important when considering vertex operators or handle operators in a global context since we might wish to transform from one frame to another via a holomorphic transition function on patch overlaps (see Sec.~\ref{sec:RS} and Sec.~\ref{sec:TPIM}). As we have seen, these holomorphic transformations generated by transition functions are typically {\it not} infinitesimal, so it is not particularly useful to adopt the traditional viewpoint (in CFT literature) and use \cite{Polchinski_v1} the energy momentum operator to generate the required change of frame on patch overlaps, since the latter generate {\it infinitesimal} changes of frame. And to generate finite changes of frame we would need to compose all such infinitesimal transformations. The approach discussed in Sec.~\ref{sec:HCNO} is much simpler when finite changes of frame are of interest, as we will illustrate next. 
\sk

So let us consider the matter and energy-momentum operators (\ref{eq:T-bosonic}), namely:
\begin{equation}\label{eq:T-bosonic2}
\boxed{
\begin{aligned}
T_{\rm m}^{(z_1)}&=-\tfrac{1}{\alpha'}\!:\!(\partial_{z_1} x_{(z_1)})^2\!:_{z_1}\\
T_{\rm g}^{(z_1)} &= \,\,:\!c^{(z_1)}(\partial_{z_1} b^{(z_1)})+2(\partial_{z_1}c^{(z_1)})b^{(z_1)}\!:_{z_1},
\end{aligned}
}
\end{equation}
where the left- and right-hand sides are evaluated at some point $p\in U_1$ in the holomorphic $z_1$ chart coordinates. It will be convenient to write the left-hand sides as,
$$
:\!T_{\rm m}^{(z_1)}\!:_{z_1},\qquad {\rm and}\qquad :\!T_{\rm g}^{(z_1)}\!:_{z_1},
$$
respectively, since this will allow us to disentangle changes in normal ordering from holomorphic coordinate transformations. We can then change normal ordering keeping coordinates fixed by making use of (\ref{eq:CNOxbc-w1}) or (\ref{eq:CNOxbc-w<->z}), according to which it is not hard to show that,
\begin{equation}\label{eq:Tm-wz cno}
\phantom{\qquad (c_{\rm m}\equiv d_{\rm cr})}\boxed{:\!T_{\rm m}^{(z_1)}\!:_{w_1} =:\!T_{\rm m}^{(z_1)}\!:_{z_1}-\frac{c_{\rm m}}{12}\Big[\frac{\partial_{z_1}^3w_1}{\partial_{z_1}w_1}-\frac{3}{2}\Big(\frac{\partial_{z_1}^2w_1}{\partial_{z_1}w_1}\Big)^2\Big]}\qquad (c_{\rm m}\equiv d_{\rm cr})
\end{equation}
where we have also taken into account the following relations. The term in the brackets is known as the {\it Schwarzian derivative} \cite{Polchinski_v1}.  This term has the property that it vanishes when $z_1\mapsto w_1(z_1)$ is generated by SL(2,$\mathbf{C}$), but we consider arbitrary holomorphic reparametrisations here. Let us briefly discuss the derivation of (\ref{eq:Tm-wz cno}).
\sk

Since $w_1(z_1)$ is by definition a {\it holomorphic} function of $z_1$ this means it has a convergent Taylor expansion. So let us Taylor expand the matter contribution in (\ref{eq:GGGw1}) around $z'=z$,
\begin{equation}
\begin{aligned}
G_{\rm m}^{(w_1;z_1)}(z',z)&=-\frac{\alpha'}{2}\ln\big|w_1(z')-w_1(z)\big|^2\\
&=G_{\rm m}^{(z_1;z_1)}(z',z)-\frac{\alpha'}{2}\ln\Big|\sum_{n=1}^{\infty}\frac{1}{n!}(z'-z)^{n-1}\partial_{z}^{n}w_1(z)\Big|^2\\
\end{aligned}
\end{equation}
We define local operators inside normal ordering by point splitting \cite{EllisMavromatosSkliros15}. 
It is then not too hard to show that for $|z'-z|$ small (the general formula is given in (\ref{eq:dndmGwz})),
\begin{equation}\label{eq:d1d1Gwz}
\begin{aligned}
&\partial_{z'}\partial_{z}\ln\Big|\sum_{n=1}^{\infty}\frac{1}{n!}(z'-z)^{n-1}\partial_{z}^{n}w_1(z)\Big|^2=\\
&\qquad=\frac{2}{12}\bigg[\frac{\partial_{z}^3w_1}{\partial_{z}w_1}-\frac{3}{2}\Big(\frac{\partial_{z}^2w_1}{\partial_{z}w_1}\Big)^2\bigg]+\frac{1}{12}\bigg[3\Big(\frac{\partial_{z}^2w_1}{\partial_{z}w_1}\Big)^3+\frac{\partial_{z}^4w_1}{\partial_{z}w_1}-4\frac{\partial_{z}^3w_1\,\partial_{z}^2w_1}{(\partial_{z}w_1)^2}\bigg]\,(z'-z)+\dots,
\end{aligned}
\end{equation}
where the `\dots' denote terms $\mathcal{O}((z'-z)^2)$. 
This follows directly by using the chain rule, taking into account that only the $(z'-z)^{n-1}$ terms depend on $z'$ and that {\it both} $(z'-z)^{n-1}$ and $\partial_{z}^{n}w_1(z)$ depend on $z$. Since only the $z'\rightarrow z$ limit is of interest we can drop all terms on the right-hand side that vanish in this limit\footnote{This is so for the energy-momentum operator, but for more general operators where there are higher derivatives appearing one needs correspondingly higher order terms in the expansion in $z'-z$. The reader might notice that the coefficient of $z'-z$ is proportional to the first derivative of the Schwarzian derivative, but it is not obvious whether the Schwarzian derivative plays a prominent role at higher order in $z'-z$.}. Now since, according to (\ref{eq:CNOxbc-w1}), the $w_1$ and $z_1$ normal-ordered energy-momentum in $z_1$ coordinates respectively are given explicitly by,
$$
:\!T_{\rm m}^{(z_1)}\!:_{w_1} =\lim_{z'\rightarrow z}\Big[-\frac{1}{\alpha'}\,\Big(\partial_{z'}x_{(z_1)}(z')\cdot \partial_{z}x_{(z_1)}(z)-\partial_{z'}\partial_{z}G_{\rm m}^{(w_1;z_1)}(z',z)\Big)\Big],
$$
and,
$$
:\!T_{\rm m}^{(z_1)}\!:_{z_1} =\lim_{z'\rightarrow z}\Big[-\frac{1}{\alpha'}\,\Big(\partial_{z'}x_{(z_1)}(z')\cdot \partial_{z}x_{(z_1)}(z)-\partial_{z'}\partial_{z}G_{\rm m}^{(z_1;z_1)}(z',z)\Big)\Big],
$$
the result (\ref{eq:Tm-wz cno}) follows immediately.
\sk

A good exercise is to show that the corresponding result for the ghost sector, on account of (\ref{eq:T-bosonic2}), is indeed precisely:
\begin{equation}\label{eq:Tg-wz cno}
\phantom{\qquad (c_{\rm g}=-26)}\boxed{:\!T_{\rm g}^{(z_1)}\!:_{w_1} =:\!T_{\rm g}^{(z_1)}\!:_{z_1}-\frac{c_{\rm g}}{12}\Big[\frac{\partial_{z_1}^3w_1}{\partial_{z_1}w_1}-\frac{3}{2}\Big(\frac{\partial_{z_1}^2w_1}{\partial_{z_1}w_1}\Big)^2\Big]}\qquad (c_{\rm g}\equiv -26)
\end{equation}
where one makes use of the defining equations for $z_1$ normal ordering (\ref{eq:CNOxbc-z1}) and that for $w_1$ normal ordering (in $z_1$ coordinates) (\ref{eq:CNOxbc-w1}), and the second relation in (\ref{eq:GGGw1}). Then, keeping normal ordering fixed one can change coordinates naively on the left-hand side of the relation (\ref{eq:Tg-wz cno}),
\begin{equation}\label{eq:Tg-wz cno2}
:\!(\partial_{z_1}w_1)^2T_{\rm g}^{(w_1)}\!\!:_{w_1} \,=\,\,:\!T_{\rm g}^{(z_1)}\!\!:_{z_1}-\frac{c_{\rm g}}{12}\Big[\frac{\partial_{z_1}^3w_1}{\partial_{z_1}w_1}-\frac{3}{2}\Big(\frac{\partial_{z_1}^2w_1}{\partial_{z_1}w_1}\Big)^2\Big],
\end{equation}
where we took into account that the energy-momentum transforms like a tensor when normal ordering is held fixed. 
So as one expects, including both matter and ghost contributions, the total energy-momentum is a tensor operator (since the total central charge vanishes in critical string theory which is the only case we consider). 
\sk

Notice that the total energy-momentum (with central charge in the critical dimension, $c_{\rm T}=c_{\rm m}+c_{\rm g}=0$) is invariant under changes in normal ordering keeping coordinates fixed,
$$
T^{(z_1)}\equiv \,\,\,:\!T^{(z_1)}\!\!:_{z_1}\,\,\equiv \,\,\,:\!T_{\rm m}^{(z_1)}+T_{\rm g}^{(z_1)}\!\!:_{z_1}\,\,=\,\,\,:\!T_{\rm m}^{(z_1)}+T_{\rm g}^{(z_1)}\!\!:_{w_1},
$$
which in turn implies one can change coordinates naively, namely:
\begin{equation}\label{eq:Tw1zw-naive}
(\partial_{z_1}w_1)^2T^{(w_1)}=T^{(z_1)},
\end{equation}
which is precisely as one expects since the total energy-momentum is expected to transform as a tensor (i.e.~conformal primary) under holomorphic transformations.
\sk

We have shown that it is the {\it change in normal ordering} that is the obstruction to the ghost or matter sector energy-momentum operator transforming as a tensor, not the change in coordinates. Conversely, the transformation rule (\ref{eq:Tw1zw-naive}) can be applied to either the matter or ghost sector independently {\it provided} we keep the normal ordering fixed, 
\begin{equation}\label{eq:Tw1zw-naive2}
(\partial_{z_1}w_1)^2:\!T_m^{(w_1)}\!:_{z_1}\,=\,:\!T^{(z_1)}_m\!:_{z_1},\qquad {\rm and}\qquad(\partial_{z_1}w_1)^2:\!T_g^{(w_1)}\!:_{z_1}\,=\,:\!T^{(z_1)}_g\!:_{z_1}.
\end{equation}

\subsubsection*{\underline{Example 3:} Ghost Current}

Another simple example of a non-primary is the {\it ghost current} defined by, 
$$
j_{\rm g}^{(z_1)}(z) =-:\!b^{(z_1)}c^{(z_1)}(z)\!:_{z_1},
$$  
with a corresponding charge, 
$$
Q_{\rm g}^{(z_1)}\dfn \frac{1}{2\pi i}\oint \big(\rmd z\,j_{\rm g}^{(z_1)}(z)-\rmd\bar{z}\,\tilde{j}_{\rm g}^{(z_1)}(\bar{z})\big).
$$
This charge depends on the choice of normal ordering and it only has the usual interpretation (where $c$ and $b$ have ghost charge $1$ and $-1$ respectively) when normal ordering is carried out in cylindrical coordinates, $w_1$, with $z_1=e^{-iw_1}$. According to the above procedure, the change in normal ordering (keeping coordinates fixed), for a general holomorphic function, $w_1(z_1)$, is given by:
\begin{equation}
\begin{aligned}
:\!j_{\rm g}^{(z_1)}(z)\!:_{w_1}  = :\!j_{\rm g}^{(z_1)}(z)\!:_{z_1}+\frac{3}{2}\frac{\partial_{z_1}^2w_1}{\partial_{z_1}w_1},
\end{aligned}
\end{equation}
So given that inside the normal ordering we can change coordinates naively and that $j_{\rm g}$ has weight $h=1$,
$$
\partial_{z_1}w_1:\!j_{\rm g}^{(w_1)}(w)\!:_{w_1}  =\,\, :\!j_{\rm g}^{(z_1)}(z)\!:_{z_1}+\frac{3}{2}\frac{\partial_{z_1}^2w_1}{\partial_{z_1}w_1}.
$$
As always, when chart coordinates and normal ordering coordinates coincide we omit the normal ordering symbol, e.g.,
$$
j_{\rm g}^{(w_1)}(w)\equiv \,\,:\!j_{\rm g}^{(w_1)}(w)\!:_{w_1},\qquad {\rm and}\qquad j_{\rm g}^{(z_1)}(z)\equiv\,\,:\!j_{\rm g}^{(z_1)}(z)\!:_{z_1},
$$
and similarly for corresponding charges, etc. 
In terms of cylindrical coordinates, the conventional ghost charge is then given by:
\begin{equation}
\begin{aligned}
Q_{\rm g}^{(w_1)} &\dfn \frac{1}{2\pi i}\oint \big(\rmd w\,j_{\rm g}^{(w_1)}(w)-\rmd\bar{w}\,\tilde{j}_{\rm g}^{(w_1)}(\bar{w})\big)\\
&=\frac{1}{2\pi i}\oint \big(\rmd z\,j_{\rm g}^{(z_1)}(z)-\rmd\bar{z}\,\tilde{j}_{\rm g}^{(z_1)}(\bar{z})\big)-3\\
&=Q_{\rm g}^{(z_1)} -3.
\end{aligned}
\end{equation}

\subsubsection*{\underline{Example 3:} Primaries}

The energy-momentum example is particularly elucidating which is why we elaborated on it in detail. Drawing from that we can, e.g., make a general statement about all primaries with {\it vanishing weights}. These are {\it invariant} under changes in normal ordering when chart coordinates are held fixed, that is:\footnote{This relation is at the heart of the physical state conditions for primary vertex operators \cite{Polchinski87}.}
$$
\boxed{:\!\mathscr{O}^{(z_1)}(p)\!:_{w_1}=\,:\!\mathscr{O}^{(z_1)}(p)\!:_{z_1},\qquad \textrm{$h=\tilde{h}=0$ primaries for $w_1(z_1)$ holomorphic}}
$$
Consequently, if such primaries are constructed to transform {\it naively} as {\it scalars} under conformal changes of coordinates (with fixed normal ordering), then these operators will automatically transform as scalars after normal ordering also:
$$
:\!\mathscr{O}^{(w_1)}(p)\!:_{w_1}\,\,=\,\,\,:\!\mathscr{O}^{(z_1)}(p)\!:_{w_1}=\,\,\,:\!\mathscr{O}^{(z_1)}(p)\!:_{z_1},
$$
implying, in turn, that they can (after translating them to integrated picture) be integrated on an arbitrary Riemann surface in a globally well-defined way. They are, in particular, guaranteed to agree on patch overlaps when we glue with holomorphic transition functions. This is ultimately why primaries are simple to work with on general Riemann surfaces. Unfortunately, handle operators are offshell and are not primaries, so one must work harder. 

\subsubsection*{\underline{Example 4:} Rescalings, Rotations and Shifts}

Building on what we have learnt, we can show immediately that {\it all} local operators, $\hat{\mathscr{A}}^{(z_1)}(p)$, (including offshell vertex operators, which might be associated to handle operators) transform as (possibly linear superpositions of) primaries (i.e.~tensors) under rigid rescalings, rigid rotations and rigid shifts of coordinates (including corresponding changes in normal ordering). In particular, making use of the notation we have developed, under:
$$
\boxed{z_1\mapsto w_1(z_1)=\lambda z_1+a}
$$ 
according to (\ref{eq:GGGw1}), the change in normal ordering with fixed coordinates is entirely encoded in the quantities:
\begin{equation}\label{eq:GGGw1=lambdaz1+a}
\begin{aligned}
G_{\rm m}^{(\lambda z_1+a\,;z_1)}(z',z)&= G_{\rm m}^{(z_1)}(z',z)-\frac{\alpha'}{2}\ln|\lambda|^2\\
G_{\rm g}^{(\lambda z_1+a\,;z_1)}(z',z)&=G_{\rm g}^{(z_1)}(z',z),\qquad {\rm with}\qquad z'=z_1(p'),\quad z=z_1(p),
\end{aligned}
\end{equation}
and derivatives thereof. 
Since $\lambda$ is a constant, the shift in the first relation in (\ref{eq:GGGw1=lambdaz1+a}) can only affect a zero mode contribution in the matter sector of a general local vertex operator. Since this zero mode usually resides in an overall exponential (with remaining factors being independent of it), according to (\ref{eq:CNOxbc-w<->z}) or (\ref{eq:CNOxbc-w1}) and (\ref{eq:GGGw1=lambdaz1+a}),
\begin{equation}\label{eq:CNOexpikx-w1}
\begin{aligned}
:\!\exp\big(ik\cdot x_{(z_1)}\big)\!:_{\lambda z_1+a} \,\,&= (\lambda\bar{\lambda})^{-\frac{\alpha'}{4}k^2} :\!\exp\big(ik\cdot x_{(z_1)}\big)\!:_{z_1}
\end{aligned}
\end{equation}
where we made use of the functional identity (\ref{eq:exp YdK}). 
Therefore, {\it any local operator} that depends on the matter zero modes as in (\ref{eq:CNOexpikx-w1}) transforms as a {\it tensor} under {\it rigid rescalings, rotations and shifts}, $z_1\mapsto \lambda z_1+a$,
\begin{equation}\label{eq:CNOAexpikx-w1}
\boxed{
\begin{aligned}
:\!\hat{\mathscr{A}}^{(z_1)}(x,b,c)e^{ik\cdot x_{(z_1)}}(p)\!:_{\lambda z_1+a} \,\,&= (\lambda\bar{\lambda})^{-\frac{\alpha'}{4}k^2} :\!\hat{\mathscr{A}}^{(z_1)}(x,b,c)e^{ik\cdot x_{(z_1)}}(p)\!:_{z_1},
\end{aligned}
}
\end{equation}
for any local operator, $\hat{\mathscr{A}}^{(z_1)}(x,b,c)$, constructed out of $x_{(z_1)}(p),b^{(z_1)}(p),c^{(z_1)}(p)$ (and derivatives thereof) that is independent of the matter zero modes (so that $x_{(z_1)}(p)$ only appears differentiated). This is because the subtractions in either of the two frames are carried out using the same propagator up to a constant shift, see (\ref{eq:GGGw1=lambdaz1+a}). Equivalently, the quantities $\Delta_{\rm m}^{(w_1;z_1)}=$const.~and $\Delta_{\rm g}^{(w_1;z_1)}=0$ in (\ref{eq:DDDw<->z}) so that the change in normal ordering with fixed coordinates (\ref{eq:CNOxbc-w<->z}) is almost trivial. 
The generalisation to local operators that are chirally split \cite{SklirosCopelandSaffin17}, where $x=x_L+x_R$, and have well-defined chiral and anti-chiral scaling dimensions is also immediate,
\begin{equation}\label{eq:CNOAexpikx-w1LR}
\begin{aligned}
:\!\hat{\mathscr{A}}^{(z_1)}&(x,b,c)e^{ik_L\cdot x^{(z_1)}_L+ik_R\cdot x^{(z_1)}_R}(p)\!:_{\lambda z_1+a} \,\,= \\
&=\lambda^{-\frac{\alpha'}{4}k_L^2}\bar{\lambda}^{-\frac{\alpha'}{4}k_R^2} :\!\hat{\mathscr{A}}^{(z_1)}(x,b,c)e^{ik_L\cdot x^{(z_1)}_L+ik_R\cdot x^{(z_1)}_R}(p)\!:_{z_1}
\end{aligned}
\end{equation}
Generalising further, if we wish to remain agnostic about the zero mode dependence and matter content, we have that for arbitrary operators (\ref{eq:CNOAexpikx-w1LR}) gets replaced by:
\begin{equation}\label{eq:CNOAexpikx-w1gen}
\boxed{
\begin{aligned}
:\!\hat{\mathscr{A}}^{(z_1/q+a)}(p)\!:_{z_1/q+a} \,\,&= q^{L_0^{(z_1)}}\bar{q}^{\tilde{L}_0^{(z_1)}} :\!\hat{\mathscr{A}}^{(z_1)}(p)\!:_{z_1}
\end{aligned}
}
\end{equation}
which we simply write as $\hat{\mathscr{A}}^{(z_1/q+a)}(p)$ and $\hat{\mathscr{A}}^{(z_1)}(p)$ respectively, so that as mentioned when the explicit normal ordering symbols are absent it is always to be understood that normal ordering is carried out in the frame depicted.
\sk

In the following section we begin to study handle operators, starting from general properties that are independent of the choice of matter CFT.

\section{Handle Operators}\label{sec:mainSCS}
In this section we apply what we have learnt in Sec.~\ref{sec:RS} and Sec.~\ref{sec:TPIM} to cutting open and gluing full string amplitudes across arbitrary (trivial or non-trivial) homology cycles. When we cut open a handle we end up with one (or two) Riemann surfaces with states on the resulting two boundaries. Assuming these boundary states can be chosen to have well-defined (total) scaling dimensions, we may quite generally (at fixed complex structure) map these to local operators using the operator-state correspondence. In particular, making use of freedom in time-slicing the path integral in different ways, we can imagine having a boundary state and then gluing in a disc with a local vertex operator at the centre. This local operator being chosen such that it reproduces the boundary state we started from. It will be efficient to work in a coherent state basis for the resulting local vertex operator(s). It is actually not at all obvious {\it a priori} that such a local offshell coherent state vertex operator exists, and if it does exist locally that it also extends to a globally well-defined operator, at least modulo U$(1)$ frame rotations. 

\subsection{Defining Properties of Coherent States}\label{sec:SCS}
An appropriate {\bf definition} for a {\bf string coherent state} that is local\footnote{The notion of {\it locality} on the worldsheet is quite subtle and potentially even ambiguous.  
To clarify, a local insertion on the worldsheet may in general be constructed out of multiple operators that are inserted and integrated over the remaining Riemann surface, at least in a subset of the full moduli space. So an insertion that is interpreted as local from one viewpoint may have a non-local origin. (This follows from properties of cutting and gluing path integrals across various cycles.) Then there is also the possibility of constructing non-local coherent states, whereby the aforementioned insertions are manifest. Both of these cases are interesting, as is the interplay between them. 
The definition given here makes worldsheet locality manifest. 
} 
on the worldsheet (but in general non-local in spacetime) was put forward in \cite{HindmarshSkliros10,Skliros11DPhil,SklirosHindmarsh11}. Various types of coherent states in the BRST formulation have been considered in the literature, e.g.~\cite{DiVecchiaNakayamaPetersenSciuto87}. Here we abstract away from the usual formulations\footnote{(But as in \cite{HindmarshSkliros10,Skliros11DPhil,SklirosHindmarsh11} we still wish to identify them with local operators with well-defined scaling dimension, even offshell.)}. Geometrically, we present a definition for a Riemann surface with fixed complex structure and we make moduli deformations explicit (from a variety of viewpoints) in forthcoming sections. When placed into  a general (background-independent and superstring) context the definition is the following:
\begin{itemize}\label{cohstatedfn}
\item[{\bf (1)}] {\bf continuity:} it depends on a {\it set of continuous} (and possibly also discrete) quantum numbers
\item[{\bf (2)}] {\bf completeness:} it produces a resolution of unity with respect to these quantum numbers, e.g.,
\begin{equation}\label{eq:completeness}
\begin{aligned}
&e^{-\chi(\Sigma)\Phi}\av[\Big]{\dots_1\dots_2\Big\rangle_{\Sigma}=\suminnt\limits_{a} \,\,e^{-\chi(\Sigma_2)\Phi}\Big\langle\dots_1\hat{\mathscr{A}}_{a}^{({\bf z}_1)}\Big\rangle_{\Sigma_1}e^{-\chi(\Sigma_2)\Phi}\Big\langle\hat{\mathscr{A}}^{a}_{({\bf z}_2)}\dots_2}_{\Sigma_2},\\
\end{aligned}
\end{equation}
where the $\hat{\mathscr{A}}_{a}^{({\bf z}_1)}$ span the full string Hilbert space of interest, inserted at the origin of a local coordinate frame labelled by\footnote{Depending on context, the label ${\bf z}_1$ represents a full set of (Grassmann-even and possibly Grassmann-odd) chiral and anti-chiral parameters required to specify a location on the corresponding (super-)Riemann surface in a local patch. There is also an {\it implicit} and context-dependent gluing condition relating ${\bf z}_1$ to ${\bf z}_2$. In general we glue with (super)conformal transformations, but the corresponding coordinates do not usually extend analytically beyond patch overlaps. In (super)string theory we also integrate over the (super)moduli space of (super-)Riemann surfaces of interest, and with some care \cite{Witten12c,LaNelson90} the corresponding measures also factorise for separating degenerations as shown in (\ref{eq:completeness}). Branch cuts associated to Ramond sector handles require particular care \cite{PolchinskiCai88,Witten12c,deLacroixErbinKashyapSenVerma17}. There are precisely analogous relations for non-separating degenerations as we discuss later.} ${\bf z}_1$ of a chart on $\Sigma_1$, and $\hat{\mathscr{A}}^{a}_{({\bf z}_2)}$ is its dual (specified below) and inserted locally at the origin of coordinate frame ${\bf z}_2$ of a chart on $\Sigma_2$. The notation, {\footnotesize{${\Sigma\!\!\!\!\!\int}_a$}}, is suggestive that we integrate over the continuous quantum numbers (including those mentioned in {\bf (1)}) and sum over any discrete ones. A `$\hat{\phantom{A}}$' on $\hat{\mathscr{A}}_a$ denotes {\it fixed-picture} vertex operators, and the corresponding integrated-picture vertex operators will be denoted by $\mathscr{A}_a$.
\item[{\bf (3)}] {\bf symmetries:} it transforms correctly under all (super)string symmetries of interest 
\end{itemize}
We next provide a very concise explanation of the various items in the above list before entering into greater depth in the remaining article. 
\sk

In {\bf (1)} the notion of continuity may best be exhibited by an example: for coherent states denoted by $\hat{\mathscr{A}}_a$ (and their duals by $\hat{\mathscr{A}}^a$) labelled by continuous quantum numbers $\{a\}$ there should exist a limit for the two-point sphere amplitude: $\lim_{a\rightarrow b}e^{-2\Phi}\langle \hat{\mathscr{A}}^a\hat{\mathscr{A}}_b\rangle_{S^2}$. This it to be contrasted with, e.g., momentum quantum numbers (associated to momentum eigenstates) where the overlap would be singular and such a limit does not exist. (Of course there can and will generically also be momentum quantum numbers characterising the states.) So momentum and position do not qualify \cite{KlauderSkagerstam85} as continuous quantum numbers. An important special case is when the sum/integral over quantum numbers $\{a\}$ is interpreted as a path integral over {\it spacetime fields}, but this will actually be a derived concept.
\sk

In {\bf (2)} the dots in (\ref{eq:completeness}) denote arbitrary sets of (usually fixed-picture \cite{Witten12c}) vertex operator insertions and the (context-dependent) path integral measures and supermoduli space are implicit (or we can think of (\ref{eq:completeness}) as a CFT axiom where one does not integrate over supermoduli space). The expectation value with subscript $\Sigma$ denotes carrying out the path integral over all elementary matter and (super-)ghost fields on (super-)Riemann surface $\Sigma$ with an appropriate (implicit) ghost measure, as well as a finite-dimensional integral over all worldsheet (super-)moduli. 
So the above definition is given for a Riemann surface with fixed complex structure. Similar remarks hold for the expectation values associated to $\Sigma_1$, $\Sigma_2$ on the right-hand side; the result and insertions factorise in the simple manner shown (with an important detail related to Ramond (R) pinches where there is also an insertion ``on the pinch'' and it is more precise to think in terms of divisors \cite{Witten12c}). The formal parameter $\Phi$ in (\ref{eq:completeness}) may, e.g., be identified with the zero mode of the dilaton (when a rigid shift symmetry, $\Phi\mapsto \Phi+c$, is present) but more generally has been included explicitly for book-keeping purposes, the quantity $\chi(\Sigma)$ being the corresponding Euler characteristic of the reduced Riemann surface $\Sigma_{\rm red}$ (obtained from $\Sigma$ by setting all odd parameters to zero \cite{Witten12c}) and one usually identifies $g_s=e^{\Phi}$ with the string coupling. For example, for closed genus-$\g $ Riemann surfaces, $\chi(\Sigma)=2-2\g $. 
\sk

Finally, item {\bf (3)} is simply to be interpreted in the usual sense: any {\it external} vertex operator should be BRST invariant if the corresponding amplitudes are to be physical or pure gauge, and any string amplitude should be invariant under phase rotations of the coordinate used to specify its location (because as we have discussed, see e.g.~Sec.~\ref{sec:WB}, this phase is not globally defined, the obstruction being the Euler number \cite{Polchinski88}). One difference compared to mass/momentum eigenstates is that coherent states are not required to transform as irreducible representations of the Poincar\'e group. As for states propagating through internal cycles these are generically not BRST-invariant unless the corresponding handle is cut (in the Cutkosky sense), but the phase of the corresponding centred chart coordinate used to specify their location will always be integrated.
\sk

It is natural to conjecture that the three conditions {\bf (1)}-{\bf (3)} must be satisfied by any coherent state construction in (super)string theory that is local on the worldsheet (but perhaps non-local in spacetime), by direct analogy with defining properties of coherent states in field theory or quantum mechanics \cite{KlauderSkagerstam85}. We state from the outset however that these definitions clearly will not determine the coherent state basis uniquely; for example, the sum/integral over $a$ includes a sum over ghost numbers, and usually only a subset of all coherent states survive due to ghost number conservation. For separating degenerations a single term in this sum survives when asymptotic states have definite ghost number (see below for further detail) whereas for non-separating degenerations the number is much larger (possibly countably infinite). In general, in handle operators we sum over ghost number. Furthermore, we can always add BRST exact terms (or powers thereof) to any given coherent state basis and these need-not be equivalent (see for example the zero momentum case in \cite{Polchinski88}).  In cuts, consistency requires that only onshell coherent states should survive, and we shall demonstrate this explicitly when we discuss the Virasoro-Shapiro amplitude derived by gluing two three-point amplitudes. We discuss gauge invariance in the handle operator formalism in Sec.~\ref{sec:BRST-AC}.

\subsection{Cutting and Gluing in General CFT's}\label{sec:CGGCFT}
It is convenient to begin by considering cutting and sewing in CFTs, where there are no worldsheet moduli to integrate over.\footnote{This equivalently applies to degenerations to which we do not associate any moduli deformations. A concrete example is the following: cut open a sphere two-point string amplitude and insert a resolution of unity --clearly there will be no moduli associated to this degeneration.} 
We will appeal to a standard conformal field theory result\cite{Polchinski88,Sonoda88a,Sonoda88b,Polchinski_v1,Sen15b}, but which may also be regarded as an axiom, sometimes referred to as the {\it gluing axiom}. For a fairly broad and modern discussion of two-dimensional CFT's see \cite{Yin17}.  Before delving into detail, let us very briefly outline the procedure. 
\sk

We consider a (trivial or non-trivial) homology cycle (or handle) (see Fig.~\ref{fig:sep-nonsep} on p.~\pageref{fig:sep-nonsep}) and cut open the path integral across this handle (i.e.~across the common boundary of the two annuli shown in each of the two figures). If we associate two charts, $(U_1',z_1)$ and $(U_2',z_2)$, to the (one or two) resulting surface(s) which now contains two holes we can perform a coordinate transformation that sets the radius of either or both circles to one, $|z_1|=|z_2|=1$, and glue with transition function $z_1z_2=1$ on patch overlaps. Incidentally, by patch or chart overlaps here we mean the chart overlaps inherited by the atlas of the original uncut surface, since after it is cut open it may incorrectly seem that such overlaps are absent. (This comment is directly related to Fig.~\ref{fig:genericdualtriangles} (on p.~\pageref{fig:genericdualtriangles}); note that Fig.~\ref{fig:sep-nonsep} (on p.~\pageref{fig:sep-nonsep}) depicts {\it cell decompositions}, which are turn analogous to the {\it solid lines} in the second diagram in Fig.~\ref{fig:genericdualtriangles}.) We then insert a sum/integral over a complete set of boundary states, each of which is replaced by a disc of unit radius with corresponding local operator insertions, $\hat{\mathscr{A}}_a^{(z_1)}$, and $\hat{\mathscr{A}}_b^{(z_2)}$, (on $\Sigma_1$ and $\Sigma_2$ respectively in the case of separating degenerations and $\Sigma$ for non-separating degenerations), and sum over all such operators. Finally, we (path) integrate out the fields that map the boundary into field space. This is a standard procedure in quantum mechanics where it is referred to as `time slicing the path integral' and the corresponding freedom in time slicing in different ways, all of which are equivalent. So by the basic principles of quantum mechanics this procedure precisely reproduces the original path integral with fixed complex structure, but we have replaced the smooth handle by a pair of local offshell vertex operators, and we sum over all such operators. 
\sk

To incorporate small moduli deformations we may then, e.g., pick a gauge slice in moduli space such that we associate translation, pinch and twist moduli to this handle (and we can proceed similarly for any remaining handles or cycles depending on the slice of interest and subject to the number of ghost zero modes). This can be incorporated, effectively, by dressing these local operators, $\hat{\mathscr{A}}_a^{(z_1)}$, and $\hat{\mathscr{A}}_b^{(z_2)}$, accordingly with the $\hat{B}_k$-ghost contributions from the path integral measure derived in Sec.~\ref{sec:TPIM} (by partitioning them appropriately). To then transition from small to finite or large moduli deformations we must ensure that the resulting integrand is globally well-defined in moduli space and modular invariant. This in turn ensures that diffeomorphism invariance (which includes diffeomorphisms continuously connected to the identity and large diffeomorphisms not continuously connected to the identity) of the full amplitude remains intact and we end up with a well-defined path integral. (Recall from above however that we fix the invariance associated to Weyl transformations when we use holomorphic normal coordinates to define the handle operators.)

\sk
Since this procedure is composed of a multitude of subtle steps, to proceed it helps to break it down into smaller chunks. Primarily, focusing on local properties of the aforementioned cutting procedure, this is most easily carried out if the operators, $\hat{\mathscr{A}}_a^{(z_1)}$, and $\hat{\mathscr{A}}_b^{(z_2)}$, have well-defined {\it scaling dimensions} (i.e.~we will take as a minimum requirement that they be eigenstates of $L_0+\tilde{L}_0$, even offshell). If we are to identify $\hat{\mathscr{A}}_a^{(z_1)}$ and $\hat{\mathscr{A}}_b^{(z_2)}$ with general offshell and local coherent state vertex operators (which is the objective) requiring that $\hat{\mathscr{A}}_a^{(z_1)}$, $\hat{\mathscr{A}}_b^{(z_2)}$ have well-defined scaling dimensions is a non-trivial constraint that we shall study in great detail below. (The fact that this is non-trivial is due to the fact that these operators correspond to an infinite superposition of mass eigenstates, each one of which {\it a priori} have different scaling dimension since they have different offshell momenta and masses.) In fact, requiring $\hat{\mathscr{A}}_a^{(z_1)}$ have well-defined scaling dimension is not absolutely essential.

\sk
Let us assume that a coherent state basis for these vertex operators can be found, and let the corresponding quantum numbers $a,b$ contain the appropriate continuous subset (in accordance with the definition in Sec.\!~\ref{sec:SCS}). In fact, the procedure we discuss holds for arbitrary bases. Every operator is inserted at the origin of their corresponding coordinate patches, namely at $z_1=z_2=0$. Then, focusing primarily on a {\it separating degeneration} for concreteness, the above procedure (described in a certain amount of detail in \cite{Polchinski88,Alvarez-GaumeGomezMooreVafa88,Sonoda88a,Polchinski_v1,Sen15b}) leads to the following factorisation formula (which is independent of basis):
\begin{equation}\label{eq:completenessc}
e^{-\chi(\Sigma)\Phi}\Big\langle \dots_1\dots_2\Big\rangle_{\Sigma}=\suminnt\limits_{a,b} \,\,e^{-\chi(\Sigma_1)\Phi}\Big\langle\dots_1\hat{\mathscr{A}}_{a}^{(z_1)}\Big\rangle_{\Sigma_1}\mathscr{G}^{ab}e^{-\chi(\Sigma_2)\Phi}\Big\langle\hat{\mathscr{A}}_{b}^{(z_2)}\dots_2\Big\rangle_{\Sigma_2},
\end{equation}
with gluing relation $z_1z_2=1$, and the labels $a,b$ span a ``complete'' set of coherent states, but they will not necessarily be BRST invariant (or primaries). The quantity $\chi$ is the Euler characteristic of the corresponding uncut surfaces, so, e.g., if $\Sigma$ is a genus $\g$ Riemann surface with $\n$ vertex operator insertions and $\Sigma_1$, $\Sigma_2$ are Riemann surfaces of genus $\g_1$, $\g_2=\g-\g_1$ with, say, $\n_1$ and $\n_2=\n-\n_1$ vertex operators respectively then $\chi(\Sigma)=2-2\g$, $\chi(\Sigma_1)=2-2\g_1$ and $\chi(\Sigma_2)=2-2\g_2$.  It is important to note that the sums over $a,b$ are {\it a priori} independent and implicitly also contain a sum over ghost numbers.\footnote{In the corresponding superstring summing also over picture number would be overcounting \cite{deLacroixErbinKashyapSenVerma17} (up to possible boundary terms in moduli space), so one normally chooses a picture number (such as the canonical one \cite{Witten12c}) and sums over states of that picture number. For offshell states one is more or less forced to pick the canonical picture number.} Ghost charge conservation (which is discussed below) will eliminate most of these states. The easiest way to keep track of this will be to glue with states of indefinite ghost number (as discussed below), because then one can use the same states for all factorisations. Furthermore, the operators $\hat{\mathscr{A}}_a^{(z)}$ span all elements of all conformal families (each of which is labelled by one primary and all of its descendants \cite{BelavinPolyakovZamolodchikov84}).  In the case of interest (where the total central charge vanishes) there are also degenerate families \cite{BelavinPolyakovZamolodchikov84} containing null states that nevertheless also satisfy the primary state conditions, so care is needed: by the `no ghost' theorem the BRST exact and unphysical states will decouple when they are expected to (up to possible boundary terms which require particular care) \cite{Sen15b,Witten12c}. We will discuss this appearance of boundary terms in moduli space in detail in Sec.~\ref{sec:BRST-AC} on p.~\pageref{sec:BRST-AC}.
\sk

There will generically also be contributions from {\it non-level-matched states}, i.e.~vertex operators, $\hat{\mathscr{A}}_a^{(z_1)}$, not annihilated by $L_0^{(z_1)}-\tilde{L}_0^{(z_1)}$, when the cycle we cut open corresponds to a {\it non-trivial homology cycle} of $\Sigma$ (due to global properties of the moduli space\footnote{The standard example where non-level-matched states are seen to contribute to the path integral is the torus amplitude, where although the relevant phase, $\tau_1$, is integrated, there does not arise a Kronecker delta since the lower limit of the $\tau_2$ integral depends on $\tau_1$ \cite{Polchinski86}. DPS thanks Edward Witten for highlighting an implication of this and Ashoke Sen for a related discussion in the string field theory context. In closed string field theory the non-level matched contributions are absorbed into string vertices (in the Feynman diagram sense).}), see Fig.~\ref{fig:sep-nonsep} on p.~\pageref{fig:sep-nonsep}. For a {\it trivial homology cycle} degenerations the integral over the relevant phase, $z_1/|z_1|$, actually does\footnote{This presumably follows from the explicit calculation of Sec.~\ref{sec:VSG} combined with a locality argument.} give rise to a Kronecker delta so that in fact only level-matched states contribute. (Incidentally, this latter comment is related to why {\it external} state vertex operators can be taken to be level-matched even when taken off the mass shell.)

\sk
Concerning the dilaton dependence in (\ref{eq:completenessc}), note that for a genus-$\g $ compact oriented Riemann surface the Euler number $\chi(\Sigma)=2-2\g $, and the powers of the string coupling, $g_s\equiv e^{\Phi}$, on the left- and right-hand sides will match if there is a factor $e^{2\Phi}$ distributed implicitly inside the operator $\hat{\mathscr{A}}_a\mathscr{G}^{ab}\hat{\mathscr{A}}_b$, which is precisely the expected dilaton coupling for {\it two} closed string vertex operators. So it is consistent to take the $\hat{\mathscr{A}}_a$ to implicitly contain a factor of $g_s$, and similarly for $\hat{\mathscr{A}}_b$, so that the quantity $\mathscr{G}^{ab}$ does {\it not} contain dilaton dependence. 
It is useful to absorb the (as of yet) undetermined quantity $\mathscr{G}^{ab}$ into a redefinition:
\begin{equation}\label{eq:A^adfn}
\boxed{\hat{\mathscr{A}}^a_{(z_2)}\dfn \suminnt\limits_b\,\,\mathscr{G}^{ab}\hat{\mathscr{A}}_b^{(z_2)}}
\end{equation}
and we will call the new object $\hat{\mathscr{A}}^a_{(z_2)}$ the {\it dual} of $\hat{\mathscr{A}}_a^{(z_2)}$ \cite{Witten12c}. Notice that we conventionally ``raise the index'' by left multiplication. The dual, $\hat{\mathscr{A}}^a_{(z_2)}$, is \cite{Zwiebach93} nor the BPZ nor the hermitian conjugate\footnote{More precisely, we are working on a Euclidean signature worldsheet so the notion of `hermitian conjugate' is replaced by `Euclidean adjoint' \cite{Polchinski_v1}.} of $\hat{\mathscr{A}}_a^{(z_2)}$, (primarily because the ghost charges of $\hat{\mathscr{A}}_a$ and $\hat{\mathscr{A}}^a$ generically differ) and we derive explicitly its precise relation to $\hat{\mathscr{A}}_a^{(z_1)}$ below. 
\sk

The matter sector in $\hat{\mathscr{A}}^a_{(z_2)}$ will be identified with the Euclidean adjoint of the matter sector in $\hat{\mathscr{A}}_a^{(z_1)}$. We will also see very explicitly why it is not necessary \cite{Witten12c} to ever take the hermitian conjugate of a Grassmann-odd variable. (The point here is that the Grassmann-odd quantum numbers associated to the ghost sector of the coherent state, $\hat{\mathscr{A}}_a^{(z_1)}$, are completely independent of the corresponding Grassmann-odd quantum numbers associated to the ghost sector of the dual coherent state, $\hat{\mathscr{A}}^a_{(z_2)}$: the two sets of quantum numbers are coupled, or entangled, only via the measure, $\Sigma_a \!\!\!\!\!\!\!\int\,\,\,$.)
\sk

Following \cite{Polchinski_v1}, in order to identify $\mathscr{G}^{ab}$ let us set $\Sigma_2=S^2$ while replacing $\dots_2$ in (\ref{eq:completenessc}) with a local operator, $\hat{\mathscr{A}}_c^{(z_3)}$, in a patch coordinatised by $z_3$ (and placed, say, at the origin $z_3=0$). We then glue to the remaining part of $S^2$ by $z_2z_3=1$, so (\ref{eq:completenessc}) reduces to:
\begin{equation}\label{eq:consistency_condition}
e^{-\chi(\Sigma)\Phi}\Big\langle \dots_1\hat{\mathscr{A}}_c^{(z_3)}\Big\rangle_{\Sigma}=\suminnt\limits_{a,b} \,\,e^{-\chi(\Sigma)\Phi}\Big\langle\dots_1\hat{\mathscr{A}}_{a}^{(z_1)}\Big\rangle_{\Sigma}\mathscr{G}^{ab}e^{-2\Phi}\Big\langle\hat{\mathscr{A}}_{b}^{(z_2)}\hat{\mathscr{A}}_c^{(z_3)}\Big\rangle_{S^2}.
\end{equation}
It is tempting to hence identify $\mathscr{G}^{ab}$ with the ``inverse'' of the standard BPZ inner product, $\mathscr{G}_{ab}$, \cite{BelavinPolyakovZamolodchikov84,Polchinski_v1,Zwiebach93},
\begin{equation}\label{eq:Gab}
\mathscr{G}_{ab}\dfn e^{-2\Phi}\big\langle \hat{\mathscr{A}}_a^{(z_2)}\hat{\mathscr{A}}_b^{(z_3)}\big\rangle_{S^2},
\end{equation}
where $z_2z_3=1$. However, because we would like to be able to identify these states with coherent states where the quantum numbers, $a,b,c,\dots$, contain a {\it continuous} subset, it is not expected to be possible for an inverse to exist in the following strict sense,
\begin{equation}\label{eq:naiveG^ab}
\phantom{\qquad \textrm{(naive inverse)}}\suminnt\limits_b\,\mathscr{G}^{ab}\mathscr{G}_{bc}\equiv \delta^a_{\phantom{a}c}\qquad \textrm{(naive inverse)}
\end{equation}
More precisely, Kronecker delta (or delta function) overlap is inconsistent with the continuity requirement of the defining property of coherent states (see {\bf (1)} on p.~\!\!\pageref{cohstatedfn}),
\begin{equation}\label{eq:deltaij}
e^{-2\Phi}\big\langle \hat{\mathscr{A}}^a_{(z_2)}\hat{\mathscr{A}}_b^{(z_3)}\big\rangle_{S^2}\neq\delta^a_{\phantom{j} b},
\end{equation}
where we have made use of (\ref{eq:A^adfn}), because a smooth or continuous limit $\lim_{a\rightarrow b}\delta^a_{\phantom{j} b}$ does not exist for Kronecker delta or Dirac delta function overlaps. 
Thankfully, we do not need to rely on such an inverse (\ref{eq:naiveG^ab}) in order for the more fundamental consistency requirement (\ref{eq:consistency_condition}) to be satisfied. Adopting the notation (\ref{eq:A^adfn}), all that is necessary for consistent factorisation (\ref{eq:consistency_condition}) is that there exists the notion of a dual, $\hat{\mathscr{A}}^{a}_{(z_2)}$, such that for every operator labelled by quantum numbers $c$:\footnote{Incidentally, the two-point amplitude appearing here is not to be confused with that discussed in \cite{ErbinMaldacenaSkliros19}. The latter and its conclusions rely crucially on the fact that the asymptotic states are {\it onshell}, whereas here the relevant two-point amplitude is offshell. Then, the implicit indefinite ghost number of the various coherent states makes this quantity non-zero.}
\begin{equation}\label{eq:consistency_condition2}
\boxed{\hat{\mathscr{A}}_c^{(z_3)}=\suminnt\limits_{a} \,\,\hat{\mathscr{A}}_{a}^{(z_3)}e^{-2\Phi}\Big\langle\hat{\mathscr{A}}^{a}_{(z_2)}\hat{\mathscr{A}}_c^{(z_3)}\Big\rangle_{S^2}}
\end{equation}
We have taken into account that the gluing conditions, $z_1z_2=1$ and $z_3z_2=1$, are such that we can extend $z_1=z_3$ throughout the sewn region. This fundamental consistency requirement (\ref{eq:consistency_condition2}) is to be interpreted as follows: inserting an operator $\hat{\mathscr{A}}_c^{(z_3)}$ into arbitrary correlation functions is equivalent to inserting the operator ${\Sigma\!\!\!\!\!\int}_{a} \,\,\hat{\mathscr{A}}_{a}^{(z_3)}e^{-2\Phi}\langle\hat{\mathscr{A}}^{a}_{(z_2)}\hat{\mathscr{A}}_c^{(z_3)}\rangle_{S^2}$. This is clearly a much more general statement than (\ref{eq:naiveG^ab}) and remains true for any basis of states including coherent states, whether primaries or (more generally) BRST invariant or not.
\sk

Given that (\ref{eq:consistency_condition2}) is independent of a chosen basis, it also applies to mass (or momentum) eigenstates. That is, denoting momentum eigenstate basis labels by $i,j,\dots$, and a coherent state basis by $a,b,\dots$ as above we can change basis (independently of whether they are orthogonal or not) in string amplitudes by using the relations:
\begin{equation}\label{eq:consistency_condition2b}
\begin{aligned}
&\hat{\mathscr{A}}_i^{(z_3)}=\suminnt\limits_{a} \,\,\hat{\mathscr{A}}_{a}^{(z_3)}e^{-2\Phi}\Big\langle\hat{\mathscr{A}}^{a}_{(z_2)}\hat{\mathscr{A}}_i^{(z_3)}\Big\rangle_{S^2}\\
&\hat{\mathscr{A}}_c^{(z_3)}=\sum_{j} \,\,\hat{\mathscr{A}}_{j}^{(z_3)}e^{-2\Phi}\Big\langle\hat{\mathscr{A}}^{j}_{(z_2)}\hat{\mathscr{A}}_c^{(z_3)}\Big\rangle_{S^2}\\
&\hat{\mathscr{A}}_i^{(z_3)}=\sum_j \,\,\hat{\mathscr{A}}_{j}^{(z_3)}e^{-2\Phi}\Big\langle\hat{\mathscr{A}}^{j}_{(z_2)}\hat{\mathscr{A}}_i^{(z_3)}\Big\rangle_{S^2}
\end{aligned}
\end{equation}

In order to get a handle on how exactly we should identify the dual vertex operators, $\hat{\mathscr{A}}^a_{(z_2)}$, given $\hat{\mathscr{A}}_a^{(z_1)}$, let us primarily consider ghost charge\footnote{Recall \cite{Witten12c} that $c,\tilde{c}$ each contribute ghost charge $N_{\rm gh}=1$, whereas $b,\tilde{b}$ contribute ghost charge $N_{\rm gh}=-1$.} conservation. 
When there exist\footnote{It is also possible to cut open a cycle to which we do not associated any moduli, and this depends on our choice of gauge slice in moduli space and on the specific diagram we cut open. For example, when we cut open a sphere 2-point amplitude there are no moduli associated to this cycle.} moduli and for genus-$\g $, total ghost charge \cite{Witten12c,Sen15b} of the inserted operators into correlation functions (including those of the corresponding $\hat{B}_k$ discussed below) must add up to $6-6\g$, and any terms for which total ghost charge does not add up to this number will give vanishing contribution. This relates the ghost charge of the terms in $\hat{\mathscr{A}}_a^{(z_1)}$ that contribute to that of the terms in the dual, $\hat{\mathscr{A}}^a_{(z_2)}$, that contribute. Recall that the operators, $\hat{\mathscr{A}}_a^{(z_1)}$ and $\hat{\mathscr{A}}^a_{(z_2)}$, need {\it not} have definite ghost charge, and what we are saying is that of the various terms in these operators the only terms that will contribute are those for which the ghost charges are related in a manner that will be explained next. 
\sk

Let us consider (\ref{eq:consistency_condition2}) (identical reasoning applies to (\ref{eq:consistency_condition2b})) and denote the total ghost charge of the terms that contribute in $\hat{\mathscr{A}}_c$, $\hat{\mathscr{A}}_a$ and $\hat{\mathscr{A}}^a$ by $n_c$, $n_a$ and $n^a$ respectively. Ghost charge conservation in (\ref{eq:consistency_condition2}) then tells us that $n^a+n_c=6$ and $n_a=n_c$, the first relation following from ghost charge conservation inside the two-point function (where $\g =0$) and the second following from the fact that once the two-point sphere amplitude is evaluated it is simply a c-number. Therefore, whatever the operators $\hat{\mathscr{A}}^a$ and $\hat{\mathscr{A}}_a$ turn out to be, the only terms that are going to contribute will be those whose total ghost charges are related by,
$$
n^a=6-n_a.
$$ 
This relation between ghost numbers of dual vertex operators is general and holds for both separating and non-separating degenerations of arbitrary genus amplitudes. If we denote the Hilbert space associated to all ghost number $n_a$ states by $\mathcal{H}_{n_a}$, (so that there exist terms in $\hat{\mathscr{A}}_a$ belonging to $\mathcal{H}_{n_a}$) then that of the dual space \cite{Witten12c} will therefore be $\mathcal{H}^*_{n_a}\cong \mathcal{H}_{n^a}=\mathcal{H}_{6-n_a}$, (so that there exist terms in $\hat{\mathscr{A}}^a$ belonging to $\mathcal{H}_{6-n_a}$). The sum/integral over $a$ in (\ref{eq:consistency_condition2}) or (\ref{eq:completenessc}) (on account of (\ref{eq:A^adfn})) therefore contains a discrete sum:
\begin{equation}\label{eq:oplusHHdual}
\oplus_{n_a\in\mathbf{Z}}\mathcal{H}_{n_a}\otimes \mathcal{H}_{6-n_a}.
\end{equation}
This justifies placing the $\hat{B}_k$ insertions associated, e.g., to pinch moduli (see below and also the discussion following (\ref{eq:BqBqbar})) on {\it either} \cite{Polchinski88} $\hat{\mathscr{A}}_a$ or $\hat{\mathscr{A}}^a$ (or indeed inside $\mathscr{G}^{ab}$ which is also a common choice \cite{Sen15b}), and this must be the case as can be seen by a contour argument to shuffle the relevant $\hat{B}_k$ around. These observations lead one to identify the two-point sphere amplitude we have been discussing with the dual (and non-degenerate \cite{Witten12c} when restricted to the BRST cohomology) pairing, $\omega: \mathcal{H}^*\times \mathcal{H}\rightarrow \mathbf{C}$:
\begin{equation}\label{eq:omega(a,c)}
\omega(\hat{\mathscr{A}}^{a},\hat{\mathscr{A}}_c)\dfn e^{-2\Phi}\big\langle\hat{\mathscr{A}}^{a}_{(u)}\hat{\mathscr{A}}_c^{(z)}\big\rangle_{S^2},\qquad {\rm with}\qquad uz=1,
\end{equation}
where recall that both operators are inserted at the origin of their respective coordinate systems, $u,z$, that are in turn glued together on patch overlaps (an equatorial band) to form a sphere by the indicated identification. 
\sk

Summarising, the basic gluing formula associated to cutting open a {\it separating handle}, (a trivial homology cycle) at fixed complex structure moduli takes the form:
\begin{equation}\label{eq:completeness-sep}
\boxed{e^{-\chi(\Sigma)\Phi}\Big\langle \dots_1\dots_2\Big\rangle_{\Sigma}=\suminnt\limits_{a} \,\,e^{-\chi(\Sigma_1)\Phi}\Big\langle\dots_1\hat{\mathscr{A}}_{a}^{(z_1)}\Big\rangle_{\Sigma_1}e^{-\chi(\Sigma_2)\Phi}\Big\langle\hat{\mathscr{A}}^{a}_{(z_2)}\dots_2\Big\rangle_{\Sigma_2}}
\end{equation}
where we glue with transition function $z_1z_2=1$. When we integrate over moduli space there is generically (and implicitly at this stage)  also a sum over permutations of external vertex operators on the right-hand side, determined by the condition that moduli space is covered once (more about which later).
\sk

\begin{figure}
\begin{center}
\includegraphics[angle=0,origin=c,width=0.55\textwidth]{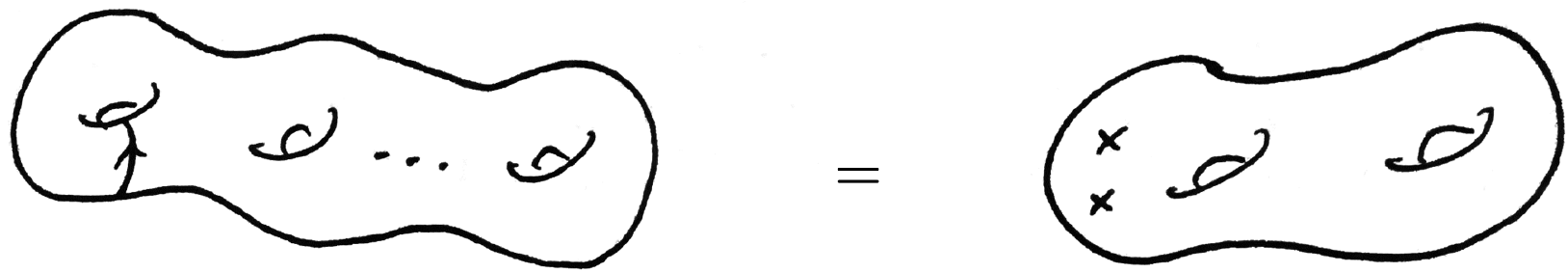}
\caption{Schematic representation underlying the idea that inserting a handle operator on a lower genus Riemann surface can represent (at fixed complex structure) a Riemann surface with one additional loop (a non-trivial homology cycle), as defined in (\ref{eq:completeness_nonsep}). There is an analogous diagram for (\ref{eq:completeness-sep}) whereby two Riemann surfaces, $\Sigma_1,\Sigma_2$, are glued by means of a handle operator across a cycle which from the viewpoint of $\Sigma$ is homologous to zero.}\label{fig:111xx}
\end{center}
\end{figure}

When cutting across a {\it non-separating handle} (associated to a non-trivial homology cycle) similar reasoning applies. The corresponding formula at fixed complex structure moduli is given by:
\begin{equation}\label{eq:completeness_nonsep}
\boxed{e^{-\chi(\Sigma_\g)\Phi}\Big\langle \dots\Big\rangle_{\Sigma_\g}= \,\,e^{-\chi(\Sigma_{\g-1})\Phi}\Big\langle\dots\suminnt\limits_{a}\,\hat{\mathscr{A}}_{a}^{(z_1)}\hat{\mathscr{A}}^{a}_{(z_2)}\Big\rangle_{\Sigma_{\g-1}}}
\end{equation}
and the gluing relation (transition function) is the same as above, $z_1z_2=1$. This corresponds to cutting open a handle and replacing it by a bi-local operator insertion. So if the correlation function on the left-hand side in (\ref{eq:completeness_nonsep}) is over a genus-$\g$ surface, on the right-hand side it is over a genus-$(\g-1)$ surface with the additional bi-local insertion. To reconstruct the entire amplitude we also need to provide the remaining transition functions and cocycle relations, so that we obtain a globally well-defined construction. We emphasise that these formulas (\ref{eq:completeness-sep}) and (\ref{eq:completeness_nonsep}) are exact. A pictorial representation of (\ref{eq:completeness_nonsep}) is shown in Fig.~\ref{fig:111xx}. 
\sk

\begin{figure}
\begin{center}
\includegraphics[width=0.6\textwidth]{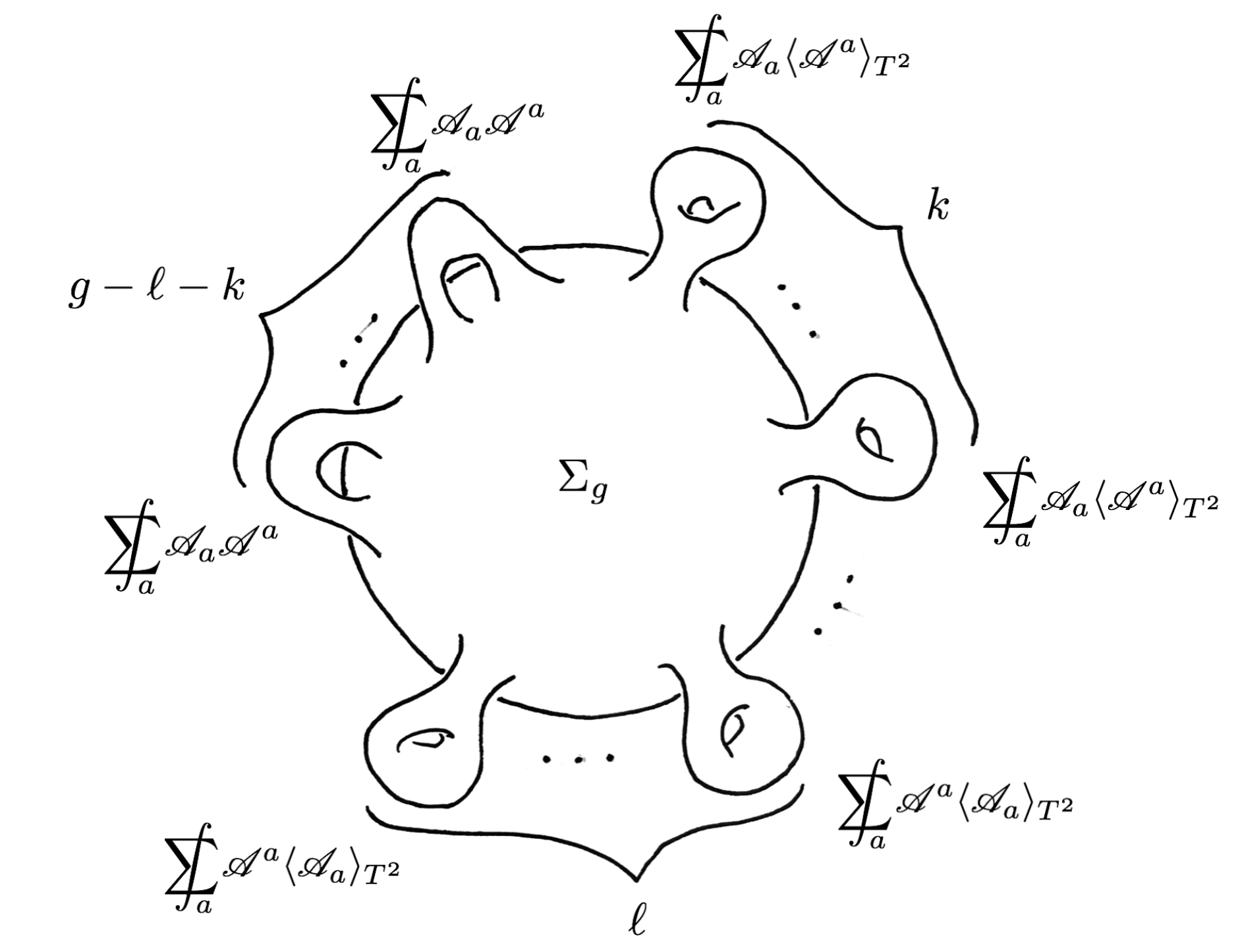}
\caption{One can cut open the path integral using handle operators in a variety of ways, many of which are related by OPE associativity and modular invariance. The insertions are schematic, e.g., path integral measure contributions, $\hat{B}_k$, are implicit.}
\label{fig:Riemann3}
\end{center}
\end{figure}

One can then imagine using such handle operators to cut open the path integral in a variety of ways related by duality (OPE associativity and modular invariance), see Fig.~\ref{fig:Riemann3}. 
The resulting path integral will be exact provided: (a) we sum over the entire basis of intermediate states; (b) we arrive at a globally well-defined construction (under complex structure deformations); and (c) we cover moduli space once. In the current document we will address steps (a) and (b). Step (c) is currently the least well understood, but tree-level worldsheet duality and one-loop modular invariance are checked in Sec.~\ref{sec:EA}. 
\sk

In Sec.~\ref{sec:GWDCS} we generalise the above to include deformations of complex structure moduli, but let us elaborate on various consistency conditions that the fixed-complex structure handle operators must satisfy, as well as a residual symmetry that this cutting procedure leaves unfixed. 

\subsection{Gluing Consistency Condition}\label{sec:GCC}
When the contour of the BRST charge encircles one of the two local operators, say $\hat{\mathscr{A}}_{a}^{(z_1)}$, that comprise the bi-local handle operator it must be possible to deform the contour without obstruction (at fixed complex structure) so that it encircles the dual local operator, $\hat{\mathscr{A}}^{a}_{(z_2)}$. This of course must be the case if the full handle operator is to indeed represent a handle of a Riemann surface. As we will elaborate much more fully in Sec.~\ref{sec:GOCS}, this condition demands that handle operators satisfy the following condition,
\begin{equation}\label{eq:BRST gluing condition}
\big(Q_B^{(z_1)}+Q_B^{(z_2)}\big)\cdot \suminnt\limits_{a}\,\,\hat{\mathscr{A}}_a^{(z_1)} \hat{\mathscr{A}}^a_{(z_2/q)}=0,
\end{equation}
which is the local operator version of the statement that we can deform a BRST-charge contour encircling $\mathscr{A}_a^{(z_1)}$ to that encircling $\mathscr{A}^a_{(z_2)}$ without obstruction, which is certainly necessary (at least at fixed complex structure) if the handle operator is to indeed represent a handle. We rescaled the $z_2$ coordinate by a complex number $q$, so that the gluing relation is now:
$$
z_1z_2=q.
$$
(The BRST charge is not affected by such rescalings, in particular $Q_B^{(a z_1+b)}=Q_B^{(z_1)}$ for any $z_1$-independent quantities $a,b$.)
Similar remarks must also hold for the relevant Virasoro generators (in the standard basis) and the corresponding condition that mode contours can be deformed from one of the two local operators to the other is:
\begin{equation}\label{eq:Ln gluing condition}
\big(L_n^{(z_1)}-L_{-n}^{(z_2/q)}\big)\cdot \suminnt\limits_{a}\,\,\hat{\mathscr{A}}_a^{(z_1)} \hat{\mathscr{A}}^a_{(z_2/q)}=0,
\end{equation}
where note that $L_{-n}^{(z_2/q)}=q^{n}L_{-n}^{(z_2)}$ and $\hat{\mathscr{A}}^a_{(z_2/q)}=q^{L_0^{(z_2)}}\bar{q}^{\tilde{L}_0^{(z_2)}}\hat{\mathscr{A}}^a_{(z_2)}$. 
There is a similar relation for the anti-chiral half. There are also similar relations for the ghost modes and matter modes -- although the matter modes depend on the choice of background. See also (\ref{eq:gluing modesbosonic}) and (\ref{eq:abcmodes1->2}), where these gluing conditions are discussed in much greater detail.

\subsection{Residual Conformal Symmetry}\label{sec:RSG}
We have used explicit holomorphic frames, $z_1,z_2$, in order to represent a handle by a bi-local operator (or handle operator),
$$
\suminnt\limits_{a}\,\,\hat{\mathscr{A}}_a^{(z_1)} \hat{\mathscr{A}}^a_{(z_2/q)},
$$
with corresponding transition functions, $z_1z_2=q$. 
\sk

Consider the set of holomorphic (for $z_1,z_2\neq0$) reparametrisations,
\begin{equation}\label{eq:z1'z2'hol}
\begin{aligned}
z_1\mapsto z_1'&=z_1+\sum_{n\in\mathbf{Z}}\varepsilon_n^1z_1^{n+1}\\
z_2\mapsto z_2'&=z_2+\sum_{n\in\mathbf{Z}}\varepsilon_n^2z_2^{n+1},\\
\end{aligned}
\end{equation}
(with $\varepsilon_n^1,\varepsilon_n^2$ infinitesimal) 
of the frame coordinates, $z_1,z_2$, that leave the gluing relation invariant, 
and hence also the summand of the (fixed-complex structure) handle operator invariant,
\begin{equation}\label{eq:A1'A2'=A1A2}
\boxed{
\begin{aligned}
\hat{\mathscr{A}}_a^{(z_1')}\hat{\mathscr{A}}^a_{(z_2'/q)}&=
\hat{\mathscr{A}}_a^{(z_1)}\hat{\mathscr{A}}^a_{(z_2/q)},\qquad {\rm when}\qquad z_1'z_2'=z_1z_2=q.
\end{aligned}
}
\end{equation}
Substituting (\ref{eq:z1'z2'hol}) into $z_1'z_2'=z_1z_2$ we see that the subset of reparametrisations (\ref{eq:z1'z2'hol}) that leave the gluing relation invariant are precisely the ``diagonal subgroup'' generated by:
\begin{equation}\label{eq:e1e2}
\varepsilon_n^2+q^{-n}\varepsilon_{-n}^1=0,\qquad \forall n\in\mathbf{Z}.
\end{equation}

To show that the fixed-complex structure handle operator is indeed invariant under this subset determined by (\ref{eq:e1e2}) as stated in (\ref{eq:A1'A2'=A1A2}) we make use of a standard result that under holomorphic reparametrisations (\ref{eq:z1'z2'hol}) subject to (\ref{eq:e1e2}), general operators transform as,
\begin{equation}\label{eq:A->A' z->z'}
\begin{aligned}
\hat{\mathscr{A}}_a^{(z_1)}\mapsto \hat{\mathscr{A}}_a^{(z_1')}&=\hat{\mathscr{A}}_a^{(z_1)}-\sum_{n\in\mathbf{Z}}\big(\varepsilon_n^1L_n^{(z_1)}+\bar{\varepsilon}_n^1\tilde{L}_n^{(z_1)}\big)\hat{\mathscr{A}}_a^{(z_1)}\\
\hat{\mathscr{A}}^a_{(z_2/q)}\mapsto \hat{\mathscr{A}}^a_{(z_2'/q)}&=\hat{\mathscr{A}}^a_{(z_2/q)}+\sum_{n\in\mathbf{Z}}\big(\varepsilon_n^1q^nL_{-n}^{(z_2)}+\bar{\varepsilon}_n^1\bar{q}^n\tilde{L}_{-n}^{(z_2)}\big)\hat{\mathscr{A}}^a_{(z_2/q)}.
\end{aligned}
\end{equation}
These relations in turn imply that to leading order in $\varepsilon_n^1,\bar{\varepsilon}_n^1$,\footnote{There is a factor of $q^n$ missing in equation (9.4.9) on p.~303 in \cite{Polchinski_v1}.}
\begin{equation}\label{eq:A1'A2'=A1A2b}
\begin{aligned}
\hat{\mathscr{A}}_a^{(z_1')}\hat{\mathscr{A}}^a_{(z_2'/q)}=
\hat{\mathscr{A}}_a^{(z_1)}\hat{\mathscr{A}}^a_{(z_2/q)}&-\sum_{n\in\mathbf{Z}}\varepsilon_n^1\Big[\big(L_n^{(z_1)}\hat{\mathscr{A}}_a^{(z_1)}\big)\hat{\mathscr{A}}^a_{(z_2/q)}-\hat{\mathscr{A}}_a^{(z_1)}\big(q^nL_{-n}^{(z_2)}\hat{\mathscr{A}}^a_{(z_2/q)}\big)\Big]\\
&- \sum_{n\in\mathbf{Z}}\bar{\varepsilon}_n^1\Big[\big(\tilde{L}_n^{(z_1)}\hat{\mathscr{A}}_a^{(z_1)}\big)\hat{\mathscr{A}}^a_{(z_2/q)}-\hat{\mathscr{A}}_a^{(z_1)}\big(\bar{q}^n\tilde{L}_{-n}^{(z_2)}\hat{\mathscr{A}}^a_{(z_2/q)}\big)\Big],
\end{aligned}
\end{equation}
and therefore taking into account the gluing consistency condition (\ref{eq:Ln gluing condition}) (and the corresponding anti-chiral counterpart) we see that indeed the handle operator remains invariant as stated in (\ref{eq:A1'A2'=A1A2}). (Recall that $L_{-n}^{(z_2/q)}=q^{n}L_{-n}^{(z_2)}$). Incidentally, the two-point amplitude (\ref{eq:omega(a,c)}) is clearly also invariant under (\ref{eq:z1'z2'hol}) subject to (\ref{eq:e1e2}) (where to apply (\ref{eq:A1'A2'=A1A2b}) we can take $q=1$). 
\sk

The residual symmetry (\ref{eq:z1'z2'hol}) subject to (\ref{eq:e1e2}) is quite special. So before moving on let us also briefly discuss yet another holomorphic change of coordinates that does {\it not} (manifestly) leave the amplitude invariant. Consider in particular the set of holomorphic  reparametrisations,
\begin{equation}\label{eq:z1'z2'holb}
\begin{aligned}
z_1\mapsto z_1'&=z_1+\sum_{n\geq0}\varepsilon_n^1z_1^{n+1}\\
z_2\mapsto z_2'&=z_2,
\end{aligned}
\end{equation}
with $\varepsilon_n^1$ infinitesimal, so that $z_2$ remains fixed. The gluing relation, $z_1z_2=q$, for $|q|\ll1$ is now changed to:
\begin{equation}
\begin{aligned}
z_1'z_2'
&=q(1+\varepsilon_0^1)+\mathcal{O}(q^2),
\end{aligned}
\end{equation}
and so the holomorphic reparametrisation (\ref{eq:z1'z2'holb}) induces a change in modulus,
$$
q\mapsto q'=q(1+\varepsilon_0^1)+\mathcal{O}(q^2).
$$
One might not expect such a reparametrisation of $q$ to be visible in string amplitudes, but it is sometimes the case that the integral over $q$ is IR divergent (as $|q|\rightarrow0$), so that (when the cancelled-propagator argument does not apply) an appropriate cutoff, $|q|>\delta$, must be introduced and then the limit $\delta\rightarrow 0$ taken. The reparametrisation  $q\mapsto q'$ then induces a corresponding change in the cutoff prescription, $\delta\mapsto \delta'$, and one must check that physical amplitudes are independent of both the cutoff and of the cutoff prescription. For example, in cases where the cancelled propagator argument does not apply, such as in the case of non-vanishing tadpoles \cite{LaNelson90}, one must deform the background (by inserting a local operator perhaps smeared over a distance $\delta$) in order to restore conformal invariance, and then demonstrate cutoff-prescription independence by showing that there is a corresponding change in normal ordering (as in (\ref{eq:A->A' z->z'})) of the background shift that precisely cancels the $\varepsilon_0^1$ dependence. 

\subsection{The Commutator $\partial_{z_{v}} \!:\!\mathscr{A}\!:-:\partial_{z_{v}} \mathscr{A}\!:$}\label{sec:comm dA-dA}
Handle operators are comprised of a pair of local operators, $\hat{\mathscr{A}}_a^{(z_1)}(p_1)$ and $\hat{\mathscr{A}}^a_{(z_2/q)}(p_2)$. In this section we will consider either one of these local operators, and in particular their derivative with respect to the base point with respect to which the frames have been defined. These base points $p_1,p_2$ might eventually be interpreted as moduli, and when one wishes to show that BRST-exact states decouple from the path integral up to a total derivative in moduli space one ends up having to be precise about whether the relevant derivative is inside or outside the normal ordering defining these local operators, since it must be outside the normal ordering in order to apply Stoke's theorem and end up with a boundary integral in moduli space.
\sk

In particular, derivatives generically do not commute with conformal normal ordering unless the Ricci scalar and its derivatives vanish. In this section we derive the precise commutator in the case that the derivative under consideration is with respect to the base point at which a general local normal-ordered (vertex) operator is inserted. Such base points are often associated to complex structure moduli. It will be efficient to take a transition function approach to shifting punctures so the prerequisite for this section is Sec.~\ref{sec:SPUTF}.
\sk

Since we need to zoom in now on charts that are related by a shift of the base point we will make the notation more explicit again and adopt an auxiliary coordinate system, $\sigma^a$. In particular, we make use of the correspondence:
\begin{equation}\label{eq:z1pzs1 corr}
\boxed{
z_1(p) \leftrightarrow z_{\sigma_1}(\sigma),\qquad z_1(p_1)=0\leftrightarrow z_{\sigma_1}(\sigma_1)=0
}
\end{equation}
So the point we have been denoting by $p_1$ will now be associated to the coordinate value $\sigma=\sigma_1$ in the auxiliary coordinate, and a generic point, $p$, of the chart will be denoted by $\sigma$, (see Sec.~\ref{sec:SPUTF}).
\sk

From the chain rule and (\ref{eq:deltazss'2}) we then learn that a derivative with respect to the modulus, $z_{\sigma_1}(\sigma_1)$, at fixed $\sigma$ can be expressed as:
\begin{equation}\label{eq:d/ds_mod}
\begin{aligned}
\frac{\partial}{\partial z_{\sigma_1}(\sigma_1)}\Big|_{\sigma} &= \frac{\partial z_{\sigma_1}(\sigma)}{\partial z_{\sigma_1}(\sigma_1)}\Big|_{\bar{z}_{\sigma_1}(\sigma)}\frac{\partial}{\partial z_{\sigma_1}(\sigma)}\Big|_{\sigma}+\frac{\partial \bar{z}_{\sigma_1}(\sigma)}{\partial z_{\sigma_1}(\sigma_1)}\Big|_{z_{\sigma_1}(\sigma)}\frac{\partial}{\partial \bar{z}_{\sigma_1}(\sigma)}\Big|_{\sigma}\\
&=\frac{\partial}{\partial z_{\sigma_1}(\sigma)}\Big|_{\sigma} -\frac{1}{4}\sum_{n=1}^\infty \frac{1}{(n+1)!}\nabla_{\bar{z}_{\sigma_1}}^{n-1}R_{(2)}(\sigma_1)\Big(\!-\bar{z}_{\sigma_1}(\sigma)^{n+1}\frac{\partial}{\partial \bar{z}_{\sigma_1}(\sigma)}\Big)\Big|_{\sigma}.
\end{aligned}
\end{equation}
We now recognise the quantity in the parenthesis in the last term on the right-hand side as being a standard basis representation of the Virasoro generator $\tilde{L}_n^{(z_{\sigma_1})}$ based at $\sigma=\sigma_1$ in chart coordinate, $z_{\sigma_1}$, while the first term is $-L_{-1}^{(z_{\sigma_1})}$, so that (\ref{eq:d/ds_mod}) can equivalently be written in the form:
\begin{equation}\label{eq:D_zs}
\boxed{
\hat{D}_{z_{v_1}}  =-L_{-1}^{(z_{\sigma_1})}-\frac{1}{4}\sum_{n=1}^\infty \frac{1}{(n+1)!}\big(\nabla_{\bar{z}_{\sigma_1}}^{n-1}R_{(2)}(\sigma)\big)\Big|_{\sigma=\sigma_1}\!\!\!\tilde{L}_n^{(z_{\sigma_1})}
}
\end{equation}
which was derived from an entirely independent viewpoint above, see (\ref{eq:hatDzv2R}) on p.~\pageref{eq:hatDzv2R}. 
When acting on operators there is a correspondence, 
\begin{equation}\label{eq:d/dzv1=-d/dzs1s1}
\hat{D}_{z_{v_1}}\sim-\frac{\partial}{\partial z_{v_1}}\equiv \frac{\partial}{\partial z_{\sigma_1}(\sigma_1)},
\end{equation}
see also (\ref{eq:dzv=ds1s1}). The correspondence in the first relation in (\ref{eq:d/dzv1=-d/dzs1s1}) requires a certain amount of care, because the derivative acts on every function of $z_v$ whereas as we see in (\ref{eq:D_zs}), acting with $\hat{D}_{z_{v_1}}$ gives a non-vanishing result only when what it acts on is not annihilated by the depicted Virasoro generators. Furthermore, the minus sign appearing in (\ref{eq:d/dzv1=-d/dzs1s1}) is convention-dependent and is included in order to be in agreement with the meaning of $z_{v_1}$ in the remaining document. Related signs can be traced back to the definition (\ref{eq:dzs*}) which leads to the minus sign in (\ref{eq:dz*=-ds*e}). 
The relation (\ref{eq:D_zs}) is essentially the result of interest, written in terms of holomorphic normal coordinates, but let us elaborate further to become entirely explicit. 
\sk

Let us also write (\ref{eq:D_zs}) in terms of $\sigma_1$ and $\sigma$ derivatives. From (\ref{eq:framefield2}) and its complex conjugate we primarily learn that,
\begin{equation}\label{eq:d/ds*}
\begin{aligned}
\frac{\partial}{\partial\sigma_1^a}&=\frac{\partial z_{\sigma_1}(\sigma_1)}{\partial\sigma_1^a}\frac{\partial}{\partial z_{\sigma_1}(\sigma_1)}+\frac{\partial \bar{z}_{\sigma_1}(\sigma_1)}{\partial\sigma_1^a}\frac{\partial}{\partial \bar{z}_{\sigma_1}(\sigma_1)}\\
&=-e_a^{z_{\sigma_1}}(\sigma_1)\frac{\partial}{\partial z_{\sigma_1}(\sigma_1)}-e_a^{\bar{z}_{\sigma_1}}(\sigma_1)\frac{\partial}{\partial \bar{z}_{\sigma_1}(\sigma_1)},
\end{aligned}
\end{equation}
Substituting (\ref{eq:D_zs}) into (\ref{eq:d/ds*}) leads to,
\begin{equation}\label{eq:d/ds*2}
\begin{aligned}
\frac{\partial}{\partial\sigma_1^a}
&=\Big(e_a^{z_{\sigma_1}}(\sigma_1)L_{-1}^{(z_{\sigma_1})}+e_a^{\bar{z}_{\sigma_1}}(\sigma_1)\tilde{L}_{-1}^{(z_{\sigma_1})}\Big)\\
&\quad+\frac{1}{4}\sum_{n=1}^\infty \frac{1}{(n+1)!}\Big(e_a^{\bar{z}_{\sigma_1}}(\sigma_1)\nabla_{z_{\sigma_1}}^{n-1}R_{(2)}(\sigma_1)L_n^{(z_{\sigma_1})}+e_a^{z_{\sigma_1}}(\sigma_1)\nabla_{\bar{z}_{\sigma_1}}^{n-1}R_{(2)}(\sigma_1)\tilde{L}_n^{(z_{\sigma_1})}\Big).\\
\end{aligned}
\end{equation}

Let us show that (\ref{eq:d/ds*2}) gives the {\it commutator} of a derivative with respect to the base point, $\sigma_1$, with the normal ordering operation \cite{Polchinski88}. Suppose that we let (\ref{eq:d/ds*2}) act on a general local normal-ordered operator, $:\!\hat{\mathscr{A}}^{(z_{\sigma_1})}(\sigma_1)\!:_{z_{\sigma_1}}$, that is inserted where the frame, $z_{\sigma_1}(\sigma)$, is based, namely at $\sigma=\sigma_1$ where $z_{\sigma_1}(\sigma_1)\equiv0$. Then what (\ref{eq:d/ds*2}) tells us is that:\footnote{Incidentally, it is perhaps useful to note that there is a sign typo in the coefficient of the Ricci scalar in equation (A.6) in \cite{Polchinski88}.}
\begin{equation}\label{eq:d/ds*3}
\begingroup\makeatletter\def\f@size{11}\check@mathfonts
\def\maketag@@@#1{\hbox{\m@th\large\normalfont#1}}%
\boxed{
\begin{aligned}
&\frac{\partial}{\partial\sigma_1^a}:\hat{\mathscr{A}}^{(z_{\sigma_1})}(\sigma_1)\!:_{z_{\sigma_1}}\,\,-\,\,
:\frac{\partial}{\partial\sigma^a}\hat{\mathscr{A}}^{(z_{\sigma_1})}(\sigma)\Big|_{\sigma=\sigma_1}\!\!\!\!\!:_{z_{\sigma_1}}=\\
&\,=\frac{1}{4}\sum_{n=1}^\infty \frac{1}{(n+1)!}\Big(e_a^{\bar{z}_{\sigma_1}}(\sigma_1)\nabla_{z_{\sigma_1}}^{n-1}R_{(2)}(\sigma_1)L_n^{(z_{\sigma_1})}+e_a^{z_{\sigma_1}}(\sigma_1)\nabla_{\bar{z}_{\sigma_1}}^{n-1}R_{(2)}(\sigma_1)\tilde{L}_n^{(z_{\sigma_1})}\Big):\!\hat{\mathscr{A}}^{(z_{\sigma_1})}(\sigma_1)\!:_{z_{\sigma_1}}
\end{aligned}
}
\endgroup
\end{equation}
We made use of the relation,
$$
:\frac{\partial}{\partial\sigma^a}\hat{\mathscr{A}}^{(z_{\sigma_1})}(\sigma)\Big|_{\sigma=\sigma_1}\!\!\!\!\!:_{z_{\sigma_1}}=\Big(e_a^{z_{\sigma_1}}(\sigma_1)L_{-1}^{(z_{\sigma_1})}+e_a^{\bar{z}_{\sigma_1}}(\sigma_1)\tilde{L}_{-1}^{(z_{\sigma_1})}\Big)
:\!\hat{\mathscr{A}}^{(z_{\sigma_1})}(\sigma_1)\!:_{z_{\sigma_1}},
$$
the first equality in (\ref{eq:framefield2}) (on p.~\pageref{eq:framefield2}), the chain rule, and also that when $L_{-1}^{(z_{\sigma_1})}$ acts on a local normal-ordered operator it yields the normal-ordered derivative of that operator \cite{Polchinski_v1},
$$
L_{-1}^{(z_{\sigma_1})}:\!\hat{\mathscr{A}}^{(z_{\sigma_1})}(\sigma_1)\!:_{z_{\sigma_1}}\,=\,\, :\!\partial_{z_{\sigma_1}}\hat{\mathscr{A}}^{(z_{\sigma_1})}(\sigma)|_{\sigma=\sigma_1}\!:_{z_{\sigma_1}}
$$

Although we adopt fairly explicit notation throughout, perhaps it is useful to also rewrite (\ref{eq:d/ds*3}) more concisely as,
\begin{equation}\label{eq:partial_z commutator}
\boxed{\,\,\partial_z:\!\mathscr{A}(\sigma_1)\!:_z\,\,=\,\,:\!\partial_z\mathscr{A}(\sigma_1)\!:_z+\frac{1}{4}\sum_{n=1}^{\infty}\frac{1}{(n+1)!}\big(\nabla_{\bar{z}}^{n-1}R_{(2)}(\sigma_1)\big)\tilde{L}_n:\!\mathscr{A}(\sigma_1)\!:_z\,\,}
\end{equation}
which (although somewhat schematic) perhaps makes the commutator more explicit. On the left-hand side the derivative is with respect to the base point whereas in the first term of the right-hand side it is with respect to the frame coordinate, but that is also evaluated at the base point. Of course, a similar relation holds for the anti-chiral derivative. 
\sk

In Sec.~\ref{sec:CPUBPS} we will derive (\ref{eq:D_zs}) and in particular the correspondence (\ref{eq:d/dzv1=-d/dzs1s1}) from yet another complementary conformal field theory viewpoint. 

\subsection{Wu-Yang Boundary Terms}\label{sec:WuYang}

Although different from the original context where these terms were perhaps first considered \cite{WuYang75}, Wu-Yang terms arise when integrals of total derivatives (in the absence of physical boundaries) do not integrate to zero after applying Stoke's theorem. An example was considered in \cite{Polchinski87,Nelson89}, and what follows gives a closely related (but much more explicit) account. These terms can arise when the integrand is not globally defined, in which case one can rather proceed by considering a cell decomposition and examine how the contributions from cell boundaries glue together to reconstruct the full integral. In conformal normal ordering this is generic, integrals of total derivatives do not vanish, whereas in Weyl normal ordering one can directly apply Stoke's theorem provided one takes into account that in the latter case derivatives do not commute with normal ordering, which in turn makes manifest the contribution from the boundary of moduli space. So Weyl normal ordering (which is to conformal normal order using holomorphic normal coordinates) in a sense localises the non-local (or global) information contained in the Wu-Yang terms, which in the general case are encoded in the right-hand side of (\ref{eq:d/ds*3}). To unravel the relevant points it will perhaps be clearest to proceed by example.
\sk

The example of interest is to integrate the result (\ref{eq:d/ds*3}) with $:\!\hat{\mathscr{A}}^{(z_{\sigma_1})}(\sigma_1)\!:_{z_{\sigma_1}}$ identified with the following local operator, 
$
:\!x^{\mu}\partial_{b}x^\nu(\sigma_1)\!:_{z_{1}},
$ 
where normal ordering is carried out using the $z_{1}$ (which is {\it shorthand} for  $z_{\sigma_1}$, so that $z_1(\sigma_1)=0$) holomorphic normal coordinate. 
A short calculation using (\ref{eq:d/ds*3}) yields (with $\alpha'=2$),
$$
\boxed{
g^{ab}(\sigma_1)\partial_a\big(\!:\!x^{\mu}\partial_{b}x^\nu(\sigma_1)\!:_{z_{1}}\!\big)=g^{ab}(\sigma_1):\!\partial_a\big(x^{\mu}\partial_{b}x^\nu(\sigma_1)\big)\!:_{z_{1}}-\frac{1}{2}\eta^{\mu\nu}R_{(2)}(\sigma_1)
}
$$
We keep the more explicit notation for derivatives given in (\ref{eq:d/ds*3}) implicit here. 
Let us integrate left- and right-hand sides over $\Sigma$ using the natural diffeomorphism-invariant measure, assuming there are no other insertions,
\begin{equation}\label{eq:int d()=(d)+R}
\begin{aligned}
\int_\Sigma \rmd^2&\sigma_1\sqrt{g(\sigma_1)}g^{ab}(\sigma_1)\partial_a\big(\!:\!x^{\mu}\partial_{b}x^\nu(\sigma_1)\!:_{z_{1}}\!\big)=\\
&=\int_\Sigma \rmd^2\sigma_1\sqrt{g(\sigma_1)}g^{ab}(\sigma_1):\!\partial_a\big(x^{\mu}\partial_{b}x^\nu(\sigma_1)\big)\!:_{z_{1}}-\frac{1}{2}\eta^{\mu\nu}\int_\Sigma \rmd^2\sigma_1\sqrt{g(\sigma_1)}R_{(2)}(\sigma_1).
\end{aligned}
\end{equation}
Next write this in terms of an integral over the base coordinates, $z_v,\bar{z}_v$, using holomorphic normal coordinates. See equations (\ref{eq:dz*=-ds*e}), (\ref{eq:dzsigma*-measure}), (\ref{eq:gabzs*}), and (\ref{eq:dzv=ds1s1}). In particular, 
$
\rmd^2\sigma_12\sqrt{g(\sigma_1)}=\rmd^2z_v,
$ 
and (for any scalars $f,h$),
$$
\frac{1}{2}g^{ab}\partial_a\big(\!:\!f\partial_bh\!:\!\big)(\sigma_1)=\partial_{z_v}\big(\!:\!f\partial_{\bar{z}_v}h\!:\!\big)(\sigma_1)+\partial_{\bar{z}_v}\big(\!:\!f\partial_{z_v}h\!:\!\big)(\sigma_1),
$$
and therefore (\ref{eq:int d()=(d)+R}) can also be written as,
\begin{equation}\label{eq:int d()=(d)+R 2}
\begin{aligned}
\int_\Sigma \rmd^2z_v&\Big[\partial_{z_v}\big(\!:\!x^\mu\partial_{\bar{z}_v}x^\nu(\sigma_1)\!:_{z_{1}}\!\big)+\partial_{\bar{z}_v}\big(\!:\!x^\mu\partial_{z_v}x^\nu(\sigma_1)\!:_{z_{1}}\!\big)\Big]=\\
&=\int_\Sigma \rmd^2z_v\Big[:\!\partial_{z_v}\big(\!x^\mu\partial_{\bar{z}_v}x^\nu(\sigma_1)\big)\!:_{z_{1}}+:\!\partial_{\bar{z}_v}\big(x^\mu\partial_{z_v}x^\nu(\sigma_1)\!\big):_{z_{1}}\Big]-2\pi\eta^{\mu\nu}\chi(\Sigma),
\end{aligned}
\end{equation}
where we took into account that the integral over the Ricci scalar yields the Euler characteristic,
$$
\chi(\Sigma)=\frac{1}{4\pi}\int_\Sigma \rmd^2\sigma_1\sqrt{g(\sigma_1)}R_{(2)}(\sigma_1).
$$
We can integrate by parts naively on the left-hand side of (\ref{eq:int d()=(d)+R 2}) (since the derivative is outside the normal ordering) and in the absence of a physical boundary the left-hand side therefore integrates to zero. So we learn that,
\begin{equation}\label{eq:int d()=(d)+R 3}
\begin{aligned}
\eta^{\mu\nu}\chi(\Sigma)=\frac{1}{2\pi}\int_\Sigma \rmd^2z_v\Big[:\!\partial_{z_v}\big(\!x^\mu\partial_{\bar{z}_v}x^\nu(\sigma_1)\big)\!:_{z_{1}}+:\!\partial_{\bar{z}_v}\big(x^\mu\partial_{z_v}x^\nu(\sigma_1)\!\big):_{z_{1}}\Big].
\end{aligned}
\end{equation}

Let us check this result. In holomorphic normal coordinates derivatives do not commute with normal ordering in the presence of non-vanishing curvature. Normal ordering using holomorphic normal coordinates (which in turn fixes most of the residual symmetry of the Beltrami equation) was termed `Weyl normal ordering' in \cite{Polchinski88}, to distinguish it from `conformal normal ordering' \cite{Polchinski87} where derivatives {\it do} commute with normal ordering. So we can also evaluate the right-hand side in (\ref{eq:int d()=(d)+R 3}) without using holomorphic normal coordinates, and instead assume the frame coordinates, $z_1$, are {\it holomorphic} in the base point, $\sigma_1$. That is we will evaluate (\ref{eq:int d()=(d)+R 3}) using {\it conformal normal ordering} \cite{Polchinski87,Nelson89}, and this serves as a consistency check of the various approaches. 
We will use the formalism developed in Sec.~\ref{sec:NB} and Sec.~\ref{sec:TFCR} and assume the reader is familiar with these sections. 
\sk

So we consider a cell decomposition of $\Sigma$, with charts $\{(U_m,z_m)\}$ and non-overlapping cells $\{V_m\}$ where every $V_m\subset U_m$, for all $m$. On patch overlaps, $U_m\cap U_n\neq\zero$, we have holomorphic transition functions, $z_m=f_{mn}(z_n)$, corresponding cocycle relations, etc., recall the formalism of Sec.~\ref{sec:TFCR}. We introduce a partition of unity, $\{\lambda_m\}$ satisfying $\sum_m\lambda_m=1$, subordinate to the open cover $\mathscr{U}=\{U_m,U_n,\dots\}$, and then make use of the result (\ref{eq:nopartitionofunity}) which in turn allows us to write (\ref{eq:int d()=(d)+R 3}) as follows,
\begin{equation}\label{eq:int d()=(d)+R 3b}
\begin{aligned}
\eta^{\mu\nu}\chi(\Sigma)
&=\frac{1}{2\pi}\sum_m\int_{V_m} \rmd^2z_m\Big[\partial_{z_m}\big(\!:\!x^\mu\partial_{\bar{z}_m}x^\nu(\sigma)\!:_{z_m}\big)-\partial_{\bar{z}_m}\big(-:\!x^\mu\partial_{z_m}x^\nu(\sigma)\!:_{z_m}\big)\Big].
\end{aligned}
\end{equation}
We took into account that in conformal normal ordering derivatives commute with normal ordering. Integrating by parts using Green's theorem (\ref{eq: Greens theorem3}), grouping together pairs of boundaries (denoted by $(mn)$) and changing variables keeping normal ordering {\it fixed}, we can rewrite (\ref{eq:int d()=(d)+R 3b}),
\begin{equation}\label{eq:int d()=(d)+R 4}
\begin{aligned}
\eta^{\mu\nu}\chi(\Sigma)
&=\frac{1}{2\pi i}\sum_{(mn)}\int_{C_{mn}} \rmd z_m\Big(\!:\!x^\mu\partial_{z_m}x^\nu(\sigma)\!:_{z_m}-\!:\!x^\mu\partial_{z_m}x^\nu(\sigma)\!:_{z_n}\Big)\\
&\quad-\frac{1}{2\pi i}\sum_{(mn)}\int_{C_{mn}} \rmd\bar{z}_m\Big(\!:\!x^\mu\partial_{\bar{z}_m}x^\nu(\sigma)\!:_{z_m}-\!:\!x^\mu\partial_{\bar{z}_m}x^\nu(\sigma)\!:_{z_n}\Big).
\end{aligned}
\end{equation}
Recall that the contour $C_{mn}$ traverses the overlap $U_m\cap U_n$ counterclockwise with respect to $U_m$, so that $C_{mn}$ may be identified with the common boundary of the sets $V_m$ and $V_n$ as depicted in Fig.~\ref{fig:segment} on p.~\pageref{fig:segment}, see also Fig.~\ref{fig:tripleoverlapsUV}.
\sk

We now make use of the results of Sec.~\ref{sec:CNO} and Sec.~\ref{sec:HCNO} where we discussed conformal (or Weyl) normal ordering and changes in normal ordering keeping coordinates fixed. Applying the result (\ref{eq:opes2}) and (\ref{eq:opes_w1fixedz1}) to the case of interest, a short calculation yields:
\begin{equation}\label{eq:xdx_zm-xdx_zn}
\boxed{:\!x^\mu\partial_{z_m}x^\nu(\sigma)\!:_{z_m}-\!:\!x^\mu\partial_{z_m}x^\nu(\sigma)\!:_{z_n} = -\frac{1}{2}\eta^{\mu\nu}\partial_{z_m}\ln\big(\partial_{z_m}f_{nm}(z_m)\big)}
\end{equation}
The term on the right-hand side is reminiscent of Wu-Yang terms and encodes the contribution to the integral (\ref{eq:int d()=(d)+R 4}) from patch overlaps. 
When we substitute (\ref{eq:xdx_zm-xdx_zn}) back into (\ref{eq:int d()=(d)+R 4}) we find,
\begin{equation}\label{eq:int d()=(d)+R 5}
\begin{aligned}
\eta^{\mu\nu}\chi(\Sigma)
&=\frac{1}{2\pi i}\sum_{(mn)}\int_{C_{mn}} \rmd z_m\Big(-\frac{1}{2}\eta^{\mu\nu}\partial_{z_m}\ln\big(\partial_{z_m}f_{nm}(z_m)\big)\Big)\\
&\quad-\frac{1}{2\pi i}\sum_{(mn)}\int_{C_{mn}} \rmd\bar{z}_m\Big(-\frac{1}{2}\eta^{\mu\nu}\partial_{\bar{z}_m}\ln\big(\partial_{\bar{z}_m}\bar{f}_{nm}(\bar{z}_m)\big)\Big).
\end{aligned}
\end{equation}
The two terms in the first and second lines on the right-hand side are actually real and equal, so in fact we can rearrange the result into the following,
\begin{equation}\label{eq:Euler2}
\boxed{
\begin{aligned}
\chi(\Sigma)
&=\frac{i}{2\pi}\sum_{(mn)}\int_{C_{mn}}\!\!\rmd z_m\partial_{z_m}\ln f'_{nm}(z_m)
\end{aligned}
}
\end{equation}
Comparing finally with (\ref{eq:Euler}) we find precise agreement. Notice that the derivation here is entirely independent of the derivation that led to (\ref{eq:Euler}) and is rather based on local composite operator insertions into the path integral. As mentioned above, we have assumed for simplicity that there are no other operators on the Riemann surface. Furthermore, although there is a potential zero mode in $x^\mu$, it actually cancels out in (\ref{eq:xdx_zm-xdx_zn}). 
\sk

The short calculation of this section demonstrates some aspects of the subtle interplay between Weyl normal ordering (with non-holomorphic base point) and conformal normal ordering (with holomorphic base point). In particular, and more generally, the right-hand side in the relation (\ref{eq:d/ds*3}) (which is {\it a priori} defined using Weyl normal ordering) encodes {\it all} Wu-Yang terms for arbitrary operators in local information, as pointed out in \cite{Polchinski88}. The same procedure discussed here can be carried out for arbitrary local operators: when we integrate both left- and right-hand sides in (\ref{eq:d/ds*3}) the first term on the left-hand side vanishes after integration by parts (up to contributions from the boundary of moduli space) and the remaining relation gives the Wu-Yang contributions associated to patch overlaps from the viewpoint of conformal or Weyl normal ordering.

\subsection{Shifting Local Operators with Transition Functions}\label{sec:CPUBPS}
Making use of the results of Sec.~\ref{sec:SPUTF} (and keeping the comments associated to (\ref{eq:z1pzs1 corr}) in mind), we next determine how an arbitrary local operator,  $\mathscr{A}^{(z_{\sigma_1})}(\sigma)$, evaluated at a point $\sigma$ transforms under a shift of {\it base point}, 
\begin{equation}\label{eq:s->s'=s+ds}
\sigma_1\mapsto \sigma_1'=\sigma_1+\delta \sigma_1,
\end{equation}
with respect to which the holomorphic normal coordinate, $z_{\sigma_1}(\sigma)$, is defined. Recall that by definition, $z_{\sigma_1}(\sigma_1)\equiv0$. 
\sk

We first consider the case where the operator, $\mathscr{A}^{(z_{\sigma_1})}(\sigma)$, under consideration is a\footnote{By `conformal primary' we will always mean Virasoro primary, and reserve the terminology `SL($2,\mathbf{C}$) primary', etc., if the object under consideration transforms as a tensor under a subgroup of the local infinite-dimensional conformal algebra.} conformal primary, and we denote it by: $\mathscr{O}^{(z_{\sigma_1})}(\sigma)$. 
Primaries are local operators that transform as {\it tensors} under holomorphic changes of frame, $z_{\sigma_1}(\sigma)\mapsto f(z_{\sigma_1}(\sigma))$, so we can perform some important consistency checks using them before discussing the transformation property of a general local operator under shifts. 
\sk

In Sec.~\ref{sec:SPUTF}, see in particular (\ref{eq:zsds(s')}), we showed that a shift of base point, $\sigma_1\mapsto \sigma_1'=\sigma_1+\delta \sigma_1$, induces a holomorphic change of frame coordinates,  $z_{\sigma_1}\mapsto z_{\sigma_1'}=z_{\sigma_1+\delta\sigma_1}$, where:
\begin{equation}\label{eq:zsds(s')2}
z_{\sigma_1+\delta\sigma_1}(\sigma)=z_{\sigma_1}(\sigma)+\delta z_{\sigma_1}(\sigma_1)+\sum_{n=1}^\infty\big(\delta \bar{z}_{\sigma_1}(\sigma_1)\frac{1}{4(n+1)!}\nabla_{z_{\sigma_1}}^{n-1}R_{(2)}(\sigma_1)\big)z_{\sigma_1}(\sigma)^{n+1},
\end{equation}
and we neglect the overall arbitrary phase. 
This change of coordinates is holomorphic with respect to $\sigma$ (more precisely $z_{\sigma_1}(\sigma)$), but it is clearly {\it not} holomorphic with respect to $\sigma_1$. The question we would like to primarily address then is: given $\mathscr{O}^{(z_{\sigma_1})}(\sigma)$ what is the corresponding quantity $\mathscr{O}^{(z_{\sigma_1'})}(\sigma)$ with $ \sigma_1'$ as in (\ref{eq:s->s'=s+ds}). 
\sk

For any given primary $\mathscr{O}^{(z)}(\sigma)$ of conformal weight $(h,\tilde{h})$ we know that under general holomorphic transformations, $z\mapsto w(z)$, 
$
\mathscr{O}^{(w)}(\sigma)=\mathscr{O}^{(z)}(\sigma)(\partial_zw)^{-h}(\partial_{\bar{z}}\bar{w})^{-\tilde{h}}.
$ 
In the case of interest, where $w=z_{\sigma_1'}$ and $z=z_{\sigma_1}$, we can (by definition) write this more explicitly as,
\begin{equation}\label{eq:Owz=Ozs}
\begin{aligned}
\mathscr{O}^{(z_{\sigma_1'})}(z_{\sigma_1'}(\sigma))&=\mathscr{O}^{(z_{\sigma_1})}(z_{\sigma_1}(\sigma))(\partial_{z_{\sigma_1}}z_{\sigma_1'}(\sigma))^{-h}(\partial_{\bar{z}_{\sigma_1}}\bar{z}_{\sigma_1'}(\sigma))^{-\tilde{h}}
\end{aligned}
\end{equation}
Before evaluating this it is convenient to rewrite (\ref{eq:zsds(s')2}) as follows,
\begin{equation}\label{eq:z->w=z+enzn.}
z_{\sigma_1}\mapsto z_{\sigma_1'}=z_{\sigma_1}+\sum_{n=-1}^\infty \varepsilon_nz_{\sigma_1}^{n+1},
\end{equation}
with identifications, 
$$
\varepsilon_{-1}=\delta z_{\sigma_1}(\sigma_1), \qquad\varepsilon_0=0,\qquad {\rm and}\qquad\varepsilon_{n\geq1}=\delta \bar{z}_{\sigma_1}(\sigma_1)\frac{1}{4(n+1)!}\nabla_{z_{\sigma_1}}^{n-1}R_{(2)}(\sigma_1),
$$
and analogous expressions for the complex conjugates.
\sk

Let us now consider the left- and right-hand sides of (\ref{eq:Owz=Ozs}) independently. Taylor expanding the left-hand side on account of (\ref{eq:z->w=z+enzn.}) yields,
\begin{equation}\label{eq:Os*'=Os*Taylor}
\begin{aligned}
\mathscr{O}^{(z_{\sigma_1'})}(\sigma)&\equiv \mathscr{O}^{(z_{\sigma_1'})}(z_{\sigma_1'}(\sigma))\\
&\simeq\mathscr{O}^{(z_{\sigma_1'})}(z_{\sigma_1}(\sigma))+\sum_{n=-1}^\infty \big(\varepsilon_nz_{\sigma_1}(\sigma)^{n+1}\partial_{z_{\sigma_1}}\
+\bar{\varepsilon}_n\bar{z}_{\sigma_1}(\sigma)^{n+1}\partial_{\bar{z}_{\sigma_1}}\big)\mathscr{O}^{(z_{\sigma_1})}(z_{\sigma_1}(\sigma))+\dots\\
\end{aligned}
\end{equation}
where the derivative $\partial_{z_{\sigma_1}}$ is with respect to $z_{\sigma_1}(\sigma)$, 
whereas for the right-hand side of (\ref{eq:Owz=Ozs}) we make use of the expansion,
\begin{equation}\label{eq:dzszs*}
\begin{aligned}
(\partial_{z_{\sigma_1}}z_{\sigma_1'}(\sigma))^{-h}&=\Big[\partial_{z_{\sigma_1}(\sigma)}\Big(z_{\sigma_1}(\sigma)+\sum_{n=-1}^\infty \varepsilon_nz_{\sigma_1}(\sigma)^{n+1}\Big)\Big]^{-h}\\
&\simeq 1-h\sum_{n=1}^\infty (n+1)\varepsilon_nz_{\sigma_1}(\sigma)^{n}+\dots\\
\end{aligned}
\end{equation}
where the `\dots' in (\ref{eq:dzszs*}) and (\ref{eq:Os*'=Os*Taylor}) denote terms of higher order in $\varepsilon_n,\bar{\varepsilon}_n$. There is an entirely analogous expression for the complex conjugate of (\ref{eq:dzszs*}) with the replacement $h\rightarrow\tilde{h}$. Substituting these results into (\ref{eq:Owz=Ozs}) we learn that at a generic point $\sigma$,
\begin{equation}\label{eq:Os*'=Os*-LL..}
\begin{aligned}
&\mathscr{O}^{(z_{\sigma_1'})}(z_{\sigma_1}(\sigma))
=\mathscr{O}^{(z_{\sigma_1})}(z_{\sigma_1}(\sigma))- \big(\delta z_{\sigma_1}(\sigma_1)\partial_{z_{\sigma_1}}+\delta \bar{z}_{\sigma_1}(\sigma_1)\partial_{\bar{z}_{\sigma_1}}\big)\mathscr{O}^{(z_{\sigma_1})}(z_{\sigma_1}(\sigma))\\
&\,\,\,\,-\delta \bar{z}_{\sigma_1}(\sigma_1)\sum_{n=1}^\infty \frac{1}{4}\frac{1}{(n+1)!}\nabla_{z_{\sigma_1}}^{n-1}R_{(2)}(\sigma_1)\,\big(h(n+1)z_{\sigma_1}(\sigma)^{n}+z_{\sigma_1}(\sigma)^{n+1}\partial_{z_{\sigma_1}}\big)\mathscr{O}^{(z_{\sigma_1})}(z_{\sigma_1}(\sigma))\\
&\,\,\,\,-\delta z_{\sigma_1}(\sigma_1)\sum_{n=1}^\infty \frac{1}{4}\frac{1}{(n+1)!}\nabla_{\bar{z}_{\sigma_1}}^{n-1}R_{(2)}(\sigma_1)\,\big(\tilde{h}(n+1)\bar{z}_{\sigma_1}(\sigma)^{n}+\bar{z}_{\sigma_1}(\sigma)^{n+1}\partial_{\bar{z}_{\sigma_1}}\big)\mathscr{O}^{(z_{\sigma_1})}(z_{\sigma_1}(\sigma))\\
\end{aligned}
\end{equation}
From a standard result in conformal field theory we can write the various terms on the right-hand side in terms of Virasoro generators,
$$
\big[L_n^{(z_{\sigma_1})},\mathscr{O}^{(z_{\sigma_1})}(z_{\sigma_1}(\sigma))\big]=\big(h(n+1)z_{\sigma_1}(\sigma)^{n}+z_{\sigma_1}(\sigma)^{n+1}\partial_{z_{\sigma_1}}\big)\mathscr{O}^{(z_{\sigma_1})}(z_{\sigma_1}(\sigma)),
$$
with an analogous relation for the anti-chiral half, 
and so in particular (\ref{eq:Os*'=Os*-LL..}) can equivalently be rewritten as follows,
\begin{equation}\label{eq:Os*'=Os*-LL}
\boxed{
\begin{aligned}
\mathscr{O}^{(z_{\sigma_1+\delta \sigma_1})}&(z_{\sigma_1}(\sigma))-\mathscr{O}^{(z_{\sigma_1})}(z_{\sigma_1}(\sigma))=\\
=&\,\,\delta z_{v_1}\Big(L_{-1}^{(z_{\sigma_1})}+\sum_{n=1}^\infty \frac{1}{4}\frac{1}{(n+1)!}\nabla_{\bar{z}_{\sigma_1}}^{n-1}R_{(2)}(\sigma_1)\,\tilde{L}_n^{(z_{\sigma_1})}\Big)\mathscr{O}^{(z_{\sigma_1})}(z_{\sigma_1}(\sigma))\\
&\,\delta\bar{z}_{v_1}\Big(\tilde{L}_{-1}^{(z_{\sigma_1})}+\sum_{n=1}^\infty \frac{1}{4}\frac{1}{(n+1)!}\nabla_{z_{\sigma_1}}^{n-1}R_{(2)}(\sigma_1)\,L_n^{(z_{\sigma_1})}\Big)\mathscr{O}^{(z_{\sigma_1})}(z_{\sigma_1}(\sigma))\\
\end{aligned}
}
\end{equation}
where we defined:
$$
\delta z_{v_1}\dfn -\delta z_{\sigma_1}(\sigma_1),\qquad \delta\bar{z}_{v_1}\dfn -\delta \bar{z}_{\sigma_1}(\sigma_1),
$$
and took into account that the generators $L_n^{(z_{\sigma_1})}$ annihilate the SL$(2,\mathbf{C})$ vacuum (i.e.~the unit operator) when $n\geq-1$. 
This relation (\ref{eq:Os*'=Os*-LL}) in turn implies that,
\begin{equation}\label{eq:D_zs2}
\frac{\partial}{\partial z_{v_1}}\Big|_{z_{\sigma_1}(\sigma)}\!\!\!\!\!\mathscr{O}^{(z_{\sigma_1})}(z_{\sigma_1}(\sigma))
=\Big(L_{-1}^{(z_{\sigma_1})}+\sum_{n=1}^\infty \frac{1}{4}\frac{1}{(n+1)!}\nabla_{\bar{z}_{\sigma_1}}^{n-1}R_{(2)}(\sigma_1)\,\tilde{L}_n^{(z_{\sigma_1})}\Big)\mathscr{O}^{(z_{\sigma_1})}(z_{\sigma_1}(\sigma)).
\end{equation}
Since the arguments of the two terms on the left-hand side in (\ref{eq:Os*'=Os*-LL}) are both evaluated at $z_{\sigma_1}(\sigma)$ it follows that the resulting derivative on the left-hand side of (\ref{eq:D_zs2}) is evaluated at fixed $z_{\sigma_1}(\sigma)$ as we indicate. In fact, we could have indicated more explicitly all of the quantities (in addition to $z_{\sigma_1}(\sigma)$) that are held fixed while taking the derivative indicted on the left-hand side, but we do not want to overload the notation any further: more fully, we are to keep $z_{\sigma_1}(\sigma),\bar{z}_{\sigma_1}(\sigma)$ fixed (which follows from the explicit arguments on the left-hand side of (\ref{eq:Os*'=Os*-LL})), and also $\bar{z}_{\sigma_1}(\sigma_1)$ (which follows from the chain rule and the right-hand side in (\ref{eq:Os*'=Os*-LL})). 
\sk

Comparing (\ref{eq:D_zs2}) to (\ref{eq:D_zs}), we have reached an alternative derivation of the latter in the specific case when the operands under consideration are conformal primaries. From the more explicit expression (\ref{eq:Os*'=Os*-LL..}) we see that inserting the operator at the origin of the holomorphic normal coordinate (which is to take $\sigma=\sigma_1$ at which $z_{\sigma_1}(\sigma_1)\equiv0$) the terms associated to the sum over $n$ {\it vanish}. The latter statement follows immediately from the defining properties of a local primary operator, namely that,
$$
L_n^{(z_{\sigma_1})}\mathscr{O}^{(z_{\sigma_1})}(\sigma_1)=0,\qquad {\rm for}\qquad n\geq1.
$$

As a consistency check let us also consider the mode expansion of a {\it chiral} primary in particular,
$$
\mathscr{O}^{(z_{\sigma_1})}(\sigma)=\sum_{m\in\mathbf{Z}}\frac{\mathscr{O}_m^{(z_{\sigma_1})}}{z_{\sigma_1}(\sigma)^{m+h}}.
$$
It is easy to check that the above expressions, in particular (\ref{eq:D_zs2}), are consistent with the mode operator results. The following standard relation is useful,
$$
\big[L_{n}^{(z_{\sigma_1})},\mathscr{O}_m^{(z_{\sigma_1})}\big]=\big[n(h-1)-m\big]\mathscr{O}_{m+n}^{(z_{\sigma_1})}.
$$

We now generalise to arbitrary local operators, $\mathscr{A}^{(z_{\sigma_1})}(\sigma)$. In fact, the result is immediate. As the derivation of (\ref{eq:D_zs}) shows this expression (\ref{eq:D_zs2}) holds for an arbitrary operator, $\mathscr{A}^{(z_{\sigma_1})}(z_{\sigma_1}(\sigma))\equiv \mathscr{A}^{(z_{\sigma_1})}(\sigma)$, inserted at an arbitrary point $\sigma$ using a frame, $z_{\sigma_1}$, based at $\sigma=\sigma_1$ within the local chart of interest,
\begin{equation}\label{eq:D_zs3}
\boxed{
\frac{\partial}{\partial z_{v_1}}\Big|_{z_{\sigma_1}(\sigma)}\!\!\!\!\!\mathscr{A}^{(z_{\sigma_1})}(\sigma)
=\Big(L_{-1}^{(z_{\sigma_1})}+\sum_{n=1}^\infty \frac{1}{4}\frac{1}{(n+1)!}\nabla_{\bar{z}_{\sigma_1}}^{n-1}R_{(2)}(\sigma_1)\,\tilde{L}_n^{(z_{\sigma_1})}\Big)\mathscr{A}^{(z_{\sigma_1})}(\sigma)
}
\end{equation}
A difference between (\ref{eq:D_zs2}) and (\ref{eq:D_zs3}) is that in the case of primaries the terms associated to the sum over $n$ (which might be thought of as ``connection terms'') vanish when $\sigma=\sigma_1$, whereas for more general operators the connection terms generically contribute non-trivially. 
\sk

Recall also equation (\ref{eq:hatDzv2R}) from p.~\pageref{eq:hatDzv2R}, which in the notation of the current section reads:
$$
\hat{D}_{z_{v_1}} =-L_{-1}^{(z_{\sigma_1})}-\sum_{n=1}^\infty \frac{1}{4}\frac{1}{(n+1)!}\Big(\nabla_{\bar{z}_{\sigma_1}}^{n-1}R_{(2)}(\sigma)\Big)\Big|_{\sigma=\sigma_1}\,\tilde{L}_n^{(z_{\sigma_1})}.
$$
What we have shown here is that when acting on operators $\mathscr{A}^{(z_{\sigma_1})}(\sigma)$ the quantity $\hat{D}_{z_{v_1}}$ generates a derivative with respect to $z_{v_1}$,
\begin{equation}\label{eq:Dzv=-dzv}
\boxed{
\hat{D}_{z_{v_1}}\mathscr{A}^{(z_{\sigma_1})}(\sigma)=-\frac{\partial}{\partial z_{v_1}}\Big|_{z_{\sigma_1}(\sigma)}\!\!\!\!\!\mathscr{A}^{(z_{\sigma_1})}(\sigma)
}
\end{equation}
Recalling also that $\big\{Q_B^{(z_1)},\hat{B}_{z_v}\big\}=\hat{D}_{z_v}$, the relation (\ref{eq:Dzv=-dzv}) plays a vital role when demonstrating that BRST-exact states decouple from string amplitudes. This is discussed further in Sec.~\ref{sec:BRST-AC}.

\subsection{Local Operators Under Weyl Rescalings}\label{sec:LOUWR}
In this section we apply what we have learnt in Sec.~\ref{sec:HTFWR} to understand the transformation of general local operators, $\hat{\mathscr{A}}_a^{(z_{\sigma_1})}(\sigma_1)$, under arbitrary Weyl rescalings of the metric, $g_{ab}(\sigma)\mapsto e^{\delta\phi(\sigma)}g_{ab}(\sigma)$, when we have adopted a slice in moduli space associated to picking holomorphic normal coordinates, $z_{\sigma_1}(\sigma)$, to define the frame with respect to which these general local operators are defined. 
\sk

Under a general infinitesimal (meromorphic) reparametrisation of the form,
\begin{equation}\label{eq:z->w Weyl}
z_{\sigma_1}(\sigma)\mapsto w_{\sigma_1}(\sigma)=z_{\sigma_1}(\sigma)+\sum_{n\in\mathbf{Z}} \varepsilon_nz_{\sigma_1}(\sigma)^{n+1}
\end{equation}
we have that,
\begin{equation}
\hat{\mathscr{A}}_a^{(z_{\sigma_1})}(\sigma_1)\mapsto \hat{\mathscr{A}}^{(w_{\sigma_1})}(\sigma_1)=\hat{\mathscr{A}}^{(z_{\sigma_1})}(\sigma_1)-\sum_{n\in\mathbf{Z}}\big(\varepsilon_nL_n^{(z_{\sigma_1})}+\bar{\varepsilon}_n\tilde{L}_n^{(z_{\sigma_1})}\big)\hat{\mathscr{A}}^{(z_{\sigma_1})}(\sigma_1),
\end{equation}
recall (\ref{eq:z'z}). 
We can read off the relevant expressions for the $\varepsilon_n$ by comparing (\ref{eq:z->w Weyl}) to (\ref{eq:dz from Weyl3}), 
\begin{equation}
\varepsilon_{-n}=0,\quad\varepsilon_0=\frac{1}{2}(\delta \phi+i\delta \beta)(\sigma_1),\quad \varepsilon_n=\frac{1}{(n+1)!}\nabla^{n}_{z_{\sigma_1}}\delta\phi(\sigma)\big|_{\sigma=\sigma_1},\quad (n\geq1)
\end{equation}
so we learn that under arbitrary local Weyl rescalings of the underlying auxiliary metric components, $g_{ab}(\sigma)$, 
$$
\boxed{
g_{ab}(\sigma)\mapsto e^{\delta\phi(\sigma)}g_{ab}(\sigma)
}
$$
a general operator, $\hat{\mathscr{A}}_a^{(z_{\sigma_1})}(\sigma_1)$, inserted at $\sigma=\sigma_1$ transforms according to:
\begin{equation}\label{eq:A(z)->A(w) Weyl}
\boxed{
\begin{aligned}
\hat{\mathscr{A}}^{(w_{\sigma_1})}(\sigma_1)&=e^{-\frac{1}{2}\delta \phi(\sigma_1)(L_0^{(z_{\sigma_1})}+\tilde{L}_0^{(z_{\sigma_1})})-\frac{i}{2}\delta \beta(\sigma_1)(L_0^{(z_{\sigma_1})}-\tilde{L}_0^{(z_{\sigma_1})})}\Big\{\hat{\mathscr{A}}^{(z_{\sigma_1})}(\sigma_1)\\
&\quad-\sum_{n=1}^\infty\frac{1}{(n+1)!}\Big[\big(\nabla^{n}_{z_{\sigma_1}}\delta\phi(\sigma_1)\big)L_n^{(z_{\sigma_1})}+\big(\nabla^{n}_{\bar{z}_{\sigma_1}}\delta\phi(\sigma_1)\big)\tilde{L}_n^{(z_{\sigma_1})}\Big]\hat{\mathscr{A}}^{(z_{\sigma_1})}(\sigma_1)\Big\}
\end{aligned}
}
\end{equation}
and $\delta \beta(\sigma_1)$ is the (effectively arbitrary) phase introduced in Sec.~\ref{sec:HTFWR}. This result can be used to study the Weyl transformation properties of general operators including handle operators in particular. 
\sk

As an aside, notice that weight-$(h,\tilde{h})$ {\it conformal primaries}, $\hat{\mathscr{O}}_a^{(z_{\sigma_1})}(\sigma_1)$, (inserted at $\sigma=\sigma_1$ where the conformal frame $z_{\sigma_1}$ is based) transform very simply under Weyl rescalings since the second term in (\ref{eq:A(z)->A(w) Weyl}) vanishes in this case,
\begin{equation}\label{eq:O(z)->O(w) Weyl}
\begin{aligned}
\hat{\mathscr{O}}^{(w_{\sigma_1})}(\sigma_1)&=e^{-\frac{1}{2}\delta \phi(\sigma_1)(h+\tilde{h})-\frac{i}{2}\delta \beta(\sigma_1)(h-\tilde{h})}\hat{\mathscr{O}}^{(z_{\sigma_1})}(\sigma_1).
\end{aligned}
\end{equation}

Incidentally, holomorphic normal coordinates are invertible around the base point, in the sense that we can interpret the argument, $\sigma$, of $\delta\phi(\sigma)$ as being a function of $z_{\sigma_1},\bar{z}_{\sigma_1}$, and $\sigma_1$, in particular 
$
\sigma=\sigma(z_{\sigma_1},\bar{z}_{\sigma_1},\sigma_1).
$

\subsection{Gluing with Deformed Complex Structure}\label{sec:GWDCS}
Let us now assemble the above ingredients in order to transition from CFT to string theory. To glue and cut string amplitudes in the presence of complex structure deformations, e.g., to apply the factorisation formula (\ref{eq:completeness-sep}) or (\ref{eq:completeness_nonsep}), we simply insert the ghost contributions,
\begin{equation}\label{eq:prodBBk2}
\prod_{k=1}^\m\hat{B}_{\tau^k}\hat{B}_{\bar{\tau}^k},
\end{equation}
associated to the path integral measure, see (\ref{eq:fullpathintegral}) and Sec.~\ref{sec:IToTF:I}, and the appropriate measures of the corresponding moduli space integrals. We have a choice as to how to partition these ghost contributions and vertex operator insertions, which corresponds to different factorisations and different slices in moduli space. An essential rule is that we distribute (or partition) these insertions such that the number of moduli is correct for each of the subamplitudes, and there is otherwise no distinguished way of carrying this out. 

\subsubsection*{Separating Degenerations}\label{sec:sepdeg}
Let us focus on a {\it separating degeneration}, namely the factorisation formula (\ref{eq:completeness-sep}). Depending on the gauge slice of interest, in the vicinity of the cut we can associate, say, three complex moduli, namely the locations of the two vertex operators on either side of the cut and the (twist and pinch) modulus of the cycle we are cutting open. Or we can instead associate, say, one complex modulus associated to pinching and twisting the cut cycle, leaving the locations of the two vertex operators fixed, etc. There is considerable freedom here, but the important point is that the number of moduli (or $\hat{B}_k$ insertions) add up to the required number determined by the Riemann-Roch index theorem. The Riemann-Roch theorem not only applies to the full amplitude, but also to the sub-amplitudes after the cut. Rather than go through every possibility, let us suppose that we indeed associate three complex moduli to a given separating degeneration, the remaining cases being entirely analogous (and simpler). So making these measure contributions in (\ref{eq:completeness-sep}) explicit yields,
\begin{equation}\label{eq:completeness-sep BBB}
\boxed{e^{-\chi(\Sigma)\Phi}\Big\langle \dots_1\dots_2\Big\rangle_{\Sigma}
=\suminnt\limits_{a} \,\,e^{-\chi(\Sigma_1)\Phi}\Big\langle\dots_1\mathscr{A}_{a}^{(z_1)}\Big\rangle_{\Sigma_1}e^{-\chi(\Sigma_2)\Phi}\Big\langle\int_{q,\bar{q}} \big[\hat{B}_q\hat{B}_{\bar{q}}\mathscr{A}^{a}_{(z_2/q)}\big]\dots_2\Big\rangle_{\Sigma_2}}
\end{equation}
where the vertex operators appearing are in {\it integrated} picture (which is why $\mathscr{A}_{a}$ and $\mathscr{A}^{a}$ appear rather than $\hat{\mathscr{A}}_{a}$ and $\hat{\mathscr{A}}^{a}$ respectively) as defined in (\ref{eq:intvertop1}). 
Recall from Sec.~\ref{sec:PH} the discussion associated to the ghost contributions to the path integral measure for pinching moduli. 
Notice that in (\ref{eq:completeness-sep}) we glued with $q=1$, but we have found it more convenient here to glue with general $|q|<1$ so we redefined $z_2\mapsto z_2/q$ to reach (\ref{eq:z1z2=q}), so that the relevant transition function gluing the two surfaces, $\Sigma_1$ and $\Sigma_2$, together is,
$z_1z_2=q$. Associating the insertions, $\hat{B}_q\hat{B}_{\bar{q}}$, to $\mathscr{A}_{(z_2/q)}^a$ rather than $\mathscr{A}^{(z_1)}_a$ is just a matter of convention, as one can see by a standard contour deformation argument. 
\sk

The explicit measure contribution in turn reads,
\begin{equation}
\begin{aligned}
\int_{q,\bar{q}}\, \hat{B}_q\hat{B}_{\bar{q}}=\int_{q,\bar{q}}\,\frac{b_0^{(z_2)}\tilde{b}_0^{(z_2)}}{q\bar{q}},\qquad{\rm with}\qquad\int_{q,\bar{q}} \equiv \int idq\wedge \rmd\bar{q},
\end{aligned}
\end{equation}
(we usually write $\rmd^2q= idq\wedge \rmd\bar{q}$) which, defining $q\equiv re^{i\theta}$, can also be equivalently written in terms of $r,\theta$ moduli,
\begin{equation}
\begin{aligned}
&\,\int_{r,\theta}\hat{B}_r\hat{B}_\theta= \int_{r,\theta}\Big(\!-\frac{1}{r}\big(b_0^{(z_2)}+\tilde{b}_0^{(z_2)}\big)\Big)\Big(\!-i\big(b_0^{(z_2)}-\tilde{b}_0^{(z_2)}\big)\Big),
\end{aligned}
\quad{\rm with}\quad \int_{r,\theta}\equiv \int i \rmd r\wedge \rmd\theta.
\end{equation}
The two expressions are equal, $\int_{r,\theta}\hat{B}_r\hat{B}_\theta=\int_{q,\bar{q}}\, \hat{B}_q\hat{B}_{\bar{q}}$. In fact, we could have just as well written $r^{h_a+\tilde{h}_a}\mathscr{A}^{a}_{(z_2e^{-i\theta})}$ instead of $\mathscr{A}^{a}_{(z_2/q)}$, but we have not yet shown that this is possible. (That it is possible is fundamental in order to be able to interpret these vertex operators as local operators.) All in all, we have associated to the cutting procedure three complex moduli. 
\sk

We have kept $q$ dependence implicit in the operator $\mathscr{A}_{(z_2/q)}^a$ for the following reason. One can show that (\ref{eq:completeness-sep BBB}) implies (and this is only true for separating degenerations) that it is possible to choose a basis, $\hat{\mathscr{A}}^a_{(z_2)}$, for gluing satisfying:
\begin{equation}\label{eq:L0L0tildea=0}
(L_0^{(z_2)}-\tilde{L}_0^{(z_2)})\hat{\mathscr{A}}^a_{(z_2)}=0,\qquad {\rm and}\qquad (L_0^{(z_2)}+\tilde{L}_0^{(z_2)})\hat{\mathscr{A}}^a_{(z_2)}=(h_a+\tilde{h}_a)\hat{\mathscr{A}}^a_{(z_2)},
\end{equation}
where $h_a+\tilde{h}_a$ is the total weight, and identical relations hold also for $\hat{\mathscr{A}}_a^{(z_1)}$. The fact that this is possible follows from the fact that for such degenerations the integral over twist moduli, $\theta=-i\ln (q/|q|)$, gives rise to a Kronecker delta that ultimately sets $h_a=\tilde{h}_a$. This remains true when $\hat{\mathscr{A}}^a_{(z_2)}$ is in a coherent state basis. So when (\ref{eq:L0L0tildea=0}) holds we may effectively extract the $q,\bar{q}$-dependence from the correlators in (\ref{eq:completeness-sep}) by choosing a basis for $\hat{\mathscr{A}}^a_{(z_2)}$ that is an eigenstate of $L_0,\tilde{L}_0$, with eigenvalues (weights) $h_a,\tilde{h}_a$, which is in turn equivalent to the statement:
\begin{equation}\label{eq:AqqbarA}
\phantom{\qquad \textrm{(separating degeneration)}}\boxed{\hat{\mathscr{A}}^a_{(z_2/q)}=q^{h_a}\bar{q}^{\tilde{h}_a}\hat{\mathscr{A}}^a_{(z_2)}}\qquad \textrm{(separating degeneration)}
\end{equation}
but we emphasise that this is only true in the context of coherent states when (\ref{eq:L0L0tildea=0}) holds. So {\it effectively} $h_a=\tilde{h}_a$, in that non-level-matched terms do not contribute. These points are automatically incorporated when we take (\ref{eq:completeness-sep BBB}) as our starting point. 
\sk

Since we integrate over the phase of $q,\bar{q}$, the range of the moduli  integration is over an open disc whose radius depends on the features on the surface outside the pinch (which in turn depends on where in moduli space we are). For example, if the closest feature is an external vertex operator then we integrate over the radius, $r$, from zero (modulo infrared divergences that must be dealt with on a case-by-case basis \cite{Polchinski88,LaNelson90,Witten12c}) up to this closest other vertex operator (and this is also the radius of convergence of the OPE).\footnote{In fact this is not quite what we do (because the range depends on the neighbouring features on {\it both} sides of the pinch via the transition function) but the idea is the same. We make a more precise statement in Sec.~\ref{sec:VSG}.} Even for a single such degeneration, (\ref{eq:completeness-sep BBB}) is strictly speaking not an equality, since the full moduli space is not covered, and there are also implicit sums over permutations of the external vertex operators. For instance, to reconstruct the Virasoro-Shapiro amplitude where there are four external vertex operators and one modulus we can cover the full moduli space by inserting two external fixed picture vertex operators on $\Sigma_1$ and $\Sigma_2$, i.e.~we glue two three-point amplitudes, integrate $q,\bar{q}$ over the indicated range (such that, say, the explicit propagator associated to the pinch provides all $s$-channel poles) and add to this term an analogous term after permuting any two of the vertex operators on opposite sides of the pinch (such that, e.g., the propagator gives rise to all $u$-channel poles). By worldsheet duality this covers the full moduli space \cite{Polchinski_v1}, and the $t$-channel poles become manifest only after summing over all states propagating through the pinches. We will demonstrate this explicitly in the case of the Virasoro-Shapiro amplitude that will be derived by gluing two 3-point amplitudes using a coherent state basis for the vertex operators in the handle operator. So this is a significant point of departure from the standard closed string field theory where all channels, $s,t,u$, are manifest, but where the price to be paid is the necessity to include a four-point vertex associated to the bulk of moduli space. (Furthermore, the integration regions for moduli integrals in closed string field theory are correspondingly much more complicated).
\sk

If the external vertex operators, $\dots_1\dots_2$, are taken to have canonical ghost number $N_{\rm gh}=2$, such as $\hat{\mathscr{V}}_b=\tilde{c}c\mathscr{V}_b$ (as may  be taken to be the case provided the vertex operator momenta are non-vanishing \cite{Witten12c} and provided the conditions spelt out in Sec.~\ref{sec:EVO} are fulfilled) with $\mathscr{V}_b$ the matter contribution, then the only terms in the sum over operators $\hat{\mathscr{A}}_a^{(z_1)}$ that will contribute (in the sum over ghost numbers, see (\ref{eq:oplusHHdual})) in (\ref{eq:completeness-sep BBB}) will also have $N_{\rm gh}=2$, while the corresponding dual, $\hat{\mathscr{A}}^a_{(z_2)}$, according to (\ref{eq:oplusHHdual}) will have \cite{Witten12c} ghost number $N_{\rm gh}=4$ in order to saturate the ghost number anomaly. We could just as well have taken the $b_0\tilde{b}_0$ factor to act on $\hat{\mathscr{A}}_a^{(z_1)}$ and the net result would be the same: in that case the sum over $a$ would have support only on the $N_{\rm gh}=4$ terms in $\hat{\mathscr{A}}_a^{(z_1)}$ and the $N_{\rm gh}=2$ terms in $\hat{\mathscr{A}}^a_{(z_2)}$.\footnote{If we indeed take the remaining asymptotic vertex operators to have ghost number $N_{\rm gh}=2$, one might wonder why the path integral is not zero: could not one use a contour argument to pull the $b_0\tilde{b}_0$ contours away from $z_2=0$ to enclose the remaining vertex operators? The answer is of course `yes', but given the $b_0\tilde{b}_0$ operators are based at $z_2=0$ they will not annihilate the external vertex operators (as one can check explicitly), so everything is consistent. This is why we include an index $b_0\rightarrow b_0^{(z_2)}$ to denote that the corresponding operator is based at the origin of the $z_2$ frame, $b_n^{(z_2)}\equiv \oint \frac{\rmd z_2}{2\pi iz_2}z_2^{n+2}b(z_2)$ (or even $b_n^{(z)}(w)\equiv \oint \frac{\rmd z}{2\pi i(z-w)}(z-w)^{n+2}b(z)$ when it is based at a point $z=w$ in the $z$ frame).} 
\sk

Applied to mass eigenstates, the factorisation property of amplitudes (\ref{eq:completeness-sep BBB}) is a standard result in string theory. That it applies even when the states $\{\hat{\mathscr{A}}_a\}$ constitute a coherent state basis however, as we have claimed, is quite subtle. This is a good point to summarise what we have learnt. From the coherent state defining properties laid out in Sec.~\ref{sec:SCS} and demanding consistent factorisation leads to the following three key differences in the case of coherent states compared to mass eigenstates: 
\begin{itemize}\label{(c)comment}
\item[{\bf (a)}] coherent states are not orthogonal, see (\ref{eq:deltaij});
\item[{\bf (b)}] consistent factorisation requires that (\ref{eq:consistency_condition2}) (and the related statements (\ref{eq:consistency_condition2b})) is satisfied;
\item[{\bf (c)}] it is desirable that coherent states have well defined scaling dimensions $h_a+\tilde{h}_a$, even offshell, in order to be able to apply the operator-state correspondence in a simple manner, but generically spin, $h_a-\tilde{h}_a$, is not well-defined.
\end{itemize} 
All these follow directly from the defining properties of coherent states (see Sec.~\ref{sec:SCS}) and general factorisation formulas. In particular, property {\bf (a)} is true by virtue of the coherent state defining property {\bf (1)}: the continuity requirement of the associated quantum numbers is not consistent with a Kronecker delta (or delta function)\footnote{This does not exclude the possibility that there are additional quantum numbers (other than the continuous ones) that {\it do} have Kronecker delta or delta function overlap.} overlap. As a result, condition {\bf (b)} associated to consistent factorisation is non-trivial for coherent states, even though it is almost trivial for momentum eigenstates where a Kronecker delta overlap is possible. 
Condition {\bf (c)}, has not yet been elaborated on. In order to understand this point we primarily need to understand the constraints on what states can propagate through a pinch on a Riemann surface. 

\subsubsection*{Non-Separating Degenerations}
Let us mention briefly a few changes in the case of non-separating degenerations. 
Cutting along a {\it non-separating cycle} (such as $A_I$ or $B_I$ homology cycles in a canonical intersection basis \cite{DHokerPhong} with $I=1,\dots,\g$, see Fig.~\ref{fig:homology})  
where the factorisation formula (\ref{eq:completeness_nonsep}) is of interest, the situation is only slightly different as compared to the separating case.  Here the best we can hope for is that $\hat{\mathscr{A}}^a_{(z_2)}$ be an eigenstate of $L_0+\tilde{L}_0$, and it will generically {\it not} be annihilated by $L_0-\tilde{L}_0$.\footnote{We thank Edward Witten for a discussion on this point. Notice this is different from the situation in string field theory where coherent state bases that are also eigenstates of $L_0,\tilde{L}_0$ can be chosen for both separating and non-separating degenerations. The reason for this distinction is that in string field theory the non-level-matched contributions are absorbed into internal vertices. We thank Ashoke Sen for a discussion on the corresponding situation in string field theory.} In general, being an eigenstate of both $L_0+\tilde{L}_0$ and $L_0-\tilde{L}_0$ as in (\ref{eq:AqqbarA}) is actually more than we need, in that to be able to apply the operator state correspondence in a simple manner the crucial requirement is that it be an eigenstate of $L_0+\tilde{L}_0$ (and not necessarily of $L_0-\tilde{L}_0$). Indeed, if we wish to use a coherent state basis for cut cycles non-homologous to zero (such as an $A_I$ homology cycle for any $I=1,\dots,\g$) the best we can hope for is that it satisfies,
\begin{equation}\label{eq:Ah1}
\hat{\mathscr{A}}^a_{(z_2/q)}=r^{h_a+\tilde{h}_a}\hat{\mathscr{A}}^a_{(z_2e^{-i\theta})},\qquad {\rm with}\qquad q=re^{i\theta},
\end{equation}
which {\it replaces} $\hat{\mathscr{A}}^a_{(z_2)}$ in (\ref{eq:completeness_nonsep}) (where we had taken $q=1$). We derive this (that (\ref{eq:Ah1}) is indeed valid for coherent states) explicitly for the case of flat backgrounds in (\ref{eq:L0Lb0A=DA}), for both separating and non-separating degenerations. 
We could also absorb the phase of $q$ into a redefinition of $z_2$ (or $z_1$), which is equivalent to gluing with transition function, $z_1z_2=r$, 
with $r$ real and positive. (If we then wish to equip the cut cycle with pinch and twist moduli we could identify these with $r$ and the phase of either $z_1$ or $z_2$ respectively). The relation (\ref{eq:Ah1}) is the minimum requirement that a coherent basis will be required to satisfy. 
We emphasise that the relation (\ref{eq:Ah1}) is more or less forced upon us if we wish to replace handles by {\it bi-local} operators, since otherwise the coherent states would change even in the absence of interaction as they propagate forward in worldsheet time \cite{JoeBigBook}.
\sk

Keeping these differences in mind, we can associate again, say, three moduli to the vicinity of the cut, namely the locations of the two vertex operators on either side and a (pinch and twist) modulus. (As mentioned above there is a variety of possibilities depending on the gauge slice of interest, and compatibility with the Riemann-Roch theorem.). The factorisation formula for non-separating degenerations (\ref{eq:completeness_nonsep}) is then,
\begin{equation}\label{eq:completeness_nonsep BBB}
\boxed{e^{-\chi(\Sigma_\g)\Phi}\Big\langle \dots\Big\rangle_{\Sigma_\g}= \,\,e^{-\chi(\Sigma_{\g-1})\Phi}\Big\langle\dots\suminnt\limits_{a}\,\mathscr{A}_{a}^{(z_1)}\int_{q,\bar{q}}\!\big[\hat{B}_q\hat{B}_{\bar{q}}\,\mathscr{A}^{a}_{(z_2/q)}\big]\Big\rangle_{\Sigma_{\g-1}}}
\end{equation}
Again, since the hats, `$\hat{\phantom{a}}$', are absent these are generically integrated vertex operators. Of course there can be situations where the underlying surface is such that some of these appear at fixed rather than integrated, but this really depends on the gauge slice of interest or the question one wishes to ask. We would rather not write down explicitly every possibility since this will most likely not add much to the discussion. 
\sk

One can show that the relation (\ref{eq:completeness_nonsep BBB}) is indeed consistent for all genera, $\g\geq1$, when the number of asymptotic states is at least three, $\n\geq3$. The remaining cases, $\n=0,1,2$, are also easy to deal with but one needs to be careful with zero modes such as conformal Killing vectors, and taking into account any unfixed automorphisms. How to deal with these exceptional cases is discussed in Sec.~\ref{sec:TEC}.

\section{Gauge Invariance}\label{sec:BRST-AC}
Spacetime gauge invariance (or target space gauge invariance in general) from a worldsheet perspective corresponds to the statement that if an external vertex operator, $\hat{\mathscr{V}}_j$, is shifted by a BRST-exact local operator, 
$$
\hat{\mathscr{V}}_j\mapsto \hat{\mathscr{V}}_j+Q_B\hat{\mathscr{W}}_j,
$$
the full path integral (\ref{eq:fullpathintegral}) (on p.~\pageref{eq:fullpathintegral}) is unchanged (when external states are BRST-invariant). In this section we demonstrate how this decoupling of BRST-exact contributions takes place when we use handle operators (associated to a non-separating cycle) to construct higher genus Riemann surfaces, or when we use handle operators to glue Riemann surfaces (associated to a separating cycle), or when we use general BRST-invariant (not necessarily $(1,1)$ conformal primary) external vertex operators. (Understanding the formalism for general BRST-invariant vertex operators for asymptotic states is also the necessary step for an offshell formulation of string theory \cite{Sen15b}.) In particular, we demonstrate that BRST-exact contributions decouple up to boundary terms in moduli space. Such boundary terms must either vanish or cancel, and there is a variety of mechanisms (such as the cancelled-propagator argument, the Fischler-Susskind mechanism and mass renormalisation) by which this cancellation occurs.
\sk

As in all previous sections (except in some parts of Sec.~\ref{sec:NOII}), also here the discussion will be general enough to apply to any 2D matter CFT provided the total central charge (matter plus ghosts) adds up to zero. From Sec.~\ref{sec:EHOC} onwards we focus on an explicit matter CFT.

\subsection{The General Case}\label{sec:TGC}
Let us primarily review the general case \cite{Zwiebach93,Sen15b}, and in particular show that the path integral measure always contributes total derivatives when BRST-exact states are inserted into the path integral. Since this is a rather formal and not particularly transparent argument we will in the following subsections focus explicitly on the various types of contributions (depending on the gauge slice of interest in moduli space) that arise from handle operators. 
\sk

We begin by considering the quantity:
\begin{equation}
\Big(\prod_{k=1}^\m \hat{B}_k\Big)Q_B,
\end{equation}
where we momentarily (implicitly) take all operators to be defined in the same chart coordinates (say $z_1$) so that we do not have to worry about changing the orientation of the BRST-charge contour as it is swept through the various patches of the Riemann surface. (In fact, the only property that we need for the following is that the BRST current is a {\it tensor}, so that the BRST charge is globally-defined and we only need to worry about flipping the contour orientation as we move the contour across the Riemann surface. In what follows we keep the contour orientation fixed, and then it can be flipped as appropriate in any given chart where we wish to evaluate the contour integral.) Having said that, we then wish to commute the BRST charge through to the left,  and this gives rise to various anticommutators,
\begin{equation}\label{eq:B1..Bk-1QBo}
\begin{aligned}
\Big(\prod_{k=1}^\m \hat{B}_k\Big)Q_B&=(-)^\m Q_B\Big(\prod_{k=1}^\m \hat{B}_k\Big)+\sum_{k=1}^\m (-)^{\m-k}\hat{B}_1\dots \hat{B}_{k-1}\big\{Q_B,\hat{B}_k\big\}\hat{B}_{k+1}\dots \hat{B}_\m.
\end{aligned}
\end{equation}
At this stage we have only made use of the fact that the $\hat{B}_k$ and $Q_B$ are Grassmann-odd. 
Since the quantity $\big\{Q_B,\hat{B}_k\big\}$ is proportional to linear superpositions of Virasoro generators, $L_n,\tilde{L}_n$, it is Grassmann-even, so we can pull it through to the left in the second term on the right-hand side which in turn leaves behind various commutators, the relevant terms according to (\ref{eq:B1..Bk-1QBo}) being (for $k>1$, the $k=1$ term being trivial),
\begin{equation}\label{eq:B1..Bk-1QB}
\begin{aligned}
\hat{B}_1&\dots \hat{B}_{k-1}\big\{Q_B,\hat{B}_k\big\} =\\
&=\big\{Q_B,\hat{B}_k\big\}\hat{B}_1\dots \hat{B}_{k-1}+\sum_{\ell=1}^{k-1}
\hat{B}_1\dots \hat{B}_{\ell-1}\big[\hat{B}_\ell,\big\{Q_B,\hat{B}_k\big\}\big]\hat{B}_{\ell+1}\dots \hat{B}_{k-1}.
\end{aligned}
\end{equation}
Now since the quantity $\big\{Q_B,\hat{B}_k\big\}$ is a linear superposition of Virasoro generators, $L_n,\tilde{L}_n$, and the $\hat{B}_\ell$ are linear superpositions of $b_n,\tilde{b}_n$ ghost modes it follows from $[L_n,b_m]=(n-m)b_{n+m}$ that $\big[\hat{B}_\ell,\big\{Q_B,\hat{B}_k\big\}\big]$ simply anticommutes with the remaining $\hat{B}_k$ in (\ref{eq:B1..Bk-1QB}). So we can pull it through to the left at the only expense of picking up $\ell-1$ minus signs,
\begin{equation}\label{eq:B1..Bk-1QB2}
\begin{aligned}
\hat{B}_1&\dots \hat{B}_{k-1}\big\{Q_B,\hat{B}_k\big\} =\\
&=\big\{Q_B,\hat{B}_k\big\}\hat{B}_1\dots \hat{B}_{k-1}- \sum_{\ell=1}^{k-1}(-)^{\ell}\big[\hat{B}_\ell,\big\{Q_B,\hat{B}_k\big\}\big]
\hat{B}_1\dots \hat{B}_{\ell-1}\hat{B}_\ell \hspace{-.38cm}/\hspace{.38cm}\!\!\hat{B}_{\ell+1}\dots \hat{B}_{k-1}.
\end{aligned}
\end{equation}
Substituting this back into (\ref{eq:B1..Bk-1QBo}) we learn that,
\begin{equation}\label{eq:B1..Bk-1QBo2}
\begin{aligned}
\Big(\prod_{k=1}^\m &\hat{B}_k\Big)Q_B
=(-)^\m Q_B\Big(\prod_{k=1}^\m \hat{B}_k\Big)+\sum_{k=1}^\m (-)^{\m-k}\big\{Q_B,\hat{B}_k\big\}\hat{B}_1\dots \hat{B}_{k-1}\hat{B}_k \hspace{-.41cm}/\hspace{.41cm}\!\!\hat{B}_{k+1}\dots \hat{B}_\m\\
&-\sum_{1\leq \ell<k\leq \m}(-)^{\m-k-\ell}\big[\hat{B}_\ell,\big\{Q_B,\hat{B}_k\big\}\big]
\hat{B}_1\dots \hat{B}_{\ell-1}\hat{B}_\ell \hspace{-.38cm}/\hspace{.38cm}\!\!\hat{B}_{\ell+1}\dots \hat{B}_{k-1}\hat{B}_k \hspace{-.41cm}/\hspace{.41cm}\!\!\hat{B}_{k+1}\dots \hat{B}_\m,
\end{aligned}
\end{equation}
and if we furthermore multiply left- and right-hand sides by $(-)^{\m-1}$ and rearrange we obtain:
\begin{equation}\label{eq:B1..Bk-1QBo3}
\boxed{
\begin{aligned}
Q_B\Big(&\prod_{k=1}^\m \hat{B}_k\Big)-(-)^\m\Big(\prod_{k=1}^\m\hat{B}_k\Big)Q_B=\\
&=\sum_{k=1}^\m (-)^{k-1}\big\{Q_B,\hat{B}_k\big\}\hat{B}_1\dots \hat{B}_{k-1}\hat{B}_k \hspace{-.41cm}/\hspace{.41cm}\!\!\hat{B}_{k+1}\dots \hat{B}_\m\\
&\quad+\sum_{1\leq \ell<k\leq \m}(-)^{k+\ell}\big[\hat{B}_\ell,\big\{Q_B,\hat{B}_k\big\}\big]
\hat{B}_1\dots \hat{B}_{\ell-1}\hat{B}_\ell \hspace{-.38cm}/\hspace{.38cm}\!\!\hat{B}_{\ell+1}\dots \hat{B}_{k-1}\hat{B}_k \hspace{-.41cm}/\hspace{.41cm}\!\!\hat{B}_{k+1}\dots \hat{B}_\m\\
\end{aligned}
}
\end{equation}
Recall from (\ref{eq:D=QB}) that $\big\{Q_B,\hat{B}_k\big\}=\hat{D}_k$, and on {\it operators}, $\hat{D}_k\sim -\frac{\partial}{\partial t^k}$, where the operator $\hat{D}_k$ associated to modulus $t^k$ is given explicitly in (\ref{eq:Tintmub8z}). Note that this generates a derivative that is {\it outside} the normal ordering of the operators on which it acts (unlike $L_{-1}$ which when acting on local operators generates a normal-ordered derivative of the operator -- see Sec.~\ref{sec:comm dA-dA}).
\sk

The relation (\ref{eq:B1..Bk-1QBo3}) is a standard result (see equation (2.20) in \cite{Sen15b}), and also \cite{VerlindeVerlinde87b} and \cite{Witten12c} for related discussions. As we have shown, see (\ref{eq:Bintmub8z}), the measure contributions, $\hat{B}_k$, can be expressed both in terms of transition functions (as in \cite{Sen15b,deLacroixErbinKashyapSenVerma17}) or in terms of a metric (as in \cite{Polchinski88}), see (\ref{eq:hatBzvR}) (the transition function approach is more efficient). The precise manner in which the gauge slice has been chosen in fact does not matter and we treat both cases on equal footing. From (\ref{eq:B1..Bk-1QBo3}), and independently of the explicit examples to be discussed in the following subsections, using the standard identity for a global formula for an exterior derivative (see, e.g., Th.~20.14, p.~233 in \cite{Tu11}) (as shown explicitly in \cite{Sen15b}) the terms on the right-hand side in (\ref{eq:B1..Bk-1QBo3}) lead precisely to a total derivative in moduli space. Combining this with the fact that the BRST operator commutes through the identity operator (\ref{eq:gluing Qbosonic2}) (up to a minus sign that arises from the change of contour orientation of the BRST charge since the BRST charge is defined with the standard orientation within each chart), it follows that as we pass the contour of the BRST charge through a handle operator it will always give rise to a total derivative in moduli space. The main objective in the current section is to make this entirely manifest, so we discuss the various explicit cases in the following subsections.

\subsection{Handle Operators with No Moduli}
Having reviewed the general argument that BRST-exact external vertex operators decouple from physical amplitudes we now zoom in more closely on the precise manifestation of this statement in terms of handle operators. We begin with a discussion of the trivial case where there are no moduli associated to a given handle operator.
\sk

As we will discuss in Sec.~\ref{sec:GOCS}, the BRST charge contour can be passed through the unit operator (\ref{eq:1intaAA}) without obstruction, the precise statement being (\ref{eq:gluing Qbosonic}):
\begin{equation}\label{eq:gluing Qbosonic2}
\big(Q_B^{(z_1)}+Q_B^{(z_2)}\big)\,\suminnt\limits_{a}\,\,\hat{\mathscr{A}}_a^{(z_1)}\hat{\mathscr{A}}^a_{(z_2/q)}=0,
\end{equation}
where we glue with $z_1z_2=q$. 
This is because mode contours can be deformed without obstruction across a handle, reflecting the fact that the transition function, $z_1z_2=q$, implementing the gluing is holomorphic on the (annular) patch overlap, $U_1\cap U_2$. For clarity, and taking into account that $Q_B^{(z_2)}=Q_B^{(z_2/q)}$, the relation (\ref{eq:gluing Qbosonic2}) is equivalent to:
\begin{equation}\label{eq:QAA=-AQA}
\suminnt\limits_{a}\,\,\big[Q_B^{(z_1)}\hat{\mathscr{A}}_a^{(z_1)}\big]\hat{\mathscr{A}}^a_{(z_2/q)}=-\suminnt\limits_{a}\,\,(-)^{|a|}\hat{\mathscr{A}}_a^{(z_1)}\big[Q_B^{(z_2/q)}\hat{\mathscr{A}}^a_{(z_2/q)}\big],
\end{equation}
and when the local operator, $\hat{\mathscr{A}}_a^{(z_1)}$ (equivalently $\hat{\mathscr{A}}^a_{(z_2/q)}$), is Grassmann-even (which is the only case we will need in the current document) $|a|=0$, whereas if it is Grassmann-odd, $|a|=1$. (Incidentally, the operators $\hat{\mathscr{A}}_a^{(z_1)}$ and $\hat{\mathscr{A}}^a_{(z_1)}$ are always Grassmann-even in the full set of offshell  coherent state basis we construct in this document, essentially because we also allow for Grassmann-odd ``quantum numbers'', see (\ref{eq:offshellA_a}); but there are bases whose elements span both Grassmann-odd and Grassmann-even elements.) It is helpful to have a visual representation in mind to make manifest that the contour of the $Q_B^{(z_1)}$ and $Q_B^{(z_2)}$ charges encircle the full operators $\hat{\mathscr{A}}_a^{(z_1)}$ and $\hat{\mathscr{A}}^a_{(z_2/q)}$ respectively (in the canonical counterclockwise sense with respect to the $z_1$ and $z_2$ frames respectively), so that, e.g., one can unwrap the contour of the charge $Q_B^{(z_2)}$ off to the remaining surface (where it might pick up additional contributions depending on whether other local operators away from this handle are BRST-invariant or not). 

\subsection{Fixed-Picture Handle Operators with Pinch \& Twists}
We next consider the case when to the handle operator of interest we wish to associate pinch and twist moduli, which corresponds to variations in the real and imaginary part of the gluing parameter, $\ln q$, respectively, where the relevant transition function is $z_1z_2=q$. The relevant handle operator with pinch and twist moduli reads:
\begin{equation}\label{eq:1+pinch}
\boxed{
\hat{H}=\suminnt\limits_a\,\,\hat{\mathscr{A}}_a^{(z_1)}\! \int \rmd^2q\big[\hat{B}_q\hat{B}_{\bar{q}}\hat{\mathscr{A}}^a_{(z_2/q)}\big]}
\end{equation}
The BRST charge does not commute with the pinching and twisting moduli measure contributions, $\hat{B}_q,\hat{B}_{\bar{q}}$, associated to this cut  (i.e.~those associated to transition functions $z_1z_2=q$ and variations $\delta q,\delta\bar{q}$). This is because of the relation $\{Q_B,b_0\}=L_0$. 
So rather than (\ref{eq:gluing Qbosonic2}) we pick up a total derivative corresponding to a boundary contribution in moduli space. 
\sk

\begin{figure}
\begin{center}
\includegraphics[angle=0,origin=c,width=0.85\textwidth]{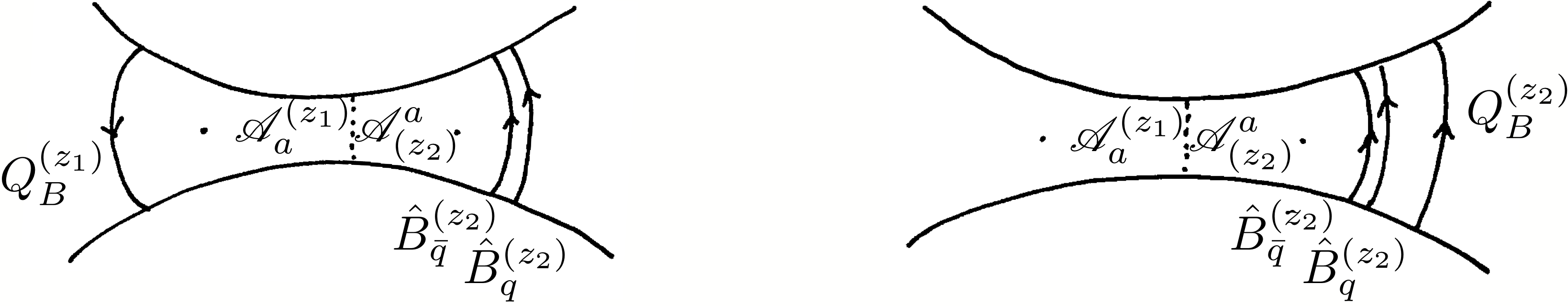}
\caption{Schematic representation of the contour deformation of the BRST charge as it is passed through a handle operator from the $z_1$ chart to the $z_2$ chart with transition function $z_1z_2=1$ (in the main text we rescale $z_2\mapsto z_2/q$ and glue with transition function $z_1(z_2/q)=1$), after deforming the latter with pinch and twist moduli. Notice that as the contour of the BRST charge, $Q_B^{(z_1)}$, is pulled through $\mathscr{A}_a^{(z_1)}$ it first encounters $\mathscr{A}^a_{(z_2)}$ and {\it then} the measure contributions, $\hat{B}_{\bar{q}}$ and finally $\hat{B}_q$. Consequently, the operator $\hat{B}_q\hat{B}_{\bar{q}}\mathscr{A}_{z_2}^a$ is encircled entirely by the contour of the BRST charge, $Q_B^{(z_2)}$, (as depicted in the sketch on the right) at the expense of giving rise to a boundary term, see (\ref{eq:QABqBqA4}). Notice also a flip in the orientation of the BRST-charge contour and the corresponding relation (on chart overlaps), $Q_B^{(z_1)}+Q_B^{(z_2)}=0$.}\label{fig:BRSTcontBqBqbar}
\end{center}
\end{figure}

Let us go through the derivation of this boundary contribution carefully, see also Fig.~\ref{fig:BRSTcontBqBqbar}. 
The starting point is the quantity:
\begin{equation}\label{eq:QABqBqA}
\suminnt\limits_{a}\,\,\,\big(Q_B^{(z_1)}\hat{\mathscr{A}}_a^{(z_1)}\big)\big(\hat{B}_q\hat{B}_{\bar{q}}\hat{\mathscr{A}}^a_{(z_2/q)}\big).
\end{equation}
Deforming the contour of the BRST charge through the handle using the transition function, $z_1z_2=q$, the fundamental relation (\ref{eq:QAA=-AQA}) implies that (\ref{eq:QABqBqA}) is equivalent to:
\begin{equation}\label{eq:QABqBqA2}
\begin{aligned}
-\suminnt\limits_{a}\,\,\,(-)^{|a|}\hat{\mathscr{A}}_a^{(z_1)}\big[\hat{B}_q\hat{B}_{\bar{q}}\big(Q_B^{(z_2/q)}\hat{\mathscr{A}}^a_{(z_2/q)}\big)\big],
\end{aligned}
\end{equation}
where we took into account that as the contour of the BRST charge passes from the $z_1$ chart to the $z_2/q$ chart it encounters $\mathscr{A}^a_{(z_2/q)}$ {\it before} it encounters the measure contributions, $\hat{B}_{\bar{q}}$ and finally $\hat{B}_q$. (More generally, if these measure contributions had odd Grassmannality there would be an additional minus sign, a comment that is relevant in (\ref{eq:Ub0b0QA}).) Recall now that,
$$
\hat{B}_q= -\frac{b_0^{(z_2/q)}}{q},\qquad \hat{B}_{\bar{q}}=-\frac{\tilde{b}_0^{(z_2/q)}}{\bar{q}},
$$
and also $b_0^{(z_2/q)}=b_0^{(z_2)}$ and $Q_B^{(z_2/q)}=Q_B^{(z_2)}$. Then, we can commute the BRST charge through to the left of the factor $\hat{B}_q\hat{B}_{\bar{q}}$ in order to ensure that it encircles the entire operator $\hat{B}_q\hat{B}_{\bar{q}}\hat{\mathscr{A}}^a_{(z_2/q)}$ at the expense of picking up a commutator, $[\hat{B}_q\hat{B}_{\bar{q}},Q_B^{(z_2)}]$. That is, the expression (\ref{eq:QABqBqA2}) is also equivalent to:
\begin{equation}\label{eq:QABqBqA3}
\begin{aligned}
\suminnt\limits_{a}\,\,\,(-)^{|a|}\hat{\mathscr{A}}_a^{(z_1)}\Big\{\big[Q_B^{(z_2)},\hat{B}_q\hat{B}_{\bar{q}}\big]\hat{\mathscr{A}}^a_{(z_2/q)}-Q_B^{(z_2)}\big(\hat{B}_q\hat{B}_{\bar{q}}\hat{\mathscr{A}}^a_{(z_2/q)}\big)\Big\}.
\end{aligned}
\end{equation}
By explicit computation, it is a simple exercise to then show that the relevant commutator is given by,
\begin{equation}
\big[Q_B^{(z_2)},\hat{B}_{q}\hat{B}_{\bar{q}}\big]=\frac{1}{q\bar{q}}\big(\tilde{b}_0^{(z_2)}L_0^{(z_2)}-b_0^{(z_2)}\tilde{L}_0^{(z_2)}\big),
\end{equation}
and therefore since,
$$
\hat{\mathscr{A}}^a_{(z_2/q)}=q^{L_0^{(z_2)}}\bar{q}^{\tilde{L}_0^{(z_2)}}\hat{\mathscr{A}}^a_{(z_2)},
$$
we see that after some elementary rearrangements,
\begin{equation}
\begin{aligned}
\big[Q_B^{(z_2)},\hat{B}_q\hat{B}_{\bar{q}}\big]\hat{\mathscr{A}}^a_{(z_2/q)}& = \frac{1}{q\bar{q}}\big(\tilde{b}_0^{(z_2)}L_0^{(z_2)}-b_0^{(z_2)}\tilde{L}_0^{(z_2)}\big)q^{L_0^{(z_2)}}\bar{q}^{\tilde{L}_0^{(z_2)}}\hat{\mathscr{A}}^a_{(z_2)}\\
&=\frac{\partial}{\partial \bar{q}}\big(\hat{B}_q\hat{\mathscr{A}}_{(z_2/q)}^a\big)-\frac{\partial}{\partial q}\big(\hat{B}_{\bar{q}}\hat{\mathscr{A}}_{(z_2/q)}^a\big).
\end{aligned}
\end{equation}
Substituting this back into (\ref{eq:QABqBqA3}) and setting the resulting quantity equal to the quantity we started from (\ref{eq:QABqBqA}), including the integral over $q,\bar{q}$ and rearranging we find that,
\begin{equation}\label{eq:QABqBqA4}
\begin{aligned}
\suminnt\limits_{a}\,\,\,\big(Q_B^{(z_1)}&\hat{\mathscr{A}}_a^{(z_1)}\big)\int \rmd^2q\big(\hat{B}_q\hat{B}_{\bar{q}}\hat{\mathscr{A}}^a_{(z_2/q)}\big)+\suminnt\limits_{a}\,\,\,(-)^{|a|}\hat{\mathscr{A}}_a^{(z_1)}\int \rmd^2q\big(Q_B^{(z_2)}\hat{B}_q\hat{B}_{\bar{q}}\hat{\mathscr{A}}^a_{(z_2/q)}\big)=\\
&=\suminnt\limits_{a}\,\,\,(-)^{|a|}\hat{\mathscr{A}}_a^{(z_1)}\int \rmd^2q\Big\{\frac{\partial}{\partial \bar{q}}\big(\hat{B}_q\hat{\mathscr{A}}_{(z_2/q)}^a\big)-\frac{\partial}{\partial q}\big(\hat{B}_{\bar{q}}\hat{\mathscr{A}}_{(z_2/q)}^a\big)\Big\}.
\end{aligned}
\end{equation}
Notice that the insertion on the $z_2$ patch is precisely an integral of a total derivative in moduli $q,\bar{q}$. 
It is natural to state the final result (\ref{eq:QABqBqA4}) in terms of the following forms,
\begin{equation}\label{eq:hatWA}
\boxed{
\begin{aligned}
\hat{\mathbb{W}}_{(z_2)}^a&=\frac{1}{i}\Big[\rmd\bar{q}\,\big(q^{L_0}\bar{q}^{\tilde{L}_0}\hat{B}_{\bar{q}}\hat{\mathscr{A}}^a_{(z_2)}\big)+\rmd q\,\big(q^{L_0}\bar{q}^{\tilde{L}_0}\hat{B}_q\hat{\mathscr{A}}^a_{(z_2)}\big)\Big]\\
\hat{\mathbb{A}}^a_{(z_2)}&
=\rmd^2q \big(q^{L_0}\bar{q}^{\tilde{L}_0}\hat{B}_q\hat{B}_{\bar{q}}\hat{\mathscr{A}}^a_{(z_2)}\big)\\
\end{aligned}
}
\end{equation}
The hats, `$\hat{\phantom{a}}$', on these quantities, $\hat{\mathbb{W}}_{(z_2)}^a$ and $\hat{\mathbb{A}}^a_{(z_2)}$, are meant to denote that they are associated to {\it fixed-picture} vertex operators since in this gauge slice we do not integrate over the location of the operators $\hat{\mathscr{A}}_a^{(z_1)}$ and $\hat{\mathscr{A}}^a_{(z_2/q)}$. There will be corresponding expressions in integrated picture below where the hats are absent. In terms of the quantities (\ref{eq:hatWA}), the result (\ref{eq:QABqBqA4}) can be written concisely as follows,
\begin{equation}\label{eq:gluing Qbosonic2withbbSD}
\boxed{
\begin{aligned}
\suminnt\limits_{a}\,\,\,\big(Q_B^{(z_1)}&\hat{\mathscr{A}}_a^{(z_1)}\big)\int \hat{\mathbb{A}}^a_{(z_2)}
&=\suminnt\limits_a\,\,(-)^{|a|}\hat{\mathscr{A}}_a^{(z_1)} \int\rmd \hat{\mathbb{W}}_{(z_2)}^a-\suminnt\limits_{a}\,\,\,(-)^{|a|}\hat{\mathscr{A}}_a^{(z_1)}\int \big(Q_B^{(z_2)}\hat{\mathbb{A}}^a_{(z_2)}\big)
\end{aligned}
}
\end{equation}
equivalently,
\begin{equation}\label{eq:gluing Qbosonic2withbbNSD}
\boxed{
\begin{aligned}
\big(Q_B^{(z_1)}+Q_B^{(z_2)}\big)\suminnt\limits_{a}\,\,\,&\hat{\mathscr{A}}_a^{(z_1)}\!\!\int \hat{\mathbb{A}}^a_{(z_2)}
&=\suminnt\limits_a\,\,(-)^{|a|}\hat{\mathscr{A}}_a^{(z_1)} \int\rmd \hat{\mathbb{W}}_{(z_2)}^a
\end{aligned}
}
\end{equation}
where on the right-hand side we wrote $\rmd=\rmd q\,\partial_q+\rmd\bar{q}\,\partial_{\bar{q}}$. 
The relations (\ref{eq:gluing Qbosonic2withbbSD}) and (\ref{eq:gluing Qbosonic2withbbNSD}) appear when these handle operators are associated to separating and non-separating degenerations respectively. In the former case the BRST-exact term on the left- and right-hand sides are inserted on the Riemann surface $\Sigma_1$ and $\Sigma_2$ respectively, whereas in the latter case they are inserted on the same Riemann surface, $\Sigma$. In both cases the obstruction to BRST-exact decoupling generated by the insertion of the handle operator is a boundary term in moduli space. The various boundary contributions are in turn generically determined by the operator product expansions of the operator $\hat{\mathbb{W}}^a_{(z_2)}$ with all other operator insertions appearing.
\sk

It is perhaps useful to also make note of the following relations: 
$$
\hat{B}_{q} \hat{\mathscr{A}}^a_{(z_2)}=\frac{\mathbi{c}_0^{(2)}}{q}\hat{\mathscr{A}}^a_{(z_2)},\qquad{\rm and}\qquad \hat{B}_{\bar{q}}\hat{\mathscr{A}}^a_{(z_2)}=\frac{\tilde{\mathbi{c}}_0^{(2)}}{\bar{q}}\hat{\mathscr{A}}^a_{(z_2)},
$$
which are obtained by explicit computation from the coherent state (\ref{eq:offshellA^a}). These relations can in turn be substituted back into (\ref{eq:hatWA}), 
which in turn allows one to integrate out $\mathbi{c}_0^{(2)},\tilde{\mathbi{c}}_0^{(2)}$ in the sum/integral over $a$ in (\ref{eq:gluing Qbosonic2withbbSD}) and (\ref{eq:gluing Qbosonic2withbbNSD}) using the measure (\ref{eq:dmua}). 
\sk

As discussed in the previous subsection, the appearance of a boundary term in the integral over moduli space as we commute the BRST operator through the ghost measure contributions is generic. All such boundary terms must cancel.  For example, when the cut produces a tadpole then the  Fischler-Susskind mechanism is at play, whereby this boundary term is cancelled by a lower-genus contribution with an insertion associated to shifting the background \cite{FischlerSusskind86a,FischlerSusskind86b}, see in particular \cite{Polchinski88,LaNelson90,Witten12c}. For an explicit analysis of the combinatorics of tadpole cancellation in a scalar field theory context (to all orders in perturbation theory) see \cite{EllisMavromatosSkliros15}. (In fact, the coherent states we have constructed do not apply at zero momentum as already discussed in Sec.~\ref{sec:GOCS}, since the change of variables (\ref{eq:lightconek}) breaks down. Notice that (\ref{eq:handle}) however {\it can} be used at zero momentum, as can the states of Sec.~\ref{sec:HOEM}.) Corresponding statements exist when the cut singles out a two-point amplitude, in which case (and when a non-renormalisation theorem does not set this contribution to zero from the outset) the boundary term is cancelled by mass (and wavefunction) renormalisation of the singled out external vertex operator \cite{Sen88}. 
\sk

Various other boundary terms can be cancelled by the standard `cancelled propagator' argument, whereby the external momenta are analytically continued off the mass shell to a regime where the boundary contribution vanishes identically and then we analytically continue back to physical external momenta. When this is possible, by a famous theorem of complex analysis such boundary contributions vanish identically. 

\subsection{Integrated-Picture Local Operators}\label{sec:IPVO}
It is well-known that  {\it physical closed-string fixed-picture vertex operators}, $\hat{\mathscr{V}}$, are defined to take values in the cohomology of the BRST charge, $Q_B$, while being annihilated by $b_0-\tilde{b}_0$ \cite{Nelson89},
\begin{equation}\label{eq:physicalVhat}
Q_B\hat{\mathscr{V}}=0\quad\mod \quad\hat{\mathscr{V}}\sim \hat{\mathscr{V}}+Q_B\hat{\mathscr{W}},\qquad {\rm and}\qquad (b_0-\tilde{b}_0)\hat{\mathscr{V}}=0.
\end{equation}
The $b_0-\tilde{b}_0=0$ constraint will be derived below (see also Sec.~\ref{sec:EVO} for related comments). We will study here the decoupling of BRST-exact contributions when we commute the BRST charge through a local vertex operator in {\it integrated} picture. Such a vertex operator can either be associated to an external state or an internal handle operator.
\sk

Let us in particular apply the general result (\ref{eq:B1..Bk-1QBo3}) to the case when we pass the BRST charge through the integrated vertex operator, $\hat{B}_{z_{v_1}}\hat{B}_{\bar{z}_{v_1}}\hat{\mathscr{A}}_a^{(z_1)}$, where $\m=2$. One finds,
\begin{equation}\label{eq:[B1B2,QB]}
\begin{aligned}
Q_B^{(z_1)}\big(&\hat{B}_{z_{v_1}}\hat{B}_{\bar{z}_{v_1}}\hat{\mathscr{A}}_a^{(z_1)}\big)-\hat{B}_{z_{v_1}}\hat{B}_{\bar{z}_{v_1}}\big(Q_B^{(z_1)}\hat{\mathscr{A}}_a^{(z_1)}\big)=\\
&=\big\{Q_B^{(z_1)},\hat{B}_{z_{v_1}}\big\}\hat{B}_{\bar{z}_{v_1}}\hat{\mathscr{A}}_a^{(z_1)}-\big\{Q_B^{(z_1)},\hat{B}_{\bar{z}_{v_1}}\big\}\hat{B}_{z_{v_1}} \hat{\mathscr{A}}_a^{(z_1)}-\big[\hat{B}_{z_{v_1}},\big\{Q_B^{(z_1)},\hat{B}_{\bar{z}_{v_1}}\big\}\big]
\hat{\mathscr{A}}_a^{(z_1)}\,\,\\
\end{aligned}
\end{equation}
Evaluating the (anti-)commutators of interest yields,
\begin{equation}\label{eq:QBQBBQB comms}
\begingroup\makeatletter\def\f@size{11}\check@mathfonts
\def\maketag@@@#1{\hbox{\m@th\large\normalfont#1}}%
\begin{aligned}
\big\{Q_B^{(z_1)},\hat{B}_{z_{v_1}}\big\} 
&= -L_{-1}^{(z_1)}(p_1)-\frac{1}{4}\sum_{n\geq1}\frac{1}{(n+1)!}\nabla_{\bar{z}_{v_1}}^{n-1}R_{(2)}(z_{v_1})\tilde{L}_n^{(z_1)}(p_1)\\
\big\{Q_B^{(z_1)},\hat{B}_{\bar{z}_{v_1}}\big\} 
&= -\tilde{L}_{-1}^{(z_1)}(p_1)-\frac{1}{4}\sum_{n\geq1}\frac{1}{(n+1)!}\nabla_{z_{v_1}}^{n-1}R_{(2)}(z_{v_1})L_n^{(z_1)}(p_1)\\
\big[\hat{B}_{z_{v_1}},\big\{Q_B,\hat{B}_{\bar{z}_{v_1}}\big\}\big] &=-
\frac{1}{4}R_{(2)}(z_{v_1})\big(b_0^{(z_1)}-\tilde{b}_0^{(z_1)}\big)+\frac{1}{4}\sum_{n\geq1}\frac{1}{(n+1)!}\partial_{\bar{z}_{v_1}}\Big(\nabla_{\bar{z}_{v_1}}^{n-1}R_{(2)}(z_v)\Big)\tilde{b}_n^{(z_1)}(p_1)\\
&\qquad -\frac{1}{4}\sum_{n\geq1}\frac{1}{(n+1)!}\partial_{z_{v_1}}\Big(\nabla_{z_{v_1}}^{n-1}R_{(2)}(z_v)\Big)b_n^{(z_1)}(p_1).
\end{aligned}
\endgroup
\end{equation}
Indeed, the right-hand sides of the first two relations in (\ref{eq:QBQBBQB comms}) were shown in (\ref{eq:D_zs3}), see also (\ref{eq:Dzv=-dzv}), to generate precisely derivatives with respect to the moduli $z_{v_1},\bar{z}_{v_1}$ respectively via the correspondence,
\begin{equation}\label{eq:QB =d}
\big\{Q_B,\hat{B}_{z_{v_1}}\big\} \leftrightarrow-\frac{\partial}{\partial z_{v_1}},\qquad \big\{Q_B,\hat{B}_{\bar{z}_{v_1}}\big\} \leftrightarrow-\frac{\partial}{\partial \bar{z}_{v_1}},
\end{equation}
but care is needed because the left-hand sides of these latter relations commute through local functions, such as $\nabla_{z_{v_1}}^{n-1}R_{(2)}(z_v)$ whereas the right-hand sides do not. In particular, derivatives of these local functions (coefficients of $b_n^{(z_1)},\tilde{b}_n^{(z_1)}$ in $\hat{B}_{\bar{z}_{v_1}}$ and $\hat{B}_{z_{v_1}}$ in particular) are generated by the {\it last} commutator in (\ref{eq:QBQBBQB comms}). All in all, the various derivatives combine to yield the following result for the commutator of interest (\ref{eq:[B1B2,QB]}),
\begin{equation}\label{eq:[B1B2,QB]2}
\begingroup\makeatletter\def\f@size{11}\check@mathfonts
\def\maketag@@@#1{\hbox{\m@th\large\normalfont#1}}%
\begin{aligned}
\big[Q_B^{(z_1)},&\hat{B}_{z_{v_1}}\hat{B}_{\bar{z}_{v_1}}\big]\hat{\mathscr{A}}_a^{(z_1)}=\frac{1}{4}R_{(2)}(z_{v_1})\big(b_0^{(z_1)}-\tilde{b}_0^{(z_1)}\big)\hat{\mathscr{A}}_a^{(z_1)}+\frac{\partial}{\partial \bar{z}_{v_1}}\big(\hat{B}_{z_{v_1}}\hat{\mathscr{A}}_a^{(z_1)}\big)-\frac{\partial}{\partial z_{v_1}}\big(\hat{B}_{\bar{z}_{v_1}}\hat{\mathscr{A}}_a^{(z_1)}\big)
\end{aligned}
\endgroup
\end{equation}
Including the integration measure associated to moduli variations, $\delta z_{v_1},\delta\bar{z}_{v_1}$, we have shown that the corresponding integral yields (up to the $b_0^{(z_1)}-\tilde{b}_0^{(z_1)}$ term) an integral of a total derivative,
\begin{equation}\label{eq:[B1B2,QB]2}
\begingroup\makeatletter\def\f@size{11}\check@mathfonts
\def\maketag@@@#1{\hbox{\m@th\large\normalfont#1}}%
\begin{aligned}
&\int_{\Sigma_1} \rmd^2z_{v_1}Q_B^{(z_1)}\big(\hat{B}_{z_{v_1}}\hat{B}_{\bar{z}_{v_1}}\hat{\mathscr{A}}_a^{(z_1)}\big)-\int_{\Sigma_1} \rmd^2z_{v_1}\hat{B}_{z_{v_1}}\hat{B}_{\bar{z}_{v_1}}\big(Q_B^{(z_1)}\hat{\mathscr{A}}_a^{(z_1)}\big)=\\
&\qquad=\int_{\Sigma_1} \rmd^2z_{v_1}R_{z_{v_1}\bar{z}_{v_1}}\big(b_0^{(z_1)}-\tilde{b}_0^{(z_1)}\big)\hat{\mathscr{A}}_a^{(z_1)}+\int_{\Sigma_1} \rmd^2z_{v_1}\Big[\frac{\partial}{\partial \bar{z}_{v_1}}\big(\hat{B}_{z_{v_1}}\hat{\mathscr{A}}_a^{(z_1)}\big)-\frac{\partial}{\partial z_{v_1}}\big(\hat{B}_{\bar{z}_{v_1}}\hat{\mathscr{A}}_a^{(z_1)}\big)\Big].
\end{aligned}
\endgroup
\end{equation}
Notice that since we are using holomorphic normal coordinates we have from (\ref{eq:R}) and $g_{z_1\bar{z}_1}(z_{v_1})=1/2$ that $R_{(2)}(z_{v_1})=4R_{z_{v_1}\bar{z}_{v_1}}(z_{v_1})$. 
\sk

Also here, as in (\ref{eq:hatWA}), the result (\ref{eq:[B1B2,QB]2}) is most naturally written in terms of the $z_1$-frame (equivalently $z_{\sigma_1}$-frame and based at $\sigma=\sigma_1$) normal-ordered forms:
\begin{equation}
\boxed{
\begin{aligned}
\mathbb{W}_a^{(z_1)}&
= \frac{1}{i}\Big[\rmd\bar{z}_{v_1}\big(\hat{B}_{\bar{z}_{v_1}}\hat{\mathscr{A}}^{(z_1)}_a\big)
+\rmd z_{v_1}\big(\hat{B}_{z_{v_1}}\hat{\mathscr{A}}^{(z_1)}_a\big)\Big]\\
\mathbb{A}_a^{(z_1)}&
=\rmd^2z_{v_1}\big(\hat{B}_{z_{v_1}}\hat{B}_{\bar{z}_{v_1}}\hat{\mathscr{A}}_a^{(z_1)}
\big)\\
\mathbb{U}_a^{(z_1)}&=\rmd^2z_{v_1}\big[R_{z_{v_1}\bar{z}_{v_1}}\big(b_0^{(z_1)}-\tilde{b}_0^{(z_1)}\big)\hat{\mathscr{A}}_a^{(z_1)}\big]\\
\end{aligned}
}
\end{equation}
but now the hats, `$\hat{\phantom{a}}$', that appeared in (\ref{eq:hatWA}) are omitted which is meant to denote that these correspond to {\it integrated-picture} vertex operators, in that the location of the insertion $\hat{\mathscr{A}}_a^{(z_1)}$ is identified with a modulus and integrated over the Riemann surface.  In terms of these quantities the relation (\ref{eq:[B1B2,QB]2}) takes the form:
\begin{equation}\label{eq:[B1B2,QB]WAU}
\boxed{
\begin{aligned}
& Q_B^{(z_1)}\mathbb{A}_a^{(z_1)}= \rmd^2z_{v_1}\hat{B}_{z_{v_1}}\hat{B}_{\bar{z}_{v_1}}\big(Q_B^{(z_1)}\hat{\mathscr{A}}_a^{(z_1)}\big)+\mathbb{U}_a^{(z_1)}+ \rmd \mathbb{W}_a^{(z_1)}
\end{aligned}
}
\end{equation}
where we omitted the integral signs, $\int_{\Sigma_1}$, since this relation also holds at the level of the integrand. 
This is of course an operator equation and as such holds inside the path integral. 
Since the only moduli dependence of $\mathbb{W}_a^{(z_1)}$ is associated to variations, $\delta z_{v_1},\delta \bar{z}_{v_1}$, the total derivative appearing in (\ref{eq:[B1B2,QB]WAU}) reads: $\rmd= \rmd z_{v_1}\partial_{z_{v_1}}+\rmd\bar{z}_{v_1}\partial_{\bar{z}_{v_1}}$.
\sk

The right-hand side of (\ref{eq:[B1B2,QB]WAU}) must not contribute to amplitudes. In fact, the first term on the right-hand side is not yet in its final form since we need to specify whether the operator $\hat{\mathscr{A}}_a^{(z_1)}$ is an {\it external fixed-picture vertex operator} or whether it is part of a {\it handle operator}. In the latter case we can pull the BRST charge contour through the handle onto the $z_2$ patch using the transition function $z_1z_2=q$ and commuting the BRST charge through the various remaining local operators from the measure and also its dual, $\mathscr{A}^a_{(z_2)}$. (We discuss this in the following section in detail.) And we can then finally pull the BRST charge contour off the Riemann surface provided external vertex operators are BRST invariant and potential anomalies (tadpoles, mass renormalisation, etc.) are dealt with. 
\sk

If this integrated-picture operator $\mathbb{A}_a^{(z_1)}$ is not part of a handle operator and is rather associated to an {\it external vertex operator} then the quantity $\hat{\mathscr{A}}_a^{(z_1)}$ is identified with a {\it fixed-picture} external vertex operator. As such, if it is physical it should be BRST-invariant, 
\begin{equation}\label{eq:QBA=0}
Q_B^{(z_1)}\hat{\mathscr{A}}_a^{(z_1)}=0,
\end{equation}
and the first term on the right-hand side in (\ref{eq:[B1B2,QB]WAU}) vanishes. 
Furthermore, the $\mathbb{U}_a^{(z_1)}$ term on the right-hand side of (\ref{eq:[B1B2,QB]WAU}) must not contribute which is guaranteed if $\hat{\mathscr{A}}_a^{(z_1)}$ is annihilated by $b_0^{(z_1)}-\tilde{b}_0^{(z_1)}$, 
\begin{equation}\label{eq:b0b0tildeA=0}
\big(b_0^{(z_1)}-\tilde{b}_0^{(z_1)}\big)\hat{\mathscr{A}}_a^{(z_1)}=0,
\end{equation}
which is a good way of seeing {\it why} external closed-string fixed-picture vertex operators are required \cite{Polchinski88,Nelson89,Sen15b} to be annihilated by $b_0^{(z_1)}-\tilde{b}_0^{(z_1)}$. (This is discussed further in Sec.~\ref{sec:EVO}.) That the factor $b_0^{(z_1)}-\tilde{b}_0^{(z_1)}$ appears multiplied by the Ricci scalar (or tensor) is indicative of the global origin of this constraint. When the two equations (\ref{eq:b0b0tildeA=0}) and (\ref{eq:QBA=0}) hold, and making use of the identity $\{Q_B,b_0-\tilde{b}_0\}=L_0-\tilde{L}_0$, we learn that the following relation is automatic
\begin{equation}\label{eq:L0L0tildeA=0}
\big(L_0^{(z_1)}-\tilde{L}_0^{(z_1)}\big)\hat{\mathscr{A}}_a^{(z_1)}=0\qquad\Rightarrow \qquad \hat{\mathscr{A}}_a^{(z_1)}=\hat{\mathscr{A}}_a^{(e^{i\theta}z_1)},
\end{equation}
for any angle $\theta$. This in turn implies that we can also take an average: 
\begin{equation}\label{eq:AveragePhase}
\boxed{ \hat{\mathscr{A}}_a^{(z_1)}=\int_0^{2\pi}\frac{\rmd\theta}{2\pi}\hat{\mathscr{A}}_a^{(z_1/e^{i\theta})}}
\end{equation}
which explicitly removes the U(1) ambiguity. 
\sk

Let us add an isolated comment here, that when local operators, $\hat{\mathscr{A}}_a^{(z_1)}$, and $\hat{\mathscr{A}}^a_{(z_2)}$ are associated to handle operators that implement a {\it trivial-homology cycle} we can always average over the corresponding U(1) phase as in (\ref{eq:AveragePhase}), because in this case we always identify this phase with a modulus and furthermore the domain of the moduli space integration is always such that only states satisfying (\ref{eq:L0L0tildeA=0}) propagate. We will make use of this in Sec.~\ref{sec:VSG} where we glue two three-point amplitudes to construct a four-point amplitude.
\sk

If  $\mathscr{A}_a^{(z_1)}$ is part of a handle operator (associated to either a trivial or a non-trivial homology cycle) the $\mathbb{U}_a^{(z_1)}$ term will not contribute if the phase of the local coordinate is identified with a modulus and taken to be integrated. Because if it is integrated then there is an additional factor $b_0^{(z_2)}-\tilde{b}_0^{(z_2)}$ from the measure (associated to identifying the phase with a modulus), recall (\ref{eq:Bintpinchtheta}), and then the $\mathbb{U}_a^{(z_1)}$ term in (\ref{eq:[B1B2,QB]WAU}) vanishes because $(b_0^{(z_1)}-\tilde{b}_0^{(z_1)})(b_0^{(z_2)}-\tilde{b}_0^{(z_2)})=0$.
\sk

The $\rmd \mathbb{W}_a^{(z_1)}$ term in  (\ref{eq:[B1B2,QB]WAU}) contributes an integral of a total derivative, as expected on general grounds, and will decouple from amplitudes up to possible boundary terms in moduli space that must be arranged to cancel (by shifting the background, by mass renormalisation, etc.). We emphasise that the resulting derivatives are {\it outside} the normal ordering, a crucial point if one is to use Stoke's theorem to integrate by parts and hence localise the potential anomaly onto the boundary of moduli space. In particular, our choice of using holomorphic normal coordinates \cite{Polchinski88} ensures there are no `Wu-Yang' type terms on patch overlaps (of the type encountered in the Appendix of \cite{Polchinski87} and also \cite{Nelson89}), see also Sec.~\ref{sec:WuYang} for further detail.
\sk

Let us briefly note that we can also rewrite (\ref{eq:[B1B2,QB]2}) as follows,\footnote{In the published version of this article there is a typo in (\ref{eq:[B1B2,QB]3}), namely there should be no factors of $R_{(2)}$ on the left-hand side of this equality (corrected in the current document).}
\begin{equation}\label{eq:[B1B2,QB]3}
\begingroup\makeatletter\def\f@size{11}\check@mathfonts
\def\maketag@@@#1{\hbox{\m@th\large\normalfont#1}}%
\begin{aligned}
&\int_{\Sigma_1} \rmd^2\sigma_12\sqrt{g}Q_B^{(z_{\sigma_1})}\big(\hat{B}_{z_{v_1}}\hat{B}_{\bar{z}_{v_1}}\hat{\mathscr{A}}_a^{(z_{\sigma_1})}\big)(\sigma_1)-\int_{\Sigma_1} \rmd^2\sigma_12\sqrt{g}\hat{B}_{z_{v_1}}\hat{B}_{\bar{z}_{v_1}}\big(Q_B^{(z_{\sigma_1})}\hat{\mathscr{A}}_a^{(z_{\sigma_1})}\big)(\sigma_1)\\
&\qquad=\frac{1}{4}\int_{\Sigma_1} \rmd^2\sigma_12\sqrt{g}R_{(2)}\big(b_0^{(z_{\sigma_1})}-\tilde{b}_0^{(z_{\sigma_1})}\big)\hat{\mathscr{A}}_a^{(z_{\sigma_1})}(\sigma_1)+\int_{\Sigma_1}\rmd \mathbb{W}_a^{(z_1)}\\
\end{aligned}
\endgroup
\end{equation}
where we made use of (\ref{eq:dzsigma*-measure}) and (\ref{eq:dzv=ds1s1}). This makes reparametrisation invariance manifest: recall that the frame coordinate $z_{\sigma_1}(\sigma_1)$ transforms as a scalar under reparametrisations, $\sigma_1\mapsto\hat{\sigma}_1(\sigma_1)$, (at least modulo a local phase) and that the operator $\hat{\mathscr{A}}_a^{(z_{\sigma_1})}(\sigma_1)$ is normal-ordered in the frame $z_{\sigma_1}(\sigma)$ and evaluated at $\sigma=\sigma_1$.
\sk

Since we are incorporating the effects of arbitrary worldsheet curvature (that the ``shape'' of the pinch changes as it is translated across the surface), and since the ill-defined U(1) phase is integrated, and since reparametrisation invariance is manifest, the resulting integrated vertex operator (whether it is identified with an asymptotic state or part of a handle operator) is globally well-defined in moduli space.
\sk

In the following section we derive the corresponding expression for the obstruction to BRST-exact decoupling associated to inserting a handle operator (to which we attach {\it three} complex moduli). Also there the conclusion will be that the resulting handle operators are globally well-defined in moduli space.

\subsection{Integrated-Picture Handle Operators with Pinch \& Twists}
\begin{figure}
\begin{center}
\includegraphics[angle=0,origin=c,width=0.4\textwidth]{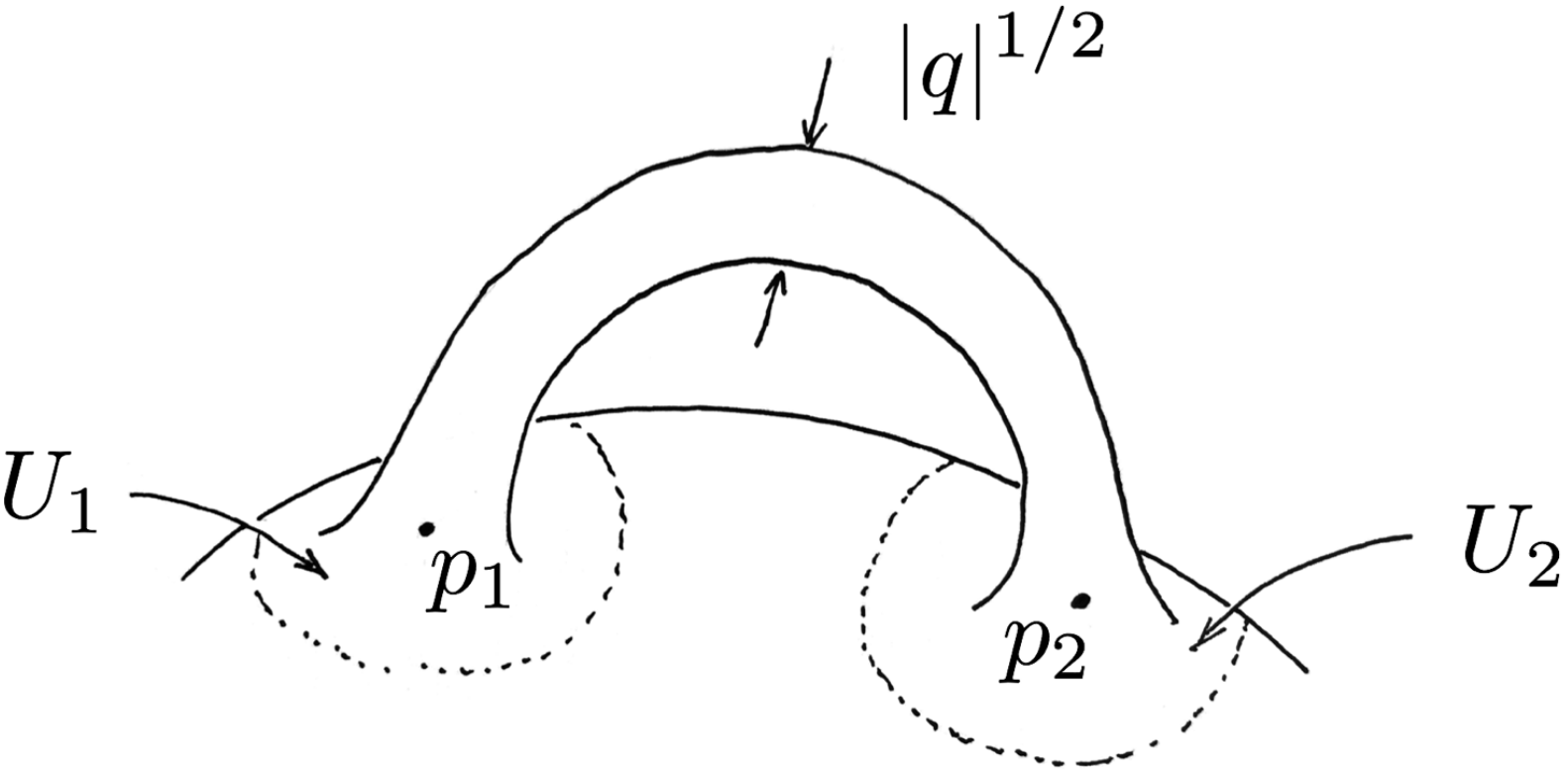}
\caption{Pictorial representation of a handle operator to which we associated three complex moduli, corresponding to the locations, $p_1,p_2$, of the two insertions and a corresponding complex modulus, $q$, for the pinch/twist of the corresponding handle.}\label{fig:3mod}
\end{center}
\end{figure}

There is a similar story for the commutator of the BRST charge with a handle operator to which we wish to associate three complex moduli, associated to shifting the two local operators (with corresponding moduli variations, $\delta z_{v_1}$ and $\delta z_{v_2}$) and also pinching/twisting the handle (with corresponding moduli variations, $\delta q$). The corresponding handle operator reads:
\begin{equation}\label{eq:handleoperatorX4}
\boxed{
\hat{H}=\int \rmd^2z_{v_1}\int \rmd^2z_{v_2}\int \rmd^2q\suminnt\limits_{a}\,\,\big[\hat{B}_{z_{v_1}}\hat{B}_{\bar{z}_{v_1}}\hat{\mathscr{A}}_a^{(z_1)}(p_1)\big]\,\big[\hat{B}_q\hat{B}_{\bar{q}}\hat{B}_{z_{v_2}}\hat{B}_{\bar{z}_{v_2}}\hat{\mathscr{A}}^a_{(z_2/q)}(p_2)\big]
}
\end{equation}
The measure $\rmd^2z_{v_1}$ (and similarly for $z_{v_2}$) might best be thought of as shorthand for the more explicitly reparametrisation-invariant measure: $ \rmd^2\sigma_12\sqrt{g(\sigma_1)}$. Recall (\ref{eq:dzsigma*-measure}) and (\ref{eq:dzv=ds1s1}). 
A corresponding sketch is shown in Fig.~\ref{fig:3mod} (for the case of cutting across a non-trivial homology cycle, but the expression (\ref{eq:handleoperatorX4}) holds also for trivial-homology cycles). 
So we would like the analogue of (\ref{eq:gluing Qbosonic2withbbNSD}) for this `handle operator'. 
The relevant term can be evaluated and one finds,\footnote{In the published version of this article in equation (\ref{eq:handleoperatorX5z}), the interpretation of the initial contours in $Q_B^{(z_1)}$ and $Q_B^{(z_2)}$ is not the relevant one for the study of gauge invariance, and consequently the subsequent analysis in that section also needs modifying. XXX Eqns (6.490), (6.494), (6.496), (6.497), (6.498) and (6.499) have all been changed, check references to these.}
\begin{equation}\label{eq:handleoperatorX5z}
\begingroup\makeatletter\def\f@size{11}\check@mathfonts
\def\maketag@@@#1{\hbox{\m@th\large\normalfont#1}}%
\begin{aligned}
\Big(Q_B^{(z_1)}&+Q_B^{(z_2)}\Big)\suminnt\limits_{a}\,\,\Big(\hat{B}_{z_{v_1}}\hat{B}_{\bar{z}_{v_1}}\hat{\mathscr{A}}_a^{(z_1)}\Big)\,\Big(\hat{B}_q\hat{B}_{\bar{q}}\hat{B}_{z_{v_2}}\hat{B}_{\bar{z}_{v_2}}\hat{\mathscr{A}}^a_{(z_2/q)}\Big)=\\
&=\suminnt\limits_a\,\,\bigg\{\Big(\Big[Q_B^{(z_1)},\hat{B}_{z_{v_1}}\hat{B}_{\bar{z}_{v_1}}\Big]\hat{\mathscr{A}}_a^{(z_1)}\Big)\,\Big(\hat{B}_q\hat{B}_{\bar{q}}\hat{B}_{z_{v_2}}\hat{B}_{\bar{z}_{v_2}}\hat{\mathscr{A}}^a_{(z_2/q)}\Big)\\
&\qquad\quad +\Big(\hat{B}_{z_{v_1}}\hat{B}_{\bar{z}_{v_1}}\hat{\mathscr{A}}_a^{(z_1)}\Big)\Big(\Big[Q_B^{(z_2)},\hat{B}_q\hat{B}_{\bar{q}}\Big]\hat{B}_{z_{v_2}}\hat{B}_{\bar{z}_{v_2}}\hat{\mathscr{A}}^a_{(z_2/q)}+\Big[Q_B^{(z_2)},\hat{B}_{z_{v_2}}\hat{B}_{\bar{z}_{v_2}}\Big]\hat{B}_q\hat{B}_{\bar{q}}\hat{\mathscr{A}}^a_{(z_2/q)}\Big)\bigg\}\\
\end{aligned}
\endgroup
\end{equation}
We made use of the relation, 
$$
\Big[\hat{B}_q\hat{B}_{\bar{q}},\big[Q_B^{(z_2)},\hat{B}_{z_{v_2}}\hat{B}_{\bar{z}_{v_2}}\big]\Big]=0
$$
as well as,
$$
\suminnt\limits_{a}\,\,\big[Q_B^{(z_1)}\hat{\mathscr{A}}_a^{(z_1)}\big]\hat{\mathscr{A}}^a_{(z_2/q)}=-\suminnt\limits_{a}\,\,\hat{\mathscr{A}}_a^{(z_1)}\big[Q_B^{(z_2)}\hat{\mathscr{A}}^a_{(z_2/q)}\big].
$$
The remaining commutators on the right-hand side in (\ref{eq:handleoperatorX5z}) are also easily evaluated (taking into account the results of the previous subsections) and we find:
\begin{equation}\label{eq:[B1B2,QB]2b}
\begingroup\makeatletter\def\f@size{11}\check@mathfonts
\def\maketag@@@#1{\hbox{\m@th\large\normalfont#1}}%
\begin{aligned}
\big[Q_B^{(z_1)},\hat{B}_{z_{v_1}}\hat{B}_{\bar{z}_{v_1}}\big]\hat{\mathscr{A}}_a^{(z_1)}&=\frac{\partial}{\partial \bar{z}_{v_1}}\big(\hat{B}_{z_{v_1}}\hat{\mathscr{A}}_a^{(z_1)}\big)-\frac{\partial}{\partial z_{v_1}}\big(\hat{B}_{\bar{z}_{v_1}}\hat{\mathscr{A}}_a^{(z_1)}\big)+
\frac{1}{4}R_{(2)}(z_{v_1})\big(b_0^{(z_1)}-\tilde{b}_0^{(z_1)}\big)\hat{\mathscr{A}}_a^{(z_1)}\\
\big[Q_B^{(z_2)},\hat{B}_{z_{v_2}}\hat{B}_{\bar{z}_{v_2}}\big]\hat{B}_q\hat{B}_{\bar{q}}\hat{\mathscr{A}}^a_{(z_2/q)}&=
\frac{\partial}{\partial \bar{z}_{v_2}}\big(\hat{B}_q\hat{B}_{\bar{q}}\hat{B}_{z_{v_2}}\hat{\mathscr{A}}^a_{(z_2/q)}\big)-\frac{\partial}{\partial z_{v_2}}\big(\hat{B}_q\hat{B}_{\bar{q}}\hat{B}_{\bar{z}_{v_2}}\hat{\mathscr{A}}^a_{(z_2/q)}\big)\\
\big[Q_B^{(z_2)},\hat{B}_q\hat{B}_{\bar{q}}\big]\hat{B}_{z_{v_2}}\hat{B}_{\bar{z}_{v_2}}\hat{\mathscr{A}}^a_{(z_2/q)}&=\frac{\partial}{\partial \bar{q}}\big(\hat{B}_q\hat{B}_{z_{v_2}}\hat{B}_{\bar{z}_{v_2}}\hat{\mathscr{A}}^a_{(z_2/q)}\big)-\frac{\partial}{\partial q}\big(\hat{B}_{\bar{q}}\hat{B}_{z_{v_2}}\hat{B}_{\bar{z}_{v_2}}\hat{\mathscr{A}}^a_{(z_2/q)}\big)
\end{aligned}
\endgroup
\end{equation}

Perhaps the last relation in (\ref{eq:[B1B2,QB]2b}) deserves some elaboration. Let us go through the various steps. 
Making use of the commutator,
$$
\big[Q_B^{(z_2)},\hat{B}_{q}\hat{B}_{\bar{q}}\big]=\frac{1}{q\bar{q}}\big(\tilde{b}_0^{(z_2)}L_0^{(z_2)}-b_0^{(z_2)}\tilde{L}_0^{(z_2)}\big),
$$
as well as the explicit expressions for $\hat{B}_{z_{v_2}},\hat{B}_{\bar{z}_{v_2}}$ and also $\hat{B}_q,\hat{B}_{\bar{q}}$, we learn that:
\begin{equation}\label{eq:QBBqBz2A}
\begin{aligned}
&\big[Q_B^{(z_2)},\hat{B}_q\hat{B}_{\bar{q}}\big]\hat{B}_{z_{v_2}}\hat{B}_{\bar{z}_{v_2}}\hat{\mathscr{A}}^a_{(z_2/q)}\\
&=\frac{1}{q\bar{q}}\big(\tilde{b}_0L_0-b_0\tilde{L}_0\big)
\big(-b_{-1}^{(z_2/q)}-\sum_{n=1}^\infty\bar{\epsilon}_n\tilde{b}_n^{(z_2/q)} \big)\big(-\tilde{b}_{-1}^{(z_2/q)}-\sum_{n=1}^\infty\epsilon_nb_n^{(z_2/q)} \big)\hat{\mathscr{A}}^a_{(z_2/q)}\\
&=\frac{1}{q\bar{q}}\big(\tilde{b}_0L_0-b_0\tilde{L}_0\big)
\big(qb_{-1}^{(z_2)}+\sum_{n=1}^\infty\bar{\epsilon}_n\bar{q}^{-n}\tilde{b}_n^{(z_2)} \big)\big(\bar{q}\tilde{b}_{-1}^{(z_2)}+\sum_{n=1}^\infty\epsilon_nq^{-n}b_n^{(z_2)} \big)q^{L_0}\bar{q}^{\tilde{L}_0}\hat{\mathscr{A}}^a_{(z_2)}\\
&=\frac{1}{q\bar{q}}\big(\tilde{b}_0L_0-b_0\tilde{L}_0\big)q^{L_0}\bar{q}^{\tilde{L}_0}
\big(b_{-1}^{(z_2)}+\sum_{n=1}^\infty\bar{\epsilon}_n\tilde{b}_n^{(z_2)} \big)\big(\tilde{b}_{-1}^{(z_2)}+\sum_{n=1}^\infty\epsilon_nb_n^{(z_2)} \big)\hat{\mathscr{A}}^a_{(z_2)}\\
\end{aligned}
\end{equation}
where in the second equality we extracted out the $q,\bar{q}$ dependence by making use of the fact that (independently of whether $\mathscr{A}_{(z_2)}^a$ is an eigenstate of $L_0,\tilde{L}_0$): 
$$
\hat{\mathscr{A}}^a_{(z_2/q)}=q^{L_0}\bar{q}^{\tilde{L}_0}\hat{\mathscr{A}}^a_{(z_2)},\qquad {\rm and}\qquad b_n^{(z_2/q)} = q^{-n}b_n^{(z_2)},
$$
whereas in the third equality we commuted the factor $q^{L_0}\bar{q}^{\tilde{L}_0}$ through to the left using $[L_0,b_n]=-nb_n$ and $[\tilde{L}_0,\tilde{b}_n]=-n\tilde{b}_n$. 
The quantities $\epsilon_n$ and $\bar{\epsilon}_n$ are local functions, but the relevant observation here is that they do not depend on $q,\bar{q}$. Combining this observation with the following,
$$
\frac{1}{q\bar{q}}\big(\tilde{b}_0L_0-b_0\tilde{L}_0\big)q^{L_0}\bar{q}^{\tilde{L}_0}
=\frac{\partial}{\partial \bar{q}}\big(\hat{B}_qq^{L_0}\bar{q}^{\tilde{L}_0}\big)-\frac{\partial}{\partial q}\big(\hat{B}_{\bar{q}}q^{L_0}\bar{q}^{\tilde{L}_0}\big),
$$
the right-hand side of the third equality in (\ref{eq:QBBqBz2A}) can be rewritten in terms of $q,\bar{q}$ derivatives with the result:
\begin{equation}\label{eq:QBBqBz2A2}
\begin{aligned}
&\big[Q_B^{(z_2)},\hat{B}_q\hat{B}_{\bar{q}}\big]\hat{B}_{z_{v_2}}\hat{B}_{\bar{z}_{v_2}}\hat{\mathscr{A}}^a_{(z_2/q)}
=\frac{\partial}{\partial \bar{q}}\big(\hat{B}_q\hat{B}_{z_{v_2}}\hat{B}_{\bar{z}_{v_2}}\hat{\mathscr{A}}^a_{(z_2/q)}\big)-\frac{\partial}{\partial q}\big(\hat{B}_{\bar{q}}\hat{B}_{z_{v_2}}\hat{B}_{\bar{z}_{v_2}}\hat{\mathscr{A}}^a_{(z_2/q)}\big),
\end{aligned}
\end{equation}
which is precisely the last equality in (\ref{eq:[B1B2,QB]2b}).
\sk

Let us now substitute the results (\ref{eq:[B1B2,QB]2b}) into (\ref{eq:handleoperatorX5z}), which yields:
\begin{equation}\label{eq:handleoperatorX5z2}
\begingroup\makeatletter\def\f@size{11}\check@mathfonts
\def\maketag@@@#1{\hbox{\m@th\large\normalfont#1}}%
\begin{aligned}
\Big(Q_B^{(z_1)}&+Q_B^{(z_2)}\Big)\suminnt\limits_{a}\,\,\Big(\hat{B}_{z_{v_1}}\hat{B}_{\bar{z}_{v_1}}\hat{\mathscr{A}}_a^{(z_1)}\Big)\otimes\Big(\hat{B}_q\hat{B}_{\bar{q}}\hat{B}_{z_{v_2}}\hat{B}_{\bar{z}_{v_2}}\hat{\mathscr{A}}^a_{(z_2/q)}\Big)=\\
&=\suminnt\limits_a\,\,\Bigg\{\bigg[\frac{\partial}{\partial \bar{z}_{v_1}}\Big(\hat{B}_{z_{v_1}}\hat{\mathscr{A}}_a^{(z_1)}\Big)-\frac{\partial}{\partial z_{v_1}}\Big(\hat{B}_{\bar{z}_{v_1}}\hat{\mathscr{A}}_a^{(z_1)}\Big)\bigg]\otimes\Big(\hat{B}_q\hat{B}_{\bar{q}}\hat{B}_{z_{v_2}}\hat{B}_{\bar{z}_{v_2}}\hat{\mathscr{A}}^a_{(z_2/q)}\Big)\\
&\qquad\quad+
\Big(\frac{1}{4}R_{(2)}(z_{v_1})\big(b_0^{(z_1)}-\tilde{b}_0^{(z_1)}\big)\hat{\mathscr{A}}_a^{(z_1)}\Big)\otimes\Big(\hat{B}_q\hat{B}_{\bar{q}}\hat{B}_{z_{v_2}}\hat{B}_{\bar{z}_{v_2}}\hat{\mathscr{A}}^a_{(z_2/q)}\Big)\\
&\qquad\quad +\Big(\hat{B}_{z_{v_1}}\hat{B}_{\bar{z}_{v_1}}\hat{\mathscr{A}}_a^{(z_1)}\Big)\otimes\bigg[\frac{\partial}{\partial \bar{q}}\Big(\hat{B}_q\hat{B}_{z_{v_2}}\hat{B}_{\bar{z}_{v_2}}\hat{\mathscr{A}}^a_{(z_2/q)}\Big)-\frac{\partial}{\partial q}\Big(\hat{B}_{\bar{q}}\hat{B}_{z_{v_2}}\hat{B}_{\bar{z}_{v_2}}\hat{\mathscr{A}}^a_{(z_2/q)}\Big)\\
&\qquad\qquad\qquad\qquad+\frac{\partial}{\partial \bar{z}_{v_2}}\Big(\hat{B}_q\hat{B}_{\bar{q}}\hat{B}_{z_{v_2}}\hat{\mathscr{A}}^a_{(z_2/q)}\Big)-\frac{\partial}{\partial z_{v_2}}\Big(\hat{B}_q\hat{B}_{\bar{q}}\hat{B}_{\bar{z}_{v_2}}\hat{\mathscr{A}}^a_{(z_2/q)}\Big)\bigg]\Bigg\}\\
\end{aligned}
\endgroup
\end{equation}

To make manifest that the resulting relation on the right-hand side is really contributing boundary terms in the integral over moduli, and to identify these boundary contributions, recall that if we define, 
$
\rmd = \rmd x^1\partial_1+\dots +\rmd x^\m\partial_\m,
$ 
and an $\m-1$ form,
$$
W = \sum_{k=1}^\m A_{1\dots \,k\!\!\!/\,\dots\m}\rmd x^1\wedge \dots \rmd x^k\!\!\!\!\!\slash\,\,\, \wedge\dots \wedge \rmd x^\m,
$$
then $\rmd W$ is the following $\m$ form,
$$
\rmd W = \rmd x^1\wedge  \dots \wedge \rmd x^\m\Big( \sum_{k=1}^\m(-)^{k-1}\partial_kA_{1\dots \,k\!\!\!/\,\dots\m}\Big).
$$
Let us apply this to the case of interest (\ref{eq:handleoperatorX5z2}). 
We define the following quantities (the first three of which were defined above and repeated here for convenience),
\begin{equation}\label{eq:WAUWA}
\boxed{
\begin{aligned}
\mathbb{W}_a^{(z_1)}&
= \frac{1}{i}\Big[\rmd\bar{z}_{v_1}\big(\hat{B}_{\bar{z}_{v_1}}\hat{\mathscr{A}}^{(z_1)}_a\big)
+\rmd z_{v_1}\big(\hat{B}_{z_{v_1}}\hat{\mathscr{A}}^{(z_1)}_a\big)\Big]\\
\mathbb{A}_a^{(z_1)}&
=\rmd^2z_{v_1}\big(\hat{B}_{z_{v_1}}\hat{B}_{\bar{z}_{v_1}}\hat{\mathscr{A}}_a^{(z_1)}
\big)\\
\mathbb{U}_a^{(z_1)}&=\rmd^2z_{v_1}\big[R_{z_{v_1}\bar{z}_{v_1}}\big(b_0^{(z_1)}-\tilde{b}_0^{(z_1)}\big)\hat{\mathscr{A}}_a^{(z_1)}\big]\\
\mathbb{W}^a_{(z_2)}&
=\frac{1}{i}\Big[\rmd\bar{q}\,\rmd^2z_{v_2}\big(q^{L_0}\bar{q}^{\tilde{L}_0}\hat{B}_{\bar{q}}\hat{B}_{z_{v_2}}\hat{B}_{\bar{z}_{v_2}}\hat{\mathscr{A}}^a_{(z_2)}\big)
+dq\,\rmd^2z_{v_2}\big(q^{L_0}\bar{q}^{\tilde{L}_0}\hat{B}_q\hat{B}_{z_{v_2}}\hat{B}_{\bar{z}_{v_2}}\hat{\mathscr{A}}^a_{(z_2)}\big)\\
&\quad+\rmd^2q\, \rmd\bar{z}_{v_2}\big(q^{L_0}\bar{q}^{\tilde{L}_0}\hat{B}_q\hat{B}_{\bar{q}}\hat{B}_{\bar{z}_{v_2}}\hat{\mathscr{A}}^a_{(z_2)}\big)
+\rmd^2q \,\rmd z_{v_2}\big(q^{L_0}\bar{q}^{\tilde{L}_0}\hat{B}_q\hat{B}_{\bar{q}}\hat{B}_{z_{v_2}}\hat{\mathscr{A}}^a_{(z_2)}\big)\Big]\\
\mathbb{A}^a_{(z_2)}&
=\rmd^2q \,\rmd^2z_{v_2}\big(q^{L_0}\bar{q}^{\tilde{L}_0}\hat{B}_q\hat{B}_{\bar{q}}\hat{B}_{z_{v_2}}\hat{B}_{\bar{z}_{v_2}}\hat{\mathscr{A}}^a_{(z_2)}\big)\\
\end{aligned}
}
\end{equation}
in terms of which the relation (\ref{eq:handleoperatorX5z2}) can be written as follows. Including also the measure contribution, $\rmd^2z_{v_1}\rmd^2q\, \rmd^2z_{v_2}$, the relation (\ref{eq:handleoperatorX5z2}) which gives the obstruction to the decoupling of BRST-exact terms generated by inserting a handle operator, $\Sigma_a\!\!\!\!\!\!\!\int\,\,
\mathbb{A}_a^{(z_1)}\mathbb{A}^a_{(z_2)}$, takes the form:
\begin{equation}\label{eq:handleoperatorX5z2xx}
\begin{aligned}
\big(Q_B^{(z_1)}+Q_B^{(z_2)}\big)\suminnt\limits_{a}\,\,
\mathbb{A}_a^{(z_1)}\mathbb{A}^a_{(z_2)}&
=\suminnt\limits_{a}\,\,
\Big[\big(\mathbb{U}_a^{(z_1)}+\rmd \mathbb{W}_a^{(z_1)}\big)\mathbb{A}^a_{(z_2)}+\mathbb{A}_a^{(z_1)}\rmd \mathbb{W}^a_{(z_2)}\Big].
\end{aligned}
\end{equation}
The quantity $\rmd$ appearing is an exterior derivative in moduli space, 
$$
\rmd = \rmd\tau^k\partial_{\tau^k}+\rmd\bar{\tau}\partial_{\bar{\tau}^k},
$$
whose explicit realisation simplifies depending on what it acts on. For example, when acting on $\mathbb{W}_a^{(z_1)}$ (which only depends on the moduli $z_{v_1},\bar{z}_{v_1}$) it reduces to 
$
\rmd = \rmd z_{v_1}\partial_{z_{v_1}}+\rmd\bar{z}_{v_1}\partial_{\bar{z}_{v_1}},
$ 
whereas when it acts on $\mathbb{W}^a_{(z_2)}$ (which only depends on the moduli $q,\bar{q},z_{v_2},\bar{z}_{v_2}$) it reduces to 
$
\rmd = dq\,\partial_q+\rmd\bar{q}\,\partial_{\bar{q}}+\rmd z_{v_2}\partial_{z_{v_2}}+\rmd\bar{z}_{v_2}\partial_{\bar{z}_{v_2}}.
$ 
\sk

Notice that all terms on the right-hand side in (\ref{eq:handleoperatorX5z2xx}) are total derivatives, and so contribute at most boundary terms when integrated, {\it except} for the following:
$$
\mathbb{U}_a^{(z_1)}\mathbb{A}^a_{(z_2)}.
$$
So we must either show that this term vanishes, or that it equals a total derivative. In fact it vanishes since it is proportional to:
\begin{equation}\label{eq:Ub0b0QA}
\begin{aligned}
\big(b_0^{(z_1)}-&\tilde{b}_0^{(z_1)}\big)\hat{\mathscr{A}}_a^{(z_1)}\big(q^{L_0}\bar{q}^{\tilde{L}_0}\hat{B}_q\hat{B}_{\bar{q}}\hat{B}_{z_{v_2}}\hat{B}_{\bar{z}_{v_2}}\hat{\mathscr{A}}^a_{(z_2)}\big)\\
&=\hat{\mathscr{A}}_a^{(z_1)}\big[q^{L_0}\bar{q}^{\tilde{L}_0}\hat{B}_q\hat{B}_{\bar{q}}\hat{B}_{z_{v_2}}\hat{B}_{\bar{z}_{v_2}}\big(b_0^{(z_2)}-\tilde{b}_0^{(z_2)}\big)\hat{\mathscr{A}}^a_{(z_2)}\big]\\
&=\hat{\mathscr{A}}_a^{(z_1)}\big[q^{L_0}\bar{q}^{\tilde{L}_0}\frac{b_0^{(z_2)}}{q}\frac{\tilde{b}_0^{(z_2)}}{\bar{q}}\big(b_0^{(z_2)}-\tilde{b}_0^{(z_2)}\big)\hat{B}_{z_{v_2}}\hat{B}_{\bar{z}_{v_2}}\hat{\mathscr{A}}^a_{(z_2)}\big]\\
&=0,
\end{aligned}
\end{equation}
where in the first equality we followed similar reasoning as that leading from (\ref{eq:QABqBqA}) to (\ref{eq:QABqBqA2}), in particular the contours of the $b_0,\tilde{b}_0$ modes first encounter $\mathscr{A}^a_{(z_2)}$ as they are passed through the handle, and subsequently $\hat{B}_{\bar{z}_{v_2}}$, then $\hat{B}_{z_{v_2}}$, and $\hat{B}_{\bar{q}}$ and finally $\hat{B}_{q}$. Since the latter two are proportional to $\tilde{b}_0$ and $b_0$ respectively, which are Grassmann-odd, the result vanishes. (Recall that on chart overlaps, $b_0^{(z_1)}=b_0^{(z_2)}$, so that the change in contour orientation is compensated by the change of frame variables.)
\sk

Taking this into account, we obtain the following result for the contribution from handle operators when considering the decoupling of BRST-exact insertions. The most useful form for {\it non-separating degenerations} is:
\begin{equation}\label{eq:QhandleoperatorNSD}
\boxed{
\begin{aligned}
\big(Q_B^{(z_1)}+Q_B^{(z_2)}\big)\suminnt\limits_{a}\,\,
\mathbb{A}_a^{(z_1)}\mathbb{A}^a_{(z_2)}
&=\suminnt\limits_{a}\,\,
\Big[\rmd \mathbb{W}_a^{(z_1)}\mathbb{A}^a_{(z_2)}+\mathbb{A}_a^{(z_1)}\rmd \mathbb{W}^a_{(z_2)}\Big]\\
\end{aligned}
}
\end{equation}
whereas the most useful form for {\it separating degenerations} is:
\begin{equation}\label{eq:QhandleoperatorSD}
\boxed{
\begin{aligned}
\suminnt\limits_{a}\,\,
\big(Q_B^{(z_1)}\mathbb{A}_a^{(z_1)}\big)\mathbb{A}^a_{(z_2)}
&=\suminnt\limits_{a}\,\,
\Big[\rmd \mathbb{W}_a^{(z_1)}\mathbb{A}^a_{(z_2)}+\mathbb{A}_a^{(z_1)}\rmd \mathbb{W}^a_{(z_2)}
-\mathbb{A}_a^{(z_1)}\,Q_B^{(z_2)}\mathbb{A}^a_{(z_2)}\Big]\\
\end{aligned}
}
\end{equation}
In the latter case, namely (\ref{eq:QhandleoperatorSD}), if $Q_B^{(z_1)}\mathbb{A}_a^{(z_1)}$ is on Riemann surface $\Sigma_1$ the quantity $Q_B^{(z_2)}\mathbb{A}^a_{(z_2)}$ on the right-hand side is inserted on Riemann surface $\Sigma_2$ (when the handle operator glues $\Sigma_1$ to $\Sigma_2$). The BRST-charge contour can be freely pulled off the operator $\mathbb{A}^a_{(z_2)}$ and we can apply the results we have already obtained in this and previous subsections to commute it through any remaining insertions on $\Sigma_2$. As we have shown this procedure will either result in further boundary contributions in moduli space, or it will give explicitly vanishing contribution.
\sk

So combining all these observations, the commutator of the full handle operator with the BRST charge indeed gives rise to a {\it total} derivative in moduli space of a local normal-ordered quantity. This remains true for all moduli since the above derivation applies also to external vertex operators, and the remaining moduli can all be associated to handle operators with the gauge slice chosen as above. In particular, if there are multiple handles the BRST charge can be passed through each one in precisely the same manner as above giving rise to corresponding boundary terms. This is because the gauge slice we have chosen (in a vague sense) ``decouples moduli as much as possible'', by allowing one to vary a subset keeping the remaining moduli fixed. In particular, it is only via the corresponding integration {\it domains} that they remain coupled. (These domains are in turn determined by requiring the resulting amplitudes be modular invariant; if we were to instead integrate over Teichm\"uller space, $\mathrm{T}_{\g,\n}$,  also the integration domains of the various handle operator moduli space integrals would be decoupled.)
\sk

The various possible boundary contributions and in particular their {\it cancellation} constitutes a large and interesting subject in its own right and we will not be able to delve deeper into this in the current document. 

\section{Explicit Handle Operators}\label{sec:EHOC}
In this section we derive an explicit expression for handle operators, using an offshell basis for string coherent states (and their duals) from defining Riemann surface data and general CFT factorisation properties. In particular, we will derive an explicit expression for a complete set of states propagating through a generic (homologically trivial or non-trivial) cycle of a genus-$\g$ Riemann surface with any set of asymptotic vertex operator insertions. When we restrict to the BRST-invariant subspace and mod out by BRST-exact contributions the resulting vertex operators can (up to normalisation that will be discussed in detail) also be interpreted as asymptotic states for string amplitudes. More generally, the coherent states we construct can also be inserted into amplitudes as {\it offshell} external state vertex operators. In the latter case there will be an explicit coordinate dependence, and inserting an external coherent state using different local coordinate systems leads to different amplitudes. Thankfully, these seemingly different amplitudes are expected to be related by field redefinitions in the corresponding field theory\footnote{We thank Ashoke Sen for a discussion along these lines.}, so this removes the apparent ambiguity. In fact, the vertex operators we construct also form a natural and explicit basis for closed string fields when we project onto the level-matched contributions. 

\subsection{The Derivation -- Fixed Complex Structure}\label{sec:GOCS}

So we now zoom in to arrive at explicit expressions for the operators, $\hat{\mathscr{A}}_a^{(z_1)}$ and $\hat{\mathscr{A}}^a_{(z_2)}$, that arise in (\ref{eq:completeness-sep BBB}) or (\ref{eq:completeness_nonsep BBB}) in the specific case of bosonic string theory in flat spacetime. It is best however to begin from the fixed-complex structure expressions  (\ref{eq:completeness-sep}) and (\ref{eq:completeness_nonsep}). 
The operators $\hat{\mathscr{A}}_a^{(z_1)}$ and $\hat{\mathscr{A}}^a_{(z_2)}$ will in general be constructed out of the elementary fields of the theory, $x(z,\bar{z})$, $b(z)$ and $c(z)$, and derivatives thereof. The chiral half of the BRST current, $j_B(z)$, that of the total stress tensor, $T_{zz}$, and the elementary fields $\sqrt{\frac{2}{\alpha'}}i\partial_z x(z)$, $b(z)$ and $c(z)$ are primaries of fixed conformal weights, $h=1,2,1,2$ and $-1$ respectively, so consider one such primary, $\mathscr{O}$, of weight $h$. In the local $z_1$ and $z_2$ frames respectively (see Appendix \ref{sec:PMO}) it has mode expansions:
\begin{equation}
\mathscr{O}^{(z_1)}(\e{z}_1) = \sum_{n}\frac{\mathscr{O}_n^{(z_1)}}{\e{z}_1^{n+h}},\qquad {\rm and}\qquad \mathscr{O}^{(z_2)}(\e{z}_2) = \sum_{n}\frac{\mathscr{O}_n^{(z_2)}}{\e{z}_2^{n+h}},
\end{equation}
where $n$ spans integers or half integers depending on context, and\footnote{Regarding notation, we can make manifest the auxiliary coordinates, $\sigma^a$, $a=1,2$, by using the correspondence (\ref{eq:z1pzs1 corr}). Some further comments on notation and conventions that are useful here are provided in Appendix \ref{sec:CFTC}.} $\e{z}_1\equiv z_1(p)$, where $p\in U_1\cap U_2$ is the point where the field is evaluated and $z_1$ is the coordinate chart used to define it. (Similar remarks hold for $\e{z}_2\equiv z_2(p)$, where notice that the corresponding local operator, $\mathscr{O}^{(z_2)}(\e{z}_2)$, is evaluated at the same point $p\in U_1\cap U_2$ but in the $z_2$ chart coordinates.) 
In order to relate the modes $\mathscr{O}_n^{(z_1)}$ in the $z_1$ frame to corresponding modes $\mathscr{O}_n^{(z_2)}$ in the $z_2$ frame we recall that on patch overlaps we identify points $z_1$ and $z_2$ when they are related by the gluing relation (or transition function), 
$$
\phantom{\qquad\qquad \textrm{(for $p\in U_1\cap U_2\neq\zero$)}}
z_1(p)z_2(p)=q,\qquad\qquad \textrm{(for $p\in U_1\cap U_2\neq\zero$)}
$$ 
But the corresponding (tensor) operators, $\mathscr{O}^{z_1}(\e{z}_1)$ and $\mathscr{O}^{(z_2)}(\e{z}_2)$, must then also be related by a conformal transformation on patch overlaps, 
$
\mathscr{O}^{(z_1)}(\e{z}_1)\rmd\e{z}_1^h=\mathscr{O}^{(z_2)}(\e{z}_2)\rmd\e{z}_2^h,
$ 
in particular,
\begin{equation}\label{eq:gluingmodes=0bosonic}
\mathscr{O}^{(z_1)}(\e{z}_1)-(-)^hq^{-h}\e{z}_2^{2h}\mathscr{O}^{(z_2)}(\e{z}_2)=0\qquad\Leftrightarrow\qquad  
\mathscr{O}_n^{(z_1)}-(-)^hq^n\mathscr{O}_{-n}^{(z_2)}=0.
\end{equation}
This is essentially BPZ conjugation, but notice the non-standard powers of the pinch parameter, $q$. As an aside, commutation relations are preserved on both sides of the pinch up to this power of $q$, which is as it must be because the modes are not invariant under dilatations. To see that commutation relations are preserved one makes use of the fact that BPZ conjugation, for which $q=1$ and $z_1\mapsto 1/z_2$, reverses the order with which the operators hit the $z_2=0$ vacuum compared to the order they hit the $z_1=0$ vacuum  \cite{Zwiebach93}, and switching back to the original ordering introduces some necessary signs in order to obtain the familiar commutation relations (for both Grassmann-even and Grassmann-odd mode operators). 
\sk

\begin{figure}
\begin{center}
\includegraphics[angle=0,origin=c,width=0.4\textwidth]{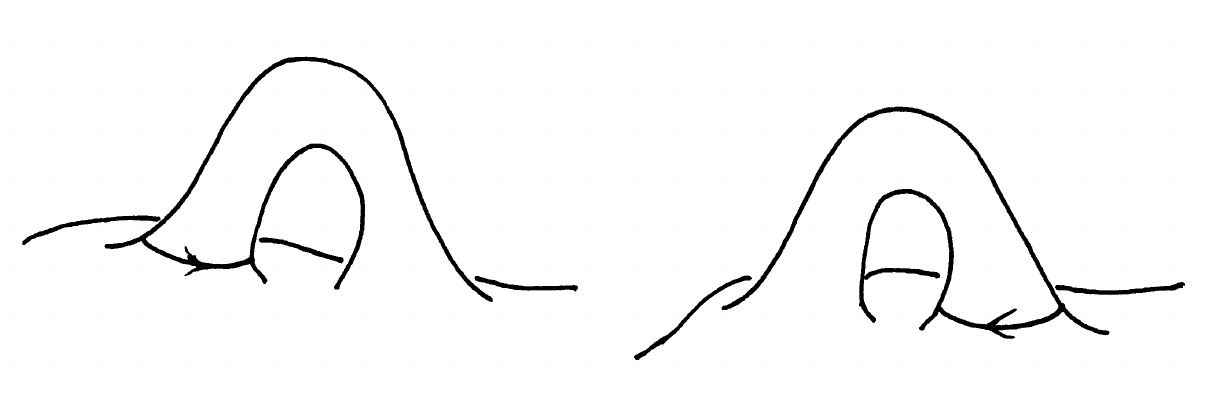}
\caption{Pictorial representation of the gluing consistency condition (\ref{eq:gluing modesbosonic}), or any of the relations in (\ref{eq:abcmodes1->2}). The two sketched configurations should in particular lead to identical amplitudes. The depicted contour in the left diagram (and that in the right diagram) is to be identified with the contours of the associated mode operators or charges on the $z_1$ and (after a change in contour orientation) on the $z_2$ charts respectively.}\label{fig:cont1}
\end{center}
\end{figure}

So we interpret the second relation in (\ref{eq:gluingmodes=0bosonic}) as an operator statement that must hold inside the path integral and consistent gluing (for fixed complex structure) then requires that:
\begin{equation}\label{eq:gluing modesbosonic}
\boxed{\Big(\mathscr{O}_n^{(z_1)}-(-)^hq^n\mathscr{O}_{-n}^{(z_2)}\Big)\cdot \suminnt\limits_{a}\,\,\hat{\mathscr{A}}_a^{(z_1)} \hat{\mathscr{A}}^a_{(z_2/q)}=0}
\end{equation}
which must be true for any basis, including a coherent state basis (since the operand is simply a resolution of unity). The case of interest here is to identify $\mathscr{O}_n^{(z)}$ with the matter modes, $\alpha_n^{(z)}$, ghost modes, $c_n^{(z)}$, $b_n^{(z)}$, as well as the stress tensor modes, $L_n^{(z)}$, and the chiral half of the BRST charge, $Q_B^{(z)}$. Identical results apply for a general matter CFT that has a free-field mode expansion, and the modes $\alpha_n^{(z)}$ are then to be replaced accordingly. (Similar reasoning applies to the anti-chiral sector.)  A pictorial representation of the gluing consistency condition (\ref{eq:gluing modesbosonic}) is shown in Fig.~\ref{fig:cont1}, where the two diagrams correspond to the two additive terms in (\ref{eq:gluing modesbosonic}).
\sk

A couple of comments: the explicit $q$ parameter is redundant since it can be absorbed into a rescaling of the frame coordinates, $z_1,z_2$, but we include it for later convenience. Secondly, the relation (\ref{eq:gluing modesbosonic}) must hold independently of whether there is a modulus associated to this factorisation, and hence there are no $b_0,\tilde{b}_0$ or $b_{-1}\tilde{b}_{-1}$, etc., insertions at this stage. 
\sk

To be completely explicit, from (\ref{eq:gluingmodes=0bosonic}) and the standard matter CFT of critical bosonic string theory we learn that (for all $n\in\mathbf{Z}$):
\begin{equation}\label{eq:abcmodes1->2}
\boxed{
\begin{aligned}
\alpha_n^{(z_1)}+q^n&\alpha_{-n}^{(z_2)}=0,\qquad
b_n^{(z_1)}-q^nb_{-n}^{(z_2)}=0,\qquad
c_n^{(z_1)}+q^nc_{-n}^{(z_2)}=0,\\
&L_n^{(z_1)}-q^nL_{-n}^{(z_2)}=0,\qquad\mathrm{and}\qquad  Q_B^{(z_1)}+Q_B^{(z_2)}=0
\end{aligned}
}
\end{equation}
and all these must hold in the sense of (\ref{eq:gluing modesbosonic}), namely as operator equations inside the path integral. 
(For the chiral half of the BRST charge we are to take $n=0$ and $h=1$ in (\ref{eq:gluing modesbosonic}).) 
\sk

A crucial point now is that these relations (\ref{eq:gluing modesbosonic}) can be used to explicitly determine what operators $\hat{\mathscr{A}}_a^{(z_1)},\hat{\mathscr{A}}^a_{(z_2)}$ can consistently propagate through the handle that has been cut open and replaced by a bi-local operator insertion. But these operators will certainly not be unique given that: (1) we can if we wish use different coordinate charts for the gluing and the resulting physical amplitudes should be independent of this (which must be checked explicitly); (2) we are free to choose a {\it basis} of states, and in the current document the basis of interest that leads to a unified and simple formalism is a {\it coherent state basis}. 
\sk

It is worth emphasising that the BRST charge contour can be smoothly deformed through the cut independently (!) of whether the states $\hat{\mathscr{A}}_a^{(z_1)},\hat{\mathscr{A}}^a_{(z_2)}$ are BRST invariant, the gluing consistency condition involving the BRST charge being:
\begin{equation}\label{eq:gluing Qbosonic}
\big(Q_B^{(z_1)}+Q_B^{(z_2)}\big)\cdot \suminnt\limits_{a}\,\,\hat{\mathscr{A}}_a^{(z_1)}\hat{\mathscr{A}}^a_{(z_2/q)}=0.
\end{equation}
So stronger statements such as $Q_B^{(z_1)}\cdot \hat{\mathscr{A}}_a^{(z_1)}=0$ and/or $Q_B^{(z_2)}\cdot \hat{\mathscr{A}}^a_{(z_2/q)}=0$ are not enforced. This is because strings propagating through loops or through intermediate propagators are offshell (i.e.~not BRST-invariant) unless these handles are pinched, or more precisely unless the associated intermediate propagators are cut (in the Cutkosky sense \cite{Cutkosky60,PiusSen16}). 
\sk

A solution to (\ref{eq:gluing modesbosonic}) and hence also (\ref{eq:gluing Qbosonic}) is well-known for {\it mass eigenstates} \cite{Polchinski88,Zwiebach93} and we write it down directly.\footnote{The general solution to $b_0^{(z_1)}-b_0^{(z_2)}=0$ in (\ref{eq:abcmodes1->2}) is a linear combination of  $|1\rangle^1\otimes |1\rangle^2$ and $(\tilde{c}_0^{(z_1)}+\tilde{c}_0^{(z_2)})(c_0^{(z_1)}+c_0^{(z_2)})|1\rangle^1\otimes |1\rangle^2$. However, the requirement for  consistent factorisation (\ref{eq:consistency_condition2}) further implies that only the terms $(\tilde{c}_0^{(z_1)}+\tilde{c}_0^{(z_2)})(c_0^{(z_1)}+c_0^{(z_2)})|1\rangle^1\otimes |1\rangle^2$ survive. Nevertheless, if one wishes to make manifest that handle operators are essentially density matrices it is clearest if one replaces $(\tilde{c}_0^{(z_1)}+\tilde{c}_0^{(z_2)})(c_0^{(z_1)}+c_0^{(z_2)})]|1\rangle^1\otimes |1\rangle^2$ by $[1+(\tilde{c}_0^{(z_1)}+\tilde{c}_0^{(z_2)})(c_0^{(z_1)}+c_0^{(z_2)})|1\rangle^1\otimes |1\rangle^2$, because then the sum, $\Sigma\!\!\!\!\!\int_{\,\,a}=1$, which roughly corresponds to the statement that the sum of probabilities equals 1. This is implemented by replacing (\ref{eq:1+c0ct0}) by (\ref{eq:1+c0ct0z}).\label{foot:density}} When each of the remaining insertions on $\Sigma_1$ and $\Sigma_2$, denoted by $\dots_1$ and $\dots_2$ in (\ref{eq:completeness-sep BBB}), have ghost number $N_{\rm gh}=2$ the sum (\ref{eq:oplusHHdual}) collapses to a single contribution, $n_a=2$ and $n^a=4$. The $b_0,\tilde{b}_0$ term then projects onto the ghost number-two part of $\hat{\mathscr{A}}^a_{(z_2)}$. In general the vertex operators $\hat{\mathscr{A}}_a^{(z_1)},\hat{\mathscr{A}}^a_{(z_2)}$ will not have definite ghost number, especially when we cut open a loop across a cycle that is not homologous to zero (a non-separating degeneration), so it will be useful to consider indefinite ghost number vertex operators that automatically cover all cases of interest. A solution to (\ref{eq:gluing modesbosonic}) that applies in all of these cases reads \cite{Polchinski88,Zwiebach93}:\footnote{\label{foot:Newton}The normalisation is chosen such that $S^2$ correlators are normalised by (\ref{eq:<..>normalisation1}) and the relation to the gravitational coupling is $\kappa_D=2\pi g_D$ where $g_D$ is the $D$-dimensional (dimensionful) string coupling, with $\kappa_D^2=8\pi G_D$ and $G_D$ Newton's constant in $D$ non-compact spacetime dimensions. E.g., in $D = 4$, $\kappa_4^{-1} = 2.4 \times 10^{18}$ GeV is the reduced Planck mass. These conventions are consistent with \cite{Polchinski_v1}. Regarding  the spacetime length dimension of the coupling, $[g_D]=L^{D/2-1}$.}
\begin{equation}\label{eq:handle}
\boxed{
\begin{aligned}
\suminnt\limits_{a}&\,\,\,\,\hat{\mathscr{A}}_a^{(z_1)}|1\rangle^1\otimes\hat{\mathscr{A}}^a_{(z_2/q)}|1\rangle^2= \\
&= \frac{\alpha'g_D^2}{8\pi i}\int \frac{\rmd^Dk}{(2\pi)^D}e^{ik\cdot (x_0^{(z_1)}-x_0^{(z_2)})}(q\bar{q})^{\frac{\alpha'}{4}k^2-1}\\
&\quad\times\exp \Big[\sum_{n=1}^{\infty}q^n\Big(\!-\frac{1}{n}\alpha_{-n}^{(z_1)}\cdot \alpha_{-n}^{(z_2)}+c_{-n}^{(z_1)}b_{-n}^{(z_2)}-b_{-n}^{(z_1)}c_{-n}^{(z_2)}\Big)\Big]\\
&\quad\times \exp\Big[\sum_{n=1}^{\infty}\bar{q}^n\Big(\!-\frac{1}{n}\tilde{\alpha}_{-n}^{(z_1)}\cdot \tilde{\alpha}_{-n}^{(z_2)}+\tilde{c}_{-n}^{(z_1)}\tilde{b}_{-n}^{(z_2)}-\tilde{b}_{-n}^{(z_1)}\tilde{c}_{-n}^{(z_2)}\Big)\Big]\\
&\quad\times(\tilde{c}_0^{(z_1)}+\tilde{c}_0^{(z_2)})(c_0^{(z_1)}+c_0^{(z_2)})\tilde{c}_1^{(z_1)}c_1^{(z_1)}\tilde{c}_1^{(z_2)}c_1^{(z_2)}|1\rangle^{1}\otimes|1\rangle^{2}.
\end{aligned}
}
\end{equation}
An interesting related and early path integral approach to factorisation of amplitudes is \cite{DHokerPhong85}. 
Let us make a few comments/clarifications before proceeding with the calculation.
\sk

Notation-wise, we could (and will below) have written `$1$' rather than $|1\rangle^{1}$ (and similarly for $|1\rangle^{2}$) to indicate that the mode contours create vertex operators out of the vacuum and that the vacuum corresponds to the absence of insertions. That is, we are really thinking of the left- and right-hand sides in (\ref{eq:handle}) as a sum/integral over bi-local vertex operators, but by the state/operator correspondence there is no real distinction. 
\sk

Secondly, strictly speaking the expression (\ref{eq:handle}) is correct as it stands only if $D=26$. However, for simple (such as toroidal) compactifications and general constant background fields the result can easily be generalised to any $D\leq26$. The essential difference is that instead of having just an integral over $k^\mu$ we would also have a (dimensionless) sum over all momentum and winding modes which are in turn implemented by splitting $x=x_L+x_R$ and introducing chiral and anti-chiral momenta as given in equation (3.48) in \cite{SklirosCopelandSaffin17}. This splitting is justified when (in a canonical intersection homology basis) we also cut along {\it all} $A_I$ cycles of the Riemann surface, i.e.~for $I=1,\dots,\g$, and introduce a corresponding resolution of unity (\ref{eq:handle}) for every such cut cycle. So in a sense the chiral-splitting theorem \cite{BelavinKnizhnik86,VerlindeVerlinde87,DijkgraafVerlindeVerlinde88,D'HokerPhong89,SklirosCopelandSaffin17} here is implicit. We neglect the case $D<26$ in order to avoid cluttering the notation further, but as mentioned the generalisation is easily implemented and essentially immediate. Also the metric, $\eta_{\mu\nu}$, associated to mode contractions gets replaced accordingly, see \cite{SklirosCopelandSaffin17} for details.\footnote{(Even the normalisation factor $\frac{\alpha'g_D^2}{8\pi i}$ survives this generalisation, and we have written $g_D$ rather than $g_c$ as defined in \cite{Polchinski_v1} precisely for this reason. The two are related by $g_{26}|_{\rm here}={g_c}|_{\rm there}$ or $g_D|_{\rm here}=g_{c,d}|_{\rm there}$, and note also the difference in convention, $D|_{\rm here}=d|_{\rm there}$.)\label{foot:normal}}
\sk

A third comment is that the similarity between (\ref{eq:handle}) and {\it boundary states} \cite{Sen05} (see also \cite{PolchinskiCai88,CallanLovelaceNappiYost88} for earlier work) is not accidental, as one can quickly convince oneself. In fact, one should be able to make use of this similarity to derive corresponding closed offshell string coherent state vertex operators that are sourced by D branes in precise analogy to the closed string {\it intermediate} state vertex operators discussed here. This is because boundary states are essentially unit operators that implement the correct boundary conditions on D-branes. Incidentally, the usual boundary states are not coherent states since they do not satisfy the defining properties (they do not depend on continuous quantum numbers). As mentioned, the correct interpretation is that they represent unit operators. But the procedure outlined below can be used to turn them into a linear superposition of coherent states, in the resolution of unity sense.
\sk

Also, we have written the result (\ref{eq:handle}) for general $q,\bar{q}$ in order to keep track of the scaling dimensions of the associated operators and also for later convenience. 
The fact that this equation solves (\ref{eq:gluing modesbosonic}) follows from a standard contour argument \cite{Polchinski_v1}, and can be verified explicitly by making use of the commutation relations. Also, we have singled out the $c_0,\tilde{c}_0$ and $c_1,\tilde{c}_1$ contributions (which lead to the variety of ghost vacua) since these will play a special role. 
\sk

Another side remark is that if we had included measure insertions in (\ref{eq:gluing modesbosonic}) associated to complex structure deformations (see Sec.~\ref{sec:TPIM}), such as $b_0^{(z_2)}\tilde{b}_0^{(z_2)}/(q\bar{q})$, then the aforementioned contour argument applied to (\ref{eq:gluing Qbosonic}) would only hold up to a boundary term in the $q,\bar{q}$ moduli space integral due to the commutation relations (\ref{eq:QbLn commutators}), which must ultimately cancel. This is a general property of all measure contributions, $\hat{B}_k$, in that inserting a BRST-exact vertex operator into an amplitude generically leads to boundary terms in moduli space due to the commutators (\ref{eq:QbLn commutators}) and (\ref{eq:QBBDzv}), this was discussed in Sec.~\ref{sec:BRST-AC}. 
\sk

The $x_0^{(z_1)}$ and $x_0^{(z_2)}$ in (\ref{eq:handle}) are the zero modes of $x(z,\bar{z})$ associated to $\hat{\mathscr{A}}_a^{(z_1)}$ and its dual, $\hat{\mathscr{A}}^a_{(z_2)}$, respectively. The superscript on $x_0^{(z_1)}$ for instance indicates that it is defined in the $z_1$ frame; with the contour convention $\frac{1}{2}\oint (\frac{\rmd\e{z}_1}{2\pi i\e{z}_1}-\frac{\rmd\bar{\e{z}}_1}{2\pi i\bar{\e{z}}_1})=1$ it can be written as:
$$
x_0^{(z_1)}=\frac{1}{2}\oint \Big(\frac{\rmd\e{z}_1}{2\pi i\e{z}_1}-\frac{\rmd\bar{\e{z}}_1}{2\pi i\bar{\e{z}}_1}\Big)x^{(z_1)}(\e{z}_1,\bar{\e{z}}_1).
$$
The vacua $|1\rangle^{1}$ and $|1\rangle^{2}$ are the unique (up to normalisation) SL(2,$\mathbf{C}$) vacua defined by choosing contours in the corresponding operators that can be contracted to a point without obstruction, 
\begin{equation}\label{eq:annihilation_operators}
\begin{aligned}
&\alpha_{n}^{(z_1)}|1\rangle^{1}=0,\qquad n\geq1\\
&c_n^{(z_1)}|1\rangle^{1}=0,\qquad n\geq 2\\
&b_n^{(z_1)}|1\rangle^{1}=0,\qquad n\geq -1\\
&L_n^{(z_1)}|1\rangle^{1}=0,\qquad n\geq-1,
\end{aligned}
\end{equation}
and similarly for quantities in the $z_2$ frame. The first three relations in (\ref{eq:annihilation_operators}) define the annihilation operators in terms of modes and the ${\rm SL}(2,\mathbf{C})$ ground state. Notice that the $b_{-1}^{(z_1)},\tilde{b}_{-1}^{(z_1)}$ correspond to annihilation operators, so normal ordering in (\ref{eq:handle}) is not yet implemented. Furthermore, given the $x(z,\bar{z})\cdot 1$ OPE is non-singular there is the correspondence,
$$
e^{ik\cdot x_0^{(z_1)}}|1\rangle^1 \leftrightarrow \,\,:\!e^{ik\cdot x^{(z_1)}}(p_1)\!:_{z_1},
$$
where $:\,:_{z_1}$ denotes normal ordering in the $z_1$ frame, and the operator on the right-hand side is inserted at $p_1$ where $z_1(p_1)=0$. The corresponding modes in (\ref{eq:annihilation_operators}) outside of the indicated ranges of the integer $n$ correspond to creation operators when acting on $|1\rangle^1$. Notice that {\it all} creation operators are contained in (\ref{eq:handle}). From (\ref{eq:abcmodes1->2})  corresponding statements exist for the BPZ conjugate vacuum $^1\langle1|$, obtained from (\ref{eq:annihilation_operators}) by $z_1\mapsto 1/z_1$, where the mode integer signs are flipped, $n\rightarrow -n$. 
\sk

The normalisation factor $\alpha'g_D^2/8\pi i$ in (\ref{eq:handle}) is convention dependent. Modulo a comment in the footnote on p.~\pageref{foot:normal}, our conventions are in agreement with Polchinski \cite{Polchinski_v1,Polchinski88}, and in particular the matter and ghost path integral measures are normalised by:\footnote{A useful equivalent statement (when again all quantities are defined in the $z_1$ frame) is the following:
\begin{equation}
 e^{-2\Phi}\Big\langle 
\tfrac{1}{2}\partial^2c\,\partial c\,c(\e{z}_1)\,\tfrac{1}{2}\bar{\partial}^2\tilde{c}\,\bar{\partial}\tilde{c}\,\tilde{c}(\bar{\e{z}}_1):\!e^{ik\cdot x(z,\bar{z})}\!:\Big\rangle_{S^2}=\frac{8\pi i}{\alpha'g_D^2}(2\pi)^D\delta^D(k).
\end{equation}
}
\begin{equation}\label{eq:<..>normalisation1}
\begin{aligned}
& e^{-2\Phi}\Big\langle \tilde{c}^{(z_1)}(\bar{\e{z}}_3)c^{(z_1)}(\e{z}_3)\,\tilde{c}^{(z_1)}(\bar{\e{z}}_2)c^{(z_1)}(\e{z}_2)\,\tilde{c}^{(z_1)}(\bar{\e{z}}_1)c^{(z_1)}(\e{z}_1)\,:\!e^{ik\cdot x^{(z_1)}(\e{z},\bar{\e{z}})}\!:_{z_1}\Big\rangle_{S^2}=\\
&\qquad \qquad=\frac{8\pi i}{\alpha'g_D^2}\big|\e{z}_{12}\e{z}_{23}\e{z}_{31}\big|^2(2\pi)^D\delta^D(k),
\end{aligned}
\end{equation}
and this is independent of whether we use a mass eigenstate or coherent state vertex operator basis (the latter being a linear superposition of mass eigenstates). In particular, the overall normalisation in (\ref{eq:handle}) has been fixed by making use of the consistency condition (\ref{eq:consistency_condition2}): we choose any vertex operator, such as $\hat{\mathscr{A}}_c^{(z)}=g_D\tilde{c}c\,e^{ip\cdot x(0,0)}$ with $p^2=4/\alpha'$ (but we can also take it offshell), and compute the right-hand side of (\ref{eq:consistency_condition2}) using (\ref{eq:handle}) (with $q=1$, although this is immaterial as it can scaled away). To compute the resulting two-point amplitude we use (\ref{eq:<..>normalisation1}) for the zero modes and ghosts (here we Taylor expand in $z_{32}$ around $z_3=z_2$ and then take $z_2\rightarrow \infty$ and $z_1\rightarrow 0$) and compute the matter contractions using Wick's theorem. For general vertex operators (which may not in general be primaries, or even SL$(2,\mathbf{C})$ primaries), in computing (\ref{eq:consistency_condition2}) one can unwrap the mode contours off one vertex operator onto the other and use commutation relations. We discuss this for general vertex operators in, e.g., Sec.~\ref{sec:S2ptA}. 
\sk

It will be convenient to rewrite the exponentials involving $q,\bar{q}$: making use of the defining relation for cycle index polynomials (of the symmetric group) (\ref{eq:CycleIndexGenFun}) in (\ref{eq:handle}) yields,
\begin{equation}\label{eq:intermbb}
\begin{aligned}
\suminnt\limits_a\,\,\,&\hat{\mathscr{A}}_a^{(z_1)}|1\rangle^{1}\otimes \hat{\mathscr{A}}^a_{(z_2/q)} |1\rangle^{2}= \\
&= \frac{\alpha'g_D^2}{8\pi i}\int \frac{\rmd^Dk}{(2\pi)^D}e^{ik\cdot (x_0^{(z_1)}-x_0^{(z_2)})}(q\bar{q})^{\frac{\alpha'}{4}k^2-1}\sum_{n,m=0}^{\infty}q^n\bar{q}^m\\
&\quad\times Z_n\big(-\alpha_{-s}^{(z_1)}\cdot \alpha_{-s}^{(z_2)}+sc_{-s}^{(z_1)}b_{-s}^{(z_2)}-sb_{-s}^{(z_1)}c_{-s}^{(z_2)}\big)\\
&\quad\times Z_m\big(-\tilde{\alpha}_{-s}^{(z_1)}\cdot \tilde{\alpha}_{-s}^{(z_2)}+s\tilde{c}_{-s}^{(z_1)}\tilde{b}_{-s}^{(z_2)}-s\tilde{b}_{-s}^{(z_1)}\tilde{c}_{-s}^{(z_2)}\big)\\
&\quad\times(\tilde{c}_0^{(z_1)}+\tilde{c}_0^{(z_2)})(c_0^{(z_1)}+c_0^{(z_2)})\tilde{c}_1^{(z_1)}c_1^{(z_1)}\tilde{c}_1^{(z_2)}c_1^{(z_2)}|1\rangle^{1}\otimes |1\rangle^{2}.
\end{aligned}
\end{equation}
We next would like to exchange the order of the sum over $n,m$ and the integral over $k^{\mu}$. The ability to do this is certainly necessary for unitarity, and in particular if one is to be able to isolate the contribution to a pinch from specific channels. This in turn will be possible if there exists an analytic continuation for the $k$ integral that leads to a finite answer, and in turn such an analytic continuation will be possible provided $k$ is not fixed by kinematics (e.g., we would not be able to continue $k$ if there appears a delta function $\delta^D(k)$, which would in turn appear if we were to consider tadpoles\footnote{For tadpole degenerations one can adopt the coherent state basis (\ref{eq:offshellA_alpha}) on p.~\pageref{eq:offshellA_alpha} which are valid at zero momentum.}). Nevertheless, the convergence of the sum over $n,m$ should be guaranteed by general statements about the convergence of the OPE (and according to Polchinski \cite{Polchinski88} it can be proven using a Cauchy-Schwarz inequality argument). So  (\ref{eq:intermbb}) takes the form,
\begin{equation}\label{eq:interm}
\begin{aligned}
\suminnt\limits_a\,\,\,&\hat{\mathscr{A}}_a^{(z_1)}|1\rangle^{1}\otimes \hat{\mathscr{A}}^a_{(z_2/q)}|1\rangle^{2}= \\
&=\sum_{n,m=0}^{\infty}\frac{\alpha'g_D^2}{8\pi i}\int \frac{\rmd^Dk}{(2\pi)^D}e^{ik\cdot (x_0^{(z_1)}-x_0^{(z_2)})}q^{\frac{\alpha'}{4}k^2+n-1}\bar{q}^{\frac{\alpha'}{4}k^2+m-1}\\
&\qquad\times Z_n\big(\!-\alpha_{-s}^{(z_1)}\cdot \alpha_{-s}^{(z_2)}+sc_{-s}^{(z_1)}b_{-s}^{(z_2)}-sb_{-s}^{(z_1)}c_{-s}^{(z_2)}\big)\\
&\qquad\times Z_m\big(\!-\tilde{\alpha}_{-s}^{(z_1)}\cdot \tilde{\alpha}_{-s}^{(z_2)}+s\tilde{c}_{-s}^{(z_1)}\tilde{b}_{-s}^{(z_2)}-s\tilde{b}_{-s}^{(z_1)}\tilde{c}_{-s}^{(z_2)}\big)\\
&\qquad\times(\tilde{c}_0^{(z_1)}+\tilde{c}_0^{(z_2)})(c_0^{(z_1)}+c_0^{(z_2)})\tilde{c}_1^{(z_1)}c_1^{(z_1)}\tilde{c}_1^{(z_2)}c_1^{(z_2)}|1\rangle^{1}\otimes |1\rangle^{2}.
\end{aligned}
\end{equation}

If we were interested in a {\it momentum eigenstate} basis the exponents of $q,\bar{q}$ would be immediately identified with the conformal weights of these operators at level numbers $n,m$,
$$
h_j=\frac{\alpha'}{4}k^2+n-1,\qquad \tilde{h}_j=\frac{\alpha'}{4}k^2+m-1,
$$
since on the left-hand side of (\ref{eq:interm}) we could also write\footnote{Recall we are using labels $i,j,\dots$ for momentum eigenstates and $a,b,\dots$ (but also $\alpha,\beta,\dots$, see below) for coherent states.} $\hat{\mathscr{A}}^j_{(z_2/q)}=q^{L_0}\bar{q}^{\tilde{L}_0}\hat{\mathscr{A}}^j_{(z_2)}$ which in turn enables us to simply read off the corresponding $L_0,\tilde{L}_0$ eigenvalues by comparing to the right-hand side. In particular, momentum eigenstates can be taken to be eigenstates of $L_0,\tilde{L}_0$. 
Coherent states however correspond to a linear superposition of mass or momentum eigenstates. Coherent states associated to cutting along a non-separating degeneration cannot in general be taken to have well-defined spin, $h_a-\tilde{h}_a$, (even though the total weight $h_a+\tilde{h}_a$ will be well-defined as we will show momentarily). 
\sk

Let us change coordinates in the integral over $k^\mu$. We first switch to lightcone {\it coordinates}, $k^{\pm}\dfn \frac{1}{\sqrt{2}}(k^0\pm k^{D-1})$, and for every $n,m$ make a change of variables  $k^{\mu}\mapsto \e{p}^{\mu}$, with:\footnote{This is the same kind of change of variables that is used in the DDF operator approach to the construction of vertex operators \cite{DelGiudiceDiVecchiaFubini72,AdemolloDelGuidiceDiVecchiaFubini74,D'HokerGiddings87,SklirosHindmarsh11}, the two differences here being that: (1) $\e{p}^2$ is {\it not} restricted to the mass shell value, $\e{p}^2=4/\alpha'$; (2) we do not have the $L_0-\tilde{L}_0=0$ constraint that would allow us to absorb $n$ and $m$ into the chiral and anti-chiral halves respectively; the latter is the obstruction to chirally splitting the matter zero modes when the handle operator implements a cut across a non-trivial homology cycle (which is in turn related to modular invariance).}
\begin{equation}\label{eq:lightconek}
k^+=\e{p}^+,\qquad k^-=\e{p}^-+\frac{n+m}{\alpha'\e{p}^+},\qquad{\rm and}
\qquad k^i=\e{p}^i,
\end{equation}
and we are in Lorentzian signature, e.g., $k^2=-2k^+k^-+{\bf k}^2$. 
We state from the outset that this change of variables (\ref{eq:lightconek}) is not well-defined in general, and especially if the path integrals on the right-hand side of (\ref{eq:completeness-sep BBB}):
\begin{itemize} 
\item[1)] produce a delta function $\delta^D(k)$ (because $\e{p}^+$ in (\ref{eq:lightconek}) appears in a denominator). In Sec.~\ref{sec:S1ptA} we will explicitly consider such a case (the one-point sphere amplitude) and show that this ambiguity is in this case harmless;
\item[2)] produce a delta function $\delta^D(k-k)$. In Sec.~\ref{sec:MI} we will explicitly consider such an example (the one-loop vacuum amplitude) and we will discuss a simple way to compute the relevant amplitude.  \label{item:deltak}
\end{itemize}
In general, one should proceed carefully after having made the change of variables (\ref{eq:lightconek}) (e.g., one should study the domain of integration, corresponding analytic continuations, scaling dimensions, in detail when integrating out $\e{p}$). Also, since $\e{p}^+$ appears in the denominator one is introducing an apparent non-locality in spacetime. The reader may wonder what is the advantage of making this change of variables. An advantage is that this change of variables (as we will see) will produce vertex operators that are {\it local} on $\Sigma$. (In Sec.~\ref{sec:HOEM} we discuss how to undo this change of variables when it is desirable to do so and obtain coherent states that are valid at all momenta, but the price will be that rather than being local on $\Sigma$ they will be ``smeared out'' over a region determined by $|q|$.)
\sk

From (\ref{eq:lightconek}) follow the relations,\footnote{As noted above, for the index contractions note that we are in Lorentzian signature where $A_\mu B^\mu=-A^-B^+-A^+B^-+\sum_{i=1}^{D-2}A^iB^i$. When explicitly integrating out momenta or moduli one should analytically continue the relevant momenta to a regime of absolute convergence before analytically continuing back to the physical regime, paying careful attention to cuts or pinch singularities (see \cite{DHokerPhong95,Witten13b,PiusSen16} and also \cite{AnSMatrix,LandauLifshitzRQF}). (Due to modular invariance there should be no UV divergences at any fixed-loop order.)}
$$
\rmd^Dk = \rmd^D\e{p},\qquad {\rm and}\qquad \frac{\alpha'}{2}k^2 = \frac{\alpha'}{2}\e{p}^2-(n+m).
$$
If we do not want to specify a direction explicitly for the lightcone we can instead introduce a null vector $\e{q}^{\mu}$ and demand that the change of coordinates takes the form:
\begin{equation}\label{eq:kpq_constr}
k^{\mu}=\e{p}^{\mu}-\frac{1}{2}(n+m)\e{q}^{\mu},\qquad {\rm with} \qquad \e{p}\cdot \e{q}=2/\alpha',\qquad \e{q}^2=0.
\end{equation}
Choosing, e.g., $\e{q}^-=-2/(\alpha'\e{p}^+)$ reproduces the above choice (\ref{eq:lightconek}), but any other choice of $\e{q}^{\mu}$ subject to the restrictions in (\ref{eq:kpq_constr}) is just as good. Implementing this change of variables in (\ref{eq:interm}) we learn primarily that,
\begin{equation}
\begin{aligned}
q^{\frac{\alpha'}{4}k^2+n-1}\bar{q}^{\frac{\alpha'}{4}k^2+m-1}
=r^{\frac{\alpha'}{2}\e{p}^2-2}e^{i(n-m)\theta},
\end{aligned}
\end{equation}
where we defined:
\begin{equation}\label{eq:q=rexpitheta}
\boxed{q\equiv r\,e^{i\theta},\qquad \bar{q} \equiv r\,e^{-i\theta}}
\end{equation}
Notice that this change of variables redefines the momentum at every mass level such that the {\it full} tower of string states scale in the same way under rescalings, $r\mapsto \lambda r$. 
In particular therefore,
\begin{equation}\label{eq:dDk=dDp..}
\begin{aligned}
\frac{\rmd^Dk}{(2\pi)^D}\,&e^{ik\cdot (x_0^{(z_1)}-x_0^{(z_2)})}(q\bar{q})^{\frac{\alpha'}{4}k^2-1}q^n\bar{q}^m =\\
&= \frac{\rmd^D\e{p}}{(2\pi)^D}\,e^{i\e{p}\cdot (x_0^{(z_1)}-x_0^{(z_2)})}\,(q\bar{q})^{\frac{\alpha'}{4}\e{p}^2-1}\,\big(e^{i\theta}\,e^{-i\e{q}\cdot \frac{1}{2}(x_0^{(z_1)}-x_0^{(z_2)})}\big)^n\big(e^{-i\theta}\,e^{-i\e{q}\cdot \frac{1}{2}(x_0^{(z_1)}-x_0^{(z_2)})}\big)^m.
\end{aligned}
\end{equation}
To see that both left- and right-hand sides in (\ref{eq:dDk=dDp..}) (and hence also in (\ref{eq:intermb})) have the same scaling dimensions under $q\mapsto \lambda q$ one must take into account the gluing relation, $z_1z_2=q$, and also the scaling dimensions of the (normal-ordered) exponentials. (There are related comments under the title `Scaling dimension' on p.~\pageref{scaldim}.)
\sk

Let us then take into account the cycle index polynomial scaling relation in (\ref{eq:Zproperties}), and also the scaling relation (\ref{eq:Ah1}), to discover that (\ref{eq:interm}) can also be written as follows:
\begin{equation}\label{eq:intermb}
\begin{aligned}
&\suminnt\limits_a\,\,r^{h_a+\tilde{h}_a}\hat{\mathscr{A}}_a^{(z_1)}|1\rangle^{1}\otimes \hat{\mathscr{A}}^a_{(z_2e^{-i\theta})}|1\rangle^{2}= \\
&\quad=\sum_{n,m=0}^{\infty} \frac{\alpha'g_D^2}{8\pi i}\int \frac{\rmd^D\e{p}}{(2\pi)^D}e^{i\e{p}\cdot (x_0^{(z_1)}-x_0^{(z_2)})}r^{\frac{\alpha'}{2}\e{p}^2-2}\\
&\qquad\times Z_n\Big(e^{is\theta}\big(\!-\alpha_{-s}^{(z_1)}\cdot \alpha_{-s}^{(z_2)}+sc_{-s}^{(z_1)}b_{-s}^{(z_2)}-sb_{-s}^{(z_1)}c_{-s}^{(z_2)}\big)e^{-is\e{q}\cdot \frac{1}{2}(x_0^{(z_1)}-x_0^{(z_2)})}\Big)\\
&\qquad \times Z_m\Big(e^{-is\theta}\big(\!-\tilde{\alpha}_{-s}^{(z_1)}\cdot \tilde{\alpha}_{-s}^{(z_2)}+s\tilde{c}_{-s}^{(z_1)}\tilde{b}_{-s}^{(z_2)}-s\tilde{b}_{-s}^{(z_1)}\tilde{c}_{-s}^{(z_2)}\big)e^{-is\e{q}\cdot \frac{1}{2}(x_0^{(z_1)}-x_0^{(z_2)})}\Big)\\
&\qquad\times(\tilde{c}_0^{(z_1)}+\tilde{c}_0^{(z_2)})(c_0^{(z_1)}+c_0^{(z_2)})\tilde{c}_1^{(z_1)}c_1^{(z_1)}\tilde{c}_1^{(z_2)}c_1^{(z_2)}|1\rangle^{1}\otimes|1\rangle^{2}.
\end{aligned}
\end{equation}

As we will see from the discussion associated to (\ref{eq:scale-phase exp}), (\ref{eq:intermb}) is already telling us something interesting: comparing left- and right-hand sides the entire $r$ dependence is factored out and explicit on both left- and right-hand sides, so the total weights of the states propagating through the pinch are, for a given momentum $\e{p}$, {\it all equal}, even offshell, namely the total scaling dimension, $\Delta_a\dfn h_a+\tilde{h}_a$, is $\Delta_a=\frac{\alpha'}{2}\e{p}^2-2$. 
\sk

We next interchange the order of the sums and integrals in (\ref{eq:intermb}) again (subject again to the assumption of existence of an analytic continuation that guarantees absolute convergence) and sum over $n,m$ by making use of the defining relation for cycle index polynomials (\ref{eq:CycleIndexGenFun}):
\begin{equation}\label{eq:intermbc}
\begin{aligned}
&\suminnt\limits_a\,\,r^{h_a+\tilde{h}_a}\hat{\mathscr{A}}_a^{(z_1)}|1\rangle^{1}\otimes \hat{\mathscr{A}}^a_{(z_2/e^{i\theta})}|1\rangle^{2}= \\
&\quad= \,\frac{\alpha'g_D^2}{8\pi i}\int \frac{\rmd^D\e{p}}{(2\pi)^D}e^{i\e{p}\cdot (x_0^{(z_1)}-x_0^{(z_2)})}r^{\frac{\alpha'}{2}\e{p}^2-2}\\
&\qquad\times\exp\Big[\sum_{n=1}^{\infty}e^{in\theta}\Big(\!-\frac{1}{n}\alpha_{-n}^{(z_1)}\cdot \alpha_{-n}^{(z_2)}+c_{-n}^{(z_1)}b_{-n}^{(z_2)}-b_{-n}^{(z_1)}c_{-n}^{(z_2)}\Big)e^{-in\e{q}\cdot \frac{1}{2}(x_0^{(z_1)}-x_0^{(z_2)})}\Big]\\
&\qquad \times \exp\Big[\sum_{n=1}^{\infty}e^{-in\theta}\Big(\!-\frac{1}{n}\tilde{\alpha}_{-n}^{(z_1)}\cdot \tilde{\alpha}_{-n}^{(z_2)}+\tilde{c}_{-n}^{(z_1)}\tilde{b}_{-n}^{(z_2)}-\tilde{b}_{-n}^{(z_1)}\tilde{c}_{-n}^{(z_2)}\Big)e^{-in\e{q}\cdot \frac{1}{2}(x_0^{(z_1)}-x_0^{(z_2)})}\Big]\\
&\qquad\times (\tilde{c}_0^{(z_1)}+\tilde{c}_0^{(z_2)})(c_0^{(z_1)}+c_0^{(z_2)})\tilde{c}_1^{(z_1)}c_1^{(z_1)}\tilde{c}_1^{(z_2)}c_1^{(z_2)}|1\rangle^{1}\otimes|1\rangle^{2}.
\end{aligned}
\end{equation}

Comparing (\ref{eq:intermbc}) to what we started with in (\ref{eq:handle}), this procedure has removed all $r$ dependence from the exponential and exchanged it with plane-wave momentum-dependent exponentials. 
What remains is to decouple the operators associated to the $z_1$ and $z_2$ frames as much as possible, since this will allow us to read off the corresponding local vertex operators. We do this by introducing integral representations for the various exponentials. For the chiral half of the matter sector,\footnote{For clarity, the integral representation of interest here is (for any Grassmann-even complex numbers $A_1,A_2,G$, and the integral is over the entire complex plane):
\begin{equation}\label{eq:GaussianIntegral}
\boxed{e^{A_1GA_2} = \int\frac{\rmd\bar{z}\wedge \rmd z}{2\pi iG}\,e^{-\bar{z}G^{-1}z}e^{zA_1}e^{\bar{z}A_2}}
\end{equation}
}
\begin{equation}\label{eq:aa chiral}
\begingroup\makeatletter\def\f@size{11}\check@mathfonts
\def\maketag@@@#1{\hbox{\m@th\large\normalfont#1}}%
\begin{aligned}
\exp\Big[&\sum_{n=1}^{\infty}e^{in\theta}\Big(\!-\frac{1}{n}\alpha_{-n}^{(z_1)}\cdot \alpha_{-n}^{(z_2)}\Big)e^{-in\e{q}\cdot \frac{1}{2}(x_0^{(z_1)}-x_0^{(z_2)})}\Big]\\
&=\bigg(\prod_{n>0,\mu}\int \frac{\rmd\mathbi{a}_{n}^{*\mu}\wedge \rmd\mathbi{a}_{n}^\mu}{2\pi in}\bigg)\exp\Big(\!-\sum_{n>0}\frac{1}{n}\mathbi{a}_n\cdot \mathbi{a}_n^*\Big)\\
&\quad\times \exp\Big(\sum_{n>0}\frac{1}{n}\mathbi{a}_n\cdot \alpha_{-n}^{(z_1)}\,e^{-in\e{q}\cdot \frac{1}{2}x_0^{(z_1)}}\Big)\times\exp\Big(-\sum_{n>0}\frac{1}{n}e^{in\theta}\mathbi{a}_n^*\cdot \alpha_{-n}^{(z_2)}\,e^{in\e{q}\cdot \frac{1}{2}x_0^{(z_2)}}\Big),
\end{aligned}
\endgroup
\end{equation}
where the $\bm{a}_n^{\mu}$ are continuous (complex) quantum numbers (recall the definition of a coherent state) and may be associated to polarisation tensors for the modes $\alpha_{-n}^{\mu}$. (There are no transversality conditions for these.) The anti-chiral half is similarly given by,
\begin{equation}\label{eq:aa antichiral}
\begingroup\makeatletter\def\f@size{11}\check@mathfonts
\def\maketag@@@#1{\hbox{\m@th\large\normalfont#1}}%
\begin{aligned}
\exp\Big[&\sum_{n=1}^{\infty}e^{-in\theta}\Big(\!-\frac{1}{n}\tilde{\alpha}_{-n}^{(z_1)}\cdot \tilde{\alpha}_{-n}^{(z_2)}\Big)e^{-in\e{q}\cdot \frac{1}{2}(x_0^{(z_1)}-x_0^{(z_2)})}\Big]\\
&=\bigg(\prod_{n>0,\mu}\int \frac{\rmd\tilde{\mathbi{a}}_{n}^{\mu*}\wedge \rmd\tilde{\mathbi{a}}_{n}^{\mu}}{2\pi in}\bigg)\exp\Big(\!-\sum_{n>0}\frac{1}{n}\tilde{\mathbi{a}}_n\cdot \tilde{\mathbi{a}}_n^*\Big)\\
&\quad\times \exp\Big(\sum_{n>0}\frac{1}{n}\tilde{\mathbi{a}}_n\cdot \tilde{\alpha}_{-n}^{(z_1)}\,e^{-in\e{q}\cdot \frac{1}{2}x_0^{(z_1)}}\Big)\times\exp\Big(-\sum_{n>0}\frac{1}{n}e^{-in\theta}\tilde{\mathbi{a}}_n^*\cdot \tilde{\alpha}_{-n}^{(z_2)}\,e^{in\e{q}\cdot \frac{1}{2}x_0^{(z_2)}}\Big).
\end{aligned}
\endgroup
\end{equation}

Notice that the second exponentials in the last line of either (\ref{eq:aa chiral}) or (\ref{eq:aa antichiral}), i.e.~the terms associated to the $z_2$ patch, are related to those in the $z_1$ patch (i.e.~the first exponential in the last lines of the same equations) by explicit conjugation of all factors or $i=\sqrt{-1}$ (recall from the operator-state correspondence the creation operators, $\alpha_{-n},\tilde{\alpha}_{-n}$, also have an explicit factor of $i$). So the matter sector $z_2$ patch vertex operators are precisely {\it Euclidean adjoints} of the matter sector $z_1$ patch vertex operators \cite{Polchinski_v1,Polchinski88}, and it is satisfying to see this notion derived directly from the defining properties of coherent states, the string path integral and Riemann surface data. (The Euclidean adjoint vertex operators actually include a subtle phase in non-trivial topologies that may not have been discussed in the literature\footnote{Based on unpublished work of one of us (DS) with Paul Saffin}.) 
\sk

We can proceed in a similar way for the chiral half of the ghost sector,\footnote{Here we make use of the following integral representation; for any Grassmann-odd quantities $A_1,B_1,A_2,B_2$, Grassmann-even numbers $G,\Delta$ (although the case $G=\Delta=1$ is of interest in (\ref{eq:bc-cb chiral}) and (\ref{eq:bc-cb antichiral})), and Grassmann-odd integration variables, $\theta^1,\theta^2,\eta^1,\eta^2$:
\begin{equation}\label{eq:Grassmann-odd ints}
\boxed{e^{A_1G^{-1}B_2+A_2\Delta^{-1}B_1} = (G\Delta)^{-1}\int [\rmd\theta^1\rmd\eta^2\rmd\theta^2\rmd\eta^1]\,e^{\eta^1\Delta\theta^2+\eta^2G\theta^1}e^{\theta^1A_1+\eta^1B_1}e^{\theta^2A_2+\eta^2B_2}}
\end{equation}
}
\begin{equation}\label{eq:bc-cb chiral}
\begingroup\makeatletter\def\f@size{11}\check@mathfonts
\def\maketag@@@#1{\hbox{\m@th\large\normalfont#1}}%
\begin{aligned}
&\exp\bigg[\sum_{n=1}^{\infty}e^{in\theta}\Big(c_{-n}^{(z_1)}b_{-n}^{(z_2)}-b_{-n}^{(z_1)}c_{-n}^{(z_2)}\Big)e^{-in\e{q}\cdot \frac{1}{2}(x_0^{(z_1)}-x_0^{(z_2)})}\bigg]\\
&=\Bigg(\prod_{n>0}\int \big[\rmd\mathbi{c}_n^{(1)}\rmd\mathbi{b}_n^{(2)} \rmd\mathbi{c}_n^{(2)}\rmd\mathbi{b}_n^{(1)} \big]\exp\Big(\mathbi{b}_n^{(1)}\mathbi{c}_n^{(2)}+\mathbi{b}_n^{(2)}\mathbi{c}_n^{(1)}\Big)\Bigg)\\
&\quad\times \exp\bigg[\sum_{n>0}\Big(\mathbi{c}_n^{(1)}c_{-n}^{(z_1)}+\mathbi{b}_n^{(1)}b_{-n}^{(z_1)}\Big)e^{-in\e{q}\cdot \frac{1}{2}x_0^{(z_1)}}\bigg] \exp\bigg[\sum_{n>0}e^{in\theta}\Big(\mathbi{c}_n^{(2)}c_{-n}^{(z_2)}+\mathbi{b}_n^{(2)}b_{-n}^{(z_2)}\Big)e^{in\e{q}\cdot \frac{1}{2}x_0^{(z_2)}}\bigg]\\
\end{aligned}
\endgroup
\end{equation}
The quantities $\mathbi{b}_n^{(i)}$, $\mathbi{c}_n^{(i)}$, for $i=1,2$, are Grassmann-odd quantum numbers, whereas the symbol $\int [\dots \rmd\theta_1\rmd\theta_2\dots]$ is to be interpreted as in \cite{Witten12a}, namely it is {\it odd} under the interchange $\rmd\theta_1\leftrightarrow \rmd\theta_2$ (despite the fact that differentials of odd variables are even). 
\sk

For the corresponding anti-chiral half we similarly have:
\begin{equation}\label{eq:bc-cb antichiral}
\begingroup\makeatletter\def\f@size{11}\check@mathfonts
\def\maketag@@@#1{\hbox{\m@th\large\normalfont#1}}%
\begin{aligned}
&\exp\bigg[\sum_{n=1}^{\infty}e^{-in\theta}\Big(\tilde{c}_{-n}^{(z_1)}\tilde{b}_{-n}^{(z_2)}-\tilde{b}_{-n}^{(z_1)}\tilde{c}_{-n}^{(z_2)}\Big)e^{-in\e{q}\cdot \frac{1}{2}(x_0^{(z_1)}-x_0^{(z_2)})}\bigg]\\
&=\Bigg(\prod_{n>0}\int \big[\rmd\tilde{\mathbi{c}}_n^{(1)}\rmd\tilde{\mathbi{b}}_n^{(2)} \rmd\tilde{\mathbi{c}}_n^{(2)}\rmd\tilde{\mathbi{b}}_n^{(1)} \big]\exp\Big(\tilde{\mathbi{b}}_n^{(1)}\tilde{\mathbi{c}}_n^{(2)}+\tilde{\mathbi{b}}_n^{(2)}\tilde{\mathbi{c}}_n^{(1)}\Big)\Bigg)\\
&\quad\times \exp\bigg[\sum_{n>0}\Big(\tilde{\mathbi{c}}_n^{(1)}\tilde{c}_{-n}^{(z_1)}+\tilde{\mathbi{b}}_n^{(1)}\tilde{b}_{-n}^{(z_1)}\Big)e^{-in\e{q}\cdot \frac{1}{2}x_0^{(z_1)}}\bigg] \exp\bigg[\sum_{n>0}e^{-in\theta}\Big(\tilde{\mathbi{c}}_n^{(2)}\tilde{c}_{-n}^{(z_2)}+\tilde{\mathbi{b}}_n^{(2)}\tilde{b}_{-n}^{(z_2)}\Big)e^{in\e{q}\cdot \frac{1}{2}x_0^{(z_2)}}\bigg].
\end{aligned}
\endgroup
\end{equation}
Bowing to standard convention, barred quantities are obtained from unbarred ones by complex conjugation whereas tilded quantities are associated to anti-chiral halves and are considered independent of untilded ones. 

\sk
Finally, we factorise the vacuum terms by means of the following integral representation:
\begin{equation}\label{eq:1+c0ct0}
\begingroup\makeatletter\def\f@size{11}\check@mathfonts
\def\maketag@@@#1{\hbox{\m@th\large\normalfont#1}}%
\begin{aligned}
&\big(\tilde{c}_0^{(z_1)}+\tilde{c}_0^{(z_2)}\big)\big(c_0^{(z_1)}+c_0^{(z_2)}\big)=\\
&=\int \big[\rmd\tilde{\mathbi{c}}_0 \rmd\mathbi{c}_0\big] \exp\big(\mathbi{c}_0c_0^{(z_1)}+\tilde{\mathbi{c}}_0\tilde{c}_0^{(z_1)}\big)\exp\big(\mathbi{c}_0c_0^{(z_2)}+\tilde{\mathbi{c}}_0\tilde{c}_0^{(z_2)}\big)\\
&=\int \big[\rmd\tilde{\mathbi{c}}_0^{(1)} \rmd\mathbi{c}_0^{(1)}\rmd\tilde{\mathbi{c}}_0^{(2)} \rmd\mathbi{c}_0^{(2)}\big]\delta(\mathbi{c}_0^{(2)}-\mathbi{c}_0^{(1)})\delta(\tilde{\mathbi{c}}_0^{(2)}-\tilde{\mathbi{c}}_0^{(1)}) \\
&\qquad\qquad\qquad\times\exp\big(\mathbi{c}_0^{(1)}c_0^{(z_1)}+\tilde{\mathbi{c}}_0^{(1)}\tilde{c}_0^{(z_1)}\big)\exp\big(\mathbi{c}_0^{(2)}c_0^{(z_2)}+\tilde{\mathbi{c}}_0^{(2)}\tilde{c}_0^{(z_2)}\big)
\end{aligned}
\endgroup
\end{equation}
Let us also note that if we replace $(\tilde{c}_0^{(z_1)}+\tilde{c}_0^{(z_2)})(c_0^{(z_1)}+c_0^{(z_2)})$ in (\ref{eq:handle}) by $1+(\tilde{c}_0^{(z_1)}+\tilde{c}_0^{(z_2)})(c_0^{(z_1)}+c_0^{(z_2)})$ then the relevant factorisation formula is:
\begin{equation}\label{eq:1+c0ct0z}
\begin{aligned}
&1+(\tilde{c}_0^{(z_1)}+\tilde{c}_0^{(z_2)})(c_0^{(z_1)}+c_0^{(z_2)})=\\
&=\int [\rmd\tilde{\mathbi{c}}_0 \rmd\mathbi{c}_0]e^{\mathbi{c}_0\tilde{\mathbi{c}}_0} \exp\big(\mathbi{c}_0c_0^{(z_1)}+\tilde{\mathbi{c}}_0\tilde{c}_0^{(z_1)}\big)\exp\big(\mathbi{c}_0c_0^{(z_2)}+\tilde{\mathbi{c}}_0\tilde{c}_0^{(z_2)}\big).
\end{aligned}
\end{equation}
This relates to a comment in the footnote of p.~\pageref{foot:density}. 
That the quantum numbers (the coefficients) of the $c_0^{(z_1)}$ and $c_0^{(z_2)}$ are the same in this case (on the right-hand side of the first equality in (\ref{eq:1+c0ct0}) or in (\ref{eq:1+c0ct0z})) can also be seen from the $b_0^{(z_1)}-b_{0}^{(z_2)}=0$ gluing constraint (and the fact that the latter (anti-)commute with the remaining modes), and similarly for the anti-chiral half. In the second equality in (\ref{eq:1+c0ct0}) we inserted two delta function constraints that enable one to fully factorise the last two exponentials and endow the two vertex operators with independent quantum numbers (that become identical only inside the resolution of unity) as shown, using $\int \rmd\mathbi{c}\,\delta(\mathbi{c}-\mathbi{c}')f(\mathbi{c})=f(\mathbi{c}')$. This will be convenient notation-wise. (Recall that $\delta(\mathbi{c}-\mathbi{c}')=(\mathbi{c}-\mathbi{c}')$ for Grassmann-odd variables, $\mathbi{c},\mathbi{c}'$.)

\sk
Let us now collect the above results and substitute them back into (\ref{eq:intermbc}),
\begin{equation}\label{eq:intermbcd}
\begingroup\makeatletter\def\f@size{11}\check@mathfonts
\def\maketag@@@#1{\hbox{\m@th\large\normalfont#1}}%
\begin{aligned}
&
\suminnt\limits_a\,r^{h_a+\tilde{h}_a}\hat{\mathscr{A}}_a^{(z_1)}|1\rangle^1\otimes\hat{\mathscr{A}}^a_{(z_2/e^{i\theta)}}|1\rangle^2= 
\\
&=\frac{\alpha'}{8\pi i}\int \frac{\rmd^D\e{p}}{(2\pi)^D}\bigg(\prod_{n>0,\mu}\int \frac{\rmd{\mathbi{a}_n^{\mu *}}\wedge \rmd\mathbi{a}_n^{\mu}}{2\pi in}\frac{\rmd{\tilde{\mathbi{a}}_n^{\mu *}}\wedge \rmd\tilde{\mathbi{a}}_n^{\mu}}{2\pi in}\bigg)\exp\bigg[\sum_{n=1}^{\infty}\Big(\!-\frac{1}{n}\mathbi{a}_n\cdot \mathbi{a}_n^*-\frac{1}{n}\tilde{\mathbi{a}}_n\cdot \tilde{\mathbi{a}}_n^*\Big)\bigg]\\
&\quad\times \int \big[\rmd\tilde{\mathbi{c}}_0^{(1)} \rmd\mathbi{c}_0^{(1)}\rmd\tilde{\mathbi{c}}_0^{(2)} \rmd\mathbi{c}_0^{(2)}\big]\delta\big(\mathbi{c}_0^{(2)}-\mathbi{c}_0^{(1)}\big)\delta\big(\tilde{\mathbi{c}}_0^{(2)}-\tilde{\mathbi{c}}_0^{(1)}\big) \\
&\quad\times
\,\bigg( \prod_{n=1}^{\infty}\int \big[\rmd\mathbi{c}_n^{(1)}\rmd\mathbi{b}_n^{(2)} \rmd\mathbi{c}_n^{(2)}\rmd\mathbi{b}_n^{(1)}\rmd\tilde{\mathbi{c}}_n^{(1)}\rmd\tilde{\mathbi{b}}_n^{(2)} \rmd\tilde{\mathbi{c}}_n^{(2)}\rmd\tilde{\mathbi{b}}_n^{(1)}  \big]\bigg)\\
&\quad\times\,\exp\bigg[\sum_{n=1}^{\infty}\Big(\mathbi{b}_n^{(1)}\mathbi{c}_n^{(2)}+\mathbi{b}_n^{(2)}\mathbi{c}_n^{(1)}+\tilde{\mathbi{b}}_n^{(1)}\tilde{\mathbi{c}}_n^{(2)}+\tilde{\mathbi{b}}_n^{(2)}\tilde{\mathbi{c}}_n^{(1)}\Big)\bigg]r^{\frac{\alpha'}{2}\e{p}^2-2}\\
&\quad\times g_D\exp \Bigg[\sum_{n>0}\Big(\frac{1}{n}\mathbi{a}_n\cdot \alpha_{-n}^{(z_1)}+\mathbi{b}_n^{(1)}b_{-n}^{(z_1)}+\mathbi{c}_n^{(1)}c_{-n}^{(z_1)}\,\Big)e^{-in\e{q}\cdot \frac{1}{2}x_0^{(z_1)}}\Bigg]\\
&\hspace{2.31cm}\times \exp\Bigg[\sum_{n>0}\Big(\frac{1}{n}\tilde{\mathbi{a}}_n\cdot \tilde{\alpha}_{-n}^{(z_1)}+\tilde{\mathbi{b}}_n^{(1)}\tilde{b}_{-n}^{(z_1)}+\tilde{\mathbi{c}}_n^{(1)}\tilde{c}_{-n}^{(z_1)}\,\Big)e^{-in\e{q}\cdot \frac{1}{2}x_0^{(z_1)}}\Bigg]\\
&\hspace{2.33cm} \times \exp\big(\mathbi{c}_0^{(1)}c_0^{(z_1)}+\tilde{\mathbi{c}}_0^{(1)}\tilde{c}_0^{(z_1)}\big) \tilde{c}_1^{(z_1)}c_1^{(z_1)}e^{i\e{p}\cdot x_0^{(z_1)}}|1\rangle^1\\
&\quad\otimes g_D\exp\Bigg[\sum_{n>0}e^{in\theta}\Big(-\frac{1}{n}\mathbi{a}_n^*\cdot \alpha_{-n}^{(z_2)}+\mathbi{b}_n^{(2)}b_{-n}^{(z_2)}+\mathbi{c}_n^{(2)}c_{-n}^{(z_2)}\,\Big)e^{in\e{q}\cdot \frac{1}{2}x_0^{(z_2)}}\Bigg]\\
&\hspace{2.36cm}\times\exp\Bigg[\sum_{n>0}e^{-in\theta}\Big(-\frac{1}{n}\tilde{\mathbi{a}}_n^*\cdot \tilde{\alpha}_{-n}^{(z_2)}+\tilde{\mathbi{b}}_n^{(2)}\tilde{b}_{-n}^{(z_2)}+\tilde{\mathbi{c}}_n^{(2)}\tilde{c}_{-n}^{(z_2)}\,\Big)e^{in\e{q}\cdot \frac{1}{2}x_0^{(z_2)}}\Bigg]\\
&\hspace{2.36cm}\times  \exp\big(\mathbi{c}_0^{(2)}c_0^{(z_2)}+\tilde{\mathbi{c}}_0^{(2)}\tilde{c}_0^{(z_2)}\big)\tilde{c}_1^{(z_2)}c_1^{(z_2)}e^{-i\e{p}\cdot x_0^{(z_2)}}|1\rangle^2\end{aligned}
\endgroup
\end{equation}
We have brought the factorisation formula into such a form that we can simply read off the full set of (generically offshell) coherent states that can propagate through the cut handles, their duals, as well as their weights and measure associated to summing over their quantum numbers that characterise them. Onshell, these include physical, unphysical and BRST exact contributions (we return to this point). 
\sk

So we can finally specify precisely what we mean by `{\it resolution of unity}' and the corresponding `{\it sum over states}' and identify an explicit coherent state basis for the vertex operators comprising handle operators. Collecting the above we can rewrite the fixed-complex structure handle operators (\ref{eq:intermbcd}) as follows:
\begin{equation}\label{eq:dpdmua}
\begin{aligned}
\boxed{\,\,\suminnt\limits_a\,\,
\hat{\mathscr{A}}_a^{(z_1)}|1\rangle^{1}\otimes q^{L_0^{(z_2)}}\bar{q}^{\tilde{L}_0^{(z_2)}}\hat{\mathscr{A}}^a_{(z_2)}|1\rangle^{2}= 
\frac{\alpha'}{8\pi i}\int \frac{\rmd^D\e{p}}{(2\pi)^D}
\int \rmd\mu_{\mathbi{a}\mathbi{b}\mathbi{c}}\,\hat{\mathscr{A}}_a^{(z_1)}|1\rangle^{1}\otimes q^{L_0^{(z_2)}}\bar{q}^{\tilde{L}_0^{(z_2)}}\hat{\mathscr{A}}^a_{(z_2)}|1\rangle^{2}\,\,}
\end{aligned}
\end{equation}
where we glue with $z_1z_2=q$, and we have defined the measure:
\begin{equation}\label{eq:dmua}
\begin{aligned}
\int \rmd\mu_{\mathbi{a}\mathbi{b}\mathbi{c}}&\dfn 
\bigg(\prod_{n>0,\mu}\int \frac{\rmd{\mathbi{a}_n^{\mu *}}\wedge \rmd\mathbi{a}_n^{\mu}}{2\pi in}\frac{\rmd{\tilde{\mathbi{a}}_n^{\mu *}}\wedge \rmd\tilde{\mathbi{a}}_n^{\mu}}{2\pi in}\bigg)\exp\bigg[\sum_{n=1}^{\infty}\Big(\!-\frac{1}{n}\mathbi{a}_n\cdot \mathbi{a}_n^*-\frac{1}{n}\tilde{\mathbi{a}}_n\cdot \tilde{\mathbi{a}}_n^*\Big)\bigg]\\
&\qquad\times \int \big[\rmd\tilde{\mathbi{c}}_0^{(1)} \rmd\mathbi{c}_0^{(1)}\rmd\tilde{\mathbi{c}}_0^{(2)} \rmd\mathbi{c}_0^{(2)}\big]\delta(\mathbi{c}_0^{(2)}-\mathbi{c}_0^{(1)})\delta(\tilde{\mathbi{c}}_0^{(2)}-\tilde{\mathbi{c}}_0^{(1)}) \\
&\qquad\times
\,\bigg(\prod_{n=1}^{\infty}\int \big[\rmd\mathbi{c}_n^{(1)}\rmd\mathbi{b}_n^{(2)} \rmd\mathbi{c}_n^{(2)}\rmd\mathbi{b}_n^{(1)}\rmd\tilde{\mathbi{c}}_n^{(1)}\rmd\tilde{\mathbi{b}}_n^{(2)} \rmd\tilde{\mathbi{c}}_n^{(2)}\rmd\tilde{\mathbi{b}}_n^{(1)}  \big]\bigg)\\
&\qquad\times\,\exp\bigg[\sum_{n=1}^{\infty}\Big(\mathbi{b}_n^{(1)}\mathbi{c}_n^{(2)}+\mathbi{b}_n^{(2)}\mathbi{c}_n^{(1)}+\tilde{\mathbi{b}}_n^{(1)}\tilde{\mathbi{c}}_n^{(2)}+\tilde{\mathbi{b}}_n^{(2)}\tilde{\mathbi{c}}_n^{(1)}\Big)\bigg].
\end{aligned}
\end{equation}
The (fixed-picture) coherent states in turn read,
\begin{equation}\label{eq:offshellA_a}
\begin{aligned}
\hat{\mathscr{A}}_a^{(z_1)}|1\rangle^{1}&\dfn 
g_D\exp \Bigg[\sum_{n>0}\Big(\frac{1}{n}\mathbi{a}_n\cdot \alpha_{-n}^{(z_1)}+\mathbi{b}_n^{(1)}b_{-n}^{(z_1)}+\mathbi{c}_n^{(1)}c_{-n}^{(z_1)}\,\Big)e^{-in\e{q}\cdot \frac{1}{2}x_0^{(z_1)}}\Bigg]\\
&\qquad\times \exp\Bigg[\sum_{n>0}\Big(\frac{1}{n}\tilde{\mathbi{a}}_n\cdot \tilde{\alpha}_{-n}^{(z_1)}+\tilde{\mathbi{b}}_n^{(1)}\tilde{b}_{-n}^{(z_1)}+\tilde{\mathbi{c}}_n^{(1)}\tilde{c}_{-n}^{(z_1)}\Big)e^{-in\e{q}\cdot \frac{1}{2}x_0^{(z_1)}}\Bigg]\\
&\qquad\times \exp\big(\mathbi{c}_0^{(1)}c_0^{(z_1)}+\tilde{\mathbi{c}}_0^{(1)}\tilde{c}_0^{(z_1)}\big) \tilde{c}_1^{(z_1)}c_1^{(z_1)}e^{i\e{p}\cdot x_0^{(z_1)}}|1\rangle^{1},\\
\end{aligned}
\end{equation}
which are to be interpreted as being inserted at $p_1$ where $z_1=0$ in the $z_1$ frame, that is, at the origin of a local centred coordinate chart in the vicinity of the cut handle. The duals are in turn given by:
\begin{equation}\label{eq:offshellA^a q=1}
\begin{aligned}
\hat{\mathscr{A}}^a_{(z_2)}|1\rangle^{2}&\dfn 
g_D\exp\Bigg[\sum_{n>0}\Big(\!-\frac{1}{n}\mathbi{a}_n^*\cdot \alpha_{-n}^{(z_2)}+\mathbi{b}_n^{(2)}b_{-n}^{(z_2)}+\mathbi{c}_n^{(2)}c_{-n}^{(z_2)}\,\Big)e^{in\e{q}\cdot \frac{1}{2}x_0^{(z_2)}}\Bigg]\\
&\qquad\times\exp\Bigg[\sum_{n>0}\Big(\!-\frac{1}{n}\tilde{\mathbi{a}}_n^*\cdot \tilde{\alpha}_{-n}^{(z_2)}+\tilde{\mathbi{b}}_n^{(2)}\tilde{b}_{-n}^{(z_2)}+\tilde{\mathbi{c}}_n^{(2)}\tilde{c}_{-n}^{(z_2)}\,\Big)e^{in\e{q}\cdot \frac{1}{2}x_0^{(z_2)}}\Bigg]\\
&\qquad\times  \exp\big(\mathbi{c}_0^{(2)}c_0^{(z_2)}+\tilde{\mathbi{c}}_0^{(2)}\tilde{c}_0^{(z_2)}\big)\tilde{c}_1^{(z_2)}c_1^{(z_2)}e^{-i\e{p}\cdot x_0^{(z_2)}}|1\rangle^{2}\\
\end{aligned}
\end{equation}
which are to be interpreted as inserted at $p_2$ where $z_2=0$ in the $z_2$ frame. 
Perhaps it is useful to display explicitly the result of twisting and rescaling the frame coordinates, $z_2\mapsto z_2/q$, which is generated by acting with $q^{L_0}\bar{q}^{\tilde{L}_0^{(z_2)}}$ on (\ref{eq:offshellA^a q=1}), 
$$
q^{L_0^{(z_2)}}\bar{q}^{\tilde{L}_0^{(z_2)}}\hat{\mathscr{A}}^a_{(z_2)}=\hat{\mathscr{A}}^a_{(z_2/q)},
$$
with:
\begin{equation}\label{eq:offshellA^a}
\begin{aligned}
\hat{\mathscr{A}}^a_{(z_2/q)}|1\rangle^{2}&\dfn 
g_D\exp\Bigg[\sum_{n>0}e^{in\theta}\Big(\!-\frac{1}{n}\mathbi{a}_n^*\cdot \alpha_{-n}^{(z_2)}+\mathbi{b}_n^{(2)}b_{-n}^{(z_2)}+\mathbi{c}_n^{(2)}c_{-n}^{(z_2)}\,\Big)e^{in\e{q}\cdot \frac{1}{2}x_0^{(z_2)}}\Bigg]\\
&\qquad\times\exp\Bigg[\sum_{n>0}e^{-in\theta}\Big(\!-\frac{1}{n}\tilde{\mathbi{a}}_n^*\cdot \tilde{\alpha}_{-n}^{(z_2)}+\tilde{\mathbi{b}}_n^{(2)}\tilde{b}_{-n}^{(z_2)}+\tilde{\mathbi{c}}_n^{(2)}\tilde{c}_{-n}^{(z_2)}\,\Big)e^{in\e{q}\cdot \frac{1}{2}x_0^{(z_2)}}\Bigg]\\
&\qquad\times   (q\bar{q})^{\frac{\alpha'}{4}\e{p}^2-1}\exp\big(\mathbi{c}_0^{(2)}c_0^{(z_2)}+\tilde{\mathbi{c}}_0^{(2)}\tilde{c}_0^{(z_2)}\big)\tilde{c}_1^{(z_2)}c_1^{(z_2)}e^{-i\e{p}\cdot x_0^{(z_2)}}|1\rangle^{2}\\
\end{aligned}
\end{equation}
So these are inserted on a local coordinate patch on the opposite or dual side of the cut handle.
\sk

Let us make some observations: 
\begin{itemize}
\item \underline{\bf Scaling dimension:}\label{scaldim} The vertex operators (\ref{eq:offshellA^a})  have well-defined scaling dimension. To see this notice primarily that although the (matter or ghost) {\it modes}, $\mathscr{O}_{-n}^{(z_2)}$, can be pulled in and out of the normal ordering freely (see e.g., eqn (2.8.6) in \cite{Polchinski_v1}) (which means that their scaling dimension inside vertex operators is the naive one determined by (\ref{eq:On-scal-rot-shif})), the exponentials are more subtle. In particular, under $z_2\mapsto z_2/q$,
\begin{equation}
\begin{aligned}
:e^{-i(\e{p}-N\e{q})\cdot x^{(z_2/q)}}\!:\, &=:(q\bar{q})^{\frac{\alpha'}{4}(\e{p}-N\e{q})^2}e^{-i(\e{p}-N\e{q})\cdot x^{(z_2)}}\!:\,\\
&= \,:\!\big((q\bar{q})^{\frac{\alpha'}{4}\e{p}^2}e^{-i\e{p}\cdot x^{(z_2)}}\big)\big((q\bar{q})^{-N}e^{iN\e{q}\cdot x^{(z_2)}}\big)\!:,
\end{aligned}
\end{equation}
where in going from the first to the second equality we made use of the last two relations in (\ref{eq:kpq_constr}), 
implying the somewhat counterintuitive {\it effective} ``rule of thumb'',
\begin{equation}\label{eq:effective rule}
\begin{aligned}
e^{-i\e{p}\cdot x^{(z_2/q)}}&=(q\bar{q})^{\frac{\alpha'}{4}\e{p}^2}e^{-i\e{p}\cdot x^{(z_2)}}\\
&=r^{\frac{\alpha'}{2}\e{p}^2}e^{-i\e{p}\cdot x^{(z_2)}}\\
e^{in\e{q}\cdot \frac{1}{2}x^{(z_2/q)}}&=(q\bar{q})^{-n/2}
e^{in\e{q}\cdot \frac{1}{2}x^{(z_2)}}\\
&=r^{-n}e^{in\e{q}\cdot \frac{1}{2}x^{(z_2)}}\\
\mathscr{O}_{-n}^{(z_2/q)}e^{in\e{q}\cdot \frac{1}{2}x^{(z_2/q)}}&=\big(q^{n}\mathscr{O}_{-n}^{(z_2)}\big)\big((q\bar{q})^{-n/2}e^{in\e{q}\cdot \frac{1}{2}x^{(z_2)}}\big)\\
&=e^{in\theta}\mathscr{O}_{-n}^{(z_2)}e^{in\e{q}\cdot \frac{1}{2}x^{(z_2)}}
\end{aligned}
\qquad\textrm{(effective rule)}
\end{equation} 
These effective rules can be used inside vertex operators, e.g., inside the arguments of exponentials containing oscillators in the coherent states. 
The scaling of the exponential is inherited from the general relation, $:e^{ik\cdot x^{(z_2/q)}}\!\!:\,\,\equiv q^{L_0^{(z_2)}}\bar{q}^{\tilde{L}_0^{(z_2)}}\!\!:\!\!e^{ik\cdot x^{(z_2)}}\!\!:\,\,=(q\bar{q})^{\frac{\alpha'}{4}k^2}\!:\!\!e^{ik\cdot x^{(z_2)}}\!\!:$. In particular, $:\!e^{ip\cdot x}e^{ik\cdot x}\!:\,\neq \,:\!e^{ip\cdot x}\!:\,\,:\!e^{ik\cdot x}\!:$, which in turn implies that the two exponentials, $e^{-i\e{p}\cdot x^{(z_2/q)}}$ and $e^{in\e{q}\cdot \frac{1}{2}x^{(z_2)}}$, do not scale independently. Crucially, we learn that $\mathscr{O}_{-n}^{(z_2)}e^{in\e{q}\cdot \frac{1}{2}x_0^{(z_2)}}$ is {\it effectively} a scale-invariant combination, as will be arbitrary integer powers of it also,
\begin{equation}\label{eq:scale-phase exp}
\begin{aligned}
r^{L_0^{(z_2)}+\tilde{L}_0^{(z_2)}}\Big(\mathscr{O}_{-n}^{(z_2)}e^{in\e{q}\cdot \frac{1}{2}x^{(z_2)}}\Big)^a&=\Big(\mathscr{O}_{-n}^{(z_2)}e^{in\e{q}\cdot \frac{1}{2}x^{(z_2)}}\Big)^a\\
e^{i\theta(L_0^{(z_2)}-\tilde{L}_0^{(z_2)})}\Big(\mathscr{O}_{-n}^{(z_2)}e^{in\e{q}\cdot \frac{1}{2}x^{(z_2)}}\Big)^a&=\Big(e^{in\theta}\mathscr{O}_{-n}^{(z_2)}e^{in\e{q}\cdot \frac{1}{2}x^{(z_2)}}\Big)^a\\
\end{aligned}
\qquad\textrm{(effective rule)}
\end{equation}
for any positive integer $a$. The phase dependence on the right-hand side of the second relation is inherited entirely from the phase dependence of the mode. 
\sk

The first relation in (\ref{eq:scale-phase exp}) is the primary reason as to why the resulting coherent state vertex operators (\ref{eq:offshellA^a}) have well-defined scaling dimension, allowing us to interpret them as local operators. The precise statement is:
\begin{equation}\label{eq:L0Lb0A=DA}
\boxed{
\hat{\mathscr{A}}^a_{(z_2/r)}=r^{L_0^{(z_2)}+\tilde{L}_0^{(z_2)}}\hat{\mathscr{A}}^a_{(z_2)}=r^{\Delta_a}\hat{\mathscr{A}}^a_{(z_2)},\qquad{\rm with}\qquad \Delta_a=\frac{\alpha'}{2}\e{p}^2-2
}
\end{equation}
with an identical relation for the $z_1$ frame vertex operators. This provides a first principles derivation of the fundamental assumption (\ref{eq:Ah1}), see also {\bf (c)} on p.~\pageref{(c)comment}. 
It is saying that independently of the fact that coherent states are not mass eigenstates, they have well-defined scaling dimension and it therefore makes sense to identify them with {\it local operators} (in the sense of \cite{Polchinski88}). The second relation in (\ref{eq:scale-phase exp}) is the obstruction to the resulting coherent states not having a well-defined phase.
\sk

It is perhaps worth emphasising that the statement (\ref{eq:L0Lb0A=DA}) is true independently of whether the states are offshell or onshell. Furthermore, requiring that {\it all} coherent state vertex operators be scale-invariant, $\Delta_a=0$, is equivalent to the single condition of placing the {\it vacuum} on which the tower of states is built onshell, $\e{p}^2=4/\alpha'$, which the reader will recognise as the tachyon onshell condition of bosonic string theory. BRST invariance however is not guaranteed if we choose to put these states onshell, because of the relation $\{Q_B,b_0+\tilde{b}_0\}=L_0+\tilde{L}_0$ does not imply that $Q_B \mathscr{A}_a^{(z_1)}=0$ when $(L_0+\tilde{L}_0)\mathscr{A}_a^{(z_1)}=0$ (or even in addition when $(b_0+\tilde{b}_0)\mathscr{A}_a^{(z_1)}=0$).
\item \underline{\bf Duals:} Notice that the {\it matter} ``polarisation tensors'' of the $z_2$ patch dual vertex operators (\ref{eq:offshellA^a q=1}) are related by complex conjugation to those of the $z_1$ patch vertex operators (\ref{eq:offshellA_a}) (and also the momenta are flipped), whereas there is no such relation for the corresponding quantities in the ghost sector (as one would expect given the absence of a notion of complex conjugation for Grassmann variables \cite{Witten12c}), and it is satisfying to see this derived. 
\item \underline{\bf Ghost number:} Although the left-hand side of the equality in (\ref{eq:bc-cb chiral}) has definite ghost number, either of the two exponentials of the mode operators on the right-hand side {\it do not} have definite ghost number. A manifestly definite ghost number is restored only after the Grassmann-odd quantum numbers have been integrated out. This is why the vertex operators (\ref{eq:offshellA_a}) and their duals (\ref{eq:offshellA^a}) do not have definite ghost number. It is also why (as claimed in Sec.~\ref{sec:EVO}) external vertex operators need not have definite ghost number either.
\end{itemize}

\subsection{Handle Operators at Exceptional Momenta}\label{sec:HOEM}
The handle operators derived above associated to the states (\ref{eq:offshellA_a}) and (\ref{eq:offshellA^a q=1}) are valid at ``sufficiently generic'' momenta. So the question arises as to what states to use at {\it exceptional momenta}, for example when we wish to study tadpole degenerations, or when the exceptional cases 1) and 2) on p.~\pageref{item:deltak} are realised.
\sk

As discussed above, the reason as to why the above basis breaks down at zero momentum is because the change of variables, $k\mapsto \e{p}$, in (\ref{eq:kpq_constr}) is not valid at $\e{p}=0$. To consider handle operators at exceptional momenta we could therefore undo this change of variables, $k\mapsto \e{p}$, which is in turn effectively implemented by the replacements:
\begin{equation}\label{eq:p->k q -> exp}
\boxed{
\begin{aligned}
&e^{i\theta}\,e^{-i\e{q}\cdot \frac{1}{2}(x_0^{(z_1)}-x_0^{(z_2)})}\qquad\rightarrow \qquad q\\
&e^{-i\theta}\,e^{-i\e{q}\cdot \frac{1}{2}(x_0^{(z_1)}-x_0^{(z_2)})}\qquad\rightarrow \qquad \bar{q}.
\end{aligned}
}
\end{equation}
These replacements leave the resolution of unity {\it invariant} when we glue with $z_1z_2=q$. The precise statement is that we can rather glue with the following handle operator,
\begin{equation}\label{eq:dpdmualpha}
\begin{aligned}
\,\,\suminnt\limits_{\alpha}\,\,
\hat{\mathscr{A}}_\alpha^{(z_1)}|1\rangle^1\otimes\hat{\mathscr{A}}^\alpha_{(z_2/q)}|1\rangle^2= 
\frac{\alpha'}{8\pi i}\int \frac{\rmd^Dk}{(2\pi)^D}
\int \rmd\mu_{\mathbi{a}\mathbi{b}\mathbi{c}}\,\hat{\mathscr{A}}_{\alpha}^{(z_1)}|1\rangle^1\otimes\hat{\mathscr{A}}^{\alpha}_{(z_2/q)}|1\rangle^2\,\,
\end{aligned}
\end{equation}
with measure $\rmd\mu_{\mathbi{a}\mathbi{b}\mathbi{c}}$ as above, repeated here for later reference,
\begin{equation}\label{eq:dmuaC}
\begin{aligned}
\int \rmd\mu_{\mathbi{a}\mathbi{b}\mathbi{c}}&\dfn 
\bigg(\prod_{n>0,\mu}\int \frac{\rmd{\mathbi{a}_n^{\mu *}}\wedge \rmd\mathbi{a}_n^{\mu}}{2\pi in}\frac{\rmd{\tilde{\mathbi{a}}_n^{\mu *}}\wedge \rmd\tilde{\mathbi{a}}_n^{\mu}}{2\pi in}\bigg)\exp\bigg[\sum_{n=1}^{\infty}\Big(\!-\frac{1}{n}\mathbi{a}_n\cdot \mathbi{a}_n^*-\frac{1}{n}\tilde{\mathbi{a}}_n\cdot \tilde{\mathbi{a}}_n^*\Big)\bigg]\\
&\qquad\times \int \big[\rmd\tilde{\mathbi{c}}_0^{(1)} \rmd\mathbi{c}_0^{(1)}\rmd\tilde{\mathbi{c}}_0^{(2)} \rmd\mathbi{c}_0^{(2)}\big]\delta(\mathbi{c}_0^{(2)}-\mathbi{c}_0^{(1)})\delta(\tilde{\mathbi{c}}_0^{(2)}-\tilde{\mathbi{c}}_0^{(1)}) \\
&\qquad\times
\,\bigg(\prod_{n=1}^{\infty}\int \big[\rmd\mathbi{c}_n^{(1)}\rmd\mathbi{b}_n^{(2)} \rmd\mathbi{c}_n^{(2)}\rmd\mathbi{b}_n^{(1)}\rmd\tilde{\mathbi{c}}_n^{(1)}\rmd\tilde{\mathbi{b}}_n^{(2)} \rmd\tilde{\mathbi{c}}_n^{(2)}\rmd\tilde{\mathbi{b}}_n^{(1)}  \big]\bigg)\\
&\qquad\times\,\exp\bigg[\sum_{n=1}^{\infty}\Big(\mathbi{b}_n^{(1)}\mathbi{c}_n^{(2)}+\mathbi{b}_n^{(2)}\mathbi{c}_n^{(1)}+\tilde{\mathbi{b}}_n^{(1)}\tilde{\mathbi{c}}_n^{(2)}+\tilde{\mathbi{b}}_n^{(2)}\tilde{\mathbi{c}}_n^{(1)}\Big)\bigg].
\end{aligned}
\end{equation}
and the following explicit expressions for the coherent state, $\hat{\mathscr{A}}_{\alpha}^{(z_1)}$, and its dual, $\hat{\mathscr{A}}^{\alpha}_{(z_2/q)}$,
\begin{equation}\label{eq:offshellA_alpha}
\begin{aligned}
\hat{\mathscr{A}}_{\alpha}^{(z_1)}|1\rangle^1&\dfn 
g_D\exp \bigg[\sum_{n>0}\Big(\frac{1}{n}\mathbi{a}_n\cdot \alpha_{-n}^{(z_1)}+\mathbi{b}_n^{(1)}b_{-n}^{(z_1)}+\mathbi{c}_n^{(1)}c_{-n}^{(z_1)}\,\Big)\bigg]\\
&\qquad\times \exp\bigg[\sum_{n>0}\Big(\frac{1}{n}\tilde{\mathbi{a}}_n\cdot \tilde{\alpha}_{-n}^{(z_1)}+\tilde{\mathbi{b}}_n^{(1)}\tilde{b}_{-n}^{(z_1)}+\tilde{\mathbi{c}}_n^{(1)}\tilde{c}_{-n}^{(z_1)}\Big)\bigg]\\
&\qquad\times \exp\big(\mathbi{c}_0^{(1)}c_0^{(z_1)}+\tilde{\mathbi{c}}_0^{(1)}\tilde{c}_0^{(z_1)}\big) \tilde{c}_1^{(z_1)}c_1^{(z_1)}e^{ik\cdot x_0^{(z_1)}}|1\rangle^1,\\
&\phantom{a}\\
\hat{\mathscr{A}}^{\alpha}_{(z_2/q)}|1\rangle^2&\dfn 
g_D\exp\bigg[\sum_{n>0}q^n\Big(\!-\frac{1}{n}\mathbi{a}_n^*\cdot \alpha_{-n}^{(z_2)}+\mathbi{b}_n^{(2)}b_{-n}^{(z_2)}+\mathbi{c}_n^{(2)}c_{-n}^{(z_2)}\,\Big)\bigg]\\
&\qquad\times\exp\bigg[\sum_{n>0}\bar{q}^n\Big(\!-\frac{1}{n}\tilde{\mathbi{a}}_n^*\cdot \tilde{\alpha}_{-n}^{(z_2)}+\tilde{\mathbi{b}}_n^{(2)}\tilde{b}_{-n}^{(z_2)}+\tilde{\mathbi{c}}_n^{(2)}\tilde{c}_{-n}^{(z_2)}\,\Big)\bigg]\\
&\qquad\times   (q\bar{q})^{\frac{\alpha'}{4}k^2-1}\exp\big(\mathbi{c}_0^{(2)}c_0^{(z_2)}+\tilde{\mathbi{c}}_0^{(2)}\tilde{c}_0^{(z_2)}\big)\tilde{c}_1^{(z_2)}c_1^{(z_2)}e^{-ik\cdot x_0^{(z_2)}}|1\rangle^2\\
\end{aligned}
\end{equation}
These states are valid at exceptional momenta. 
We are using  labels `$\alpha$' on the left-hand sides of these relations and in (\ref{eq:dpdmualpha}) rather than `$a$' to distinguish this set from the set defined in (\ref{eq:offshellA^a q=1}) and (\ref{eq:offshellA^a}). The measure, $\rmd\mu_{\mathbi{a}\mathbi{b}\mathbi{c}}$, is as defined in (\ref{eq:dmua}). These states are not yet normal-ordered (since $b_{-1},\tilde{b}_{-1}$ appears which annihilates the SL(2,$\mathbf{C}$) vacuum) -- the normal-ordered expressions are discussed in Sec.~\ref{sec:NOHO}.
\sk

These coherent states do not have well-defined scaling dimension, and (roughly speaking) $\hat{\mathscr{A}}^{\alpha}_{(z_2/q)}$ is effectively smeared out over a region $|z_2|\leq |q|$, the point being that the ``size'', $|q|$, of the pinched region cannot be disentangled from both coherent states on both sides of the pinch, so there is an associated non-locality on $\Sigma$ associated to this. The price of demanding locality on the worldsheet, as we can see in (\ref{eq:offshellA_a}) and (\ref{eq:offshellA^a}) (to which there are associated local operators on $\Sigma$) is non-locality in {\it spacetime}.  Because recall that $\e{q}\cdot\e{p}=2/\alpha'$, so that (in a convenient coordinate system) $\e{q}^-=-2/(\alpha'\e{p}^+)$. The fact that $\e{p}^+$ appears in the denominator in the argument of the exponential in (\ref{eq:offshellA_a}) or (\ref{eq:offshellA^a}) is suggestive of non-locality in spacetime over distances $L\sim \mathcal{O}(\alpha'\e{p}^+)$. This type of non-locality is complementary to that associated to the extended nature of strings in spacetime. One should keep in mind that these are coordinate-dependent statements. 
\sk

\begin{figure}
\begin{center}
\includegraphics[angle=0,origin=c,width=0.55\textwidth]{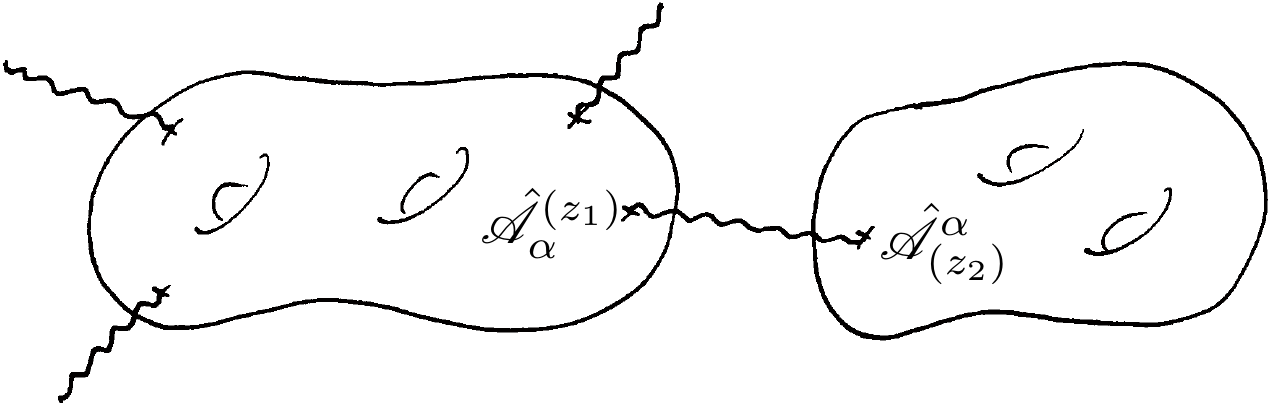}
\caption{The handle operator associated to the basis (\ref{eq:offshellA_alpha}) is valid at zero momentum and can therefore also be used to study tadpole degenerations (and corresponding background shifts that cancel the corresponding BRST anomaly \cite{Polchinski88,LaNelson90}).}\label{fig:tadpoles}
\end{center}
\end{figure}

The choice of basis depends on the question one wishes to ask. For example, the interchange of the sum and integral in going from (\ref{eq:intermbb}) to (\ref{eq:interm}) (which required an analytic continuation in momenta) has not been carried out here. Therefore, the basis (\ref{eq:offshellA_alpha}) can also be used at exceptional momenta, for example for handle operators associated to {\it tadpole} degenerations, see Fig.~\ref{fig:tadpoles},  which corresponds to cutting the path integral across a cycle that pinches of a Riemann surface with no external vertex operators. This is in turn useful in studies of background shifts in relation to the Fischler-Susskind mechanism \cite{FischlerSusskind86a,FischlerSusskind86b,Polchinski88,LaNelson90,Tseytlin90b,PiusRudraSen14}.
\sk

In what follows we usually make explicit reference to the handle operators of the previous section, but the replacement (\ref{eq:p->k q -> exp}) can be carried out when it is desirable to change basis to that of the current section. It is worth pointing out that there are cases where we can work with the handle operators of the previous section even at zero momentum as we demonstrate with an explicit example in Sec.~\ref{sec:S1ptA}, because there the obstruction associated to the constraint $\e{p}\cdot \e{q}=2/\alpha'$ does not play a role.

\subsection{Target Space Interpretation of Handle Operators}
In this section we derive yet another explicit basis for handle operators, a basis in which their target-space interpretation is immediate.
\sk

In (\ref{eq:handle}) we adopted a momentum space approach to the construction of handle operators, but this is not necessary, and in fact obscures some underlying physics that becomes clearer in position space. There is a long history about how position space conformal vertex operators have difficulties with satisfying physical state conditions when these are associated to asymptotic states. This is because in order to transition to position space one performs a Fourier transform, integrating over all momenta. But the physical state conditions single out a specific subset of allowed momenta, so the resulting vertex operators typically cannot easily be placed onshell \cite{Polchinski_v1}. 
\sk

Handle operators however are intrinsically offshell objects, so the above obstruction is absent. Furthermore, if one wishes to understand how loop corrections might manifest themselves in terms of target space intuition it is useful to change basis and somewhat reorganise the computation. Let us be specific.
\sk

Consider again the explicit expression from which fixed-picture handle operators were derived, namely (\ref{eq:handle}),
\begin{equation}\label{eq:handle_position}
\begin{aligned}
\suminnt\limits_{a}&\,\,\,\,\hat{\mathscr{A}}_a^{(z_1)}\hat{\mathscr{A}}^a_{(z_2/q)}= \\
&= \frac{\alpha'g_D^2}{8\pi i}\int \frac{\rmd^Dk}{(2\pi)^D}(q\bar{q})^{\frac{\alpha'}{4}k^2-1}e^{ik\cdot (x_0^{(z_1)}-x_0^{(z_2)})}\\
&\quad\times\exp \Big[\sum_{n=1}^{\infty}q^n\Big(\!-\frac{1}{n}\alpha_{-n}^{(z_1)}\cdot \alpha_{-n}^{(z_2)}+c_{-n}^{(z_1)}b_{-n}^{(z_2)}-b_{-n}^{(z_1)}c_{-n}^{(z_2)}\Big)\Big]\\
&\quad\times \exp\Big[\sum_{n=1}^{\infty}\bar{q}^n\Big(\!-\frac{1}{n}\tilde{\alpha}_{-n}^{(z_1)}\cdot \tilde{\alpha}_{-n}^{(z_2)}+\tilde{c}_{-n}^{(z_1)}\tilde{b}_{-n}^{(z_2)}-\tilde{b}_{-n}^{(z_1)}\tilde{c}_{-n}^{(z_2)}\Big)\Big]\\
&\quad\times(\tilde{c}_0^{(z_1)}+\tilde{c}_0^{(z_2)})(c_0^{(z_1)}+c_0^{(z_2)})\tilde{c}_1^{(z_1)}c_1^{(z_1)}|1\rangle^{1}\tilde{c}_1^{(z_2)}c_1^{(z_2)}|1\rangle^{2},
\end{aligned}
\end{equation}
and then focus on the spacetime momentum-dependent terms,
\begin{equation}\label{eq:intk->x}
\begin{aligned}
\int \frac{\rmd^Dk}{(2\pi)^D}\,(q\bar{q})^{\frac{\alpha'}{4}k^2}&:e^{ik\cdot x_{(z_1)}}:_{z_1}\,:e^{-ik\cdot x_{(z_2)}}:_{z_2}
\end{aligned}
\end{equation}
We want to primarily integrate out $k$ in this expression, but we wish to do so in a manner that preserves the tensor product structure of the two local operators. The additional oscillator and ghost contributions are to be considered implicitly present, but they will not add to the story at this stage. The trick will be to integrate out $k$ and then introduce a new integral which (after some changes of variables) will physically correspond to the ``centre of mass'' (target space) position of the two local operators. 
The reader can check that this can be achieved by primarily rewriting (\ref{eq:intk->x}) as follows (in Euclidean space),
\begin{equation}\label{eq:intk->x2}
\begin{aligned}
&\int \frac{\rmd^Dk}{(2\pi)^D}\,(q\bar{q})^{\frac{\alpha'}{4}k^2}:e^{ik\cdot x_{(z_1)}}:_{z_1}\,:e^{-ik\cdot x_{(z_2)}}:_{z_2}\\
&\qquad =(\alpha'\tau_2)^{D/2}\int \prod_{\mu=0}^{D-1}\frac{\rmd\bar{\lambda}_\mu \wedge \rmd\lambda_{\mu}}{2\pi i}\,e^{-(2\pi\alpha'\tau_2)\lambda\cdot \bar{\lambda}}:e^{-(4\pi\alpha'\tau_2)^{-1}x_{(z_1)}^2 +\lambda\cdot x_{(z_1)}}:_{z_1}\\
&\hspace{7.7cm}\times :e^{-(4\pi\alpha'\tau_2)^{-1}x_{(z_2)}^2 +\bar{\lambda}\cdot x_{(z_2)}}:_{z_2},
\end{aligned}
\end{equation}
where $q=e^{2\pi i\tau}$ and $\tau=\tau_1+i\tau_2$, and then (taking into account that $\tau_2>0$) defining new variables of integration (corresponding to spacetime ``centre of mass'' position and momentum respectively),
$$
Y_\mu\dfn 2\pi\alpha'{\rm Re}(\lambda_\mu),\qquad k_\mu\dfn {\rm Im}(\lambda_\mu),\qquad\mu=0,1,\dots,D-1,
$$
in terms of which the relation (\ref{eq:intk->x2}) takes the following guise,
\begin{equation}\label{eq:intk->x3}
\boxed{
\begin{aligned}
&\int \frac{\rmd^Dk}{(2\pi)^D}\,(q\bar{q})^{\frac{\alpha'}{4}k^2}:e^{ik\cdot x_{(z_1)}}:_{z_1}\,:e^{-ik\cdot x_{(z_2)}}:_{z_2}=\\
&\qquad =\int \rmd^DY\int \frac{\rmd^Dk}{(2\pi)^D}\,e^{-2\pi\alpha'\tau_2k^2}(\pi^2\alpha'\tau_2)^{-D/2}\,\\
&\quad\qquad\times:\exp\Big[-\frac{(x_{(z_1)}-Y)^2}{4\pi\alpha'\tau_2} +ik\cdot x_{(z_1)}\Big]:_{z_1}\,\,:\exp\Big[-\frac{(x_{(z_2)}-Y)^2}{4\pi\alpha'\tau_2} -ik\cdot x_{(z_2)}\Big]:_{z_2}
\end{aligned}
}
\end{equation}
Although this is perhaps a subtle operator to work with (the two operators which produce the handle do not have well-defined scaling dimension), the physics becomes nevertheless particularly transparent and so offers some additional physical intuition. In particular, (\ref{eq:intk->x3}) is telling us that the dominant contribution associated to a handle insertion (at fixed worldsheet complex structure) will come from regions in the path integral over $x$ (after inclusion of matter and ghost oscillators and ghost zero modes) where the two local operators are ``close by'' in spacetime (with respect to $\sqrt{\alpha'\tau_2}$), where $x_{(z_1)}\sim x_{(z_2)}\sim Y$, and this position, $Y$, is in turn integrated over. On the other hand, large $k^2$ (i.e.~the UV contribution) is exponentially suppressed (after analytic continuation to Euclidean signature). Furthermore, contributions coming from regimes where these operators are far apart in spacetime are exponentially suppressed. Another point which provides additional intuition is the similarity of the $Y$ dependence in (\ref{eq:intk->x3}) with a corresponding exponential factor appearing in the one-loop partition function in the presence of D-branes \cite{Polchinski94,Polchinski95}; this is not accidental. Finally, regarding worldsheet moduli space it is evident that the dominant contribution (for fixed $x_{(z_1)},x_{(z_2)},Y$) will come from large $\tau_2$, whereby the corresponding (trivial or non-trivial) homology cycle of the handle degenerates. These properties should broadly remain qualitatively true after including the tower of oscillators and for any finite number of handle operators insertions (e.g., for any finite number of string loops).

\subsection{Normal-Ordered Handle Operators}\label{sec:NOHO}
Normal ordering and changes in normal ordering are discussed in general terms in Sec.~\ref{sec:NOII}, as is the link to Polchinski's notion of `conformal/Weyl normal ordering' \cite{Polchinski87,Polchinski88}. Here we are interested in normal ordering the vertex operators used to construct handle operators. One can show that conformal or Weyl (when we use holomorphic normal coordinates to fix the frame) normal-ordered vertex operators in radial coordinates are equivalent to creation-annihilation normal-ordered vertex operators where all annihilation operators are to the right of creation operators in each of the two local operators. Since creation-annihilation normal ordering is more efficient we focus on this here.
\sk

We begin with the fixed-picture coherent states (\ref{eq:offshellA_a}),
\begin{equation}\label{eq:offshellA_a2}
\begin{aligned}
|\hat{\mathscr{A}}_a^{(z_1)}\rangle^1&\dfn 
g_D
\exp \Bigg[\sum_{n>0}\Big(\frac{1}{n}\mathbi{a}_n\cdot \alpha_{-n}^{(z_1)}+\mathbi{b}_n^{(1)}b_{-n}^{(z_1)}+\mathbi{c}_n^{(1)}c_{-n}^{(z_1)}\,\Big)e^{-in\e{q}\cdot \frac{1}{2}x_0^{(z_1)}}\Bigg]\\
&\,\,\,\,\quad\times \exp\Bigg[\sum_{n>0}\Big(\frac{1}{n}\tilde{\mathbi{a}}_n\cdot \tilde{\alpha}_{-n}^{(z_1)}+\tilde{\mathbi{b}}_n^{(1)}\tilde{b}_{-n}^{(z_1)}+\tilde{\mathbi{c}}_n^{(1)}\tilde{c}_{-n}^{(z_1)}\Big)e^{-in\e{q}\cdot \frac{1}{2}x_0^{(z_1)}}\Bigg]\\
&\,\,\,\,\quad\times \exp\big(\mathbi{c}_0^{(1)}c_0^{(z_1)}+\tilde{\mathbi{c}}_0^{(1)}\tilde{c}_0^{(z_1)}\big) \tilde{c}_1^{(z_1)}c_1^{(z_1)}e^{i\e{p}\cdot x_0^{(z_1)}}|1\rangle^{1},\\
\end{aligned}
\end{equation}
where it is manifest that the exponentials contain not only creation operators but also the annihilation operators $b_{-1}^{(z_1)}$ and $\tilde{b}_{-1}^{(z_1)}$ (which annihilate the SL(2,$\mathbf{C}$) vacuum, $|1\rangle^1$, not the ghost vacuum), so the states are not explicitly normal-ordered.\footnote{Incidentally, the reason these annihilation operators are present in (\ref{eq:offshellA_a2}) in the first place is because in the resolution of unity (\ref{eq:handle}) we see that if we are to include all creation operators, in particular $c_{-1}$, we must also include a corresponding $b_{-1}$ since the two are coupled by the gluing conditions (\ref{eq:abcmodes1->2}).} In making use of the operator-state correspondence, that will in turn enable us to read off the corresponding vertex operators from the states we may like to bring the annihilation operators into plain view and commute them through to their respective vacua, so that the corresponding vertex operators come out in normal-ordered form. We can accomplish this in various ways and we will mention two. 
\sk

For the first of the two, let us return to (\ref{eq:intermbcd}) and integrate out the various ghost quantum numbers associated to the level number $n=1$ terms, $\bm{b}_1^{(1)},\tilde{\bm{b}}_1^{(1)},\bm{c}_1^{(1)},\tilde{\bm{c}}_1^{(1)},\bm{b}_1^{(2)},\tilde{\bm{b}}_1^{(2)},\bm{c}_1^{(2)},\tilde{\bm{c}}_1^{(2)}$. One finds:
\begin{equation}\label{eq:16terms}
\begingroup\makeatletter\def\f@size{11}\check@mathfonts
\def\maketag@@@#1{\hbox{\m@th\large\normalfont#1}}%
\begin{aligned}
\int \big[\rmd\mathbi{c}_1^{(1)}&\rmd\mathbi{b}_1^{(2)} \rmd\mathbi{c}_1^{(2)}\rmd\mathbi{b}_1^{(1)} \rmd\tilde{\mathbi{c}}_1^{(1)}\rmd\tilde{\mathbi{b}}_1^{(2)} \rmd\tilde{\mathbi{c}}_1^{(2)}\rmd\tilde{\mathbi{b}}_1^{(1)} \big]\exp\Big(\mathbi{b}_1^{(1)}\mathbi{c}_1^{(2)}+\mathbi{b}_1^{(2)}\mathbi{c}_1^{(1)}+\tilde{\mathbi{b}}_1^{(1)}\tilde{\mathbi{c}}_1^{(2)}+\tilde{\mathbi{b}}_1^{(2)}\tilde{\mathbi{c}}_1^{(1)}\Big)\\
&\hspace{-1cm}\times \exp\bigg[\Big(\mathbi{b}_1^{(1)}b_{-1}^{(z_1)}+\mathbi{c}_1^{(1)}c_{-1}^{(z_1)}\Big)e^{-i\e{q}\cdot \frac{1}{2}x_0^{(z_1)}}+\Big(\tilde{\mathbi{b}}_1^{(1)}\tilde{b}_{-1}^{(z_1)}+\tilde{\mathbi{c}}_1^{(1)}\tilde{c}_{-1}^{(z_1)}\Big)e^{-i\e{q}\cdot \frac{1}{2}x_0^{(z_1)}}\bigg]\tilde{c}_1^{(z_1)}c_1^{(z_1)}|1\rangle^{1}\\
&\hspace{-1cm}\times \exp\bigg[e^{i\theta}\Big(\mathbi{b}_1^{(2)}b_{-1}^{(z_2)}+\mathbi{c}_1^{(2)}c_{-1}^{(z_2)}\Big)e^{i\e{q}\cdot \frac{1}{2}x_0^{(z_2)}}+e^{-i\theta}\Big(\tilde{\mathbi{b}}_1^{(2)}\tilde{b}_{-1}^{(z_2)}+\tilde{\mathbi{c}}_1^{(2)}\tilde{c}_{-1}^{(z_2)}\Big)e^{i\e{q}\cdot \frac{1}{2}x_0^{(z_2)}}\bigg]\tilde{c}_1^{(z_2)}c_1^{(z_2)}|1\rangle^{2}\\
=\,\,&\Big(\tilde{c}_1^{(z_1)}+e^{-i\theta}\tilde{c}_{-1}^{(z_2)}e^{-i\e{q}\cdot   \frac{1}{2}(x_0^{(z_1)}-x_0^{(z_2)})}\Big)\Big(c_1^{(z_1)}+e^{i\theta}c_{-1}^{(z_2)}e^{-i\e{q}\cdot   \frac{1}{2}(x_0^{(z_1)}-x_0^{(z_2)})}\Big)\\
&\times \Big(\tilde{c}_1^{(z_2)}+e^{-i\theta}\tilde{c}_{-1}^{(z_1)}e^{-i\e{q}\cdot   \frac{1}{2}(x_0^{(z_1)}-x_0^{(z_2)})}\Big)\Big(c_1^{(z_2)}+e^{i\theta}c_{-1}^{(z_1)}e^{-i\e{q}\cdot   \frac{1}{2}(x_0^{(z_1)}-x_0^{(z_2)})}\Big)|1\rangle^{1}\otimes |1\rangle^{2}
\end{aligned}
\endgroup
\end{equation}
As a side remark, when proceeding in this manner (i.e.~integrating out quantum numbers characterising the states propagating through pinches) one generically ends up with effective vertex operator descriptions, and the more quantum numbers one integrates out the less information is retained in the remaining expressions (although there are large redundancies associated to gauge invariance and the corresponding ability to add BRST-exact terms). In the extreme case where one integrates over {\it  all} quantum numbers one ends up with the identity operator (which is precisely analogous to a boundary state in the presence of D-branes, but in the closed string context that we are focusing on in this document there are no true boundaries). 

\sk
As one sees from (\ref{eq:16terms}) there are sixteen (distinct) terms, and one can read off the various normal ordered vertex operators (and their duals)  directly from (\ref{eq:16terms}) and (\ref{eq:intermbcd}). This approach is useful for low mass levels (it was also adopted in \cite{Polchinski88} where the massless case was examined in detail). E.g., when cutting across separating cycles it is often the case \cite{Polchinski88} that only one of these sixteen terms will be contributing to the amplitude due to ghost charge conservation. For massive states and/or for non-separating degenerations ghost charge conservation is much less constraining and generically all terms will contribute, so this way of proceeding is not efficient in the generic case.
\sk

A more useful approach to obtain explicitly normal-ordered vertex operators will be to {\it not} integrate out the quantum numbers, $\bm{b}_1^{(1)},\tilde{\bm{b}}_1^{(1)},\bm{c}_1^{(1)},\tilde{\bm{c}}_1^{(1)},\bm{b}_1^{(2)},\tilde{\bm{b}}_1^{(2)},\bm{c}_1^{(2)},\tilde{\bm{c}}_1^{(2)}$. This is the second of the two approaches mentioned above and leads to the most compact and explicitly normal-ordered vertex operators, $\hat{\mathscr{A}}_a^{(z_1)}$ and $\hat{\mathscr{A}}^a_{(z_2)}$. 
\sk

Focusing on the $n=1$ ghost contributions in (\ref{eq:offshellA_a2}), we expand the exponentials, shrink the mode operator contours around $z_1=\bar{z}_1=0$ and evaluate the associated (singular and non-singular) OPE's. One finds (with the definitions (\ref{eq:ABCcondensed_notation})):
\begin{equation}\label{eq:b-1c-1terms}
\begingroup\makeatletter\def\f@size{11}\check@mathfonts
\def\maketag@@@#1{\hbox{\m@th\large\normalfont#1}}%
\begin{aligned}
&\exp\Big(\e{B}_1^{(1)}b_{-1}^{(z_1)}+\e{C}_1^{(1)}c_{-1}^{(z_1)}+\tilde{\e{B}}_1^{(1)}\tilde{b}_{-1}^{(z_1)}+\tilde{\e{C}}_1^{(1)}\tilde{c}_{-1}^{(z_1)}\Big)\tilde{c}_1^{(z_1)}c_1^{(z_1)}|1\rangle^{1}\\
&\quad = \exp\Big(\tilde{\e{C}}^{(1)}_1\tilde{c}_{-1}^{(z_1)}+\e{C}_1^{(1)}c_{-1}^{(z_1)}\Big)\Big(\tilde{c}_1^{(z_1)}+\tilde{\e{B}}_1^{(1)}\Big)\Big(c_1^{(z_1)}+\e{B}_1^{(1)}\Big)|1\rangle^1
\end{aligned}
\endgroup
\end{equation}
Substituting this into the vertex operator expression (\ref{eq:offshellA_a2}) leads to the following explicitly normal-ordered coherent state basis:
\begin{equation}\label{eq:offshellA_a3}
\boxed{
\begin{aligned}
&|\hat{\mathscr{A}}_a^{(z_1)}\rangle^1\dfn 
g_D\tilde{\e{U}}_{(1)}\e{U}_{(1)}\,e^{i\e{p}\cdot x_0^{(z_1)}}|1\rangle^{1}
\end{aligned}
}
\end{equation}
where, noting that various terms exponentiate, we have defined,
\begin{equation}\label{eq:offshellA_a_}
\boxed{
\begin{aligned}
\tilde{\e{U}}_{(1)}&\dfn \exp \bigg(\sum_{n\geq1}\frac{1}{n}\tilde{\e{A}}_n^{(1)}\!\cdot \!\tilde{\alpha}_{-n}^{(z_1)}+\sum_{n\geq2}\tilde{\e{B}}_n^{(1)}\tilde{b}_{-n}^{(z_1)}+\sum_{n\geq0}\tilde{\e{C}}_n^{(1)}\tilde{c}_{-n}^{(z_1)}\,\bigg)\big(\tilde{c}_1^{(z_1)}+\tilde{\e{B}}_1^{(1)}\big)
\\
\e{U}_{(1)}&\dfn  \exp \bigg(\sum_{n\geq1}\frac{1}{n}\e{A}_n^{(1)}\!\cdot \!\alpha_{-n}^{(z_1)}+\sum_{n\geq2}\e{B}_n^{(1)}b_{-n}^{(z_1)}+\sum_{n\geq0}\e{C}_n^{(1)}c_{-n}^{(z_1)}\bigg)\big(c_1^{(z_1)}+\e{B}_1^{(1)}\big)
\end{aligned}
}
\end{equation}
We have found it convenient to condense the notation slightly and define the following  quantities (with the obvious Grassmannality):
\begin{equation}\label{eq:ABCcondensed_notation}
\begin{aligned}
&\e{A}_n^{(1)}\dfn \mathbi{a}_ne^{-inq\cdot \frac{1}{2}x_0^{(z_1)}},\qquad \tilde{\e{A}}_n^{(1)}\dfn \tilde{\mathbi{a}}_ne^{-inq\cdot \frac{1}{2}x_0^{(z_1)}}\\
&\e{B}_n^{(1)}\dfn \mathbi{b}_n^{(1)}e^{-inq\cdot \frac{1}{2}x_0^{(z_1)}},\qquad \tilde{\e{B}}_n^{(1)}\dfn \tilde{\mathbi{b}}_n^{(1)}e^{-inq\cdot \frac{1}{2}x_0^{(z_1)}}\\
&\e{C}_n^{(1)} \dfn \mathbi{c}_n^{(1)}e^{-inq\cdot \frac{1}{2}x_0^{(z_1)}},\qquad \tilde{\e{C}}^{(1)}_n \dfn \tilde{\mathbi{c}}_n^{(1)}e^{-inq\cdot \frac{1}{2}x_0^{(z_1)}}
\end{aligned}
\end{equation}
We will abuse the notation slightly and use the same symbols (\ref{eq:ABCcondensed_notation}) for the corresponding vertex operators, but note that in the latter context we are to interpret the factors $e^{-inq\cdot \frac{1}{2}x_0^{(z_1)}}$ as $e^{-inq\cdot \frac{1}{2}x^{(z_1)}}$ (with of course similar remarks for the $z_2$ frame vertex operator).
\sk

As a side remark, in Sec.~\ref{sec:VSG} we will average over the phase of the local operator since we will be cutting across a trivial homology cycle. This can be implemented by primarily replacing $z_1$ by $z_1/e^{i\phi}$, and integrating over $\phi$ with measure $\int_0^{2\pi}\rmd\phi/(2\pi)$. This replacement can in turn also be implemented by leaving $z_1$ unchanged and rather replacing the quantum numbers:
\begin{equation}\label{eq:abc phase repl}
(\mathbi{a}_n,\mathbi{b}_n,\mathbi{c}_n,\tilde{\mathbi{a}}_n,\tilde{\mathbi{b}}_n,\tilde{\mathbi{c}}_n)\quad\rightarrow \quad(e^{in\phi}\mathbi{a}_n,e^{in\phi}\mathbi{b}_n,e^{in\phi}\mathbi{c}_n,e^{-in\phi}\tilde{\mathbi{a}}_n,e^{-in\phi}\tilde{\mathbi{b}}_n,e^{-in\phi}\tilde{\mathbi{c}}_n),
\end{equation}
in (\ref{eq:ABCcondensed_notation}), and to be precise:
\begin{equation}\label{eq:Aphasez1}
\begin{aligned}
\hat{\mathscr{A}}_a^{(z_{\sigma_1}/e^{i\phi})}(\sigma_1) =\hat{\mathscr{A}}_a^{(z_{\sigma_1})}(\sigma_1)\Big|_{\substack{
         (\mathbi{a}_n,\mathbi{b}_n,\mathbi{c}_n)\,\,\,\rightarrow \,\,\,e^{in\phi}(\mathbi{a}_n,\mathbi{b}_n,\mathbi{c}_n)\\
       \, \, \,\,(\tilde{\mathbi{a}}_n,\tilde{\mathbi{b}}_n,\tilde{\mathbi{c}}_n)\,\,\,\rightarrow \,\,\,e^{-in\phi}(\tilde{\mathbi{a}}_n,\tilde{\mathbi{b}}_n,\tilde{\mathbi{c}}_n)}
        }
\end{aligned}
\end{equation}

Gathering the above, a complete (or rather over-complete) set of  local offshell fixed-picture normal-ordered coherent state vertex operators can be written down immediately, 
\begin{equation}\label{eq:offshellA_a_VO}
\boxed{
\begin{aligned}
&\hat{\mathscr{A}}_a^{(z_1)}(p_1)\dfn 
g_D:\!\tilde{\e{U}}_{(1)}^{(z_1)}\e{U}_{(1)}^{(z_1)}\,e^{i\e{p}\cdot x^{(z_1)}}(p_1)\!:_{z_1}
\end{aligned}
}
\end{equation}
where, taking into account the operator-state correspondence relations (\ref{eq:annihil.ops,op-state}):
\begin{equation}\label{eq:offshellA_a_VOUU}
\begingroup\makeatletter\def\f@size{11}\check@mathfonts
\def\maketag@@@#1{\hbox{\m@th\large\normalfont#1}}%
\boxed{
\begin{aligned}
\tilde{\e{U}}_{(1)}^{(z_1)}& \dfn\exp\Bigg[\sum_{n\geq2}\Big(\frac{\tilde{\mathbi{b}}_n^{(1)}}{(n-2)!}\bar{\nabla}^{n-2}\tilde{b}\,\Big)e^{-in\e{q}\cdot \frac{1}{2}x^{(z_1)}}+\sum_{n\geq0}\Big(\frac{\tilde{\mathbi{c}}_n^{(1)}}{(n+1)!}\bar{\nabla}^{n+1}\tilde{c}\,\Big)e^{-in\e{q}\cdot \frac{1}{2}x^{(z_1)}}\Bigg]\\
&\qquad\times\exp \Bigg[\sum_{n\geq1}\frac{1}{n}\!\frac{i\tilde{\mathbi{a}}_n}{(n-1)!}\cdot\sqrt{\frac{2}{\alpha'}}\bar{\nabla}^nx\,e^{-in\e{q}\cdot \frac{1}{2}x^{(z_1)}}\Bigg]\big(\tilde{c}+\tilde{\mathbi{b}}_1^{(1)}\,e^{-i\e{q}\cdot \frac{1}{2}x^{(z_1)}}\big)(p_1)\\
\e{U}_{(1)}^{(z_1)}&\dfn \exp\Bigg[\sum_{n\geq2}\Big(\frac{\mathbi{b}_n^{(1)}}{(n-2)!}\nabla^{n-2}b\,\Big)e^{-in\e{q}\cdot \frac{1}{2}x^{(z_{1})}}+\sum_{n\geq0}\Big(\frac{\mathbi{c}_n^{(1)}}{(n+1)!}\nabla^{n+1}c\,\Big)e^{-in\e{q}\cdot \frac{1}{2}x^{(z_{1})}}\Bigg]\\
&\qquad\times\exp \Bigg[\sum_{n\geq1}\frac{1}{n}\!\frac{i\mathbi{a}_n}{(n-1)!}\cdot\sqrt{\frac{2}{\alpha'}}\nabla^nx\,e^{-in\e{q}\cdot \frac{1}{2}x^{(z_{1})}}\Bigg]\big(c+\mathbi{b}_1^{(1)}\,e^{-i\e{q}\cdot \frac{1}{2}x^{(z_{1})}}\big)(p_1)
\end{aligned}
}
\endgroup
\end{equation}
We have kept it implicit that the arguments of the various terms are to be evaluated in the $z_1$ frame and at $p_1$ (where $z_1(p_1)=0$). Furthermore, we have made use of the fact that this vertex operator is defined using holomorphic normal coordinates where partial derivatives can equivalently be replaced by covariant derivatives at $p_1$, see Sec.~\ref{sec:HNC}. 
\sk

The vertex operators (\ref{eq:offshellA_a_VO}) can be used to write down explicit expressions for fixed-picture handle operators associated to cutting either trivial or non-trivial homology cycles of any given Riemann surface associated to the standard matter CFT of bosonic string theory \cite{Polchinski_v1}. We will use them in explicit amplitude computations in Sec.~\ref{sec:EA}.

\subsection{Integrated Picture Local Operators}
Let us briefly discuss how the above fixed-picture vertex operator, $\hat{\mathscr{A}}_a^{(z_1)}$, is modified when we translate it to integrated picture, $\mathscr{A}_a^{(z_1)}$, which corresponds to identifying a complex structure deformation with a shift in the location of either one or both vertex operators out of which any given handle operator is constructed. Adopting an auxiliary coordinate system (see Sec.~\ref{sec:SPUTF} and Sec.~\ref{sec:SHUTF}) the map from fixed- to integrated-picture is furnished by (\ref{eq:intvertop1sigma_1}), namely:
\begin{equation}\label{eq:integrvertop}
\boxed{\hat{\mathscr{A}}_a^{(z_{\sigma_1})}(\sigma_1)\qquad\rightarrow\qquad \mathscr{A}_a^{(z_{\sigma_1})}= \int \rmd^2\sigma_12\sqrt{g(\sigma_1)}\hat{B}_{z_v}\hat{B}_{\bar{z}_v}\hat{\mathscr{A}}_a^{(z_{\sigma_1})}(\sigma_1)}
\end{equation}
where $g(\sigma_1)\equiv \det g_{ab}(\sigma_1)$ and $B_{z_v}\tilde{B}_{\bar{z}_v}$ is the contribution from the path integral measure in holomorphic normal coordinates, derived using a metric viewpoint in Sec.~\ref{sec:TP} and a holomorphic transition function viewpoint in Sec.~\ref{sec:SHUTF} (the two are equivalent but the latter makes reparametrisation invariance manifest). 
\sk

We emphasise from the outset that the operator (\ref{eq:integrvertop}) makes sense as a path integral insertion only if it is insensitive to the phase of the local frame coordinate, $z_{\sigma_1}$, which will in turn be the case either if it is part of a handle operator (in which case there will also be a factor $\hat{B}_\theta$ that elevates the phase, $\theta$, to a modulus that is to be integrated), or if $\hat{\mathscr{A}}_a^{(z_{\sigma_1})}(\sigma_1)$ is level-matched from the outset, which will be the case if it is associated to an asymptotic state. For further elaboration see Sec.~\ref{sec:EVO} and the discussion below. 
\sk

The explicit expression for the measure that translates fixed-picture vertex operators to integrated-picture using holomorphic normal coordinates is given, e.g., in (\ref{eq:Bintmub8z2xxx}), reproduced here for convenience,
\begin{equation}\label{eq:Bintmub8z2xxxs}
\begin{aligned}
\hat{B}_{z_v}&=-b_{-1}^{(z_{\sigma_1})}(\sigma_1)-\frac{1}{4}\sum_{n=1}^\infty\frac{1}{(n+1)!}\big(\nabla_{\bar{z}_{\sigma_1}}^{n-1}R_{(2)}(\sigma)\big)\Big|_{\sigma=\sigma_1}\!\!\!\tilde{b}_n^{(z_{\sigma_1})}(\sigma_1)\\
\hat{B}_{\bar{z}_v}&=-\tilde{b}_{-1}^{(z_{\sigma_1})}(\sigma_1)-\frac{1}{4}\sum_{n=1}^\infty\frac{1}{(n+1)!}\big(\nabla_{z_{\sigma_1}}^{n-1}R_{(2)}(\sigma)\big)\Big|_{\sigma=\sigma_1}\!\!\!b_n^{(z_{\sigma_1})}(\sigma_1)
\end{aligned}
\end{equation}
Explicitly evaluating the operator $\mathscr{A}_a^{(z_{\sigma_1})}$ making use of the normal-ordered fixed picture vertex operator given in (\ref{eq:offshellA_a_VO}), the result exponentiates and we find,
\begin{equation}\label{eq:IntPictVertOp}
\begingroup\makeatletter\def\f@size{11}\check@mathfonts
\def\maketag@@@#1{\hbox{\m@th\large\normalfont#1}}%
\boxed{
\begin{aligned}
B_{z_v}&B_{\bar{z}_v}\hat{\mathscr{A}}_a^{(z_{\sigma_1})}(\sigma_1)=\\
&=g_D:\!\exp\Bigg[\sum_{n\geq2}\Big(\frac{\mathbi{b}^{(1)}_n}{(n-2)!}\nabla_{z_{\sigma_1}}^{n-2}b\,\Big)e^{-in\e{q}\cdot \frac{1}{2}x^{(z_{\sigma_1})}}+\mathbi{c}^{(1)}_0\nabla_{z_{\sigma_1}}c\\
&\qquad\quad+\sum_{n\geq1}\mathbi{c}^{(1)}_n\Big(\frac{1}{(n+1)!}\nabla_{z_{\sigma_1}}^{n+1}c-\frac{1}{4}\frac{1}{(n+1)!}\nabla_{z_{\sigma_1}}^{n-1}R_{(2)}(\sigma_1)\big(\tilde{c} +\tilde{\mathbi{b}}^{(1)}_1\,e^{-i\e{q}\cdot \frac{1}{2}x^{(z_{\sigma_1})}}\big)\Big)e^{-in\e{q}\cdot \frac{1}{2}x^{(z_{\sigma_1})}}\\
&\quad\qquad\qquad+\sum_{n\geq1}\frac{1}{n}\!\frac{i\mathbi{a}_n}{(n-1)!}\cdot\sqrt{\frac{2}{\alpha'}}\nabla_{z_{\sigma_1}}^nx\,e^{-in\e{q}\cdot \frac{1}{2}x^{(z_{\sigma_1})}}\Bigg]e^{i\e{p}\cdot\frac{1}{2}x^{(z_{\sigma_1})}}(\sigma_1)\\
&\qquad\times \exp\Bigg[\sum_{n\geq2}\Big(\frac{\tilde{\mathbi{b}}^{(1)}_n}{(n-2)!}\nabla_{\bar{z}_{\sigma_1}}^{n-2}\tilde{b}\,\Big)e^{-in\e{q}\cdot \frac{1}{2}x^{(z_{\sigma_1})}}+\tilde{\mathbi{c}}^{(1)}_0\nabla_{\bar{z}_{\sigma_1}}\tilde{c}\\
&\qquad\quad+\sum_{n\geq1}\tilde{\mathbi{c}}^{(1)}_n\Big(\frac{1}{(n+1)!}\nabla_{\bar{z}_{\sigma_1}}^{n+1}\tilde{c}-\frac{1}{4}\frac{1}{(n+1)!}\nabla_{\bar{z}_{\sigma_1}}^{n-1}R_{(2)}(\sigma_1)\big(c +{\bf b}_1\,e^{-i\e{q}\cdot \frac{1}{2}x^{(z_{\sigma_1})}}\big)\Big)e^{-in\e{q}\cdot \frac{1}{2}x^{(z_{\sigma_1})}}\\
&\quad\qquad\qquad+\sum_{n\geq1}\frac{1}{n}\!\frac{i\tilde{\mathbi{a}}_n}{(n-1)!}\cdot\sqrt{\frac{2}{\alpha'}}\nabla_{\bar{z}_{\sigma_1}}^nx\,e^{-in\e{q}\cdot \frac{1}{2}x^{(z_{\sigma_1})}}\Bigg]e^{i\e{p}\cdot \frac{1}{2}x^{(z_{\sigma_1})}}(\sigma_1)\!:_{z_{\sigma_1}}
\end{aligned}
}
\endgroup
\end{equation}
Notice that the Ricci scalar contribution and matter zero modes provide the obstruction to chiral splitting as expected on general grounds, and that the full vertex operator is at most quadratic in $R_{(2)}$ since $c,\tilde{c}$ are Grassmann-odd. 
\sk

When the above vertex operators are used to cut open trivial homology cycles only the level-matched subset will propagate . So in this case we can also project onto the level-matched subset from the outset by integrating over the phase of the frame coordinate,
$$
B_{z_v}B_{\bar{z}_v}\hat{\mathscr{A}}_a^{(z_{\sigma_1})}(\sigma_1)\qquad\rightarrow\qquad \int_0^{2\pi}\frac{\rmd\phi}{2\pi}B_{z_v}B_{\bar{z}_v}\hat{\mathscr{A}}_a^{(z_{\sigma_1}/e^{i\phi})}(\sigma_1).
$$
We can in turn make this phase explicit in (\ref{eq:IntPictVertOp}) by noting that rotating the frame coordinate by a phase, $z_{\sigma_1}\mapsto e^{i\phi}z_{\sigma_1}$, can be undone by replacing the quantum numbers, $\mathbi{a}_n$, $\mathbi{b}^{(1)}_n$ and $\mathbi{c}^{(1)}_n$ by $e^{in\phi}\mathbi{a}_n$, $e^{in\phi}\mathbi{b}^{(1)}_n$ and $e^{in\phi}\mathbi{c}^{(1)}_n$ respectively (and similarly for the anti-chiral halves where $-\phi$ rather than $\phi$ appears in the exponents). To be precise,
\begin{equation}
\begin{aligned}
B_{z_v}&B_{\bar{z}_v}\hat{\mathscr{A}}_a^{(z_{\sigma_1}/e^{i\phi})}(\sigma_1) =\\ &=B_{z_v}B_{\bar{z}_v}\hat{\mathscr{A}}_a^{(z_{\sigma_1})}(\sigma_1)\big|_{(\mathbi{a}_n,\mathbi{b}_n,\mathbi{c}_n,\tilde{\mathbi{a}}_n,\tilde{\mathbi{b}}_n,\tilde{\mathbi{c}}_n)\rightarrow (e^{in\phi}\mathbi{a}_n,e^{in\phi}\mathbi{b}_n,e^{in\phi}\mathbi{c}_n,e^{-in\phi}\tilde{\mathbi{a}}_n,e^{-in\phi}\tilde{\mathbi{b}}_n,e^{-in\phi}\tilde{\mathbi{c}}_n)}
\end{aligned}
\end{equation}
As we have demonstrated on general grounds in Sec.~\ref{sec:BRST-AC}, when the above integrated vertex operator is used to construct a handle operator the resulting amplitudes are gauge invariant up to a total derivative on moduli space provided this phase, $\phi$, is identified with a modulus and is integrated (which always attaches a ghost contribution $(b_0-\tilde{b}_0)$ to either the $z_{\sigma_1}$ or the $z_{\sigma_2}$ frame vertex operator).
\sk

Finally, as we have already shown in (\ref{eq:zsigma* scalar}) on p.~\pageref{eq:zsigma* scalar}, the quantity $z_{\sigma_1}(\sigma_1)$ transforms as a scalar under reparametrisations, $\sigma_1\mapsto \hat{\sigma}_1(\sigma_1)$, (modulo a local U(1) phase that is ambiguous) and since all quantities appearing in (\ref{eq:IntPictVertOp}) depend on $\sigma_1$ only implicitly via $z_{\sigma_1}(\sigma_1)$ (or $z_{\sigma_1}(\sigma)$ evaluated at $\sigma=\sigma_1$) it follows that the full integrated vertex operator transforms as a scalar under reparametrisations of the underlying auxiliary coordinate on patch overlaps, and is hence (modulo U(1) frame rotations) globally well-defined on the entire Riemann surface on which it is inserted. Since we always average over this phase (we always take this U(1) to be integrated) \cite{Polchinski88} the corresponding ambiguity will not appear in physical information.

\subsection{Fixed-Picture with Pinch and Twists}\label{sec:GWDCSII}
When we wish to associate pinch and twist moduli to the cycle we cut open using a handle operator, 
\begin{equation}\label{eq:1intaAA}
\suminnt\limits_a\,\,\hat{\mathscr{A}}_a^{(z_1)}\hat{\mathscr{A}}^a_{(z_2/q)},
\end{equation}
then this bi-local operator insertion is dressed by path integral measure contributions,
\begin{equation}\label{eq:BqBqbarX}
\hat{B}_q\hat{B}_{\bar{q}}=\frac{b_0^{(z_2/q)}\tilde{b}_0^{(z_2/q)}}{q\bar{q}},
\end{equation}
acting on either of the two insertions in (\ref{eq:1intaAA}) since $b_0^{(z_1)}=b_0^{(z_2)}$ (note also that $b_0^{(z_2/q)}=b_0^{(z_2)}$) and the handle operator (\ref{eq:1intaAA}) is therefore replaced by:
\begin{equation}\label{eq:handleoperatorX1}
\boxed{\hat{H}= \int \rmd^2q\,\,\suminnt\limits_{a}\,\,\hat{\mathscr{A}}_a^{(z_1)}(p_1)\,\big[\hat{B}_q\hat{B}_{\bar{q}}\,\hat{\mathscr{A}}^a_{(z_2/q)}(p_2)\big]}
\end{equation}
Since we are moving the ghost contributions in pairs there are no additional minus signs, the precise reasoning was explained very clearly in \cite{Sen15b} and also applies here, the only difference being that our local operators have indefinite ghost number (but definite Grassmannality). Elaborating briefly on this last point, provided we insert the correct number of path integral measure contributions the result will always project onto the correct number of ghost insertions. When cutting non-trivial homology cycles this number is summed over, but this is automatic when using the coherent state basis for the bi-local operator (\ref{eq:1intaAA}).
\sk

Let us now consider the quantity: 
$$
b_0^{(z_2)}\tilde{b}_0^{(z_2)}\hat{\mathscr{A}}^a_{(z_2/e^{i\theta})}(p_2).
$$
It is a simple exercise to check that when acting with the anti-ghosts, $b_0^{(z_2)}\tilde{b}_0^{(z_2)}$, on the coherent state basis the sole effect is to remove the $c_0,\tilde{c}_0$ dependence from {\it both} vertex operators. To see how this works in coherent state language notice primarily that applying $b_0^{(z_2)}\tilde{b}_0^{(z_2)}$ to the relevant factor in the explicit coherent state in (\ref{eq:offshellA^a}) gives,
\begin{equation}\label{eq:b0bto;de0expc0c0..}
b_0^{(z_2)}\tilde{b}_0^{(z_2)}\cdot \exp\big(\mathbi{c}_0^{(2)}\partial c+\tilde{\mathbi{c}}_0^{(2)}\bar{\partial}\tilde{c}\big)\tilde{c}c\,e^{-i\e{p}\cdot x(0,0)}=\mathbi{c}_0^{(2)}\tilde{\mathbi{c}}_0^{(2)}\tilde{c}c\,e^{-i\e{p}\cdot x(0,0)}.
\end{equation}
Since this is the only factor on which the $b_0^{(z_2)}\tilde{b}_0^{(z_2)}$ term acts non-trivially in the vertex operator $\hat{\mathscr{A}}_{(z_2)}^a$ in (\ref{eq:offshellA^a}), and since there are integrals over the Grassmann-odd quantum numbers setting $\mathbi{c}_0^{(1)}=\mathbi{c}_0^{(2)}$ and $\tilde{\mathbi{c}}_0^{(1)}=\tilde{\mathbi{c}}_0^{(2)}$ in the sum over states (\ref{eq:handleoperatorX1}), seen explicitly in (\ref{eq:dpdmua}) and (\ref{eq:dmua}), only the $\mathbi{c}_0^{(1)}=\tilde{\mathbi{c}}_0^{(1)}=0$ terms can contribute in $\hat{\mathscr{A}}_a^{(z_1)}$. So the overall effect of the pinch and twist ghost moduli insertions is to effectively set $\mathbi{c}_0^{(1)}=\tilde{\mathbi{c}}_0^{(1)}=\mathbi{c}_0^{(2)}=\tilde{\mathbi{c}}_0^{(2)}=0$ in the vertex operators and multiply the overall result by $\mathbi{c}_0^{(2)}\tilde{\mathbi{c}}_0^{(2)}$ which can then freely be integrated out.
\sk

To be precise, if we focus on the sum/integral in (\ref{eq:handleoperatorX1}) over $\tilde{\mathbi{c}}_0^{(1)} ,\mathbi{c}_0^{(1)},\tilde{\mathbi{c}}_0^{(2)},\mathbi{c}_0^{(2)}$, according to (\ref{eq:b0bto;de0expc0c0..}) and (\ref{eq:dmua}) we encounter the following integral,
\begin{equation}
\begin{aligned}
&\int \big[\rmd\tilde{\mathbi{c}}_0^{(1)} \rmd\mathbi{c}_0^{(1)}\rmd\tilde{\mathbi{c}}_0^{(2)} \rmd\mathbi{c}_0^{(2)}\big]\delta(\mathbi{c}_0^{(2)}-\mathbi{c}_0^{(1)})\delta(\tilde{\mathbi{c}}_0^{(2)}-\tilde{\mathbi{c}}_0^{(1)})\mathbi{c}_0^{(2)}\tilde{\mathbi{c}}_0^{(2)}=1,\\
\end{aligned}
\end{equation}
so we see that we reach the same result for the handle operator, $\hat{H}$, if we set $\mathbi{c}_0^{(1)}=\tilde{\mathbi{c}}_0^{(1)}=0$
in (\ref{eq:offshellA_a_VO}), likewise set $\mathbi{c}_0^{(2)}=\tilde{\mathbi{c}}_0^{(2)}=0$ in $\hat{\mathscr{A}}^a_{(z_2/q)}$, and also replace 
the corresponding measure (\ref{eq:dmua}) by,
\begin{equation}\label{eq:dmua3}
\begin{aligned}
\int \rmd\mu_{\mathbi{a}\mathbi{b}\mathbi{c}}'
&=\int \rmd\mu_{\mathbi{a}\mathbi{b}\mathbi{c}}\,\,\mathbi{c}_0^{(2)}\tilde{\mathbi{c}}_0^{(2)}\\
&=\bigg(\prod_{n>0,\mu}\int \frac{\rmd{\mathbi{a}_n^{\mu *}}\wedge \rmd\mathbi{a}_n^{\mu}}{2\pi in}\frac{\rmd{\tilde{\mathbi{a}}_n^{\mu *}}\wedge \rmd\tilde{\mathbi{a}}_n^{\mu}}{2\pi in}\bigg)\exp\bigg[\sum_{n=1}^{\infty}\Big(\!-\frac{1}{n}\mathbi{a}_n\cdot \mathbi{a}_n^*-\frac{1}{n}\tilde{\mathbi{a}}_n\cdot \tilde{\mathbi{a}}_n^*\Big)\bigg]\\
&\qquad\times
\,\bigg(\prod_{n=1}^{\infty}\int \big[\rmd\mathbi{c}_n^{(1)}\rmd\mathbi{b}_n^{(2)} \rmd\mathbi{c}_n^{(2)}\rmd\mathbi{b}_n^{(1)}\rmd\tilde{\mathbi{c}}_n^{(1)}\rmd\tilde{\mathbi{b}}_n^{(2)} \rmd\tilde{\mathbi{c}}_n^{(2)}\rmd\tilde{\mathbi{b}}_n^{(1)}  \big]\bigg)\\
&\qquad\times\,\exp\bigg[\sum_{n=1}^{\infty}\Big(\mathbi{b}_n^{(1)}\mathbi{c}_n^{(2)}+\mathbi{b}_n^{(2)}\mathbi{c}_n^{(1)}+\tilde{\mathbi{b}}_n^{(1)}\tilde{\mathbi{c}}_n^{(2)}+\tilde{\mathbi{b}}_n^{(2)}\tilde{\mathbi{c}}_n^{(1)}\Big)\bigg]\\
\end{aligned}
\end{equation}
in terms of which the handle operator (\ref{eq:handleoperatorX1}) reads,
\begin{equation}\label{eq:handleoperatorX3}
\boxed{
\begin{aligned}
\hat{H}=\,\,{\suminnt\limits_{a}}'\,\,
\hat{\mathscr{A}}_a^{'(z_1)}(p_1)\int \rmd^2q\,q^{L_0-1}\bar{q}^{\tilde{L}_0-1}\hat{\mathscr{A}}^{'a}_{(z_2)}(p_2)
\end{aligned}
}
\end{equation}
where the prime in the sum/integral over $a$ reminds us to replace $\rmd\mu_{\mathbi{a}\mathbi{b}\mathbi{c}}$ by $\rmd\mu_{\mathbi{a}\mathbi{b}\mathbi{c}}'$ in (\ref{eq:dpdmua}), and the prime on the fixed-picture vertex operators reminds us to set $\mathbi{c}_0^{(1)}=\tilde{\mathbi{c}}_0^{(1)}=\mathbi{c}_0^{(2)}=\tilde{\mathbi{c}}_0^{(2)}=0$, that is,
\begin{equation}\label{eq:eq:Ahatprime}
\hat{\mathscr{A}}_a^{'(z_1)}(p_1)\equiv \hat{\mathscr{A}}_a^{(z_1)}(p_1)\big|_{\mathbi{c}_0^{(1)}=\tilde{\mathbi{c}}_0^{(1)}=0},\qquad \hat{\mathscr{A}}^{'a}_{(z_2)}(p_2)\equiv \hat{\mathscr{A}}^{a}_{(z_2)}(p_2)\big|_{\mathbi{c}_0^{(2)}=\tilde{\mathbi{c}}_0^{(2)}=0}.
\end{equation}

\subsection{Integrated Picture with Pinch and Twists}
Generalising the handle operator (\ref{eq:handleoperatorX1}) to the case where the corresponding vertex operators are in integrated- rather than fixed-picture follows immediately from the above so we will be brief. The handle operator of interest is now:
\begin{equation}\label{eq:handleoperatorX1s}
\begin{aligned}
&\hat{H}= \int \rmd^2z_{v_1}\int \rmd^2z_{v_2}\int \rmd^2q\,\,\suminnt\limits_{a}\,\,\big[\hat{B}_{z_{v_1}}\hat{B}_{\bar{z}_{v_1}}\hat{\mathscr{A}}_a^{(z_1)}(p_1)\big]\,\big[\hat{B}_{z_{v_2}}\hat{B}_{\bar{z}_{v_2}}\hat{B}_q\hat{B}_{\bar{q}}\,\hat{\mathscr{A}}^a_{(z_2/q)}(p_2)\big].
\end{aligned}
\end{equation}
The derivation that led to (\ref{eq:handleoperatorX3}) applies here also, in particular comparing (\ref{eq:handleoperatorX1s}) to (\ref{eq:handleoperatorX1}) we see primarily that (\ref{eq:handleoperatorX1s}) can also be written as,
\begin{equation}\label{eq:handleoperatorX3int}
\begin{aligned}
\hat{H}=\int \rmd^2z_{v_1}\int \rmd^2z_{v_2}\int \rmd^2q\,\,{\suminnt\limits_{a}}'\,\,
\big[\hat{B}_{z_{v_1}}\hat{B}_{\bar{z}_{v_1}}\hat{\mathscr{A}}_a^{'(z_{\sigma_1})}(\sigma_1)\big]\,\big[\,q^{L_0-1}\bar{q}^{\tilde{L}_0-1}\hat{B}_{z_{v_2}}\hat{B}_{\bar{z}_{v_2}}\hat{\mathscr{A}}^{'a}_{(z_{\sigma_2})}(\sigma_2)\big],
\end{aligned}
\end{equation}
where again recall that $\rmd^2z_{v_1}$ is really shorthand for $\rmd^2\sigma_12\sqrt{g(\sigma_1)}$, and similarly for the $z_{v_2}$ measure, and the prime was defined in (\ref{eq:eq:Ahatprime}). We can now simply write down the answer for the action of the remaining measure contributions on the local operators by reading off the result obtained in (\ref{eq:IntPictVertOp}). The only difference is that we are to set $\mathbi{c}_0=\tilde{\mathbi{c}}_0=0$ in both the $z_{\sigma_1}$ and $z_{\sigma_2}$ patch local operators, in particular,
\begin{equation}\label{eq:IntPictVertOpprime}
\begingroup\makeatletter\def\f@size{11}\check@mathfonts
\def\maketag@@@#1{\hbox{\m@th\large\normalfont#1}}%
\begin{aligned}
B_{z_v}&B_{\bar{z}_v}\hat{\mathscr{A}}_a^{'(z_{\sigma_1})}(\sigma_1)=\\
&=g_D:\!\exp\Bigg[\sum_{n\geq2}\Big(\frac{\mathbi{b}^{(1)}_n}{(n-2)!}\nabla_{z_{\sigma_1}}^{n-2}b\,\Big)e^{-in\e{q}\cdot \frac{1}{2}x^{(z_{\sigma_1})}}\\
&\qquad\qquad+\sum_{n\geq1}\mathbi{c}^{(1)}_n\Big(\frac{1}{(n+1)!}\nabla_{z_{\sigma_1}}^{n+1}c-\frac{1}{4}\frac{1}{(n+1)!}\nabla_{z_{\sigma_1}}^{n-1}R_{(2)}(\sigma_1)\big(\tilde{c} +\tilde{\mathbi{b}}^{(1)}_1\,e^{-i\e{q}\cdot \frac{1}{2}x^{(z_{\sigma_1})}}\big)\Big)e^{-in\e{q}\cdot \frac{1}{2}x^{(z_{\sigma_1})}}\\
&\quad\qquad\qquad+\sum_{n\geq1}\frac{1}{n}\!\frac{i\mathbi{a}_n}{(n-1)!}\cdot\sqrt{\frac{2}{\alpha'}}\nabla_{z_{\sigma_1}}^nx\,e^{-in\e{q}\cdot \frac{1}{2}x^{(z_{\sigma_1})}}\Bigg]e^{i\e{p}\cdot \frac{1}{2}x^{(z_{\sigma_1})}}(\sigma_1)\\
&\qquad\times \exp\Bigg[\sum_{n\geq2}\Big(\frac{\tilde{\mathbi{b}}^{(1)}_n}{(n-2)!}\nabla_{\bar{z}_{\sigma_1}}^{n-2}\tilde{b}\,\Big)e^{-in\e{q}\cdot \frac{1}{2}x^{(z_{\sigma_1})}}\\
&\qquad\qquad+\sum_{n\geq1}\tilde{\mathbi{c}}^{(1)}_n\Big(\frac{1}{(n+1)!}\nabla_{\bar{z}_{\sigma_1}}^{n+1}\tilde{c}-\frac{1}{4}\frac{1}{(n+1)!}\nabla_{\bar{z}_{\sigma_1}}^{n-1}R_{(2)}(\sigma_1)\big(c +{\bf b}_1\,e^{-i\e{q}\cdot \frac{1}{2}x^{(z_{\sigma_1})}}\big)\Big)e^{-in\e{q}\cdot \frac{1}{2}x^{(z_{\sigma_1})}}\\
&\quad\qquad\qquad+\sum_{n\geq1}\frac{1}{n}\!\frac{i\tilde{\mathbi{a}}_n}{(n-1)!}\cdot\sqrt{\frac{2}{\alpha'}}\nabla_{\bar{z}_{\sigma_1}}^nx\,e^{-in\e{q}\cdot \frac{1}{2}x^{(z_{\sigma_1})}}\Bigg]e^{i\e{p}\cdot \frac{1}{2}x^{(z_{\sigma_1})}}(\sigma_1)\!:_{z_{\sigma_1}}
\end{aligned}
\endgroup
\end{equation}
The corresponding explicit expression for the $z_2$-frame vertex operator is obtained by replacing the $z_1$-frame labels by $z_2$, the $z_1$-frame vertex operator quantum numbers by the corresponding $z_2$-frame quantum numbers (which are in turn obtained by comparing (\ref{eq:offshellA_a}) to (\ref{eq:offshellA^a})).
\sk

In the following section we present some explicit amplitude computations, the simplest examples in particular that probe worldsheet duality, modular invariance, and normalisation. These explicit computations provide additional consistency checks of the formalism.

\section{Explicit Amplitudes}\label{sec:EA}
In this section we will compute all sphere one-point and two-point amplitudes, we will show that cutting open the path integral by cutting out a disc (with no insertions) and gluing in a handle operator (to reattach the disc to the surface) produces the SL(2,$\mathbf{C}$) vacuum, we show that the handle operators constructed (at fixed complex structure) produce a resolution of unity, and finally demonstrate that one-loop modular invariance and tree-level worldsheet duality all remain intact in the handle operator formalism.

\subsection{Sphere 1-Point Amplitudes}\label{sec:S1ptA}
We begin by computing all sphere one-point amplitudes using the offshell vertex operators (\ref{eq:offshellA_a_VO}) or (\ref{eq:offshellA_a_}) with all quantum numbers generic. The quantity of interest is then:
\begin{equation}
e^{-2\Phi}\big\langle \hat{\mathscr{A}}^a_{(z_2)}\big\rangle_{S^2}.
\end{equation}
Taking into account the normalisation (\ref{eq:S2-norm-modes}) that we use throughout (and which is in turn equivalent to Polchinski's convention \cite{Polchinski_v1}), we insert (\ref{eq:offshellA_a_VO}), integrate out the matter and ghosts (including zero modes), and find:
\begin{equation}\label{eq:<A>S2}
\boxed{
e^{-2\Phi}\big\langle \hat{\mathscr{A}}^a_{(z_2)}\big\rangle_{S^2}=\frac{8\pi i}{\alpha'g_D}(2\pi)^D\delta^D(\e{p}-\e{q})\big(\tilde{\mathbi{c}}_0^2\mathbi{c}_0^2\tilde{\mathbi{c}}_1^2\mathbi{c}_1^2\big)}
\end{equation}
The explicit argument of the delta function, $\e{p}-\e{q}$, is reminiscent of the fact that the one-point amplitude only gets contributions from onshell massless states.
\sk

To see that the result (\ref{eq:<A>S2}) not only makes sense but that it is also fundamental for overall consistency, let us next consider a generic string amplitude and focus on any patch of the underlying Riemann surface, $\Sigma$, where there are no insertions (i.e.~a region diffeomorphic to a disc). Then associate to that patch a chart $(U_1,z_1)$. We then cut out a disc, say of radius $|z_1|=1-\epsilon$ (with $0<\epsilon\ll1$), centred at $z_1=0$, so that we end up with two disconnected surfaces, namely the original Riemann surface but now with a boundary, $\Sigma \setminus {\rm D}$, and a disc, ${\rm D}$. Let us think of the disc as a sphere with a ``tiny'' hole cut out, and cover this surface with two charts. The first of these two charts, $(U_2,z_2)$, is centred at the ``centre'' of the ``hole'' (whose radius is $|z_2|=1-\epsilon$) and extends beyond the equator, and the second chart, $(U_3,z_3)$, covers the complement, but also contains an overlap $U_2\cap U_3$ which is an equatorial band (and topologically an annulus) with holomorphic transition function, $z_2z_3=1$. 
\sk

\begin{figure}
\begin{center}
\includegraphics[angle=0,origin=c,width=0.8\textwidth]{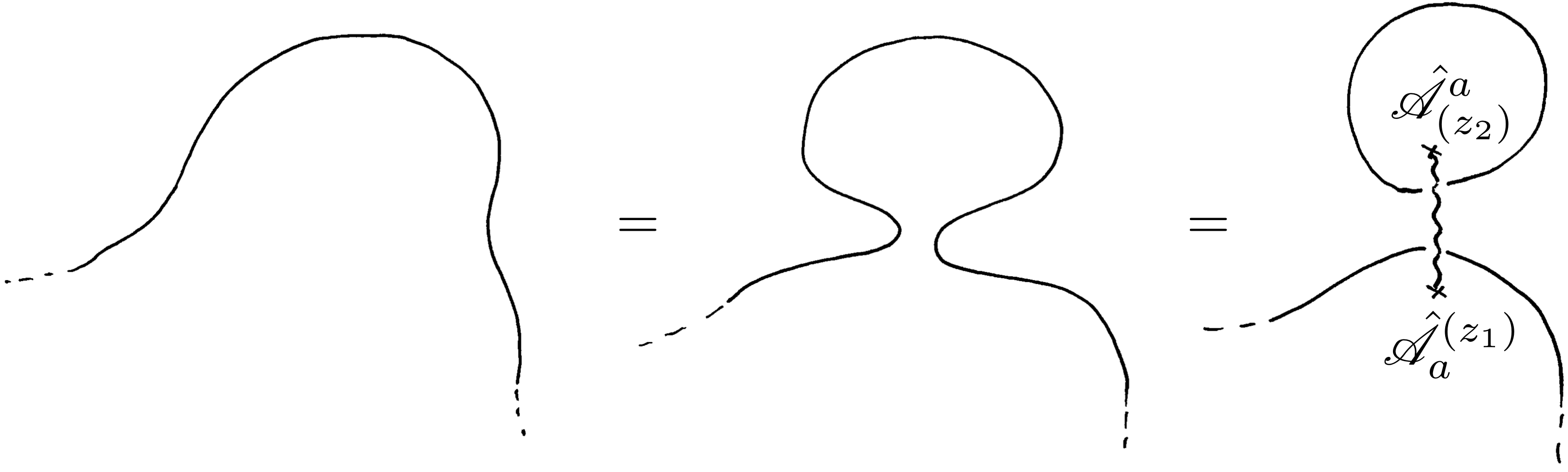}
\caption{Schematic representation of (\ref{eq:identitycut}), equivalently (\ref{eq:1=A<A>}), namely the fact that when we insert a handle operator that implements cutting open the path integral across a cycle that is homotopic to the identity the resulting amplitudes are unchanged. }\label{fig:1ptpinch}
\end{center}
\end{figure}

We next insert a handle operator that glues the two boundaries. In particular, we can implicitly imagine that we attach discs with punctures to the two boundaries and then we insert the {\it local} operators, $\mathscr{A}_a^{(z_1)}$, and $ \hat{\mathscr{A}}^a_{(z_2)}$ at these two punctures (associated to the punctured $\Sigma$ and $S^2$ respectively). The relevant transition function that connects the two local operators of the handle operator will be taken to be $z_1z_2=1$, and the sewn region corresponds to $1-\epsilon<|z_1|<1+\epsilon$. Since the disc (or once-punctured sphere) has no other insertions, it must therefore be the case that the following equality holds:
\begin{equation}\label{eq:identitycut}
e^{-\chi(\Sigma)\Phi}\big\langle\dots \big\rangle_{\Sigma} =e^{-\chi(\Sigma)\Phi}\big\langle\dots \suminnt\limits_{a}\,\,\hat{\mathscr{A}}_a^{(z_1)}\big\rangle_{\Sigma} 
e^{-2\Phi}\big\langle \hat{\mathscr{A}}^a_{(z_2)}\big\rangle_{S^2}
\end{equation}
The dots, `$\dots$', denote operator insertions that are ``sufficiently far'' from the operator insertion, $\hat{\mathscr{A}}_a^{(z_1)}$, which is in turn inserted $z_1=0$. 
Indeed, it is a simple exercise to show that substituting (\ref{eq:<A>S2}) into (\ref{eq:1=A<A>}) and tracing back the various definitions, see (\ref{eq:dpdmua}), (\ref{eq:dmua}) and (\ref{eq:offshellA_a_VO}) or (\ref{eq:offshellA_a_}), yields: 
\begin{equation}\label{eq:1=A<A>}
\boxed{1=\suminnt\limits_{a}\,\,\hat{\mathscr{A}}_a^{(z_1)}
e^{-2\Phi}\big\langle \hat{\mathscr{A}}^a_{(z_2)}\big\rangle_{S^2}}
\end{equation}
This is simply the SL(2,$\mathbf{C}$) vacuum, which is really the `number' 1 (as opposed to the unit operator, $\mathds{1}$, in the Hilbert space). The underlying geometrical configuration is depicted in Fig.~\ref{fig:1ptpinch}.
\sk

This elementary consistency check ensures that the vertex operator normalisation, the explicit expression for the summation symbol over $a$, and the normalisation of correlators (\ref{eq:S2-norm-modes}) are all consistent. Furthermore, it ensures that pinching off a sphere with no insertions leaves amplitudes invariant, as must clearly be the case.

\subsection{Sphere 2-Point Amplitudes}\label{sec:S2ptA}
\begin{figure}
\begin{center}
\includegraphics[angle=0,origin=c,width=0.75\textwidth]{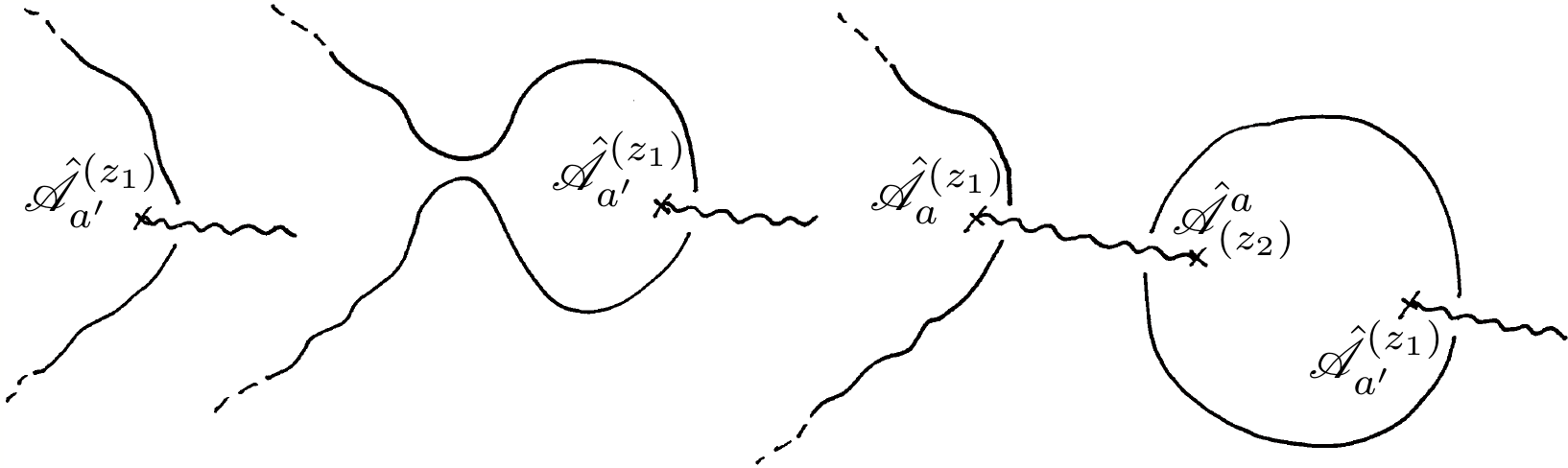}
\caption{Schematic representation of three equivalent configurations, namely the gluing consistency condition (\ref{eq:consistency_condition3}), which corresponds to the statement that we can cut open the path integral across a given cycle which contains one (or more) puncture(s) provided we sum/integrate over all boundary conditions (i.e.~over all quantum numbers of the inserted vertex operators which create the resolution of unity). The resulting amplitudes are identical before (left-most diagram) and after the cut (right-most diagram). One can also use this to change basis. The pinched-off two-point sphere amplitude is given precisely by (\ref{eq:2ptamp}) when all insertions are in a coherent state basis.}\label{fig:2ptpinch}
\end{center}
\end{figure}

In this subsection we compute all sphere 2-point amplitudes using the offshell vertex operators (\ref{eq:offshellA_a_VO}) or (\ref{eq:offshellA_a_}) with all quantum numbers generic:
\begin{equation}\label{eq:2pt}
e^{-2\Phi}\Big\langle\hat{\mathscr{A}}^{a}_{(z_2)}\hat{\mathscr{A}}_{a'}^{(z_1)}\Big\rangle_{S^2}.
\end{equation}
Armed with the resolution of unity (\ref{eq:intermbcd}), this will then be used to explicitly check the fundamental factorisation requirement (\ref{eq:consistency_condition2}),
\begin{equation}\label{eq:consistency_condition3}
\hat{\mathscr{A}}_{a'}^{(z_1)}=\suminnt\limits_{a} \,\,\hat{\mathscr{A}}_{a}^{(z_1)}e^{-2\Phi}\Big\langle\hat{\mathscr{A}}^{a}_{(z_2)}\hat{\mathscr{A}}_{a'}^{(z_1)}\Big\rangle_{S^2},\qquad {\rm with}\qquad z_1z_2=1,
\end{equation}
and its cousins (\ref{eq:consistency_condition2b}). Given that coherent states can also be regarded as generating functions for momentum eigenstates it is in fact sufficient to establish (\ref{eq:consistency_condition3}) in order to also establish the remaining relations in (\ref{eq:consistency_condition2b}). In this section we will confirm that (\ref{eq:consistency_condition3}) is indeed satisfied by the offshell coherent states (\ref{eq:offshellA_a}) and (\ref{eq:offshellA^a}) when the sum/integral $\sum\!\!\!\!\!\int\limits_{a}$ is interpreted as in (\ref{eq:dpdmua}) and (\ref{eq:dmua}). The underlying geometrical configuration is depicted in Fig.~\ref{fig:2ptpinch}.
\sk

The operators in the two-point sphere amplitude (\ref{eq:2pt}) are inserted at the origin of their respective coordinate charts $(U_1,z_1)$ and $(U_2,z_2)$, with transition function $z_1z_2=1$ on the patch overlap, $U_1\cap U_2$, which corresponds to an equatorial band with the topology of an annulus. 
We could imagine the operators $\hat{\mathscr{A}}^{a}_{(z_2)}$ and $\hat{\mathscr{A}}_{a'}^{(z_1)}$ inserted at the north $(N)$ and south $(S)$ poles respectively of a sphere, such that $z_1(S)=z_2(N)=0$.  We insert the normal-ordered $z_1$-frame offshell state (\ref{eq:offshellA_a3}) (with quantum numbers, $\{\mathbi{a}_n',\mathbi{b}_n^{'(1)},\mathbi{c}_n^{'(1)},\tilde{\mathbi{a}}_n',\tilde{\mathbi{b}}_n^{'(1)},\tilde{\mathbi{c}}_n^{'(1)},\e{p}',\e{q}'\}$ and frame modes $\{\alpha_{-n}^{(z_1)},b_{-n}^{(z_1)},\dots\}$) and its dual in the $z_2$ frame (obtained from the former by replacing the aforementioned quantum numbers by $\{-\mathbi{a}_n^*,\mathbi{b}_n^{(2)},\mathbi{c}_n^{(2)},-\tilde{\mathbi{a}}_n^*,\tilde{\mathbi{b}}_n^{(2)},\tilde{\mathbi{c}}_n^{(2)},-\e{p},-\e{q}\}$  while also replacing the frame labels of all modes, $\{\alpha_{-n}^{(z_1)},b_{-n}^{(z_1)},\dots\}\rightarrow \{\alpha_{-n}^{(z_2)},b_{-n}^{(z_2)},\dots\}$) into the two-point amplitude. Similar remarks hold for the zero modes. We then unwrap the various mode contours off the sphere (using that on patch overlaps coordinate frames are related by the transition function $z_1z_2=1$ so that (\ref{eq:gluingmodes=0bosonic}) applies with $q=1$) and use commutation relations,
\begin{equation}\label{eq:z1comms}
[\alpha_n^{(z_1)},\alpha_m^{(z_1)}]=n\delta_{n+m,0},\qquad \{b_n^{(z_1)},c_m^{(z_1)}\}=\delta_{n+m,0}.
\end{equation}
to bring the annihilation operators onto their respective SL($2,\mathbf{C}$) vacua where we in turn make use of (\ref{eq:annihilation_operators}). 
(The spacetime indices and metric $\eta^{\mu\nu}$ on the left- and right-hand sides respectively are of course implicit in the first relation, and there are similar relations for the anti-chiral halves). We then translate all zero modes to the same frame (say $z_1$) and use the universal normalisation  (\ref{eq:<..>normalisation1}) that we use throughout. 
The result is very simple:
\begin{equation}\label{eq:2ptamp}
\boxed{
\begin{aligned}
e^{-2\Phi}\Big\langle\hat{\mathscr{A}}^{a}_{(z_2)}\hat{\mathscr{A}}_{a'}^{(z_1)}\Big\rangle_{S^2}
&=\frac{8\pi i}{\alpha'}(2\pi)^D\delta^D(\e{p}-\e{p}')\delta\big(\mathbi{c}_0^{(2)}+\mathbi{c}_0^{'(1)}\big)
\delta\big(\tilde{\mathbi{c}}_0^{(2)}+\tilde{\mathbi{c}}_0^{'(1)}\big)\\
&\quad\times\exp\bigg\{\sum_{n=1}^{\infty}\Big(\frac{1}{n}\tilde{\mathbi{a}}_n^*\cdot\tilde{\mathbi{a}}_n'+\tilde{\mathbi{c}}_n^{(2)}\tilde{\mathbi{b}}_n^{'(1)}-\tilde{\mathbi{b}}_n^{(2)}\tilde{\mathbi{c}}_n^{'(1)}\Big)\bigg\}\\
&\quad
\times\exp\bigg\{\sum_{n=1}^{\infty}\Big(\frac{1}{n}\mathbi{a}_n^*\cdot\mathbi{a}_n'+\mathbi{c}_n^{(2)}\mathbi{b}_n^{'(1)}-\mathbi{b}_n^{(2)}\mathbi{c}_n^{'(1)}\Big)\bigg\}
\end{aligned}
}
\end{equation}
This holds for arbitrary offshell local operators of the form (\ref{eq:offshellA_a_}) (with generic quantum numbers). 
\sk

Let us work through the various steps leading to (\ref{eq:2ptamp}) in further detail. 
The zero modes are fairly subtle, because coherent states correspond to an infinite superposition of momentum eigenstates, and in computing (\ref{eq:2ptamp}) we therefore encounter an infinite superposition of delta functions. 
Remarkably,  these can all be rearranged so that the full two-point amplitude is indeed proportional to $\delta^D(\e{p}-\e{p}')$ as indicated in (\ref{eq:2ptamp}). Let us elaborate on how this comes about next.
\sk

Focusing initially on the matter contributions\footnote{The mechanism underlying the ghost contribution is (modulo zero modes) identical, see below.}, recall (\ref{eq:offshellA_a_}) and (\ref{eq:ABCcondensed_notation}), we will consider the quantity:
\begin{equation}\label{eq:<p|aa..|p'>}
e^{-2\Phi}\Big\langle 
:\!e^{-\sum_{n=1}^\infty\frac{1}{n}(\e{A}_n^{*}\cdot \alpha_{-n}^{(z_2)}+\tilde{\e{A}}_{n}^{*}\cdot \tilde{\alpha}_{-n}^{(z_2)})}e^{-i\e{p}\cdot x^{(z_2)}}(p_2)\!:_{z_2}\,
\,:e^{\sum_{m=1}^{\infty}\frac{1}{m}(\e{A}_m^{'}\cdot \alpha_{-m}^{(z_1)}+\tilde{\e{A}}_{m}^{'}\cdot \tilde{\alpha}_{-{m}}^{(z_1)})}e^{i\e{p}'\cdot x^{(z_1)}}(p_1)\!:_{z_1}\Big\rangle_{S^2},
\end{equation}
with ghosts implicit. 
Notice there are exponentials contained in:
\begin{equation}\label{eq:A'A'A*A*}
\begin{aligned}
&\e{A}_m^{'}\dfn \mathbi{a}_m'e^{-imq'\cdot \frac{1}{2}x^{(z_1)}},\qquad \tilde{\e{A}}_{\bar{m}}^{'}\dfn \tilde{\mathbi{a}}_{\bar{m}}'e^{-i\bar{m}q'\cdot \frac{1}{2}x^{(z_1)}}\\
&\e{A}_n^{*}\dfn \mathbi{a}_n^*e^{inq\cdot \frac{1}{2}x^{(z_2)}},\qquad \tilde{\e{A}}_{\bar{n}}^{*}\dfn \tilde{\mathbi{a}}_{\bar{n}}^*e^{i\bar{n}q\cdot \frac{1}{2}x^{(z_2)}}
\end{aligned}
\end{equation}
which as the reader may recall from the discussion in Sec.~\ref{sec:HOEM} are the price of requiring these coherent states have well-defined scaling dimension. 
Unwrapping the modes off the $z_2$ patch and into the $z_1$ patch (while preserving the definitions and frames in (\ref{eq:A'A'A*A*})) implies (\ref{eq:<p|aa..|p'>}) is also equal to,
\begin{equation}\label{eq:<p|aa..|p'>2}
e^{-2\Phi}\Big\langle 
:\!e^{\sum_{n=1}^\infty\frac{1}{n}(\e{A}_n^{*}\cdot \alpha_{n}^{(z_1)}+\tilde{\e{A}}_{n}^{*}\cdot \tilde{\alpha}_{n}^{(z_1)})}e^{-i\e{p}\cdot x^{(z_2)}}(p_2)\!:_{z_2}\,
\,:e^{\sum_{m=1}^{\infty}\frac{1}{m}(\e{A}_m^{'}\cdot \alpha_{-m}^{(z_1)}+\tilde{\e{A}}_{m}^{'}\cdot \tilde{\alpha}_{-{m}}^{(z_1)})}e^{i\e{p}'\cdot x^{(z_1)}}(p_1)\!:_{z_1}\Big\rangle_{S^2}
\end{equation}
where we made use of the fact that we can freely pull oscillator modes outside the normal ordering, and therefore that we can also change their frame while formally ``depicting'' them as being inside the original normal ordering. 
We then expand the exponentials and make use of the definitions (\ref{eq:A'A'A*A*}). Since the oscillators are now all in the $z_1$ frame we can make use of the commutators (\ref{eq:z1comms}) to pass all annihilation operators to the right and onto the SL(2,$\mathbf{C}$) vacuum at $p_1$, so that (\ref{eq:<p|aa..|p'>2}) is equivalent to:
\begin{equation}\label{eq:<p|aa..|p'>2}
\begin{aligned}
\sum_{a=0}^\infty\frac{1}{a!}&\sum_{\{n_1,\dots,n_a\}=1}^\infty\Big(\prod_{i=1}^a\frac{1}{n_i}\mathbi{a}_{n_i}^*\cdot \mathbi{a}_{n_i}'\Big)\,\,\sum_{b=0}^\infty\frac{1}{b!}\sum_{\{m_1,\dots,m_b\}=1}^\infty\Big(\prod_{j=1}^b\frac{1}{m_j}\tilde{\mathbi{a}}_{m_j}^*\cdot \tilde{\mathbi{a}}_{m_j}'\Big)\\
&\qquad\times
e^{-2\Phi}\Big\langle 
:\!e^{-i(\e{p}-\frac{N+\tilde{N}}{2}\e{q})\cdot x^{(z_2)}}(p_2)\!:_{z_2}\,
\,:\!e^{i(\e{p}'-\frac{N+\tilde{N}}{2}\e{q}')\cdot x^{(z_1)}}(p_1)\!:_{z_1}\Big\rangle_{S^2},
\end{aligned}
\end{equation}
where we defined the positive integers $N\dfn \sum_{i=1}^an_i$ and $\tilde{N}\dfn \sum_{j=1}^bm_j$. Notice that the same quantities, $N,\tilde{N}$, appear in {\it both} exponentials in the remaining correlator as a consequence of the commutation relations (\ref{eq:z1comms}). 
\sk

Focus next on the remaining correlator in (\ref{eq:<p|aa..|p'>2}). Since the inserted exponentials are primaries of conformal weight $\frac{\alpha'}{4}k^2$ where $k=\e{p}-\frac{N+\tilde{N}}{2}\e{q}$ (and similarly for the insertion at $p_2$ with $k'=\e{p}'-\frac{N+\tilde{N}}{2}\e{q}'$), we can transform the $z_2$ frame insertion to the $z_1$ frame using the transition function, $z_1(p)z_2(p)=1$,
\begin{equation}\label{eq:expkz2->z1}
\begin{aligned}
:\!e^{-ik\cdot x^{(z_2)}}(p_2)\!:_{z_2} &= \Big(\frac{\partial z_1}{\partial z_2}\frac{\partial \bar{z}_1}{\partial \bar{z}_2}(p_2)\Big)^{\frac{\alpha'}{4}k^2}\!\!\!\!\!:\!e^{-ik\cdot x^{(z_1)}}(p_2)\!:_{z_1} \\
&=|z_1(p_2)|^{\alpha'k^2}\!\!:\!e^{-ik\cdot x^{(z_1)}}(p_2)\!:_{z_1}
\end{aligned}
\end{equation}
A subtlety here is that strictly speaking the two charts overlap at neither $p_1$ nor $p_2$, but only on an annular equatorial band excluding these points, $p\neq p_1,p_2$. So the transition function, $z_1(p)z_2(p)=1$, also does not apply at precisely $p_2$ where the relation (\ref{eq:expkz2->z1}) is evaluated. 
The relevant implication is that the quantity $|z_1(p_2)|$ is formally infinite, but this is harmless since the pre-factor, $|z_1(p_2)|^{\alpha'k^2}$, appearing in (\ref{eq:expkz2->z1}) cancels out of the correlator in (\ref{eq:<p|aa..|p'>2}) as we will see momentarily. (If one prefers one can use that these insertions transform as tensors under shifts to construct new charts that do overlap at $p_1,p_2$, but the result is the same.) 
\sk

We then insert (\ref{eq:expkz2->z1}) into the correlator, and decompose $x^{(z_1)}$ into zero modes and fluctuations (in a standard fashion \cite{Polchinski_v1}). Integrating out the fluctuations leads to the following result for the remaining correlator in (\ref{eq:<p|aa..|p'>2}),
\begin{equation}\label{eq:<expexp>}
\begin{aligned}
e^{-2\Phi}\Big\langle &
:\!e^{-ik\cdot x^{(z_2)}}(p_2)\!:_{z_2}\,
\,:\!e^{ik'\cdot x^{(z_1)}}(p_1)\!:_{z_1}\Big\rangle_{S^2}\\
&\quad=e^{-2\Phi}|z_1(p_2)|^{\alpha'k^2}\Big\langle |z_1(p_2)|^{-\alpha'k^2}
\!\!:\!e^{-ik\cdot x^{(z_1)}}(p_2)e^{ik'\cdot x^{(z_1)}}(p_1)\!:_{z_1}\,\Big\rangle_{S^2}\\
&\quad=e^{-2\Phi}\Big\langle\,:\!e^{-ik\cdot x^{(z_1)}}(p_2)e^{ik'\cdot x^{(z_1)}}(p_1)\!:_{z_1}\,
\Big\rangle_{S^2},
\end{aligned}
\end{equation}
where for the Wick contractions of the fluctuations (after translating all insertions to the $z_1$ frame) we used the free-boson propagator, $\langle x_{(z_1)}^\mu(p_2)x_{(z_1)}^\nu(p_1)\rangle = -\frac{\alpha'}{2}\eta^{\mu\nu}\ln|z_1(p_2)-z_1(p_1)|^2$, taking into account that by definition $z_1(p_1)=0$. (We could also make use of the fact that the zero mode piece is independent of the location, $p_2$, of the insertion, and therefore take $p_2\rightarrow p_1$ in the remaining correlator. Since the remaining insertion is inside the one remaining normal ordering it only receives zero mode contributions.) 
\sk

Let us now include also the ghost contributions in (\ref{eq:<p|aa..|p'>2}). The following rearrangement formula is useful,
\begin{equation}\label{eq:ghostrearrange}
\begingroup\makeatletter\def\f@size{11}\check@mathfonts
\def\maketag@@@#1{\hbox{\m@th\large\normalfont#1}}%
\begin{aligned}
\exp &\big(\tilde{\e{C}}_0^{(2)}\tilde{c}_{0}^{(z_2)}+\tilde{\e{C}}_1^{(2)}\tilde{c}_{-1}^{(z_2)}\,\big)\big(\tilde{c}_1^{(z_2)}+\tilde{\e{B}}_1^{(2)}\big)\exp \big(\e{C}_0^{(2)}c_{0}^{(z_2)}+\e{C}_1^{(2)}c_{-1}^{(z_2)}\big)\big(c_1^{(z_2)}+\e{B}_1^{(2)}\big)\\
&\times\exp \big(\tilde{\e{C}}_0^{'(1)}\tilde{c}_{0}^{(z_1)}+\tilde{\e{C}}_1^{'(1)}\tilde{c}_{-1}^{(z_1)}\,\big)\big(\tilde{c}_1^{(z_1)}+\tilde{\e{B}}_1^{'(1)}\big)\exp \big(\e{C}_0^{'(1)}c_{0}^{(z_1)}+\e{C}_1^{'(1)}c_{-1}^{(z_1)}\big)\big(c_1^{(z_1)}+\e{B}_1^{'(1)}\big)\\
&\quad=\delta\big(\e{C}_0^{(2)}+\e{C}_0^{'(1)}\big)\delta\big(\tilde{\e{C}}_0^{(2)}+\tilde{\e{C}}_0^{'(1)}\big)\exp\big(-\tilde{\e{B}}_1^{(2)}\tilde{\e{C}}_1^{'(1)}+\tilde{\e{C}}_1^{(2)}\tilde{\e{B}}_1^{'(1)}\big)\\
&\qquad\times\exp\big(-\e{B}_1^{(2)}\e{C}_1^{'(1)}+\e{C}_1^{(2)}\e{B}_1^{'(1)}\big)c_{-1}^{(z_1)}c_{0}^{(z_1)}c_1^{(z_1)}\tilde{c}_{-1}^{(z_1)}\tilde{c}_{0}^{(z_1)}\tilde{c}_1^{(z_1)},
\end{aligned}
\endgroup
\end{equation}
where we unwrapped the $z_2$ frame modes onto the $z_1$ patch (using that on patch overlaps, $z_1z_2=1$). Recall also (\ref{eq:ABCcondensed_notation}) for the definitions of the various ``capital-letter'' quantities. The remaining ghost commutators are precisely analogous to the matter commutators that led to (\ref{eq:<p|aa..|p'>2}), which (taking (\ref{eq:ghostrearrange}) into account) in turn generalises to,
\begin{equation}\label{eq:<p|aa..|p'>2withghosts}
\begin{aligned}
&g_D^2\,\delta\big(\mathbi{c}_0^{(2)}+\mathbi{c}_0^{'(1)}\big)
\delta\big(\tilde{\mathbi{c}}_0^{(2)}+\tilde{\mathbi{c}}_0^{'(1)}\big)\\
&\times\sum_{a=0}^\infty\frac{1}{a!}\sum_{\{n_1,\dots,n_a\}=1}^\infty\prod_{i=1}^a\Big(\frac{1}{n_i}\mathbi{a}_{n_i}^*\cdot \mathbi{a}_{n_i}'+\mathbi{c}_n^{(2)}\mathbi{b}_n^{'(1)}-\mathbi{b}_n^{(2)}\mathbi{c}_n^{'(1)}\Big)\\
&\times\sum_{b=0}^\infty\frac{1}{b!}\sum_{\{m_1,\dots,m_b\}=1}^\infty\prod_{j=1}^b\Big(\frac{1}{m_j}\tilde{\mathbi{a}}_{m_j}^*\cdot \tilde{\mathbi{a}}_{m_j}'+\tilde{\mathbi{c}}_n^{(2)}\tilde{\mathbi{b}}_n^{'(1)}-\tilde{\mathbi{b}}_n^{(2)}\tilde{\mathbi{c}}_n^{'(1)}\Big)\\
&\times
e^{-2\Phi}\Big\langle 
:\!e^{-i(\e{p}-\frac{N+\tilde{N}}{2}\e{q})\cdot x^{(z_1)}}(p_2)c_{-1}^{(z_1)}c_{0}^{(z_1)}c_1^{(z_1)}\tilde{c}_{-1}^{(z_1)}\tilde{c}_{0}^{(z_1)}\tilde{c}_1^{(z_1)}e^{i(\e{p}'-\frac{N+\tilde{N}}{2}\e{q}')\cdot x^{(z_1)}}(p_1)\!:_{z_1}\Big\rangle_{S^2},
\end{aligned}
\end{equation}
where we also made use of (\ref{eq:<expexp>}), and included two powers of the string coupling, $g_D^2$, associated to the two vertex operators. This is now equal to the full two-point amplitude (\ref{eq:2pt}) we are aiming for. 
\sk

The remaining correlator in (\ref{eq:<p|aa..|p'>2withghosts}) is entirely associated to matter and ghost zero modes (since the entire insertion is inside the $z_1$-frame  normal ordering there are no remaining matter fluctuations). We can compute this correlator by appealing to the normalisation of the measure (\ref{eq:<..>normalisation1}) used throughout, which is in particular equivalent to:
\begin{equation}\label{eq:S2-norm-modes}
 e^{-2\Phi}\Big\langle 
\,:\!e^{-ik\cdot x^{(z_1)}}\!(p_2)\,c_{-1}^{(z_1)}c_{0}^{(z_1)}c_{1}^{(z_1)}\tilde{c}_{-1}^{(z_1)}\tilde{c}_{0}^{(z_1)}\tilde{c}_1^{(z_1)}
e^{ik'\cdot x^{(z_1)}}(p_1)\!:_{z_1}\Big\rangle_{S^2}=\frac{8\pi i}{\alpha'g_D^2}(2\pi)^D\delta^D(k-k').
\end{equation}
This in turn tells us that the two-point amplitude (\ref{eq:<p|aa..|p'>2withghosts}) is proportional to:
$$
\delta^D(k-k')=\delta^D\big(\e{p}-M\e{q}-\e{p}'+M\e{q}'\big),\qquad {\rm with}\quad M=\tfrac{1}{2}(N+\tilde{N}).
$$
Taking into account that the relations $\e{q}^2=0$ and $\e{p}\cdot \e{q}=2/\alpha'$ are valid offshell as well as onshell, and making use of the fact that we can therefore choose lightcone coordinates such that, 
$$
(\e{p}-M\e{q})^{\mu}=
\left\{\begin{aligned}
&\e{p}^--M\e{q}^-,\qquad{\rm with}\quad \e{q}^-=-\frac{2}{\alpha'}\frac{1}{\e{p}^+}\\
&\e{p}^+\\
&\e{p}^i
\end{aligned}
\right.
$$
it follows that:
\begin{equation}\label{eq:deltakk'deltapp'}
\begin{aligned}
\delta^D(k-k')&= \delta(\e{p}^--\e{p}^{'-}-M\e{q}^-+M\e{q}^{'-})\delta(\e{p}^+-\e{p}^{'+})\delta^{D-2}({\bf p}-{\bf p}') \\
&=\delta(\e{p}^--\e{p}^{'-})\delta(\e{p}^+-\e{p}^{'+})\delta^{D-2}({\bf p}-{\bf p}') \\
&=\delta^D(\e{p}-\e{p}'),
\end{aligned}
\end{equation}
where in the second line we used the $\e{p}^+=\e{p}^{'+}$ momentum conservation constraint. So the $\e{q},\e{q}'$ dependence cancels out, entirely because of momentum conservation and independently of any onshell conditions. 
\sk

Incidentally, this is a fairly remarkable observation, because one might not expect to see an overall $D$-dimensional momentum delta function unless the states under consideration are momentum eigenstates, but clearly these coherent states are not momentum eigenstates and yet they still have such an overlap. The fact that this is possible is because $\e{p}$ is not the actual momentum of either one state (there is also $\e{q}$). This property (\ref{eq:deltakk'deltapp'}) is in fact crucial for there to exist an offshell resolution of unity, as we will discuss momentarily.
\sk

Substituting (\ref{eq:deltakk'deltapp'}) into (\ref{eq:S2-norm-modes}), which is in turn substituted back into the full result for the two-point amplitude (\ref{eq:<p|aa..|p'>2withghosts}), we see that the various remaining sums and products can all be carried out and exponentiate as shown in the final answer for the two-point amplitude (\ref{eq:2ptamp}).
\sk

It is worth pointing out that we would have obtained precisely the same result (\ref{eq:2ptamp}) for the two-point amplitude had we instead used the basis of states given in Sec.~\ref{sec:HOEM} (with $q=1$), which serves as a good consistency check.
\sk

Having computed the two-point sphere amplitudes (\ref{eq:2ptamp}) we next make use of the result to check the basic factorisation requirement (\ref{eq:consistency_condition3}). Recall now that the interpretation for the sum/integral over states, $\Sigma\!\!\!\!\!\int_{\,\,a}$, for this basis was derived above, see (\ref{eq:dpdmua}) and (\ref{eq:dmua}). So we have the following representation for the right-hand side of (\ref{eq:consistency_condition3}),
\begin{equation}
\begin{aligned}
&\suminnt\limits_{a} \,\,\hat{\mathscr{A}}_{a}^{(z_1)}e^{-2\Phi}\Big\langle\hat{\mathscr{A}}^{a}_{(z_2)}\hat{\mathscr{A}}_{a'}^{(z_1)}\Big\rangle_{S^2}\\
&\qquad=\int \rmd\mu_{\mathbi{a}\mathbi{b}\mathbi{c}}\,\,\hat{\mathscr{A}}_{a}^{(z_1)}
\delta\big(\mathbi{c}_0^{(2)}+\mathbi{c}_0^{'(1)}\big)\delta\big(\tilde{\mathbi{c}}_0^{(2)}+\tilde{\mathbi{c}}_0^{'(1)}\big)\\
&
\qquad\quad\times\exp\bigg\{\sum_{s=1}^{\infty}\Big(\frac{1}{s}\mathbi{a}_s^*\cdot\mathbi{a}_s'+\mathbi{c}_s^{(2)}\mathbi{b}_s^{'(1)}-\mathbi{b}_s^{(2)}\mathbi{c}_s^{'(1)}\Big)\bigg\}\\
&\qquad\quad\times\exp\bigg\{\sum_{s=1}^{\infty}\Big(\frac{1}{s}\tilde{\mathbi{a}}_s^*\cdot\tilde{\mathbi{a}}_s'+\tilde{\mathbi{c}}_s^{(2)}\tilde{\mathbi{b}}_s^{'(1)}-\tilde{\mathbi{b}}_s^{(2)}\tilde{\mathbi{c}}_s^{'(1)}\Big)\bigg\}\\
\end{aligned}
\end{equation}
where we used the momentum-conserving delta function to integrate out the momenta and cancelled the factors $\alpha'/(8\pi i)$. The remaining integrals may be trivially carried out on account of (\ref{eq:GaussianIntegral}) and (\ref{eq:Grassmann-odd ints}), with $\hat{\mathscr{A}}_a^{(z_1)}$ given in either (\ref{eq:offshellA_a2}) or (\ref{eq:offshellA_a_VO}), leading precisely to the stated factorisation expression (\ref{eq:consistency_condition3}), namely,
$$
\suminnt\limits_{a} \,\,\hat{\mathscr{A}}_{a}^{(z_1)}e^{-2\Phi}\Big\langle\hat{\mathscr{A}}^{a}_{(z_2)}\hat{\mathscr{A}}_{a'}^{(z_1)}\Big\rangle_{S^2}=\hat{\mathscr{A}}_{a'}^{(z_1)},\qquad {\rm with}\qquad z_1z_2=1.
$$ 

This demonstrates that both vertex operators and our interpretation for the sum/integral over states are correctly normalised and the coherent state basis indeed provides a resolution of unity. In fact, as we discuss in the following section, the two-point function (\ref{eq:2ptamp}) (more precisely a close cousin of it) also plays a crucial role in demonstrating that the one-loop vacuum amplitude derived using the handle-operator formalism is {\it modular invariant}, which is a highly non-trivial (but easily-derived) result.

\subsection{One-Loop Modular Invariance}\label{sec:MI}
Computing string amplitudes using handle operators obscures {\it manifest} modular invariance, because we single out specific cycles to cut open. 
So we should check that the resulting amplitudes are indeed modular invariant in the presence of handle operators. Rather than present a general proof at arbitrary genus (which requires some additional ingredients we have not included in the current article) we will focus on the simplest example where modular invariance plays a crucial role, namely the torus, $T^2$, and in particular the {\it one-loop vacuum amplitude}. Regarding the generalisation of the proof of modular invariance to arbitrary genus, there are general reasons to believe \cite{Polchinski_v1} that if one can show that the one-loop one-point amplitude is modular invariant, so will arbitrary amplitudes constructed by gluing be modular invariant. It is important to check this statement explicitly, but we will nevertheless here focus on the simpler case of the one-loop vacuum amplitude.
\sk

Let us think about the relevant worldsheet geometry. (The setup is similar to that described in Sec.~\ref{sec:PH} with some minor changes.)
\sk

We primarily consider two coordinate charts $(U_1,z_1)$ and $(U_2,z_2)$ labelled by holomorphic frame coordinates, $z_1$ and $z_2$.  As usual then, to construct a handle we remove discs $|z_1|<(1-\epsilon)|q|^{1/2}$ and $|z_2|<(1-\epsilon)|q|^{1/2}$ with $\epsilon$ a small parameter and glue by identifying pairs of points $z_1,z_2$ if they are related by:
\begin{equation}\label{eq:z1z2=q2}
z_1z_2=q,\qquad {\rm on}\qquad U_1\cap^h U_2,
\end{equation}
as in (\ref{eq:z1z2=q}). The overlap $U_1\cap^h U_2$ is in turn taken to correspond to the annular region:\footnote{We have placed a letter `$h$' (for {\it h}andle) on the set overlap symbol, $\cap^h$, for a reason to become clear momentarily.}
\begin{equation}\label{eq:sewingregion2}
U_1\cap^hU_2=\left\{z_1,z_2\,\bigg|\,
\begin{aligned}
&(1-\epsilon)|q|^{1/2}<|z_1|<(1+\epsilon)|q|^{1/2}\\
&(1-\epsilon)|q|^{1/2}<|z_2|<(1+\epsilon)|q|^{1/2}
\end{aligned}
\right\}
\end{equation}
as in (\ref{eq:sewingregion}). 
The complex parameter $q=e^{2\pi i\tau}$ (with $\tau=\tau_1+i\tau_2$) will be identified with the modulus of the torus, but we are not done yet. In order to create a torus we must specify that the thus created handle is to be inserted, e.g., onto a sphere, so that we obtain an object that is topologically a torus. This can be done, e.g., by pulling back the two charts onto a sphere, or onto a plane (using stereographic projection), and subsequently pulling back the two discs onto the plane with SL(2,$\mathbf{C}$) transformations.  We can then remove the two discs as specified above and glue with (\ref{eq:z1z2=q2}) to create a handle. (This is essentially a Schottky parametrisation of the torus.) Since handle operators are not invariant under SL(2,$\mathbf{C}$) we will also have to change coordinates which will also induce a corresponding change in normal ordering as discussed above, see in particular (\ref{eq:CNOxbc-w<->z}) (on p.~\pageref{eq:CNOxbc-w<->z}) with the choice (\ref{eq:DDDw<->z SL2C}) for the subtractions. 
\sk

\begin{figure}
\begin{center}
\includegraphics[angle=0,origin=c,width=0.42\textwidth]{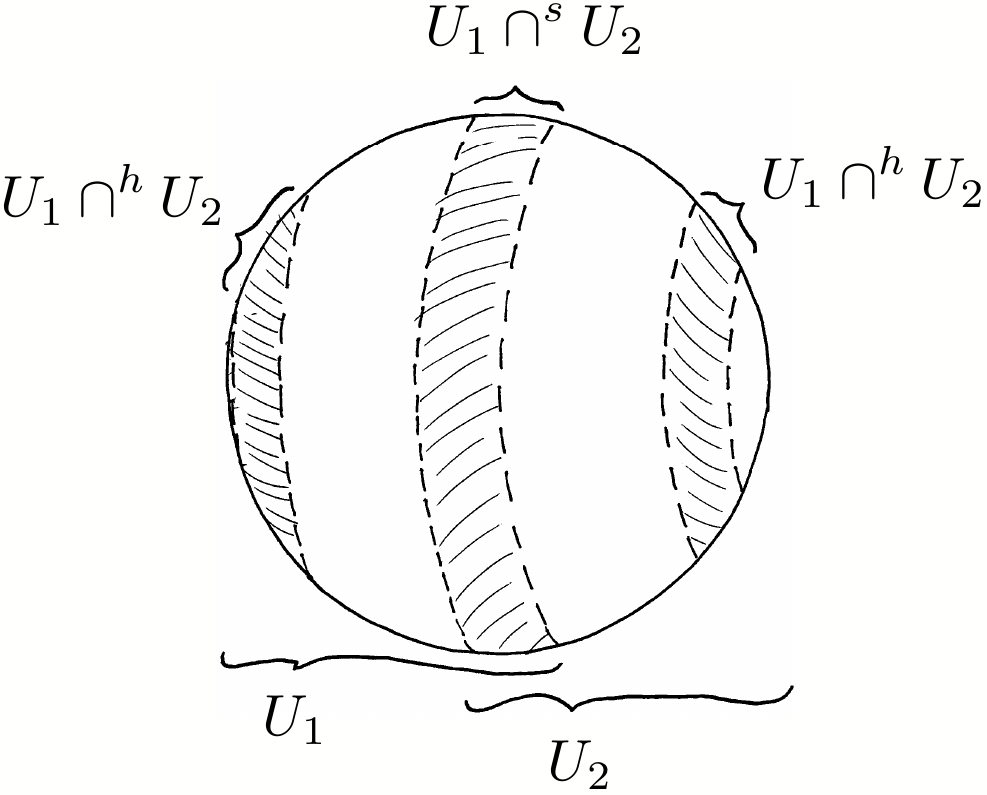}
\caption{Construction of a torus by using a cover with only two holomorphic charts, $(U_1,z_1)$ and $(U_2,z_2)$. On $U_1\cap^hU_2$ we glue with $z_1z_2=q$ whereas on $U_1\cap^sU_2$ we glue with $z_1z_2=1$. }\label{fig:T2patches}
\end{center}
\end{figure}

But in fact there is a shortcut, because the specific calculation we are aiming to carry out does not have any operator insertions on the remaining surface. We can use the fact that the chart overlap (\ref{eq:sewingregion2}) is restricted to the vicinity of the pinch, and therefore allow the two charts to also overlap in a {\it second} annular region,
\begin{equation}\label{eq:sewingregion3}
U_1\cap^s U_2=\left\{z_1,z_2\,\bigg|\,
\begin{aligned}
&(1+\epsilon)<|z_1|<(1+2\epsilon)\\
&(1+\epsilon)<|z_2|<(1+2\epsilon)
\end{aligned}
\right\}
\end{equation}
and then glue on this chart overlap to create the sphere with:
$$
z_1z_2=1,\qquad {\rm on}\qquad U_1\cap^s U_2,
$$
which is to be interpreted that chart coordinates $z_1$ and $z_2$ are to be identified if on the chart overlap (\ref{eq:sewingregion3}) they are related by $z_1z_2=1$. 
(Here the letter `$s$' (for {\it s}phere) on $\cap^s$ represents the fact that from one viewpoint this patch overlap and gluing constructs a sphere. Of course, due to the wonders of conformal field theory there is no real distinction as to which part corresponds to the sphere and which corresponds to the handle.) The setup is shown in Fig.~\ref{fig:T2patches}. 
So since the full chart overlap, 
$$
U_1\cap U_2=\big\{U_1\cap^h U_2,U_1\cap^s U_2\big\}
$$
is composed of two {\it disconnected} pieces, $U_1\cap^h U_2$ and $U_1\cap^s U_2$, and since furthermore neither of these is diffeomorphic to a disc, the resulting cover is not a `good cover'. But we do not need a good cover, and in fact with some care the above setup will be the most efficient way to construct a torus using a handle operator insertion. (We could if we like introduce additional charts but this will not be necessary here.) So we have completely specified the relevant geometry.
\sk

The torus has one conformal Killing vector (associated to invariance under rigid coordinate shifts in the two directions or homology cycles of the torus), so in the usual formulation to compute a one-loop $\n$-point amplitude we would insert one fixed-picture vertex operator and $\n-1$ integrated-picture vertex operators. For the vacuum amplitude however there are no vertex operator insertions, so we should specify how to fix the aforementioned invariance under shifts.\footnote{See also \cite{ErbinMaldacenaSkliros19,SekiTakahashi19} for a related discussion.} In particular, we cannot use the gauge-fixed path integral of Sec.~\ref{sec:FPIFP} where all vertex operators are in `fixed picture' (at least as it stands), but we {\it can} instead use the `integrated-picture' path integral of Sec.~\ref{sec:TEC}. The latter explicitly unfixes the conformal Killing group symmetries and in particular it also applies when the number, $\n$, of external states is zero, $\n=0$.
As discussed there, a good starting point for evaluating the one-loop vacuum amplitude in the BRST formalism is (\ref{eq:fullpathintegralX4}), namely:
\begin{equation}\label{eq:fullpathintegralX4z}
\begin{aligned}
Z_{T^2}=\int_{\mathcal{M}_{1,0}}\frac{\rmd^{2}q}{4\tau_2} \Big\langle\hat{B}_{q}\hat{B}_{\bar{q}}\tilde{c}^{(w_{1})} c^{(w_{1})}\Big\rangle_{T^2}.
\end{aligned}
\end{equation}
where we temporarily display the amplitude in terms of, $q=e^{2\pi i\tau}$ (and $\tau=\tau_1+i\tau_2$), since it is $q$ that appears naturally in handle operators and the transition function (\ref{eq:z1z2=q2}), and $\mathcal{M}_{1,0}$ is a fundamental domain of SL(2,$\mathbf{Z})/\mathbf{Z}_2$, such as,
\begin{equation}\label{eq:FundDomT2}
\mathcal{M}_{1,0} = \big\{\tau,\bar{\tau}\,\,\big|-\tfrac{1}{2}\leq \tau_2\leq\tfrac{1}{2},|\tau|\geq1\big\},
\end{equation}
which corresponds to integrating over all tori with distinct complex structures.\footnote{This is equivalent to Polchinski's trick of averaging over translations, see (7.3.3) in \cite{Polchinski_v1} (but in our conventions the area of the torus is $2\tau_2$ rather than $2(2\pi)^2\tau_2$).}
\sk

Using a handle operator approach enables us to compute this torus partition function, $Z_{T^2}$, by computing a correlation function on a sphere, $S^2$. 
The precise statement  is provided by the general prescription (\ref{eq:completeness_nonsep}), in particular,
\begin{equation}\label{eq:completeness_nonsepX}
\big\langle \dots\big\rangle_{T^2}= \,\,e^{-2\Phi}\Big\langle\dots\suminnt\limits_{a}\,\hat{\mathscr{A}}_{a}^{(z_1)}\hat{\mathscr{A}}^{a}_{(z_2/q)}\Big\rangle_{S^2},
\end{equation}
where the handle operator is constructed by gluing with $z_1z_2=q$ on the patch overlap $U_1\cap^h U_2$. Regarding the ghost insertions, these are evaluated with the transition function (\ref{eq:z1z2=q2}) leading to,
\begin{equation}\label{eq:BqBqbarz2}
\hat{B}_{q}\hat{B}_{\bar{q}}=b_0^{(z_2)}\tilde{b}_0^{(z_2)}/(q\bar{q}),
\end{equation}
where the mode contours encircle the annulus $U_1\cap^h U_2$ (and note that $b_0^{(z_2/q)}\tilde{b}_0^{(z_2/q)}=b_0^{(z_2)}\tilde{b}_0^{(z_2)}$). This was discussed in detail in Sec.~\ref{sec:PH} and Sec.~\ref{sec:GWDCSII}. Finally, we need to also choose where to insert the ghost operators, $\tilde{c}^{(w_{1})} c^{(w_{1})}(p)$, appearing in (\ref{eq:fullpathintegralX4z}) and then change frame coordinates as appropriate. We take $p\in U_1\cap^hU_2$. By looking back to the derivation of (\ref{eq:fullpathintegralX4z}) in (\ref{eq:fullpathintegralX4}) we see that the torus identification in terms of the $w_1$ coordinate is $w_1\sim w_1+1$ and $w_1\sim w_1+\tau$, and therefore the relevant holomorphic change of frame is\footnote{The coefficient $-i$ in the exponential is conventional \cite{Polchinski_v1} and we could also omit it without loss of generality. }, $w_1(p)\mapsto z_2(p)=e^{-iw_1}(p)$, so that writing $z\equiv z_2(p)$,
\begin{equation}\label{eq:ccwlnz}
\tilde{c}^{(w_{1})} c^{(w_{1})} (p)= \frac{\tilde{c}^{(z_{2})}(\bar{z}) c^{(z_{2})}(z)}{z\bar{z}}.
\end{equation}

Let us then substitute the results (\ref{eq:ccwlnz}), (\ref{eq:BqBqbarz2}) and (\ref{eq:completeness_nonsepX}) into (\ref{eq:fullpathintegralX4z}),
\begin{equation}\label{eq:fullpathintegralX4zz}
\begin{aligned}
Z_{T^2}=\,\,\int_{\mathcal{M}_{1,0}}\frac{\rmd^{2}q}{4\tau_2} e^{-2\Phi} \Big\langle\suminnt\limits_{a}\,\hat{\mathscr{A}}_{a}^{(z_1)}\Big(\frac{b_0^{(z_2)}\tilde{b}_0^{(z_2)}}{q\bar{q}}\frac{\tilde{c}^{(z_{2})}(\bar{z}) c^{(z_{2})}(z)}{z\bar{z}}\Big)\hat{\mathscr{A}}^{a}_{(z_2/q)}\Big\rangle_{S^2}.
\end{aligned}
\end{equation}
It is convenient to pull the $b_0,\tilde{b}_0$ modes through the $c,\tilde{c}$ operators, and in particular evaluating the relevant contour integrals yields,
$$
b_0^{(z_2)}\tilde{b}_0^{(z_2)}\frac{\tilde{c}^{(z_{2})}(\bar{z}) c^{(z_{2})}(z)}{z\bar{z}}=1-\frac{c^{(z_{2})}(z)}{z}b_0^{(z_2)}-\frac{\tilde{c}^{(z_{2})}(\bar{z})}{\bar{z}}\tilde{b}_0^{(z_2)}+\frac{\tilde{c}^{(z_{2})}(\bar{z})c^{(z_{2})}(z)}{z\bar{z}}b_0^{(z_2)}\tilde{b}_0^{(z_2)}.
$$
Noting furthermore that,
$$
b_0^{(z_2)}\hat{\mathscr{A}}^{a}_{(z_2/q)}=-\mathbi{c}_0^{(2)}\hat{\mathscr{A}}^{a}_{(z_2/q)},\qquad \tilde{b}_0^{(z_2)}\hat{\mathscr{A}}^{a}_{(z_2/q)}=-\tilde{\mathbi{c}}_0^{(2)}\hat{\mathscr{A}}^{a}_{(z_2/q)},
$$
it is seen that the effect of acting with the ghosts on $\hat{\mathscr{A}}^{a}_{(z_2/q)}$ in (\ref{eq:fullpathintegralX4zz}) leads to an exponentiation, in particular,
\begin{equation}\label{eq:fullpathintegralX4zX}
\begin{aligned}
Z_{T^2}=\,\,\int_{\mathcal{M}_{1,0}}\frac{\rmd^{2}q}{4\tau_2} \frac{e^{-2\Phi}}{q\bar{q}} \Big\langle\suminnt\limits_{a}\,\hat{\mathscr{A}}_{a}^{(z_1)}\Big(e^{-\mathbi{c}_0^{(2)}c^{(z_2)}(z)/z-\tilde{\mathbi{c}}_0^{(2)}\tilde{c}^{(z_2)}(\bar{z})/\bar{z}}\Big)\hat{\mathscr{A}}^{a}_{(z_2/q)}\Big\rangle_{S^2}.
\end{aligned}
\end{equation}
Since the amplitude is independent of $z,\bar{z}$ (recall the derivation in Sec.~\ref{sec:TEC}) we can simplify the calculation further if we also average their location across the $A$-cycle:
\begin{equation}\label{eq:cctildeaverage}
\frac{c^{(z_{2})}(z)}{z}\qquad\rightarrow\qquad \oint_A \frac{\rmd z}{2\pi iz}\frac{c^{(z_{2})}(z)}{z}=\oint_A \frac{\rmd z}{2\pi iz}\sum_{n\in\mathbf{Z}}\frac{c_n^{(z_{2})}}{z^n}=c_0^{(z_2)},
\end{equation}
and similarly for the anti-chiral half (note that the $A$ cycle traverses the patch overlap $U_1\cap^hU_2$). Therefore, (\ref{eq:fullpathintegralX4zX}) reduces to,
\begin{equation}\label{eq:fullpathintegralX4zX2}
\begin{aligned}
Z_{T^2}=\,\,\int_{\mathcal{M}_{1,0}}\frac{\rmd^{2}q}{4\tau_2} \frac{e^{-2\Phi}}{q\bar{q}} \Big\langle\suminnt\limits_{a}\,\hat{\mathscr{A}}_{a}^{(z_1)}\Big(e^{-\mathbi{c}_0^{(2)}c_0^{(z_2)}-\tilde{\mathbi{c}}_0^{(2)}\tilde{c}_0^{(z_2)}}\Big)\hat{\mathscr{A}}^{a}_{(z_2/q)}\Big\rangle_{S^2},
\end{aligned}
\end{equation}
and by looking back at the explicit expression for the local operator $\hat{\mathscr{A}}^{a}_{(z_2/q)}$ (also displayed in (\ref{eq:offshellA_alpha3})) it is seen why the specific averaging (\ref{eq:cctildeaverage}) was chosen: the explicit exponential in the correlator in (\ref{eq:fullpathintegralX4zX2}) precisely cancels a corresponding exponential in $\hat{\mathscr{A}}^{a}_{(z_2/q)}$. In particular,
\begin{equation}\label{eq:fullpathintegralX4zX3}
\begin{aligned}
Z_{T^2}=\,\,\int_{\mathcal{M}_{1,0}}\frac{\rmd^{2}q}{4\tau_2} \frac{1}{q\bar{q}} \suminnt\limits_{a}\,\bigg[e^{-2\Phi}\Big\langle\hat{\mathscr{A}}_{a}^{(z_1)}\hat{\mathscr{A}}^{a}_{(z_2/q)}\Big\rangle_{S^2}\Big|_{\tilde{\mathbi{c}}_0^{(2)}=\mathbi{c}_0^{(2)}=0}\bigg].
\end{aligned}
\end{equation}

Let us now compute the correlator in (\ref{eq:fullpathintegralX4zX3}). This is {\it almost} the same two-point amplitude (\ref{eq:2ptamp}) that was computed in Sec.~\ref{sec:S2ptA}, but let us in this section rather use the basis of states given in (\ref{eq:offshellA_alpha}) on p.~\pageref{eq:offshellA_alpha}, which after normal ordering and an appropriate rescaling $z_2\mapsto z_2/q$ read:
\begin{equation}\label{eq:offshellA_alpha3}
\begin{aligned}
\hat{\mathscr{A}}_{\alpha}^{(z_1)}&\dfn 
g_D\exp\bigg[\sum_{n\geq1}\frac{1}{n}\tilde{\mathbi{a}}_n\cdot \tilde{\alpha}_{-n}^{(z_1)}+\sum_{n\geq2}\tilde{\mathbi{b}}_n^{(1)}\tilde{b}_{-n}^{(z_1)}+\sum_{n\geq0}\tilde{\mathbi{c}}_n^{(1)}\tilde{c}_{-n}^{(z_1)}\bigg]\\
&\qquad\times 
\exp \bigg[\sum_{n\geq1}\frac{1}{n}\mathbi{a}_n\cdot \alpha_{-n}^{(z_1)}+\sum_{n\geq2}\mathbi{b}_n^{(1)}b_{-n}^{(z_1)}+\sum_{n\geq0}\mathbi{c}_n^{(1)}c_{-n}^{(z_1)}\bigg]\\
&\qquad\times \big(\tilde{c}_1^{(z_1)}+\tilde{\mathbi{b}}_1^{(1)}\big)\big(c_1^{(z_1)}+\mathbi{b}_1^{(1)}\big)e^{ik\cdot x^{(z_1)}},\\
&\phantom{a}\\
\hat{\mathscr{A}}^{\alpha}_{(z_2/q)}&\dfn 
g_D \exp\bigg[\!-\sum_{n\geq1}\bar{q}^n\frac{1}{n}\tilde{\mathbi{a}}_n^*\cdot \tilde{\alpha}_{-n}^{(z_2)}+\sum_{n\geq2}\bar{q}^n\tilde{\mathbi{b}}_n^{(2)}\tilde{b}_{-n}^{(z_2)}+\sum_{n\geq0}\bar{q}^n\tilde{\mathbi{c}}_n^{(2)}\tilde{c}_{-n}^{(z_2)}\,\Big)\bigg]\\
&\qquad\times\exp\bigg[\!-\sum_{n\geq1}q^n\frac{1}{n}\mathbi{a}_n^*\cdot \alpha_{-n}^{(z_2)}+\sum_{n\geq2}q^n\mathbi{b}_n^{(2)}b_{-n}^{(z_2)}+\sum_{n\geq0}q^n\mathbi{c}_n^{(2)}c_{-n}^{(z_2)}\bigg]
\\
&\qquad\times   (q\bar{q})^{\frac{\alpha'}{4}k^2}\big(\bar{q}^{-1}\tilde{c}_1^{(z_2)}+\tilde{\mathbi{b}}_1^{(2)}\big)\big(q^{-1}c_1^{(z_2)}+\mathbi{b}_1^{(2)}\big)e^{-ik\cdot x^{(z_2)}}
\end{aligned}
\end{equation}
The same calculation that led to (\ref{eq:2ptamp}) in the two-point amplitude computation of the previous section can also be carried out here. One subtle point is that even though $q\neq1$ in the present section we unwrap the modes through the overlap $U_1\cap^sU_2$, where the relevant transition function is $z_1z_2=1$ (rather than through the handle, $U_1\cap^hU_2$). This is just saying that having inserted the handle operator we are really to think of the resulting object as a two-point {\it sphere} amplitude, which is in turn constructed by gluing with $z_1z_2=1$ on $U_1\cap^sU_2$. So, e.g., when we unwrap the modes we make use of $\mathscr{O}_{-n}^{(z_1)}=(-)^h\mathscr{O}_{n}^{(z_2)}$. 
\sk

Evaluating the two-point amplitude with the above insertions and taking into account the above comments we find:
\begin{equation}\label{eq:2ptampT2qq}
\begin{aligned}
e^{-2\Phi}\Big\langle\hat{\mathscr{A}}_{\alpha}^{(z_1)}\hat{\mathscr{A}}^{\alpha}_{(z_2/q)}\Big\rangle_{S^2}
&=\frac{8\pi i}{\alpha'}V_D\delta\big(\mathbi{c}_0^{(1)}+\mathbi{c}_0^{(2)}\big)
\delta\big(\tilde{\mathbi{c}}_0^{(1)}+\tilde{\mathbi{c}}_0^{(2)}\big)(q\bar{q})^{\frac{\alpha'}{4}k^2-1}\\
&\quad\times\exp\bigg\{\sum_{n=1}^{\infty}\bar{q}^n\Big(\frac{1}{n}\tilde{\mathbi{a}}_n^*\cdot\tilde{\mathbi{a}}_n+\tilde{\mathbi{c}}_n^{(2)}\tilde{\mathbi{b}}_n^{(1)}-\tilde{\mathbi{b}}_n^{(2)}\tilde{\mathbi{c}}_n^{(1)}\Big)\bigg\}\\
&\quad
\times\exp\bigg\{\sum_{n=1}^{\infty}q^n\Big(\frac{1}{n}\mathbi{a}_n^*\cdot\mathbi{a}_n+\mathbi{c}_n^{(2)}\mathbi{b}_n^{(1)}-\mathbi{b}_n^{(2)}\mathbi{c}_n^{(1)}\Big)\bigg\},
\end{aligned}
\end{equation}
where $V_D\dfn (2\pi)^D\delta^D(k-k)$, and we made use of the usual normalisation of the measure given in (\ref{eq:S2-norm-modes}). Notice that according to (\ref{eq:fullpathintegralX4zX3}) we are to set $\tilde{\mathbi{c}}_0^{(2)}=\mathbi{c}_0^{(2)}=0$.
\sk

Incidentally, it is also possible to carry out this computation with the handle operator associated to local vertex operators (\ref{eq:offshellA_a}) and (\ref{eq:offshellA^a}) (or in fact their normal-ordered counterparts), but it {\it is} more subtle because one must correctly account for the scaling under $z_2\mapsto z _2/q$ of the zero modes in relation to the change of variables, $k\mapsto \e{p}$, that was discussed on p.~\pageref{item:deltak}.
\sk

Substituting (\ref{eq:2ptampT2qq}) into (\ref{eq:fullpathintegralX4zX3}), we then integrate out the various quantum numbers using the definition for the sum/integral over $\alpha$ given in (\ref{eq:dpdmualpha}) and (\ref{eq:dmuaC}). The calculation is very simple, in fact the formulas (\ref{eq:GaussianIntegral}) and (\ref{eq:bc-cb chiral}) make the computation immediate. From the latter one finds that for every integer $n$ the matter integrals produce $D$ factors of $(1-q^n)^{-1}$, whereas the ghost integrals produce correspondingly a factor of $(1-q^n)^2$ respectively. Setting $D=26$ (which is the only case where the explicit results are valid since we have not explicitly included a compactification manifold or an internal CFT more generally), we find the result:
\begin{equation}\label{eq:torusvacuum4}
\boxed{
\begin{aligned}
Z_{T^2}
&=iV_{26}\int_{\mathcal{M}_{1,0}}\frac{\rmd^{2}\tau}{4\tau_2}\,(4\pi^2\alpha'\tau_2)^{-13}|\eta(\tau)|^{-48},
\end{aligned}
}
\end{equation}
where we have written the result in terms of $\tau=\tau_1+i\tau_2$ (recall $q=e^{2\pi i\tau}$), and $\eta(\tau)$ is the Dedekind eta function, $\eta(\tau)=q^{\frac{1}{24}}\prod_{n=1}^{\infty}(1-q^n)$. Note that the overall factor of $i=\sqrt{-1}$ can be attributed to Wick rotating the time-like component of the momentum integral (the factor of $i$ appearing explicitly in (\ref{eq:2ptampT2qq}) is cancelled by a corresponding factor in the sum/integral over $a$). 
\sk

The relation (\ref{eq:torusvacuum4}) is of course the standard result for the one-loop partition function of bosonic string theory. Since this is explicitly modular invariant (i.e.~invariant under $\tau\mapsto \tau+1$ and $\tau\mapsto -1/\tau$), the integration domain, $\mathcal{M}_{1,0}$, should indeed be restricted to a fundamental domain in order to avoid overcounting, as advertised already in (\ref{eq:FundDomT2}).
\sk

This is an important and non-trivial consistency check. 
We have shown that although handle operators break {\it manifest} invariance under the modular group, they do in fact lead to modular-invariant amplitudes. Presumably this remains true at arbitrary loop order, but this has not been shown. Furthermore, the tree-level normalisation we have employed throughout the document {\it automatically} leads to the correct normalisation for one-loop amplitudes, which is also a highly non-trivial consistency check. (Recall that the traditional way to normalise string amplitudes \cite{Polchinski_v1} is by demanding consistent factorisation or unitarity, on a case by case basis. String amplitude normalisation in the handle operator approach should automatically be correct at arbitrary genus\footnote{At higher genus however it remains an open problem to unravel the action of the mapping class group, $\mathrm{Mod}_{\g,\n}$, on Teichm\"uller space, $\mathrm{T}_{\g,\n}$, in a ``handle operator gauge slice'' and account for any unfixed discrete symmetries.} once sphere amplitudes are correctly normalised, and this is indeed what we are finding here.)

\subsection{Sphere 4-Point Amplitude}\label{sec:VSG}
In this section we will perform yet another elementary consistency check of the formalism we have developed associated to gluing string amplitudes using handle operators. The focus will be on using a handle operator to cut across a trivial homology cycle (which is complementary to the one-loop partition function computed in Sec.~\ref{sec:MI} where we cut across a non-trivial homology cycle). The simplest non-trivial amplitude (in particular one that probes worldsheet duality) is the Virasoro-Shapiro amplitude which corresponds to a sphere 4-point amplitude associated to tachyon-tachyon scattering with onshell asymptotic states. We will construct this amplitude by gluing two 3-point amplitudes using the handle operators we have constructed. 
\sk

Although somewhat tangential, let us also make contact with closed string field theory (CSFT). In closed string field theory one constructs a four-point amplitude by decomposing the moduli space integrals in such a way that the full amplitude is expressed as a sum of $s$-, $t$- and $u$-channel contributions plus a contribution from a four-point vertex \cite{SonodaZwiebach90}, see also \cite{Sen19}. The correspondence with field theory is then immediate. Such a decomposition is however not necessary -- arguments based on associativity of the operator-product expansion (OPE), on bootstrap CFT, or worldsheet duality, all suggest that one should be able to reconstruct arbitrary string amplitudes by gluing the following elementary building blocks \cite{Polchinski88,Polchinski_v1}: sphere 3-pt amplitudes and torus 1-pt amplitudes. 
\sk

In fact, handle operator insertions into sphere amplitudes should presumably be sufficient to reconstruct arbitrary closed string amplitudes also. In the previous section we demonstrated one-loop modular invariance (where the cut was across a {\it non-trivial homology cycle}), so in order to be able to suggest that handle operator insertions are sufficient we should also demonstrate that cutting across a {\it trivial homology cycle} using a handle operator reproduces the correct result. Showing this will be the main objective of the current section.
\sk

In particular, focusing on a specific example of reconstructing a four-point amplitude by gluing two three-point amplitudes, we should be able \cite{Polchinski_v1} to reconstruct the full amplitude by adding {\it only} ``$s$- and $u$-channel'' contributions, the $t$-channel contribution then being implicitly contained in the infinite sums over intermediate states in the $s$- and $u$-channel terms. Similarly, the four-point vertex of CSFT is then also implicitly contained in the sum of $s$- and $u$-channel terms. In the traditional string perturbation theory approach one can either show this directly by working backwards from the known result (in terms of gamma functions) for the Virasoro-Shapiro amplitude, or one can show it directly by explicitly decomposing the moduli space integrals in a appropriate manner. We will discuss both of these before getting to the main objective which is to derive the full four-point amplitude by adding $s$- and $u$-channel contributions that are in turn constructed by gluing two three-point amplitudes using a handle operator for gluing. We will then see that the expression obtained from gluing does indeed reproduce the full amplitude non-ambiguously, the coordinate dependence used for gluing cancels out of the full amplitude, BRST-exact states decouple as shown on general grounds in Sec.\ref{sec:BRST-AC}, and the handle operators constructed are correctly normalised and consistent with unitarity.
\sk

\subsubsection{The Standard Story}
Let us now briefly recall some standard results regarding the Virasoro-Shapiro amplitude \cite{Polchinski_v1} since this will allow us to compare to the result obtained from gluing. 
The full tree-level S matrix for the scattering of four onshell tachyon primaries (without kinematic factors \cite{SklirosHindmarsh11}) reads (using standard notation \cite{Polchinski_v1}):\footnote{(In momentum space, the dimensionless S-matrix elements are obtained by supplying a factor $1/\sqrt{2k^0_jV_{D-1}}$ for every one of the external vertex operators, $j=1,\dots,4$ and $V_{D-1}\dfn (2\pi)^D\delta^D(0)$ is the formal non-compact spatial volume which cancels out of observables, allowing one to take the physical limit $V_{D-1}\rightarrow \infty$. With these kinematic factors included, the sum symbol over states is dimensionless also.)}
\begin{equation}\label{eq:S-VirasoroShapiro}
\begin{aligned}
S_{S^2}(1;2;3;4)&= g_D^4e^{-2\Phi}\int_{\mathbf{C}} \rmd^2z_4 \Big\langle \prod_{j=1}^3:\!\tilde{c}ce^{ik_j\cdot x}(z_j,\bar{z}_j)\!:\,:\!e^{ik_4\cdot x}(z_4,\bar{z}_4)\!:\Big\rangle_{S^2}\\
& = \frac{8\pi ig_D^2}{\alpha'}(2\pi)^D\delta^D(k)\frac{2\pi \Gamma(-1-\frac{\alpha'}{4}s)\Gamma(-1-\frac{\alpha'}{4}t)\Gamma(-1-\frac{\alpha'}{4}u)}{\Gamma(2+\frac{\alpha'}{4}s)\Gamma(2+\frac{\alpha'}{4}t)\Gamma(2+\frac{\alpha'}{4}u)},
\end{aligned}
\end{equation}
where we defined $k\dfn k_1+\dots+k_4$. So the standard approach is to use the ${\rm SL}(2,\mathbf{C})$ invariance of the amplitude to fix three of the four vertex operator locations using the $c$-ghosts delta function support, with the single complex structure modulus captured by integrating over the position of the one remaining vertex operator. After computing the contractions, one evaluates the resulting integral by analytically continuing the Mandelstam variables to a regime of absolute convergence (which is the region, ${\rm Re}(s)<-4/\alpha'$, ${\rm Re}(t)<-4/\alpha'$ and ${\rm Re}(u)<-4/\alpha'$, or, e.g., ${\rm Re}(s)<-4/\alpha'$, ${\rm Re}(t)<-4/\alpha'$ and ${\rm Re}(s+t)<-12/\alpha'$, where our conventions for $s,t,u$ are given below) and then define the physical amplitude by analytic continuation to Lorentzian signature physical momenta.

\sk
We note primarily that constructing amplitudes by gluing breaks manifest worldsheet duality, since one makes a choice as to which cycle to cut open (and there is no explicit symmetrising as in CSFT \cite{SonodaZwiebach90}). Consequently, e.g., the amplitude one obtains from gluing is not obviously invariant under arbitrary permutations of the $(s,t,u)$ Mandelstam variables. To make contact with the results obtained from gluing it is therefore useful to primarily make the pole structure of (\ref{eq:S-VirasoroShapiro}) manifest. We will do this in two ways: first make explicit the $s$ and $u$ pole structure manifest on the right-hand side of the second equality in (\ref{eq:S-VirasoroShapiro}); secondly, we decompose the moduli space integral on the right-hand side of the first equality in (\ref{eq:S-VirasoroShapiro}) explicitly in terms of $s$ plus $u$ channel contributions (with the $t$ channel implicit), hence showing that one can cover moduli space exactly once by gluing two three-point amplitudes and associating the one complex modulus of the four-point sphere amplitude with the plumbing fixture pinch modulus, $q$, used to glue the two surfaces. 

\sk
So we start from the gamma function representation on the right-hand side of the second equality in (\ref{eq:S-VirasoroShapiro}). 
 It is convenient to work in terms of the {\it invariant amplitude}, $\mathcal{A}_{S^2}(s,t,u)$,
\begin{equation}\label{eq:invariant amplitude}
S_{S^2}(1;2;3;4)=i(2\pi)^D\delta^D(k)\mathcal{A}_{S^2}(s,t,u).
\end{equation}

\sk
We will now expose the $s$- and $u$-channel poles and write the full amplitude correspondingly as a sum of two terms. The $t$-channel poles will be {\it implicit} in the infinite sum over intermediate states. To make this explicit, we start off by presenting our convention for the Mandelstam variables (recall we adopt $(-++\dots)$ spacetime metric signature):
\begin{equation}\label{eq:Mandelstam}
s=-(k_1+k_2)^2,\qquad t=-(k_1+k_3)^2,\qquad u=-(k_1+k_4)^2,
\end{equation}
with the momentum conservation ($k_1+\dots+k_4=0$) and onshell conditions ($k_j^2=4/\alpha'$, $j=1,\dots,4$) implying: 
$$
s+t+u=-\frac{16}{\alpha'}.
$$
We now make use of the following gamma function identity,
\begin{equation}
\frac{\Gamma(z)}{\Gamma(z+w)} = \sum_{n=0}^{\infty}\frac{(-)^n}{n!\Gamma(w-n)}\Big(\frac{1}{z+n}\Big),
\end{equation}
in (\ref{eq:S-VirasoroShapiro}), twice, once with $z=-1-\frac{\alpha'}{4}s$ and $w=-1-\frac{\alpha'}{4}t$ (with $z+w=2+\frac{\alpha'}{4}u$), and once with $z=-1-\frac{\alpha'}{4}u$ and $w=-1-\frac{\alpha'}{4}t$ (with $z+w=2+\frac{\alpha'}{4}s$). Upon using some additional standard gamma function identities\footnote{In particular, for integers $n$ and complex numbers $z$,
\begin{equation}\label{eq:GammaIdentities}
\Gamma(z+n)\Gamma(1-z-n)=(-)^{n+1}\Gamma(-z)\Gamma(1+z),\qquad \Gamma(1-z)\Gamma(z)=\frac{\pi}{\sin \pi z}.
\end{equation}
} 
the full amplitude can be written as a sum of a term containing only explicit $s$-channel poles and a term containing only explicit $u$-channel poles, 
\begin{equation}\label{eq:A_S2^(s+u)}
\mathcal{A}_{S^2}(s,t,u)=\mathcal{A}_{S^2}^s(s,t,u)+\mathcal{A}_{S^2}^u(s,t,u),
\end{equation}
with:
\begin{subequations}\label{eq:As,Au}
\begin{align}
\mathcal{A}_{S^2}^s(s,t,u)&=\Big(\frac{8\pi g_D}{\alpha'}\Big)^2\sum_{n,m=0}^{\infty} \Big(\frac{\Gamma(2+n+\frac{\alpha'}{4}t)}{\Gamma(n+1)\Gamma(2+\frac{\alpha'}{4}t)}\Big)\Big(\frac{\Gamma(2+m+\frac{\alpha'}{4}t)}{\Gamma(m+1)\Gamma(2+\frac{\alpha'}{4}t)}\Big)\nonumber\\
&\quad\times
(-)^{n+m}\frac{\sin[\pi(-1+m-\frac{\alpha'}{4}u)]}{\pi(-\frac{\alpha'}{4}u+m-1)}\cdot\frac{\cos[\pi(-1+n-\frac{\alpha'}{4}s)]}{-s+\frac{4}{\alpha'}(n-1)-i\epsilon}\label{eq:A_S2^s}\\
\mathcal{A}_{S^2}^u(s,t,u)&=\Big(\frac{8\pi g_D}{\alpha'}\Big)^2\sum_{n,m=0}^{\infty} \Big(\frac{\Gamma(2+n+\frac{\alpha'}{4}t)}{\Gamma(n+1)\Gamma(2+\frac{\alpha'}{4}t)}\Big)\Big(\frac{\Gamma(2+m+\frac{\alpha'}{4}t)}{\Gamma(m+1)\Gamma(2+\frac{\alpha'}{4}t)}\Big)\nonumber\\
&\quad\times
(-)^{n+m}\frac{\sin[\pi(-1+m-\frac{\alpha'}{4}s)]}{\pi(-\frac{\alpha'}{4}s+m-1)}\cdot\frac{\cos[\pi(-1+n-\frac{\alpha'}{4}u)]}{-u+\frac{4}{\alpha'}(n-1)-i\epsilon}\label{eq:A_S2^u}
\end{align}
\end{subequations}
where the $u$-channel contributions, $\mathcal{A}_{S^2}^u$, are obtained from $\mathcal{A}_{S^2}^s$ by interchanging $s\leftrightarrow u$, 
$
\mathcal{A}_{S^2}^u(s,t,u) = \mathcal{A}_{S^2}^s(u,t,s).
$ 
Of course (\ref{eq:As,Au}) is not the most economical way of presenting the result since either one or both of the sums over $n,m$ can be carried out, and also further gamma function identities can be used to write it in a more compact form even for fixed $n,m$. But we have presented the result in the form (\ref{eq:As,Au}) to make the pole structure {\it explicit}, i.e.~that the summand in $\mathcal{A}_{S^2}^s$ for given $n,m$ only has  explicit poles in $s$,
\begin{equation}\label{eq:s-poles}
\alpha's=-4,0,4,8,\dots,
\end{equation}
and likewise for $\mathcal{A}_{S^2}^u$ where the only explicit poles occur at $\alpha'u=-4,0,4,8,\dots$. We also included the Feynman prescription appropriate for a Minkowski process. 
\sk

Now that the pole structure has been brought to plain view it is useful to sum over $m$ in both $\mathcal{A}_{S^2}^s$ and $\mathcal{A}_{S^2}^u$ in (\ref{eq:As,Au}). On account also of the identities in (\ref{eq:GammaIdentities}) one finds:
\begin{subequations}\label{eq:As,Au2}
\begin{align}
\mathcal{A}_{S^2}^s(s,t,u)
&=\Big(\frac{8\pi g_D}{\alpha'}\Big)^2\sum_{n=0}^{\infty} \Big(\frac{\Gamma(2+n+\frac{\alpha'}{4}t)}{\Gamma(n+1)\Gamma(2+\frac{\alpha'}{4}t)}\Big)^2
\nonumber\\
&\quad\times\frac{\Gamma(-1-n-\frac{\alpha'}{4}t)\Gamma(1+n)}{\Gamma(-2-\frac{\alpha'}{4}s-\frac{\alpha'}{4}t)\Gamma(2+\frac{\alpha'}{4}s)}
\frac{\cos[\pi(-1+n-\frac{\alpha'}{4}s)]}{-s+\frac{4}{\alpha'}(n-1)-i\epsilon}\label{eq:A_S2^s2}\\
\mathcal{A}_{S^2}^u(s,t,u)
&=\Big(\frac{8\pi g_D}{\alpha'}\Big)^2\sum_{n=0}^{\infty} \Big(\frac{\Gamma(2+n+\frac{\alpha'}{4}t)}{\Gamma(n+1)\Gamma(2+\frac{\alpha'}{4}t)}\Big)^2\nonumber\\
&\quad\times\frac{\Gamma(-1-n-\frac{\alpha'}{4}t)\Gamma(1+n)}{\Gamma(-2-\frac{\alpha'}{4}u-\frac{\alpha'}{4}t)\Gamma(2+\frac{\alpha'}{4}u)}
\frac{\cos[\pi(-1+n-\frac{\alpha'}{4}u)]}{-u+\frac{4}{\alpha'}(n-1)-i\epsilon}\label{eq:A_S2^u2}
\end{align}
\end{subequations}
It is an elementary exercise to check that the sum of these two ($s$ and $u$ channel) expressions reproduces the gamma function representation on the right-hand side of the second equality in (\ref{eq:S-VirasoroShapiro}) upon including the momentum-conserving delta function.
\sk

Let us state from the outset that the decomposition into $s$- and $u$-channel contributions depicted in (\ref{eq:As,Au2}) is certainly not unique, and in fact there are various ways of representing the same amplitude $\mathcal{A}_{S^2}$ as a sum over $s$- and $u$-channels, e.g., $\mathcal{A}_{S^2}=\mathcal{A}_{S^2}^{'s}+\mathcal{A}_{S^2}^{'u}$, where $\mathcal{A}_{S^2}^{'s}\neq \mathcal{A}_{S^2}^{s}$ (and similarly for $u$) but nevertheless the sum still adds up to yield the same expression\footnote{DS thanks Ashoke Sen for an extensive discussion and for emphasising this crucial fact.} for the full amplitude $\mathcal{A}_{S^2}$. This will be important when we compare the result for the VS amplitude obtained from gluing to the expression derived directly from the known result.
\sk

We will also be interested in extracting the {\it imaginary} part of the amplitude, since by unitarity this should only get contributions from {\it onshell} propagating states through the intermediate propagator and BRST exact contributions should also decouple. 
Let us consider the case where taking the imaginary part commutes with the sum over $n$ in $\mathcal{A}_{S^2}$ expressed in terms of (\ref{eq:As,Au2}). A necessary and sufficient condition for these two operations to commute is that $\alpha't\neq -4,0,4,8,\dots$ since according to (\ref{eq:S-VirasoroShapiro}) the {\it sums} in (\ref{eq:As,Au2}) must diverge on the $t$-channel poles because the summands for fixed $n$ are holomorphic in $t$. So away from the $t$-channel poles, to extract the corresponding imaginary parts, denoted by ${\rm Im}\,\mathcal{A}_{S^2}^s(s,t,u)$ and ${\rm Im}\,\mathcal{A}_{S^2}^u(s,t,u)$, we can use the above observations (that the only poles in (\ref{eq:A_S2^s2}) and (\ref{eq:A_S2^u2}) for fixed $n$ are associated to the explicit propagators with resonances at $s=\frac{4}{\alpha'}(n-1)$ and $u=\frac{4}{\alpha'}(n-1)$ respectively), and hence make use of the standard identity $2\pi i\delta(x)=\frac{1}{x-i\epsilon}-\frac{1}{x+i\epsilon}$ according to which ${\rm Im}\,\frac{1}{x-i\epsilon}=\pi\delta(x)$. So the imaginary parts for $\alpha't\neq -4,0,4,8,\dots$ take the form:
\begin{subequations}\label{eq:ImAs,ImAu2}
\begin{align}
{\rm Im}\,\mathcal{A}_{S^2}^s(s,t,u)
&=\Big(\frac{8\pi g_D}{\alpha'}\Big)^2\sum_{n=0}^{\infty} \Big(\frac{\Gamma(2+n+\frac{\alpha'}{4}t)}{\Gamma(n+1)\Gamma(2+\frac{\alpha'}{4}t)}\Big)^2
\pi\delta\big(s-\tfrac{4}{\alpha'}(n-1)\big)\label{eq:ImA_S2^s}\\
{\rm Im}\,\mathcal{A}_{S^2}^u(s,t,u)
&=\Big(\frac{8\pi g_D}{\alpha'}\Big)^2\sum_{n=0}^{\infty} \Big(\frac{\Gamma(2+n+\frac{\alpha'}{4}t)}{\Gamma(n+1)\Gamma(2+\frac{\alpha'}{4}t)}\Big)^2\pi\delta\big(u-\tfrac{4}{\alpha'}(n-1)\big)
\label{eq:ImA_S2^u}
\end{align}
\end{subequations}
These relations have been derived here from the known gamma function representation of the VS amplitude. In the following section we carry out the corresponding computation of VS by gluing two 3-point amplitudes using offshell coherent states for the gluing.
\sk

Having shown that it is indeed possible to express the full VS amplitude as a sum of only $s$- and $u$-channel pole contributions we will next show this directly by decomposing the moduli space integral starting from an integral representation such as that given in the first equality in (\ref{eq:S-VirasoroShapiro}), which for four generic primary fixed-picture (of ghost number 2) or integrated-picture vertex operators, $\hat{\mathscr{V}}_j=\tilde{c}c\mathscr{V}_j$ or $\mathscr{V}_j$ respectively (where the latter is constructed solely out of matter contributions), for $j=1,\dots,4$, generalises to,\footnote{Note that the integration domain is written (somewhat imprecisely) as $\mathbf{C}$ rather than $\mathbf{C}\cup \{\infty\}$. In the usual approach to computing this amplitude the distinction is immaterial, but we will make a more precise statement below.}
\begin{equation}\label{eq:S-VirasoroShapiroX}
\begin{aligned}
S_{S^2}(1;2;3;4)
&= g_D^4e^{-2\Phi}\int_{\mathbf{C}} \rmd^2z_4 \Big\langle \hat{\mathscr{V}}^{(z)}_1(z_1,\bar{z}_1)\,\hat{\mathscr{V}}^{(z)}_2(z_2,\bar{z}_2)\,\hat{\mathscr{V}}^{(z)}_3(z_3,\bar{z}_3)\,\mathscr{V}^{(z)}_4(z_4,\bar{z}_4)\Big\rangle_{S^2}\\
\end{aligned}
\end{equation}
The superscript $z$ indicates that all vertex operators appearing are defined in the same coordinate chart, $(U_z,z)$, in particular: $z_j\dfn z(p_j)$, where $p_j\in U_z\subset S^2$ for $j=1,\dots,4$ denote marked points in $S^2$. 
\sk

The expression (\ref{eq:S-VirasoroShapiroX}) is a standard way of representing a four-point closed string amplitude. However, there is now a small detail that is hard to avoid, namely that there does not exist a single coordinate system for $S^2$. This is a minor point for the above sphere amplitude, and can often be ignored without trouble, but since we will eventually be cutting open the path integral across a specific cycle making explicit the various coordinate systems will be inevitable. So let us write down a slightly more precise expression for $S_{S^2}(1;2;3;4)$. 
\sk

It will be convenient to primarily translate all vertex operators to fixed-picture,
\begin{equation}\label{eq:S-VirasoroShapiroX2}
\begin{aligned}
S_{S^2}(1;2;3;4)
&= g_D^4e^{-2\Phi}\int_{\mathbf{C}} \rmd^2z_4 \Big\langle \hat{\mathscr{V}}^{(z)}(z_1,\bar{z}_1)\,\hat{\mathscr{V}}^{(z)}(z_2,\bar{z}_2)\,\frac{b_0^{(z)}\tilde{b}_0^{(z)}(p_3)}{z_4\bar{z}_4}\hat{\mathscr{V}}^{(z)}(z_3,\bar{z}_3)\,\hat{\mathscr{V}}^{(z)}(z_4,\bar{z}_4)\Big\rangle_{S^2}\\
\end{aligned}
\end{equation}
where $b_0^{(z)}(p_3)$ is a mode constructed in the $z$ coordinate chart and centred at the vertex operator insertion at $p_3$, the contour being such that it encloses {\it both} $p_3$ and $p_4$. Identical remarks apply to the anti-chiral half, $\tilde{b}_0^{(z)}(p_3)$. 
\sk

Let us cover $S^2$ with two coordinate charts, $(U_w,w)$ and $(U_v,v)$, constructed such that they overlap on an equatorial band, so that $U_w\cap U_v$ is topologically an annulus. We can extend the patch overlap to include almost the entire sphere except for the north and south poles, $p_N,p_S\in S^2$, at which the $w$ and $v$ coordinates respectively are centred,
$$
w(p_N)\dfn0,\qquad v(p_S)\dfn 0.
$$
Away from the poles, $p_N,p_S\in S^2$, the two charts overlap and we can adopt a transition function of the form:
$$
wv=1,\qquad \textrm{on $U_w\cap U_v$}.
$$
That is, any point $p\in U_w\cap U_v$ (which excludes $p_N$ and $p_S$) now has two coordinate representations $w(p)$ and $v(p)$ related via $w(p)v(p)=1$. (Notice that the transition function is holomorphic since the marked points $p_N,p_S\in S^2$ are omitted, in accordance with the general analysis of Sec.~\ref{sec:TFCR}.) At any such  point $p$ a given fixed-picture primary vertex operator defined in the $w$ chart, $\hat{\mathscr{V}}_j^{(w)}(p)$, of conformal weight $(h_j,\tilde{h}_j)$ transforms as a tensor, namely:
$$
\hat{\mathscr{V}}_j^{(w)}(w(p),\overline{w}(p)) = \hat{\mathscr{V}}_j^{(v)}(v(p),\overline{v}(p)) (\partial_wv)^{h_j}(\partial_{\bar{w}}\bar{v})^{\tilde{h}_j}.
$$
When all external vertex operators in a given string amplitude are onshell and primaries then $(h_j,\tilde{h}_j)=(0,0)$, and we have that on patch overlaps,
$$
\hat{\mathscr{V}}_j^{(w)}(w(p),\overline{w}(p)) = \hat{\mathscr{V}}_j^{(v)}(v(p),\overline{v}(p)).
$$

\subsubsection{Outline of the Gluing Approach}

We wish to derive the Virasoro-Shapiro amplitude (\ref{eq:S-VirasoroShapiro}) by gluing two three-point amplitudes using the handle operators derived above for the gluing. If we cut along the $s$-channel we (naively) expect to recover (\ref{eq:A_S2^s}), whereas if we cut along the $u$-channel we (naively) expect to recover (\ref{eq:A_S2^u}). 

\sk
For the external states we use fixed picture primary vertex operators (of ghost number $N_{\rm gh}=2$), and in particular: 
$$
\hat{\mathscr{V}}_j(z_j,\bar{z}_j)=\,g_D\!:\!\tilde{c}ce^{ik_j\cdot x}(z_j,\bar{z}_j)\!\!:,\qquad j=1,\dots,4.
$$
Given a 4-point sphere amplitude has one complex modulus we can associate that modulus with a separating degeneration. The $s$-channel contribution (in particular) obtained from gluing is then given by: 
\begin{equation}\label{eq:S-VirasoroShapiro-gluing}
\begin{aligned}
S_{S^2}^{\infty_s}(1;\dots;4)&= g_D^4\suminnt\limits_{a} \,\,e^{-2\Phi}\Big\langle \prod_{j=1,2}\!\!:\!\tilde{c}ce^{ik_j\cdot x}(z_j,\bar{z}_j)\!:\hat{\mathscr{A}}_{a}^{(z)}\Big\rangle_{S^2}\\
&\quad\times\int \rmd^2q\,q^{h_a-1}\bar{q}^{\tilde{h}_a-1} e^{-2\Phi}\Big\langle \big[b_0^{(u)}\tilde{b}_0^{(u)}\cdot \hat{\mathscr{A}}^{a}_{(u)}\big]\prod_{j=3,4}\!\!:\!\tilde{c}ce^{ik_j\cdot x}(u_j,\bar{u}_j)\!:\Big\rangle_{S^{'2}}\\
\end{aligned}
\end{equation}
The integral over the coherent state quantum numbers appearing in (\ref{eq:S-VirasoroShapiro-gluing}) has been defined in (\ref{eq:dpdmua}) and (\ref{eq:dmua}). 
The conformal weights, $(h_a,\tilde{h}_a)$, of the coherent state vertex operators are given in (\ref{eq:L0Lb0A=DA}), with $\Delta_a=h_a=\tilde{h}_a$ on account of (\ref{eq:AveragePhase}) (and the comment following the latter), see also below. It is also convenient to define the analogue of the invariant amplitude associated explicitly to the $s$-channel (\ref{eq:invariant amplitude}):
\begin{equation}\label{eq:invariant amplitude_infty_s}
S_{S^2}^{\infty_s}(1;\dots;4)=i(2\pi)^D\delta^D(k_1+\dots+k_4)\mathcal{A}_{S^2}^{\infty_s}(s,t,u).
\end{equation}
Adopting notation similar to that introduced in \cite{Vafa87}, a superscript `$^{\infty}$' indicates a result inherited from gluing two otherwise disconnected amplitudes (a separating degeneration) to distinguish it from $\mathcal{A}_{S^2}^{s}(s,t,u)$ in (\ref{eq:As,Au}) that was derived directly from the full amplitude (\ref{eq:S-VirasoroShapiro}). (Non-separating degenerations might be denoted correspondingly by `$^8$'.) 
\sk

Regarding the specifics of the worldsheet coordinates in the gluing construction, recall primarily that the gluing condition in one convenient set of coordinates is:
\begin{equation}\label{eq:zu=q}
zu=q,
\end{equation}
we place the vertex operators on $S^2$ at $z_1\equiv z(p_1)$, $z_2\equiv z(p_2)$ and the coherent state at $z(p_0)\equiv 0$, and the $p_1,p_2$ and $p_0$ indicate marked points on $U_z\subset S^2$. (We initially denote the coordinate at which the coherent state is inserted by $z_3=z(p_0)$ and subsequently set $z_3=0$.) So we place {\it all} vertex operators in the (first of the two) 3-point sphere path integrals in (\ref{eq:S-VirasoroShapiro-gluing}) in a {\it single} coordinate chart\footnote{Even though we cannot cover the whole sphere with a single coordinate chart we nevertheless have the freedom to place the vertex operators on a single chart since their locations are fixed.} $(U_z,z)$. It is to be understood that the two external vertex operators are placed outside of the sewing circle, $|z_1|,|z_2|>|q|^{1/2}$, and given these are fixed vertex operators we may take advantage of the SL$(2,\mathbf{C})$ invariance of $S^2$ which allows us to take $z_1\rightarrow \infty$ and $z_2\rightarrow 1$, but we will keep $z_2$ explicit in order to track some of the coordinate (in)dependence of the result. The coordinates for the $S^{'2}$ 3-point amplitude in (\ref{eq:S-VirasoroShapiro-gluing}) are as follows: we place the vertex operators at\footnote{The choice of subscripts is such that they correspond to the corresponding subscripts of the external vertex operator momenta.} $u_3\equiv u(p_3')$, $u_4\equiv u(p_4')$ and the coherent state at $u(p_0')\equiv 0$. (Similarly here, we initially denote the coordinate at which this coherent state is inserted by $u_2$ and subsequently set $u_2\equiv 0$.) Again, we choose to place {\it all} three vertex operators on a single coordinate chart $(U_u,u)$ without loss of generality, and $p_3',p_4'$ and $p_0'$ indicate marked points on $U_u\subset S^{'2}$. As in the case of $S^2$, also here it is to be understood that the external vertex operators are placed outside of the sewing circle, $|u_3|,|u_4|>|q|^{1/2}$, and we may take in particular $u_3\rightarrow \infty$ and $u_4\rightarrow 1$ (although again we keep $u_4$ explicit throughout) by appealing to the SL$(2,\mathbf{C})$ invariance of $S^{'2}$. 
\sk

Concerning the range of the $q,\bar{q}$ moduli integrals in (\ref{eq:S-VirasoroShapiro-gluing}), given the above comments, the natural integration domain for the $q$ integral is 
\begin{equation}\label{eq:|q| range VS}
|q|<|z_2u_4|,\qquad \big({\rm or}\qquad |q|<|z_2|^2=|u_4|^2 \qquad \textrm{if we set \,\,\,\,$|z_2|=|u_4|$}\big)
\end{equation}
This says that the pinch can only grow to a radius as large as the closest other vertex operator. 
If $|q|$ violated this inequality then the coordinates $z_2,u_4$ of the two vertex operators would be identified (because we have the gluing condition, $q=zu$) and the sewing would be inconsistent. So the indicated range is the largest consistent range for the $|q|$ integral and precisely \cite{Polchinski_v1,JoeBigBook} covers the full moduli space when the corresponding $u$-channel contribution is included as well. Notice that we can always absorb the factor $|z_2u_4|$ into rescalings of the frame coordinates, e.g.,
\begin{equation}\label{eq:zz'uu'qq'}
z\mapsto z'=\frac{z}{|z_2|},\qquad u\mapsto u'=\frac{u}{|u_4|}, \qquad{\rm with}\qquad z'u'=q',\qquad q'\dfn q/|z_2u_4|.
\end{equation}
Taking also into account that the local operators used to construct the handle operator have well-defined scaling dimension, the integration range (\ref{eq:|q| range VS}) is then replaced by,
$$
|q'|<1,
$$
in which case we are to use handle operators, $\hat{\mathscr{A}}_{a}^{(z')}$ and $\hat{\mathscr{A}}^{a}_{(u')}$, defined in the $z'$ and $u'$ frames respectively in (\ref{eq:S-VirasoroShapiro-gluing}). 
This is a general result, and (at least for separating degenerations) this can always be taken to be the range of integration of the gluing parameter. 
But we will work with $q$ instead of $q'$ in order to see explicitly how the corresponding scale cancels out of the final answer for the amplitude.
\sk

The natural range of the corresponding phase, $\theta$, where $q=|q|e^{i\theta}$, is $[0,2\pi)$. (Note that ${\rm SL}(2,\mathbf{C})$ invariance allows us to set $|z_2|=|u_4|$, in which case we have the second inequality in (\ref{eq:|q| range VS}).) Therefore, given (\ref{eq:|q| range VS}) the integral over $\theta$ will project onto level-matched contributions only, which in turn implies we can project onto level-matched contributions in the vertex operators associated to the handle operator without loss of generality. This is generic when one cuts along trivial homology cycles (whereas for non-trivial homology cycles non-level matched contributions also contribute). This will be implemented by taking the phases of the $z$ and $u$ coordinates to be integrated (denoted by the $\phi$ integral in (\ref{eq:S-VirasoroShapiro-3pt_a}), see also  (\ref{eq:L0Lb0A=DA})). 

\subsubsection{The Two 3-Point Amplitudes}
A complete set of offshell normal-ordered coherent states was derived above, see (\ref{eq:offshellA_a3}). We primarily compute the sphere 3-point amplitude appearing in (\ref{eq:S-VirasoroShapiro-gluing}),
\begin{equation}\label{eq:S-VirasoroShapiro-3pt_a}
\begin{aligned}
g_D^2&e^{-2\Phi}\Big\langle \prod_{j=1,2}\!\!:\!\tilde{c}ce^{ik_j\cdot x}(z_j,\bar{z}_j)\!:\hat{\mathscr{A}}_{a}^{(z)}(z_3,\bar{z}_3)\Big\rangle_{S^2}=\\
&=g_D^3\int_0^{2\pi}\frac{\rmd\phi}{2\pi}e^{-2\Phi}\Big\langle :\!\tilde{c}ce^{ik_1\cdot x}(z_1,\bar{z}_1)\!:\,:\!\tilde{c}ce^{ik_2\cdot x}(z_2,\bar{z}_2)\!:\,\tilde{\e{U}}_{(1)}\e{U}_{(1)}\,e^{i\e{p}\cdot x}(z_3,\bar{z}_3)\Big\rangle_{S^2}\\
\end{aligned}
\end{equation}
where the integral over $\phi$ is to be interpreted as arising from averaging over the U(1) phase of the local operator, $\hat{\mathscr{A}}_{a}^{(z)}$, as indicated in (\ref{eq:AveragePhase}). (We will also include a corresponding average for the dual operator, $\hat{\mathscr{A}}^{a}_{(u)}$.) So in fact the operators $\tilde{\e{U}}_{(1)}\e{U}_{(1)}$ appearing on the right-hand side of (\ref{eq:S-VirasoroShapiro-3pt_a}) should be interpreted as in (\ref{eq:offshellA_a3}) but with the quantum numbers appearing in (\ref{eq:ABCcondensed_notation}) rescaled accordingly as indicated in (\ref{eq:abc phase repl}). This of course does not affect the value of the full four-point amplitude we are aiming for since these phases can all be reabsorbed into redefinitions of $\theta$, the phase of the pinch modulus $q=|q|e^{i\theta}$ (see above). 
\sk

Given the coherent state is offshell the quantity $\e{p}^{\mu}$ does not satisfy any mass-shell condition (but given any vector $\e{p}^{\mu}$ we are to choose $\e{q}^{\mu}$ that is contained in $\tilde{\e{U}}_{(1)},\e{U}_{(1)}$ such that $\e{q}^2=0$ and $\e{p}\cdot \e{q}=2/\alpha'$ as discussed on p.~\pageref{eq:kpq_constr}). Let us primarily discuss the ghost contributions to the above path integral.

\subsubsection*{Ghost Contributions}

The ghost path integrals on $S^2$ that we will be needing (or higher point/genus when the latter are constructed by gluing) can be derived from repeated functional differentiation (with respect to Grassmann-odd sources $J(z,\bar{z}),I(z,\bar{z})$) of the generating function:
\begin{equation}\label{eq:bc-pathintegral}
\begin{aligned}
\e{Z}^{\rm g}(I,J;\tilde{I},\tilde{J})&\dfn\int^{S^2}\!\! \mathcal{D}(b,c,\tilde{b},\tilde{c})\,e^{-\frac{1}{2\pi}\int \,(b\bar{\partial}c+\tilde{b}\partial\tilde{c})+\int \,(Ib+Jc)+\int \,(\tilde{I}\tilde{b}+\tilde{J}\tilde{c})}\\
&\,\,=C_{S^2}^{\rm g} \,\int J\psi_1\int J\psi_2\int J\psi_3\,e^{-\int I(z,\bar{z})J(w,\bar{w})\frac{1}{z-w}}\\
&\qquad\times \int \tilde{J}\bar{\psi}_1\int \tilde{J}\bar{\psi}_2\int \tilde{J}\bar{\psi}_3\,e^{-\int \tilde{I}(z,\bar{z})\tilde{J}(w,\bar{w})\frac{1}{\bar{z}-\bar{w}}}
\end{aligned}
\end{equation}
provided the sources $I,J,\tilde{I},\tilde{J}$ only have compact support in a single coordinate chart. The path integral determinant/normalisation, $C_{S^2}^{\rm g}$, in (\ref{eq:bc-pathintegral}) is precisely equivalent to that in \cite{Polchinski_v1}, and only its combined value with the matter normalisation and dilaton (when it has a zero mode) is determined and required. The effective worldsheet measure $\rmd^2z=i\rmd z\wedge \rmd\bar{z}$ in (\ref{eq:bc-pathintegral}) is kept implicit (and the exponents on the right-hand side of the second equality are double integrals with respect to $z$ and $w$). The $\psi_a(z)$, $a=1,2,3$, are the conformal Killing vectors on $S^2$, i.e.~the three linearly-independent globally defined holomorphic vectors on $S^2$,
\begin{equation}\label{eq:CKVs_S2}
\psi_1(z) = 1,\qquad \psi_2(z) = z,\qquad {\rm and}\qquad \psi_3(z) = z^2,
\end{equation}
where we have picked an orthonormal basis with unit norm with respect to the natural inner product. These are in turn related to the $c$-ghost zero modes via the orthogonal decomposition,
\begin{equation}\label{eq:c(z)dec}
c(z) = \sum_ac_a\psi_a(z)+c'(z),
\end{equation}
where the $\{c_a\}$ are Grassmann-odd and a prime denotes quantum fluctuations. The $b$-ghosts have no zero modes on $S^2$ so the path integral measure decomposes as $\mathcal{D}(b,c)= \mathcal{D}(b',c')\prod_a\rmd c_a$. Quantities with an overbar, such as $\bar{\psi}_a$ and $\bar{z}$, are obtained by complex conjugation from $\psi_a$ and $z$, whereas a tilde, such as $\tilde{I},\tilde{J}$, denotes anti-chiral quantities that are independent of the corresponding un-tilded quantities, $I,J$. In the case of coherent states the sources, $I,J,\dots$, are physical and can be read off from equations such as (\ref{eq:offshellA_a_VOUU}) or some combination thereof.
\sk

We first make contact with standard results \cite{Polchinski_v1}. The $bc$ path integral (\ref{eq:bc-pathintegral}) chirally factorises \cite{VerlindeVerlinde87}, so it is convenient to define chiral and anti-chiral contributions,
\begin{equation}\label{eq:Z=CZZ^gg}
\e{Z}^{\rm g}(I,J;\tilde{I},\tilde{J})\equiv C_{S^2}^{\rm g}\big\langle e^{\int (Ib+Jc)}\big\rangle_{bc}\big\langle e^{\int (\tilde{I}\tilde{b}+\tilde{J}\tilde{c})}\big\rangle_{\tilde{b}\tilde{c}}
\end{equation}
where:
\begin{equation}\label{eq:Zgchiral}
\big\langle e^{\int (Ib+Jc)}\big\rangle_{bc}\equiv \int J\psi_1\int J\psi_2\int J\psi_3\,e^{-\int_z\int_w I(z,\bar{z})J(w,\bar{w})\frac{1}{z-w}},
\end{equation}
and a corresponding equation for the anti-chiral half (that is read off from (\ref{eq:bc-pathintegral})). Notice in particular that this effective chiral correlator computes expectation values of the full ghost fields (i.e.~including zero modes). Furthermore, (\ref{eq:Zgchiral}) defines the notation  $\langle \dots\rangle_{bc}$ for chiral path integrals (equivalently, chiral correlators). We could write more fully $\langle \dots\rangle_{S^2,bc}$ for the same quantity (\ref{eq:Zgchiral}) to emphasise this is a correlator on $S^2$, but this will not be necessary here.
\sk

As mentioned above, we have implicitly assumed that the sources, $I,J$, only have support in a {\it single} coordinate chart. This is actually all one needs, in that when more charts are required one may use the corresponding (holomorphic) transition functions on chart overlaps in order to place operators on different charts. Sometimes one has to work a little harder and add `Wu-Yang'-type terms depending on whether the corresponding integrands are globally well-defined (which in the current context means that they transform as tensors under conformal transformations) or not. We will not analyse this point in detail here, but rather only consider the cases we will be needing. We refer the interested reader to  \cite{WuYang75,Polchinski80,Alvarez84,Polchinski87,VerlindeH87,VerlindeHphd,SenWitten15} for some discussion of the relevant points (albeit, except for \cite{Polchinski87}, in different contexts).
\sk

As an example and to make contact with standard results, the simplest application of (\ref{eq:Zgchiral}) is the basic ghost correlator:
\begin{equation}\label{eq:<ccc>}
\begin{aligned}
\big\langle c(z_1)c(z_2)c(z_3)\big\rangle_{bc} &=\delta_{J_1}\delta_{J_2}\delta_{J_3}\e{Z}^{\rm g}(I,J)\big|_{I=J=0}\\
&= {\rm det}\,\psi_a(z_b)\\
&=z_{12}z_{13}z_{23},
\end{aligned}
\end{equation}
where $z_2,z_2,z_3$ are all considered to be in the $(U_z,z)$ chart and $\delta_{J_a}\equiv \delta/\delta J(z_a,\bar{z}_a)$ is odd and denotes functional differentiation. Note that this result is consistent  (as is the full generating function, $\e{Z}^{\rm g}(I,J;\tilde{I},\tilde{J})$) with our normalisation condition (\ref{eq:<..>normalisation1}) that is adopted throughout. 
Now, we need two patches to cover $S^2$, call them $U_z,U_u$ and corresponding charts $(U_z,z)$ and $(U_u,u)$, such that on patch overlaps, $U_u\cap U_z$, we have the gluing relation $zu=1$. The above chiral path integral could be written more precisely as $\big\langle c^{(z)}(z_1)c^{(z)}(z_2)c^{(z)}(z_3)\big\rangle_{S^2}$, so that $c^{(z)}(z_1)$ is inserted at the point $z=z_1$ in the holomorphic coordinate chart $(U_z,z)$, with $z:U_z\rightarrow \mathbf{C}$. 
\sk

Let us use the above to compute the following ghost path integral:
\begin{equation}\label{eq:<ccCS(c+B)>}
\Big\langle c(z_1)c(z_2)\,:\!e^{\int Ib+Jc}(c+\e{B}_1)(z_3)\!:\Big\rangle_{bc},
\end{equation}
where all insertions are placed on a single chart $(U_z,z)$. When we regard the local insertion at $z_3$ as being associated to a coherent state, see (\ref{eq:offshellA_a_VOUU}), we can read off the relevant sources from the latter directly:
\begin{equation}\label{eq:IJbcexp}
\begin{aligned}
I(z,\bar{z}) &= \delta^2(z-z_3)\sum_{n\geq2}\frac{\e{B}_n}{(n-2)!}\partial_z^{n-2}\\
J(z,\bar{z}) &= \delta^2(z-z_3)\sum_{n\geq1}\frac{\e{C}_n}{(n+1)!}\partial_z^{n+1},
\end{aligned}
\end{equation}
and we are using the short-hand notation (\ref{eq:ABCcondensed_notation}) for $\e{B}_n,\e{C}_n$ (and we have dropped the superscripts that distinguish one coherent state from another as there is only one coherent state in this amplitude). 

\sk
We use standard field theory techniques to evaluate (\ref{eq:<ccCS(c+B)>}). Namely, we rewrite it as an operator acting on the generating function given in (\ref{eq:Zgchiral}), $\delta_{J_1}\delta_{J_2}(\delta_{J_3}+\e{B}_1)\langle e^{\int Ib+Jc}\rangle_{bc}|_{(\ref{eq:IJbcexp})}$. 
It is to be understood that this is an equality only when we drop self-contractions (since the coherent state is normal-ordered). Carrying out the functional derivatives and substituting the explicit expressions (\ref{eq:IJbcexp}) for the sources into the final result leads to:
\begin{equation}\label{eq:Example1-bc}
\begin{aligned}
\Big\langle &c(z_1)c(z_2):\!e^{\int Ib+Jc}(c+\e{B}_1)(z_3)\!:\!\Big\rangle_{bc}=z_{12}z_{13}z_{23}\,\exp\Big[\frac{1}{z_{12}}\,\sum_{n\geq1}\Big(\frac{\e{C}_1\e{B}_n}{z_{13}^{n}}-\frac{\e{C}_1\e{B}_n}{z_{23}^n}\Big)\Big]
\end{aligned}
\end{equation}
Recall that the quantities (\ref{eq:ABCcondensed_notation}) contain matter contributions that are to be integrated out when we carry out the matter path integral (see below). 
\sk

Referring back to (\ref{eq:offshellA_a_}), we have derived a  useful and general result:
\begin{equation}\label{eq:<ccU>_S2}
\begingroup\makeatletter\def\f@size{11}\check@mathfonts
\def\maketag@@@#1{\hbox{\m@th\large\normalfont#1}}%
\begin{aligned}
\Big\langle c^{(z)}(z_1)c^{(z)}(z_2) \e{U}_{(1)}^{(z)}(z_3)\Big\rangle_{bc} =z_{12}z_{13}z_{23}\,\exp\bigg[\,\sum_{n\geq1}\Big(\frac{1}{n}\e{A}_n^{(1)}\!\cdot \!\alpha_{-n}^{(z)}(z_3)+\frac{\e{C}_1^{(1)}\e{B}_n^{(1)}}{z_{12}z_{13}^{n}}-\frac{\e{C}_1^{(1)}\e{B}_n^{(1)}}{z_{12}z_{23}^n}\Big)\bigg].
\end{aligned}
\endgroup
\end{equation}

\subsubsection*{Matter Contributions}
We next bring the amplitude to a form appropriate to make use of the ghost path integral  (\ref{eq:<ccU>_S2}),
\begin{equation}\label{eq:S-VirasoroShapiro-3pt_a}
\begin{aligned}
&g_D^2e^{-2\Phi}\Big\langle \prod_{j=1,2}\!\!:\!\tilde{c}ce^{ik_j\cdot x}(z_j,\bar{z}_j)\!:\hat{\mathscr{A}}_{a}^{(z)}(z_3,\bar{z}_3)\Big\rangle_{S^2}=g_D^3C_{S^2}^{\rm g}\int_0^{2\pi}\frac{\rmd\phi}{2\pi}e^{-2\Phi}\Big\langle :\! e^{ik_1\cdot x}(z_1,\bar{z}_1)\!:\\
&\quad\qquad\times :\!e^{ik_2\cdot x}(z_2,\bar{z}_2)\!:\,:\!\big\langle c(z_1)c(z_2)\e{U}_{(1)}(z_3)\big\rangle_{bc}\big\langle \tilde{c}(\bar{z}_1)\tilde{c}(\bar{z}_2)\tilde{\e{U}}_{(1)}(\bar{z}_3)\big\rangle_{\tilde{b}\tilde{c}}e^{i\e{p}\cdot x}(z_3,\bar{z}_3)\!:\Big\rangle_x,
\end{aligned}
\end{equation}
where $\langle \dots\rangle_x$ denotes a path integral over the matter fields, $x^{M}(z,\bar{z})$, including zero modes, the normalisation (and path integral determinant that will be absorbed into an overall normalisation). According to (\ref{eq:<ccU>_S2}) and general properties of string path integrals \cite{BelavinKnizhnik86,VerlindeVerlinde87,DijkgraafVerlindeVerlinde88,D'HokerPhong89,SklirosCopelandSaffin17} the path integral (\ref{eq:S-VirasoroShapiro-3pt_a}) factorises into chiral and anti-chiral halves when we factor out the zero mode integral (including the factor of $i=\sqrt{-1}$ from the analytic continuation of $x_0^0$ back to Minkowski spacetime),
\begin{equation}\label{eq:S-VirasoroShapiro-3pt_ab}
\begin{aligned}
&g_D^2e^{-2\Phi}\Big\langle \prod_{j=1,2}\!\!:\!\tilde{c}ce^{ik_j\cdot x}(z_j,\bar{z}_j)\!:\hat{\mathscr{A}}_{a}^{(z)}(z_3,\bar{z}_3)\Big\rangle_{S^2}=g_D^3e^{-2\Phi}C_{S^2}^{x\phantom{g}}C_{S^2}^{\rm g}|z_{12}z_{13}z_{23}|^2\int_0^{2\pi}\frac{\rmd\phi}{2\pi}i\int \rmd^Dx_0\\
&\quad\Big\langle \!:\! e^{ik_1\cdot x_L}(z_1)\!:\,:\!e^{ik_2\cdot x_L}(z_2)\!:\,:\!\,\exp \sum_{n\geq1}\Big(\frac{1}{n}\e{A}_n^{(1)}\!\cdot \!\alpha_{-n}^{(z)}(z_3)+\frac{\e{C}_1^{(1)}\e{B}_n^{(1)}}{z_{12}z_{13}^{n}}-\frac{\e{C}_1^{(1)}\e{B}_n^{(1)}}{z_{12}z_{23}^n}\Big)e^{i\e{p}\cdot x_L}(z_3)\!:\!\Big\rangle_{+}\\
&\quad\Big\langle \!:\! e^{ik_1\cdot x_R}(\bar{z}_1)\!:\,:\!e^{ik_2\cdot x_R}(\bar{z}_2)\!:\,:\!\, \exp \sum_{n\geq1}\Big(\frac{1}{n}\tilde{\e{A}}_n^{(1)}\!\cdot \!\tilde{\alpha}_{-n}^{(\bar{z})}(\bar{z}_3)+\frac{\tilde{\e{C}}_1^{(1)}\tilde{\e{B}}_n^{(1)}}{\bar{z}_{12}\bar{z}_{13}^{n}}-\frac{\tilde{\e{C}}_1^{(1)}\tilde{\e{B}}_n^{(1)}}{\bar{z}_{12}\bar{z}_{23}^n}\Big)e^{i\e{p}\cdot x_R}(\bar{z}_3)\!:\!\Big\rangle_{-}\\
\end{aligned}
\end{equation}
and we denote the aforementioned normalisation by \cite{Polchinski_v1,SklirosCopelandSaffin17} $C_{S^2}^{x\phantom{g}}$. 
The following decomposition is to be understood, $x(z,\bar{z})=x_L(z)+x_R(\bar{z})$, with:
\begin{equation}\label{eq:xLxR=x_0+x+-}
x_L(z)=\frac{1}{2}x_0+x_+(z),\qquad x_R(\bar{z})= \frac{1}{2}x_0+x_-(\bar{z}),
\end{equation}
in which case the path integral (associated to non-zero modes) is carried out using the {\it effective} rule \cite{BelavinKnizhnik86,D'HokerPhong89,SklirosCopelandSaffin17}: compute all Wick contractions using the following correlators,
$$
\langle x_+^{\mu}(z_1)x_+^{\nu}(z_2)\rangle_+=-\frac{\alpha'}{2}\eta^{\mu\nu}\ln z_{12},\qquad \langle x_-^{\mu}(\bar{z}_1)x_-^{\nu}(\bar{z}_2)\rangle_-=-\frac{\alpha'}{2}\eta^{\mu\nu}\ln \bar{z}_{12}.
$$

We adopt the standard \cite{Polchinski_v1} notation and result for the overall normalisation:
\begin{equation}\label{eq:C_S^2}
C_{S^2}\equiv e^{-2\Phi}C_{S^2}^{x\phantom{g}}C_{S^2}^{\rm g}=\frac{8\pi}{\alpha'g_D^2},
\end{equation}
which holds for path integrals involving coherent states as well as mass eigenstates. In particular, tracing through the relevant definitions one confirms that the second equality in (\ref{eq:C_S^2}) is fixed by our normalisation condition (\ref{eq:<..>normalisation1}) that we adopt throughout. 

\sk
Let us primarily consider the following term in the chiral half on the right-hand side of (\ref{eq:S-VirasoroShapiro-3pt_ab}) in detail. A variation of a standard path integral trick that makes its evaluation immediate is to rewrite it as follows,
\begin{equation}\label{eq:<eeeAe>_+}
\begin{aligned}
&\Big\langle \!:\! e^{ik_1\cdot x_L}(z_1)\!:\,:\!e^{ik_2\cdot x_L}(z_2)\!:\,:\!\,\exp \sum_{n\geq1}\Big(\frac{1}{n}\e{A}_n^{(1)}\!\cdot \!\alpha_{-n}^{(z)}(z_3)\Big)e^{i\e{p}\cdot x_L}(z_3)\!:\!\Big\rangle_{+}\\
&\qquad=\sum_{n=0}^{\infty}e^{in\phi}e^{i(k_1+k_2+p-nq)\cdot \frac{1}{2}x_0}\oint_0\frac{\rmd u}{2\pi iu}\,u^{-n}\Big\langle 
\exp\Big(i\int \rmd^2z\,J(z,\bar{z})\cdot x_+(z)\Big)
\Big\rangle_{+}\\
\end{aligned}
\end{equation}
with the definition,
$$
J(z,\bar{z})\dfn \delta^2(z-z_1)k_1+\delta^2(z-z_2)k_2+\delta^2(z-z_3)\Big((p-nq)+\sum_{s=1}^{\infty}\frac{u^s}{s}{\bf a}_s \sqrt{\frac{2}{\alpha'}}\frac{1}{(s-1)!}\partial_z^s\Big)
$$
In arriving at this representation for the correlator we made use of the explicit expression for $\e{A}_n^{(1)}$ in (\ref{eq:ABCcondensed_notation}), the decompositions (\ref{eq:xLxR=x_0+x+-}) and the cycle index polynomial relations (\ref{eq:CycleIndexGenFun}) and (\ref{eq:S_n(a)a}) in order to rearrange the exponential and extract out all zero modes. Finally, to make sense of (\ref{eq:<eeeAe>_+}) it is to be understood that we (implicitly) subtract all self contractions (all vertex operators are normal-ordered), and it is only then that the depicted equality holds. 

\sk
The remaining Wick contractions are readily calculated, see, e.g.,  \cite{SklirosCopelandSaffin17}:
\begin{equation}
\begin{aligned}
&\Big\langle \exp\Big(i\int \rmd^2z\,J(z,\bar{z})\cdot x_+(z)\Big)\Big\rangle_{+}=\exp\Big(\frac{\alpha'}{4}\int \rmd^2z\int \rmd^2z' J(z,\bar{z})\cdot J(z',\bar{z}')\ln (z-z')\Big).
\end{aligned}
\end{equation}
Substituting the explicit expression for the source, $J$, rearranging, plugging the result back into (\ref{eq:<eeeAe>_+}) and making use again of the cycle index polynomial relations (\ref{eq:CycleIndexGenFun}) and (\ref{eq:S_n(a)a}) leads to,
\begin{equation}\label{eq:<eeeAe>_+2}
\begin{aligned}
&\Big\langle \!:\! e^{ik_1\cdot x_L}(z_1)\!:\,:\!e^{ik_2\cdot x_L}(z_2)\!:\,:\!\,\exp \sum_{n\geq1}\Big(\frac{1}{n}\e{A}_n^{(1)}\!\cdot \!\alpha_{-n}^{(z)}(z_3)\Big)e^{i\e{p}\cdot x_L}(z_3)\!:\!\Big\rangle_{+}\\
&\quad =\sum_{n=0}^{\infty}e^{in\phi}e^{i(k_1+k_2+p-nq)\cdot \frac{1}{2}x_0}
z_{12}^{\frac{\alpha'}{2}k_1\cdot k_2}z_{13}^{\frac{\alpha'}{2}k_1\cdot (p-nq)}z_{23}^{\frac{\alpha'}{2}k_2\cdot (p-nq)}Z_n\Big(\sqrt{\tfrac{\alpha'}{2}}\tfrac{k_1\cdot {\bf a}_s}{z_{13}^s}+\sqrt{\tfrac{\alpha'}{2}}\tfrac{k_2\cdot {\bf a}_s}{z_{23}^s}\Big)\\
\end{aligned}
\end{equation}

A similar calculation leads to the following expression for the terms involving ghost quantum numbers,
\begin{equation}\label{eq:<eeeACBe>_+2}
\begin{aligned}
&\Big\langle \!:\! e^{ik_1\cdot x_L}(z_1)\!:\,:\!e^{ik_2\cdot x_L}(z_2)\!:\,:\!\,\exp \sum_{n\geq1}\Big(\frac{1}{n}\e{A}_n^{(1)}\!\cdot \!\alpha_{-n}^{(z)}(z_3)\Big)\sum_{m\geq1}\Big(\frac{\e{C}_1^{(1)}\e{B}_m^{(1)}}{z_{12}z_{13}^m}-\frac{\e{C}_1^{(1)}\e{B}_m^{(1)}}{z_{12}z_{23}^m}\Big)e^{i\e{p}\cdot x_L}(z_3)\!:\!\Big\rangle_{+}\\
&\qquad = \sum_{m\geq1}\sum_{n=0}^{\infty}e^{i(n+m+1)\phi}e^{i(k_1+k_2+\e{p}-(n+m+1)\e{q})\cdot \frac{1}{2}x_0}\Big(\frac{{\bf c}_1^{(1)}{\bf b}_m^{(1)}}{z_{12}z_{13}^m}-\frac{{\bf c}_1^{(1)}{\bf b}_m^{(1)}}{z_{12}z_{23}^m}\Big)\\
&\qquad\qquad\times 
z_{12}^{\frac{\alpha'}{2}k_1\cdot k_2}z_{13}^{\frac{\alpha'}{2}k_1\cdot (\e{p}-(n+m+1)\e{q})}z_{23}^{\frac{\alpha'}{2}k_2\cdot (\e{p}-(n+m+1)\e{q})}
Z_n\Big(\sqrt{\tfrac{\alpha'}{2}}\tfrac{k_1\cdot {\bf a}_s}{z_{13}^s}+\sqrt{\tfrac{\alpha'}{2}}\tfrac{k_2\cdot {\bf a}_s}{z_{23}^s}\Big)
\end{aligned}
\end{equation}
The full chiral half of the path integral, denoted by $\langle \dots\rangle_+$ in (\ref{eq:S-VirasoroShapiro-3pt_ab}), is equal to the sum of (\ref{eq:<eeeACBe>_+2}) and (\ref{eq:<eeeAe>_+2}); recall that $\e{C}_1^{(1)}$ (in particular ${\bf c}_1^{(1)}$) squares to zero. 

\sk
The evaluation and result for the anti-chiral correlator, denoted by $\langle\dots\rangle_-$ in (\ref{eq:S-VirasoroShapiro-3pt_ab}), is entirely analogous to the evaluation of the full chiral correlator, $\langle\dots\rangle_+$, we just discussed. In particular, to obtain the anti-chiral correlator we are to replace all chiral quantities visible in (\ref{eq:<eeeACBe>_+2}) to (\ref{eq:<eeeAe>_+2}) by the corresponding anti-chiral quantities while also flipping the sign of $\phi$. 

\sk
We may proceed directly from (\ref{eq:<eeeAe>_+2}) and (\ref{eq:<eeeACBe>_+2}), but it is useful to perform some elementary manipulations first that will make the structure of the full amplitude more transparent, while also allowing us to introduce some additional tricks that are useful in more general coherent state path integrals. 
Let us first substitute (\ref{eq:<eeeAe>_+2}) and (\ref{eq:<eeeACBe>_+2}) into (\ref{eq:S-VirasoroShapiro-3pt_ab}). After some tedious but straightforward manipulations we arrive at the following expression for the full three-point amplitude:
\begin{equation}\label{eq:S-VirasoroShapiro-3pt_abc}
\begin{aligned}
&g_D^2e^{-2\Phi}\Big\langle \prod_{j=1,2}\!\!:\!\tilde{c}ce^{ik_j\cdot x}(z_j,\bar{z}_j)\!:\hat{\mathscr{A}}_{a}^{(z)}(z_3,\bar{z}_3)\Big\rangle_{S^2}
\\
&=\Big(\frac{8\pi ig_D}{\alpha'}\Big)(2\pi)^D\delta^D(\mathsf{k})\int_0^{2\pi}\frac{\rmd\phi}{2\pi}\int_0^{2\pi}\frac{\rmd\theta}{2\pi}\,e^{-i\theta\frac{\alpha'}{2}(s+\e{p}^2)}\,\Big(\frac{z_{12}}{z_{13}z_{23}}\Big)^{-\frac{\alpha'}{4}s-1}\Big(\frac{\bar{z}_{12}}{\bar{z}_{13}\bar{z}_{23}}\Big)^{-\frac{\alpha'}{4}s-1}\\
&\quad \exp\sum_{n=1}^{\infty}\Bigg[
\frac{e^{in(\phi+\theta)}}{n}\sqrt{\frac{\alpha'}{2}}\Big(\frac{k_1\cdot {\bf a}_n}{z_{13}^n}+\frac{k_2\cdot {\bf a}_n}{z_{23}^n}\Big)+e^{i(n+1)(\phi+\theta)}\Big(\frac{{\bf c}_1^{(1)}{\bf b}_n^{(1)}}{z_{12}z_{13}^n}-\frac{{\bf c}_1^{(1)}{\bf b}_n^{(1)}}{z_{12}z_{23}^n}\Big)\Bigg]\\
&\quad
\exp\sum_{\tilde{n}=1}^{\infty}\Bigg[\frac{e^{i\tilde{n}(\theta-\phi)}}{\tilde{n}}\sqrt{\frac{\alpha'}{2}}\Big(\frac{k_1\cdot \tilde{\bf a}_{\tilde{n}}}{\bar{z}_{13}^{\tilde{n}}}+\frac{k_2\cdot \tilde{\bf a}_{\tilde{n}}}{\bar{z}_{23}^{\tilde{n}}}\Big)+ e^{i(\tilde{n}+1)(\theta-\phi)}\Big(\frac{\tilde{\bf c}_1^{(1)}\tilde{\bf b}_{\tilde{n}}^{(1)}}{\bar{z}_{12}\bar{z}_{13}^{\tilde{n}}}-\frac{\tilde{\bf c}_1^{(1)}\tilde{\bf b}_{\tilde{n}}^{(1)}}{\bar{z}_{12}\bar{z}_{23}^{\tilde{n}}}\Big)\Bigg]\\
\end{aligned}
\end{equation}

For clarity purposes we have displayed the result (\ref{eq:S-VirasoroShapiro-3pt_abc}) for generic $z_3$, but when we substitute this into the factorisation formula our gluing condition, $zu=q$, demands that we take $z_3=0$, so we take $z_3=0$ in what follows.
\sk

The only step in arriving at (\ref{eq:S-VirasoroShapiro-3pt_abc}) that does not come down to straightforward algebraic manipulations is the matter zero mode contribution which is dealt with as follows. The integral over zero modes leads to an infinite superposition of delta functions (for given sets of integers, $N,\tilde{N}$, that are summed over, and in fact the $\phi$ integral sets $N=\tilde{N}$ for each term in the summand) which can be rewritten in the form:
\begin{equation}\label{eq:deltatrick}
\delta^D\big(k_1+k_2+\e{p}-\tfrac{1}{2}(N+\tilde{N})\e{q}\big)=\delta^D(\mathsf{k})\int_0^{2\pi}\frac{\rmd\theta}{2\pi}\,e^{-i\theta\frac{\alpha'}{2}(s+\e{p}^2)}e^{i\theta(N+\tilde{N})},
\end{equation}
\begin{equation}\label{eq:k_dfn}
\mathsf{k}\dfn k_1+k_2+\e{p}-\frac{\alpha'}{4}(s+\e{p}^2)\e{q}.
\end{equation}
Notice the $\theta$ integral is a Kronecker delta constraint, and we have made use of the kinematic constraints (valid both onshell and offshell): $\e{q}^2=0$ and $\e{p}\cdot \e{q}=2/\alpha'$. Recall also that $\e{p}^2$ is unconstrained offshell, whereas onshell (in particular, for all onshell coherent states) $\e{p}^2=4/\alpha'$. This enables us to factor out a single overall delta function, $\delta^D(\mathsf{k})$, which is independent of $N,\tilde{N}$, as indicated in (\ref{eq:S-VirasoroShapiro-3pt_abc}). (Incidentally, these kinds of momentum-conserving delta function manipulations are very convenient for coherent state amplitudes.) Note that we may rewrite this relation by introducing an integral representation for this delta function on the left-hand side,
\begin{equation*}
\int \rmd^Dx_0\,e^{i(k_1+k_2+\e{p}-\tfrac{1}{2}(N+\tilde{N})\e{q})\cdot x_0}=(2\pi)^D\delta^D(\mathsf{k})\int_0^{2\pi}\frac{\rmd\theta}{2\pi}\,e^{-i\theta\frac{\alpha'}{2}(s+\e{p}^2)}e^{i\theta(N+\tilde{N})},
\end{equation*}
and now it becomes manifest that provided we ultimately integrate over $x_0$ we can effectively replace:
\begin{equation}
e^{-i(N+\tilde{N})\e{q}\cdot \frac{1}{2}x_0}\rightarrow e^{-i\frac{\alpha'}{4}(s+\e{p}^2)\e{q}\cdot x_0}\int_0^{2\pi}\frac{\rmd\theta}{2\pi}\,e^{-i\theta\frac{\alpha'}{2}(s+\e{p}^2)}e^{i\theta(N+\tilde{N})}.
\end{equation}
This is the only non-trivial step that has been adopted in deriving the result for the full 3-point path integral (\ref{eq:S-VirasoroShapiro-3pt_abc}). The various factors of $e^{i\theta N}$, $e^{i\theta\tilde{N}}$ combine with remaining terms and exponentiate as seen. (A note of caution: when we make use of these results in the factorisation relation (\ref{eq:S-VirasoroShapiro-gluing}) there will be an integral over $\e{p}^{\mu}$, and it is important that we integrate out $\theta$ {\it before} integrating out $\e{p}^{\mu}$ -- the two integrals do not commute. Nevertheless, the prescription presented in this paragraph is useful (also for more general amplitudes) in that it allows for compact manipulations in intermediate steps leading up to the computation of the 4-point amplitude.)
\sk

As a quick consistency check notice that if we set the continuous quantum numbers of the coherent state to zero in (\ref{eq:S-VirasoroShapiro-3pt_abc}) we precisely recover the 3-point tachyon amplitude with the correct normalisation when we use the onshell condition $\e{p}^2=4/\alpha'$. Furthermore, the fact that there remains worldsheet coordinate dependence is as expected: this is the amplitude for two onshell tachyons and one {\it offshell} and generic coherent state vertex operator. Nevertheless, this coordinate dependence will drop out when asymptotic states are onshell as expected on general grounds \cite{Sen15b} and as we shall demonstrate momentarily in the gluing formula (\ref{eq:S-VirasoroShapiro-gluing}) that is going to reproduce the Virasoro-Shapiro amplitude. 
\sk

We next integrate out $\theta$ and $\phi$ in (\ref{eq:S-VirasoroShapiro-3pt_abc}). We can proceed in various ways and we choose the most efficient that is to use the SL$(2,\mathbf{C})$ invariance of $S^2$ to take one of the primary tachyon vertex operators to infinity, say $z_1\rightarrow \infty$, while keeping the coherent state vertex operator offshell. Then, integrating out $\theta$ and $\phi$ in (\ref{eq:S-VirasoroShapiro-3pt_abc}) becomes a simple exercise when one rewrites the exponentials in terms of cycle index polynomials using the defining generating function (\ref{eq:CycleIndexGenFun}). Note that after taking the limit $z_1\rightarrow \infty$ only the arguments $\sqrt{\frac{\alpha'}{2}}k_2\cdot {\bf a}_nz_{2}^{-n}$ and $\sqrt{\frac{\alpha'}{2}}k_2\cdot \tilde{\bf a}_{\tilde{n}}\bar{z}_{2}^{-\tilde{n}}$ survive in the exponents. After integrating out $\theta,\phi$ we make use of the scaling relation for cycle index polynomials in (\ref{eq:Zproperties}). Taking these considerations into account the full 3-point amplitude reduces to:
\begin{equation}\label{eq:S-VirasoroShapiro-3pt_abbla}
\begin{aligned}
&g_D^2e^{-2\Phi}\Big\langle \prod_{j=1,2}\!\!:\!\tilde{c}ce^{ik_j\cdot x}(z_j,\bar{z}_j)\!:\hat{\mathscr{A}}_{a}^{(z)}
\Big\rangle_{S^2}
=\Big(\frac{8\pi ig_D}{\alpha'}\Big)\\
&\quad\times\sum_{n=0}^{\infty}(2\pi)^D\delta^D(k_1+k_2+\e{p}-n\e{q})(z_{2}\bar{z}_{2})^{\frac{\alpha'}{4}s-n+1}
 Z_n\big(\sqrt{\tfrac{\alpha'}{2}}k_2\cdot {\bf a}_s\big)
Z_{n}\big(\sqrt{\tfrac{\alpha'}{2}}k_2\cdot \tilde{\bf a}_s\big)\\
\end{aligned}
\end{equation}

Incidentally, this result is only simple because we have adopted a {\it coherent state basis}. This is important to emphasise: the amplitude (\ref{eq:S-VirasoroShapiro-3pt_abbla}) encodes the interaction of two onshell tachyons and {\it any} other offshell vertex operator in the string spectrum on $S^2$.  The corresponding expression for a generic mass eigenstate replacing the coherent state is prohibitively difficult to write down explicitly, since the rank of the associated polarisation tensors would be countable but increases all the way up to infinity, and one also has to prescribe a prescription for summing over such polarisation tensors. The expression (\ref{eq:S-VirasoroShapiro-3pt_abbla}) is also more general than a corresponding mass eigenstate amplitude, in that we may also view (\ref{eq:S-VirasoroShapiro-3pt_abbla}) as a generating function (associated to tachyon$+$tachyon$\rightarrow$ mass eigenstate) by differentiating with respect to the coherent state quantum numbers and using the derivative relation in (\ref{eq:Zproperties}). 
\sk

The amplitude computation associated to the $S^{'2}$ sphere in (\ref{eq:S-VirasoroShapiro-gluing}) proceeds in an entirely analogous manner. Recall that here we are to place all vertex operators in the $(U_u,u)$ chart, the coherent state being at the origin. The result analogous to (\ref{eq:S-VirasoroShapiro-3pt_abc}) is:
\begin{equation}\label{eq:3pt-VirSh_u3=0}
\begin{aligned}
&g_D^2\,e^{-2\Phi}\Big\langle \big[b_0\tilde{b}_0\cdot \hat{\mathscr{A}}^{a}_{(u)}\big]\prod_{j=3,4}\!\!:\!\tilde{c}ce^{ik_j\cdot x}(u_j,\bar{u}_j)\!:\Big\rangle_{S^{'2}}\\
&=\Big(\frac{8\pi ig_D}{\alpha'}\Big)(2\pi)^D\delta^D(\mathsf{k}')\int_0^{2\pi}\frac{\rmd\varphi}{2\pi}\int_0^{2\pi}\frac{\rmd\theta'}{2\pi}\,e^{-i\theta'\frac{\alpha'}{2}(s+\e{p}^2)}\,\Big(\frac{u_{34}}{u_{3}u_{4}}\Big)^{-\frac{\alpha'}{4}s-1}\Big(\frac{\bar{u}_{34}}{\bar{u}_{3}\bar{u}_{4}}\Big)^{-\frac{\alpha'}{4}s-1}\mathbi{c}_0^{(2)}\tilde{\mathbi{c}}_0^{(2)}\\
&\quad \exp\sum_{n=1}^{\infty}\Bigg[
\frac{e^{in(\theta'+\varphi)}}{n}\sqrt{\frac{\alpha'}{2}}\Big(-\frac{k_3\cdot {\bf a}_n^*}{u_{3}^n}-\frac{k_4\cdot {\bf a}_n^*}{u_{4}^n}\Big)+e^{i(n+1)(\varphi+\theta')}\Big(\frac{{\bf c}_1^{(2)}{\bf b}_n^{(2)}}{u_{34}u_{3}^n}-\frac{{\bf c}_1^{(2)}{\bf b}_n^{(2)}}{u_{34}u_{4}^n}\Big)\Bigg]\\
&\quad
\exp\sum_{\tilde{n}=1}^{\infty}\Bigg[\frac{e^{i\tilde{n}(\theta'-\varphi)}}{\tilde{n}}\sqrt{\frac{\alpha'}{2}}\Big(-\frac{k_3\cdot \tilde{\bf a}_{\tilde{n}}^*}{\bar{u}_{3}^{\tilde{n}}}-\frac{k_4\cdot \tilde{\bf a}_{\tilde{n}}^*}{\bar{u}_{4}^{\tilde{n}}}\Big)+ e^{i(\tilde{n}+1)(\theta'-\varphi)}\Big(\frac{\tilde{\bf c}_1^{(2)}\tilde{\bf b}_{\tilde{n}}^{(2)}}{\bar{u}_{34}\bar{u}_{3}^{\tilde{n}}}-\frac{\tilde{\bf c}_1^{(2)}\tilde{\bf b}_{\tilde{n}}^{(2)}}{\bar{u}_{12}\bar{u}_{4}^{\tilde{n}}}\Big)\Bigg],
\end{aligned}
\end{equation}
wit the definition 
\begin{equation}\label{eq:k'_dfn}
\mathsf{k}\dfn k_3+k_4-\e{p}+\frac{\alpha'}{4}(s+\e{p}^2)\e{q},
\end{equation}
and going through the same manipulations as above (taking $u_3\rightarrow \infty$) we integrate out $\varphi$ and $\theta'$ and obtain the analogue of (\ref{eq:S-VirasoroShapiro-3pt_abbla}),
\begin{equation}\label{eq:S-VirasoroShapiro-3pt_final_u}
\begin{aligned}
&g_D^2\,e^{-2\Phi}\Big\langle \big[b_0\tilde{b}_0\cdot \hat{\mathscr{A}}^{a}_{(u)}\big]\prod_{j=3,4}\!\!:\!\tilde{c}ce^{ik_j\cdot x}(u_j,\bar{u}_j)\!:\Big\rangle_{S^{'2}}=\Big(\frac{8\pi ig_D}{\alpha'}\Big)\mathbi{c}_0^{(2)}\tilde{\mathbi{c}}_0^{(2)}\\
&\times\sum_{m=0}^{\infty}(2\pi)^D\delta^D\big(k_3+k_4-\e{p}+m\e{q}\big) (u_{4}\bar{u}_{4})^{\frac{\alpha'}{4}s-m+1}Z_m\big(\!\!-\sqrt{\tfrac{\alpha'}{2}}k_4\cdot {\bf a}_s^*\big)Z_m\big(\!\!-\sqrt{\tfrac{\alpha'}{2}}k_4\cdot \tilde{\bf a}_{s}^*\big)
\end{aligned}
\end{equation}
Of course we may use momentum conservation to write the result more symmetrically in $k_3$ and $k_4$ (following standard manipulations \cite{Polchinski_v1}) but this just introduces unnecessary clutter that does not affect the result for the amplitude $\mathcal{A}_{S^2}^{\infty_s}$ that we are heading towards.

\subsubsection{Gluing 3-Point Amplitudes} 
We next substitute the 3-point sphere amplitude results (\ref{eq:S-VirasoroShapiro-3pt_abbla}) and (\ref{eq:S-VirasoroShapiro-3pt_final_u}) into the amplitude gluing equation (\ref{eq:S-VirasoroShapiro-gluing}),
\begin{equation}\label{eq:S-VirasoroShapiro-gluing3an}
\begin{aligned}
&S_{S^2}^{\infty_s}(1;\dots;4)= g_D^4\suminnt\limits_{a} \,\,e^{-2\Phi}\Big\langle \prod_{j=1,2}\!\!:\!\tilde{c}ce^{ik_j\cdot x}(z_j,\bar{z}_j)\!:\hat{\mathscr{A}}_{a}^{(z)}\Big\rangle_{S^2}\\
&\quad\times\int \rmd^2q\,q^{h_a-1}\bar{q}^{\tilde{h}_a-1} e^{-2\Phi}\Big\langle \big[b_0\tilde{b}_0\cdot \hat{\mathscr{A}}^{a}_{(u)}\big]\prod_{j=3,4}\!\!:\!\tilde{c}ce^{ik_j\cdot x}(u_j,\bar{u}_j)\!:\Big\rangle_{S^{'2}}\\
&=\Big(\frac{8\pi ig_D}{\alpha'}\Big)^2\sum_{n,m=0}^{\infty}\int \rmd^D\e{p}(2\pi)^D\delta^D(k_1+k_2+\e{p}-n\e{q})\delta^D\big(k_3+k_4-\e{p}+m\e{q}\big)\\
&\quad\times \frac{\alpha'}{8\pi i}\int \rmd^2q \,q^{(\alpha'\e{p}^2/4-1)-1}\bar{q}^{(\alpha'\e{p}^2/4-1)-1}(z_{2}\bar{z}_{2}u_{4}\bar{u}_{4})^{-\frac{\alpha'}{4}\e{p}^2+1}\\
&\quad\times \int \rmd\mu_{\mathbi{abc}}\,\mathbi{c}_0^{(2)}\tilde{\mathbi{c}}_0^{(2)}Z_n\big(\sqrt{\tfrac{\alpha'}{2}}k_2\cdot {\bf a}_s\big)Z_n\big(\sqrt{\tfrac{\alpha'}{2}}k_2\cdot \tilde{\bf a}_{s}\big)Z_m\big(\!-\sqrt{\tfrac{\alpha'}{2}}k_4\cdot {\bf a}_s^*\big)Z_m\big(\!-\sqrt{\tfrac{\alpha'}{2}}k_4\cdot \tilde{\bf a}_{s}^*\big),
\end{aligned}
\end{equation}
where we made use of the definitions (\ref{eq:dpdmua}) and (\ref{eq:dmua}). 
\sk

The ghost quantum numbers in the measure, $\rmd\mu_{\mathbi{abc}}$, (including the factor $\mathbi{c}_0^{(2)}\tilde{\mathbi{c}}_0^{(2)}$) can immediately be integrated out trivially and give unit contribution. The matter quantum numbers can be integrated out by making use of the cycle index polynomial expression (\ref{eq:S_n(a)a}) for all cycle index polynomials appearing in (\ref{eq:S-VirasoroShapiro-gluing3an}). The explicit integrals simply boil down to a product of Gaussian integrals of the form:
\begin{equation}\label{eq:GaussianIntegral2}
\int\frac{\rmd\bar{z}\wedge \rmd z}{2\pi iG}\,e^{-\bar{z}G^{-1}z}e^{zA_1}e^{\bar{z}A_2}=e^{A_1GA_2}.
\end{equation}
Using the scaling relation in (\ref{eq:Zproperties}) and repackaging the result back into cycle index polynomials on account of (\ref{eq:S_n(a)a}) leads to,
\begin{equation}\label{eq:ZnZnZmZm_ints}
\begin{aligned}
&\int \rmd\mu_{\mathbi{abc}}\,\mathbi{c}_0^{(2)}\tilde{\mathbi{c}}_0^{(2)}Z_n\big(\sqrt{\tfrac{\alpha'}{2}}k_2\cdot {\bf a}_s\big)Z_n\big(\sqrt{\tfrac{\alpha'}{2}}k_2\cdot \tilde{\bf a}_{s}\big) Z_m\big(-\sqrt{\tfrac{\alpha'}{2}}k_4\cdot {\bf a}_s^*\big)Z_m\big(-\sqrt{\tfrac{\alpha'}{2}}k_4\cdot \tilde{\bf a}_{s}^*\big)\\
&\qquad = \delta_{n,m}Z_n\big(-\tfrac{\alpha'}{2}k_2\cdot k_4\big)^2 = \delta_{n,m}\Big(\frac{\Gamma(2+\frac{\alpha'}{4}t+n)}{\Gamma(2+\frac{\alpha'}{4}t)\Gamma(n+1)}\Big)^2,
\end{aligned}
\end{equation}
where in the last equality we made use of the gamma function relation in (\ref{eq:Zproperties}), and according to our convention (\ref{eq:Mandelstam}), $-\tfrac{\alpha'}{2}k_2\cdot k_4=2+\tfrac{\alpha'}{4}t$. 
\sk

For the moduli integral, recalling the discussion associated to the range of $q,\bar{q}$ integration (\ref{eq:|q| range VS}), analytically continuing $\e{p}^2$ from a domain where the integral is finite we immediately have the following result (including the Feynman $i\epsilon$ prescription):
\begin{equation}\label{eq:FeynmannProp}
\begin{aligned}
\frac{\alpha'}{8\pi i}\int\limits_{|q|<|z_2u_4|} &\rmd^2q \,q^{(\alpha'\e{p}^2/4-1)-1}\bar{q}^{(\alpha'\e{p}^2/4-1)-1}(z_{2}u_{4}\bar{z}_{2}\bar{u}_{4})^{-\frac{\alpha'}{4}\e{p}^2+1} =\frac{-i}{\e{p}^2-\frac{4}{\alpha'}-i\epsilon}.
\end{aligned}
\end{equation}

The zero mode momentum integrals are a little bit subtle, because one must take into account the constraints, $\e{p}\cdot \e{q}=2/\alpha'$ and $\e{q}^2=0$, but the result is simple also: for any function $f(\e{p}^2)$,
\begin{equation}
\begin{aligned}
&\int \rmd^D\e{p}(2\pi)^D\delta^D(k_1+k_2+\e{p}-n\e{q})\delta^D(k_3+k_4-\e{p}+m\e{q})f(\tfrac{\alpha'}{4}\e{p}^2)\delta_{n,m}=\\
&\qquad=(2\pi)^D\delta^D(k_1+k_2+k_3+k_4)\,f(-\tfrac{\alpha'}{4}s+n)\delta_{n,m}.
\end{aligned}
\end{equation}

Now let us substitute these results back into (\ref{eq:S-VirasoroShapiro-gluing3an}), leading to the following expression for the $s$-channel contribution to the Virasoro-Shapiro invariant amplitude ({\it when} we glue with a general basis of states):
\begin{equation}\label{eq:As-VirasoroShapiro-glued}
\begin{aligned}
\mathcal{A}_{S^2}^{\infty_s}(s,t,u)=\Big(\frac{8\pi  g_D}{\alpha'}\Big)^2\sum_{n=0}^{\infty}\Big(\frac{\Gamma(2+\frac{\alpha'}{4}t+n)}{\Gamma(2+\frac{\alpha'}{4}t)\Gamma(n+1)}\Big)^2\frac{1}{-s+\tfrac{4}{\alpha'}(n-1)-i\epsilon}
\end{aligned}
\end{equation}
The corresponding result for the $u$-channel contribution is simply obtained by interchanging $s\leftrightarrow u$ in (\ref{eq:As-VirasoroShapiro-glued}):
\begin{equation}\label{eq:Au-VirasoroShapiro-glued}
\begin{aligned}
\mathcal{A}_{S^2}^{\infty_u}(s,t,u)=\Big(\frac{8\pi  g_D}{\alpha'}\Big)^2\sum_{n=0}^{\infty}\Big(\frac{\Gamma(2+\frac{\alpha'}{4}t+n)}{\Gamma(2+\frac{\alpha'}{4}t)\Gamma(n+1)}\Big)^2\frac{1}{-u+\tfrac{4}{\alpha'}(n-1)-i\epsilon}
\end{aligned}
\end{equation}
These expressions are the best starting points for extracting the imaginary part of the amplitude (away from $t$-channel poles). Indeed, it is immediate that:
$$
{\rm Im}\,\mathcal{A}_{S^2}^{\infty_s}(s,t,u)= {\rm Im}\,\mathcal{A}_{S^2}^s(s,t,u),\qquad{\rm and}\qquad {\rm Im}\,\mathscr{A}_{S^2}^{\infty_u}(s,t,u)= {\rm Im}\,\mathcal{A}_{S^2}^u(s,t,u),
$$
where we made use of (\ref{eq:ImAs,ImAu2}) and the standard result ${\rm Im}\,\frac{1}{x-i\epsilon}=\pi\delta(x)$. 
\sk

In order to demonstrate that also the real part of the amplitude is captured correctly by this gluing procedure let us manipulate these expressions further. We primarily simply rewrite (\ref{eq:As-VirasoroShapiro-glued}) and (\ref{eq:Au-VirasoroShapiro-glued}) (dropping the $i\epsilon$ since this is irrelevant for the real part) as follows,
\begin{equation}\label{eq:Asu-VirasoroShapiro-glued}
\begin{aligned}
\mathcal{A}_{S^2}^{\infty_s}(s,t,u)&=\Big(\frac{8\pi  g_D}{\alpha'}\Big)^2\frac{\alpha'}{4}\sum_{n=0}^{\infty}\Big(\frac{\Gamma(2+\frac{\alpha'}{4}t+n)}{\Gamma(2+\frac{\alpha'}{4}t)\Gamma(n+1)}\Big)^2\frac{1}{1-(2+\tfrac{\alpha'}{4}s)+n}\\
\mathcal{A}_{S^2}^{\infty_u}(s,t,u)&=\Big(\frac{8\pi  g_D}{\alpha'}\Big)^2\frac{\alpha'}{4}\sum_{n=0}^{\infty}\Big(\frac{\Gamma(2+\frac{\alpha'}{4}t+n)}{\Gamma(2+\frac{\alpha'}{4}t)\Gamma(n+1)}\Big)^2\frac{1}{1-(2+\tfrac{\alpha'}{4}u)+n}
\end{aligned}
\end{equation}
We then multiply and divide by $\Gamma(1-2-\frac{\alpha'}{4}t-n)$ in both expressions in  (\ref{eq:Asu-VirasoroShapiro-glued}) and make use of the first gamma function identity in (\ref{eq:GammaIdentities}) with the identification $z=2+\frac{\alpha'}{4}t$, according to which (taking into account also that $\Gamma(v+1)=v\Gamma(v)$):
\begin{equation}\label{eq:gammarelations}
\begin{aligned}
\Big(\frac{\Gamma(2+\frac{\alpha'}{4}t+n)}{\Gamma(2+\frac{\alpha'}{4}t)\Gamma(n+1)}\Big)^2
&=\frac{\Gamma(-1-\frac{\alpha'}{4}t)}{\Gamma(2+\frac{\alpha'}{4}t)}\frac{(-)^{n}\Gamma(2+\frac{\alpha'}{4}t+n)}{\Gamma(n+1)^2\Gamma(1-2-\frac{\alpha'}{4}t-n)}.
\\
\end{aligned}
\end{equation}
Since the first fraction on the right-hand side is precisely that found in the full result for the VS amplitude in (\ref{eq:S-VirasoroShapiro}), the remaining terms will most naturally be written entirely in terms of $s$ and $u$. 
It proves efficient to work in terms of the following quantities:
\begin{equation}\label{eq:abc}
a\dfn 2+\frac{\alpha'}{4}t,\qquad b\dfn 2+\frac{\alpha'}{4}s,\qquad{\rm and}\qquad c\dfn 2+\frac{\alpha'}{4}u,
\end{equation}
in terms of which, taking into account that $s+t+u=-16/\alpha'$ (or $a+b+c=2$),  (\ref{eq:gammarelations}) reads,
\begin{equation}\label{eq:gammarelations2}
\begin{aligned}
\Big(\frac{\Gamma(2+\frac{\alpha'}{4}t+n)}{\Gamma(2+\frac{\alpha'}{4}t)\Gamma(n+1)}\Big)^2
&=\frac{\Gamma(-1-\frac{\alpha'}{4}t)}{\Gamma(2+\frac{\alpha'}{4}t)}\frac{(-)^{n}\Gamma(2-2-\frac{\alpha'}{4}s-2-\frac{\alpha'}{4}u+n)}{\Gamma(n+1)^2\Gamma(-1+2+\frac{\alpha'}{4}s+2+\frac{\alpha'}{4}u-n)}
\\
&=\frac{\Gamma(1-a)}{\Gamma(a)}\frac{(-)^{n}\Gamma(2-b-c+n)}{\Gamma(n+1)^2\Gamma(-1+b+c-n)}.
\end{aligned}
\end{equation}
Adding the resulting $s$ and $u$ channel contributions in (\ref{eq:Asu-VirasoroShapiro-glued}), denoting the resulting quantity by $\mathcal{A}_{S^2}^{\infty}(s,t,u)=\mathcal{A}_{S^2}^{\infty_s}(s,t,u)+\mathcal{A}_{S^2}^{\infty_u}(s,t,u)$, and taking (\ref{eq:abc}) and (\ref{eq:gammarelations2}) into account yields,
\begin{equation}\label{eq:Asu-VirasoroShapiro-glued2}
\begin{aligned}
\mathcal{A}_{S^2}^{\infty}(s,t,u)
&=\Big(\frac{8\pi g_D^2}{\alpha'}\Big)2\pi \frac{\Gamma(1-a)}{\Gamma(a)}\sum_{n=0}^{\infty}\frac{(-)^{n}\Gamma(2-b-c+n)}{\Gamma(n+1)^2\Gamma(-1+b+c-n)}\Big[\frac{1}{1-b+n}+\frac{1}{1-c+n}\Big]\\
\end{aligned}
\end{equation}
The remaining sum over $n$ can now be performed and yields generalised hypergeometric functions,
\begin{equation}\label{eq:Asu-VS-sum_n}
\begin{aligned}
&\sum_{n=0}^{\infty}\frac{(-)^{n}\Gamma(2-b-c+n)}{\Gamma(n+1)^2\Gamma(-1+b+c-n)}\Big[\frac{1}{1-b+n}+\frac{1}{1-c+n}\Big]\\
&\qquad=
\frac{\Gamma(2-b-c)}{\Gamma(-1+b+c)}\Big(\frac{1}{1-b}\,\,\,\,F_{\!\!\!\!\!\!\!\!\!3\,\,\,\,\,\,\,2}\big(\begin{smallmatrix} 
      1-b ,2-b-c,2-b-c \\
      1, 2-b \\
   \end{smallmatrix};1\big)+
   \frac{1}{1-c}\,\,\,\,F_{\!\!\!\!\!\!\!\!\!3\,\,\,\,\,\,\,2}\big(\begin{smallmatrix}
      1-c ,2-b-c,2-b-c \\
      1, 2-c \\
   \end{smallmatrix};1\big)\Big) 
\end{aligned}
\end{equation}
and although we have not managed to show this analytically,\footnote{Since the residues of the full amplitude have been shown to be identical it suffices to show that the asymptotic expansions of (\ref{eq:Asu-VS-sum_n}) and (\ref{eq:Gammabc}) are identical. One might then appeal to uniqueness theorems of analytic continuation to deduce that (\ref{eq:Asu-VS-sum_n}) and (\ref{eq:Gammabc}) are equal. But this will not be pursued here any further.} according to Mathematica numerical evaluation of the right-hand side of the latter equation yields precise agreement with:
\begin{equation}\label{eq:Gammabc}
\frac{\Gamma(1-b)\Gamma(1-c)}{\Gamma(b)\Gamma(c)}.
\end{equation}
That is, the full result for the tree-level S-matrix for four tachyons (\ref{eq:Asu-VirasoroShapiro-glued2}) obtained by gluing two three-point amplitudes is,
\begin{equation}\label{eq:Asu-VirasoroShapiro-glued4}
\begin{aligned}
\mathcal{A}_{S^2}^{\infty}(s,t,u)
&=\Big(\frac{8\pi g_D^2}{\alpha'}\Big) \frac{2\pi\Gamma(1-a)\Gamma(1-b)\Gamma(1-c)}{\Gamma(a)\Gamma(b)\Gamma(c)},
\end{aligned}
\end{equation}
which according to the definitions (\ref{eq:abc}) is equivalent to:
\begin{equation}\label{eq:Asu-VirasoroShapiro-glued4}
\begin{aligned}
\mathcal{A}_{S^2}^{\infty}(s,t,u)
&=\Big(\frac{8\pi g_D^2}{\alpha'}\Big)\frac{2\pi \Gamma(-1-\frac{\alpha'}{4}s)\Gamma(-1-\frac{\alpha'}{4}t)\Gamma(-1-\frac{\alpha'}{4}u)}{\Gamma(2+\frac{\alpha'}{4}s)\Gamma(2+\frac{\alpha'}{4}t)\Gamma(2+\frac{\alpha'}{4}u)},
\end{aligned}
\end{equation}
which is indeed precisely the full standard result for the Virasoro-Shapiro amplitude \cite{Polchinski_v1}, derived here by using a handle operator insertion to glue together two three-point amplitudes.

\section{Discussion}
We have discussed how to cut Riemann surfaces and corresponding string amplitudes across trivial and non-trivial homology cycles. This is accomplished by appropriate insertion of bi-local operators called handle operators, an idea that can be traced back to work by Ooguri and Sakai \cite{OoguriSakai87} and Tseytlin \cite{Tseytlin90b,Tseytlin90} among others. Our overall underlying motivation for the construction is the hope that these handle operators can now be used to sum loops and perhaps provide the necessary link to a non-perturbative definition of string theory. 
\sk

Looking ahead regarding this last point, what remains is to understand how the modular group acts on amplitudes with handle operator insertions after which one can ask whether it is possible to sum string loops in a fashion consistent with modular invariance. It is conceivable that the old arguments regarding (super)string perturbation theory as an asymptotic series might not apply in the handle operator approach, essentially because handle operators might provide a means to sum loops before computing the path integral. (Asymptotic series for an otherwise perfectly well-defined finite quantity can arise by incorrectly interchanging a conditionally-convergent sum and integral, so this might also be the case in string theory.) This leads one to speculate that a non-perturbative string theory might take the form of an interacting two-dimensional quantum field theory compactified on a sphere, and a corresponding landscape of fixed points at which conformal field theories live, some of which will have a spacetime interpretation. It is conceivable that handle operators capture all non-perturbative effects one expects to see in a bulk definition of string theory. Of course we also need to go beyond the bosonic theory that we have explored here, the corresponding superstring construction will be discussed by one of us elsewhere \cite{Skliros20}.
\sk

In the current document we have followed Polchinski's approach \cite{Polchinski88}, where one takes full advantage of the BRST machinery, including manifest gauge invariance for arbitrary worldsheet Ricci curvature (subject to the topological Euler number constraint). This allows one, e.g., to glue string amplitudes together, or to represent higher genus string amplitudes as lower-genus amplitudes with appropriate handle operator insertions. E.g., all loop amplitudes may thus be reduced to sphere amplitudes, and in particular OPE associativity or worldsheet duality and modular invariance should suffice to ensure one can cover the entire moduli space. To ensure one covers moduli space {\it once} one needs to unravel the various domains of integration of the moduli space integrals, and we have demonstrated this explicitly in some simple examples. The construction is globally well-defined in moduli space and exact. 
\sk

We have demonstrated explicitly that the resulting amplitudes are gauge invariant (at arbitrary loop order), in that BRST-exact insertions decouple up to boundary terms in moduli space. Although modular invariance is not manifest, we have also demonstrated explicitly that the resulting amplitudes are modular-invariant at one loop. It will be important to demonstrate modular invariance at arbitrary loop order. We have also paid careful attention to various boundary terms that arise when one cuts open a handle, which are required in order for the resulting amplitudes (or moduli integrands) to depend only on worldsheet complex structure and not on a specific choice of coordinates or metric. 
\sk

The first six sections are (modulo parts in Sec.~\ref{sec:NOII} and Sec.~\ref{sec:WuYang}) entirely background-independent, whereas in the remaining Sec.~\ref{sec:EHOC} and Sec.~\ref{sec:EA} we focused on vanilla bosonic string theory to demonstrate a simple and explicit realisation of handle operators. 
\sk

Most of the results generalise immediately to the superstring. Two additional complications worth pointing out will come from specifying a globally well-defined gauge slice for the worldsheet gravitino (so that one can avoid Wu-Yang type boundary contributions associated to patch overlaps \cite{Sen15b,SenWitten15}), and also the Ramond sector we expect will present some additional subtleties due to branch cuts \cite{Witten12c}, although in principle it is understood how to proceed. Regarding the worldsheet gravitino, the price to pay for a suitably covariantised slice will be to depart from the traditional PCO approach \cite{Sen15b,SenWitten15}. But this might be necessary if one wishes to go beyond perturbation theory. The approach discussed in this document can indeed easily be generalised to superstring theory and the corresponding supermoduli space. In the superstring context the handle operator approach discussed here automatically implements a globally-defined smooth gauge slice in supermoduli space and there is no obstruction of the type discussed by Donagi and Witten \cite{DonagiWitten15}, and hence there is also no need for `vertical integration' \cite{Sen15b,SenWitten15}. As mentioned above, details for the supermoduli generalisation of the smooth gauge slice constructed in the current document using handle operators will be discussed in detail by one of us in forthcoming work.
\sk

Since handle operators (and their complex structure deformations) represent string loops or separating handles, they are comprised of offshell operators. In order to ensure that the resulting operators can be patched together consistently we have adopted Polchinski's suggestion \cite{Polchinski88} which is to normal order them using holomorphic normal coordinates. These coordinates are required to solve the Beltrami equation, and moreover the residual invariance under holomorphic reparametrisations of the Beltrami equation is fixed. Consequently, using holomorphic normal coordinates to normal order handle operators gauge fixes invariance under Weyl transformations. These coordinates also have the property that they preserve covariance, and they transform as scalars with respect to reparametrisations of an underlying auxiliary coordinate system while allowing one to proceed without making an explicit choice for the local Ricci curvature. A crucial property is that operators constructed out of holomorphic normal coordinates transform as scalars on patch overlaps modulo a U(1) ambiguity. This ambiguity is in turn removed \cite{Polchinski88} by considering combinations of operators that are independent of this phase, or by taking this phase to be integrated (by identifying it with a modulus). 
\sk

Although holomorphic normal coordinates are not holomorphic with respect to shifts in the base point at which they are centred, a significant virtue is that Wu-Yang boundary terms are not required, in that the global information contained in Wu-Yang terms is in a sense localised. So in particular one can use Stoke's theorem for integrals of total derivatives leading to contributions coming entirely from the boundary of moduli space (as opposed to boundary terms arising from patch overlaps). Since the entire construction has been developed for surfaces with generic Ricci curvature, one can also partition and distribute the curvature as desirable (such as at a point or uniformly, etc., subject to the topological Euler number constraint). Since the notion of a worldsheet metric is not necessary to define a Riemann surface it is also possible to allow for the worldsheet metric to be discontinuous or singular across various cycles provided the defining properties of transition functions are satisfied.
\sk

We have found that it is particularly efficient to adopt a coherent state basis for the construction of handle operators. These coherent states have, remarkably, well-defined total scaling dimension, even offshell. They are also local on the worldsheet, and can be considered to be a complete (or over-complete) basis for string fields when non-level-matched contributions are projected out. But since we are not adopting the string field theory partitioning of moduli space also non-level matched terms contribute when the corresponding handle operator is associated to a non-trivial homology cycle. These vertex operators can also be placed onshell whereby they can be interpreted as vertex operators associated to asymptotic states or to states produced in cuts.
\sk

One-loop modular invariance was checked by explicitly computing a one-loop vacuum amplitude by appropriate insertion of a handle operator onto a sphere, and showing perfect agreement with the standard result \cite{Polchinski86}. This checks the consistency of the use of handle operators for cutting across non-trivial homology cycles. To check consistency of the use of handle operators when cutting across trivial homology cycles we have demonstrated how to glue two three-point amplitudes and reconstruct the expected four-point amplitude. The result is again precisely in agreement with the standard result \cite{Polchinski_v1} and worldsheet duality is present. In particular, gluing across $s$- and $u$-channels implicitly contains the infinite sum over $t$-channel poles and (in contrast to string field theory \cite{SonodaZwiebach90}) a four-point vertex is not required in this formulation. A further consistency check has been to show that the coherent states out of which handle operators (at fixed-complex structure) are constructed indeed provide a resolution of unity, with the related computation of all one- and two-point sphere amplitudes. Again, the normalisation is precisely that required by unitarity, and in fact we have been particularly careful throughout to ensure that all normalisations are in agreement with factorisation and unitarity \cite{Polchinski_v1}.

\section*{Acknowledgements} 
DS gratefully acknowledges support from the Max Planck Institute for Physics in Munich where essentially all of this work was carried out. The work of DL is supported by the Origins Excellence Cluster. 
DS would like to thank Daniel Friedan, Ashoke Sen and Edward Witten for helpful discussions and remarks, the late Joe Polchinski for sharing an early unpublished version of his book, and especially Ed Copeland, Gia Dvali, Mark Hindmarsh, Joe Polchinski, Paul Saffin, John Ellis, Nick Mavromatos, for numerous insights, discussions and support that seeded the present work. He also thanks Tasos Avgoustidis for 3D-printing structures that helped visualise some Riemann surface geometry. We would also like to thank Costas Bachas, Lorenz Schlechter and Arkady Tseytlin for valuable feedback on the draft. DS has also benefited from discussions with Dmitri Bykov, Harold Erbin, Cesar Gomez, Ivo Sachs, Lorenz Schlechter and Matthias Traube.

\appendix
\newpage
\section{Notation and CFT Conventions}\label{sec:CFTC}
In this Appendix we collect some conventions on notation and elementary CFT results that we adopt throughout the text. 

\subsection{Charts and Coordinates}
Let us consider a Riemann surface with fixed complex structure and focus on some of the local data on a single chart, $(U_1,z_1)$, (see Sec.~\ref{sec:RS}). 
So given any two points $p,p'\in U_1$ we have the coordinate representations, $z_1(p),z_1(p')$, with $z_1:p\mapsto \mathbf{C}$, etc. We refer to $z_1$ as the {\it holomorphic coordinate} or the {\it frame coordinate} of the chart $(U_1,z_1)$. The choice of frame coordinate also depends on the point at which it vanishes (i.e.~at which it is centred). We consider centred coordinates throughout and place the centre of our coordinate system at $p_1\in U_1$, we have by definition:
$
z_1(p_1)\dfn 0
$. 
So the subscript on $z_1$, in addition to labelling the chart, is really a reminder that holomorphic coordinate $z_1(p)$ is based or centred  at $p_1$, and we could more fully write:
$$
\boxed{z_1(p)\equiv z_{p_1}(p),\qquad {\rm and}\qquad z_1(p_1)\equiv z_{p_1}(p_1)\equiv 0}
$$
Analogous relations are adopted in all charts, $\{(U_m,z_m)\}$, in particular, $z_m(p_m)\equiv z_{p_m}(p_m)\equiv0$, for all $m$. But note that a point $p$ can have multiple holomorphic coordinate representations, e.g., if $p\in U_m\cap U_n\neq\zero$, then $z_m(p)$ and $z_n(p)$ both represent the same point in different chart coordinates (with appropriate transition functions relating them, more about which below). The points $p,p_1,\dots$ denote abstract points on the underlying Riemann surface. 
\sk

It will occasionally be useful to introduce an auxiliary real coordinate system, $\sigma^a$, $a=1,2$, with $\sigma:p\mapsto \mathbf{R}^2$, and perhaps also endow the underlying Riemann surface with a metric, $g_{ab}(\sigma)\rmd\sigma^a \rmd\sigma^b$, thus turning it into a Riemannian manifold. In this case we denote $z_{1}(p)$ by $z_{\sigma_1}(\sigma)$, with $z_{\sigma_1}(\sigma_1)\dfn0$, etc. 
\sk

We will adopt a phrasing appropriate for the $(U_1,z_1)$ chart in the remaining subsection, but everything can immediately be generalised to any chart $(U_m,z_m)$ by simply replacing $z_1\rightarrow z_m$.

\subsection{Local Operators}\label{sec:LO}
Let us denote a generic local field or operator by: 
$$
\mathscr{A}^{(z_1)}(p),
$$
where $z_1$ in the superscript denotes the frame or chart coordinate used to define the field and $p\in U_1$ is the location where the field is evaluated. So according to the above comments, the operator $\mathscr{A}^{(z_1)}(p)$ is defined using frame coordinate, $z_1=z_{p_1}$, and so depends implicitly also on the point $p_1\in U_1$ where the frame is based (in the radial quantisation sense). When the explicit argument of a local operator is omitted it will always mean that the operator is evaluated at $p=p_1$ (where recall that $z_1(p_1)=0$):
$$
\mathscr{A}^{(z_1)}\equiv \mathscr{A}^{(z_1)}(p_1)\equiv \mathscr{A}^{(z_{p_1})}(p_1)\equiv \mathscr{A}^{(z_{p_1})}.
$$
Occasionally, instead of making manifest the location $p\in U_1$ we sometimes make manifest its coordinate representation instead, so that:
$$
\mathscr{A}^{(z_1)}(z,\bar{z})\equiv \mathscr{A}^{(z_1)}(p),\qquad {\rm with}\qquad z=z_1(p),\qquad \bar{z}=\bar{z}_1(p),
$$
and therefore when inserted at $p=p_1$ we might also write,
$$
\mathscr{A}^{(z_1)}(0,0)\equiv \mathscr{A}^{(z_1)}(p_1).
$$

In terms of the auxiliary real coordinate system, $\sigma^a$, $a=1,2$, with $\sigma:p\mapsto \mathbf{R}^2$, introduced above, we can also write $\mathscr{A}^{(z_{\sigma_1})}(\sigma)$ for the same object $\mathscr{A}^{(z_1)}(p)$, when the frame is centred at $\sigma=\sigma_1$ (where $z_{\sigma_1}(\sigma)|_{\sigma=\sigma_1}=0$) and in particular,
$$
\mathscr{A}^{(z_1)}\equiv \mathscr{A}^{(z_1)}(\sigma_1)\equiv \mathscr{A}^{(z_{\sigma_1})}(\sigma_1)\equiv \mathscr{A}^{(z_{\sigma_1})},
$$
and,
$$
\mathscr{A}^{(z_1)}(p)\equiv \mathscr{A}^{(z_1)}(\sigma)\equiv \mathscr{A}^{(z_{\sigma_1})}(\sigma).
$$
Working in terms of an auxiliary underlying coordinate system make certain aspects of reparametrisation invariance, complex structure and deformations thereof (such as shifts of the base point, $\sigma_1\mapsto \sigma_1'$) entirely manifest.
\sk

We denote generic (not necessarily primary) local operators by $\mathscr{A}$ throughout (usually with subscripts/superscripts to distinguish one such operator or frame from another as well as with appropriate arguments to denote where the operator is evaluated). {\it When} these operators have well-defined chiral and anti-chiral scaling dimension, denote it by $(h,\tilde{h})$, the corresponding OPE with the chiral half of the total energy momentum tensor, $T^{(z_1)}(z)$, takes the generic form:
\begin{equation}
\begin{aligned}
T^{(z_1)}&(z' )\mathscr{A}^{(z_1)}(z,\bar{z})  =\\
&=\dots+ \frac{h}{(z'-z)^2}\mathscr{A}^{(z_1)}(z,\bar{z})+\frac{1}{z'-z}\partial_{z}\mathscr{A}^{(z_1)}(z,\bar{z})+:\!T^{(z_1)}(z')\mathscr{A}^{(z_1)}(z,\bar{z})\!:_{z_1}.
\end{aligned}
\end{equation}
For primaries the `$\dots$' are absent whereas for more general local operators they correspond to more singular contributions in the limit $|z'-z|\rightarrow 0$.
\sk

Incidentally, this OPE with the energy-momentum tensor gives the transformation of the local operator $\mathscr{A}^{(z_1)}(z,\bar{z})$ under a change of coordinates and implicitly includes a corresponding {\it change of normal ordering}. Sometimes it is useful to disentangle these two notions, since all operators (in particular primary and non-primary operators) transform under changes of coordinates whereas {\it only primaries with vanishing weights} are invariant under changes of normal ordering (when coordinates are held fixed). This is a subtle but important point, so we will discuss it in detail in Sec.~\ref{sec:HCNO}. 

\subsection{Primaries and Mode Operators}\label{sec:PMO}
Remarks similar to the above for operators $\mathscr{A}^{(z_1)}(p)$ apply to {\it primaries} and {\it (anti-)chiral primaries} which we denote throughout by $\mathscr{O}^{(z_1)}(p)$. 
Suppose $\mathscr{O}^{(z_1)}(p)$ is a chiral primary (so that it depends on $z_1(p)$ but not the complex conjugate, $\bar{z}_1(p)$) and its associated conformal weight is $h$. That it is a chiral primary is equivalent to the statement that under conformal changes of frame, $z_1\mapsto w_1(z_1)$,
\begin{equation}\label{eq:chiralprimary}
\mathscr{O}^{(z_1)}(p)\rmd z_1(p)^h = \mathscr{O}^{(w_1)}(p)\rmd w_1(p)^h,\qquad{\rm or}\qquad \mathscr{O}^{(z_1)}(z)\rmd z^h = \mathscr{O}^{(w_1)}(w)\rmd w^h, 
\end{equation}
for the same quantities, with $z=z_1(p)$ and $w=w_1(p)$. These transformations implicitly include a change in normal ordering (discussed in Sec.~\ref{sec:HCNO}). 
When $p=p_1$ (where $z_1(p_1)=0$ and if also $w_1(p_1)=0$) we might write instead,
\begin{equation}\label{eq:chiralprimary2}
\mathscr{O}^{(z_1)}\rmd z_1^h = \mathscr{O}^{(w_1)}\rmd w_1^h.
\end{equation}
For any given chiral primary, $\mathscr{O}^{(z_1)}(p)$, we define an infinite set of modes, $\mathscr{O}_n^{(z_1)}$, as the Laurent coefficients in the expansions around the base point, $p_1$, in the $z_1$ frame:
\begin{equation}\label{eq:phi_n modes Laurent}
\mathscr{O}^{(z_1)}(z) = \sum_{n\in\mathbf{Z}}\frac{\mathscr{O}_n^{(z_1)}}{z^{n+h}},\qquad {\rm with}\qquad z=z_1(p),
\end{equation}
and we express the corresponding modes as contour integrals,
\begin{equation}\label{eq:phi_n modes}
\mathscr{O}_n^{(z_1)}\dfn \oint \frac{\rmd z}{2\pi i z}z^{n+h}\mathscr{O}^{(z_1)}(z),\qquad n\in\mathbf{Z}.
\end{equation}
We emphasise that also the mode operators depend on the coordinate frame ($z_1$) and hence also the base point ($p_1$) around which they are expanded. 
In the standard string theory literature \cite{Polchinski_v1,Witten12c} these quantities are often denoted by $\mathscr{O}(z)$ and $\mathscr{O}_n$ respectively, but we need to be more explicit in order to be able to transition from a local to a global construction that works for generic Riemann surfaces. (The contour integrals circle $z=z_1(p_1)=0$ counterclockwise and we adopt the standard normalisation, 
$\oint \rmd z/z=2\pi i$, throughout.) Occasionally, when it is not necessary to exhibit the specific frame with respect to which local operators are defined, and when we would rather make manifest the tensor transformation properties of tensor components we might write, e.g., $b=b_{zz}\rmd z^2$ or $\tilde{b}=\tilde{b}_{\bar{z}\bar{z}}\rmd\bar{z}^2$, where in a specific frame $z_1$ the quantities $b_{zz}$ and $\tilde{b}_{\bar{z}\bar{z}}$ might otherwise be denoted by $b^{(z_1)}(z)$ and $\tilde{b}^{(z_1)}(\bar{z})$ respectively.
\sk

The various chiral primary operators, modes and conformal weights relevant to the bosonic string in flat spacetime are listed, respectively, in Table \ref{tab:weightsmodes}.
\begin{table}[htbp]
   \centering
   \begin{tabular}{@{} lcrr @{}} 
      \toprule
      \quad\,$\mathscr{O}^{(z_1)}(z)$    & $\mathscr{O}_n^{(z_1)}$ & $h$\phantom{a} \\
      \midrule
      $\sqrt{\frac{2}{\alpha'}}i\partial_{z_1} x^{\mu}_{(z_1)}(z)$& $\alpha_n^{(z_1)}$&$1\phantom{a}$\\
      \quad\,$c^{(z_1)}(z)$&$c_n^{(z_1)}$&$-1$\\
      \quad\,$b^{(z_1)}(z)$&$b_n^{(z_1)}$&$2\phantom{a}$&\\
      \quad\,$T^{(z_1)}(z)$&$L_n^{(z_1)}$&$2\phantom{a}$\\
      \quad\,$j_{B}^{(z_1)}(z)$&$Q_B^{(z_1)}$&$1\phantom{a}$\\
      \bottomrule
   \end{tabular}
   \caption{The elementary chiral primary operators that appear in the bosonic string, their corresponding modes and conformal weights respectively. The $\sqrt{\frac{2}{\alpha'}}i\partial x^{\mu}(z)$ denote standard matter fields, $b(z),c(z)$ reparametrisation ghosts, $T(z)$ the energy-momentum tensor (including matter and ghost contributions such that the total central charge vanishes), and $j_B(z)$ the BRST current.  Any one of these primaries is denoted generically by $\mathscr{O}^{(z_1)}(z)$. They are related as in (\ref{eq:phi_n modes Laurent}) and (\ref{eq:phi_n modes}). (For the BRST charge which does not depend on $n$ one should take $n=0$ in (\ref{eq:phi_n modes}), see (\ref{eq:QB}).)}\label{tab:weightsmodes}
\end{table}

The corresponding operator product expansions (OPE) for the case of bosonic strings in flat spacetime read:
\begin{equation}\label{eq:opes}
\begingroup\makeatletter\def\f@size{11}\check@mathfonts
\def\maketag@@@#1{\hbox{\m@th\large\normalfont#1}}%
\begin{aligned}
&b^{(z_1)}(p')c^{(z_1)}(p) = \frac{1}{z_1(p')-z_1(p)}+:\!b^{(z_1)}(p')c^{(z_1)}(p)\!:_{z_1}\\
&x^{\mu}_{(z_1)}(p')x^{\nu}_{(z_1)}(p) = -\frac{\alpha'}{2}\eta^{\mu\nu}\ln|z_1(p')-z_1(p)|^2+:\!x^{\mu}_{(z_1)}(p')x^{\nu}_{(z_1)}(p)\!:_{z_1}\\
\end{aligned}
\endgroup
\end{equation}
These relations {\it define} what we mean by `$z_1$ normal ordering in $z_1$ frame coordinates', and $z_1$ normal ordering (independently of which coordinates are used) is denoted by $:\,\,:_{z_1}$, a concept that was made precise in \cite{Polchinski87}. This corresponds to subtracting precisely the singular terms in (\ref{eq:opes}). There are also  important related notions that we introduce in Sec.~\ref{sec:HCNO}, which is to `{\it change normal ordering keeping coordinates fixed}', and conversely to `{\it change coordinates keeping normal ordering fixed}'. 
\sk

Under rigid rescalings, rotations and shifts, $z_1\mapsto w_1=\lambda z_1+z_v$, according to (\ref{eq:phi_n modes}) and (\ref{eq:chiralprimary}) the modes transform as follows,
\begin{equation}\label{eq:On-scal-rot-shif}
\begin{aligned}
&\mathscr{O}_n^{(\lambda z_1+z_v)}(p_1)= \lambda^{n}\mathscr{O}^{(z_1)}_{n}(p_1)\qquad\Rightarrow \qquad \mathscr{O}_n^{(\lambda z_1+z_v)}(z_v)= \lambda^{n}\mathscr{O}^{(z_1)}_{n}(0)\\
&\tilde{\mathscr{O}}_n^{(\lambda z_1+z_v)}(p_1)= \bar{\lambda}^{n}\tilde{\mathscr{O}}^{(z_1)}_{n}(p_1)\qquad\Rightarrow \qquad \tilde{\mathscr{O}}_n^{(\lambda z_1+z_v)}(z_v)= \bar{\lambda}^{n}\tilde{\mathscr{O}}^{(z_1)}_{n}(0)
\end{aligned}
\end{equation}
where in the second line we also wrote down the corresponding expressions for the anti-chiral halves. These follow from the fact that the corresponding local fields transform as conformal tensors, and are therefore derived from (\ref{eq:chiralprimary}) and (\ref{eq:phi_n modes}). The relations on the right-hand side in (\ref{eq:On-scal-rot-shif}) take into account that when evaluated at $p_1\in U_1$ the coordinates read $z_1(p_1)\equiv 0$ and $w_1(p_1)=\lambda z_1(p_1)+z_v\equiv z_v$, and we will use the two notations (corresponding to the two ``columns'' in (\ref{eq:On-scal-rot-shif})) interchangeably depending on what property we wish to emphasise.

\sk
The quantity $T^{(z_1)}(z)$ in the previous subsection is the chiral component of the matter plus ghost energy-momentum tensor, $T=T^{(z_1)}(z)\rmd z^2+\tilde{T}^{(z_1)}(\bar{z})\rmd\bar{z}^2$, in particular $T = T_{\rm m}+T_{\rm g}$. In the case of strings in Minkowski spacetime (or toroidal compactifications thereof),
\begin{equation}\label{eq:T-bosonic}
\begin{aligned}
T_{\rm m}^{(z_1)}&=-\tfrac{1}{\alpha'}\!:\!(\partial_{z_1} x_{(z_1)})^2\!:_{z_1}\\
T_{\rm g}^{(z_1)} &= \,\,:\!c^{(z_1)}(\partial_{z_1} b^{(z_1)})+2(\partial_{z_1} c^{(z_1)})b^{(z_1)}\!:_{z_1}
\end{aligned}
\end{equation}

Since \cite{Polchinski87,Polchinski88} normal ordering often depends on the frame (see Sec.~\ref{sec:HCNO}) with respect to which the subtractions are carried out, one must take (usually finite) changes in normal ordering into account when gluing together charts to construct global objects that can be integrated over a Riemann surface. (Indeed, requiring that local vertex operators be independent of the normal ordering prescription leads precisely to the usual Virasoro conditions. Notice these conditions are stronger than simply requiring BRST invariance.) 
\sk

Our convention for the BRST charge (based at $p_1$ in the above sense and as indicated by the superscript, `$\phantom{i}^{(z_1)}$') is:
\begin{equation}\label{eq:QB}
Q_{\rm B}^{(z_1)} = \frac{1}{2\pi i}\oint \big(\rmd z\,j_{\rm B}^{(z_1)}-\rmd\bar{z}\,\tilde{j}_{\rm B}^{(z_1)}\big),
\end{equation}
where the chiral half of the (weight $(h,\tilde{h})=(1,1)$ primary) BRST current reads,
\begin{equation}\label{eq:jB}
j_{\rm B}^{(z_1)}(z) = \,:\!c^{(z_1)}(z)\big(T_{\rm m}^{(z_1)}(z)+\frac{1}{2}T_{\rm g}^{(z_1)}(z)\big)+\frac{3}{2}\partial_z^2c^{(z_1)}(z)\!:_{z_1}
\end{equation}
with a similar expression for the anti-chiral half. 
Let us make note of two relations that are useful for demonstrating decoupling of BRST-exact states from string amplitudes:
\begin{equation}\label{eq:QbLn commutators}
\begin{aligned}
\big\{Q_{\rm B}^{(z_1)},b_n^{(z_1)}\big\} &= L_n^{(z_1)}\\
\big[L_n^{(z_1)},\mathscr{O}_m^{(z_1)}\big]&=\big[n(h-1)-m\big]\mathscr{O}_{n+m}^{(z_1)}
\end{aligned}
\qquad
\begin{aligned}
\big\{Q_{\rm B}^{(z_1)},\tilde{b}_n^{(z_1)}\big\} &= \tilde{L}_n^{(z_1)},\\
\big[\tilde{L}_n^{(z_1)},\tilde{\mathscr{O}}_m^{(z_1)}\big]&=\big[n(\tilde{h}-1)-m\big]\tilde{\mathscr{O}}_{n+m}^{(z_1)},
\end{aligned}
\end{equation}
for any primary operator modes (\ref{eq:phi_n modes}). These follow directly from the above and the usual \cite{DiFrancescoMatheuSenechal97} CFT interpretation for (anti-)commutators of generic modes, 
$$
\{A,B] = \oint_0\rmd w\oint_w\rmd z\,a(z)b(w), \qquad{\rm with}\qquad A=\oint \rmd z\,a(z),\qquad B=\oint \rmd z\,b(z).
$$

A crucial property that follows immediately from the fact that (\ref{eq:jB}) transforms as a $(1,0)$ primary is that the BRST operator is invariant under holomorphic changes of coordinates. To be more precise, it is invariant up to a sign since within a given coordinate chart the orientation of the contour in (\ref{eq:QB}) is always the canonical one and we might need to flip the orientation to map the BRST charge of one chart to that of another. For example, we can cover a sphere with two coordinate charts, $(U_1,z_1)$ and $(U_2,z_2)$ (centred at $p_1$ and $p_2$ respectively), such that $U_1\cap U_2$ consists of an equatorial band on which $z_1$ and $z_2$ represent the same point provided they are related by the transition function $z_1z_2=1$. On either chart we can define a BRST operator, namely $Q_B^{(z_1)}$ and $Q_B^{(z_2)}$, which are centred at $p_1$ and $p_2$ respectively. When the corresponding contour of the $Q_B^{(z_1)}$ charge is deformed into the region $U_1\cap U_2$ we can change coordinates using the transition function, but since the orientation of the contour of $Q_B^{(z_2)}$ is opposite to that of $Q_B^{(z_1)}$ we pick up a minus sign: $Q_B^{(z_1)}=-Q_B^{(z_2)}$. Similar reasoning holds for any Riemann surface defined by a set of holomorphic transition functions and corresponding cocycle relations.
\sk

We next list some useful mode operator results (annihilation operators and state-operator correspondence, all of which follow directly from (\ref{eq:phi_n modes})):
\begin{subequations}\label{eq:annihil.ops,op-state}
\begin{align}
&\alpha_{n}^{(z_1)}(p_1)\cdot 1 = 0,\qquad n\geq0; \qquad\qquad\,\alpha_{-n}^{(z_1)}(p_1)\cdot 1 = \sqrt{\frac{2}{\alpha'}}\frac{i}{(n-1)!}\partial_{z_1}^nx_{(z_1)}(p_1),
\qquad n\geq1\label{eq:an-annihilation-creation}\\
&b_{n}^{(z_1)}(p_1)\cdot 1 = 0,\qquad n\geq -1;  \qquad\quad\,\,\, b_{-n}^{(z_1)}(p_1)\cdot 1 =  \frac{1}{(n-2)!}\partial_{z_1}^{n-2}b^{(z_1)}(p_1),\qquad n\geq2\label{eq:bn-annihilation-creation}\\
&c_n^{(z_1)}(p_1)\cdot 1 = 0,\qquad n\geq2;  \qquad\qquad\,\, c_{-n}^{(z_1)}(p_1)\cdot 1 =  \frac{1}{(n+1)!}\partial_{z_1}^{n+1}c^{(z_1)}(p_1),\qquad n\geq-1\label{eq:cn-annihilation-creation}\\
&L_n^{(z_1)}(p_1)\cdot 1 = 0,\qquad n\geq-1;  \qquad\quad\, L_{-n}^{(z_1)}(p_1)\cdot 1=\frac{1}{(n-2)!}\partial_{z_1}^{n-2}T^{(z_1)}(p_1),\qquad n\geq2\label{eq:Ln-annihilation-creation}
\end{align}
\end{subequations}
where `1' indicates the unique (up to normalisation) ${\rm SL}(2,\mathbf{C})$ vacuum. In the main text we write $|1\rangle$ for the corresponding state representation for the ${\rm SL}(2,\mathbf{C})$ vacuum. 

\section{Cycle Index Polynomials}\label{sec:CIP}
Cycle index polynomials play a fundamental role in studies of coherent states in string theory. 
In this section we provide a brief review of cycle index polynomials, starting from their definition, after which we present various useful properties and identities.

\subsection{Definition}\label{sec:CIP dfn and prop}
A fundamental role in discussions of string coherent states is played by {\it cycle index polynomials}, $Z_N(a_s)\equiv Z(S_N;a_1,\dots,a_N)$, of the {\it symmetric group}, $S_N$, defined (for any arguments $a_s=\{a_1,a_2,\dots\}$) by the generating function:
\begin{equation}\label{eq:CycleIndexGenFun}
\boxed{\sum_{n=0}^{\infty}Z_n(a_s)u^n\equiv \exp\Big(\sum_{s=1}^{\infty}\frac{1}{s}a_su^s\Big)}
\end{equation}
The $\{a_s\}$ are by definition independent of $u$ so we may take a contour integral of both sides leading to explicit expressions, 
\begin{subequations}\label{eq:S_m(a)}
\begin{align}
Z_N(a_s)&=\oint_0\frac{\rmd u}{2\pi iu}\,u^{-N} \exp \Big(\sum_{s=1}^{N}\frac{1}{s}a_su^s\Big),\qquad \forall N\in \mathbf{Z}\label{eq:S_n(a)a}\\
&=\sum_{k_1+2k_2+\dots+Nk_N=N}\frac{1}{k_1!}\Big(\frac{a_1}{1}\Big)^{k_1}\dots\frac{1}{k_N!}\Big(\frac{a_N}{N}\Big)^{k_N},\qquad \forall N\in \mathbf{Z}^+\label{eq:S_n(a)b},
\end{align}
\end{subequations}
where the sum in the second equality is over all partitions of $N$, that is over non-negative integers, $\{k_1,\dots,k_N\}$, subject to $k_1+2k_2+\dots+Nk_N=N$. 
The expression (\ref{eq:S_n(a)b}) can be derived directly from (\ref{eq:CycleIndexGenFun}) by taking the $N^{\rm th}$ derivative of both sides with respect to $u$, making use of Fa\`a di Bruno's formula for derivatives of composite functions \cite{Riordan58} for the right-hand side, and finally setting $u=0$. 
From (\ref{eq:S_n(a)a}) it follows immediately that $Z_{N<0}(a_s)=0$ (so that the definition of $Z_N(a_s)$ can be extended to negative integers, $N$, by means of (\ref{eq:S_n(a)a})) and $Z_0(a_s)=1$, $Z_N(0)=\delta_{N,0}$, whereas from (\ref{eq:S_n(a)b}) or (\ref{eq:S_n(a)a}), the first few explicit expressions are: 
\begin{equation}\label{eq:Zn(as)-explicit}
\begin{aligned}
Z_0(a_s)&=1\\
Z_1(a_s)&=a_1\\
Z_2(a_s)&=\frac{1}{2}a_1^2+\frac{1}{2}a_2\\
Z_3(a_s)&=\frac{1}{3!}a_1^3+\frac{1}{2}a_1a_2+\frac{1}{3}a_3\\
Z_4(a_s)&=\frac{1}{4!}a_1^4+\frac{1}{4}a_1^2a_2+\frac{1}{8}a_2^2+\frac{1}{3}a_1a_3+\frac{1}{4}a_4\\
Z_5(a_s)&=\frac{1}{5!}a_1^5+\frac{1}{12}a_1^3a_2+\frac{1}{8}a_1a_2^2+\frac{1}{6}a_1^2a_3+\frac{1}{6}a_2a_3+\frac{1}{4}a_1a_4+\frac{1}{5}a_5
\end{aligned}
\end{equation}
Notice that the coefficients in each of these explicit expressions add up to 1, reminiscent of the general result $Z_N(1)=1$ (which follows, e.g., from the integer shift relation derived below and $Z_N(0)=\delta_{N,0}$). 
\sk

\subsection{Essential Relations}\label{sec:ER}
In most applications in string theory, for fixed $N$ all terms in the explicit sum in (\ref{eq:S_n(a)b}) contribute at the same order so we must keep them all. 
This is easily dealt with as there are a number of useful relations that follow from the above definition (for $N>0$), to the extent that one hardly ever needs to use the explicit expansion (\ref{eq:S_n(a)b}):
\begin{equation}\label{eq:Zproperties}
\begin{aligned}
&Z_N(b^sa_s)=b^NZ_N(a_s)\qquad \textrm{(scaling relation)}\\
&Z_N(a_s)=\frac{1}{N}\sum_{m=1}^Na_mZ_{N-m}(a_s)\qquad \textrm{(recursion relation)}\\
&Z_N(a_s+b_s)=\sum_{m=0}^NZ_{N-m}(a_s)Z_m(b_s)\qquad \textrm{(multiplication theorem)}\\
&Z_N(a_s)=\frac{1}{N!}B_N(0!a_1,1!a_2,\dots,(N-1)!a_N),\qquad \textrm{(complete Bell polynomial relation)}\\
&Z_N(a)
=\frac{\Gamma(a+N)}{\Gamma(a)\Gamma(N+1)}\qquad \textrm{($\Gamma$-function relation)}\\
&\frac{\partial}{\partial z}Z_N(a_s)=\sum_{m=1}^N\frac{1}{m}Z_{N-m}(a_s)\frac{\partial a_m}{\partial z}\qquad \textrm{(derivative relation)}\\
&Z_N(a_s+1)=\sum_{m=0}^NZ_{N-m}(a_s)\qquad\textrm{(integer shift relation)}\\
&Z_{ng}(\delta_{s,n}a_s)=\frac{1}{g!}\Big(\frac{a_n}{n}\Big)^g\qquad \textrm{(Kronecker delta relation)}
\end{aligned}
\end{equation}
where 
 $B_N(c_1,\dots,c_N)$ is a complete Bell polynomial \cite{Riordan58,Andrews}, and $a_s,b_s,c_s,b$ are arbitrary.  Given the importance of the properties (\ref{eq:Zproperties}) in string amplitude computations, we briefly discuss their derivations next. 
\sk

The {\bf scaling relation} follows immediately from (\ref{eq:S_n(a)b}). Differentiating the scaling relation with respect to $b$ and making use of (\ref{eq:S_n(a)a}) on the left-hand side then yields the {\bf recursion relation} in (\ref{eq:Zproperties}). The {\bf multiplication theorem} of (\ref{eq:Zproperties}) follows upon replacing $a_s\rightarrow a_s+b_s$ in (\ref{eq:S_n(a)a}), replacing the resulting two exponentials in the integrand using the generating function (\ref{eq:CycleIndexGenFun}) and carrying out the contour integral.  The {\bf complete Bell polynomial relation} in (\ref{eq:Zproperties}) follows from inspecting the defining series expansion \cite{Riordan58,Andrews} of complete Bell polynomials and (\ref{eq:S_n(a)b}). To derive the {\bf $\Gamma$-function relation} of (\ref{eq:Zproperties}) we consider the quantity $Z_N(a)$, use the complete Bell polynomial relation of (\ref{eq:Zproperties}) leading to $Z_N(a)=\frac{1}{N!}B_N(0!a,1!a,\dots,(N-1)!a)$. This is then decomposed in terms of Bell polynomials using the relation, 
$
B_N(a_1,\dots,a_N)=\sum_{k=1}^NB_{N,k}(a_1,\dots,a_{N-k+1}).
$ 
The latter polynomial has the scaling property $B_{N,k}(ca_1,\dots,ca_{N-k+1})=B_{N,k}(a_1,\dots,a_{N-k+1})c^k$, which when applied to the case of interest yields: 
$
Z_N(a)=\frac{1}{N!}\sum_{k=1}^NB_{N,k}(0!,1!,\dots,(N-k)!)a^k.
$ 
But the coefficient $B_{N,k}(0!,1!,\dots,(N-k)!)$ is in turn precisely the unsigned Stirling number of the first kind \cite{Riordan58}, $c(N,k)\equiv (-)^{N-k}s(N,k)$, 
and so noting that the resulting sum over $k$ is by definition the rising factorial\footnote{Recall that the equality: 
$$
\sum_{k=1}^Nc(N,k)a^k\equiv\frac{\Gamma(N+a)}{\Gamma(a)}
$$ is the defining generating function  of unsigned Stirling numbers of the first kind \cite{Riordan58}.} \cite{Riordan58}, the stated $\Gamma$-function relation of (\ref{eq:Zproperties}) follows immediately. The {\bf derivative relation} follows immediately from (\ref{eq:S_n(a)a}). The {\bf integer shift relation} follows immediately from the multiplication theorem and the result $Z_m(1)=1$ (for any integer $m\geq0$) which in turn follows easily from the integral representation (\ref{eq:S_n(a)a}) with $a_s=1$. Finally, the {\bf Kronecker delta relation} follows immediately from the integral representation (\ref{eq:S_n(a)a}).
\sk

Let us also list a couple of complete Bell polynomial relations,
\begin{equation}
\begin{aligned}
\exp\Big(\sum_{n=1}^{\infty}\frac{1}{n!}a_nz^n\Big)&=\sum_{n=0}^{\infty}\frac{1}{n!}B_n(a_1,\dots,a_n)z^n\\
\delta B_N(a_1,\dots,a_N)&=\sum_{n=1}^{\infty}\binom{N}{n}B_{N-n}(a_1,\dots,a_{N-n})\delta a_n.
\end{aligned}
\end{equation}

\subsection{Additional Identities}
Let us consider the factorisation relation in (\ref{eq:Zproperties}), replace $a_s\rightarrow a_se^{is\theta}$, $b_s\rightarrow b_se^{-is\theta}$ and integrate out $\theta$ with measure $\int_0^{2\pi} \frac{\rmd\theta}{2\pi}$. Making use of the generating function (\ref{eq:CycleIndexGenFun}) on the left-hand side and the scaling relation on the right-hand side gives:
\begin{equation}
\int_0^{2\pi} \frac{\rmd\theta}{2\pi}\exp\sum_{n=1}^\infty \Big(\frac{1}{n}a_ne^{in\theta}+\frac{1}{n}b_ne^{-in\theta}\Big)=\sum_{n=0}^\infty Z_n(a_s)Z_n(b_s).
\end{equation}

There is also a useful relation for $n^{\rm th}$ derivatives of an exponential of a function in terms of cycle index polynomials,
$\frac{1}{n!}e^{-g(z)}\partial_z^ne^{g(z)}=Z_n(\tfrac{\partial_z^sg(z)}{(s-1)!})$, 
which follows from Faa di Bruno's formula and the complete Bell polynomial relation given above. This relation is a useful representation for Taylor series; suppose $e^{g(z)}$ has a Taylor series expansion around $z=0$ with a non-zero radius of convergence.  Then,
$$
e^{g(z)}=e^{g(0)}\sum_{n=0}^\infty Z_n\big(\tfrac{\partial_z^sg(0)}{(s-1)!}\big)z^n.
$$

Some additional identities that follow from the above relations in Appendices \ref{sec:ER} and \ref{sec:CIP dfn and prop} are the following,
\begin{equation}
\begin{aligned}
&\sum_{n=0}^{\infty}Z_n(ba_s)=\Big(\sum_{n=0}^{\infty}Z_n(a_s)\Big)^b\qquad \textrm{(sum-power relation)}\\
&\sum_{n=0}^\infty Z_n(a_s+b_s)=\Big(\sum_{n=0}^\infty Z_n(a_s)\Big)\Big(\sum_{m=0}^\infty Z_m(b_s)\Big)\qquad\textrm{(factorisation relation)}\\
&Z_n(\tfrac{1}{2}a_{2s})=Z_{2n}\big(\tfrac{1+(-)^s}{2}a_s\big)\qquad\textrm{(projection relation)}\\
&Z_{2n+1}\big(\tfrac{1+(-)^s}{2}a_s\big)=0\qquad\textrm{(vanishing theorem)}\\
&Z_n\big(\!-\sum_{j=1}^m\lambda_j^s\big)=e_n(-\lambda_1,\dots,-\lambda_m)\qquad \textrm{(elementary symmetric polynomial relation)}\\
\end{aligned}
\end{equation}
and we leave their derivation to the reader.

\subsection{Matrix Identities}
Let us also list a number of fundamental identities involving matrices, traces of matrices and cycle index polynomials. Suppose that $A$ and $B$ are $N\times N$ matrices. We have the following identities:
\begin{equation}\label{eq:Zmatrixproperties}
\begin{aligned}
&\sum_{n=0}^{\infty}Z_n({\rm Tr}A^s)={\det}^{-1}(\mathbb{1}-A) \qquad \textrm{(1$^{\rm st}$ determinant relation)}\\
&\sum_{n=0}^{\infty}Z_n(-{\rm Tr}B^s)={\det}(\mathbb{1}-B) \qquad \textrm{(2$^{\rm nd}$determinant relation)}\\
&Z_{n}(-{\rm Tr}A^s)=0\qquad \textrm{(when $n>N$)}\\
&Z_{N}(-{\rm Tr}A^s)=(-)^N{\det}A\\
&Z_{N-n}(-{\rm Tr}A^s)=Z_{n}(-{\rm Tr}A^{-s})Z_{N}(-{\rm Tr}A^s)\\
&{\det}\big(\tfrac{B}{A}\big)=\sum_{n=0}^{\infty}Z_n({\rm Tr}(\mathbb{1}+A)^s-{\rm Tr}(\mathbb{1}+B)^s)\\
&\lim_{\ell\rightarrow \infty}\sum_{n=0}^{\infty}Z_n({\rm Tr}A^s-{\rm Tr}B^s)\ell^n=\sum_{n=0}^{\infty}Z_n({\rm Tr}(\mathbb{1}+A)^s-{\rm Tr}(\mathbb{1}+B)^s)
\end{aligned}
\end{equation}
$A,B$ are general matrices of any finite rank.
\sk

It is not hard to derive the above identities\footnote{E.g., the vanishing of $Z_{n}(-{\rm Tr A^s})$ for all $n>N$ and $N\times N$ matrices, $A$, follows from the $2^{\rm nd}$ determinant relation.} and we leave most of the derivations to the reader. Let us though elaborate on the last two identities. Multiply left- and right-hand sides respectively of the $1^{\rm st}$ and $2^{\rm nd}$ determinant relations in (\ref{eq:Zmatrixproperties}),
$$
{\det}\big(\tfrac{\mathbb{1}-B}{\mathbb{1}-A}\big)=\big(\sum_{n=0}^{\infty}Z_n({\rm Tr}A^s)\big)\big(\sum_{m=0}^{\infty}Z_m(-{\rm Tr}B^s)\big),
$$
and make use of the factorisation relation,
\begin{equation}\label{eq:det(1-B)/(1-A)}
\begin{aligned}
{\det}\big(\tfrac{\mathbb{1}-B}{\mathbb{1}-A}\big)&=\sum_{n=0}^{\infty}Z_n({\rm Tr}A^s-{\rm Tr}B^s).
\end{aligned}
\end{equation}
Next rescale $A,B$ by a $c$-number, $\ell$, and use the scaling relation,
\begin{equation}\label{eq:det(1-B)/(1-A)ell}
\begin{aligned}
{\det}\big(\tfrac{\mathbb{1}-\ell B}{\mathbb{1}-\ell A}\big)&=\sum_{n=0}^{\infty}Z_n({\rm Tr}A^s-{\rm Tr}B^s)\ell^n.
\end{aligned}
\end{equation}
On the other hand, if we shift $B\mapsto B+\mathbb{1}$ and $A\mapsto A+\mathbb{1}$ in (\ref{eq:det(1-B)/(1-A)}) we learn that,
\begin{equation}\label{eq:det(1-B)/(1-A)elllimit}
\begin{aligned}
{\det}\big(\tfrac{B}{A}\big)&=\sum_{n=0}^{\infty}Z_n({\rm Tr}(\mathbb{1}+A)^s-{\rm Tr}(\mathbb{1}+B)^s),
\end{aligned}
\end{equation}
which is the second to last relation in (\ref{eq:Zmatrixproperties}). Also, by continuity of the determinant,
\begin{equation}
\lim_{\ell\rightarrow \infty}{\det}\big(\tfrac{\mathbb{1}-\ell B}{\mathbb{1}-\ell A}\big)={\det}\big(\tfrac{B}{A}\big),
\end{equation}
which enables us to restate this in terms of the right-hand sides of (\ref{eq:det(1-B)/(1-A)ell}) and (\ref{eq:det(1-B)/(1-A)elllimit}) which yields the last relation in (\ref{eq:Zmatrixproperties}):
\begin{equation}
\lim_{\ell\rightarrow \infty}\sum_{n=0}^{\infty}Z_n({\rm Tr}A^s-{\rm Tr}B^s)\ell^n=\sum_{n=0}^{\infty}Z_n({\rm Tr}(\mathbb{1}+A)^s-{\rm Tr}(\mathbb{1}+B)^s).
\end{equation}

\subsection{Integer Arguments}
There are a number of useful relations for cycle index polynomials of integer arguments, beginning from:
\begin{equation}\label{eq:Z_n-r(-n) apx}
Z_{n-r}(-n) = (-)^{n-r}\binom{n}{r},\qquad{\rm and}\qquad Z_{m+r}(-m)=0,\qquad\textrm{$\forall$ integers $n,m,r\geq1$},
\end{equation}
which follow immediately from the gamma function relation (and elementary manipulations using the identity $\Gamma(z)\Gamma(1-z)=(\sin \pi z)/\pi$); 
we also require the following double sum,
\begin{equation}\label{eq:sumZZ=ndelta apx}
\begin{aligned}
\sum_{r=1}^{n'}\sum_{s=1}^{n}&\frac{(r+s-1)!}{(r-1)!(s-1)!}Z_{n'-r}(-n')Z_{n-s}(-n)\\
&=(-)^{n'+n}\sum_{r=1}^{n'}\sum_{s=1}^{n}\frac{(-)^{r+s}(r+s-1)!}{(r-1)!(s-1)!}\binom{n'}{r}\binom{n}{s}=n\delta_{n',n}.
\end{aligned}
\end{equation}
This double sum was computed numerically for explicit integers, $n,n'$, and (conjecturally) extrapolated to all positive integers. 
Furthermore there are useful relations such as,
\begin{equation}\label{eq:sumZ_n-r(-n)}
\begin{aligned}
&\sum_{r=1}^nrZ_{n-r}(-n) = \delta_{n,1}\\
&\sum_{r=1}^nZ_{n-r}(-n)=(-)^{n-1}\\
&\sum_{r=1}^nrZ_{m+r}(m)Z_{n-r}(n)=\frac{2}{m+n}\frac{\Gamma(2n)}{\Gamma(n)^2}\frac{\Gamma(2m)}{\Gamma(m)^2}.
\end{aligned}
\end{equation}

\section{Some Elementary Complex Analysis}

\subsection{Green's Theorem}\label{sec:GT}
\begin{figure}
\begin{center}
\includegraphics[angle=0,origin=c,width=0.8\textwidth]{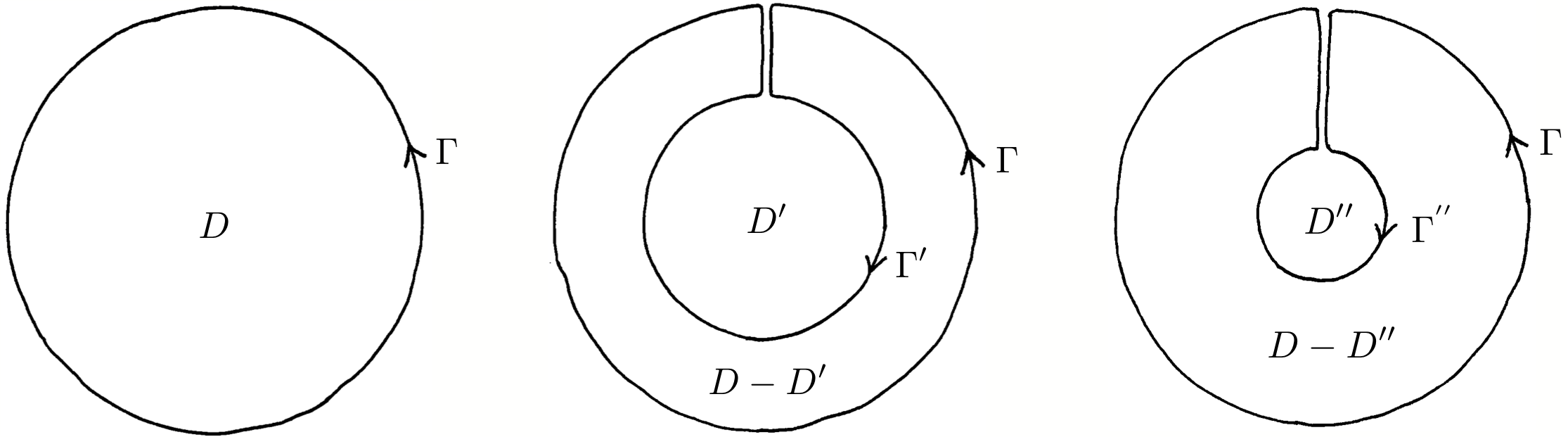}
\caption{Integration contours associated to various simply-connected domains.}\label{fig:Greens2}
\end{center}
\end{figure}

Suppose $D$ is a simply connected region in $\Sigma$ bounded by a sufficiently smooth curve $\Gamma=\partial D$ with the standard orientation as depicted in the first diagram in Fig.~\ref{fig:Greens2}.  
The Green's (or 2-$d$ Stoke's) theorem then reads,
\begin{equation}\label{eq: GreensTheorem_real}
\oint_{\Gamma}\rmd x\,v+\rmd y\,u = \int_{D}\rmd x\wedge \rmd y \,\big{(}\partial_x u-\partial_yv\big{)},
\end{equation}
where $u(x,y)$ and $v(x,y)$ real $C^1$ functions.
\sk

Let us go to complex coordinates, $z,\bar{z}$, by defining the quantities:
\begin{equation}\label{eq:zzbar xy}
\begin{aligned}
&z=x+iy,\qquad \bar{z}=x-iy\\
&\partial_z=\frac{1}{2}(\partial_x-i\partial_y),\qquad \partial_{\bar{z}}=\frac{1}{2}(\partial_x+i\partial_y)\\
&w(z,\bar{z})=u(x,y)+iv(x,y),\qquad \bar{w}(z,\bar{z})=u(x,y)-iv(x,y).
\end{aligned}
\end{equation}
Writing furthermore $\rmd^2z\dfn i\rmd z\wedge \rmd\bar{z}$, this change of variables in (\ref{eq: GreensTheorem_real}) leads to:
\begin{equation}\label{eq: Greens theorem}
\oint_{\Gamma}\rmd z\,w+\rmd\bar{z}\bar{w}=i\int_D\rmd^2z\,\big{(}\partial_{\bar{z}}w-\partial_z\bar{w}\big{)}.
\end{equation}
Note that in this derivation the quantities $w(z,\bar{z})$ and $\bar{w}(z,\bar{z})$ are $C^1$ complex functions related by complex conjugation. However, we can also make the stronger statement:
\begin{equation}\label{eq: Greens theorem2a}
\boxed{\oint_{\Gamma}\rmd z\, w(z,\bar{z})=i\int_D\rmd^2z\,\partial_{\bar{z}}w(z,\bar{z})}
\end{equation}
To see this we again make use of (\ref{eq:zzbar xy}) in (\ref{eq: Greens theorem2a}) and equate real and imaginary parts. This now yields {\it two} relations:
\begin{equation}
\begin{aligned}
&\oint_\Gamma (u\rmd x-v\rmd y)=-\int_D\rmd x\wedge \rmd y(\partial_yu+\partial_xv)\\
&\oint_\Gamma (v\rmd x+u\rmd y)=\int_D\rmd x\wedge \rmd y(\partial_xu-\partial_yv),
\end{aligned}
\end{equation}
either of which is precisely Green's theorem (\ref{eq: GreensTheorem_real}) in real coordinates, and so (\ref{eq: Greens theorem2a}) holds true independently of the complex conjugate appearing in (\ref{eq: Greens theorem}). Therefore, we can replace $\bar{w}(z,\bar{z})$ by a new complex function $\tilde{w}(z,\bar{z})$ (which is {\it not} necessarily the complex conjugate of $w(z,\bar{z})$) and hence obtain a more general version of Green's theorem (\ref{eq: Greens theorem}),\footnote{Of course, this is in agreement with the more general expression, $\int_{\partial D}\varphi=\int_{D}\rmd \varphi$, if we take $\varphi=w\rmd z+\tilde{w}\rmd\bar{z}$ and note that $\rmd = \rmd z\partial_z+\rmd\bar{z}\partial_{\bar{z}}$.}
\begin{equation}\label{eq: Greens theorem3}
\boxed{\oint_{\Gamma}\rmd z\,w+\rmd\bar{z}\tilde{w}=i\int_D\rmd^2z\,\big{(}\partial_{\bar{z}}w-\partial_z\tilde{w}\big{)}}
\end{equation}

\subsection{Generalised Cauchy Integral Formula}\label{sec:GCIF}
In this subsection we prove the `generalised Cauchy integral formula', see (\ref{eq:grt}). 
We first recall the `Cauchy integral theorem'. Consider Green's theorem (\ref{eq: Greens theorem2a}) with $w(z,\bar{z})$ replaced by a holomorphic function, $w(z)$. Green's theorem then reduces immediately to the {\it Cauchy integral theorem}:
\begin{equation}\label{eq:Cauchy}
\boxed{\oint_{\Gamma}\rmd z \,w(z)=0}
\end{equation}
We apply this to the closed contour $\Gamma+\Gamma'$ enclosing $D-D'$ shown in the second diagram in Fig.~\ref{fig:Greens} with integrand $w(z)\mapsto w(z)/(z-\zeta)$ (which is holomorphic in $D-D'$). After taking the radius of the disc $D'$ centred at $z=\zeta$ to zero and using holomorphicity of $w(z)$ to Taylor expand it, the Cauchy integral theorem (\ref{eq:Cauchy}) yields the {\it Cauchy integral formula}:
\begin{equation}\label{eq:holomresidueformula}
w(\zeta)=\frac{1}{2\pi i}\oint_{\Gamma}\rmd z\frac{w(z)}{z-\zeta}.
\end{equation}
We took the radius of $D'$ to zero but the result (\ref{eq:Cauchy}) is independent of this radius because $w(z)$ is holomorphic in $D$. 
\sk

In the present subsection we generalise the Cauchy integral formula (\ref{eq:holomresidueformula}) and show that any $C^1$ function $w(\zeta,\bar{\zeta})$ (not necessarily holomorphic) in a domain $D$ bounded by $\Gamma$ (as in\footnote{Note that the cross cuts in the second and third diagrams cancel out and hence give vanishing contribution.} the first diagram in Fig.~\ref{fig:Greens2}) 
can be given the integral representation:
\begin{equation}\label{eq:grt}
\boxed{w(\zeta,\bar{\zeta})=\frac{1}{2\pi i}\oint_\Gamma \,\rmd z\frac{w(z,\bar{z})}{z-\zeta}-\frac{1}{2\pi}\int_D\rmd^2z\frac{\partial_{\bar{z}}w(z,\bar{z})}{z-\zeta}
}
\end{equation}
We refer to this as the {\it generalised Cauchy integral formula}. Notice that the first term in (\ref{eq:grt}) is {\it analytic} in $\zeta$ and the second encodes the departure from analyticity. Let us also display the conjugate relation, obtained from (\ref{eq:grt}) by taking the complex conjugate and replacing $\overline{w(z,\bar{z})}$ by $w(z,\bar{z})$,
\begin{equation}\label{eq:grtcc}
w(\zeta,\bar{\zeta})=-\frac{1}{2\pi i}\oint_\Gamma \,\rmd\bar{z}\frac{w(z,\bar{z})}{\bar{z}-\bar{\zeta}}-\frac{1}{2\pi}\int_D\rmd^2z\frac{\partial_{z}w(z,\bar{z})}{\bar{z}-\bar{\zeta}}
\end{equation}

Following Bers \cite{BersRiemannSurfaces} or Chern \cite{Chern}, to prove (\ref{eq:grt}) we consider the version of Green's theorem given in (\ref{eq: Greens theorem2a}) applied to the domain $D-D'$ shown in Fig.~\ref{fig:Greens} and with the replacement:
$$
w(z,\bar{z})\rightarrow \frac{1}{2\pi i}\frac{w(z,\bar{z})}{z-\zeta}.
$$
We choose the inner disc $D'$ such that the point $z=\zeta$ is at the centre of this disc in Fig.~\ref{fig:Greens} and denote the radius of $D'$ by $r$. 
Then consider the following integral,
\begin{equation}\label{eq:D-D'int}
\begin{aligned}
\int\limits_{D-D'}\frac{\rmd^2z}{2\pi}\,\frac{\partial_{\bar{z}}w(z,\bar{z})}{z-\zeta}&=\int\limits_{D-D'}\frac{\rmd^2z}{2\pi}\,\partial_{\bar{z}}\Big(\frac{w(z,\bar{z})}{z-\zeta}\Big)\\
&=\frac{1}{2\pi i}\oint_{\partial D}\rmd z\frac{w(z,\bar{z})}{z-\zeta}-\frac{1}{2\pi i}\oint_{\partial D'}\rmd z\frac{w(z,\bar{z})}{z-\zeta}
\end{aligned}
\end{equation}
where in the first equality we took into account that the pole at $z=\zeta$ is absent from the region of integration allowing us to pull out the derivative $\partial_{\bar{z}}$. In the second equality we applied Green's theorem (\ref{eq: Greens theorem2a}) to the annulus $D-D'$. Let us now consider the second integral on the right-hand side of (\ref{eq:D-D'int}) in the limit $r\rightarrow 0$,
\begin{equation}\label{eq:D-D'int2}
\begin{aligned}
\lim_{r\rightarrow 0}\frac{1}{2\pi i}\oint_{\partial D'}\rmd z\frac{w(z,\bar{z})}{z-\zeta}&=\lim_{r\rightarrow 0}\frac{1}{2\pi i}\oint_{\partial D'}\rmd z\frac{w(z+\zeta,\bar{z}+\bar{\zeta})}{z}\\
&=\lim_{r\rightarrow 0}\frac{1}{2\pi i}\oint_{\partial D'}d(r e^{i\theta})\frac{w(re^{i\theta}+\zeta,re^{-i\theta}+\bar{\zeta})}{re^{i\theta}}\\
&=\lim_{r\rightarrow 0}\frac{1}{2\pi }\int_0^{2\pi}\rmd\theta\, w(re^{i\theta}+\zeta,re^{-i\theta}+\bar{\zeta})\\
&= w(\zeta,\bar{\zeta}),
\end{aligned}
\end{equation}
where in the first and second equality we changed variables of integration, and the last equality follows from the assumption of continuity of $w(z,\bar{z})$ in $D$. (Recall that every $C^1$ function is continuous, and $w(z,\bar{z})$ is by assumption at least $C^1$ in $D$.) 
Therefore, in the limit $r\rightarrow 0$ the relation (\ref{eq:D-D'int}) takes the form,
\begin{equation}\label{eq:D-D'int3}
\begin{aligned}
\lim_{r\rightarrow0}\frac{1}{2\pi}\int\limits_{D-D'}\rmd^2z\,\frac{\partial_{\bar{z}}w(z,\bar{z})}{z-\zeta}&=\frac{1}{2\pi i}\oint_{\partial D}\rmd z\frac{w(z,\bar{z})}{z-\zeta}-w(\zeta,\bar{\zeta}).
\end{aligned}
\end{equation}
Since the right-hand side is non-singular in this limit so must the left-hand side be  and we learn that (since $\partial D=\Gamma$):
\begin{equation}\label{eq:D-D'int3}
\begin{aligned}
\frac{1}{2\pi}\int_{D}\rmd^2z\frac{\partial_{\bar{z}}w(z,\bar{z})}{z-\zeta}&=\frac{1}{2\pi i}\oint_{\partial D}\rmd z\frac{w(z,\bar{z})}{z-\zeta}-w(\zeta,\bar{\zeta}).
\end{aligned}
\end{equation}
which is (\ref{eq:grt}), what we set out to show. That the left-hand side in (\ref{eq:D-D'int3}) is non-singular in the limit $r\rightarrow 0$ is in any case guaranteed since the pole of the integrand is suppressed by the integration measure. 

\subsection{Dolbeault-Grothendieck Lemma}\label{sec:DGlemma}
The Dolbeault-Grothendieck lemma states the following. Consider $\mathbf{C}^m$ with coordinates $z^k$, $1\leq k\leq m$, and let $D$ be the polydisc $|z^k|<r^k$ and $D'$ the smaller polydisc $|z^k|<{r'}^k$ such that ${r'}^k<r^k$. Let $\mu$ be a $(p,q)$-form (with $q\geq1$) in $D$ such that 
$
\bar{\partial}\mu=0.
$ 
Then there exists a $(p,q-1)$-form $v$ in $D$ such that 
$
\mu=\bar{\partial}v \quad \textrm{in $D'$}.
$ 
That is,
\begin{equation}
\boxed{\bar{\partial}\mu=0\quad \Leftrightarrow \quad \mu=\bar{\partial}v \quad \textrm{in $D'$}}
\end{equation}

We present the proof for the special case $q=m=1$, since this is the case of interest in this document (the general result then follows by induction \cite{BersRiemannSurfaces,Chern,GunningRossi,MorrowKodaira}). So we consider the case where $\mu$ is a (possibly vector-valued) $(0,1)$-form,
$$
\mu = f(z,\bar{z})\rmd\bar{z}.
$$
For a given $f(z,\bar{z})$ let us define the quantity $v(z,\bar{z})$ via,
\begin{equation}\label{eq:gzzbar-dfn}
v(z,\bar{z})\dfn \frac{1}{2\pi i}\int_{D'}\rmd\zeta\wedge \rmd\bar{\zeta}\,\frac{f(\zeta,\bar{\zeta})}{z-\zeta},
\end{equation}
and in what follows we will (for notational simplicity) write $f(z)$ and $v(z)$ for the same quantities (i.e.~we omit the anti-chiral coordinates from the notation). 
Consider now:
\begin{equation}
\begin{aligned}
\rmd \big(\rmd\bar{\zeta}f(\zeta)\ln|\zeta-z|^2\big)&=(\rmd\zeta\partial_\zeta+\rmd\bar{\zeta}\partial_{\bar{\zeta}})\big(\rmd\bar{\zeta}f(\zeta)\ln|\zeta-z|^2\big)\\
&=\rmd\zeta\wedge \rmd\bar{\zeta}\,\partial_\zeta f(\zeta)\ln|\zeta-z|^2+\rmd\zeta\wedge \rmd\bar{\zeta}f(\zeta)\partial_{\zeta}\ln|\zeta-z|^2,
\end{aligned}
\end{equation}
and introduce a disc $\Delta(z,\epsilon)\subset D'$ centred at $\zeta=z$ of radius $\epsilon$ and integrate both sides of this equation over $D-\Delta(z,\epsilon)$,
\begin{equation}\label{eq:D'Deltdinte1}
\begin{aligned}
&\int_{D'-\Delta(z,\epsilon)}\rmd \big(\rmd\bar{\zeta}f(\zeta)\ln|\zeta-z|^2\big)\\
&\quad=\int_{D'-\Delta(z,\epsilon)}\rmd\zeta\wedge \rmd\bar{\zeta}\,\partial_\zeta f(\zeta)\ln|\zeta-z|^2+\int_{D'-\Delta(z,\epsilon)}\rmd\zeta\wedge \rmd\bar{\zeta}\frac{f(\zeta)}{\zeta-z}\\
\end{aligned}
\end{equation}
For the left-hand side we can use Green's theorem (\ref{eq: Greens theorem2a}) (in fact its complex conjugate), according to which,
\begin{equation}\label{eq:D'Deltdinte}
\begin{aligned}
\int_{D'-\Delta(z,\epsilon)}&\rmd \big(\rmd\bar{\zeta}f(\zeta)\ln|\zeta-z|^2\big)\\
&=\int_{D'-\Delta(z,\epsilon)} \rmd\zeta\wedge \rmd\bar{\zeta}\partial_\zeta\big(f(\zeta)\ln|\zeta-z|^2\big)\\
&=\oint_{\partial D'}\rmd\bar{\zeta}f(\zeta)\ln|\zeta-z|^2-\underbrace{\oint_{\partial\Delta(z,\epsilon)}\rmd\bar{\zeta}f(\zeta)\ln|\zeta-z|^2}_{\dfn I_\epsilon}.
\end{aligned}
\end{equation}
Next focus on the second term on the right-hand side, call it $I_\epsilon$. We want to explore the limit $\epsilon\rightarrow 0$ and for this purpose assume that within $D'$ there exists a $\zeta$-independent quantity $B$ such that
\begin{equation}\label{eq:|f|<B}
|f(\zeta)|<B.
\end{equation}
Then,
\begin{equation}
\begin{aligned}
|I_\epsilon|&=\Big|\oint_{\partial\Delta(z,\epsilon)}\rmd\bar{\zeta}f(\zeta)\ln|\zeta-z|^2\Big|\\
&=\Big|-\int_0^{2\pi}\epsilon i\rmd\theta e^{-i\theta} f(z+\epsilon e^{i\theta})\ln \epsilon^2\Big|\\
&=2\epsilon \ln \epsilon\Big|\int_0^{2\pi}\rmd\theta\,e^{-i\theta} f(z+\epsilon e^{i\theta})\Big|\\
&\leq 2\epsilon \ln \epsilon\int_0^{2\pi}\rmd\theta \big|f(z+\epsilon e^{i\theta})\big|\\
&< 2\epsilon \ln \epsilon\int_0^{2\pi}\rmd\theta B\\
&= 4\pi B\epsilon \ln \epsilon
\end{aligned}
\end{equation}
That is, 
$$
\lim_{\epsilon\rightarrow 0}|I_\epsilon| < \lim_{\epsilon\rightarrow 0}4\pi B\epsilon \ln\epsilon=0,
$$ 
and (\ref{eq:D'Deltdinte}) then implies that:
\begin{equation}\label{eq:D'Deltdinte3}
\begin{aligned}
\lim_{\epsilon\rightarrow 0}\int_{D'-\Delta(z,\epsilon)}&\rmd \big(\rmd\bar{\zeta}f(\zeta)\ln|\zeta-z|^2\big)=\oint_{\partial D'}\rmd\bar{\zeta}f(\zeta)\ln|\zeta-z|^2.
\end{aligned}
\end{equation}
According to (\ref{eq:D'Deltdinte1}) we may rewrite the left-hand side of (\ref{eq:D'Deltdinte3}) to obtain
\begin{equation}\label{eq:D'Deltdinte4}
\begin{aligned}
\lim_{\epsilon\rightarrow 0}&\Bigg(\!\int\limits_{\,\,\,D'-\Delta(z,\epsilon)}\!\!\!\rmd\zeta\wedge \rmd\bar{\zeta}\,\partial_\zeta f(\zeta)\ln|\zeta-z|^2+\!\!\!\int\limits_{D'-\Delta(z,\epsilon)}\!\!\!\rmd\zeta\wedge \rmd\bar{\zeta}\frac{f(\zeta)}{\zeta-z}\Bigg)=\oint_{\partial D'}\rmd\bar{\zeta}f(\zeta)\ln|\zeta-z|^2.
\end{aligned}
\end{equation}
Now since the right-hand side has a finite limit as $\epsilon \rightarrow 0$ so must the left-hand side have a finite limit. Indeed, since the integrands have at most a simple pole any such singularity is suppressed by the measure, allowing us to take the limit:
\begin{equation}\label{eq:D'Deltdinte5}
\begin{aligned}
\int\limits_{D'}\rmd\zeta\wedge \rmd\bar{\zeta}\,\partial_\zeta f(\zeta)\ln|\zeta-z|^2+\int_{D'}\rmd\zeta\wedge \rmd\bar{\zeta}\frac{f(\zeta)}{\zeta-z}=\oint_{\partial D'}\rmd\bar{\zeta}f(\zeta)\ln|\zeta-z|^2,
\end{aligned}
\end{equation}
Now recall the definition of $v(z)$ in (\ref{eq:gzzbar-dfn}), on account of which (\ref{eq:D'Deltdinte5}) takes the form,
\begin{equation}\label{eq:D'Deltdinte6}
\begin{aligned}
2\pi iv(z)=-\oint_{\partial D'}\rmd\bar{\zeta}f(\zeta)\ln|\zeta-z|^2+\int\limits_{D'}\rmd\zeta\wedge \rmd\bar{\zeta}\,\partial_\zeta f(\zeta)\ln|\zeta-z|^2.
\end{aligned}
\end{equation}
We then take a $\partial_{\bar{z}}$ derivative of both sides, and we may pull the derivative into the integral (since the resulting integrals exist) with the result:
\begin{equation}\label{eq:D'Deltdinte7}
\begin{aligned}
2\pi i\partial_{\bar{z}}v(z)=-\oint_{\partial D'}\rmd\bar{\zeta}\frac{f(\zeta)}{\bar{z}-\bar{\zeta}}+\int\limits_{D'}\rmd\zeta\wedge \rmd\bar{\zeta}\,\frac{\partial_\zeta f(\zeta)}{\bar{z}-\bar{\zeta}}.
\end{aligned}
\end{equation}
According to (\ref{eq:grtcc}) however, we know that we can express any $C^1$ complex-valued $f(z)$ as:
\begin{equation}\label{eq:grtcc2}
2\pi if(z)=-\oint_{\partial D'}\rmd\bar{\zeta}\frac{f(\zeta)}{\bar{z}-\bar{\zeta}}+\int\limits_{D'}\rmd\zeta\wedge \rmd\bar{\zeta}\,\frac{\partial_\zeta f(\zeta)}{\bar{z}-\bar{\zeta}},
\end{equation}
and therefore from (\ref{eq:D'Deltdinte7}) and (\ref{eq:grtcc2}) there exists a function $v(z)$ for any given $f(z)$ satisfying the properties stated above such that
\begin{equation}
\boxed{f(z)=\partial_{\bar{z}}v(z)}
\end{equation}
which is what we set out to show. A special case of this result is when $\mu$ is a vector-valued $(0,1)$ form, that is, a {\it Beltrami differential}, 
\begin{equation}\label{eq:beltrami}
\mu=\mu_{\bar{z}}^{\phantom{z}z}\rmd\bar{z}\otimes \partial_z,
\end{equation}
in which case $v(z)=v^z(z)\partial_z$ and we have shown that $\mu_{\bar{z}}^{\phantom{a}z}$ is locally of the form:
\begin{equation}\label{eq:mu=dv}
\boxed{\mu_{\bar{z}}^{\phantom{z}z}=\partial_{\bar{z}}v^z}
\end{equation}

Unfortunately, this derivation excludes some interesting cases. For example, it is sometimes useful to consider the case where $\mu_{\bar{z}}^{\phantom{a}z}$ becomes singular along a contour or at isolated points. Such cases are excluded by (\ref{eq:|f|<B}). We will treat one such case separately below.

\subsection{Delta Function Singularities}
As pointed out at the end of the last subsection, the proof of the Dolbeault-Grothendieck lemma presented above excludes some interesting cases, where (the component of) a Beltrami differential $\mu_{\bar{z}}^{\phantom{a}z}$ becomes singular along a certain contour or at isolated points. We will treat one of these special cases here in detail, in order to expose underlying assumptions.
\sk

The case of interest is when:
$$
\boxed{\mu_{\bar{z}}^{\phantom{a}z}=2\pi\delta^2(z-\zeta)}
$$
for some fixed complex number $\zeta$. (There is something analogous that happens in the superstring for the usual choice of worldsheet gravitino gauge slice.) This corresponds to taking the worldsheet metric to become singular at some isolated point $z=\zeta$. It is a standard result that there indeed exists a function $v^z$ such that $\mu_{\bar{z}}^{\phantom{a}z}=\partial_{\bar{z}}v^z$, namely:
$$
\boxed{v^z = \frac{1}{z-\zeta}+\textrm{holomorphic}}
$$
but let us consider the derivation of this standard result in detail, if only to highlight the underlying assumptions.
\sk

The proof is again based on the generalised Cauchy integral formula (\ref{eq:grt}), repeated here for convenience,
\begin{equation}\label{eq:D-D'int8}
\begin{aligned}
w(\zeta,\bar{\zeta})&=\frac{1}{2\pi i}\oint_{\Gamma}\rmd z\frac{w(z,\bar{z})}{z-\zeta}-\frac{1}{2\pi}\int\limits_{D}\rmd^2z\,\frac{\partial_{\bar{z}}w(z,\bar{z})}{z-\zeta}.
\end{aligned}
\end{equation}
Since the relevant integrals appearing are convergent (and taking into account that the first term on the right-hand side of (\ref{eq:D-D'int8}) is analytic),
\begin{equation}\label{eq:dirac-first}
\begin{aligned}
\partial_{\bar{\zeta}}w(\zeta,\bar{\zeta})&=\frac{1}{2\pi}\partial_{\bar{\zeta}}\int_D\rmd^2z\frac{\partial_{\bar{z}}w(z,\bar{z})}{\zeta-z}\\
&=\frac{1}{2\pi}\int_D\rmd^2z\partial_{\bar{z}}w(z,\bar{z})\Big(\partial_{\bar{\zeta}}\frac{1}{\zeta-z}\Big).
\end{aligned}
\end{equation}
The term in the parenthesis is analytic everywhere except at $\zeta=z$, so the form of the left- and right-hand sides of (\ref{eq:dirac-first}) suggests the identification:
\begin{equation}\label{eq:dbar1/z=delta(z)}
\boxed{\partial_{\bar{\zeta}}\frac{1}{\zeta-z}=2\pi \delta^2(\zeta-z)}
\end{equation}
which is a standard and important result. Since we assumed that we can pull the $\partial_{\bar{\zeta}}$ derivative through the integral in going from the first to the second equality in (\ref{eq:dirac-first}) we will also provide a more careful derivation of (\ref{eq:dbar1/z=delta(z)}) next. 
\sk

To see why it is justified to pull the derivative through the integral consider again (\ref{eq:D-D'int8}),
\begin{equation}\label{eq:D-D'int6}
\begin{aligned}
w(\zeta,\bar{\zeta})-\frac{1}{2\pi i}\oint_{\Gamma}\rmd z\frac{w(z,\bar{z})}{z-\zeta}&=-\frac{1}{2\pi}\int\limits_{D}\rmd^2z\,\frac{\partial_{\bar{z}}w(z,\bar{z})}{z-\zeta}\\
&=-\frac{1}{2\pi}\int\limits_{D}\rmd^2z\,\partial_{\bar{z}}\Big(\frac{w(z,\bar{z})}{z-\zeta}\Big)+\frac{1}{2\pi}\int\limits_{D}\rmd^2z\,w(z,\bar{z})\Big(\partial_{\bar{z}}\frac{1}{z-\zeta}\Big)\\
\end{aligned}
\end{equation}
Since the first integral on the right-hand side is, by Green's theorem (\ref{eq: Greens theorem2a}),
$$
-\frac{1}{2\pi}\int\limits_{D}\rmd^2z\,\partial_{\bar{z}}\Big(\frac{w(z,\bar{z})}{z-\zeta}\Big)=-\frac{1}{2\pi i}\oint_{\Gamma}\rmd z\frac{w(z,\bar{z})}{z-\zeta},
$$
substituting this into (\ref{eq:D-D'int6}) we are left with:
\begin{equation}\label{eq:D-D'int7}
\begin{aligned}
w(\zeta,\bar{\zeta})&=\frac{1}{2\pi}\int\limits_{D}\rmd^2z\,w(z,\bar{z})\Big(\partial_{\bar{z}}\frac{1}{z-\zeta}\Big)\\
\end{aligned}
\end{equation}
We also have by definition of the Dirac delta function, $\delta^2(z-\zeta)$, that for $\zeta$ in $D$:
$$
w(\zeta,\bar{\zeta})=\int\limits_{D}\rmd^2z\,w(z,\bar{z})\delta^2(z-\zeta).
$$
Combining the preceding two equations yields:
\begin{equation}\label{eq:deltaderivation}
\int\limits_{D}\rmd^2z\,w(z,\bar{z})\Big(\partial_{\bar{z}}\frac{1}{z-\zeta}-2\pi\delta^2(z-\zeta)\Big)=0.
\end{equation}
Since the quantity $\partial_{\bar{z}}(\frac{1}{z-\zeta})$ vanishes everywhere except possibly at $z=\zeta$, and equation (\ref{eq:deltaderivation}) must hold for any $C^1$ function $w(z,\bar{z})$, we learn that:
$$
\partial_{\bar{z}}\frac{1}{z-\zeta}=2\pi\delta^2(z-\zeta),
$$
which is precisely (\ref{eq:dbar1/z=delta(z)}). Therefore, the assumption that we can pull the derivative $\partial_{\bar{\zeta}}$ through the integral that led to the first (quick) derivation of (\ref{eq:dbar1/z=delta(z)}) is fully justified. One can also argue that this is nevertheless justified since the resulting integrals exist, and this latter comment may be viewed as a consistency check.

\section{Complex Analysis on Curved Surfaces}\label{sec:CT}
Any two-dimensional Riemannian manifold is {\it locally} conformally flat, see e.g.~\cite{ChernIsothCoords55,BersRiemannSurfaces}. In this section we consider a local chart such that the corresponding conformally-flat metric takes the form $g_I=\rho(z,\bar{z})\rmd z\rmd\bar{z}$, with $\rho(z,\bar{z})\dfn 2g_{z\bar{z}}$. Notice that $g^{z\bar{z}}g_{z\bar{z}}=1$ and $\sqrt{g}=g_{z\bar{z}}$, and we use the following convention throughout: $\rmd^2z \equiv i\rmd z\wedge \rmd\bar{z}$. 

\subsection{Complex Tensors}\label{sec:CTa}
Consider a Riemann surface, $\Sigma$, and decompose the corresponding complexified cotangent bundle into chiral and anti-chiral halves, $T_{\mathbb{C}}^*\Sigma=T_L^*\Sigma\oplus T_R^*\Sigma$, so that the summands correspond to the spaces of $(1,0)$- and $(0,1)$-forms on $\Sigma$. 
We can use a corresponding basis, $\rmd z$, of $T_L^*\Sigma$ and $\rmd\bar{z}$ of $T_R^*\Sigma$ to define a complex tensor $\varphi$ of conformal weight $(h,\tilde{h})$ by:
\begin{equation}\label{eq: tensor V}
\varphi=\varphi_{z\dots \bar{z}\dots}(\rmd z)^{h}(\rmd\bar{z})^{\tilde{h}}\in K^{(h,\tilde{h})},
\end{equation}
where it is to be understood that the components, $\varphi_{z\dots \bar{z}\dots}$, carry $h$ lower $z$ indices and $\tilde{h}$ lower $\bar{z}$ indices. (This convention is really only useful for ordinary Riemann surfaces, but since this is the focus of the current document it will suffice.) The quantity $K^{(h,\tilde{h})}$ is the space of complex tensors of weight $(h,\tilde{h})$. 
It is convenient to define $K^{n}\dfn K^{(n,0)}$. The space $K^1$ (which is usually denoted by $K=T_L^*\Sigma$) is the {\it canonical bundle} whose local sections are spanned by (1,0)-forms. Using the metric $g_{z\bar{z}}$ to raise and lower indices there is an isomorphism $K^{n-m}\cong K^{(n,m)}$ and one may therefore express all tensors in terms of holomorphic indices only, e.g. we may write, $\varphi_{\phantom{a}\bar{z}z}^z=g_{z\bar{z}}\varphi^{zz}_{\phantom{aa}z}$ in which case the corresponding tensor $\varphi^{zz}_{\phantom{aa}z}\partial_z$ takes values in $K^{-1}$, and we could have equivalently written $\varphi^{zz}_{\phantom{aa}z}\partial_z$ for the same object.\footnote{(Incidentally, one advantage of using $\rmd z^{-1}$ over $\partial_z$ for basis elements is that it is convenient to use the same symbol $\rmd z^n$ for both $n>0$ and $n<0$. But this is not always convenient, e.g., commutators of tangent vectors are less transparent. So we use the notation $\partial_z$ and $\rmd z^{-1}$ interchangeably depending on context.)} We will denote tensors in $K^n$ by,
\begin{equation}\label{eq: tensor varphi}
\boxed{\varphi=\varphi_{zz\dots}\rmd z^{n}\in K^{n}}
\end{equation}
where $\varphi_{zz\dots}$ carries $n$ lower $z$ indices.
\sk

To tensors of weight $(1,1)$ there correspond top forms, $\varphi_{z\bar{z}}(z,\bar{z})\rmd z\wedge \rmd\bar{z}$, that can be integrated over $\Sigma$,
$$
\int_{\Sigma} \rmd^2z\,\varphi_{z\bar{z}}(z,\bar{z})
$$
since this object is invariant under local conformal transformations $z\mapsto w(z)$. It is hence intrinsically defined and can be integrated over the entire surface. This latter point follows from the fact that we can cover the entire Riemann surface with conformal coordinate charts such that on patch overlaps the transition functions are holomorphic, and then use a partition of unity \cite{GunningRossi} to glue together the various contributions into a globally well-defined integral. Using the metric to raise the anti-holomorphic index we can also rewrite the above integral,
$$
\int_{\Sigma} \rmd^2z\sqrt{g}\,\varphi_{z}^{\phantom{a}z}(z,\bar{z}),
$$
and indeed any tensor with an equal number of upper and lower $z$ indices (i.e.~any tensor in $K^0$ with components $\varphi_{z\dots}^{\phantom{a}z\dots}(z,\bar{z})$) is a scalar and can therefore be integrated with measure $\rmd^2z\sqrt{g}$ over the entire Riemann surface.

\subsection{Covariant Derivatives}\label{sec:CD}
To define covariant derivatives note that there is enough freedom to identify the Cauchy-Riemann operator $\partial_{\bar{z}}$ (when acting on tensors in $K^n$) with the covariant derivative $\nabla_{\bar{z}}^{(n)}$ without use of a connection; in particular, $\nabla_{\bar{z}}^{(n)}: K^{n}\rightarrow K^{n,1}$, so that:
\begin{equation}\label{eq: Cauchy-Riemann operator}
\rmd\bar{z}\nabla^{(n)}_{\bar{z}}\varphi = \rmd\bar{z}\partial_{\bar{z}}\varphi\equiv \bar{\partial}\varphi.
\end{equation}
According to the above convention (Appendix~\ref{sec:CTa}) for raising and lowering indices we could also have written the Cauchy-Riemann operator as $\nabla^z_{(n)}: K^{n}\rightarrow K^{n-1}$, with
\begin{equation}\label{eq: Cauchy-Riemann operator2}
\boxed{ (\rmd z)^{-1}\nabla^z_{(n)}\varphi = (\rmd z)^{-1}g^{z\bar{z}}\partial_{\bar{z}}\varphi}
\end{equation}
which transforms as a tensor of rank $n-1$.
\sk

Taking the complex conjugate of (\ref{eq: Cauchy-Riemann operator}), replacing $n\rightarrow -n$, and lowering (or raising if $n<0$) the resulting $n$ indices using the metric as discussed above, it follows that covariant derivatives are maps $\nabla^{(n)}_z: K^{n}\rightarrow K^{n+1}$, given by:
\begin{equation}\label{eq: nabla V}
\boxed{
\begin{aligned}
\rmd z\nabla^{(n)}_z\varphi 
&= \rmd z(\partial_z-n\Gamma_{zz}^z)\varphi\\
&=\rmd z(g_{z\bar{z}})^n\,\partial_z \,\big[(g_{z\bar{z}})^{-n}\,\varphi\big],
\end{aligned}
}
\end{equation}

Indeed, the choice (\ref{eq: Cauchy-Riemann operator}) together with a conformally-flat metric  $g_I=\rho(z,\bar{z})\rmd z\rmd\bar{z}$ defines the connection uniquely:
\begin{equation}\label{eq:Gammazzz}
\omega\dfn \rmd z\Gamma_{zz}^z=\partial \ln \rho(z,\bar{z}),\qquad \bar{\omega}\dfn \rmd\bar{z}\Gamma_{\bar{z}\bar{z}}^{\bar{z}}=\bar{\partial}\ln \rho(z,\bar{z}),
\end{equation}
where recall that $\rho=2g_{z\bar{z}}$. The remaining components of the connection not appearing explicitly in (\ref{eq:Gammazzz}) vanish. 
\sk

We usually drop the index $(n)$ from covariant derivatives when there is no ambiguity about the type of tensor it acts on. Also, we shall not in general display the differentials, $\rmd z$, $\rmd\bar{z}$, but include them in the definitions for concreteness.

\subsection{Laplacians}
We can also construct two, in general distinct, Laplacians using the differential operators (\ref{eq: nabla V}) and (\ref{eq: Cauchy-Riemann operator}),
\begin{equation}\label{eq: Laplacians}
\begin{aligned}
\Delta_{(n)}^{+}&=-2\nabla_{(n+1)}^{z}\nabla_z^{(n)}\\
\Delta_{(n)}^{-}&=-2\nabla^{(n-1)}_{z}\nabla^z_{(n)}.
\end{aligned}
\end{equation}
Explicit calculation yields:
\begin{equation}\label{eq:Dpm_n}
\begin{aligned}
\Delta_{(n)}^{+}&=-2g^{z\bar{z}}\partial_z\partial_{\bar{z}}-n\big(\!-2g^{z\bar{z}}\partial_z\partial_{\bar{z}}\ln g_{z\bar{z}}\big)+2n\Gamma^z_{zz}g^{z\bar{z}}\partial_{\bar{z}}\\
\Delta_{(n)}^{-}&=-2g^{z\bar{z}}\partial_z\partial_{\bar{z}}+2n\Gamma^z_{zz}g^{z\bar{z}}\partial_{\bar{z}}
\end{aligned}
\end{equation}
These two Laplacians are equal when acting on scalars, in which case $n=0$, so we define:
$$
\Delta_{(0)}\equiv\Delta_{(0)}^{+}=\Delta_{(0)}^{-}=-2g^{z\bar{z}}\partial_z\partial_{\bar{z}}.
$$
The factor of $-2$ in the definitions (\ref{eq: Laplacians}) is conventional and is included so as to agree with the definition of the conventional Laplacian $\Delta_{(0)}=-\frac{1}{\sqrt{g}}\partial_{\alpha}(\sqrt{g}g^{\alpha\beta}\partial_{\beta})$, which according to (\ref{eq:Dpm_n}) indeed reduces to $-2g^{z\bar{z}}\partial_z\partial_{\bar{z}}$ in the conformal metric $g_I=\rho(z,\bar{z})\rmd z\rmd\bar{z}$. Finally, as discussed below, the quantity $-2g^{z\bar{z}}\partial_z\partial_{\bar{z}}\ln g_{z\bar{z}}$ appearing in (\ref{eq:Dpm_n}) is the Ricci scalar, $R_{(2)}$. 

\subsection{Curvature}\label{sec:C}
On two-dimensional manifolds the Riemannian curvature has only one independent component and so it will be completely specified by the Ricci scalar, $R_{(2)}$. We define the latter by:
\begin{equation}\label{eq:Riccidfn}
\begin{aligned}
&[\nabla^z,\nabla_z]_{(n)}\varphi\equiv \frac{n}{2}R_{(2)}\varphi,\\
\end{aligned}
\end{equation} 
for $\varphi\in K^n$, and the commutator, when acting on elements in $K^n$, is defined by,
$$
[\nabla^z,\nabla_z]_{(n)}\dfn \nabla^z_{(n+1)}\nabla_z^{(n)}-\nabla_z^{(n-1)}\nabla^z_{(n)}.
$$
Explicit calculation using the definitions in Sec.~\ref{sec:CD} yields $R_{(2)}=-2g^{z\bar{z}}\partial_z\partial_{\bar{z}}\ln g_{z\bar{z}}$, and when $\rho=2g_{z\bar{z}}$,
\begin{equation}\label{eq:RicciScalar-z}
\boxed{ R_{(2)}=-4\rho^{-1}\partial_z\partial_{\bar{z}}\ln\rho}
\end{equation}

From (\ref{eq:Dpm_n}) it follows that we can also express the Ricci scalar in terms of the Laplacians defined in (\ref{eq: Laplacians}):
\begin{equation}\label{eq:Laplacian pm}
\Delta_{(n)}^{+}-\Delta_{(n)}^{-}=-nR_{(2)}.
\end{equation}

Our conventions are such that there is a close parallel with the notion of curvature in real manifolds. In particular, in the standard ($+\,+\,+$) conventions in the classification of Misner, Thorne and Wheeler \cite{MisnerThorneWheeler74}:
\begin{equation}\label{eq:R}
\begin{aligned}
R_{(2)} &= g^{\alpha\beta}R_{\alpha\beta}
=2g^{z\bar{z}}R_{z\bar{z}}
=2g^{z\bar{z}}R^{\alpha}_{\phantom{a}z\alpha \bar{z}}=2g^{z\bar{z}}R^{z}_{\phantom{a}zz \bar{z}},
\end{aligned}
\end{equation}
but we are in Euclidean space. 
In particular, in the $(+\,+\,+)$ conventions the components of Riemann curvature tensor in terms of the Christoffel symbol read,
$$
R^{\alpha}_{\phantom{a}\beta\gamma\delta} = \partial_{\gamma}\Gamma^{\alpha}_{\beta\delta}-\partial_{\delta}\Gamma^{\alpha}_{\beta\gamma}+\Gamma^{\alpha}_{\sigma\gamma}\Gamma^{\sigma}_{\beta\delta}-\Gamma^{\alpha}_{\sigma\delta}\Gamma^{\sigma}_{\beta\gamma},
$$
and taking into account that the only non-vanishing Christofel symbols are those depicted in (\ref{eq:Gammazzz}),
\begin{equation}\label{eq:Rzzzzbar}
\begin{aligned}
R^{z}_{\phantom{a}zz \bar{z}}& = -\partial_{\bar{z}}\Gamma^z_{zz}\\
&=-\partial_{\bar{z}}\partial_z\ln g_{z\bar{z}},
\end{aligned}
\end{equation}
the other non-vanishing components of the Riemann curvature tensor being related to $R^{z}_{\phantom{a}zz \bar{z}}$ by the symmetries of the indices and complex conjugation. According to (\ref{eq:R}) the component $R^{z}_{\phantom{a}zz \bar{z}}$ is also identified with the corresponding Ricci tensor component,
\begin{equation}\label{eq:RicciCurvature}
\boxed{R_{z\bar{z}}=-\partial_z\partial_{\bar{z}}\ln g_{z\bar{z}}}
\end{equation}
so that indeed all information about the curvature of a Riemann surface can be derived from the Ricci scalar $R_{(2)}$ and the metric component $g_{z\bar{z}}$, and there is a clear parallel between complex and real manifolds in two dimensions. 
\sk

The curvature tensor, $\mathcal{R}$, is defined in terms of the connection one-form $\omega$ in (\ref{eq:Gammazzz}) for a rank-1 complex vector bundle $E=K^{-1}$ over a local patch $U$:
\begin{equation}
\begin{aligned}
\mathcal{R} &\dfn \rmd\omega-\omega\wedge\omega \\
&=(\partial+\bar{\partial})\wedge\partial \ln g_{z\bar{z}}-(\partial \ln g_{z\bar{z}})\wedge (\partial \ln g_{z\bar{z}})\\
&=\rmd z\wedge \rmd\bar{z}(-\partial_{\bar{z}}\partial_z \ln g_{z\bar{z}}).
\end{aligned}
\end{equation}

As an example let us consider a sphere $\Sigma=S^2$ and search for a $g_{z\bar{z}}$ such that,
\begin{equation}\label{eq:R2=k}
R_{(2)}=+k,\quad\textrm{with $k>0$ a constant}
\end{equation}
Setting (\ref{eq:R}) equal to (\ref{eq:R2=k}) with $R_{z\bar{z}}$ given by (\ref{eq:RicciCurvature}) yields a differential equation for $g_{z\bar{z}}$,
$$
k=2g^{z\bar{z}}(-\partial_{\bar{z}}\partial_z \ln g_{z\bar{z}}),
$$
whose solution reads,
\begin{equation}\label{eq:g_zzbar}
g_{z\bar{z}} = \frac{4}{k}\frac{1}{(1+z\bar{z})^2}\quad \Leftrightarrow \quad R_{(2)}=+k.
\end{equation}
This metric is normalised such that it is equivalent to that of the standard metric on the sphere, 
$$
\rmd s^2 = r^2(\rmd\theta^2 + \sin^2\theta \rmd\varphi^2),
$$
which is seen by defining $z = e^{i\varphi}\tan \frac{\theta}{2}$, and taking $k=2/r^2$. The reparametrisation-invariant area of the worldsheet reads 
$$
{\rm Area}=\int_{S^2} \rmd^2z\sqrt{g}=8\pi/k=4\pi r^2,
$$ 
where the relevant integral may be computed, e.g., by exponentiating the denominator of the integrand using the gamma function integral representation, $\frac{1}{A^a}= \frac{1}{\Gamma(a)}\int_0^{\infty}\frac{\rmd y}{y}y^ae^{-Ay}$, with $A=1+z\bar{z}$ and $a=2$ (recalling that $\rmd^2z=i\rmd z\wedge \rmd\bar{z}$). 

\subsection{Inner Product and Adjoint}
The natural inner product between tensors $\varphi_{1,2}\in K^n$ with respect to the metric $g$ is
\begin{equation}\label{eq: (V1,V2)}
\boxed{(\varphi_1,\varphi_2)_g = \int_{\Sigma}\rmd^2z\sqrt{g}\,(g^{z\bar{z}})^n\,\varphi_1^*\,\varphi_2}
\end{equation}
and we define the adjoint operators $\nabla^{(n)\dagger}_z$ and $\nabla^{z\dagger}_{(n)}$ with respect to this, 
$$
\boxed{(\varphi_1,\nabla^{(n)\dagger}_z\varphi_2)_g\equiv (\nabla^{(n)}_z\varphi_1, \varphi_2)_g}
$$

We use the inner product (\ref{eq: (V1,V2)}) to define a norm: $\| \varphi\|_g^2=(\varphi,\varphi)_g$. Using the definitions it follows that
\begin{equation}
\nabla^{(n)\dagger}_z=-\nabla_{(n+1)}^z,\qquad \nabla_{(n)}^{z\dagger}=-\nabla^{(n-1)}_z.
\end{equation}

\bibliographystyle{JHEP}
\bibliography{spi-o}

\providecommand{\href}[2]{#2}\begingroup\raggedright\begin{thebibliography}{100}

\bibitem{DonagiWitten15}
R.~Donagi and E.~Witten, {\it {Supermoduli Space Is Not Projected}}, {{\it
  Proc. Symp. Pure Math.} {\bfseries 90} (2015) 19}
  [\href{https://arxiv.org/abs/1304.7798}{{\ttfamily 1304.7798}}].

\bibitem{Maldacena98}
J.~M. Maldacena, {\it {The large N limit of superconformal field theories and
  supergravity}}, {{\it Adv. Theor. Math. Phys.} {\bfseries 2} (1998) 231}.

\bibitem{RyuTakayanagi06}
S.~Ryu and T.~Takayanagi, {\it {Holographic derivation of entanglement entropy
  from AdS/CFT}}, \href{https://doi.org/10.1103/PhysRevLett.96.181602}{{\it
  Phys. Rev. Lett.} {\bfseries 96} (2006) 181602}
  [\href{https://arxiv.org/abs/hep-th/0603001}{{\ttfamily hep-th/0603001}}].

\bibitem{VanRaamsdonk10}
M.~Van~Raamsdonk, {\it {Building up spacetime with quantum entanglement}},
  \href{https://doi.org/10.1007/s10714-010-1034-0,
  10.1142/S0218271810018529}{{\it Gen. Rel. Grav.} {\bfseries 42} (2010) 2323}
  [\href{https://arxiv.org/abs/1005.3035}{{\ttfamily 1005.3035}}].

\bibitem{MaldacenaSusskind13}
J.~Maldacena and L.~Susskind, {\it {Cool horizons for entangled black holes}},
  \href{https://doi.org/10.1002/prop.201300020}{{\it Fortsch. Phys.} {\bfseries
  61} (2013) 781} [\href{https://arxiv.org/abs/1306.0533}{{\ttfamily
  1306.0533}}].

\bibitem{MaldacenaQi18}
J.~Maldacena and X.-L. Qi, {\it {Eternal traversable wormhole}},
  \href{https://arxiv.org/abs/1804.00491}{{\ttfamily 1804.00491}}.

\bibitem{SaadShenkerStanford18}
P.~Saad, S.~H. Shenker and D.~Stanford, {\it {A semiclassical ramp in SYK and
  in gravity}},  \href{https://arxiv.org/abs/1806.06840}{{\ttfamily
  1806.06840}}.

\bibitem{StanfordWitten19}
D.~Stanford and E.~Witten, {\it {JT Gravity and the Ensembles of Random Matrix
  Theory}},  \href{https://arxiv.org/abs/1907.03363}{{\ttfamily 1907.03363}}.

\bibitem{OoguriSakai87}
H.~Ooguri and N.~Sakai, {\it {String Loop Corrections From Fusion of Handles
  and Vertex Operators}},
  \href{https://doi.org/10.1016/0370-2693(87)90351-0}{{\it Phys. Lett.}
  {\bfseries B197} (1987) 109}.

\bibitem{Das88}
S.~R. Das, {\it {Renormalizing Handles and Holes in String Theory}},
  \href{https://doi.org/10.1103/PhysRevD.38.3105}{{\it Phys. Rev.} {\bfseries
  D38} (1988) 3105}.

\bibitem{Polchinski88}
J.~Polchinski, {\it {Factorization of Bosonic String Amplitudes}},
  \href{https://doi.org/10.1016/0550-3213(88)90522-6}{{\it Nucl. Phys.}
  {\bfseries B307} (1988) 61}.

\bibitem{Tseytlin90b}
A.~A. Tseytlin, {\it {Renormalization group and string loops}},
  \href{https://doi.org/10.1142/S0217751X90000301}{{\it Int. J. Mod. Phys.}
  {\bfseries A5} (1990) 589}.

\bibitem{Tseytlin90}
A.~A. Tseytlin, {\it {On 'Macroscopic String' Approximation in String Theory}},
  \href{https://doi.org/10.1016/0370-2693(90)90792-5}{{\it Phys.Lett.}
  {\bfseries B251} (1990) 530}.

\bibitem{Sen15b}
A.~Sen, {\it {Off-shell Amplitudes in Superstring Theory}},
  \href{https://doi.org/10.1002/prop.201500002}{{\it Fortsch. Phys.} {\bfseries
  63} (2015) 149} [\href{https://arxiv.org/abs/1408.0571}{{\ttfamily
  1408.0571}}].

\bibitem{Skliros20}
D.~P. Skliros, {\it {A Globally-Defined Slice for Supermoduli Space (to
  appear)}}, .

\bibitem{AtickMooreSen88}
J.~J. Atick, G.~W. Moore and A.~Sen, {\it {Some Global Issues in String
  Perturbation Theory}},
  \href{https://doi.org/10.1016/0550-3213(88)90043-0}{{\it Nucl. Phys. B}
  {\bfseries 308} (1988) 1}.

\bibitem{Witten12b}
E.~Witten, {\it {Notes On Super Riemann Surfaces And Their Moduli}},
  \href{https://arxiv.org/abs/1209.2459v5}{{\ttfamily 1209.2459v5}}.

\bibitem{Witten12c}
E.~Witten, {\it {Superstring Perturbation Theory Revisited}},
  \href{https://arxiv.org/abs/1209.5461v3}{{\ttfamily 1209.5461v3}}.

\bibitem{SenWitten15}
A.~Sen and E.~Witten, {\it {Filling the gaps with PCO's}},
  \href{https://doi.org/10.1007/JHEP09(2015)004}{{\it JHEP} {\bfseries 09}
  (2015) 004} [\href{https://arxiv.org/abs/1504.00609}{{\ttfamily
  1504.00609}}].

\bibitem{Polchinski_v1}
J.~Polchinski, {\it String Theory. Vol. 1: An Introduction to the Bosonic
  String}. Cambridge Univ. Pr., UK, 1998.

\bibitem{Nelson89}
P.~C. Nelson, {\it {Covariant Insertion of General Vertex Operators}},
  \href{https://doi.org/10.1103/PhysRevLett.62.993}{{\it Phys. Rev. Lett.}
  {\bfseries 62} (1989) 993}.

\bibitem{Polchinski87}
J.~Polchinski, {\it {Vertex Operators in the Polyakov Path Integral}},
  \href{https://doi.org/10.1016/0550-3213(87)90389-0}{{\it Nucl. Phys.}
  {\bfseries B289} (1987) 465}.

\bibitem{deLacroixErbinKashyapSenVerma17}
C.~de~Lacroix, H.~Erbin, S.~P. Kashyap, A.~Sen and M.~Verma, {\it {Closed
  Superstring Field Theory and its Applications}},
  \href{https://arxiv.org/abs/1703.06410}{{\ttfamily 1703.06410}}.

\bibitem{MoosavianZhou19}
S.~F. Moosavian and Y.~Zhou, {\it {On the Existence of Heterotic-String and
  Type-II-Superstring Field Theory Vertices}},
  \href{https://arxiv.org/abs/1911.04343}{{\ttfamily 1911.04343}}.

\bibitem{PiusRudraSen14c}
R.~Pius, A.~Rudra and A.~Sen, {\it {Mass Renormalization in String Theory:
  Special States}}, \href{https://doi.org/10.1007/JHEP07(2014)058}{{\it JHEP}
  {\bfseries 07} (2014) 058} [\href{https://arxiv.org/abs/1311.1257}{{\ttfamily
  1311.1257}}].

\bibitem{PiusRudraSen14b}
R.~Pius, A.~Rudra and A.~Sen, {\it {Mass Renormalization in String Theory:
  General States}}, \href{https://doi.org/10.1007/JHEP07(2014)062}{{\it JHEP}
  {\bfseries 07} (2014) 062} [\href{https://arxiv.org/abs/1401.7014}{{\ttfamily
  1401.7014}}].

\bibitem{LaNelson90}
H.-S. La and P.~C. Nelson, {\it {Effective Field Equations for Fermionic
  Strings}}, \href{https://doi.org/10.1016/0550-3213(90)90031-8}{{\it Nucl.
  Phys.} {\bfseries B332} (1990) 83}.

\bibitem{Ahlfors66}
L.~Ahlfors, {\it COMPLEX ANALYSIS}, University Lecture Series.

\bibitem{BersRiemannSurfaces}
L.~Bers, {\it {Riemann Surfaces, 1957-1958}}. Courant Institute of Mathematical
  Sciences, New York University, US, 1958.

\bibitem{Bers81}
L.~Bers, {\it {Finite dimensional Teichm\"uller spaces and generalizations}},
  \href{https://doi.org/10.1090/S0273-0979-1981-14933-8}{{\it Bull. Amer. Math.
  Soc.} {\bfseries 5} (1981) 131}.

\bibitem{Chern}
S.-S. Chern, {\it {Complex Manifolds without Potential Theory}}.
  \href{https://doi.org/10.1007/978-1-4684-9344-3}{10.1007/978-1-4684-9344-3}.

\bibitem{GunningRossi}
R.~Gunning and H.~Rossi, {\it Analytic Functions of Several Complex Variables},
  Ams Chelsea Publishing.

\bibitem{MorrowKodaira}
J.~Morrow and K.~Kodaira, {\it Complex Manifolds}, AMS Chelsea Publishing
  Series.

\bibitem{Tu11}
L.~Tu, {\it An Introduction to Manifolds}, Universitext. Springer New York,
  second~ed., 2011.

\bibitem{FrolicherNijenhuisI}
A.~Frolicher and A.~Nijenhuis, {\it Theory of vector-valued differential forms.
  i: Derivations in the graded ring of differential forms}, {{\it Nederlandse
  Akademie van Wetenschappen. Proceedings. Series A. Indagationes Mathematicae}
  (1956) }.

\bibitem{FrolicherNijenhuis}
A.~Frolicher and A.~Nijenhuis, {\it Invariance of vector form operations under
  mappings}, \href{https://doi.org/10.1007/BF02565938}{{\it Commentarii
  Mathematici Helvetici} {\bfseries 34} (1960) 227}.

\bibitem{KodairaNirenbergSpencer}
K.~Kodaira, L.~Nirenberg and D.~Spencer, {\it {On the Existence of Deformations
  of Complex Analytic Structures}}, {{\it Ann. Math.} {\bfseries 68} (1958)
  450?459}.

\bibitem{D'HokerPhong15b}
E.~D'Hoker and D.~H. Phong, {\it {Higher Order Deformations of Complex
  Structures}}, \href{https://doi.org/10.3842/SIGMA.2015.047}{{\it SIGMA}
  {\bfseries 11} (2015) 047}
  [\href{https://arxiv.org/abs/1502.03673}{{\ttfamily 1502.03673}}].

\bibitem{Tu17}
L.~Tu, {\it Differential Geometry: Connections, Curvature, and Characteristic
  Classes}, Graduate Texts in Mathematics.

\bibitem{KobayashiNomizu-II}
S.~Kobayashi and K.~Nomizu, {\it Foundations of Differential Geometry},
  vol.~II.

\bibitem{DHokerPhong}
E.~D'Hoker and D.~H. Phong, {\it {The Geometry of String Perturbation Theory}},
  \href{https://doi.org/10.1103/RevModPhys.60.917}{{\it Rev. Mod. Phys.}
  {\bfseries 60} (1988) 917}.

\bibitem{Alhfors}
L.~Ahlfors, {\it Lectures on Quasiconformal Mappings, $2^{\rm nd}$ ed.},
  vol.~38 of {\it University Lecture Series}.

\bibitem{Abikoff}
W.~Abikoff, {\it {The Real Analytic Theory of Teichm\"uller Space}}.
  Springer-Verlag, Berlin Heidelberg, 1980.

\bibitem{ChernIsothCoords55}
S.-S. Chern, {\it {An Elementary Proof of the Existence of Isothermal
  Parameters on a Surface}}, \href{https://doi.org/doi:10.2307/2032933}{{\it
  Proceedings of the American Mathematical Society} {\bfseries 6} (1955) 771}.

\bibitem{Milnor}
J.~Milnor and D.~Weaver, {\it Topology from the Differentiable Viewpoint},
  Princeton Landmarks in Mathematics.

\bibitem{AhlforsSario}
L.~Ahlfors and L.~Sario, {\it Riemann Surfaces}. Princeton University Press,
  1960.

\bibitem{Kodaira}
K.~Kodaira, {\it Complex Manifolds and Deformation of Complex Structures}.
  Springer-Verlag, New York, 1986.

\bibitem{VerlindeHphd}
H.~L. Verlinde, {\it {Path-Integral Formulation of Supersymmetric String
  Theory}}, Ph.D. thesis, Rijksuniversiteit Utrecht (Netherlands), 1988.

\bibitem{Friedan82}
D.~Friedan, {\it {Introduction to Polyakov's String Theory}},  in {\it {Les
  Houches Summer School in Theoretical Physics: Recent Advances in Field Theory
  and Statistical Mechanics Les Houches, France, August 2-September 10, 1982}},
  1982.

\bibitem{Huybrechts}
D.~Huybrechts, {\it Complex Geometry}. Springer-Verlag, Germany, 2005.

\bibitem{FrolicherNijenhuis57}
A.~Frolicher and A.~Nijenhuis, {\it A theorem on stability of complex
  structures}, {{\it Proc.~Natl.~Acad.~Sci.~USA} {\bfseries 43(2)} (1957) 239}.

\bibitem{BottTu}
R.~Bott and L.~Tu, {\it Differential Forms in Algebraic Topology}, Graduate
  Texts in Mathematics.

\bibitem{GiddingsMartinec86}
S.~B. Giddings and E.~J. Martinec, {\it {Conformal Geometry and String Field
  Theory}}, \href{https://doi.org/10.1016/0550-3213(86)90108-2}{{\it Nucl.
  Phys.} {\bfseries B278} (1986) 91}.

\bibitem{D'HokerGiddings87}
E.~D'Hoker and S.~B. Giddings, {\it {Unitarity of the Closed Bosonic Polyakov
  String}}, \href{https://doi.org/10.1016/0550-3213(87)90466-4}{{\it Nucl.
  Phys.} {\bfseries B291} (1987) 90}.

\bibitem{Eisenhardt}
L.~P. Eisenhardt, {\it Riemannian Geometry}. Princeton University Press, 1997.

\bibitem{Friedan80}
D.~H. Friedan, {\it {Nonlinear Models in Two + Epsilon Dimensions}},
  \href{https://doi.org/10.1016/0003-4916(85)90384-7}{{\it Annals Phys.}
  {\bfseries 163} (1985) 318}.

\bibitem{Alvarez-GaumeFreedmanMukhi81}
L.~Alvarez-Gaume, D.~Z. Freedman and S.~Mukhi, {\it {The Background Field
  Method and the Ultraviolet Structure of the Supersymmetric Nonlinear Sigma
  Model}}, \href{https://doi.org/10.1016/0003-4916(81)90006-3}{{\it Annals
  Phys.} {\bfseries 134} (1981) 85}.

\bibitem{FradkinTseytlin85}
E.~Fradkin and A.~A. Tseytlin, {\it {Quantum String Theory Effective Action}},
  \href{https://doi.org/10.1016/0550-3213(85)90559-0}{{\it Nucl.Phys.}
  {\bfseries B261} (1985) 1}.

\bibitem{Mukhi86}
S.~Mukhi, {\it {The Geometric Background Field Method, Renormalization and the
  Wess-Zumino Term in Nonlinear Sigma Models}},
  \href{https://doi.org/10.1016/0550-3213(86)90502-X}{{\it Nucl.Phys.}
  {\bfseries B264} (1986) 640}.

\bibitem{CallanGan86}
J.~Callan, Curtis~G. and Z.~Gan, {\it {Vertex Operators in Background Fields}},
  \href{https://doi.org/10.1016/0550-3213(86)90238-5}{{\it Nucl. Phys.}
  {\bfseries B272} (1986) 647}.

\bibitem{Tseytlin87}
A.~A. Tseytlin, {\it {Sigma Model Weyl-Invariance Conditions and String
  Equations of Motion}},
  \href{https://doi.org/10.1016/0550-3213(87)90588-8}{{\it Nucl. Phys.}
  {\bfseries B294} (1987) 383}.

\bibitem{HowePapadopoulosStelle88}
P.~S. Howe, G.~Papadopoulos and K.~S. Stelle, {\it {The Background Field Method
  and the Nonlinear $\sigma$ Model}},
  \href{https://doi.org/10.1016/0550-3213(88)90379-3}{{\it Nucl. Phys.}
  {\bfseries B296} (1988) 26}.

\bibitem{Osborn90}
H.~Osborn, {\it {General Bosonic $\sigma$ Models and String Effective
  Actions}}, \href{https://doi.org/10.1016/0003-4916(90)90241-F}{{\it Annals
  Phys.} {\bfseries 200} (1990) 1}.

\bibitem{Riordan58}
J.~Riordan, {\it {An Introduction to Combinatorial Analysis}}, .

\bibitem{Andrews}
G.~E. Andrews, {\it {The Theory of Partitions}}. Cambridge University Press,
  UK, 1984.

\bibitem{Chern67}
S.-S. Chern, {\it {Curves and Surfaces in Euclidean Space}}, vol.~4. 1967.

\bibitem{WuYang75}
T.~T. Wu and C.~N. Yang, {\it {Concept of Nonintegrable Phase Factors and
  Global Formulation of Gauge Fields}},
  \href{https://doi.org/10.1103/PhysRevD.12.3845}{{\it Phys. Rev.} {\bfseries
  D12} (1975) 3845}.

\bibitem{ChernChenLam}
S.~Chern, W.~Chen and K.~Lam, {\it Lectures on Differential Geometry}, Series
  on University Mathematics.

\bibitem{FriedanShenker87}
D.~Friedan and S.~H. Shenker, {\it {The Analytic Geometry of Two-Dimensional
  Conformal Field Theory}},
  \href{https://doi.org/10.1016/0550-3213(87)90418-4}{{\it Nucl. Phys.}
  {\bfseries B281} (1987) 509}.

\bibitem{Martinec86b}
E.~J. Martinec, {\it {Conformal Field Theory on a (Super)Riemann Surface}},
  \href{https://doi.org/10.1016/0550-3213(87)90252-5}{{\it Nucl. Phys.}
  {\bfseries B281} (1987) 157}.

\bibitem{Mansfield87}
P.~Mansfield, {\it {Nilpotent {BRST} Invariance of the Interacting Polyakov
  String}}, \href{https://doi.org/10.1016/0550-3213(87)90286-0}{{\it Nucl.
  Phys.} {\bfseries B283} (1987) 551}.

\bibitem{MooreNelson86}
G.~W. Moore and P.~C. Nelson, {\it {Absence of Nonlocal Anomalies in the
  Polyakov String}}, \href{https://doi.org/10.1016/0550-3213(86)90177-X}{{\it
  Nucl. Phys.} {\bfseries B266} (1986) 58}.

\bibitem{DHokerPhong86}
E.~D'Hoker and D.~H. Phong, {\it {Multiloop Amplitudes for the Bosonic Polyakov
  String}}, \href{https://doi.org/10.1016/0550-3213(86)90372-X}{{\it Nucl.
  Phys.} {\bfseries B269} (1986) 205}.

\bibitem{Alvarez83}
O.~Alvarez, {\it {Theory of Strings with Boundaries: Fluctuations, Topology,
  and Quantum Geometry}},
  \href{https://doi.org/10.1016/0550-3213(83)90490-X}{{\it Nucl. Phys.}
  {\bfseries B216} (1983) 125}.

\bibitem{Knizhnik86}
V.~G. Knizhnik, {\it {Analytic Fields on Riemann Surfaces}},
  \href{https://doi.org/10.1016/0370-2693(86)90304-7}{{\it Phys. Lett.}
  {\bfseries B180} (1986) 247}.

\bibitem{Knizhnik89}
V.~G. Knizhnik, {\it {Multiloop amplitudes in the theory of quantum strings and
  complex geometry}},
  \href{https://doi.org/10.1070/PU1989v032n11ABEH002775}{{\it Sov. Phys. Usp.}
  {\bfseries 32} (1989) 945}.

\bibitem{ErbinMaldacenaSkliros19}
H.~Erbin, J.~Maldacena and D.~Skliros, {\it {Two-Point String Amplitudes}},
  \href{https://doi.org/10.1007/JHEP07(2019)139}{{\it J. High Energ. Phys.}
  {\bfseries 07} (2019) 139}
  [\href{https://arxiv.org/abs/1906.06051}{{\ttfamily 1906.06051}}].

\bibitem{SekiTakahashi19}
S.~Seki and T.~Takahashi, {\it {Two-Point String Amplitudes Revisited by
  Operator Formalism}},  \href{https://arxiv.org/abs/1909.03672}{{\ttfamily
  1909.03672}}.

\bibitem{Weinberg85}
S.~Weinberg, {\it {Coupling Constants and Vertex Functions in String
  Theories}}, \href{https://doi.org/10.1016/0370-2693(85)91615-6}{{\it Phys.
  Lett.} {\bfseries B156} (1985) 309}.

\bibitem{Seiberg87}
N.~Seiberg, {\it {Anomalous Dimension and Mass Renormalization in String
  Theory}}, \href{https://doi.org/10.1016/0370-2693(87)90071-2}{{\it Phys.
  Lett.} {\bfseries B187} (1987) 56}.

\bibitem{Sen88}
A.~Sen, {\it {Mass Renormalization and {BRST} Anomaly in String Theories}},
  \href{https://doi.org/10.1016/0550-3213(88)90634-7}{{\it Nucl. Phys.}
  {\bfseries B304} (1988) 403}.

\bibitem{FischlerSusskind86a}
W.~Fischler and L.~Susskind, {\it {Dilaton Tadpoles, String Condensates and
  Scale Invariance}}, \href{https://doi.org/10.1016/0370-2693(86)91425-5}{{\it
  Phys.Lett.} {\bfseries B171} (1986) 383}.

\bibitem{FischlerSusskind86b}
W.~Fischler and L.~Susskind, {\it {Dilaton Tadpoles, String Condensates and
  Scale Invariance. 2.}},
  \href{https://doi.org/10.1016/0370-2693(86)90514-9}{{\it Phys.Lett.}
  {\bfseries B173} (1986) 262}.

\bibitem{PiusRudraSen14}
R.~Pius, A.~Rudra and A.~Sen, {\it {String Perturbation Theory Around
  Dynamically Shifted Vacuum}},
  \href{https://doi.org/10.1007/JHEP10(2014)070}{{\it JHEP} {\bfseries 10}
  (2014) 70} [\href{https://arxiv.org/abs/1404.6254}{{\ttfamily 1404.6254}}].

\bibitem{Weinberg85b}
S.~Weinberg, {\it {Radiative Corrections in String Theories}}, .

\bibitem{Sen20}
A.~Sen, {\it {D-instanton Perturbation Theory}},
  \href{https://arxiv.org/abs/2002.04043}{{\ttfamily 2002.04043}}.

\bibitem{Zwiebach93}
B.~Zwiebach, {\it {Closed string field theory: Quantum action and the B-V
  master equation}}, \href{https://doi.org/10.1016/0550-3213(93)90388-6}{{\it
  Nucl. Phys.} {\bfseries B390} (1993) 33}
  [\href{https://arxiv.org/abs/hep-th/9206084}{{\ttfamily hep-th/9206084}}].

\bibitem{Fay}
D.~Fay, John, {\it {Theta Functions on Riemann Surfaces}}. 1973.

\bibitem{BelavinKnizhnik86}
A.~Belavin and V.~Knizhnik, {\it {Algebraic Geometry and the Geometry of
  Quantum Strings}}, \href{https://doi.org/10.1016/0370-2693(86)90963-9}{{\it
  Phys.Lett.} {\bfseries B168} (1986) 201}.

\bibitem{Sen19}
A.~Sen, {\it {String Field Theory as World-sheet UV Regulator}},
  \href{https://arxiv.org/abs/1902.00263}{{\ttfamily 1902.00263}}.

\bibitem{FarbMargalit12}
B.~Farb and D.~Margalit, {\it A Primer on Mapping Class Groups}, Princeton
  Mathematical Series. Princeton University Press, 2012.

\bibitem{Vafa87}
C.~Vafa, {\it {Conformal Theories and Punctured Surfaces}},
  \href{https://doi.org/10.1016/0370-2693(87)91358-X}{{\it Phys. Lett.}
  {\bfseries B199} (1987) 195}.

\bibitem{SklirosCopelandSaffin17}
D.~P. Skliros, E.~J. Copeland and P.~M. Saffin, {\it {Highly Excited Strings I:
  Generating Function}},
  \href{https://doi.org/10.1016/j.nuclphysb.2016.12.022}{{\it Nucl. Phys.}
  {\bfseries B916} (2017) 143}
  [\href{https://arxiv.org/abs/1611.06498}{{\ttfamily 1611.06498}}].

\bibitem{D'HokerPhong89}
E.~D'Hoker and D.~H. Phong, {\it {Conformal Scalar Fields and Chiral Splitting
  on Super-Riemann Surfaces}}, \href{https://doi.org/10.1007/BF01218413}{{\it
  Commun. Math. Phys.} {\bfseries 125} (1989) 469}.

\bibitem{DijkgraafVerlindeVerlinde88}
R.~Dijkgraaf, E.~P. Verlinde and H.~L. Verlinde, {\it {$c = 1$ Conformal Field
  Theories on Riemann Surfaces}},
  \href{https://doi.org/10.1007/BF01224132}{{\it Commun. Math. Phys.}
  {\bfseries 115} (1988) 649}.

\bibitem{VerlindeVerlinde87}
E.~P. Verlinde and H.~L. Verlinde, {\it {Chiral Bosonization, Determinants and
  the String Partition Function}},
  \href{https://doi.org/10.1016/0550-3213(87)90219-7}{{\it Nucl. Phys.}
  {\bfseries B288} (1987) 357}.

\bibitem{Cutkosky60}
R.~E. Cutkosky, {\it {Singularities and Discontinuities of Feynman
  Amplitudes}}, \href{https://doi.org/10.1063/1.1703676}{{\it J. Math. Phys.}
  {\bfseries 1} (1960) 429}.

\bibitem{PiusSen16}
R.~Pius and A.~Sen, {\it {Cutkosky Rules for Superstring Field Theory}},
  \href{https://doi.org/10.1007/JHEP10(2016)024}{{\it JHEP} {\bfseries 10}
  (2016) 024} [\href{https://arxiv.org/abs/1604.01783}{{\ttfamily
  1604.01783}}].

\bibitem{Pompeiu}
D.~Pompeiu, {\it {Sur une classe de fonctions d'une variable complexe}},
  \href{https://doi.org/10.1007/BF03015292}{{\it Rend. Circ. Matem. Palermo}
  {\bfseries 33} (1912) 108}.

\bibitem{Polchinski86}
J.~Polchinski, {\it {Evaluation of the One Loop String Path Integral}},
  \href{https://doi.org/10.1007/BF01210791}{{\it Commun. Math. Phys.}
  {\bfseries 104} (1986) 37}.

\bibitem{Zimmermann73a}
W.~Zimmermann, {\it {Normal Products and the Short Distance Expansion in the
  Perturbation Theory of Renormalizable Interactions}},
  \href{https://doi.org/10.1016/0003-4916(73)90430-2}{{\it Annals Phys.}
  {\bfseries 77} (1973) 570}.

\bibitem{Zimmermann73b}
W.~Zimmermann, {\it {Composite Operators in the Perturbation Theory of
  Renormalizable Interactions}},
  \href{https://doi.org/10.1016/0003-4916(73)90429-6}{{\it Annals Phys.}
  {\bfseries 77} (1973) 536}.

\bibitem{EllisMavromatosSkliros15}
J.~Ellis, N.~E. Mavromatos and D.~P. Skliros, {\it {Complete Normal Ordering 1:
  Foundations}}, \href{https://doi.org/10.1016/j.nuclphysb.2016.05.024}{{\it
  Nucl. Phys.} {\bfseries B909} (2016) 840}
  [\href{https://arxiv.org/abs/1512.02604}{{\ttfamily 1512.02604}}].

\bibitem{SklirosPhysStackEx}
Wakabaloola, ``Polchinski string theory p 105 eq (3.6.18) weyl transformation
  of the massless vertex operator; https://physics.stackexchange.com/q/531862
  (version: 2020-02-21).'' Physics Stack Exchange.

\bibitem{Friedan86}
D.~Friedan, {\it {Notes on String Theory and Two-Dimensional Conformal Field
  Theory}}, {{\it Proceedings of Workshop on Unified String Theories, Santa
  Barbara, CA, Jul 29 - Aug 16, 1985} }.

\bibitem{FriedanShenkerMartinec85}
D.~Friedan, S.~H. Shenker and E.~J. Martinec, {\it {Covariant Quantization of
  Superstrings}}, \href{https://doi.org/10.1016/0370-2693(85)91466-2}{{\it
  Phys. Lett.} {\bfseries B160} (1985) 55}.

\bibitem{HindmarshSkliros10}
M.~Hindmarsh and D.~Skliros, {\it {Covariant Closed String Coherent States}},
  \href{https://doi.org/10.1103/PhysRevLett.106.081602}{{\it Phys. Rev. Lett.}
  {\bfseries 106} (2011) 1602}
  [\href{https://arxiv.org/abs/1006.2559}{{\ttfamily 1006.2559}}].

\bibitem{Skliros11DPhil}
D.~Skliros, {\it {Vertex Operators for Cosmic Strings}}, .

\bibitem{SklirosHindmarsh11}
D.~Skliros and M.~Hindmarsh, {\it {String Vertex Operators and Cosmic
  Strings}}, \href{https://doi.org/10.1103/PhysRevD.84.126001}{{\it Phys.Rev.}
  {\bfseries D84} (2011) 126001}
  [\href{https://arxiv.org/abs/1107.0730}{{\ttfamily 1107.0730}}].

\bibitem{SklirosCopelandSaffin16bb}
D.~P. Skliros, E.~J. Copeland and P.~M. Saffin, {\it {Highly Excited Strings
  II: Vertex Operators}}, {{\it (to appear)} }.

\bibitem{DiVecchiaNakayamaPetersenSciuto87}
P.~Di~Vecchia, R.~Nakayama, J.~L. Petersen and S.~Sciuto, {\it {Properties of
  the Three Reggeon Vertex in String Theories}},
  \href{https://doi.org/10.1016/0550-3213(87)90678-X}{{\it Nucl. Phys.}
  {\bfseries B282} (1987) 103}.

\bibitem{PolchinskiCai88}
J.~Polchinski and Y.~Cai, {\it {Consistency of Open Superstring Theories}},
  \href{https://doi.org/10.1016/0550-3213(88)90382-3}{{\it Nucl. Phys.}
  {\bfseries B296} (1988) 91}.

\bibitem{KlauderSkagerstam85}
J.~R. Klauder and B.-S. Skagerstam, {\it {Coherent States -- Applications in
  Physics and Mathematical Physics}}. World Scientific Publishing, Singapore,
  1985.

\bibitem{Sonoda88a}
H.~Sonoda, {\it {Sewing Conformal Field Theories}},
  \href{https://doi.org/10.1016/0550-3213(88)90066-1}{{\it Nucl. Phys.}
  {\bfseries B311} (1988) 401}.

\bibitem{Sonoda88b}
H.~Sonoda, {\it {Sewing Conformal Field Theories. 2.}},
  \href{https://doi.org/10.1016/0550-3213(88)90067-3}{{\it Nucl. Phys.}
  {\bfseries B311} (1988) 417}.

\bibitem{Yin17}
X.~Yin, {\it {Aspects of Two-Dimensional Conformal Field Theories}},
  \href{https://doi.org/10.22323/1.305.0003}{{\it PoS} {\bfseries TASI2017}
  (2017) 003}.

\bibitem{Alvarez-GaumeGomezMooreVafa88}
L.~Alvarez-Gaume, C.~Gomez, G.~W. Moore and C.~Vafa, {\it {Strings in the
  Operator Formalism}},
  \href{https://doi.org/10.1016/0550-3213(88)90391-4}{{\it Nucl. Phys.}
  {\bfseries B303} (1988) 455}.

\bibitem{BelavinPolyakovZamolodchikov84}
A.~A. Belavin, A.~M. Polyakov and A.~B. Zamolodchikov, {\it {Infinite conformal
  symmetry in two-dimensional quantum field theory}},
  \href{https://doi.org/10.1016/0550-3213(84)90052-X}{{\it Nucl. Phys.}
  {\bfseries B241} (1984) 333}.

\bibitem{JoeBigBook}
J.~Polchinski, {\it Joe's Big Book of String}. Unpublished Notes, (year
  unknown, perhaps 1995).

\bibitem{VerlindeVerlinde87b}
E.~P. Verlinde and H.~L. Verlinde, {\it {Multiloop Calculations in Covariant
  Superstring Theory}},
  \href{https://doi.org/10.1016/0370-2693(87)91148-8}{{\it Phys. Lett.}
  {\bfseries B192} (1987) 95}.

\bibitem{DHokerPhong85}
E.~D'Hoker and D.~H. Phong, {\it {Length Twist Parameters in String Path
  Integrals}}, \href{https://doi.org/10.1103/PhysRevLett.56.912}{{\it Phys.
  Rev. Lett.} {\bfseries 56} (1985) 912}.

\bibitem{Sen05}
A.~Sen, {\it {Tachyon Dynamics in Open String Theory}},
  \href{https://doi.org/10.1142/S0217751X0502519X}{{\it Int. J. Mod. Phys.}
  {\bfseries A20} (2005) 5513}
  [\href{https://arxiv.org/abs/hep-th/0410103}{{\ttfamily hep-th/0410103}}].

\bibitem{CallanLovelaceNappiYost88}
C.~G. Callan, Jr., C.~Lovelace, C.~R. Nappi and S.~A. Yost, {\it {Loop
  Corrections to Superstring Equations of Motion}},
  \href{https://doi.org/10.1016/0550-3213(88)90565-2}{{\it Nucl. Phys.}
  {\bfseries B308} (1988) 221}.

\bibitem{DelGiudiceDiVecchiaFubini72}
E.~Del~Giudice, P.~Di~Vecchia and S.~Fubini, {\it {General properties of the
  dual resonance model}},
  \href{https://doi.org/10.1016/0003-4916(72)90272-2}{{\it Ann. Phys.}
  {\bfseries 70} (1972) 378}.

\bibitem{AdemolloDelGuidiceDiVecchiaFubini74}
M.~Ademollo, E.~Del~Giudice, P.~Di~Vecchia and S.~Fubini, {\it {Couplings of
  three excited particles in the dual-resonance model}},
  \href{https://doi.org/10.1007/BF02801846}{{\it Nuovo Cim.} {\bfseries A19}
  (1974) 181}.

\bibitem{DHokerPhong95}
E.~D'Hoker and D.~Phong, {\it {The Box Graph in Superstring Theory}},
  \href{https://doi.org/10.1016/0550-3213(94)00526-K}{{\it Nucl.Phys.}
  {\bfseries B440} (1995) 24}.

\bibitem{Witten13b}
E.~Witten, {\it {The Feynman $i \epsilon$ in String Theory}},
  \href{https://doi.org/10.1007/JHEP04(2015)055}{{\it JHEP} {\bfseries 04}
  (2015) 055} [\href{https://arxiv.org/abs/1307.5124}{{\ttfamily 1307.5124}}].

\bibitem{AnSMatrix}
R.~J. Eden, P.~V. Landshoff, D.~I. Olive and J.~C. Polkinghorne, {\it The
  Analytic S-Matrix}. Cambridge, UK: Univ. Pr., UK, 1966.

\bibitem{LandauLifshitzRQF}
E.~M. Berestetskii, V.~B.~Lifshitz and L.~P. Pitaevskii, {\it Quantum
  Electrodynamics}, vol.~4 of {\it Landau and Lifshitz Course of Theoretical
  Physics}. Butterworth-Heinmann, United Kingdom, 1982.

\bibitem{Witten12a}
E.~Witten, {\it {Notes On Supermanifolds and Integration}},
  \href{https://arxiv.org/abs/1209.2199}{{\ttfamily 1209.2199}}.

\bibitem{Polchinski94}
J.~Polchinski, {\it {Combinatorics of Boundaries in String Theory}}, {{\it
  Phys. Rev.} {\bfseries D50} (1994) 6041}.

\bibitem{Polchinski95}
J.~Polchinski, {\it {Dirichlet-Branes and Ramond-Ramond Charges}}, {{\it Phys.
  Rev. Lett.} {\bfseries 75} (1995) 4724}.

\bibitem{SonodaZwiebach90}
H.~Sonoda and B.~Zwiebach, {\it {Covariant Closed String Field Theory Cannot be
  Cubic}}, \href{https://doi.org/10.1016/0550-3213(90)90108-P}{{\it Nucl.
  Phys.} {\bfseries B336} (1990) 185}.

\bibitem{Polchinski80}
J.~G. Polchinski, {\it {Vortex Operators in Gauge Field Theories}}, Ph.D.
  thesis, LBL, Berkeley, 1980.

\bibitem{Alvarez84}
O.~Alvarez, {\it {Topological Quantization and Cohomology}},
  \href{https://doi.org/10.1007/BF01212452}{{\it Commun. Math. Phys.}
  {\bfseries 100} (1985) 279}.

\bibitem{VerlindeH87}
H.~L. Verlinde, {\it {A Note on the Integral over the Fermionic Supermoduli}},
  {{\it (unpublished)} }.

\bibitem{DiFrancescoMatheuSenechal97}
P.~Di~Francesco, P.~Mathieu and D.~Senechal, {\it {Conformal Field Theory}}.

\bibitem{MisnerThorneWheeler74}
C.~W. Misner, K.~Thorne and J.~Wheeler, {\it {Gravitation}}. 1974.

\end{thebibliography}\endgroup




\end{document}